\documentclass[a4paper,11pt]{scrbook}
\pdfoutput=1
\usepackage[normalem]{ulem}
\usepackage[utf8]{inputenc}
\usepackage{cite}
\usepackage[left=2cm,right=2cm,top=2cm,bottom=3cm]{geometry}
\usepackage{xcolor}
\usepackage{float}

\usepackage{epsf}
\usepackage{graphicx}

\usepackage{amsmath}
\usepackage{amssymb}
\usepackage{mathrsfs}
\usepackage{slashed}
\usepackage{mathtools}
\usepackage{bbold}

\usepackage{mathrsfs}

\usepackage{mathbbol}

\usepackage{amsfonts}
\usepackage{bbding}

\usepackage{amsfonts}
\usepackage{xfrac} 
\usepackage{booktabs} 
\usepackage{soul} 

\usepackage{tikz}
\usetikzlibrary{shapes,arrows,positioning,automata,backgrounds,calc,er,patterns}
\usepackage[compat=1.1.0]{tikz-feynman}

\usepackage{color}
\definecolor{cred}{RGB}{180,50,40}
\definecolor{purple}{RGB}{180,90,180}
\definecolor{darkgreen}{RGB}{0, 100, 0}

\usepackage{lscape}

\usepackage{minitoc}

\usepackage[
colorlinks=true
,urlcolor=black
,anchorcolor=black
,citecolor=black
,filecolor=black
,linkcolor=black
,menucolor=black
,linktocpage=true
,pdfproducer=medialab
,pdfa=true
]{hyperref}
\usepackage{braket}
\allowdisplaybreaks[1]

\title{Indirect searches for new physics via flavour observables}
\author{Jonathan Kriewald}
\begin{document}

\begin{titlepage}
\pagestyle{empty}
\vspace*{-35mm}
\begin{center}

\renewcommand{\familydefault}{cmss}

\vspace*{18mm}
\mbox{\Large \textbf{UNIVERSIT\'E CLERMONT AUVERGNE}}

\vspace*{5mm}
{\Large ECOLE DOCTORALE DES SCIENCES FONDAMENTALES}

\vspace*{12mm}
\mbox{\hspace*{-0mm}\textbf{\huge TH\`ESE}}

\vspace*{12mm}
{\large pr\'esent\'ee pour obtenir le grade de} 

\vspace*{5mm}
{\Large DOCTEUR D'UNIVERSIT\'E en physique}

\vspace*{5mm}
{\large Sp\'ecialit\'e : Physique des Particules} 

\vspace*{8mm}
{\Large Par \textbf{Jonathan Kriewald}}

\vspace*{5mm}
{\large Laboratoire de Physique de Clermont, CNRS/IN2P3}

\vspace*{15mm}
\textbf{\huge Indirect searches for New Physics via flavour observables}

\vspace*{15mm}
{\large soutenu publiquement le 13 d\'ecembre 2021, devant la commission d'examen :}

\vspace*{8mm}
{\large 
\renewcommand{\arraystretch}{1.2}
\begin{tabular}{lll}
M. & S. MONTEIL & \hspace*{8mm}Président et Examinateur \\
Mme. & S. FAJFER & \hspace*{8mm}Rapporteuse\\
M. \hspace*{1mm}& E. FERN\'ANDEZ-MART\'INEZ &  \hspace*{8mm}Rapporteur\\
Mme. & A. ABADA & \hspace*{8mm}Examinatrice \\
Mme. & G. HILLER & \hspace*{8mm}Examinatrice\\
M. & V. MOR\'ENAS &  \hspace*{8mm}Examinateur \\
Mme. & I. RIPP-BAUDOT & \hspace*{8mm}Examinatrice \vspace*{2mm}\\
Mme. & A. M. TEIXEIRA & \hspace*{8mm}Directrice de thèse\\
M. & J. ORLOFF & \hspace*{8mm}Directeur de thèse
\end{tabular}
\renewcommand{\arraystretch}{1.}

}

\vspace*{15mm}
\mbox{\hspace*{-3mm}

\raisebox{-6.5mm}{\includegraphics[width=32mm]{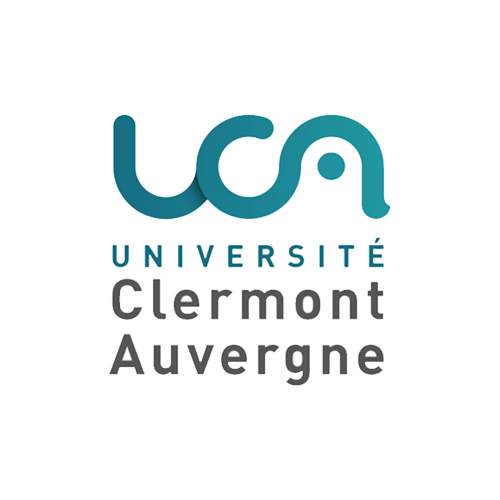}} \hspace*{12mm}
\includegraphics[width=35mm]{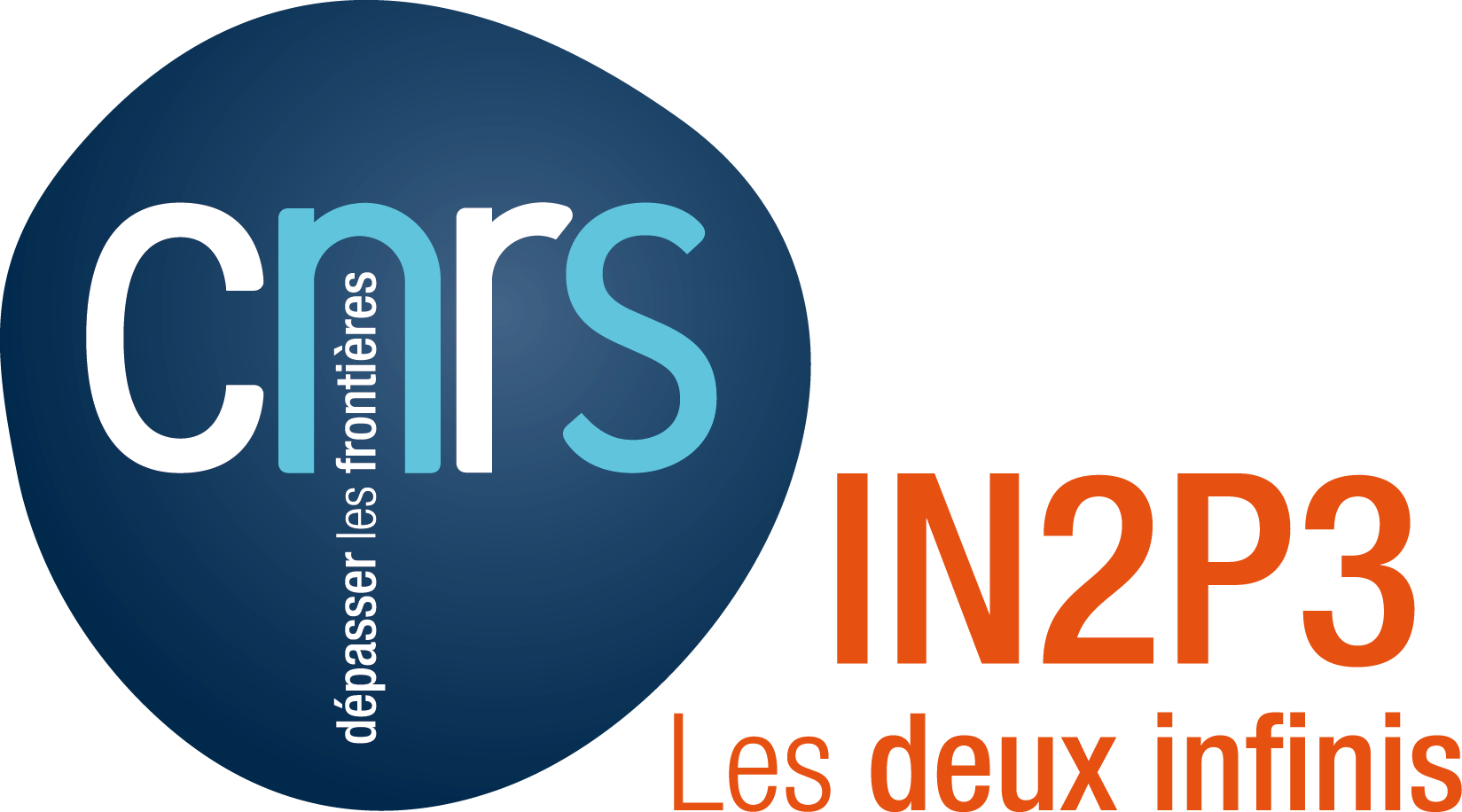}\hspace{12mm}
\includegraphics[width=50mm]{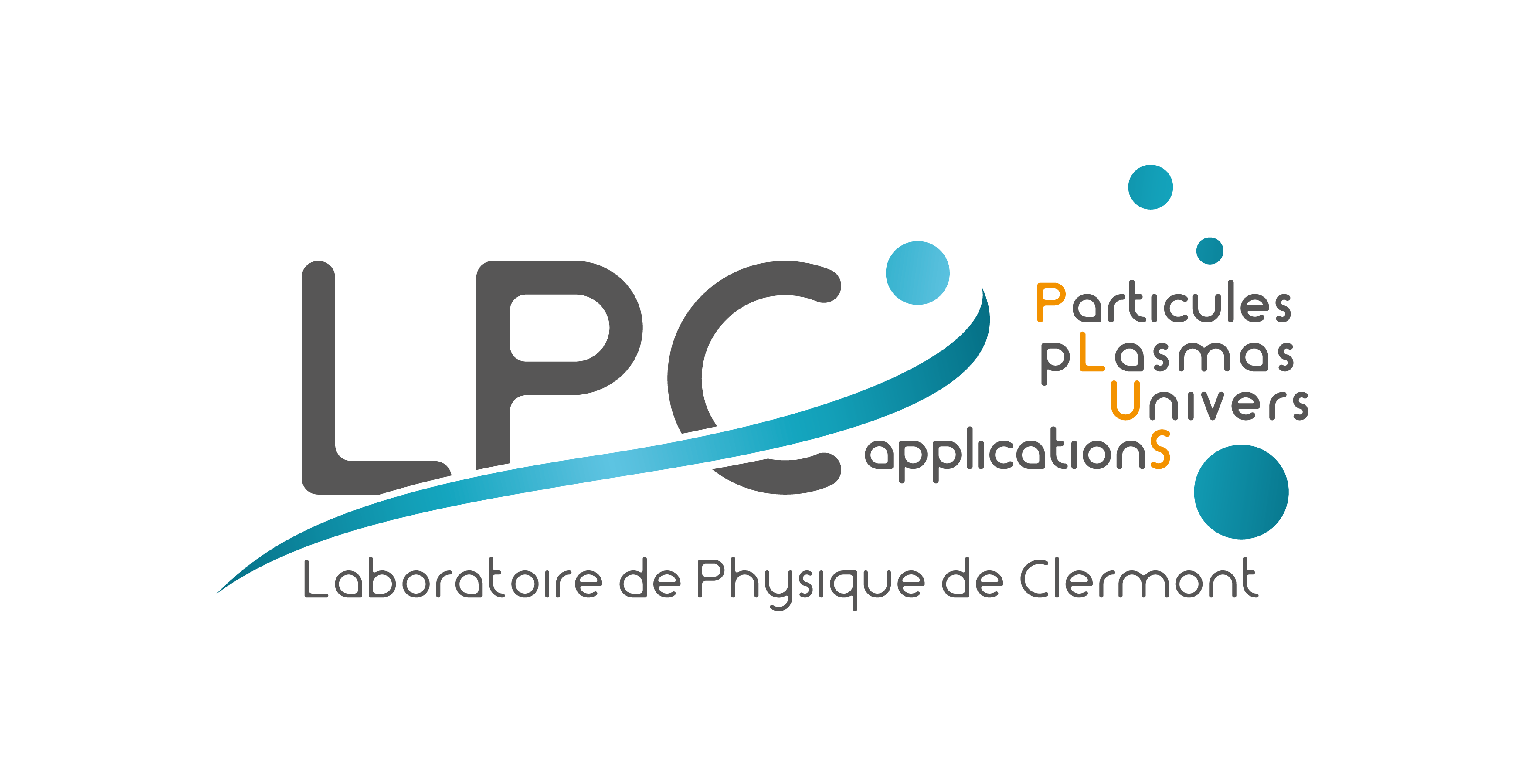}

}
\renewcommand{\familydefault}{}
\end{center}

\end{titlepage}
\frontmatter
\addchap{Acknowledgements}
\adjustmtc
First and foremost I want to express my wholehearted gratitude to my advisers Jean and Ana, for your incredible professional and personal support. 
You taught me so much about physics and life (for example skiing), and spent countless nights and weekends with me at the lab working together.
None of this would have been possible without you.
Thank you for everything!

I would  also like to thank the CNRS/IN2P3, the Université Clermont Auvergne and Dominique Pallin as representative of the lab for the financial support for the various projects of this doctorate.
Moreover I am grateful to the TH team for welcoming me so kindly into the group.

Furthermore I am greatly indebted to my family and friends for their unlimited support.
Thank you Chandan, for teaching me so much about the flavours of particle physics and the flavours of Indian cuisine, spontaneously travelling to Corsica together and countless discussions about physics and life.
Thank you Kevin \& Janina for the many vacations in Geneva, preventing me from working all the weekends.
Thank you Innes for your friendship and many therapeutic discussions from the other side of the world.
A huge thank you to all my friends in Clermont, Sofia, Halime, Emanuelle, Andreas, Théo, Mike, Guillaume, Ioan, Lennart and Nazlim for so much time spent together at the lab and in the mountains, eating inconceivable amounts of food and surviving through the pandemic together.

Many thanks also to Svjetlana Fajfer and Enrique Fernández-Martínez for refereeing this thesis as well as the other jury members Stéphane Monteil,  Asmaa Abada, Gudrun Hiller, Vincent Morénas and Isabelle Ripp-Baudot for taking the time to read my work and taking part in my defence.

Finally, I wholeheartedly want to thank my parents and my brother for their unconditional love and support.

\chapter*{Abstract}
Precision measures of electroweak and flavour observables, at both low and high energies, are highly complementary to direct searches for New Physics at high-energy colliders. Despite the discovery of the Higgs boson at the Large Hadron Collider, and of the overwhelming successes of the Standard Model, several observational and theoretical problems remain to be addressed. In addition to neutrino oscillation phenomena, the Standard Model fails to explain the baryon asymmetry of the Universe, and does not offer a viable dark matter candidate. In recent years, numerous deviations between the Standard Model prediction and experimental measurements have been identified; interestingly, most are closely connected to lepton flavours.

In this thesis we have explored several aspects of flavour physics, focusing on the phenomenological implications of models of massive neutrinos, and of several Standard Model extensions capable of accommodating current tensions on anomalous magnetic moments of charged leptons and several $B$-meson decay observables.

Following a brief introduction to the Standard Model and an overview of phenomenological aspects of flavour in the lepton sector, numerous phenomenological aspects of massive neutrinos (with particular attention to the interplay of leptonic CP violation and lepton flavour violation) are explored. The very appealing inverse seesaw mechanism (endowed with flavour and CP symmetries) is analysed, focusing on phenomenological consequences.
The tensions in the anomalous magnetic moments of the electron and the muon are also investigated, especially in view of further peculiar excesses in nuclear transitions.

Finally, $B$-meson decay anomalies, which have been the object of increasing attention, are considered, relying on model-independent analysis (using effective field theories) which pave the way to the phenomenological study of simplified leptoquark models.

\chapter*{R\'esum\'e}
Les mesures de précision d’observables électrofaibles et de saveurs, à la fois à basse et à haute énergie, sont hautement complémentaires aux recherches directes de nouvelle physique au près des collisionneurs à haute énergie. Malgré la découverte du boson de Higgs au LHC, et les succès remarquables du Modèle Standard, il reste néanmoins quelques problèmes observationnels et théoriques à résoudre. 
Outre que le phénomène d'oscillation des neutrinos, le Modèle Standard ne parvient pas à expliquer l'asymétrie baryonique de l'Univers et n'offre pas de candidat viable au problème de la matière noire. Ces dernières années, de nombreuses déviations entre les prédictions du Modèle Standard et les mesures expérimentales ont été identifiées ; il est intéressant de remarquer que la plupart d’entre elles sont étroitement liées aux saveurs leptoniques.

Dans cette thèse, nous avons exploré plusieurs aspects de la physique des saveurs (aussi bien leptonique qu’hadronique), en nous concentrant sur les implications phénoménologiques de modèles de neutrinos massifs, ainsi que sur certaines extensions du Modèle Standard capables d'accommoder les tensions actuelles en ce qui concerne les moments magnétiques anormaux des leptons chargés et aussi de plusieurs observables liées à la désintégration des mésons $B$.

Après une brève introduction au Modèle Standard et un aperçu des aspects phénoménologiques de la saveur dans le secteur leptonique, de nombreux aspects phénoménologiques des neutrinos massifs (avec une attention particulière à l'interaction entre la violation de CP leptonique et la violation de la saveur leptonique) sont explorés. Une réalisation du mécanisme du ``Inverse Seesaw", qui demeure une des possibilités les plus envisagées actuellement pour expliquer les masses et mélanges des neutrinos, est analysée, en particulier en ce qui concerne les conséquences phénoménologiques d’une telle réalisation dotée de symétries de saveur et de CP. Les tensions dans les moments magnétiques anormaux de l'électron et du muon sont également étudiées, notamment en vue d'autres excès observés dans des transitions nucléaires. 

Enfin, les anomalies dans les désintégrations des mésons $B$, qui font l'objet d'une attention croissante ces toutes dernières années, sont examinées, en s'appuyant sur une analyse indépendante du modèle (à l'aide des théories effectives), et qui ouvre la voie à l'étude phénoménologique de modèles simplifiés de leptoquarks.

\dominitoc
\tableofcontents
\mainmatter

\addchap{Introduction}
\minitoc

\noindent
With the 2012 discovery of a scalar boson at LHC, in good agreement with the properties of a Higgs boson, the electroweak sector of the Standard Model was finally completed~\cite{ATLAS:2015yey}.
Albeit being a massive breakthrough, this discovery was well anticipated, as it was well known from the LEP and Tevatron results of electroweak precision measurements that, should the Standard Model be an accurate description of Nature, LHC should discover a scalar boson with a mass of $\sim 100\:\mathrm{GeV}$.
Furthermore, in the past, precision measurements of electroweak decays, for instance the muon decay, led to strong lower bounds on the electroweak gauge boson masses long before they were directly discovered.
History tells a similar story in the flavour sector of the Standard Model.
After the discovery of the strange quark, the ``three quark model'' with a $SU(3)$ flavour symmetry first led to the prediction of many new bound states in the form of baryons and mesons, which were subsequently discovered.
This ``model'' had however the striking problem that it would predict flavour changing neutral currents (FCNC) at tree-level, which were however not confirmed by experimental data.
This led to the hypothesis of the charm quark, in order to suppress FCNC transitions via a generalised Glashow-Iliopoulos-Maiani (GIM) mechanism~\cite{Glashow:1970gm}.
Moreover, the discovery of CP violation in Kaon decays led to the hypothesis of a third quark generation; CP violation is only possible if there are at least three families~\cite{Kobayashi:1973fv}.
Subsequent precision measurements of neutral $K^0 - \bar K^0$ meson mixing allowed establishing stringent lower bounds on the top-quark mass, suggesting that it was far heavier than the other quarks.
The vast combined effort of experimental direct and indirect searches, as well as phenomenological studies to interpret the data during the last sixty years has carried particle physics into an unprecedented era: precision measurements in almost all sectors of high-energy physics corroborate the predictions of the Standard Model with great accuracy.

Despite what became a clear theoretical and phenomenological success story, it also rapidly became clear that the Standard Model cannot be the end of the line,
the most obvious reason being that it does not account for a quantum theory of gravity, and thus cannot describe all known fundamental interactions.
Furthermore, the description of the Higgs sector is far from satisfactory, without a fundamental principle behind it.
Understanding the exact mechanism of electroweak symmetry breaking is also related to the flavour problem; why are the fermion masses so hierarchical? Why are there three generations of fermions?
Theoretical and aesthetical issues aside, the Standard Model lacks a viable Dark Matter candidate and cannot account for the baryon asymmetry of the Universe.
Furthermore, and most strikingly, with the discovery of neutrino oscillations and their successful description via the ``Pontecorvo-Maki-Nakagawa-Sakata'' mechanism~\cite{Pontecorvo:1957cp,Pontecorvo:1957qd,Maki:1962mu}, necessarily implying that neutrinos are massive, the first irrefutable evidence of New  Physics in a laboratory has been found.

To tackle the aforementioned problems, many models and frameworks have been proposed in the past, often suggesting New Physics states present at the TeV-scale.
Until the present day, direct signals of such new states have so far eluded discovery at LHC.
However, measurements of flavour observables and electroweak precision tests indirectly impose stringent constraints on the parameter space and mass scale of New Physics models.
In particular, precision measurements of hadron flavour observables during the last twenty years have allowed to stringently constrain the Cabibbo-Kobayashi-Maskawa quark mixing matrix, overwhelmingly pointing towards the unitarity of the same.
Consequently, any additional fermion content that mixes with the Standard Model quarks cannot have large mixings, and a fourth quark generation has been effectively ruled out.
Moreover, precision measurements of rare decays (such as $B_s\to \mu\mu$ strikingly consistent with the Standard Model prediction) have allowed to almost rule out many models aiming at addressing the puzzle of electroweak symmetry breaking; in other words, ``flavour is the usual graveyard of  beyond the Standard Model electroweak theories''.

Contrary to the quark sector, the lepton (flavour) sector is far from being mastered.
While the entries of the quark mixing matrix have been determined with great precision, the experimental effort of measuring the lepton mixing parameters has just started to reach its ``precision era''.
By itself, the neutrino sector is at the source of many open questions; neither the absolute scale nor the mechanism behind the origin of neutrino masses are presently known.
Moreover, while the quark mixing pattern is very hierarchical (as is the spectrum of quark masses), the PMNS is not, further worsening the overall ``flavour problem''.
Since neutrino oscillation phenomena necessarily imply that neutrinos are massive, one expects the accidental lepton flavour symmetries of the Standard Model to be violated in Nature.
These consist of individual lepton flavour conservation and lepton flavour universality.
Thus, experimental tests of these symmetries seem to be particularly appealing to achieve a better understanding of the lepton sector, to constrain New Physics contributions and possibly to discover indirect hints of New Physics effects that manifest at low energies in lepton phenomena.

While it is clear that neutral lepton flavour is violated in Nature, searches for charged lepton flavour violating processes have so far only returned negative results.
In turn, this allows to place tight constraints on models aiming at providing a viable mechanism of neutrino mass generation.
In the same manner, measurements of precision observables sensitive to the violation of lepton flavour universality  at high ($W\to\ell\nu$ and $Z\to \ell\ell$ decays) and low energies ((semi-) leptonic $K$ and $\pi$ decays) seem to be consistent with the Standard Model predictions, leading to stringent bounds on the unitarity of the lepton mixing matrix and therefore on the presence of (hypothetical) additional neutral fermions that possibly mix with the Standard Model neutral leptons.
It is however important to notice that lepton flavour universality violation and charged lepton flavour violation can also occur in New Physics models without any (direct) connection to massive neutrinos.

Although the overwhelming majority of flavour observables so far measured appears to be consistent with the Standard Model paradigm of flavour, in recent years several observables related to lepton flavour started exhibiting significant deviations from their respective Standard Model predictions.
Among them are the anomalous magnetic moments of the electron and the muon.
In particular, measurements of the anomalous magnetic moment of the muon persistently remain in tension with the Standard Model prediction\footnote{Recent lattice QCD evaluations of the leading-order hadronic vacuum polarisation lead to a far milder tensions between Standard Model prediction and the measurements. This is further discussed in the subsequent chapters.}.
Combining the measurements at Brookhaven and Fermilab, the tension currently amounts to $+4.2\,\sigma$ (standard deviations).
More recently, due to the availability of competitive independent measurements of the electromagnetic fine-structure constant $\alpha_e$ (using Caesium atoms\footnote{A recent measurement of $\alpha_e$ using Rubidium atoms exhibits a $>5\,\sigma$ tensions with the Caesium result and leads to a milder tension in the electron anomalous magnetic moment. This is further discussed in subsequent chapters.}), the anomalous magnetic moment of the electron shows a tension with the Standard Model prediction as well, leading to a deviation of $-2.5\,\sigma$.
Interestingly, the sign (and the size) of the respective deviations could point towards the presence of lepton flavour universality violating New Physics interactions.

Lepton flavour ratios of decay widths of semi-leptonic charged and neutral current $B$-meson decays are directly sensitive to lepton flavour universality violation.
During the last decade, measurements of the ratios $R_{D^{(\ast)}}\equiv B\to D^{(\ast)}\tau\nu / B\to D^{(\ast)}\ell\nu$ and $R_{K^{(\ast)}}\equiv B\to K^{(\ast)}\mu\mu/B\to K^{(\ast)}ee$ exhibit persistent tensions with their respective Standard Model predictions, most recently reaching $3.1\,\sigma$ for the measurement of $R_K$.
Moreover, measurements of the differential branching fractions in $B\to K^\ast\mu\mu$ and $B_s\to \phi\mu\mu$, as well as measurements of the angular coefficients in the decay $B\to K^\ast (\to K\pi)\mu\mu$ show (local) deviations reaching $>3\,\sigma$.
If interpreted in terms of a presence of New Physics, the so-called $B$-anomalies, particularly in neutral current $b\to s\ell\ell$ transitions, draw a consistent picture: there seems to be a ``force that takes muons away''.
While the discovery and measurements of neutrino oscillations are the first irrefutable evidence of New Physics, the flavour anomalies in the anomalous magnetic moments and $B$-meson decays are certainly interesting \textit{indirect} hints on New Physics.

In order to find Standard Models extensions that accommodate the anomalous data, one needs to understand the ``big picture'' of low-energy flavour physics.
In addition to the apparent deviations from the Standard Model predictions, negative results, so-called null results, provide crucial input for understanding the low-energy effects of potential Standard Model extensions.
While a specific New Physics field is only irrefutably discovered by direct searches, indirect searches offer the invaluable guiding principle where and what to look for.
Since there are numerous hints that there could be New Physics at the TeV-scale, or more importantly above the electroweak scale, one can use \textit{effective field theories} to parametrise generic New Physics effects in low-energy observables.
This allows to find requirements on an ultra-violet (UV) New Physics model, channelling the power of low-energy data from indirect searches.
Indirect searches are more than complementary to high-energy data; they can offer valuable guiding principles to identify suitable New Physics candidates and clarify potential manifestation, thus providing the crucial input that is needed to eventually uncover a more complete description of Nature.

\medskip
The present thesis is organised as follows:
After a brief introduction to the Standard Model, in which we discuss its theoretical and observational shortcomings, we review phenomenological aspects of flavour in the lepton sector, with particular emphasis on lepton flavour observables that are well suited to test the Standard Model and its symmetries, and possibly search for New Physics.
Chapters~\ref{sec:massivenu} and~\ref{chap:CPV} are devoted to a review of phenomenological aspects of massive neutrinos, with particular attention to the interplay of leptonic CP violation and lepton flavour violation induced by the presence of heavy neutral leptons.
In Chapter~\ref{chap:ISS}, we study the inverse seesaw mechanism endowed with a flavour symmetry $G_f$, a CP symmetry, and its phenomenological consequences.
The remaining Chapters are then devoted to the indirect hints on New Physics in flavour observables; in Chapter~\ref{sec:g-2paper}, motivated by a peculiar excess in nuclear transitions, we study a low-scale $U(1)_{B-L}$ model as a solution for the $g-2$ anomalie(s), while Chapters~\ref{chap:bphysics} and~\ref{chap:lq} are devoted to the $B$-meson decay anomalies: starting from a model-independent analysis using effective field theories, we study simplified leptoquark models with particular emphasis on its future observability.
Some final remarks and an outlook are given in Chapter~\ref{chap:concs}.

\bigskip
This thesis relies on the following original scientific contributions\footnote{All results presented here (in particular plots) which are not explicitly referred to a publication have been originally prepared for this thesis.}:
\begin{itemize}
    \item C.~Hati, \textbf{J.~Kriewald}, J.~Orloff and A.~M.~Teixeira,
    ``A nonunitary interpretation for a single vector leptoquark combined explanation to the $B$-decay anomalies,''
    JHEP \textbf{12} (2019), 006
    \href{https://arxiv.org/abs/1907.05511}{[arXiv:1907.05511 [hep-ph]]}.
    
    \item C.~Hati, \textbf{J.~Kriewald}, J.~Orloff and A.~M.~Teixeira,
    ``Anomalies in $^8$Be nuclear transitions and $(g-2)_{e,\mu}$: towards a minimal combined explanation,''
    JHEP \textbf{07} (2020), 235
    \href{https://arxiv.org/abs/2005.00028}{[arXiv:2005.00028 [hep-ph]].}
    
    \item C.~Hati, \textbf{J.~Kriewald}, J.~Orloff and A.~M.~Teixeira,
    ``The fate of vector leptoquarks: the impact of future flavour data,''
    Eur. Phys. J. C \textbf{81} (2021) no.12, 1066
    \href{https://arxiv.org/abs/2012.05883}{[]arXiv:2012.05883 [hep-ph]]}.
    
    \item \textbf{J.~Kriewald}, C.~Hati, J.~Orloff and A.~M.~Teixeira,
    PoS \textbf{ICHEP2020} (2021), 258
    \href{https://arxiv.org/abs/2012.06315}{[arXiv:2012.06315 [hep-ph]]}.
    
    \item \textbf{J.~Kriewald}, C.~Hati, J.~Orloff and A.~M.~Teixeira,
    ``Leptoquarks facing flavour tests and $b\to s\ell\ell$ after Moriond 2021,''
    \href{https://arxiv.org/abs/2104.00015}{arXiv:2104.00015 [hep-ph]},
    Contribution to \textbf{Moriond EW 2021}.
    
    \item A.~Abada, \textbf{J.~Kriewald} and A.~M.~Teixeira,
    ``On the role of leptonic CPV phases in cLFV observables,''
    Eur. Phys. J. C \textbf{81} (2021) no.11, 1016
    \href{https://arxiv.org/abs/2107.06313}{[arXiv:2107.06313 [hep-ph]]}.
    
    \item C.~Hagedorn, \textbf{J.~Kriewald}, J.~Orloff and A.~M.~Teixeira,
    ``Flavour and CP symmetries in the inverse seesaw,''
    \href{https://arxiv.org/abs/2107.07537}{arXiv:2107.07537 [hep-ph]},
    accepted by EPJC.
    
    \item \textbf{J.~Kriewald}, A.~Abada, A.~M.~Teixeira, 
    ``The role of leptonic CPV phases in cLFV observables,''  \href{https://arxiv.org/abs/2110.15177}{arXiv:2110.15177 [hep-ph]}, Contribution to TAUP and NuFact 2021.
\end{itemize}

\adjustmtc

\chapter{The Standard Model}
\label{chap:SM}
\minitoc

The Standard Model of Particle Physics~\cite{Weinberg:1967tq,Glashow:1961tr,Salam:1968rm} provides an extraordinarily successful and yet simple description of Nature at its smallest scales; it offers a common framework to describe elementary particles and their electroweak and strong interactions.
Despite its exceptional success, it is now firmly established that the Standard Model (SM) cannot account for a certain number of observations, and one must thus envisage theoretical constructions - including new degrees of freedom (new particles and/or new interactions), capable of of accounting for experimental data.
Moreover, a strong theoretical interest also fuels the study of ``New Physics beyond the SM (BSM)'', as the the latter might provide a solution, or at least ameliorate, some of the theoretical puzzles of the SM.

In this chapter, we briefly describe the building blocks of the Standard Model, and enumerate its observational and theoretical caveats.
At the end, we also give a brief overview 
of how low-energy imprints of New Physics can be studied in the absence of concrete models (i.e. well identified BSM constructions) by means of effective field theories.

\section{Flavour in the Standard Model}

The Standard Model is a renormalisable quantum field theory, invariant under both the Poincaré group and the semi-simple (local) gauge group $SU(3)_c\times SU(2)_L\times U(1)_Y$.
In addition to the associated gauge bosons, the SM comprises three families of quarks and leptons, as well as a single fundamental scalar field.
Their representations under the non-abelian groups $SU(3)_c$ and $SU(2)_L$, as well as their charge under the abelian gauge group $U(1)_Y$, are listed in Table~\ref{tab:SM}. The convention of the $U(1)_Y$ (hyper)charge is such that $Q_f^\text{em} = Y_f^{U(1)}+T_{3\,f}^{SU(2)}$.
\renewcommand{\arraystretch}{1.3}
\begin{table}[ht!]
\begin{center}
  \begin{tabular}{|c|c|c|c|}
  \hline
      Field & $SU(3)_c$ & $SU(2)_L$ & $U(1)_Y$\\
  \hline
  \hline
  $Q = \left(u_L, \, d_L\right)^T$ & $ \mathbf{3} $ & $ \mathbf{2} $ &
  $ \frac{1}{6} $ \\ 
  $ \ell = \left(\nu_L, \, e_L\right)^T $ & $ \mathbf{1} $ & $
  \mathbf{2} $ & $ -\frac{1}{2} $ \\ 
  $ u_R $ & $ \mathbf{3} $ & $ \mathbf{1} $ & $ \frac{2}{3} $ \\ 
  $ d_R $ & $ \mathbf{3} $ & $ \mathbf{1} $ & $ -\frac{1}{3} $\\ 
   $ e_R $ & $ \mathbf{1} $ & $ \mathbf{1} $ & $ -1 $ \\
  \hline
  $ H = (H^+, H^0)^T$ & $ \mathbf{1} $ & $ \mathbf{2} $ & $ \frac{1}{2} $ \\
  \hline
    $G$ & $\mathbf{8}$ & $\mathbf{1}$ & 0\\
    $W$ & $\mathbf{1}$ & $\mathbf{3}$ & 0\\
    $B$ & $\mathbf{1}$ & $\mathbf{1}$ & 0\\ 
  \hline
  \end{tabular}
  \end{center}
\caption{Field content of the Standard Model and
  the corresponding representations under the gauge group $SU(3)_c\times SU(2)_L$, as well as their charge under the abelian gauge group $U(1)_Y$.} 
  \label{tab:SM}
\end{table}
\renewcommand{\arraystretch}{1.}
Once the gauge group and matter content have been defined, the renormalisable SM Lagrangian is fully determined,
\begin{eqnarray}
    \mathcal L_\text{SM} &=&  -\frac{1}{4} B_{\mu\nu}B^{\mu\nu} - \frac{1}{4}W_{\mu\nu}^a W_a^{\mu\nu} - \frac{1}{4} G_{\mu\nu}^a G_a^{\mu\nu}\nonumber\\
    &\phantom{=}& + i \bar Q_L^i \slashed D Q_L^i + i \bar u_R^i \slashed D u_R^i + i \bar L_L^i \slashed D L_L^i + i \bar e_R^i \slashed D e_R^i\nonumber\\
    &\phantom{=}& + Y_{ij}^u \bar Q_L^i \tilde H u_R^j + Y_{ij}^d \bar Q_L^i H d_R^j + Y_{ij}^\ell \bar L_L^i H e_R^j + \text{H.c.}\nonumber\\
    &\phantom{=}& + \left|D_\mu H\right|^2 + \mu^2 \left|H\right|^2 - \lambda \left|H\right|^4\,.
\end{eqnarray}
In the above, $i,j = 1,2,3$ are family indices, $\slashed D = D_\mu \gamma^\mu$, and $\tilde H = i \sigma_2 H$; $Y^f$ denotes the Yukawa couplings, $\lambda$ the quartic Higgs self-coupling and $\mu$ the Higgs mass term. With the exception $\mu$, all of the previous couplings are dimensionless, so that theoretically the SM can be extrapolated to a wide range of energies (or scales). 
Furthermore, the interactions between gauge fields and the fermions are encoded in the gauge covariant derivative, given by
\begin{equation}
   D_\mu   = \partial_{\mu} + i g_s G_\mu^a T^{SU(3)}_a + i g_w W_\mu^a T^{SU(2)}_a + i g' Y B_\mu\,,
\end{equation}
in which the couplings $g_s, g_w, g'$ denote the different gauge couplings of $SU(3)_c$, $SU(2)_L$ and $U(1)_Y$, and $T^{(\mathcal G)}_a$ are the generators of the (non-abelian) gauge group $\mathcal G$ in the representation of the fermion (or boson) the derivative is acting on.
The gauge kinetic terms of the gauge fields $F$ are written in terms of their field strength tensors, which are defined as
\begin{equation}
    F_{\mu\nu}^a \equiv \partial_\mu F_\nu^a - \partial_\nu F_\mu^a + i g_{(\mathcal{G})} f^{abc}_{(\mathcal G)}  F_\mu^b F_\nu^c\,,
\end{equation}
in which $g_{(\mathcal{G})}$ is the associated gauge coupling and $f^{abc}_{(\mathcal G)}$ are the structure constants of the corresponding lie-algebra spanned by the gauge group $\mathcal G$ (for the abelian group all $f_{U(1)}^{abc}=0$).

Provided that the parameters of the Higgs sector (which are necesarily input by hand) fulfil certain conditions ($\mu^2,\lambda > 0$), the Higgs field develops a vacuum expectation value (vev) $\langle H\rangle = (0,\frac{v}{\sqrt{2}})^T$ with $v= \sqrt{\mu^2/\lambda}\simeq 246~\mathrm{GeV}$, and (spontaneously) breaks the SM gauge group to $SU(3)_c\times U(1)_\text{em}$. This phenomenon is the so-called Brout-Englert-Higgs (BEH) mechanism~\cite{Englert:1964et,Higgs:1964ia,Higgs:1964pj}.
After electroweak symmetry breaking (EWSB), three massive gauge bosons emerge, the $Z^0$ and $W^\pm$, while the gluons and a linear combination of $B$ and $W$, the photon $A$ (or $\gamma$), remain massless.
In the broken phase, the physical (electroweak) gauge fields can be written in terms of the original gauge fields $W$ and $B$ as
\begin{align}
    W_\mu^\pm &= \frac{1}{\sqrt{2}}(W_\mu^1 \mp i W_\mu^2)\quad&\text{with mass}\quad &M_W = \frac{g_w v}{2}\,,\nonumber\\
    Z_\mu^0 &= \frac{1}{\sqrt{g_w^2 + g^{\prime 2}}}(g W_\mu^3 - g' B_\mu)\quad&\text{with mass}\quad &M_Z = \frac{v}{2}\sqrt{g_w^2 + g^{\prime 2}}\,,\nonumber\\
    A_\mu &= \frac{1}{\sqrt{g_w^2 + g^{\prime 2}}} (g' W_\mu^3 + g B_\mu)\quad&\text{with mass}\quad &M_A = 0\,.
\end{align}
Furthermore, the fermions acquire masses 
\begin{equation}
    m^f = Y^f \langle H\rangle
\end{equation}
via their couplings to the Higgs.

In the SM fermion dynamics is governed by the interactions with the gauge bosons and by the Yukawa couplings.
The Yukawa couplings (and thus the mass matrices $m^f$) are in general not diagonal.
In order to obtain the physical (massive) fermion fields, the fermion Yukawa couplings have to be diagonalised.
After EWSB, the quark mass terms ($q=u,d$) in the Lagrangian can be recast as
\begin{equation}
    \mathcal L_\text{mass}^q \sim \bar q_{L}^i M_{ij}^q q_R^j = \bar q_{L}^i V_{L}^{q\dagger}\, V_L^q\, M_{ij}^q\, V_{R}^{q\dagger}\, V_R^q\, q_R^j = \hat{\bar q}_L^i m_i^q \hat{q}_R^i\,,
\end{equation}
in which the physical (mass) basis is denoted by $\hat{\phantom{q}}$.
The unitary matrices $V_{L,R}^q$ diagonalise the mass matrices and relate the interaction basis with the mass basis via
\begin{equation}
    m_\text{diag}^q = V_L^q\, M_{ij}^q\, V_R^{q\dagger}\,,\quad\text{and}\quad\hat{q}_{L,R} = V_{L,R}^q q_{L,R}\,,
\end{equation}
and equivalently for the charged leptons.

In the quark sector\footnote{For the lepton sector, since neutrinos are massless due to the absence of right-handed neutrinos and/or Higgs triplets, one can without loss of generality choose to work in a basis in which the charged lepton Yukawa couplings are diagonal, leading to strict lepton flavour conserving charged current interactions.}, inserting the above transformations into the interaction Lagrangian leads to flavour violation in charged currents (cc), parametrised by the Cabibbo-Kobayashi-Maskawa (CKM) quark mixing matrix ($V_\text{CKM}$)
\begin{equation}
    \mathcal L_\text{cc}^q \sim - \frac{g_w}{\sqrt{2}} V_\text{CKM}^{ij} W_\mu^+ \bar u_{L i} \gamma^\mu d_{L j}\,,\quad V_\text{CKM} = V_L^u V_L^{d\dagger}\,.
\end{equation}
The CKM matrix is a complex and (special) unitary $3\times3$ matrix, and thus fully parametrised by 4 real (physical) parameters\footnote{A special unitary $3\times3$ matrix has in general eight free parameters. However, four of the complex phases can be re-absorbed into field re-definitions of the quark fields (which are Dirac particles) and are thus unphysical, leaving four physical parameters.}.
These parameters are usually cast, in the so-called standard parametrisation, in terms of three real mixing angles and one phase.
Due to the misalignment between the mass bases of up- and down-type quarks, the CKM matrix is in general non-trivial and thus leads to hadronic flavour violation, which has been experimentally observed in a number of meson and baryon decays.
Moreover, the Kobayashi-Maskawa mechanism~\cite{Kobayashi:1973fv}, via its single (physical) phase, naturally provides a source of CP violation.
Flavour changing neutral current (FCNC) interactions remain absent at tree-level and are, at higher order, naturally suppressed via a generalised Glashow-Iliopoulos-Maiani (GIM) mechanism~\cite{Glashow:1970gm}.

All in all, the SM has 18 free parameters: the three gauge couplings ($g_s, g_w, g'$), the two parameters of the Higgs potential, 10 parameters in the quark sector (quark masses, three CKM angles and one CP violating phase), and the three charged lepton masses.

Although not a priori imposed, the SM Lagrangian exhibits a number of ``accidental'' symmetries, which allow to understand and explain certain properties and phenomena. The exact accidental symmetries of the SM correspond to: 
\begin{itemize}
    \item Baryon number conservation (global $U(1)_B$): each quark carries $B_q = 1/3$, while leptons do not ($B_\ell = 0$). This symmetry is only broken at the quantum level in so-called sphaleron processes, which however conserve $B-L$, with $L$ being the total lepton number.
    \item Individual lepton number conservation (global $U(1)_{L_e}\times U(1)_{L_\mu}\times U(1)_{L_\tau}$): among other consequences, these symmetries forbid lepton flavour violating processes such as charged lepton flavour violating decays (e.g. $\mu\to e\gamma$) as well as neutrino oscillations. Furthermore, the conservation of individual lepton number renders all couplings of SM gauge bosons to leptons flavour universal.
    This naturally implies global lepton number conservation, a global $U(1)_L$ symmetry, which is also only broken at the quantum level in the aforementioned sphaleron processes.
\end{itemize}
Interestingly, these accidental symmetries might also hint on paths towards (preferred) extensions of the SM, which will be discussed in several of the subsequent chapters.

The SM Lagrangian also possesses several approximate accidental symmetries, corresponding to global exact symmetries only broken by small couplings.
In the limit of vanishing Yukawa couplings and $g' = 0$, the SM has an additional global $SU(2)$ symmetry (under which the Higgs transforms as a doublet).
The Higgs vev breaks this to the so-called ``custodial'' $SU(2)_C$, under which the massive $W^\pm$ and $Z^0$ bosons form a triplet with degenerate mass $M_C = M_W = M_Z$.
The small coupling $g'$ then breaks the custodial symmetry, and in the limit of vanishing Yukawa couplings one finds
\begin{equation}
    \frac{M_W^2}{M_Z^2\cos^2\theta_w} \equiv \rho = 1\,,
\end{equation}
with the weak mixing angle $\tan\theta_w = g'/g_w$.
Corrections due to non-vanishing Yukawa couplings (dominated by the top Yukawa $y_t = \frac{m_t \sqrt{2}}{v}\simeq 1$) only arise at loop-level; this accidental symmetry thus gives rise to a non-trivial prediction of the SM, $\rho \simeq 1$, which has been very well tested experimentally.

In the limit of vanishing Yukawa couplings $Y^f = 0$ the SM has five additional global $U(3)$ symmetries, associated to the three families of $Q_L, u_R, d_R, L_L, e_R$.
The latter symmetries (and their subgroups) allow to understand many properties in (quark) flavour physics, and pave the way to study New Physics models of flavour (e.g with flavour symmetries, which we will discuss later on).
We note here that all these accidental symmetries are a direct consequence of only including renormalisable (dimension $\leq 4$) operators in the SM Lagrangian.
Extensions of the SM Lagrangian via non-renormalisable operators will be also discussed later on.

\medskip
The SM constitutes one of the most successful theories in modern physics: based on an elegant theoretical framework, it describes \textit{almost all} experimental observations in particle physics with high precision.
Following the discovery of the $Z$ and $W$ gauge bosons, the new boson discovered at the LHC increasingly complies with the SM requirements of a ``Higgs boson'' (in particular concerning its spin/parity).
The measurements of the ATLAS and CMS collaborations, mutually in good agreement, give the average value of~\cite{ATLAS:2015yey}
\begin{equation}
    m_H = (125.09 \pm 0.24)\:\mathrm{GeV}\,.
    \label{eqn:higgsdirect}
\end{equation}
The electroweak sector of the SM has been tested via an impressive amount of measurements.
Electroweak precision observables (EWPO) have been systematically used to probe the predictions of the SM, and to constrain its unknown parameters (related to the Higgs sector).
Although EWPOs comprise a huge set of  possible measurements, these have been reduced by the LEP and Tevatron working groups, and include for instance the mass and width of the $W$ boson, and various $Z$-pole observables such as the weak mixing angle $\sin^2\theta_w$, decay widths to SM fermions, amongst many more.
In addition to determining the electroweak properties of the SM, these measurements allow to perform extensive consistency checks of the SM\footnote{For a comprehensive overview of the current experimental status see e.g.~\cite{ParticleDataGroup:2020ssz}.}.
For instance, a global fit of all EWPOs to experimental data leads to a Higgs mass of~\cite{ParticleDataGroup:2020ssz} 
\begin{equation}
    m_H^\text{EWPO} = 90^{+18}_{-16}\:\mathrm{GeV}\,,
\end{equation}
in agreement with the direct mass measurement (cf. Eq.~\eqref{eqn:higgsdirect}) at $1.8\,\sigma$.

\begin{figure}
    \centering
    \includegraphics[width=0.6\textwidth]{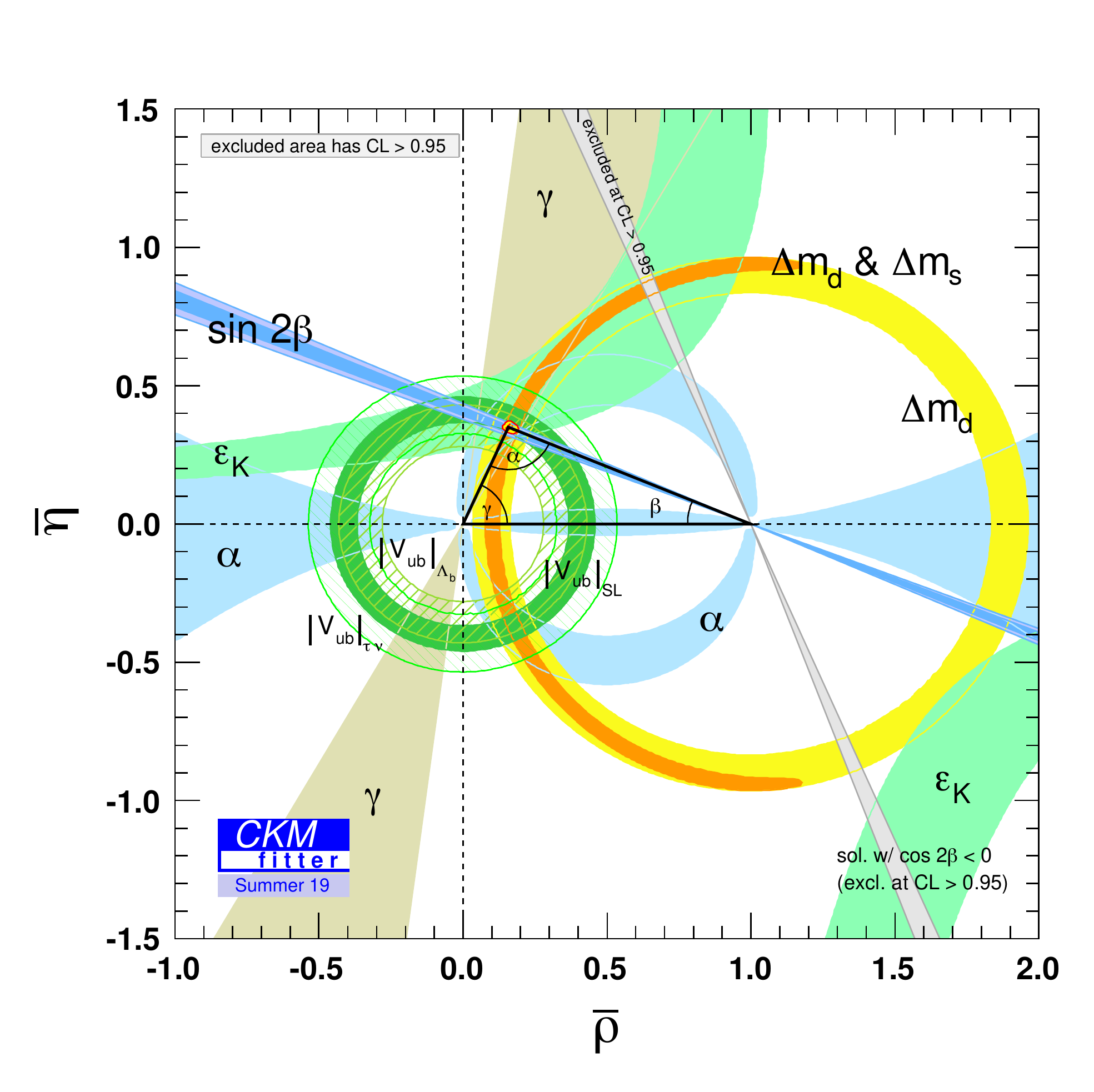}
    \caption{Constraints on the unitarity of the CKM matrix inferred from a large number of experimental measurements of (quark) flavour violating hadron processes. Figure taken from~\cite{Charles:2004jd}.}
    \label{fig:UTCKM}
\end{figure}

This ``success story'' continues in the realm of (quark) flavour physics.
The experimentally measured quark mixing and CP violation observables are in general well accounted for by the CKM paradigm of flavour, rooted in the unitary CKM matrix and the GIM mechanism; the CKM matrix exhibits a strongly hierarchical structure, and is furthermore highly consistent with unitarity.
The fundamental CKM parameters, including possible deviations from unitarity, have been constrained by a large series of observables to impressive precision~\cite{Charles:2004jd}, as can be seen in Fig.~\ref{fig:UTCKM}.

Finally, the study of (hadronic) flavour transitions and decays has gone through significant developments in recent years.
While on the experimental side more and more experimental data has been accumulated in a number of facilities, on the theory side important progress has been achieved, with higher-order effects increasingly under control.
Strong interactions between quarks, mediated by the gluons, are described by Quantum Chromodynamics (QCD), whose non-perturbative nature at low-energies (large distances) can be successfully treated via lattice QCD (LQCD).

As of today, the ensemble of past and current experimental facilities has allowed to draw a very clear and consistent picture of the SM status: with some remarkable exceptions, Nature at the elementary level can be described by Quantum Chromodynamics and electroweak interactions, which can be modelled via a Higgs mechanism for EWSB.
Additionally, one has the CKM paradigm precisely describing quark flavour transitions and CP violation, with a generalised GIM mechanism suppressing unwanted FCNCs.
In other words, the Standard Model provides an outstanding description of Nature and works far better than what could have been initially expected.

\section{The need for New Physics}
Despite its simplicity and remarkable phenomenological success, the SM is plagued with several theoretical and observational issues.
While the theoretical caveats can be somehow accepted at the expense of aesthetics - perhaps Nature is ``just so'' - the observational and experimental problems remain and urgently call for New Physics.

\paragraph{Theoretical caveats of the SM}
The theoretical formulation of the SM has several important issues that should be addressed and either solved or at least improved upon.
Some of them deal with emergent patterns, others are related to a disparity of the existing energy scales.
In the following we highlight some of the most curious and pressing issues.
\begin{itemize}
    \item \textbf{Quantum theory of gravity.} There are four known fundamental interactions of Nature that elementary particles are subject to.
    While the SM provides a quantum description of the strong, weak and electromagnetic interactions, Einstein's theory of General Relativity offers a framework to explain gravitational phenomena at large distances. In the energy range accessible to human experiments, gravity is indeed much weaker than the other fundamental interactions and therefore completely negligible in what regards studying phenomena of elementary particles.
    At very high energies, above the Planck scale $\Lambda_\text{Pl} \sim 10^{19}\:\mathrm{GeV}$ (or equivalently at distances below the Planck length $r_\text{Pl}\sim 10^{-34}\mathrm{m}$), gravitational interactions become relevant.
    Although the SM as such was never meant to constitute a theory of everything, a quantum theory of gravity is nevertheless required.
    
    \item \textbf{Strong CP problem.} The QCD Lagrangian allows for an additional parameter of the SM, $\theta_\text{QCD}$, encompassing another source of CP violation via a non-vanishing phase. CP violation arises from the term $\int dx^4 \theta_\text{QCD} \epsilon^{\mu\nu\rho\sigma} G_{\mu\nu}^a G_{\rho\sigma}^a$ (often omitted in the Lagrangian). This CP-odd term has physical effects if quarks are massive, leading to non-vanishing electric dipole moments (EDM) of hadrons. Upper bounds on the neutron EDM however imply that the strong CP phase must be vanishingly small, $\theta_\text{QCD}\lesssim 10^{-9}$. Since in the limit $\theta_\text{QCD}\to 0$ CP is not restored, this is a technically unnatural regime for a dimensionless parameter in the sense of 't Hooft~\cite{tHooft:1979rat}.
    
    \item \textbf{Hierarchy problem.} The interactions described by the the SM are rooted in energy scales below the TeV, with the QCD confinement scale $\Lambda_\text{QCD}$ at a few hundred MeV, and the electroweak scale $\Lambda_\text{EW}$ at the order of a hundred GeV.
    However, in its symmetric phase, the SM could in principle remain valid up to the Planck scale, at which it must be necessarily modified to include the effects of quantum gravity.
    This suggests a ``desert'' between the known fundamental scales, spanning over thirteen orders of magnitude.
    Without a fundamental principle (a symmetry), the light SM scales, particularly $\Lambda_\text{EW}$, cannot be generically protected from quantum corrections from the larger scale, and thus cannot be stabilised.
    While quantum corrections to the SM fermions and (massive) gauge bosons are regulated by their underlying symmetries (gauge symmetry and chiral symmetry in the case of fermions), the bare mass term in the Higgs potential exhibits a disturbing sensitivity to New Physics: be it in the form of additional fields present in almost any possible SM extension, the Planck scale directly, or even to scales lying beyond.
    
    At leading order, the Higgs mass is simply given by the bare mass term in the Higgs potential, $M_H^2 = \mu^2$.
    Considering only the SM field content, the dominant one-loop corrections are given by
    \begin{equation}
        \Delta M_H^2|_\text{1-loop}^\text{SM} = \frac{3\Lambda_\text{UV}^2}{8 \pi^2 v^2}[M_H^2 + 2 M_W^2 + M_Z^2 - 4 m_t^2 + ...]\,,
    \end{equation}
    in which $\Lambda_\text{UV}^2$ is an ultraviolet (UV) momentum cutoff, introduced to regulate the loop integral.
    The UV cutoff scale can be interpreted as the (minimum) energy scale at which New Physics must be be manifest to alter the high-energy behaviour of the SM.
    Since no new (heavy) states have been discovered thus far, the only known scale of New Physics is the Planck scale, and one can only hypothesise that the SM remains valid up to the Planck scale.
    If indeed $\Lambda_\text{UV}\sim \Lambda_\text{Pl}$, the quantum corrections to the Higgs mass are 30 orders of magnitude larger then the required value.
    One thus needs a very fine cancellation between the bare Higgs mass and the radiative corrections, to have a physical Higgs mass in the range determined by experiments.
    Moreover, such a cancellation would have to occur at all orders in perturbation theory, which would then imply that the entire mass spectrum of the SM would be sensitive to the cutoff scale.
    
    One straightforward way to overcome this problem is to postulate $\Lambda_\text{UV}$ to be far below the Planck scale, in turn necessitating the presence of new degrees of freedom not too far from the electroweak scale.
    The additional states then in turn also contribute to the Higgs mass corrections, a problem that persists in dimensional regularisation as well\footnote{However, as it was recently shown for scalar field theories~\cite{Mooij:2021ojy}, if one resorts to a finite renormalisation scheme based on the Callan-Symanzik equations~\cite{Callan:1970yg,Symanzik:1970rt}, neither a cutoff scale nor a departure from four spacetime dimensions (dimensional regularisation) is needed to compute (and renormalise) the radiative corrections, and the Higgs mass is manifestly finite.
    In other words, the ``hierarchy problem'' being first of all an aesthetical problem of fine-tuning, might not be a problem after all.}; the Higgs mass will always be sensitive to the masses of the heaviest particles that it  couples to, and moreover, the sensitivity of the SM to these new scales does not disappear.
    This leads to the hypothesis that there should exist new heavy states at the TeV-scale, to ameliorate the amount of fine-tuning required to stabilise the Higgs mass.
    
    \item \textbf{Flavour puzzle.} In the SM, all fermions appear to be organised in three families without any particular underlying reason. The dynamics of flavour transitions is governed by a set of complex matrices in flavour (or family) space, the Yukawa couplings $Y^f$, that allow for a parametrisation of the fermion masses.
    However, there is no explanation behind the very hierarchical structure of fermion masses (spanning many orders of magnitude), nor to the patterns of fermion mixing. Furthermore, there is no theoretical necessity for three generations of fermions, since, for instance, the gauge anomalies of the SM gauge group are cancelled generation-wise.
\end{itemize}
Moreover, in what regards EWSB, the parameters of the Higgs potential are basically just put in by hand, unrelated to a fundamental principle or symmetry. Despite the discovery of a scalar boson at LHC that strongly resembles a SM-like Higgs, a deeper understanding of the mechanism behind EWSB is desirable.

Finally, one might ask the question of whether the gauge couplings, and therefore the gauge interactions, might unify at a higher energy scale.
With the SM field content and the experimentally determined boundary conditions, this is not possible.

\paragraph{Observational problems}
Although the issues discussed in the above paragraph challenge the theoretical foundation of the SM, none invalidates its phenomenological success nor its viability as a simple description of Nature.
Even if the SM does not provide an explanation to many phenomena, it nevertheless offers a self-consistent parametrisation of these, and one could accept that Nature is ``just so''.

Independently of the theoretical caveats, the SM insufficiency is revealed by its inability to account for three major observations:
\begin{itemize}
    \item \textbf{Oscillating neutrinos.} Upon comparison of the expected and experimentally measured solar and atmospherical neutrino fluxes, the observed discrepancy provided the first unambiguous evidence for New Physics beyond the SM.
    The first discrepancies arose already in 1968, when first measurements of the solar neutrino flux conducted by the Homestake experiment~\cite{Cleveland:1998nv} revealed a significant deficit between the observed and theoretically expected flux of solar electron neutrinos.
    This deficit, $N_\nu^\text{exp} \simeq \frac{1}{3}N_\nu^\text{theo}$, was subsequently confirmed by a large number of independent experiments~\cite{Cleveland:1998nv,Kaether:2010ag,SAGE:2009eeu,Super-Kamiokande:2010tar,SNO:2011hxd,Bellini:2011rx,BOREXINO:2014pcl}, and it became clear that it could not be explained by modifying the Solar Standard Model~\cite{Bahcall:1987jc}.

    The hypothesis of massive neutrino oscillations, first suggested by Pontecorvo in 1958~\cite{Pontecorvo:1957cp,Pontecorvo:1957qd}, and subsequently modelled by Maki, Nakagawa and Sakata in 1962~\cite{Maki:1962mu}, became eventually irrefutable, marking the first failure of the SM in describing a low-energy particle physics phenomenon; neutrino oscillation necessarily imply that neutrinos are massive (and that there are non-trivial mixings in the lepton sector).
    
    By construction, the SM does not contain right-handed neutrinos (nor Higgs triplets), so that neutrino masses cannot be generated by the Higgs mechanism. Although the SM can be minimally extended by three right-handed neutrinos in order to generate neutrino masses by coupling to the Higgs, the SM symmetries would not prevent a potentially large Majorana mass term for the right-handed fields, leading to the violation of lepton number, and rendering neutrinos Majorana fermions\footnote{The potential implications of having massive Majorana neutrinos are further discussed in the following chapters.}.
    
    Irrespective of their nature, neutrino oscillations, and therefore lepton mixing, can be described with a leptonic  analogue of the CKM matrix~\cite{Maki:1962mu}, the so-called ``Pontecorvo-Maki-Nakagawa-Sakata'' (PMNS) matrix.
    While the CKM matrix, as experimentally determined, is very hierarchical, the PMNS is not, which thus deepens the aforementioned flavour problem.
    The neutrino masses by themselves, although not (yet) individually determined, lie below the eV-scale, and thus further increase the flavour problem in what concerns the vastly different fermion mass scales.
    
    \item \textbf{Baryon asymmetry of the Universe.} Astrophysical and cosmological observations strongly suggest that the Universe is dominated by matter. This matter dominance can be cast in the so-called baryon asymmetry of the Universe (BAU), defined as
    \begin{equation}
        \eta \equiv \frac{n_B}{n_\gamma} = \frac{n_b - n_{\bar b}}{n_\gamma}\,,
    \end{equation}
    in which $n_b$, $n_{\bar b}$ and $n_\gamma$ denote the baryon, anti-baryon and photon number densities.
    This number can be determined independently by measurements of the abundance of light elements as predicted by Big-Bang-Nucleosynthesis (BBN), and by measurements of power spectrum of the cosmic microwave background (CMB).
    Interestingly, both measurements are in remarkable agreement with each other, leading to a value of the order $\eta \sim 6\times 10^{-10}$.
    
    One could interpret this slight overabundance of matter as a primordial asymmetry, 
    but the inflation of the universe would exponentially dilute it, 
    and it would corresponds to extremely fine-tuned initial conditions of adding one quark to $10^9$ quark-antiquark pairs. One possible circumvention of this problem relies in the hypothesis that the BAU is generated dynamically, the so-called baryogenesis mechanism.
    In 1967~\cite{Sakharov:1967dj} Sakharov noticed that in order to obtain a baryon asymmetry from a primordial baryon symmetric state, three requirements must be satisfied, the so-called ``Sakharov conditions''.
    These are the presence of baryon number violation, C and CP violation, and a departure from thermal (and kinetic) equilibrium.
    Although the SM in principle fulfils these conditions~\cite{Kuzmin:1985mm}, the amount of CP violation is not sufficient to create a sufficiently large BAU, and the Higgs is too heavy to generate a strong enough electroweak phase transition, leading to a weak departure from thermal equilibrium.
    
    Thus, in order to generate the observed BAU, a New Physics model with a stronger source of CP violation and a stronger departure from thermal equilibrium is necessary.
    Interestingly, the existence of oscillating massive neutrinos opens the door to new sources of CP violation, and therefore to the possibility to explain the observed BAU via the dynamical generation of a lepton asymmetry, which is subsequently converted into a baryon asymmetry via electroweak sphalerons - the so-called mechanism of leptogenesis.
    
    \item \textbf{Dark matter.} Astronomical and cosmological observations accumulated unequivocal evidence suggesting that the majority of the matter coupled to gravity in the Universe, observed in galaxies and large scale structures, is non-luminous, or ``dark''.
    The observations supporting the evidence include studies of the motion of galaxy clusters, galactical rotation curves, and (weak and strong) gravitational lensing, amongst others.
    Studies of the CMB anisotropies in conjunction with baryon acoustic oscillation data have led to a (cold) dark matter density of~\cite{ParticleDataGroup:2020ssz}
    \begin{equation}
        \Omega_\text{CDM} h^2 = 0.1198 \pm 0.0026\,,
    \end{equation}
    in which $h$ is the Hubble constant.
    Despite the amount of evidence of the existence of gravitational interactions of dark matter, its precise nature is unknown. One of the most simple solutions consists in hypothesising that it might be a fundamental particle, which the SM fails to provide\footnote{Baryonic matter (in the form of macroscopic objects, e.g. brown dwarfs, black holes or MACHOS), and SM-like light massive neutrinos can only account for a fraction of the observed dark matter density.}.
\end{itemize}

\begin{figure}
    \centering
    \includegraphics[width=0.5\textwidth]{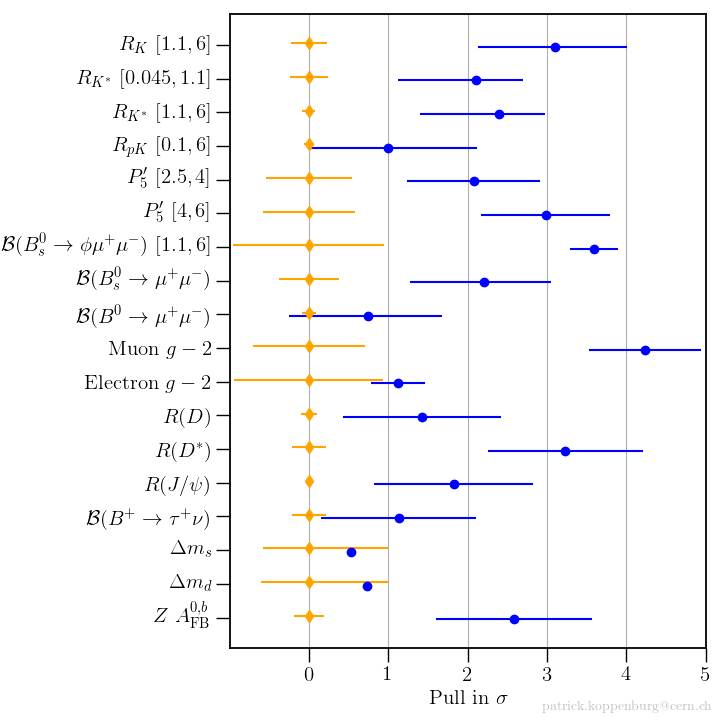}
    \caption{Compilation of several flavour observables currently exhibiting tensions between experimental data and their associated theoretical predictions. From~\cite{koppenburgwebsite}.}
    \label{fig:pkoppenburg}
\end{figure}

In addition, although not as firmly established, in recent years more and more data on low-energy flavour processes has been accumulated leading to numerous discrepancies with the SM predictions.
Many of these tensions between theory and experiment lie around the $3-4\,\sigma$ level, and almost all of them are related to (final state) lepton flavours.
These include, among others, precision observables such as the anomalous magnetic moments of the muon and the electron, lepton flavour universality tests in charged and neutral current semi-leptonic $B$-meson decays, and differential branching fraction and angular distributions of rare $b\to s\ell\ell$ decays.
Even though hadronic uncertainties can still be improved, current data draws a consistent picture of hints pointing towards New Physics effects in the lepton sector.
In Fig.~\ref{fig:pkoppenburg} we show an overview of several (lepton) flavour-related observables currently exhibiting tensions between their SM prediction and their experimentally determined values~\cite{koppenburgwebsite}.

\section{Effective field theory}
\label{sec:eft}

As discussed in the above, motivations to extend the SM are abundant and there are several reasons to believe that there is New Physics, in the form of new particles and/or interactions, between the electroweak and the Planck scale.
Although we do not know how a UV completion of the SM should look like to successfully explain all low-energy phenomena that the SM presently cannot account for, we can parametrise its effects with the help of effective field theories (EFT).
If one accepts that there is indeed New Physics at a certain energy scale $\Lambda_\text{NP}$, with $\Lambda_\text{EW} < \Lambda_\text{NP} < \Lambda_\text{Pl}$, one can extend the SM Lagrangian via non-renormalisable operators (with mass dimension $d>4$) composed of the SM fields.
The New Physics scale naturally serves as a UV-cutoff with a physical meaning, and thus allows to regulate the otherwise divergent radiative corrections.
Schematically, the effective Lagrangian can be written as
\begin{equation}
    \mathcal L_\text{eff} = \mathcal L_\text{SM} + \mathcal L_\text{NR}^{d>4} = \mathcal L_\text{SM} + \sum_{n\geq 1} \frac{\mathcal C_{ij}^{d = 4 + n}}{\Lambda^n_\text{NP}}\mathcal O_{ij}^{d = 4+n}\,.
    \label{eqn:eft}
\end{equation}
In the above, $\mathcal O$ are the effective operators of mass dimension $d > 4$, which are gauge- and Lorentz-invariant combinations of SM fields. Associated to these are the dimensionless effective couplings $\mathcal C$, the so-called Wilson coefficients, which are suppressed by $d-4$ powers of the cutoff scale $\Lambda_\text{NP}$. This EFT is called the Standard Model effective field theory or SMEFT.
Once the new heavy fields present in the UV-complete theory have been ``integrated out'', all New Physics effects are then encoded in the Wilson coefficients.
This can be understood as follows:
at energies (or more precisely momenta) much smaller than the New Physics scale we can Taylor-expand the fundamental amplitudes in powers of $p^2/\Lambda_\text{NP}^2$, effectively removing heavy propagators from the amplitude calculation, and thus yielding effective four-fermion diagrams. Moreover, the Taylor expansion is usually done to zeroth order, thus yielding local operators that do not depend on internal momenta.
This is completely analogous to the four Fermi theory (a low energy effective field theory with the electroweak interaction as its UV completion). As an illustrative example, let us consider the muon decay. Since $M_W \gg m_\mu$, the available momentum transfer is tiny compared to $M_W$ and it can thus be integrated out, by Taylor expanding the $W$ propagator.
In the unitary gauge we then have 
\begin{eqnarray}
    \mathcal A(\mu\to e\bar\nu_e\nu_\mu) &=& \frac{g_w^2}{2}(\bar e_L \gamma_\mu \nu_{eL}) \frac{-i\left(g^{\mu\nu} - \frac{p^\mu p^\nu}{M_W^2}\right)}{p^2 - M_W^2}(\bar\nu_{\mu L} \gamma_\nu \mu_L)\nonumber\\
    &\stackrel{p^2\ll M_W^2}{\approx}& \frac{i g_w^2}{2 M_W^2}(\bar e_L \gamma_\mu \nu_{eL})(\bar\nu_{\mu L} \gamma^\mu \mu_L)\,,
\end{eqnarray}
from which we can define the Fermi constant $\frac{4 G_F}{\sqrt{2}}\equiv \frac{g_w^2}{2 M_W^2}$ as the coefficient of the dimension-6 four-fermion operator. This also suggests that the natural cutoff scale is the $W$ mass, giving the cutoff scale a physical meaning: for sufficiently large momenta, $p^2\to M_W^2$, the effective theory loses its validity and becomes non-renormalisable.
More formally, since the amplitudes computed in the non-renormalisable theory are suppressed by powers of $(p/\Lambda_\text{NP})^{d-4}$, one can renormalise the theory provided $p\ll \Lambda_\text{NP}$.

\smallskip
The expansion to higher-dimensional operators in Eq.~\eqref{eqn:eft} can in principle be done up to an arbitrarily large mass dimension, and some processes only receive contributions at dimensions larger than six (e.g. neutrinoless double-beta decay or neutron-antineutron oscillations).

Taking the full SM as an effective theory, we have at dimension-5 only a single operator, the so-called Weinberg operator
\begin{equation}
    \mathcal L_{d=5} = \frac{C_{ij}}{2\Lambda}(\overline{L_i^c}\widetilde H^\ast)(\widetilde H^\dagger L_j)\,,
\end{equation}
which violates total lepton number by two units and gives rise to a Majorana mass for the left-handed neutrinos $\nu_L$.
Specific realisations of the Weinberg operator (the seesaw mechanisms) will be discussed in more detail in Section~\ref{sec:numassgen}.
At dimension-6, and if total baryon number conservation is imposed\footnote{If total baryon number conservation is relaxed, there are 4 additional operators.}, there is a total of 59 non-redundant operators~\cite{Grzadkowski:2010es}, containing 0, 2, and 4 fermion fields. Taking into account flavour, the amount of baryon number conserving operators is given by $2499$.

Since the Wilson coefficients are effective \textit{coupling constants}, they are subject to quantum corrections and thus run under renormalisation group evolution (RGE), analogously to the gauge and Yukawa couplings in the (renormalisable) SM.
Due to radiative (QCD and QED) corrections, the Wilson coefficients do not only run under renormalisation, but RGE can also lead to operator mixing.
These effects are of paramount importance, and have to be consistently taken into account, in order to obtain accurate predictions of low-energy observables within a given EFT.
In order to study New Physics effects (from a high-scale UV theory), the practical procedure is thus as follows.
One computes the Wilson coefficients (or a relevant subset thereof) at the New Physics (or rather matching) scale by matching the full UV theory to the effective Lagrangian\footnote{This can be done in several ways, one of which consists in requiring that amplitudes computed in the UV theory, and subsequently Taylor expanded for small momenta, (order by order) match the same amplitudes computed in the effective theory.}.
The matching scale is usually set to the mass of the lightest New Physics field that has been integrated out.
The Wilson coefficients are subsequently run down to the observable scale, for instance $\mu_\text{obs}\simeq M_W$; due to operator mixing under RGE, Wilson coefficients that are vanishing at the matching scale can nevertheless receive a non-vanishing (and often non-negligible) contribution at the observable scale.
If the observable scale is lower, the next lightest SM fields are integrated out at the usual RGE thresholds, that is at $\mu\sim M_W$ one performs the matching to the weak effective theory (WET), in which the Higgs, the weak gauge bosons and the top quark have been integrated out.
Fortunately, the matching conditions from SMEFT to WET as well as the anomalous dimension matrices have been calculated up to NLO precision for both SMEFT and WET~\cite{Jenkins:2013zja,Jenkins:2013wua,Alonso:2013hga,Jenkins:2017jig,Aebischer:2017gaw,Jenkins:2017dyc}.
For certain applications, for instance rare $B$-meson decays, the matching of the SM to WET and the anomalous dimension matrices for the subsequent RGE have been calculated to even higher precision, as outlined in Chapter~\ref{chap:bphysics}.

The approach of effective field theories offer two significant advantages:
\begin{itemize}
    \item \textbf{Simplifying calculations.} Calculating physical observables within the appropriate EFT (e.g. WET for $B$-meson decays), already reduces the amount of counter-term diagrams that have to be computed for the renormalisation of (QED and QCD) radiative corrections. Once the observables and the running of the Wilson coefficients have been computed, one has a model-independent result. To study a given New Physics model all that is left to do is to compute the relevant Wilson coefficients at the matching scale $\Lambda_\text{NP}\geq \mu_\text{obs}$.
    Furthermore, the running and operator mixing also leads to non-trivial correlations between observables that could otherwise be easily missed. This effect is discussed in more detail in Chapters~\ref{chap:bphysics} and~\ref{chap:lq}.
    
    \item \textbf{Constraining the unknown.} In addition to the above, even in the absence of a concrete UV model, one can quantitatively probe New Physics effects in low-energy observables using EFT.
    In particular, for a given set of observables (e.g. $b\to s\ell\ell$ observables), one can constrain New Physics contributions to Wilson coefficients on top of the SM ones, or even to operators which are absent in the SM, by fitting sets of Wilson coefficients to experimental data.
    This allows to estimate the inherent New Physics scale that can be indirectly probed by a given observable.
    Depending on the observable (or sets there of), experiments probe the ratio $\mathcal C/\Lambda_\text{NP}^{d-4}$.
    Restricting the discussion to dimension-6 operators for simplicity, one can then argue, that a given indirect experimental upper bound (U.L.) on the ratio $\mathcal C/\Lambda_\text{NP}^{2}$ probes New Physics scales of the order
    \begin{equation}
        \Lambda_\text{NP} = \sqrt{\frac{\mathcal C}{\mathrm{U.L.}}} \simeq \sqrt{\frac{1}{\mathrm{U.L.}}}\,,
    \end{equation}
    in which the last approximation is obtained if one requires the New Physics couplings generating a given Wilson coefficient to be natural $\mathcal C\sim \mathcal O(1)$.
    This definition has however several caveats: on the one hand, especially regarding flavour changing processes, the couplings might not be exactly natural, as is  the case of the electroweak interactions.
    On the other hand, $\mathcal C$ might be naturally suppressed by one or multiple loop-factors, if it is only generated at higher order.
    Nevertheless, throughout this thesis, we will resort to the requirement of $\mathcal C\sim \mathcal O(1)$ to determine the inherent New Physics scales, keeping in mind that it should not be directly interpreted as the mass of a New Physics field.
    
    Should one observe a tension between experimental data and the SM prediction, as is currently the case for several $b\to s\ell\ell$ and $b\to c\tau\nu$ observables, one can furthermore use EFT fits to the data in question in order to derive a ``guide'' for model building.
    In other words, one can fit model-building motivated ``hypotheses'', i.e. sets of Wilson coefficients, to the data to establish requirements a candidate model must fulfill, which can be done at the observable scale; the resulting patterns of New Physics and the inherent New Physics scale can then be re-interpreted in SMEFT at a high matching scale.
    The results of such an EFT analysis often allow to single out candidate models that could potentially accommodate a given excess in the low-energy data. This approach will be applied in Chapters~\ref{chap:bphysics} and ~\ref{chap:lq}.
\end{itemize}

To summarise, the approach of effective field theories to study New Physics is a powerful tool allowing to constrain (or identify) New Physics contributions to low-energy observables in a model-independent way.
Results of EFT analyses then often provide crucial data-driven bottom-up input for model building, allowing to identify and develop, step-by-step, a suitable extension of the SM.

\chapter{New Physics searches via lepton observables}
\label{chap:lepflav}
\minitoc

\noindent
Up to the present moment, massive neutrinos remain the only confirmed evidence for New Physics observed in a laboratory. 
Despite numerous decades of dedicated studies, both on the  theoretical and on the experimental fronts, the lepton sector of the SM remains far from being mastered.
Testing the SM and its (accidental) symmetries in the lepton sector is thus an appealing and promising pathway to indirectly search for New Physics interactions.

The present and following chapters are dedicated to an overview of some of the most important lepton flavour observables and how these can deviate from their SM expectations.
In the next chapter we focus on SM extensions via heavy neutral leptons (well motivated SM extensions allowing to accommodate oscillation data),  and discuss their phenomenological impact on lepton flavour observables.

The goal is to emphasis the role of lepton flavour observables as powerful probes of New Physics, acting in a complementary way to direct searches at colliders; despite indirect in nature, these searches at the ``high-intensity'' frontier have the potential to explore much higher energy scales than those directly accessible in high-energy experiments.

In what follows, we thus begin by considering neutrino oscillations, then discuss observables sensitive to the violation of lepton flavour universality, followed by a brief introduction to anomalous magnetic lepton moments $(g-2)_\ell$, and finally provide an overview of the most promising lepton flavour violating observables.
\section{Flavour violation in the neutral lepton sector: neutrino oscillations}
\label{sec:osci}
The so-far observed neutral fermions, $\nu_e,\, \nu_\mu,\, \nu_\tau$, are weak interaction eigenstates, forming $SU(2)_L$ doublets with the corresponding charged leptons, the so-called ``active'' neutrinos\footnote{The existence of additional neutral, weakly interacting light fermions is severely constrained by LEP data~\cite{ParticleDataGroup:2020ssz} on the invisible $Z$-boson decay width, thus strongly disfavouring further ``active'' neutrinos with mass $m < M_Z/2$.}.
In the SM, the flavour of neutrinos is defined by their production in weak charged current interactions; if for example a $W$-boson decay produces an electron, the associated (anti-) neutrino is defined as $\bar\nu_e$. 
This labelling defines the weak interaction (or flavour) basis of $\nu_\alpha \equiv (\nu_e,\, \nu_\mu,\, \nu_\tau)$. Consequently, this also allows to identify neutrino flavours in direct detection by the associated interaction with a charged lepton.
Since neutrinos are massless in the SM, their weak interaction basis is identical to their (physical) mass basis, leading to the conservation of flavour in weak charged current interactions.
However, the discovery of neutrino oscillations implies that at least two of the 3 active neutrinos are massive.
Lepton mass and interactions bases are no longer equivalent and this misalignment, just as what occurs in the quark sector, consequently leads to the violation of lepton flavour in charged current interactions, and lepton flavour non-conservation is encoded in the PMNS mixing matrix~\cite{Pontecorvo:1957cp,Pontecorvo:1957qd,Maki:1962mu}.
After EWSB the leptonic charged current is given by
\begin{equation}
    \mathcal L_W^{\ell\nu} = -\frac{g_w}{\sqrt{2}} W_\mu^- \bar\ell V_L^{\ell\dagger} \gamma^\mu P_L V_L^\nu \nu + \text{H.c.}\,,
    \label{eqn:wlnu}
\end{equation}
in which $g_w$ is the weak coupling constant (as before) and $P_L = (1-\gamma_5)/2$ denotes the left-handed chirality projector.
The presence of the diagonalisation matrices $V_L^\ell$ and $V_L^\nu$ now allows to define the PMNS mixing matrix as
\begin{equation}
    U_\text{PMNS} = V_L^{\ell\dagger}V_L^\nu\,. 
\end{equation}
For simplicity, and without loss of generality, one can always choose a basis in which the charged lepton Yukawa couplings are diagonal, and thus $V_L^\ell = \mathbb{1}$.
In this case the PMNS mixing matrix is simply given by $U_\text{PMNS} = V_L^\nu$.

Analogously to the CKM matrix in the quark sector, the PMNS mixing matrix can be cast in its standard parametrisation as
\begin{equation}
    U_\text{PMNS} = \begin{pmatrix}
                        c_{12}c_{13} & s_{12} c_{13} & s_{13} e^{-i \delta}\\
                        -s_{12}c_{23} - c_{12} s_{23} s_{13} e^{i\delta} & c_{12} c_{23} - s_{12} s_{23} s_{13} e^{i\delta} & s_{23} c_{13}\\
                        s_{12} s_{23} -c_{12} c_{23} s_{13} e^{i\delta} & -c_{12} s_{23} -s_{12} c_{23} s_{13} e^{i\delta} & c_{23} c_{13}
                    \end{pmatrix} \times \mathrm{diag}\left(1, e^{i\varphi_2}, e^{i\varphi 3}\right)\,,
\end{equation}  
in which $c_{ij} = \cos \theta_{ij}$ and $s_{ij} = \sin \theta_{ij}$; $\theta_{ij}$ are the (real) neutrino mixing angles, $\delta$ is the CP violating Dirac phase, and $\varphi_{2, 3}$ are the CP violating Majorana phases.
If neutrinos are Majorana particles, the latter phases cannot be re-absorbed via field re-definitions and are thus physical.
By themselves, neutrino oscillations are however not sensitive to the neutrino nature nor to the Majorana phases.

In analogy with the quark sector, one can also obtain a basis and parametrisation-independent measure of CP violation using the leptonic Jarlskog invariant~\cite{Jarlskog:1985ht} defined as 
\begin{equation}
    J_\text{CP}^\ell = \mathrm{Im}\left[U_{\alpha i} U_{\alpha j}^\ast U_{\beta i}^\ast U_{\beta j}\right]\,,\quad i<j\,,\:\alpha\neq\beta \,,
    \label{eqn:jarlskog}
\end{equation}
in which $U_{\alpha i}$ are the elements of $U_\text{PMNS}$; in the standard parametrisation, and for the ``$e\mu23$'' quartet one finds
\begin{equation}
    J_\text{CP}^\ell = \cos\theta_{12}\sin\theta_{12}\cos\theta_{23}\sin\theta_{23}\cos^2\theta_{13}\sin\theta_{13}\sin\delta_\text{CP}\,.
    \label{eqn:jarlskog_standard}
\end{equation}
As one can see in Eqs.~\eqref{eqn:jarlskog} and \eqref{eqn:jarlskog_standard}, the convention-independent measure of CP violation is not sensitive to the Majorana phases (since these cancel in the product of PMNS elements).

In order to determine the PMNS elements, different flavour transitions using neutrino oscillation experiments must be measured.
The transition probability, or oscillation probability, for a neutrino of energy $E$ created with flavour $\alpha$ to to be detected as flavour $\beta$ after propagating a distance $L$ in vacuum is given by~\cite{Giunti:2007ry}
\begin{eqnarray}
    P_{\nu_\alpha \to \nu_\beta}(L, E) = \delta_{\alpha \beta} &-& 4\sum_{k>j} \mathrm{Re}[U_{\alpha k}^\ast U_{\beta k} U_{\alpha j} U_{\beta j}^\ast] \sin^2\left(\frac{\Delta m_{kj}^2 L}{4 E}\right)\nonumber\\
    &+& 2\sum_{k>j}\mathrm{Im}[U_{\alpha k}^\ast U_{\beta k} U_{\alpha j} U_{\beta j}^\ast]\sin\left(\frac{\Delta m_{kj}^2 L}{2E}\right)\,,
    \label{eqn:osci}
\end{eqnarray}
in which $\Delta m_{kj}^2 = m_k^2 - m_j^2$, $m_k$ denotes the mass of the neutrino mass eigenstate $\nu_k$, $E$ is the kinetic energy and $L$ is the propagation length, that is the distance between the source and the detector.
In Eq.~\eqref{eqn:osci} one can see that, depending on the propagation length $L$ and the neutrino energy $E$, experimental measurements of the transition probabilities allow to determine the entries of $U$ and to measure the squared mass differences $\Delta m_{kj}^2$, but do not offer a direct access to the individual masses. 
Furthermore, it is evident that the second part in Eq.~\eqref{eqn:osci} is directly proportional the amount of CP violation.
However, notice that vacuum oscillations alone are not sufficient to determine the signs of $\Delta m_{kj}^2$ (independent of the amount of CP violation). This can be overcome if matter effects, such as the Mikheev-Smirnov-Wolfenstein (MSW) effect~\cite{Wolfenstein:1977ue, Mikheev:1986wj}, are consistently taken account. For example, measuring oscillations of solar neutrinos allowed to establish $\Delta m_{21}^2>0$, while the sign of $\Delta m_{31}^2$ is still unclear, and could be determined by the JUNO~\cite{JUNO:2015zny} and DUNE~\cite{DUNE:2015lol} experiments.
Therefore, there are currently two orderings\footnote{In principle, neutrino masses could be quasi-degenerate if the absolute neutrino mass scale is much larger than $\Delta m_{kj}^2$. However, cosmological measurements and bounds on $0\nu\beta\beta$ currently disfavour such a scenario~\cite{Lattanzi:2020iik}.} of the light neutrino spectrum which are still compatible with data, commonly referred to as \textit{normal ordering} (NO), where $m_{\nu_1}<m_{\nu_2}<m_{\nu_3}$ and \textit{inverted ordering} (IO), where $m_{\nu_3}<m_{\nu_1}<m_{\nu_2}$.

\medskip
With a worldwide experimental effort, different neutrino oscillation probabilities have been measured from solar neutrinos~\cite{Cleveland:1998nv,Kaether:2010ag,SAGE:2009eeu,Super-Kamiokande:2010tar,SNO:2011hxd,Bellini:2011rx,BOREXINO:2014pcl} produced in the solar fusion cycles~\cite{Vinyoles:2016djt}, atmospheric neutrinos~\cite{IceCube:2014flw} produced in pion decays from cosmic rays in the upper atmosphere~\cite{Honda:2015fha}, reactor neutrinos~\cite{KamLAND:2013rgu,DayaBay:2016ssb,DoubleChooz:2011ymz} at nuclear power reactors and accelerator neutrinos~\cite{T2K:2011ypd} produced at particle accelerators.
A global analysis~\cite{Esteban:2020cvm} of most of the available oscillation data resulting in a global fit of the neutrino mixing angles $\theta_{ij}$, the mass squared differences $\Delta m_{kj}^2$ and the Dirac phase $\delta$ for both possible orderings, has been performed by the NuFIT collaboration\footnote{For other global analyses see e.g.~\cite{deSalas:2020pgw}.}.
Their latest results (October 2021) are displayed in Table~\ref{tab:nufit}.

\renewcommand{\arraystretch}{1.3}
\begin{table}[h!]
    \centering
    \begin{tabular}{|l|cc|cc|}
        \hline
         & \multicolumn{2}{c|}{Normal ordering}  & \multicolumn{2}{c|}{Inverted ordering}\\
         \hline
         & b.f. $\pm 1\,\sigma$ & $3\,\sigma$ range & b.f. $\pm 1\,\sigma$ & $3\,\sigma$ range\\
         \hline
         $\sin^2 \theta_{12}$ & $0.304_{-0.012}^{+0.013}$ & $0.269\to0.343$ & $0.304_{-0.012}^{+0.013}$ & $0.269\to0.343$\\
         $\sin^2\theta_{23}$ & $0.573_{-0.023}^{+0.018}$ & $0.405\to 0.620$ & $0.578_{-0.021}^{+0.017}$ & $0.410 \to 0.623$\\
         $\sin^2\theta_{13}$ & $0.02220_{-0.00062}^{+0.00068}$ & $0.02034\to0.02430$ & $0.02238_{-0.00062}^{+0.00064}$ & $0.02053\to0.02434$\\
         $\delta/^\circ$ & $195_{-25}^{+52}$ & $105\to 405$ & $287_{-32}^{+27}$ & $192\to361$\\
         $\Delta m_{21}^2\,/\,10^{-5}\,\mathrm{eV}^2$ & $7.42_{-0.20}^{+0.21}$ & $6.82\to8.04$ & $7.42_{-0.20}^{+0.21}$ & $6.82 \to 8.04$\\
         $\Delta m_{3\ell}^2\,/\,10^{-3}\,\mathrm{eV}^2$ & $+2.515_{-0.028}^{+0.028}$ & $+2.431 \to +2.599$ & $-2.498_{-0.029}^{+0.028}$ & $-2.584 \to -2.413$\\
         \hline
         
    \end{tabular}
    \caption{Global fit of neutrino oscillation parameters as determined by the NuFIT collaboration~\cite{Esteban:2020cvm}, not taking into account the atmospheric data of the Superkamiokande experiment. The squared mass difference for normal ordering is defined as $\Delta m_{32}^2$, whereas for inverted ordering it is $\Delta m_{31}^2$.}
    \label{tab:nufit}
\end{table}
\renewcommand{\arraystretch}{1.}

Normal ordering is slightly preferred over inverted ordering by experimental data.
Since the absolute masses of the light neutrinos are unknown, one usually parametrises the masses depending on the lightest neutrino mass $m_0$ and the ordering of the light spectrum as
\begin{eqnarray}
	m_{\nu_1}^\text{NO} &=& m_0\,,\quad\quad m_{\nu_1}^\text{IO} = \sqrt{|\Delta m^2_{32} + \Delta m^2_{21} - m_0^2|}\,,\\
	m_{\nu_2}^\text{NO} &=& \sqrt{m_0^2 + \Delta m^2_{21}}\,,\quad\quad m_{\nu_2}^\text{IO} = \sqrt{|\Delta m^2_{32} - m_0^2|}\,,\\
	m_{\nu_3}^\text{NO} &=& \sqrt{m_0^2 + \Delta m^2_{31}}\,,\quad\quad m_{\nu_3}^\text{IO} = m_0\,.
\end{eqnarray}

\medskip
As previously mentioned, measuring the oscillation probabilities does not allow to determine the absolute mass scale of the light (active) neutrinos.
However, experiments measuring the kinematical spectrum of Tritium $\beta$ decays~\cite{Kraus:2004zw, Troitsk:2011cvm, KATRIN:2019yun} are sensitive to the so called ``effective electron-neutrino mass'' $m_\beta$, which is defined as~\cite{Abada:2018qok}
\begin{equation}
    m_\beta = \sqrt{\sum_{i} m_i^2 |U_{e i}|^2}\,,
\end{equation}
and which modifies the endpoint of the spectrum.
The most recent upper limit established by the KATRIN experiment~\cite{KATRIN:2019yun} leads to $m_\beta \leq 1.1 \:\mathrm{eV}$.
In the future, the KATRIN experiment aims at reaching a sensitivity of $\sim0.2\:\mathrm{eV}$.

Another bound on the sum of light degrees of freedom (assuming only SM fermions) can be established by cosmological measurements of the cosmic microwave background and baryon acoustic oscillations, but these require further assumptions on the underlying cosmological model.
A conservative bound from cosmological observations has been obtained by the Planck Collaboration in~\cite{Planck:2018vyg}, as
\begin{equation}
    m_{\nu_1} + m_{\nu_2} + m_{\nu_3} \lesssim 0.12\:\mathrm{eV}\quad\text{ corresponding to }\quad m_0 \lesssim 0.04 \: \mathrm{eV}\
    \label{eq:sumnuexp}
\end{equation}
for the lightest neutrino mass.

\medskip
Should neutrinos be Majorana fermions, they can mediate lepton number violating (LNV) processes such as neutrinoless double-beta decay ($0\nu\beta\beta$) (in certain radioactive isotopes which allow for double beta decays).
The amplitude of $0\nu\beta\beta$ decays is directly proportional to the effective Majorana mass $m_{ee}$, which can be written as~\cite{Blennow:2010th,Abada:2014nwa}
\begin{eqnarray}
    m_{ee} &\simeq&  \sum_{i=1}^{3} \, U_{e i}^2 \, p^2 \, \frac{m_{ i}}{p^2-m_{i}^2}\,,
    \label{eqn:0nubb_chap2}
\end{eqnarray}
in which $p^2$ corresponds to the virtual momentum of the neutrino, with $p^2 \simeq -(100 \, \mathrm{MeV})^2$ (an average estimate over
different values depending on the decaying nucleus). 
Also for $0\nu\beta\beta$ decays there is a worldwide experimental search going on; a signal confirming the Majorana hypothesis would be groundbreaking.
So far, no neutrinoless double-beta decay has been observed, but upper bounds on $m_{ee}$ have been established.
For instance the KamLAND-ZEN experiment~\cite{KamLAND-Zen:2016pfg} sets an upper limit on $m_{ee} \lesssim  (61 \div 165)\:\mathrm{MeV}$ obtained using the isotope $^{136}\mathrm{Xe}$. Similar limits have been obtained by other collaborations, for distinct choices of isotopes: 
$m_{ee} < (78 \div 239) \, \mathrm{meV}$ also for $^{136} \mathrm{Xe}$, by EXO-200~\cite{EXO-200:2019rkq};
$m_{ee} < (79 \div 180) \, \mathrm{meV}$ for $^{76}\mathrm{Ge}$, as derived by GERDA~\cite{GERDA:2020xhi};
$m_{ee} < (200 \div 433) \, \mathrm{meV}$ also for $^{76} \mathrm{Ge}$ by the Majorana Demonstrator~\cite{Majorana:2019nbd};  
$m_{ee} < (75 \div 350) \, \mathrm{meV}$ for $^{130} \mathrm{Te}$, obtained by CUORE~\cite{CUORE:2019yfd}.
The ranges result from different (very challenging) computations of the nuclear matrix elements.

\begin{figure}
    \centering
    \includegraphics[width=0.6\textwidth]{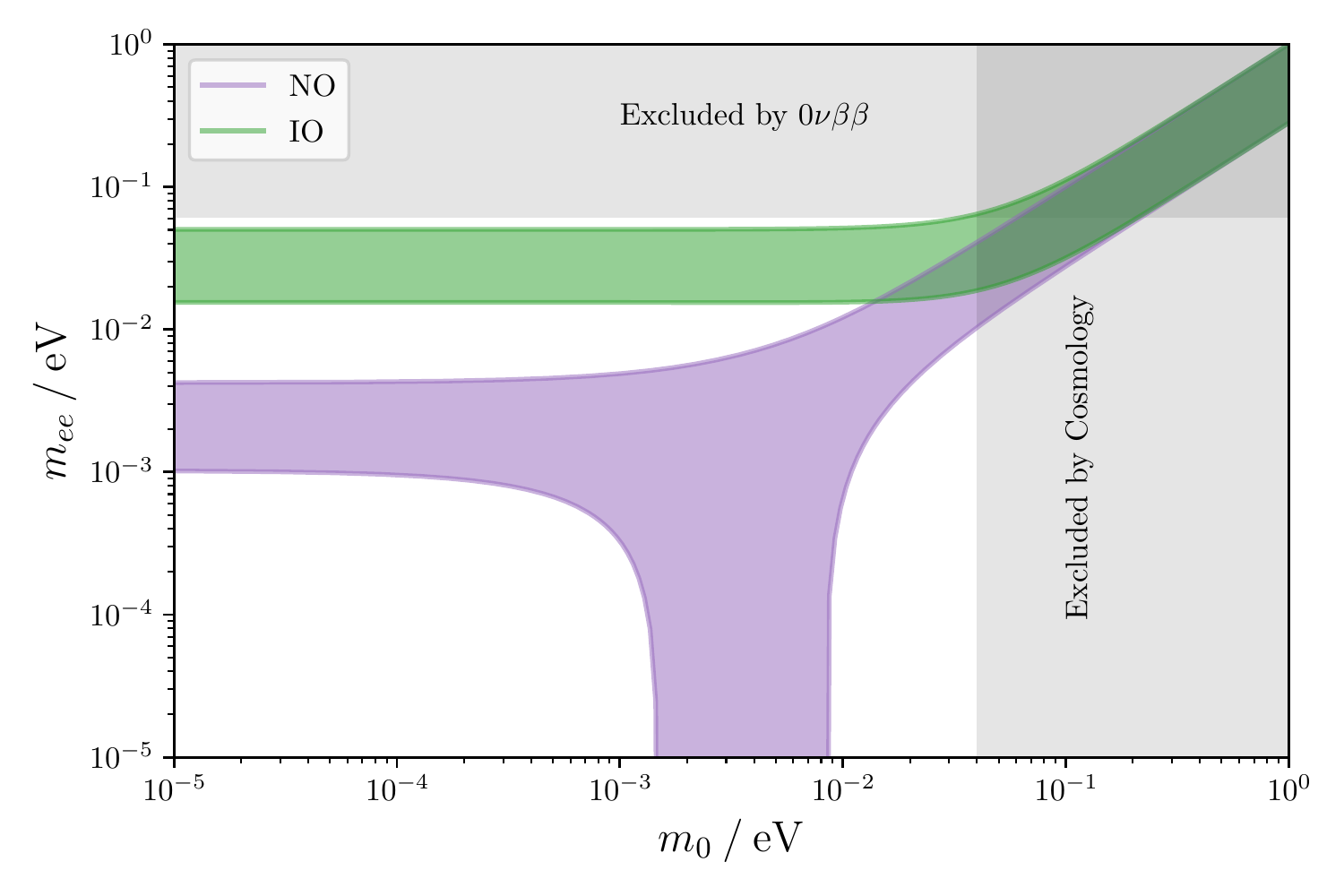}
    \caption{Predictions for the effective mass of neutrinoless double-beta decay $m_{ee}$ at $95\%\,\mathrm{C.L.}$ depending on the lightest neutrino mass $m_0$, under the assumption of 3 light Majorana neutrinos.
    The grey regions correspond to upper bounds on the half-life of $0\nu\beta\beta$ as determined by KamLAND-ZEN~\cite{KamLAND-Zen:2016pfg} and to cosmological bounds on the sum of light neutrino masses as obtained by Planck~\cite{Planck:2018vyg}.
    The purple region corresponds to a normal ordered spectrum, the green to inverted ordering.}
    \label{fig:sm_0nubb}
\end{figure}

Assuming that the three light neutrinos are Majorana, current oscillation data (at $95\%\,\mathrm{C.L.}$) allows for a wide range for $m_{ee}$, depending on the lightest neutrino mass $m_0$ and on the ordering of the light spectrum.
This can be seen in Figure~\ref{fig:sm_0nubb}, where we show a plot of $m_{ee}$ depending on $m_0$ for the two orderings. The widths of the bands correspond to varying the oscillation parameters and the CP violating phases (within the experimentally preferred $3\,\sigma$ regions).
As can be seen, on the one hand, improved upper bounds on $m_{ee}$ will probe (and potentially rule out) an inverted ordering of the light Majorana neutrinos.
On the other hand, the predictions for $m_{ee}$ of normal ordered light Majorana neutrinos has a strong suppression in the region $10^{-3} < m_0 < 10^{-2}$. 
This is due to interference effects of the Majorana phases present in $U_\text{PMNS}$.

\section{Lepton flavour universality violation}
\label{sec:LFUVnu}
In the SM, all lepton families have the same quantum charges, such that the couplings of the photon and the $Z$-boson are blind to lepton flavour.
Furthermore, and as emphasised before, in the SM neutrinos are massless and lepton flavour is strictly conserved.
Apart from lepton flavour conservation, this leads to another accidental symmetry of the SM called lepton flavour universality (LFU), only broken by the scalar sector (i.e. Yukawa interactions), since charged leptons have different masses.

As previously discussed in Section~\ref{sec:osci}, neutrino oscillations imply that neutrinos are necessarily massive, and lepton flavour is violated in charged current interactions. One thus expects that LFU is violated as well. 

The violation of LFU in charged current interactions can, for instance, be tested with $W\to\ell\nu$ decays and leptonic pseudo-scalar meson decays ($P\to \ell\nu$), where the presence of massive neutrinos leads to a modification of the decay rates already at tree-level.
Moreover, depending on the specific BSM construction, additional fermion or boson fields can contribute to the (leptonic) decays of the electroweak gauge bosons, weak decays of charged leptons and light mesons, either at tree-level or at higher order.
Thus, observables sensitive to LFU violation are very appealing laboratories to test the SM and to search for New Physics.
In the following we will briefly comment on a few selected observables of relevance to the remainder of this thesis.

\subsection{Decays of weak bosons}
As previously mentioned, the couplings of the $Z$-boson to charged leptons are by construction strictly flavour-universal, such that apart from (small) phase space corrections the partial widths $\Gamma(Z\to\ell^+\ell^-)$ are expected to be universal as well.
The current experimental bounds~\cite{ParticleDataGroup:2020ssz} on the (non-)universality of $Z$-decays yield
\begin{equation}
    \frac{\Gamma(Z\to \mu^+\mu^-)^\text{exp}}{\Gamma(Z\to e^+ e^-)^\text{exp}} = 1.0001 \pm 0.0024\,;\quad \frac{\Gamma(Z\to \tau^+\tau^-)^\text{exp}}{\Gamma(Z\to e^+ e^-)^\text{exp}} = 1.0020 \pm 0.0032\,,
\end{equation}
in excellent agreement with the SM LFU expectation, and thus leading to tight constraints on BSM contributions.

Similarly, the rates of the leptonic $W$-boson decays are also expected to be universal for all final charged lepton flavours.
The experimental measurements from LEP~\cite{ALEPH:2013dgf} of the individual rates and the SM predictions~\cite{Kniehl:2000rb} (including next-to-leading order corrections) of the branching ratios presently exhibit a mild tension,
\begin{eqnarray}
    \mathrm{BR}(W\to e\nu)^{\text{SM}} &=& 0.108383\,; \quad \mathrm{BR}(W\to e\nu)^{\text{exp}} = 0.1071 \pm 0.0016\,,\\
    \mathrm{BR}(W\to \mu\nu)^{\text{SM}} &=& 0.108383\,; \quad \mathrm{BR}(W\to \mu\nu)^{\text{exp}} = 0.1063 \pm 0.0015\,,\\
    \mathrm{BR}(W\to \tau\nu)^{\text{SM}} &=& 0.108306\,; \quad \mathrm{BR}(W\to \tau\nu)^{\text{exp}} = 0.1138 \pm 0.0021\,.
\end{eqnarray}
As can be clearly seen, the SM expectation for the $W\to\ell\nu$ branching fractions (with $\ell = e,\,\mu$) are lepton flavour universal while the experimental data for $W\to\tau\nu$ slightly deviates from that.
A deviation from LFU can be seen more prominently in the ratios of branching ratios~\cite{ALEPH:2013dgf}
\begin{eqnarray}
    \frac{\mathrm{BR}(W\to \tau\nu)^{\text{SM}}}{\mathrm{BR}(W\to e\nu)^{\text{SM}}} &=& 0.9993\,; \quad  \frac{\mathrm{BR}(W\to \tau\nu)^{\text{exp}}}{\mathrm{BR}(W\to e\nu)^{\text{exp}}} = 1.063 \pm 0.027\,,\\
    \frac{\mathrm{BR}(W\to \tau\nu)^{\text{SM}}}{\mathrm{BR}(W\to \mu\nu)^{\text{SM}}} &=& 0.9993\,; \quad  \frac{\mathrm{BR}(W\to \tau\nu)^{\text{exp}}}{\mathrm{BR}(W\to \mu\nu)^{\text{exp}}} = 1.070 \pm 0.06\,,
\end{eqnarray}
in which the experimental covariances have been consistently taken into account.

Recently, the ATLAS experiment performed a more precise measurement~\cite{ATLAS:2021icw} of the ratio $R_{\tau/\mu}^W = \frac{\mathrm{BR}(W\to \tau\nu)}{\mathrm{BR}(W\to \mu\nu)}$ directly, resulting in
\begin{equation}
    R_{\tau\mu}^W = 0.992 \pm 0.013(\text{tot.}) [\pm 0.007 (\text{stat.}) \pm 0.011 (\text{syst.})]\,,
\end{equation}
in which the LEP results on the $\tau\to\mu\nu\bar\nu$ branching fraction were used as an input.
In contrast to the LEP result, this measurement is in very good agreement with LFU as predicted by the SM.

\mathversion{bold}
\subsection{$\tau$-lepton decays}
\mathversion{normal}
The lepton universality of the $W\ell\nu$ vertex can be further investigated using decays of the $\tau$-lepton.
Considering only decays into final state leptons, one can construct another ratio of decay widths 
\begin{equation}
    R_\tau \equiv \frac{\Gamma(\tau^-\to \mu^-\nu\bar\nu)}{\Gamma (\tau^-\to e^-\nu\bar\nu)}\,,
    \label{eqn:Rtau1}
\end{equation}
which is sensitive to the presence of New Physics via deviation from LFU.
In the SM (with vanishing neutrino masses), the ratio is predicted to be $R_\tau = 0.973$~\cite{Pich:2009zza}. 
Combining experimental data from the ARGUS~\cite{ARGUS:1991zhv}, CLEO~\cite{CLEO:1996oro} and BaBar~\cite{BaBar:2009lyd} experiments, the HFLAV collaboration finds in their global fit\cite{HFLAV:2019otj}
\begin{equation}
    R_\tau = 0.9761 \pm 0.0028\,,
\end{equation}
consistent with the SM prediction at less than $2\,\sigma$, and thus placing a strong constraint on New Physics models.

Due to its comparatively large mass, the $\tau$-lepton can also decay into hadronic final states, which can be sensitive probes of LFU violation as well.
In particular, it is interesting to consider the ratios
\begin{eqnarray}
R_K^{\ell\tau} \equiv \frac{\Gamma(\tau\to K\nu)}{\Gamma(K\to \ell\nu)}\quad\text{and}\quad R_\pi^{\ell\tau}\equiv \frac{\Gamma(\tau\to \pi\nu)}{\Gamma(\pi\to \ell\nu)}\,,
\end{eqnarray}
with $\ell = e, \mu$. 
These observables are indirect probes of the universality of the $\tau$-coupling, mostly free from hadronic uncertainties (since these cancel in the ratio).
However, there are no available direct measurements of the ratios and a combination of experimental data without taking into account potential systematic effects cannot be done unambiguously. Therefore, we will not consider these decays here.

\subsection{Decays of light mesons}
Leptonic decays of charged mesons also constitute powerful probes of LFU, since their tree-level decays are also mediated by a $W$-boson exchange.
In order to minimise the impact of hadronic uncertainties, one can consider ratios of the form
\begin{equation}
    R_P \equiv \frac{\Gamma(P^+ \to \ell_{\alpha}^+\nu)}{\Gamma(P^+\to \ell_{\beta}^+\nu)}\,,
    \label{eqn:RP1}
\end{equation}
so that the SM predictions can be computed with a very high precision.
In order to compare experimental data with the SM prediction, or with the prediction of a given New Physics model, it is convenient to parametrise possible deviations ($\Delta r_P$) from the SM expectation as
\begin{equation}
    R_P = R_P^\text{SM}(1 + \Delta r_P)\,.
    \label{eqn:deltarp}
\end{equation}
In the past, due to experimental accessibility, attention was mostly devoted to the ratios\footnote{The ratio $R_K$ do not correspond to the lepton universality ratios of rare $B$-meson decays, which are commonly denoted by the same symbol.}
\begin{equation}
    R_K \equiv \frac{\Gamma(K^+\to e^+\nu)}{\Gamma( K^+\to \mu^+\nu)}\quad\text{and}\quad R_\pi \equiv \frac{\Gamma(\pi^+\to e^+\nu)}{\Gamma (\pi^+\to \mu^+\nu)}\,.
\end{equation}
Comparing the SM predictions~\cite{Cirigliano:2007xi} with experimental measurements~\cite{Girrbach:2012km, ParticleDataGroup:2020ssz}
\begin{eqnarray}
    R_K^{\:\text{SM}} &=& (2.477 \pm 0.001)\times 10^{-5}\,, \quad R_K^{\:\text{exp}} = (2.488 \pm 0.010)\times 10^{-5}\,,\\
    R_\pi^{\:\text{SM}} &=& (1.2354 \pm 0.0002)\times 10^{-4}\,, \quad R_\pi^{\:\text{exp}} = (1.230 \pm 0.004)\times 10^{-4}\,,
\end{eqnarray}
thus implies for $\Delta r_K$ and $\Delta r_\pi$
\begin{equation}
    \Delta r_K = (4\pm 4) \times 10^{-3}\,, \quad \Delta r_\pi = (-4 \pm 3) \times 10^{-3}\,,
\end{equation}
suggesting that observation agrees with the SM predictions at the $1\:\,\sigma$ level, and thus again providing tight constraints on LFU-violating New Physics models.

Neutral and charged current decays of heavy mesons into (semi-) leptonic final states also offer powerful probes of New Physics; however we postpone their discussion to Chapter~\ref{chap:bphysics}.

\section{Magnetic moments of charged leptons}
\label{sec:g-2section}
The magnetic (dipole) moment of a charged particle is a measure of the particle's tendency to align with a magnetic field.
For a fermion, or in particular a charged lepton with spin $\vec S$ and mass $m_\ell$, the magnetic moment is given by
\begin{equation}
    \vec M = g_\ell \,\frac{e}{2 m_\ell}\, \vec S\,,
\end{equation}
in which $g_\ell$ is the ``coupling strength'' of the lepton to a magnetic field, the so-called ``Landé factor''.
The Dirac equation implies $g_\ell = 2$, but this result is susceptible to quantum corrections.
In quantum electrodynamics (QED), a charged lepton coupled to an external magnetic field is described by a gauge-invariant lepton current coupled to an off-shell photon.
The gauge-invariant electromagnetic lepton current can in general be parametrised as
\begin{equation}
    \mathcal J_\mu = \bar \ell (p')\left[F_1(q^2) \gamma_\mu + \frac{i}{2m_\ell} F_2(q^2) \,\sigma_{\mu\nu} q^\nu - F_3(q^2)\gamma_5 \,\sigma_{\mu\nu}q^\nu + F_4(q^2)(q^2\gamma_\mu - 2 m_\ell q_\mu)\gamma_5\right]\ell(p)\,,
\end{equation}
in which $q$ is the momentum of the photon and $F_i$ are the electromagnetic form factors.
The Landé factor is then given by
\begin{equation}
    g_\ell = 2(F_1(0) + F_2(0))\,.
\end{equation}
At tree-level in the SM, we have $F_1(0) = 1$ and $F_{2,3,4}(0) = 0$, leading to $g_\ell = 2 = g_\text{Dirac}$.
Higher order corrections in perturbation theory to $F_1$ only modify the original coupling to the photon and thus give the scale dependence of the electron charge $e$, such that corrections to $g_\ell$ can only come from higher order contributions to $F_2(0)$.
The other form factor $F_3(0)$ induces the electric dipole moment $d_\ell$, while $F_4$ is only relevant for short distance virtual photon exchanges, often called ``anapole''.

The higher order corrections contributing to $F_2(0)$ and therefore to $g_\ell$, are conveniently captured in the so-called \textit{anomalous magnetic moment} defined as
\begin{equation}
    a_\ell \equiv \frac{g_\ell - g_\text{Dirac}}{g_\text{Dirac}} = \frac{g_\ell - 2}{2} = F_2(0)\,,
\end{equation}
commonly referred to as $(g-2)_\ell$.
The first correction at next-to-leading order (NLO) in QED was first calculated in 1948, resulting in $a_\ell = \frac{\alpha_e}{2\pi}$, where $\alpha_e = \frac{e^2}{4\pi}$ is the  electromagnetic fine structure constant.
Since then, a lot of progress has been made.
In general, the quantum corrections to the anomalous magnetic moment can be divided into three categories. There are contributions from pure QED, that only depend on the charged lepton mass and $\alpha_e$, and which have been fully perturbatively calculated up to 5-loop accuracy. 
For the anomalous magnetic moment of the muon $a_\mu$, corrections from weak interactions have also been calculated up to NLO precision (2-loop).
Furthermore, QCD corrections due to hadronic light-by-light scattering~\cite{Colangelo:2015ama,Green:2015sra,Gerardin:2016cqj,Blum:2016lnc,Colangelo:2017qdm,Colangelo:2017fiz,Blum:2017cer,Hoferichter:2018dmo,Hoferichter:2018kwz}, hadronic vacuum
polarisation~\cite{Chakraborty:2016mwy,Jegerlehner:2017lbd,DellaMorte:2017dyu,Davier:2017zfy,Borsanyi:2017zdw,Blum:2018mom,Keshavarzi:2018mgv,Colangelo:2018mtw,Davier:2019can}, and higher-order hadronic
corrections~\cite{Kurz:2014wya,Colangelo:2014qya} have to be taken into account to achieve a sufficiently accurate SM prediction.
Prior to the most recent lattice QCD-based computation of the leading order hadronic vacuum polarisation (LO HVP) contribution\footnote{Due to the comparatively large uncertainties in past lattice QCD computations, another method to determine the LO HVP relies on a data-driven approach using data of hadron production from virtual photons in $e^+e^-$ scattering. For a review of these evaluations see~\cite{Aoyama:2020ynm}.} by the BMW collaboration~\cite{Borsanyi:2020mff}, the SM prediction recently
compiled by the ``Muon $g-2$ Theory Inititative''~\cite{Aoyama:2020ynm} was found to be
\begin{equation}\label{eq:amu:SMwhite}
    a_\mu^\text{SM}\, =\, 116\, 591\, 810 \, (43) \times 10^{-11}\,,
\end{equation}
where the bulk of the uncertainty is associated with hadronic contributions.

Following the recently disclosed first results from the ``g-2'' E989 experiment at FNAL~\cite{Muong-2:2021ojo}, which are in good agreement with the previous findings of the BNL E821 experiment~\cite{Bennett:2006fi}, the current experimental average for the muon anomalous magnetic moment~\cite{Muong-2:2021ojo} is given by
\begin{equation}\label{eq:amu:exp}
    a_\mu^\text{exp}\, =\, 116\, 592\, 061\, (41) \times 10^{-11}\,,
\end{equation}
which should be compared to its SM expectation (cf. Eq.~\eqref{eq:amu:SMwhite}), leading to the following $4.2\,\sigma$ tension between  
theory and observation
\begin{equation}\label{eq:amu:delta}
    \Delta a_\mu\, \equiv a_\mu^\text{SM}\,- \, 
    a_\mu^\text{exp}\,
    \, =\, 251 \, (59) \times 10^{-11}\,.
\end{equation}
The impressive accuracy of the theoretical prediction and the experimental measurements renders $a_\mu$ a high-precision observable, extremely sensitive to contributions of New Physics.

\begin{figure}
    \centering
    \includegraphics[width=0.6\textwidth]{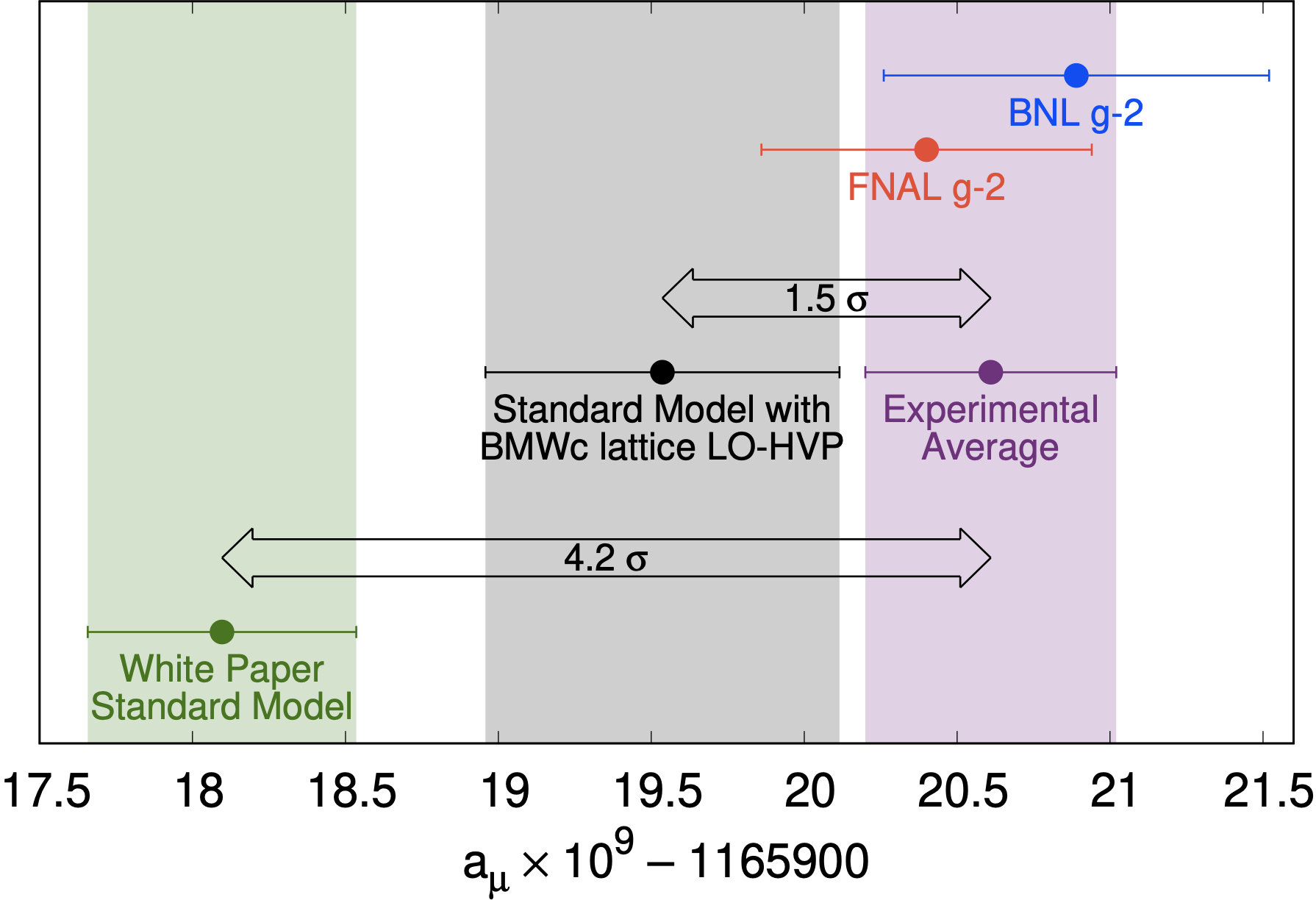}
    \caption{Overview of the current averages of the SM predictions and experimental measurements of $a_\mu$. The green region denotes the SM prediction compiled in~\cite{Aoyama:2020ynm}, the grey region denotes the SM prediction taking into account the lattice QCD determination of HVP as obtained in~\cite{Borsanyi:2020mff}, and the purple region denotes the experimental average of the BNL~\cite{Bennett:2006fi} and FNAL~\cite{Muong-2:2021ojo} measurements.
    Figure taken from~\cite{Lellouchplot}.}
    \label{fig:g-2overview}
\end{figure}

The value obtained taking into account the BMW collaboration computation ($a_\mu^\text{SM}=116\,591\,954\, (57)\times 10^{-11}$) would suggest $\Delta a_\mu= 107\, (70)\times 10^{-11}$, corresponding to a $1.5\,\sigma$ tension between  
theory and observation. 
While waiting for further confirmation\footnote{In~\cite{Crivellin:2020zul} it was pointed out that such hadronic
  vacuum polarisation contributions could potentially lead to conflicts
  with electroweak fits, inducing tensions in other relevant
  observables (hitherto in good agreement with the SM).} of lattice QCD based computations of the LO HVP contributions, in what follows we will rely on  $\Delta a_\mu$ obtained from the SM value as given in Eq.~(\ref{eq:amu:SMwhite}).
An overview of the averages of the SM predictions and the experimental measurements is shown in Fig.~\ref{fig:g-2overview}~\cite{Lellouchplot}.

Under the assumption of a significant tension between theory and observation, as given by Eq.~(\ref{eq:amu:delta}), the need for New Physics capable of accounting for such a sizeable discrepancy is manifest; 
several minimal, as well as more complete NP models, have been thoroughly explored in the light of the recent experimental results (for a recent review see, for example,~\cite{Athron:2021iuf} and references therein).  

\begin{figure}
    \centering
    \includegraphics[width=0.8\textwidth]{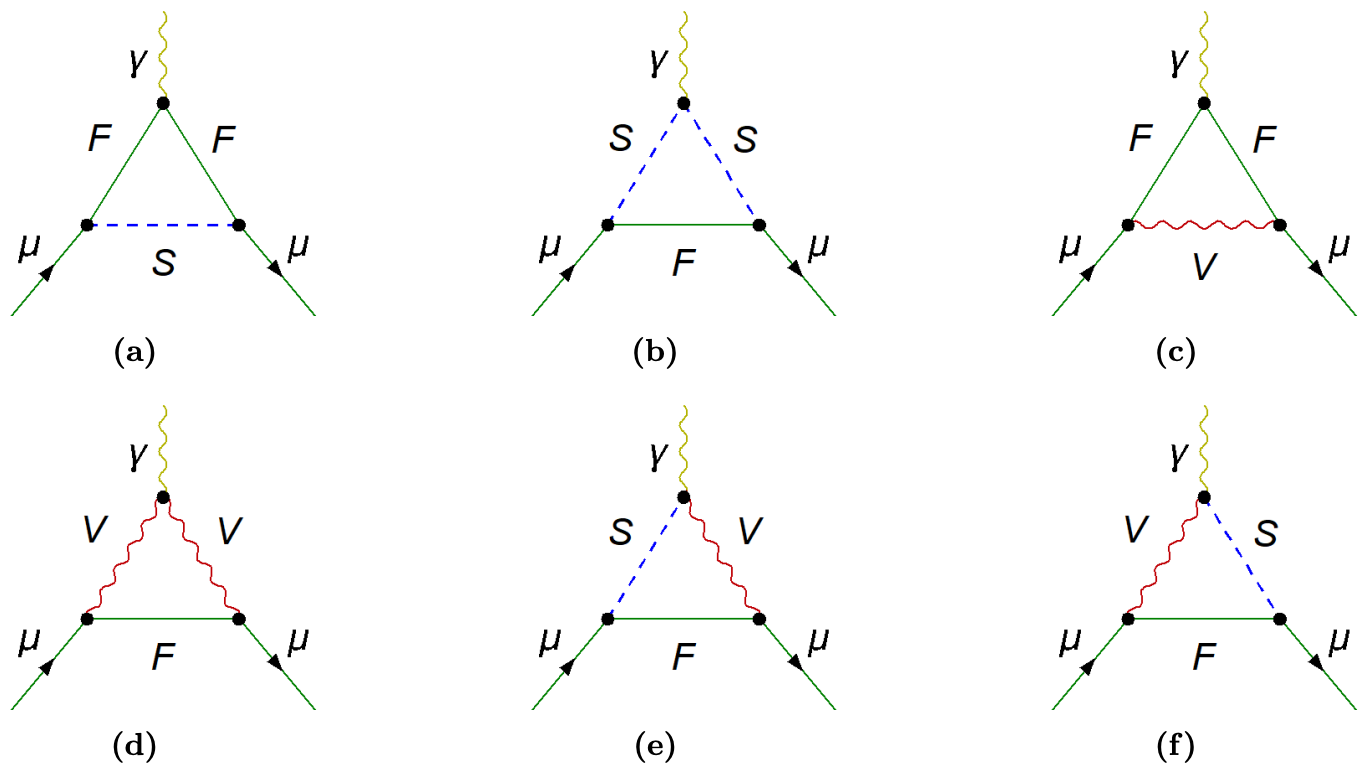}
    \caption{Illustrative examples of New Physics contributions to the anomalous magnetic moment of the muon $a_\mu$ at the one-loop level. Here, $F$ denotes a virtual fermion, $V$ a virtual vector boson and $S$ a virtual scalar.
    Figure taken from~\cite{Athron:2021iuf}.}
    \label{fig:g-2diagrams}
\end{figure}

\smallskip
In order to accommodate the tension in $a_\mu$, New Physics contributions are expected to appear at the one-loop level. 
In Fig.~\ref{fig:g-2diagrams} one has a general overview of additional scalar ($S$), vector ($V$) and fermion ($F$) fields contributing to the anomalous magnetic moments of the muon $a_\mu$.
Depending on the ratio of masses inside the loop $m_F/m_S$ or $m_F/m_V$, and on the sizes of the relevant (chiral) couplings, important New Physics contributions are possible.
In general, the tension $\Delta a_\mu$ can be explained with comparatively light BSM fields and sizeable couplings responsible for chiral enhancements.
New Physics explanations involving light BSM fields are however subject to an extensive array of other indirect constraints from LHC, flavour factories, and dark matter searches.
For a comprehensive survey of candidate models involving up to three BSM fields, that explain the tension in $a_\mu$ (with possible connections to e.g. dark matter), see~\cite{Athron:2021iuf}.

\begin{figure}
    \centering
    \includegraphics[width=0.6\textwidth]{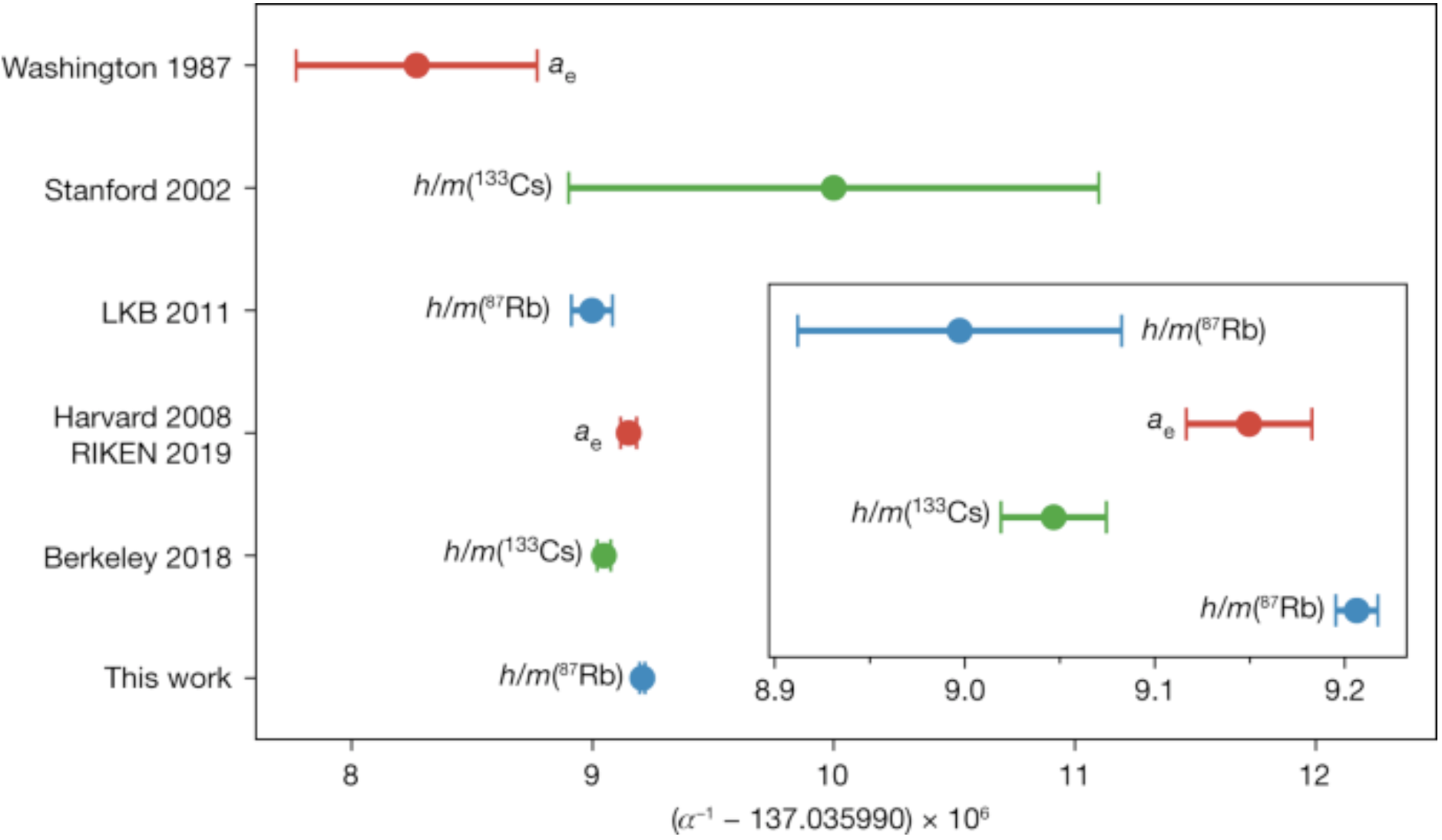}
    \caption{Overview of different experimental measurements of the electromagnetic finestructure constant $\alpha_e$ at low energies. Notice the large tension between the most recent determinations using Caesium and Rubidium atoms. Figure taken from~\cite{Morel:2020dww}.}
    \label{fig:alphaeoverview}
\end{figure}

\bigskip
The anomalous magnetic moment of the electron $a_e$ has been calculated in QED to an impressive 4-loop accuracy. 
From the experimental side, until recently, measurements of $a_e$ have been used to infer the low-energy value of $\alpha_e$.
Interestingly, a precise measurement of $\alpha_e$ 
using $\mathrm{Cs}$ atoms~\cite{Parker:2018vye,Yu:2019gdh}, is at the source of yet another discrepancy, this time
concerning the electron's anomalous magnetic moment. The 
experimental measurement of the electron anomalous magnetic moment
$a_e$~\cite{Hanneke:2008tm} 
\begin{equation}
a_e^\text{exp}\, =\, 1\,159\,652\,180.73(28)\times 10^{-12}\,
\end{equation}
currently exhibits a $2.5\,\sigma$ deviation from 
the SM prediction (relying on $\alpha_e$ from Caesium atoms), 
\begin{equation}
\label{Delta_aeCS}
\Delta a_e^\text{Cs}\,=\,a_e^\text{exp} - a_e^\text{SM}\,\sim\, 
-0.88 (0.36)\times 10^{-12}\,.
\end{equation}
In~\cite{Morel:2020dww}, a more recent estimation of $\alpha_e$ was obtained, this time relying on Rubidium atoms; the new determination of $\alpha_e$ (implying an overall deviation above the $5\,\sigma$ level for $\alpha_e$) now suggests milder tensions between observation and theory prediction, 
\begin{equation}\label{eq:ae:deltaRb}
    \Delta a_e^\text{Rb} \, =\, 0.48 \, (0.30)\times 10^{-12}\,,
\end{equation}
corresponding to $\mathcal{O}(1.7\,\sigma)$ deviation.
An overview of different measurements of $\alpha_e$ is shown in Fig.~\ref{fig:alphaeoverview}.
Other than signalling deviations from the SM expectation, 
it is interesting to notice the potential impact of \textit{both}
$\Delta a_e$ and $\Delta a_\mu$: other than having an opposite sign, the ratio $\Delta a_\mu/\Delta a_e$ does not exhibit the 
na\"ive scaling $\sim m^2_\mu/ m^2_e$ (expected  from the magnetic dipole operator, in which a mass insertion of the SM lepton is responsible for the required chirality flip~\cite{Giudice:2012ms}). This behaviour renders a common explanation of both tensions quite challenging, calling upon a departure from a minimal flavour violation (MFV) hypothesis, or from single new particle extensions of the SM (coupling to charged leptons~\cite{Davoudiasl:2018fbb,Kahn:2016vjr,Crivellin:2018qmi,Dorsner:2020aaz}). 
Notice that the pattern in both $\Delta a_e$ and $\Delta a_\mu$ could be also perceived as suggestive of a violation of flavour universality.
In Chapter~\ref{sec:g-2paper} we will attempt at constructing such a combined explanation relying on a minimal SM extension.

Finally, regarding the anomalous magnetic moment of the $\tau$-lepton, the experimental precision~\cite{DELPHI:2003nah} is still very poor compared with the theoretical uncertainty~\cite{Eidelman:2007sb},
\begin{eqnarray}
    a_\tau^\text{SM} &=& (117721\pm 5)\times 10^{-8}\,,\nonumber\\
    -0.052 < a_\tau^\text{exp} &<& 0.013\,,
\end{eqnarray}
so that unfortunately this observable cannot yet be used to infer useful information on possible New Physics contributions.

\section{Charged lepton flavour violation}
\label{sec:clfv_intro}
After the unexpected discovery of cosmic ray muons in 1937, it was first believed that such a state corresponded to an excited electron, such that it could radiatively decay into an electron and a photon ($e^\ast\to e \gamma$).
Experimental searches for this process, using artificial muons produced at accelerators, returned negative results with upper limits on the branching fraction of the order $\mathcal O(10^{-5}$.
In turn, this gave rise to the hypothesis of the muon neutrino $\nu_\mu$ whose presence would allow for a (GIM-) cancellation of the otherwise large neutral current loop effects (arising in a single neutrino scenario).
The subsequent discovery of $\nu_\mu$ then led to the introduction of separate lepton flavours ($e$ and $\mu$).

This introduction of $\nu_\mu$ to suppress the unwanted FCNC in the lepton sector was in essence analogous to the the GIM suppression of FCNC in the quark sector.
However, contrary to the quark sector, in the absence of a mechanism leading to non-vanishing neutrino masses, individual lepton flavours are strictly conserved. 
Formally, this leads to an accidental symmetry of the (lepton) SM Lagrangian which is then invariant under global $U(1)_e\times U(1)_\mu\times U(1)_\tau$ lepton field transformations.

As extensively discussed in Section~\ref{sec:osci}, the discovery of neutrino oscillations implies that the SM formulation of the lepton sector is at least incomplete - neutrinos have masses and their oscillations are a direct manifestation that neutral lepton flavours are not conserved, so that the accidental $U(1)^3$ symmetry of the SM is broken in Nature.
This implies, analogously to the quark sector, that charged lepton currents violate lepton flavour, and thus opens the door to cLFV transitions, unless accidental cancellations are at work.
Just like massive neutrinos constitute an irrefutable signal of New Physics, the same can be said of cLFV. 

The most minimal SM extension that accommodates neutrino oscillation data, as described in Section~\ref{sec:numassgen}, would in principle allow for lepton flavour violating transitions.
However, due to the unitarity of the PMNS matrix, and to the tiny differences of the neutrino masses, there is a strong GIM cancellation, so that the expected rates are vanishingly small.
For instance, the prediction for $\mu \to e \gamma$ in this framework, using the current experimental constraints on neutrino mixing, is approximately given by~\cite{Petcov:1976ff,Bilenky:1977du} 
\begin{equation}
    \mathrm{BR}(\mu\to e \gamma) \simeq \frac{3\alpha_e}{32\pi}\left|\sum_{i=1}^3 U_{ei} U_{\mu i}^\ast \frac{m_{\nu_i}^2}{M_W^2}\right|^2\simeq \mathcal O (10^{-55})\,,
\end{equation}
clearly lying beyond the reach of any experimental sensitivity.
Similar (extremely small) values are found for processes such as $\mu\to eee$ decays and the analogous lepton flavour violating $\tau$ decays.
The observation of such cLFV signals would thus imply that more involved BSM extensions are needed in order to simultaneously explain the origin of neutrino masses and to interpret a possible cLFV signal.
Any observation of cLFV would imply new degrees of freedom: the SM must be non-trivially extended.

It is however important to stress that although neutrino oscillations imply that lepton flavour is violated in Nature, a possible observation of charged lepton flavour violating processes is not necessarily associated with neutrino oscillation phenomena; cLFV can emerge as an independent process, without any connection to the mechanism of neutrino mass generation.
Furthermore, it is important to stress the strong difference of leptonic FCNC transitions and FCNC in the quark sector in what regards contributions of New Physics.
Albeit strongly suppressed, the SM does lead to observable rates in quark FCNC transitions such as $b\to s \ell\ell$, which are currently subject to extensive experimental investigation (see Chapter~\ref{chap:bphysics}). 
Here, New Physics contributions are either invoked to address possible tensions, or are strongly constrained by current experimental data.
In stark contrast is the lepton sector - there is in essence no SM contribution and any confirmed observation of cLFV is necessarily an indisputable signal of New Physics, as it cannot be interpreted or explained in terms of SM theoretical uncertainties.

\begin{figure}
    \centering
    \includegraphics[width=0.8\textwidth]{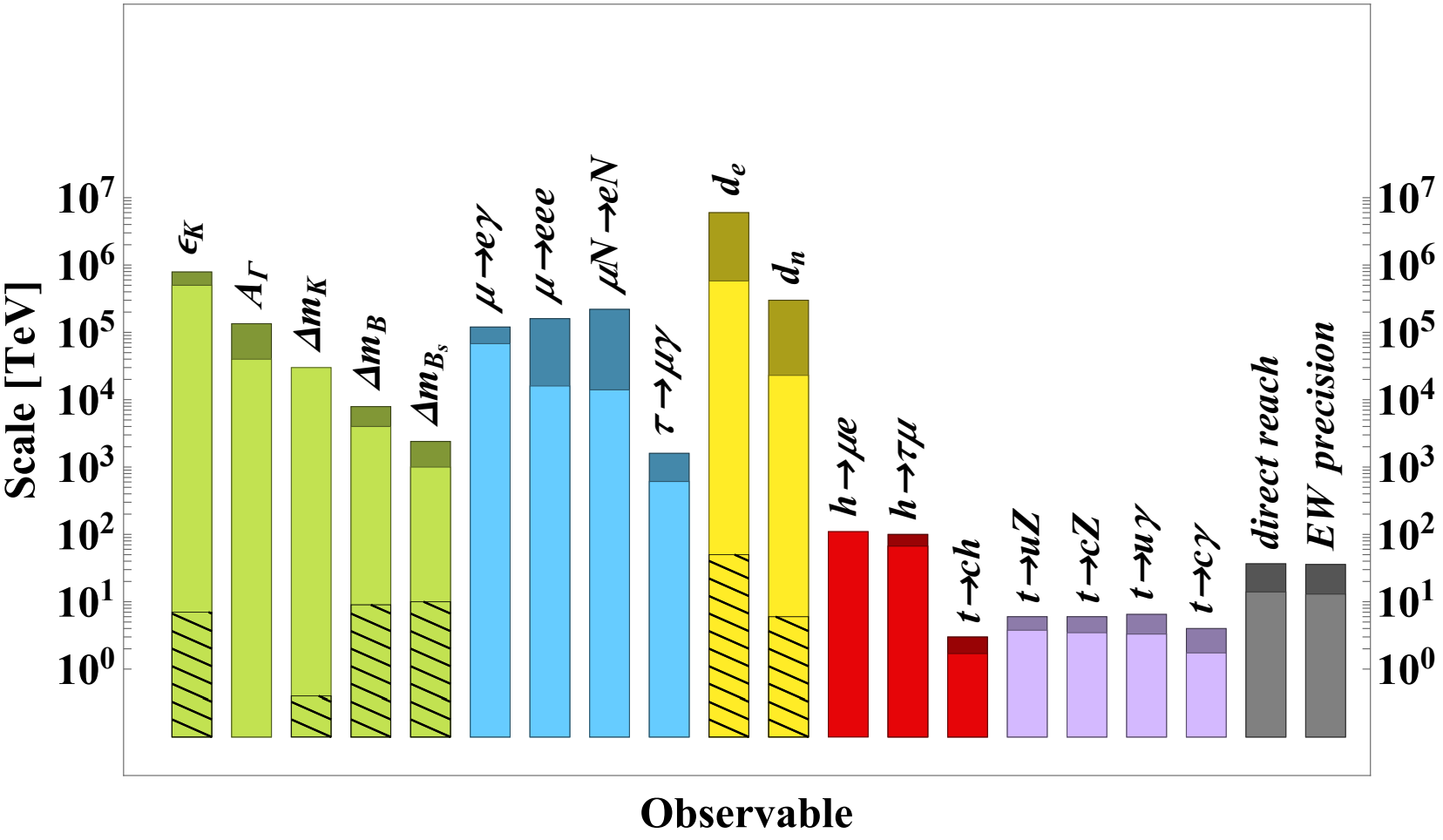}
    \caption{New Physics scales to be indirectly probed by the indicates observables. The darkend areas are the ``na\"ive'' New Physics scales by assuming the Wilson coefficients of order one, the coloured bars indicate the inherent New Physics scales assuming weak interaction strengths, while the hatched areas account for loop-suppression due to higher order effects.  Figure taken from Ref.~\cite{EuropeanStrategyforParticlePhysicsPreparatoryGroup:2019qin}.}
    \label{fig:EPPSU_flavour}
\end{figure}

Obviously, the non-observation of such signals and the implied experimental upper bounds on the associated processes, consequently lead to tight constraints on the parameters of New Physics models that could in principle predict sizeable rates for cLFV transitions.
From a model-independent perspective, one can argue that the inherent scale of New Physics that can be probed with current and future cLFV dedicated experiments is up to thousands of $\mathrm{TeV}$, far beyond the direct reach of current and future colliders~\cite{EuropeanStrategyforParticlePhysicsPreparatoryGroup:2019qin} (cf. Chapter~\ref{sec:eft}).
An overview of this is shown in Fig.~\ref{fig:EPPSU_flavour}, where one has the inherent New Physics scales to be indirectly probed by several flavour observables\footnote{As can be seen in Fig.~\ref{fig:EPPSU_flavour}, the electric dipole moments (EDM) of the electron $d_e$ and neutron $d_n$ probe a very high New Physics scale as well, due to the fact that they occur in the SM only at the four-loop level. In New Physics models they can already be generated at the one-loop level, are however not of relevance for this thesis.}.

In summary, an observation of cLFV might constitute the first (albeit indirect) discovery of New Physics, without having directly observed a signal of a new fermion or boson at colliders.
Furthermore, searches for signals of cLFV observables might offer complementary information to a potential direct discovery of new states at high-energy colliders, such that cLFV observables will potentially prove to be crucial in disentangling New Physics scenarios in the lepton sector.
This re-inforces the importance of experimental searches for cLFV processes.

\paragraph{Muon cLFV}
Muons are possibly the best laboratory to look for cLFV, since they can be abundantly produced and have a comparatively long lifetime.
Furthermore, due to their low mass, the number of kinematically allowed decay channels, flavour violating or not, is relatively small and the final states can be studied with great precision.
Very high intensity muon beams are possible (obtained at meson factories and proton accelerators), allowing for a great variety of muon dedicated experiments with extremely high sensitivities.
In view of this, it comes with no surprise that the best available experimental sensitivities, and consequently the best available bounds on cLFV processes, arise from rare muon processes.

In addition to the radiative and three-body decays ($\mu^+\to e^+\gamma$ and $\mu^+\to e^+e^-e^+$), several facilities are dedicated to studying muonic atoms.
Muonic atoms are formed when a muon is ``stopped'' in some target material, usually very pure elements. After cascading down the energy levels of the atom, the muon becomes bound in the $1s$ ground state
\begin{equation}
    \mu^- + (A, Z) \to [\mu^-(A,Z)^+]_{1s}\,,
\end{equation}
where $A$ and $Z$ respectively denote the mass and atomic number of the target nucleus.
While cascading down the energy levels, the muon emits a characteristic X-ray spectrum, which allows identifying the process.
Since a muon is an unstable particle, it will eventually decay despite being in a bound state.
The bound muon decays via interactions with the target nucleus, either exchanging a virtual photon, or, in the presence of New Physics, undergoing some non-electromagnetic interaction.
In the SM, there are two possible outcomes. 
Either the muon decays in orbit (DIO) into an electron and two neutrinos, or it is captured by the target nucleus via inverse $\beta$ decay.
In the presence of New Physics, the exotic process of neutrinoless muon capture can occur
\begin{equation}
    [\mu^-(A, Z)^+]_{1s} \to (A, Z) + e^-\,,
\end{equation}
in which the electron is produced with sufficient kinetic energy to escape the Coulomb potential of the target nucleus, which can be left in the ground state, or in an excited one.
Usually dominating, and from an experimental point of view the most advantageous, is the first case, called ``coherent capture''.
The rate of coherent captures with respect to other captures strongly depends on the chosen target nucleus~\cite{Kitano:2002mt}.
In addition to being usually dominant, the coherent capture also leads to a cleaner experimental signature, since the the energy of the final state electron is monochromatic and typically lies well above the kinematical endpoint\footnote{The energy of the escaping electron depends on the binding energy and therefore on the nucleus.} of the spectrum of the SM muon decay $\mu\to e \bar \nu_e \nu_\mu$.
This process is usually referred to as ``$\mu-e$ conversion'' and the associated observable is defined as
\begin{equation}
    \mathrm{CR}(\mu-e, \mathrm{N}) = \frac{\Gamma(\mu^- + \mathrm{N} \to e^- + \mathrm{N})}{\Gamma(\mu^- + \mathrm{N}\to \text{all captures})}\,,
\end{equation}
which from a theoretical point of view has the additional advantage that most of the nuclear form factors cancel out, only the overlap integrals between the nuclear and leptonic wave function remain to be computed~\cite{Kitano:2002mt}.

In the presence of lepton number violating interactions, another neutrinoless $\mu- e$ conversion can take place,  given by
\begin{equation}
    \mu^- \to (A,Z) \to e^+ (A, Z-2)^{(\ast)}\,,
\end{equation}
in which the the final state nucleus can be in its ground state or an excited one.
Here, contrary to the $\mu^- - e^-$ conversion, no coherent enhancement is possible since the final and initial state nuclei are necessarily different from each other.
Due to its LNV nature, this process is closely related to neutrinoless double-beta decay.
From the theoretical perspective there is however a caveat; all but one of the nuclear form factors are unknown~\cite{Domin:2004tk,Geib:2016atx,Geib:2016daa}, and therefore we do not consider this process here.

Another cLFV process in muonic atoms was proposed in~\cite{Koike:2010xr}. It consists of a bound $1s$ muon and a bound $1s$ electron converting into a pair of electrons, and has been identified as potentially complementary to other cLFV muon processes:
\begin{equation}
    \mu^- e^- \to e^- e^-\,.
\end{equation}
As it has been pointed out in~\cite{Koike:2010xr}, it offers several experimental advantages. On the one hand, the experimental signal consists of two (almost) back-to-back emitted electrons with the same energy. On the other hand, this process is enhanced by the Coulomb potential of the nucleus, with respect to other observables in muonic atoms.
So far, this process has not been experimentally investigated, but discussions are underway so that it might be studied at COMET.

\smallskip
Further interesting observables concern Muonium ($\mathrm{Mu}$). Muonium is a Coulomb bound state consisting of an electron and an anti-muon ($e^-\mu^+$) which is formed when a $\mu^+$ slows down inside matter and captures an electron.
Being free of hadronic uncertainties, this hydrogen-like bound state is well described by electroweak interactions and is used to study fundamental constants of the SM, or search for deviations from the SM induced by the presence of possible New Physics interactions.

Concerning cLFV transitions, one can study the spontaneous conversion of Muonium into anti-Muonium ($\overline{\text{Mu}}= e^+\mu^-$) and the cLFV decay of Muonium, $\mathrm{Mu}\to e^+ e^-$. An observation of these would again be a clear signal of New Physics.
The Muonium cLFV observables are further discussed in Section~\ref{sec:muonicatoms} in the context of SM extensions via heavy neutral leptons.

\paragraph{\pmb{$\tau$}-lepton cLFV}
Due to their large mass and consequently their large phase space, $\tau$-leptons offer a vast array of cLFV signatures.
Besides the radiative and three-body cLFV decays in full analogy to the muon sector, there are also numerous semi-leptonic cLFV decays into a lighter lepton and one or two mesons\footnote{Searches for lepton and baryon number violating $\tau$ decays have also been conducted, for example $\tau\to p\mu^+\mu^-$.}.
Studying cLFV decays across all lepton families is paramount to the understanding of the underlying New Physics flavour structures.

In addition to the radiative decays ($\tau\to e\gamma$ and $\tau\to\mu\gamma$) and same-lepton three-body decays ($\tau\to eee$ and $\tau\to \mu\mu\mu$), four other fully leptonic cLFV final states are possible: 
\begin{eqnarray}
    \tau^- \to \mu^- e^+ e^-\,, &\phantom{=}& \tau^-\to e^- \mu^+\mu^-\,,\\
    \tau^-\to \mu^- e^+ \mu^-\,, &\phantom{=}& \tau^- \to e^- \mu^+ e^-\,,
\end{eqnarray}
in which the decays of the second row correspond to a ``double'' flavour violation.
Thus, depending on the underlying New Physics framework, the different (charge) signatures can have very distinct amplitudes.
Furthermore, the different semi-leptonic channels can offer very distinct probes of New Physics.
Assuming there is only one meson in the final state, $\tau$ decays into $q\bar q$ and a lighter lepton are of particular interest, for instance $\tau \to \phi \mu$, because in this case there can be a resonant enhancement of the cLFV process.

From the experimental side, due to their much larger mass and much shorter lifetime, $\tau$-leptons are not as readily available as muons.
However, they can be produced at the so-called $B$-factories, as for example Belle (II) and BaBar.
These experiments are situated at $e^+ e^-$ colliders running at energies to abundantly produce the $\Upsilon(nS)$ $b\bar b$ resonances.
Usually they operate at the energy of the $\Upsilon(4S)$ meson (i.e. $\sqrt{s} = 10.58\:\mathrm{GeV}$) which almost exclusively decays into $B \bar B$ pairs ($\mathrm{BR}(\Upsilon(4S)\to B\bar B) \geq 96\%$\cite{ParticleDataGroup:2020ssz}), hence the name ``$B$-factories''. Runs at the lower $b\bar b$ resonances can however also produce copious amounts of $\tau^+\tau^-$ pairs; the consequently high luminosities enable exhaustive studies of cLFV $\tau$-lepton decays.
For instance, the Belle experiment has searched for 46 distinct cLFV $\tau$ decay modes, and Belle II is expected to significantly improve the obtained bounds~\cite{Belle-II:2018jsg}.

\paragraph{cLFV meson decays}
Many experiments have searched for signals of cLFV in the decays of an extensive array of neutral and charged mesons.
These processes probe
\begin{equation}
    q\to q^{(\prime)} \ell_\alpha\ell_\beta
\end{equation}
contact interactions, possibly accompanied by another final state meson.
The most stringent bounds have been obtained for neutral $K_L$ decays, but results for heavy meson decays have nevertheless reached an impressive level.
Of particular interest are neutral and charged decays of mesons containing a $b$ quark, due to their abundant production at the aforementioned $B$ factories and to  their large phase space leading to a plethora of possible final states.
From a phenomenological point of view, these decays are interesting due to several hints of New Physics possibly coupled to second and/or third generation fermions, as will be discussed in detail in Chapters~\ref{chap:bphysics} and~\ref{chap:lq}.
As we will discuss, several cLFV $B$-meson decays can be directly connected to recently experimentally measured deviations from SM.

\paragraph{cLFV at high energies}
In addition to cLFV searches at the ``intensity frontier'', which include the aforementioned low-energy lepton and meson decays, one can also look for distinct types of cLFV signals at higher energies, which are a consequence of the (on-shell) production of certain states. 
The most interesting channels are perhaps those of lepton flavour violating $Z$ and Higgs decays, as well as lepton flavour violating di-lepton tails in $pp$ collisions.

Similarly to cLFV lepton decays, lepton flavour violating decays of the $Z$-boson are forbidden in the SM and, even if the SM is minimally extended to accommodate neutrino oscillation data (cf. Section~\ref{sec:numassgen}), these are highly GIM-suppressed, leading to extremely small rates:
\begin{equation}
    \mathrm{BR}(Z\to \mu^\pm \tau^\mp) \lesssim 10^{-54}\,,\quad\mathrm{BR}(Z\to e^\pm \tau^\mp)\sim\mathrm{BR}(Z\to e^\pm \mu^\mp) \lesssim 10^{-60}\,.
\end{equation}
Sizeable rates for cLFV $Z$-boson decays then reflect a non-trivial BSM construction. 
Furthermore, cLFV decays of charged leptons are often (depending on the underlying model) at least partly mediated via lepton flavour violating $Z$-penguin diagrams, so that studying cLFV $Z$-decays offers important complementary information, and might help disentangling New Physics scenarios.

Experimentally, stringent upper bounds on the different decay channels have been obtained, especially at LEP which performed as a ``$Z$-factory''.
These bounds are expected to be significantly improved at a future FCC-ee running as a $Z$ and Higgs factory, i.e. respectively at the $Z$ and Higgs poles, or at energies at which resonant $Zh$ production is possible.

Of particular interest are also cLFV decays of the Higgs boson, 
\begin{equation}
    h\to e^\pm \mu^\mp\,,\quad h\to e^\pm \tau^\mp\,,\quad h\to  \mu^\pm \tau^\mp\,.
\end{equation}
In addition to probing the presence of cLFV in general, the above decays offer unique access to possible flavour violating (effective) Yukawa couplings and may thus offer insight about BSM mechanisms of mass generation in the lepton sector, and possible connections to the scalar sector of a given New Physics framework.

Finally, as recently pointed out in~\cite{Angelescu:2020uug}, searches for high-$p_T$ lepton flavour violating di-lepton tails in $pp$ collisions
\begin{equation}
    pp \to e\mu\,,\quad pp \to e\tau\,,\quad pp\to\mu\tau\,,
\end{equation}
offer important model-independent complementary probes to semi-leptonic cLFV $\tau$ decays and (semi-) leptonic cLFV meson decays,
since they allow to derive indirect upper bounds on effective operators that encode $qq^{(\prime)};\ell_\alpha\ell_\beta$ contact interactions that might be responsible for these decays\footnote{Further model-dependent channels can also be studied, should $\sqrt{s}$ be sufficiently large in order to allow for the production of New Physics states. In this case, the di-lepton tails exhibit a resonant enhancement via e.g. $pp\to X\to e\mu$.}.
Consequently, this also allows to derive indirect upper bounds on the associated semi-leptonic cLFV decays.
In some cases, the indirect bounds derived from high-$p_T$ data already give more stringent upper bounds than direct searches for the decays~\cite{Angelescu:2020uug}.
With increasing statistics at the LHC, these probes are expected to become more and more relevant.

\paragraph{Overview}
As extensively argued, the observation of one (or several) cLFV processes would be a clear signal of physics beyond the SM.
Currently, there is a vast world-wide 
array of dedicated experiments and searches, at different energy scales, aiming at discovering 
cLFV transitions.
In Table~\ref{tab:cLFVdata} we list current experimental bounds and future sensitivities\footnote{Note that for the Mu3e experiment~\cite{Blondel:2013ia} we display a second more optimistic future sensitivity, reflecting the potential of having a very high intensity muon beam available; in our (numerical) analyses throughout this thesis we use the latter (optimal) sensitivity.} for some of the ``purely leptonic'' observables here considered.
In Table~\ref{tab:semicLFV} we display current upper bounds and future sensitivities for semi-leptonic cLFV processes (semi-leptonic decays of $\tau$-leptons and (semi-) leptonic meson decays) of particular interest for this thesis.

\renewcommand{\arraystretch}{1.3}
\begin{table}[h!]
    \centering
    \hspace*{-7mm}{\small\begin{tabular}{|c|c|c|}
    \hline
    Observable & Current bound & Future Sensitivity  \\
    \hline\hline
    $\text{BR}(\mu\to e \gamma)$    &
    \quad $<4.2\times 10^{-13}$ \quad (MEG~\cite{TheMEG:2016wtm})   &
    \quad $6\times 10^{-14}$ \quad (MEG II~\cite{Baldini:2018nnn}) \\
    $\text{BR}(\tau \to e \gamma)$  &
    \quad $<3.3\times 10^{-8}$ \quad (BaBar~\cite{Aubert:2009ag})    &
    \quad $3\times10^{-9}$ \quad (Belle II~\cite{Kou:2018nap})      \\
    $\text{BR}(\tau \to \mu \gamma)$    &
     \quad $ <4.4\times 10^{-8}$ \quad (BaBar~\cite{Aubert:2009ag})  &
    \quad $10^{-9}$ \quad (Belle II~\cite{Kou:2018nap})     \\
    \hline
    $\text{BR}(\mu \to 3 e)$    &
     \quad $<1.0\times 10^{-12}$ \quad (SINDRUM~\cite{Bellgardt:1987du})    &
     \quad $10^{-15(-16)}$ \quad (Mu3e~\cite{Blondel:2013ia})   \\
    $\text{BR}(\tau \to 3 e)$   &
    \quad $<2.7\times 10^{-8}$ \quad (Belle~\cite{Hayasaka:2010np})&
    \quad $5\times10^{-10}$ \quad (Belle II~\cite{Kou:2018nap})     \\
    $\text{BR}(\tau \to 3 \mu )$    &
    \quad $<3.3\times 10^{-8}$ \quad (Belle~\cite{Hayasaka:2010np})  &
    \quad $5\times10^{-10}$ \quad (Belle II~\cite{Kou:2018nap})     \\
    & & \quad$5\times 10^{-11}$\quad (FCC-ee~\cite{Abada:2019lih})\\
        $\text{BR}(\tau^- \to e^-\mu^+\mu^-)$   &
    \quad $<2.7\times 10^{-8}$ \quad (Belle~\cite{Hayasaka:2010np})&
    \quad $5\times10^{-10}$ \quad (Belle II~\cite{Kou:2018nap})     \\
    $\text{BR}(\tau^- \to \mu^-e^+e^-)$ &
    \quad $<1.8\times 10^{-8}$ \quad (Belle~\cite{Hayasaka:2010np})&
    \quad $5\times10^{-10}$ \quad (Belle II~\cite{Kou:2018nap})     \\
    $\text{BR}(\tau^- \to e^-\mu^+e^-)$ &
    \quad $<1.5\times 10^{-8}$ \quad (Belle~\cite{Hayasaka:2010np})&
    \quad $3\times10^{-10}$ \quad (Belle II~\cite{Kou:2018nap})     \\
    $\text{BR}(\tau^- \to \mu^-e^+\mu^-)$   &
    \quad $<1.7\times 10^{-8}$ \quad (Belle~\cite{Hayasaka:2010np})&
    \quad $4\times10^{-10}$ \quad (Belle II~\cite{Kou:2018nap})     \\
    \hline
    $\text{CR}(\mu- e, \text{N})$ &
     \quad $<7 \times 10^{-13}$ \quad  (Au, SINDRUM~\cite{Bertl:2006up}) &
    \quad $10^{-14}$  \quad (SiC, DeeMe~\cite{Nguyen:2015vkk})    \\
    & &  \quad $2.6\times 10^{-17}$  \quad (Al, COMET~\cite{Krikler:2015msn,Adamov:2018vin,KunoESPP19})  \\
    & &  \quad $8 \times 10^{-17}$  \quad (Al, Mu2e~\cite{Bartoszek:2014mya})\\
    \hline
    \hline 
    $\mathrm{BR}(Z\to e^\pm\mu^\mp)$ & \quad$< 4.2\times 10^{-7}$\quad (ATLAS~\cite{Aad:2014bca}) & \quad$\mathcal O (10^{-10})$\quad (FCC-ee~\cite{Abada:2019lih}\\
    $\mathrm{BR}(Z\to e^\pm\tau^\mp)$ & \quad$< 5.2\times 10^{-6}$\quad (OPAL~\cite{Akers:1995gz}) & \quad$\mathcal O (10^{-10})$\quad (FCC-ee~\cite{Abada:2019lih}\\
    $\mathrm{BR}(Z\to \mu^\pm\tau^\mp)$ & \quad$< 5.4\times 10^{-6}$\quad (OPAL~\cite{Akers:1995gz}) & \quad $\mathcal O (10^{-10})$\quad (FCC-ee~\cite{Abada:2019lih}\\
     $\mathrm{BR}(h\to e^\pm\mu^\mp)$ & \quad$< 6.1\times 10^{-5}$\quad\cite{ParticleDataGroup:2020ssz}  & ---\\
    $\mathrm{BR}(h\to e^\pm\tau^\mp)$ & \quad$< 4.7\times 10^{-3}$\quad\cite{ParticleDataGroup:2020ssz}  & ---\\
    $\mathrm{BR}(h\to \mu^\pm\tau^\mp)$ & \quad$< 2.5\times 10^{-3}$\quad\cite{ParticleDataGroup:2020ssz}  & ---\\
    \hline
    \end{tabular}}
    \caption{Current experimental bounds and future sensitivities on cLFV observables considered in this work. All limits are given at $90\%\:\mathrm{C.L.}$, and the Belle II sensitivities correspond to an integrated luminosity of $50\:\mathrm{ab}^{-1}$.}
    \label{tab:cLFVdata}
\end{table}
\renewcommand{\arraystretch}{1.}

\renewcommand{\arraystretch}{1.3}
\begin{table}[h!]
\begin{center}
	\begin{tabular}{|c|c|c|}
	\hline
	Observable & Current bound & Future Sensitivity\\
	\hline
	\hline
	$\mathrm{BR}(\tau\to\pi e)$ & $<8\times10^{-8}$\quad Belle~\cite{Miyazaki:2007jp} & $<4\times10^{-10}\quad$ Belle II~\cite{Kou:2018nap}\\
	$\mathrm{BR}(\tau\to\pi \mu)$ & $<1.1\times10^{-7}$\quad Belle~\cite{Miyazaki:2007jp} & $<5\times10^{-10}\quad$ Belle II~\cite{Kou:2018nap}\\
	$\mathrm{BR}(\tau\to\phi e)$ & $<3.1\times10^{-8}$\quad Belle~\cite{Miyazaki:2011xe} & $<5\times10^{-10}\quad$ Belle II~\cite{Kou:2018nap}\\
	$\mathrm{BR}(\tau\to\phi \mu)$ & $<8.4\times10^{-8}$\quad Belle~\cite{Miyazaki:2011xe} & $<2\times10^{-9}\quad$ Belle II~\cite{Kou:2018nap}\\
	$\mathrm{BR}(\tau\to\rho e)$ & $<1.8\times10^{-8}$\quad Belle~\cite{Miyazaki:2011xe} & $<3\times10^{-10}\quad$ Belle II~\cite{Kou:2018nap}\\
	$\mathrm{BR}(\tau\to\rho \mu)$ & $<1.2\times10^{-8}$\quad Belle~\cite{Miyazaki:2011xe} & $<2\times10^{-10}\quad$ Belle II~\cite{Kou:2018nap}\\
	\hline
	$\mathrm{BR}(B^+\to K^+\tau^+e^-)$ & $<1.5\times10^{-5}$\quad BaBar~\cite{Lees:2012zz} & $<2.1\times10^{-6}\quad$ Belle II~\cite{Kou:2018nap}\\
	$\mathrm{BR}(B^+\to K^+\tau^-e^+)$ & $<4.3\times10^{-5}$\quad BaBar~\cite{Lees:2012zz} &\\
	$\mathrm{BR}(B^+\to K^+\tau^+\mu^-)$ & $<2.8\times10^{-5}$\quad BaBar~\cite{Lees:2012zz} & $<3.3\times10^{-6}\quad$ Belle II~\cite{Kou:2018nap}\\
	$\mathrm{BR}(B^+\to K^+\tau^-\mu^+)$ & $<4.5\times10^{-5}$\quad BaBar~\cite{Lees:2012zz} &\\
	\hline
	$\mathrm{BR}(B^0\to e^\pm\tau^{\mp})$ & $<2.8\times10^{-5}$\quad BaBar~\cite{Aubert:2008cu} & $ <1.6\times10^{-5}\quad$ Belle II~\cite{Kou:2018nap}\\
	$\mathrm{BR}(B^0\to \mu^\pm\tau^{\mp})$ & $<1.4\times10^{-5}$\quad LHCb~\cite{Aaij:2019okb} & $ <1.3\times10^{-5}\quad$ Belle II~\cite{Kou:2018nap}\\
	$\mathrm{BR}(B_s\to\mu^\pm\tau^{\mp})$ & $<4.2\times10^{-5}\quad$ LHCb~\cite{Aaij:2019okb} & --- \\
	$\mathrm{BR}(B^+\to K^+\tau^+\mu^-)$ & $< 2.8\times 10^{-5}\quad$ BaBar~\cite{Lees:2012zz} & $<3.3\times 10^{-6}\quad$ Belle II~\cite{Kou:2018nap}\\
	$\mathrm{BR}(B_s\to\phi\mu^\pm\tau^\mp)$ & $<4.3\times10^{-5}$\cite{Tanabashi:2018oca} & ---\\
	\hline
	$\text{BR}(K_L \to \mu^\pm e^\mp)$ & $< 4.7\times 10^{-12}\quad$~\cite{Tanabashi:2018oca} & ---\\
	\hline
	\end{tabular}
	\caption{Current upper bounds and future sensitivities (at $90\%$~C.L.) for semi-leptonic cLFV processes (decay channels) of particular interest for this thesis.}
	\label{tab:semicLFV}
\end{center}
\end{table}
\renewcommand{\arraystretch}{1.}

All these limits, regarding many cLFV observables, are by themselves impressive, and most of them are expected to be improved in the coming years at upcoming and future facilities.
Should a signal of New Physics be observed in a certain channel, the vast array of experimental searches will help to cross-check the observations and disentangle the underlying mechanism responsible for cLFV, and therefore identify the theoretical construction that could be behind it.
An example of this, concerning neutrino mass generation, will be the topic of the following chapter.
In Chapter~\ref{chap:lq} we will discuss how semi-leptonic cLFV processes might be paramount in cross-checking possible explanations of indirect hints of New Physics in $b\to s\ell\ell$ transitions.

\chapter{Massive neutrinos}
\label{sec:massivenu}
\minitoc

\noindent
The discovery of neutrino oscillations constituted the first irrefutable laboratory evidence of New Physics.
Moreover, massive neutrinos open the door to the violation of lepton flavour: by themselves, neutrino oscillations signal the violation of neutral lepton flavour and in the absence of a fundamental principle (imposed symmetry), any SM extension accommodating massive and mixing neutrinos is expected to also allow for charged lepton flavour violating processes, such as $\mu\to e\gamma$ decays.

As discussed in Section~\ref{sec:osci}, despite the vast worldwide experimental effort in determining the oscillation parameters of massive neutrinos, several open questions remain, suggesting that our understanding of the neutral lepton sector is far from complete.
First of all, being electrically neutral, neutrinos can be described as Dirac or Majorana fermions, meaning they could be their own anti-particle.
Furthermore, the absolute mass scale and the ordering of the light neutrino spectrum is still unknown. 

Measurements of the invisible $Z$-boson decay width furthermore confirm the existence of $3$ light neutral states (with masses smaller than the mass of the $Z$-boson), the 3 active neutrinos. Nevertheless, the existence of additional neutral fermions, the so-called sterile states (without SM gauge interactions), remains a viable and appealing possibility.
In particular, heavy neutral leptons (HNL) are often invoked in SM extensions that aim at accommodating oscillation data, and offer an appealing mechanism for neutrino mass generation.

\medskip
In this chapter we briefly review some phenomenological aspects of massive neutrinos.

\section{Neutrino mass generation}
\label{sec:numassgen}
In order to accommodate neutrino oscillation data, the SM has to be extended.
If one imposes lepton number conservation, neutrinos are Dirac fermions and we can minimally extend the SM field content by three right-handed  neutrinos $\nu_R$, This allows to directly write a Yukawa interaction term between the SM lepton doublet and $\nu_R$, $Y^\nu H \bar L_L \nu_R$, in full analogy to the other fermions (quarks and charged leptons).
After EWSB, this leads to a mass term of the form
\begin{equation}
    \mathcal L_\text{mass}^\text{Dirac} = \bar \ell_L m_\ell \ell_R + \bar \nu_L m_D\nu_R + \text{H.c.}\,,
\end{equation}
in which $m_\ell$ is the mass matrix of the charged leptons, $m_D = Y^\nu v/\sqrt{2}$ and $v$ is the SM Higgs vev.
Similarly to the quark sector, the charged lepton and neutrino mass terms can then be diagonalised by bi-unitary transformations
\begin{equation}
    m_\nu^\text{diag} = V_L^\nu m_D V_R^{\nu\dagger}\,,\quad m_\ell^\text{diag} = V_L^\ell m_\ell V_R^{\ell\dagger}\,,
\end{equation}
with the transformations from the interaction basis into the mass basis (denoted by $\hat{\phantom\ell}$) given by
\begin{equation}
    \hat\nu_{L,R} = V_{L,R}^\nu \nu_{L,R}\,,\quad \hat\ell_{L,R} = V_{L,R}^\ell \ell_{L,R}\,.
\end{equation}
In the mass basis, we can then define the physical Dirac spinor $\psi_\nu = \nu_L + \nu_R$ which fulfils the Dirac equation.
Consequently, the PMNS matrix is then given as $U_{\text{PMNS}} = V_L^{\ell\dagger}V_L^\nu$, or if we choose to work the weak basis in which the charged lepton Yukawa couplings are diagonal (as mentioned before), simply as $U_\text{PMNS} = V_L^\nu$.  

Although this ad-hoc extension provides a working explanation of the oscillation data, ensuring compatibility with experimental bounds on the absolute mass scale of neutrinos ($m_\nu\lesssim 0.1\,\mathrm{eV}$) would require the Yukawa couplings $Y^\nu$ to be extremely small, $Y^\nu\lesssim 10^{-12}$. This begs the question why there is such a huge hierarchy in the Yukawa couplings between the charged and neutral lepton sectors (or even worse, if one considers all fermions) and consequently raises the issue of naturalness.
A more problematic aspect however is that due to the true singlet nature of $\nu_R$ (no electric nor colour charge, and an $SU(2)_L$-singlet), the SM gauge symmetry in principle allows for a potentially large Majorana mass term of the form $m_{RR}\bar\nu_R\nu_R^c$. Unless a symmetry is enforced, such a term would lead to the violation of total lepton number $L$ by two units.

Despite its shortcomings, this Dirac neutrino ad-hoc SM extension can be advocated as adding only additional spin states to the SM field content; thus it is appealing due to its minimality.

\medskip
At the expense of either breaking gauge invariance or losing renormalisability, the SM field content allows for a Majorana mass term of the form $m_{LL}\overline{\nu_L}\nu_L^c$.
In contrast to fermions carrying a gauge charge, the spinors $\psi$ and $\psi^c$ of a neutral fermion to which no globally conserved charge is associated, do not necessarily correspond to different fields, but rather to different helicity states, and thus obey the same equation of motion.
This implies one could have $\psi = \psi^c$, which is commonly called the ``Majorana condition''.
A Majorana bispinor can then be constructed out of a single chiral component, $\psi_\text{M} = \psi_L + C\bar\psi_L^T$, giving rise to a mass term of the form
\begin{equation}
    \mathcal L_\text{mass}^\text{Majorana} = \frac{1}{2} m_\text{M} \left(\overline{\psi_L^c}\psi_L + \overline{\psi_L} \psi_L^c\right)\,.
\end{equation}
In principle, this type of mass term could be realised in Nature for neutrinos, since these are neutral particles; as such, neutrinoless double-beta decays and other LNV interactions would be possible. 
However, with the SM neutrinos, a mass term of the form $m_\text{M} \overline{\nu_L}\nu_L^c$ violates $SU(2)_L$ gauge invariance, since it transforms as an $SU(2)_L$ triplet.

\begin{figure}
    \centering
\mbox{    
\includegraphics[width=0.25\textwidth]{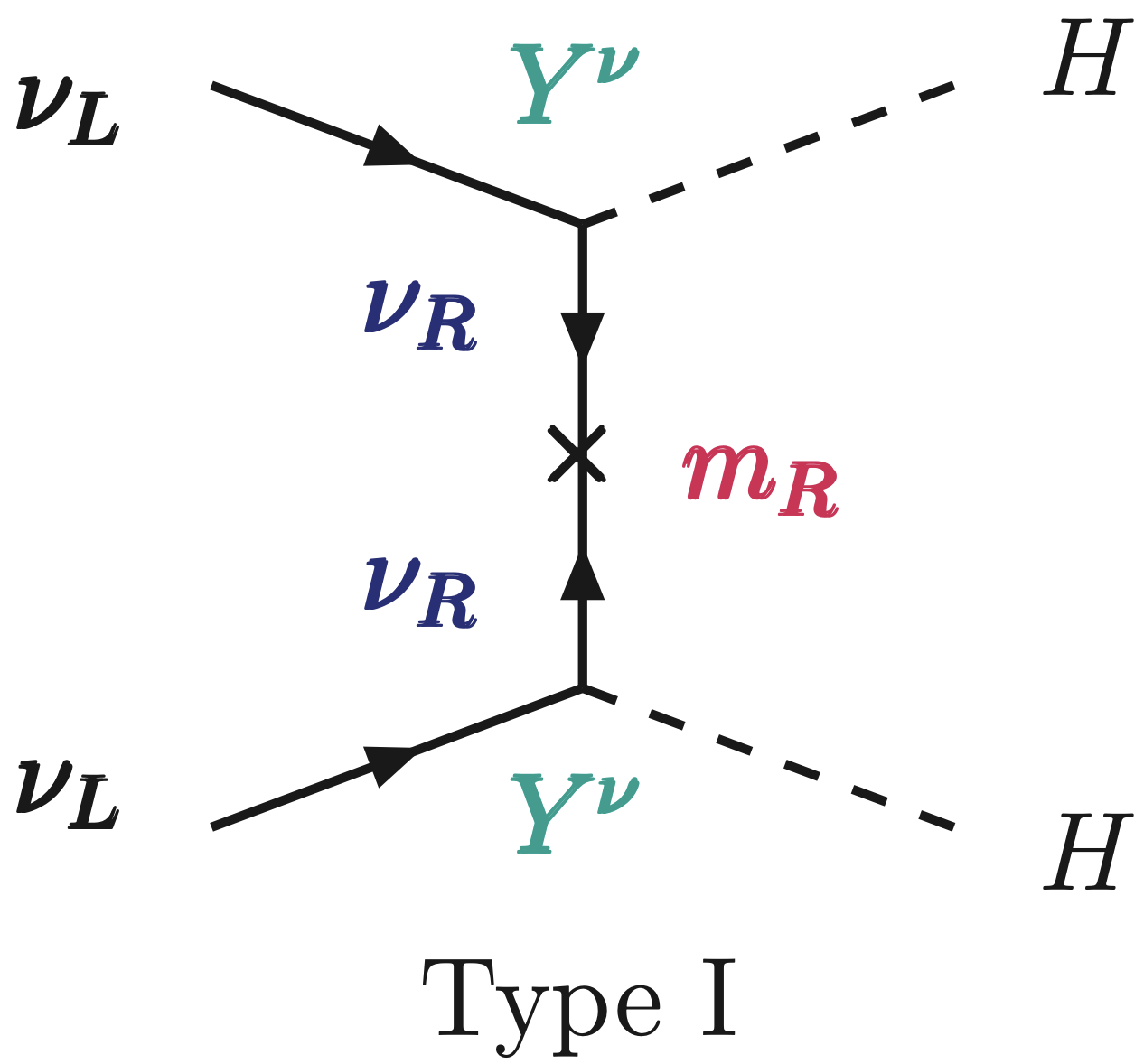}
\hspace{5mm}\includegraphics[width=0.35\textwidth]{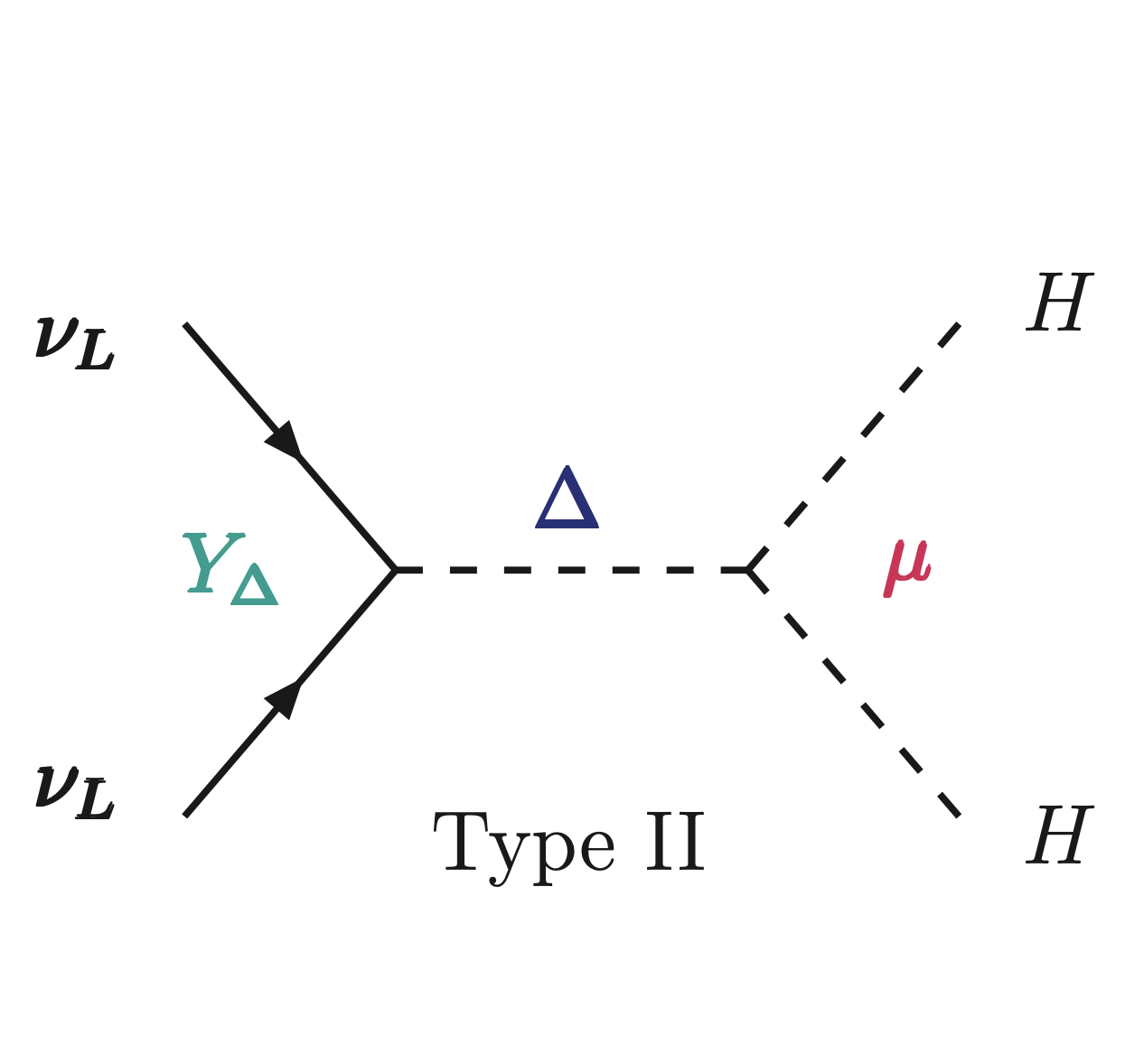}
\hspace{5mm}\includegraphics[width=0.28\textwidth]{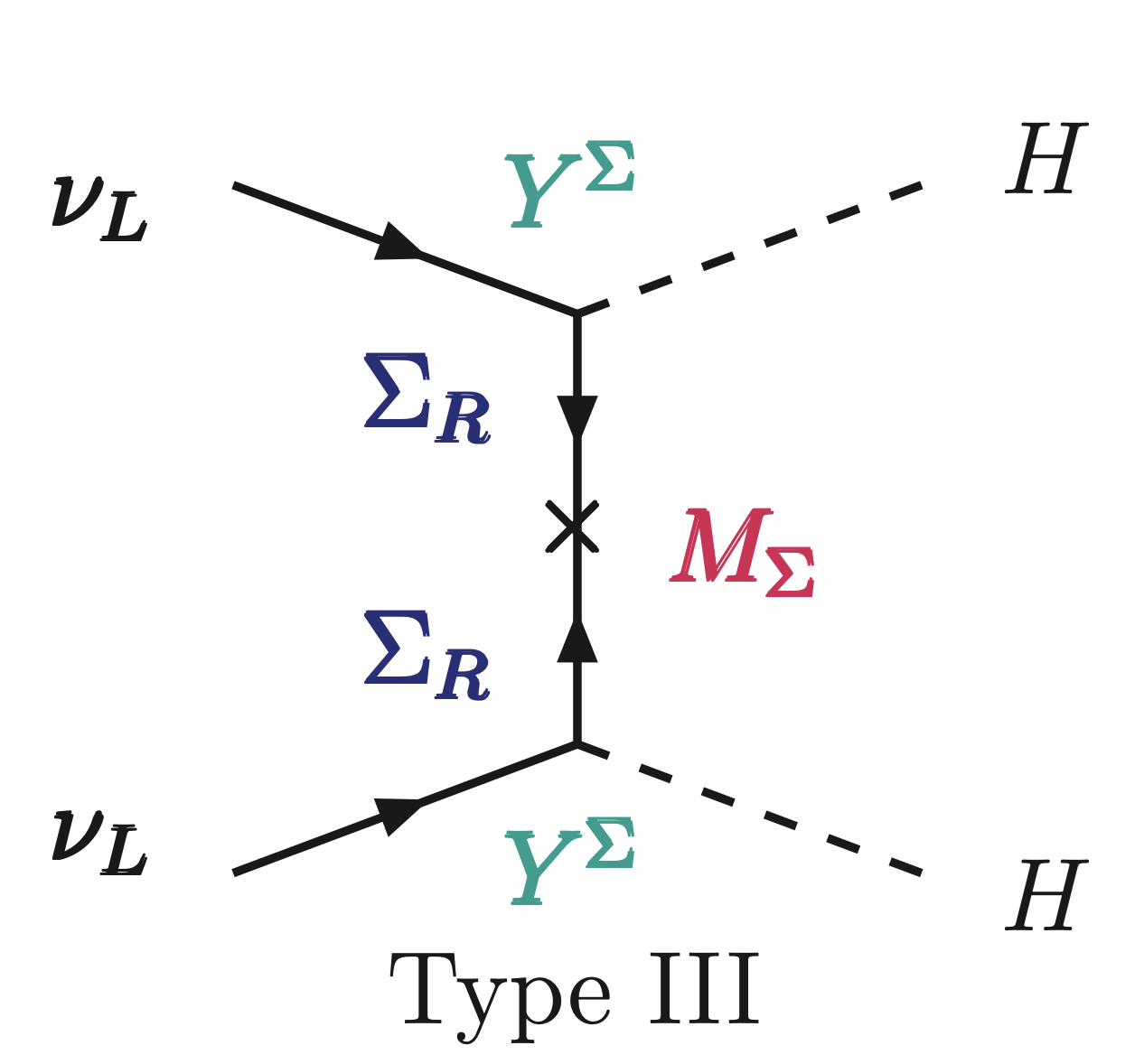}

}
    \caption{Tree-level realisations of the Weinberg operator represented by Feynman diagrams.
    From left to right we depict the type I, type II and type III seesaw mechanisms.}
    \label{fig:seesaws}
\end{figure}

Gauge invariance can be recovered if one assumes that this term arises from a non-renormalisable dimension-5 operator, the so-called Weinberg operator, which is the only gauge invariant dimension 5 operator that can be constructed out of SM fields (cf. Chapter~\ref{sec:eft}. 
It is given by
\begin{equation}
    \mathcal L_{d=5} = \frac{C_{ij}}{2\Lambda}(\overline{L_i^c}\widetilde H^\ast)(\widetilde H^\dagger L_j)\,,
\end{equation}
$\Lambda$ is the New Physics scale at which lepton number is broken.
Here, the Weinberg operator transforms under $SU(2)_L$ as a fermion singlet, suggesting that it can be generated at tree-level by singlet fermions such as RH neutrinos $\nu_R$, which is the case of a type I seesaw mechanism.
Gauge invariance allows for two additional realisations of the Weinberg operator, via scalar triplets or fermion triplets.
In Fig.~\ref{fig:seesaws} we illustrate the three seesaw types by the associated tree-level diagrams giving rise to a realisation of the Weinberg operator.
Assuming a tree-level realisation, this respectively leads to the type II~\cite{Barbieri:1979ag,Cheng:1980qt,Magg:1980ut,Lazarides:1980nt,Schechter:1980gr,Mohapatra:1980yp} (SM extensions via a scalar triplet) and to the type III~\cite{Foot:1988aq,Ma:1998dn} (SM extensions via a fermion tiplet) seesaw mechanisms.
Independent of the realisation, after EWSB, the Weinberg operator gives rise to an effective Majorana mass term for the left-handed neutrinos as
\begin{equation}
    \mathcal L_{d=5} = \frac{v^2 C_{ij}}{2 \Lambda} (\overline{\nu_{i L}^c} \nu_{L j}) + \text{H.c.}\,,
\end{equation}
where the suppression $\Lambda_\text{EW}/\Lambda$ is manifest. 
Depending on the specific realisation, the dimensionless coefficients $C_{ij}$ contain the full combination of couplings, loop factors etc..
If $C_{ij}\sim \mathcal O(1)$, compatibility with current experimental data implies for the New Physics scale $\Lambda \sim \mathcal O(10^{16})\:\mathrm{GeV}$, interestingly close to the GUT scale. This is the case of the ``vanilla type I seesaw mechanism''.
In fact, the first proposals of a type I seesaw were actually done in the framework of GUT $SO(10)$ models~\cite{Minkowski:1977sc,Yanagida:1979as,Glashow:1979nm,Gell-Mann:1979vob,Mohapatra:1979ia}.

However, depending on the underlying UV model, the effective couplings $C_{ij}$ can also be (very) small; be it due to loop suppression if the Weinberg operator is not realised at the tree-level, or due to arguments based on symmetry, which is the case of many low-scale seesaw variants, such as the Inverse 
Seesaw (ISS)~\cite{Schechter:1980gr,Gronau:1984ct,Mohapatra:1986bd},  
the Linear Seesaw (LSS)~\cite{Barr:2003nn,Malinsky:2005bi}  and the 
$\nu$-MSM~\cite{Asaka:2005an,Asaka:2005pn,Shaposhnikov:2008pf}.
Furthermore, neutrino masses can also be generated by higher-dimensional operators. For an exhaustive classification see e.g.~\cite{Gargalionis:2020xvt,Gargalionis:2019drk}.

As an illustrative example, we briefly review the high-scale type I seesaw mechanism and one of its many low-scale variants, the inverse seesaw mechanism; both rely on minimal SM extensions via right-handed neutrinos $\nu_R$ (and other fermion singlets), commonly referred to as heavy neutral leptons.

\subsection{Type I seesaw}
Since right-handed neutrinos $\nu_R$ are true gauge singlets (also referred to as ``sterile''), the SM symmetries allow for Yukawa interactions with the SM Higgs field and for a potentially large Majorana mass term. If we assign lepton number $L = +1$ to $\nu_R$, the Yukawa interaction with the SM Higgs field also preserves lepton number.
Furthermore, due to their singlet nature, the number of right-handed neutrinos $\nu_R$ is in principle unconstrained, since these do not contribute to gauge anomalies.
In particular, the number of right-handed states does not need to replicate the number of families of the SM.
However, for realistic models that accommodate experimental data, at least two right-handed states are necessary, one for each observed non-zero neutrino squared mass difference, thus implying $n_{\nu_R}\geq 2$.
In the following we will assume $n_{\nu_R} = 3$.
We can thus write the (type I seesaw) Lagrangian as
\begin{equation}
    \mathcal L_\text{mass}^\nu = -Y_\nu^{ij} \overline{L_i} \tilde H \nu_{Rj} - \frac{1}{2} m_R^{ij} \overline{\nu_{Ri}^c}\nu_{Rj} + \text{H.c.}\,,
\end{equation}
in which $Y_\nu$ is a complex Yukawa matrix and $m_R$ is a complex symmetric Majorana mass matrix, which can be taken as diagonal without loss of generality.
After EWSB, a Dirac mass term of the form $m_D = Y_\nu v/\sqrt{2}$ is generated.
The previous Lagrangian can be rewritten in matrix form,
\begin{equation}
    \mathcal L_\text{mass}^\nu = -\frac{1}{2} \left(\overline{\nu_L^c}, \overline{\nu_R} \right) \begin{pmatrix}\mathbb{0} & m_D^T \\ m_D  & m_R\end{pmatrix} \begin{pmatrix}\nu_L^c\\\nu_R\end{pmatrix} \equiv -\frac{1}{2}\left(\overline{\nu_L^c}, \overline{\nu_R} \right) M_\nu\begin{pmatrix}\nu_L^c\\\nu_R\end{pmatrix}\,,
\end{equation}
with $\nu_R = \left(\nu_{R1}, \nu_{R2}, \nu_{R3}\right)$ and $\nu_L = \left(\nu_e, \nu_\mu, \nu_\tau\right)_L$.
We stress here that the indices of the right-handed neutrinos merely correspond to generation indices and not to flavour.
If $m_R \gg m_D$, we can perturbatively block-diagonalise $M_\nu$ via a unitary transformation $\mathcal V$.
At leading order\footnote{For higher order terms in the expansion and their phenomenological impact, see e.g.~\cite{Grimus:2000vj,Hettmansperger:2011bt}.} in $m_D/m_R$ we then have
\begin{equation}
    \mathcal V^T\, M_\nu \,\mathcal V \simeq \begin{pmatrix}- m_D^T m_R^{-1} m_D & \mathbb{0} \\ \mathbb{0} & m_R\end{pmatrix}\,,
\end{equation}
where the transformation is given (also at leading order) by
\begin{equation}
    \mathcal V \simeq \begin{pmatrix} \mathbb{1} - \frac{1}{2} B B^\dagger & B\\ -B^\dagger & \mathbb{1} - \frac{1}{2} B^\dagger B \end{pmatrix}\,,
\end{equation}
in which $B = m_D^\dagger (m_R^{-1})^\ast$ (at leading order).
In the so-called seesaw limit (i.e. $m_R \gg m_D$), the light (Majorana) neutrino mass matrix is consequently given by 
\begin{equation}
m_\nu \simeq -m_D^T m_R^{-1} m_D\,,
\label{eqn:seesawI}
\end{equation}
which is subsequently diagonalised by the unitary matrix $U_\text{PMNS}$ (with $V_L^\ell = \mathbb{1}$) as
\begin{equation}
    m_\nu^\text{diag} = - U_\text{PMNS}^T \,m_D^T m_R^{-1} m_D \,U_\text{PMNS}\,.
\end{equation}
The coefficient $C_{ij}$ of the Weinberg operator is then given by a combination of the (Dirac) Yukawa couplings and the Majorana mass term.
As one can see from the mass term of the light neutrinos (cf. Eq.~\eqref{eqn:seesawI}), the light neutrino masses are ``suppressed'' by the (heavy) masses of the right-handed neutrinos, thus potentially allowing for natural values of the Yukawa couplings, $Y_\nu\sim\mathcal O(1)$.
This also suggests that the scale of New Physics, which is also the scale of lepton number violation, is $\Lambda \simeq m_R\simeq 10^{16}\:\mathrm{GeV}$.

Assuming that $m_R$ is diagonal, the full diagonalisation matrix is thus given by
\begin{equation}
    \mathcal U = \mathcal V \begin{pmatrix}U_\text{PMNS} & \mathbb{0}\\ \mathbb{0} & \mathbb{1}\end{pmatrix}\,.
\end{equation}
As before (cf. Eq.~\eqref{eqn:wlnu}), the mixing also leads to a modification of the charged weak current which can be cast in the mass basis as
\begin{equation}
    \mathcal{L}_{W^\pm}\, =\, -\frac{g_w}{\sqrt{2}} \, W^-_\mu \,
\sum_{\alpha=1}^{3} \sum_{j=1}^{6} \mathcal{U}_{\alpha j} \bar \ell_\alpha 
\gamma^\mu P_L \nu_j \, + \, \text{H.c.}\,,
\end{equation}
assuming that the charged lepton Yukawa couplings are diagonal.
In the above equation $j = 1,...,6$ corresponds to the neutrino mass eigenstates, while $\alpha=1,2,3$ (or $\alpha=e, \mu,\tau$) to the charged lepton flavours.
Thus, the phenomenologically relevant part of $\mathcal U$ is encoded in the upper $3\times6$ block of $\mathcal U$ 
\begin{equation}
\left((\mathbb{1} - \frac{1}{2}B B^\dagger)\,U_\text{PMNS}, B\right) \equiv \left(\tilde U_\text{PMNS}, B\right)\,.
\end{equation}
Consequently, the left $3\times3$ block that describes the mixing between the mostly active (SM-like) neutrinos, $\tilde U_\text{PMNS}$, is no longer unitary.
Notice that for $m_R\gg m_D$ the heavy states effectively decouple and unitarity is restored, that is $\tilde U_\text{PMNS} \simeq U_\text{PMNS}$.
Constraints on the unitarity of $\tilde U_\text{PMNS}$ will be discussed in Section~\ref{sec:unitaritynu}.

\medskip
In order to ensure compatibility with neutrino oscillation data, a convenient way to parametrise the neutrino Yukawa couplings is the so-called ``Casas-Ibarra'' parametrisation~\cite{Casas:2001sr}.
The Dirac masses can be cast as (assuming $m_R = m_R^\text{diag}$)
\begin{equation}
    Y_\nu v = m_D = i \sqrt{m_R^\text{diag}} R\sqrt{m_\nu^\text{diag}} U_\text{PMNS}^\dagger\,,
    \label{eqn:ynuci}
\end{equation}
in which the complex orthogonal matrix $R$ can be parametrised as
\begin{equation}
    R = \begin{pmatrix} c_2 c_3 & -c_1 s_3 - s_1 s_2 c_3 & s_1 s_3 - c_1 s_2 c_3\\
                        c_2 s_3 & c_1 c_3 - s_1 s_2 s_3 & -s_1 c_3 - c_1 s_2 s_3\\
                        s_2 & s_1 c_2 & c_1 c_2
        \end{pmatrix}\,,
        \label{eqn:Rmatrix}
\end{equation}
with $c_i \equiv \cos\theta_i\,,\: s_i \equiv \sin \theta_i$, and $\theta_i$ are arbitrary complex angles.

The complex orthogonal matrix $R$ is a priori not constrained by oscillation data; however, it is of paramount importance in processes in which the heavy neutrinos contribute directly, either virtually, or as initial/final state particles.

\subsection{Inverse seesaw}
\label{sec:ISSintro}
The Inverse Seesaw mechanism (ISS)~\cite{Schechter:1980gr,Gronau:1984ct,Mohapatra:1986bd} is a particularly appealing SM extension via right-handed neutrinos and additional sterile states.
Contrary to the ``vanilla'' type I seesaw mechanism, it allows to accommodate massive neutrinos (and neutrino oscillation data) with natural values of the Yukawa couplings for comparatively low masses of the additional fermions, and low scales of lepton number violation.
In the ISS, $n_R\geq 2$ generations of RH neutrinos $\nu_R$ and $n_X$ generations\footnote{The number of RH and sterile neutrinos can in principle be different from each other (notice however that not all combinations successfully allow to accommodate oscillation data), leading to an interesting phenomenology with possible connections to dark matter and leptogenesis. For a systematic study of minimal inverse seesaw scenarios see~\cite{Abada:2014vea,Abada:2014zra,Abada:2015rta,Abada:2017ieq}.} of extra $SU(2)_L$-singlet fermions $X$, both carrying lepton number $L=+1$, are added to the field content of the SM.
For simplicity, here we focus on (3,3) ISS realisations, corresponding to $n_R = n_X = 3$ generations of extra fermions.
Both the fields $X$ and $\nu_R$ can acquire a Majorana mass and the specific assignment of $L$ to $\nu_R$ and $X$ furthermore allows for gauge invariant and lepton number conserving Yukawa couplings between $\nu_L$ and $\nu_R$.
We can write the ISS Lagrangian as
\begin{equation}
\mathcal L_\text{ISS} = -Y_\nu^{ij} \,\overline{L_i^c}\,\widetilde H \,\nu_{Rj}^c - m_R^{ij}\, \overline{\nu_{Ri}}\, X_j - \frac{1}{2}\mu_R^{ij}\, \overline{\nu_{Ri}^c}\,\nu_{Rj} - \frac{1}{2} \mu_X^{ij}\, \overline{X_i^c}\, X_j + \text{H.c.}\,,
\end{equation}
in which only the terms proportional to $\mu_X$ and $\mu_R$ break lepton number.
In the limit in which $\mu_{X,R}\to 0$, lepton number is restored, and the active neutrinos remain massless to all orders in perturbation theory.
Consequently, since in this limit the symmetry of the Lagrangian is enhanced (lepton number conservation is recovered), small values for both $\mu_X$ and $\mu_R$ are technically natural in the sense of 't Hooft~\cite{tHooft:1979rat}.
After EWSB, analogously to the type I seesaw, a Dirac mass term of the form $m_D = Y_\nu v/\sqrt{2}$ is generated.
Again, we can rewrite the Lagrangian in matrix form and obtain
\begin{equation}
    \mathcal L_\text{ISS} = -\frac{1}{2} \left(\overline{\nu_L^c}, \overline{\nu_R}, \overline{X^c}\right) 
    \begin{pmatrix}
        \mathbb{0} & m_D & \mathbb{0}\\
        m_D^T & \mu_R & m_R\\
        \mathbb{0} & m_R^T & \mu_X
    \end{pmatrix}
    \begin{pmatrix}
    \nu_L\\\nu_R^c\\X
    \end{pmatrix}\,.
\end{equation}
We note here that $\mu_R$ does not contribute to the active neutrino masses at leading order, and we therefore neglect it in the subsequent discussion.
In the limit of $\mu_X \ll m_D\ll m_R$, we can perturbatively diagonalise the ($9\times 9$) mass matrix which leads to an approximate expression for the 3 light (mostly active) neutrino masses given by
\begin{equation}
    m_\nu \simeq  m_D \, \left( m_{R}^{-1} \right)^{T} \, \mu_X \, m_{R}^{-1} \, m_D^T\, \equiv U_\text{PMNS}^\ast\: m_\nu^\text{diag}\: U_\text{PMNS}^\dagger.
    \label{eqn:mnuISS}
\end{equation}
By introducing a new $3\times3$ mass matrix as
\begin{equation}
	M \,= \,m_R \,\mu_X^{-1}\,m_R^T\,,
\end{equation}
one recovers an expression for the light neutrino masses 
strongly resembling that of the type I seesaw model, 
\begin{equation}
	m_\nu \simeq m_D\, M^{-1}\, m_D^T\,.
\end{equation}
The heavier 6 states form heavy pseudo-Dirac pairs with masses $\propto M_R \pm \mu_X$, such that the small lepton number breaking parameter $\mu_X$ controls the smallness of the active neutrino masses and the non-degeneracy (or mass splitting) of the heavy states.
From Eq.~\eqref{eqn:mnuISS} it can be seen that, in order to accommodate the light neutrino masses, two scales are relevant - the small lepton number breaking scale $\mu_X$ and the heavy scale $m_R$.
For illustrative purposes, let us notice that assuming natural Yukawa couplings $Y_\nu\sim 1$ and TeV-scale masses $m_R\sim 1\:\mathrm{TeV}$, one is led to a lepton number breaking scale of order $1\:\mathrm{eV}$.

In order to accommodate oscillation data, several useful parametrisations are possible.
Firstly, we can consider a modified Casas-Ibarra parametrisation~\cite{Casas:2001sr}, thus encoding the flavour structure of the active neutrinos in the Yukawa couplings $Y_\nu$.
The Dirac mass term can then be written as
\begin{equation}
    m_D^T = V^\dagger\sqrt{M^\text{diag}}R\sqrt{m_\nu^\text{diag}} U_\text{PMNS}^\dagger\,,
    \label{eqn:ISSCI}
\end{equation}
where $V$ is a unitary matrix that diagonalises $M = V^\dagger M^\text{diag}V^\ast$ and $R$ is a complex orthogonal matrix as given in Eq.~\eqref{eqn:Rmatrix}.
Another parametrisation has been suggested in~\cite{Arganda:2014dta}, in which the flavour structure of the light sector is encoded in $\mu_X$, leading to
\begin{equation}
    \mu_X = m_R^T m_D^{-1} U_\text{PMNS}^\ast m_\nu^\text{diag} U_\text{PMNS}^\dagger (m_D^T)^{-1} m_R\,,
\end{equation}
assuming that the Dirac mass matrix $m_D$ is invertible.
In the most simple case, both $m_R$ and $m_D$ can be chosen to be diagonal (as emphasised in~\cite{Arganda:2014dta}, non-minimal textures in $m_D$ can have significant impact on the associated phenomenology). 

A scenario like the ISS is extremely appealing from a phenomenological point of view, since sizeable Yukawa couplings and comparatively low masses of the heavy neutral leptons lead to a very rich (flavour) phenomenology.
In turn, such realisations are also in general subject to abundant experimental constraints.

\section{SM extensions via heavy neutral leptons: modified currents}

As previously mentioned, in models featuring heavy neutral leptons that mix with the (mostly active) light neutrinos, the standard PMNS mixing scheme is modified.
The modified leptonic mixings 
are consequently parametrised by a $(3+n_S)\times(3+n_S)$ unitary mixing matrix, $\mathcal U$;  its upper left $3\times3$ block corresponds to the left-handed leptonic mixing matrix, the would-be PMNS, $\tilde{U}_\text{PMNS}$. 
We can $\tilde U_\text{PMNS}$ as
\begin{equation}
    U_\text{PMNS} \to \tilde U_\text{PMNS} = (\mathbb{1} - \eta)U_\text{PMNS}\,,
    \label{eqn:Utildeeta}
\end{equation}
in which the matrix $\eta$ contains the deviation from unitariy.
The ensuing non-unitarity of 
$\tilde{U}_\text{PMNS}$ and the enlargement of the mixing matrix will lead to modified charged and neutral currents, which can be cast in the physical basis as
\begin{align}\label{eq:lagrangian:WGHZ}
& \mathcal{L}_{W^\pm}\, =\, -\frac{g_w}{\sqrt{2}} \, W^-_\mu \,
\sum_{\alpha=1}^{3} \sum_{j=1}^{3 + n_S} \mathcal{U}_{\alpha j} \bar \ell_\alpha 
\gamma^\mu P_L \nu_j \, + \, \text{H.c.}\,, \nonumber \\
& \mathcal{L}_{Z^0}^{\nu}\, = \,-\frac{g_w}{2 \cos \theta_w} \, Z_\mu \,
\sum_{i,j=1}^{3 + n_S} \bar \nu_i \gamma ^\mu \left(
P_L {C}_{ij} - P_R {C}_{ij}^* \right) \nu_j\,, \nonumber \\
& \mathcal{L}_{Z^0}^{\ell}\, = \,-\frac{g_w}{4 \cos \theta_w} \, Z_\mu \,
\sum_{\alpha=1}^{3}  \bar \ell_\alpha \gamma ^\mu \left(
\mathbf{C}_{V} - \mathbf{C}_{A} \gamma_5 \right) \ell_\alpha\,, \nonumber \\
& \mathcal{L}_{H^0}\, = \, -\frac{g_w}{2 M_W} \, H  \,
\sum_{i\ne j= 1}^{3 + n_S}  {C}_{ij}  \bar \nu_i\left(
P_R m_i + P_L m_j \right) \nu_j + \, \text{H.c.}\ , \nonumber \\
& \mathcal{L}_{G^0}\, =\,\frac{i g_w}{2 M_W} \, G^0 \,
\sum_{i,j=1}^{3 + n_S} {C}_{ij}  \bar \nu_i  
\left(P_R m_j  - P_L m_i  \right) \nu_j\,+ \, \text{H.c.}, \nonumber  \\
& \mathcal{L}_{G^\pm}\, =\, -\frac{g_w}{\sqrt{2} M_W} \, G^- \,
\sum_{\alpha=1}^{3}\sum_{j=1}^{3 + n_S} \mathcal{U}_{\alpha j} 
\bar \ell_\alpha\left(
m_i P_L - m_j P_R \right) \nu_j\, + \, \text{H.c.}\,, 
\end{align}
with $n_S$ being the number of sterile states in the neutrino spectrum, and in which we recall that  
\begin{equation}
    {C}_{ij} = \sum_{\rho = 1}^3
  \mathcal{U}_{i\rho}^\dagger \,\mathcal{U}_{\rho j}^{\phantom{\dagger}}\:. 
  \label{eqn:Cij}
\end{equation}
In the above, the indices 
$\alpha, \rho = 1, \dots, 3$ denote the flavour of the charged leptons, while $i, j = 1, \dots, 3+n_S$ correspond to the physical (massive) 
neutrino states; 
as before, $g_w$ denotes the weak coupling constant, and
$\cos^2 \theta_w =  M_W^2 /M_Z^2$.
The coefficients $\mathbf{C}_{V}$ and $\mathbf{C}_{A}$ 
parametrise the SM vector and axial-vector currents 
for the interaction of neutrinos with charged leptons, respectively given by 
$\mathbf{C}_{V} = \frac{1}{2} + 2 \sin^2\theta_w$ and 
$\mathbf{C}_{A} = \frac{1}{2}$.

An immediate constraint of theoretical nature can be derived by imposing that decays of the HNL comply with perturbative unitarity~\cite{Chanowitz:1978mv,Durand:1989zs,Korner:1992an,Bernabeu:1993up,Fajfer:1998px,Ilakovac:1999md}, which gives a direct bound on their decay width $\frac{\Gamma_{\nu_i}}{m_{\nu_i}} < \frac{1}{2} (i \geq 4)$. 
Since the dominant contribution arises from $W$-boson exchanges, one can obtain a bound on the sterile masses and their couplings to active states, which can be written as
\begin{equation}
    m_{\nu_i}^2 C_{ii} < 2\ \frac{M_W^2}{\alpha_w}\quad(i\geq4)\,.
\end{equation}

Furthermore, the active-sterile mixings and the departure from unitarity of $\tilde U_\text{PMNS}$ can have an impact on several observables, inducing deviations from SM predictions, such as the violation of lepton flavour universality, enhanced charged lepton flavour violating processes and new contributions to many other precision observables and collider processes.
This leads to (indirect) bounds on the entries of $\eta$, particularly on the diagonal elements, which we proceed to discuss.

\section{Constraints on heavy neutral leptons from precision observables}
The modification of the interaction Lagrangian and therefore of the weak interaction vertices can consequently lead to a breaking of lepton flavour universality, lepton flavour conservation and lepton number conservation, all of which are accidental symmetries of the SM, so that any experimental measurement signalling their breaking is a clear hint of New Physics.
Thus far, no signal has been observed.
The null results and the experimental bounds (discussed in the previous chapter) in turn lead to tight constraints on models with heavy neutral leptons.
\subsection{Lepton flavour universality}

In the presence of HNL, at leading order, the modified decay rate of the $W$-boson can be written as~\cite{Abada:2013aba} 
\begin{eqnarray}
    \Gamma(W\to\ell_\alpha\nu) = \sum_{j=1}^{N_{\text{max}}^{(\ell_\alpha)}} \frac{\lambda^{\frac{1}{2}}(M_W, m_{\ell_\alpha}, m_{\nu_j})}{48 \pi M_W}\sqrt{8} G_F \left|\mathcal U_{\alpha j}\right|^2 \left(2 M_W^2 - m_{\ell_\alpha}^2 - m_{\nu_j}^2 - \frac{(m_{\ell_\alpha}^2 - m_{\nu_j}^2)^2}{M_W^2}\right)\,,
\end{eqnarray}
in which the kinematical function $\lambda(a, b, c)$ is defined as
\begin{equation}
    \lambda(a, b, c) = (a^2 - b^2 - c^2)^2 - 4b^2c^2
    \label{eqn:kallenlambda}
\end{equation}
and $N_{\text{max}}^{(\ell_\alpha)}$ denotes the heaviest neutrino that is kinematically allowed as a final state, depending on the charged lepton $\ell_\alpha$.

\medskip
Apart from the $W\to\ell\nu$ decays, weak leptonic decays of charged pseudo-scalar mesons are sensitive to a modified $W\ell\nu$ vertex.
Their decay rates can be cast (at leading order) as~\cite{Abada:2013aba}
\begin{equation}
    \Gamma(P\to\ell_\alpha\nu) = \sum_{j=1}^{N_{\text{max}}^{(\ell_\alpha)}}\frac{G_F^2 f_{P}^2}{8 \pi m_P^3}\left|\mathcal U_{\alpha j}\right|^2|V_{q_u q_d}^\text{CKM}|^2\lambda^{\frac{1}{2}}(m_P, m_{\ell_\alpha}, m_{\nu_j})\left[m_P^2(m_{\nu_j}^2 + m_{\ell_\alpha}^2) - (m_{\ell_\alpha}^2 - m_{\nu_j}^2)^2\right]\,,
    \label{eqn:Pellnu}
\end{equation}
where $f_P$ and $m_P$ are the decay constant and the mass of the meson $P$, and $V_{q_u q_d}^\text{CKM}$ is the CKM element relevant in view of the quark content of the meson.
Since the decay constant $f_P$ is plagued by hadronic uncertainties and the CKM elements lead to further uncertainties, it is useful to construct ratios of decay widths sensitive to LFUV, so that the hadronic uncertainties and CKM elements approximately cancel.
We can for instance consider the ratios 
\begin{equation}
    R_P \equiv \frac{\Gamma(P^+ \to \ell_{\alpha}^+\nu)}{\Gamma(P^+\to \ell_{\beta}^+\nu)}\,,
    \label{eqn:RP2}
\end{equation}
with $m_{\ell_\beta}>m_{\ell_\alpha}$. 
At tree-level, the expression for $R_P$ in the SM extended by sterile neutrinos is given by~\cite{Abada:2012mc}
\begin{equation}
    R_P = \frac{\sum_{j=1}^{N_{\text{max}}^{(\ell_\alpha)}} F^{\alpha j} G^{\alpha j}}{\sum_{k=1}^{N_{\text{max}}^{(\ell_\beta)}} F^{\beta k} G^{\beta k}}
\end{equation}
with 
\begin{eqnarray}
    F^{\alpha j} = |\mathcal U_{\alpha j}|^2 \quad\text{and}\quad G^{\alpha j} = \left[m_P^2(m_{\nu_j}^2 + m_{\ell_\alpha}^2) - (m_{\nu_j}^2 - m_{\ell_\alpha}^2)^2\right]\lambda^{\frac{1}{2}}(m_P, m_{\ell_\alpha}, m_{\nu_j})\,.
\end{eqnarray}
Nevertheless, higher order radiative corrections, that are different for each lepton flavour, are not incorporated in Eq.~\eqref{eqn:Pellnu}, so it is further useful to consider the deviation of Eq.~\eqref{eqn:RP2} from the SM predictions, which in some cases have been obtained with very high precision~\cite{Cirigliano:2007xi}.
As mentioned in Chapter~\ref{chap:lepflav} (see Eq.~\eqref{eqn:deltarp}), it therefore proves convenient to parametrise the deviation from the SM prediction as
\begin{equation}
    R_P = R_P^\text{SM}(1 + \Delta r_P)\,.
\end{equation}
In the limit of vanishing neutrino masses and no lepton mixing (i.e. $m_{\nu_j} = 0$ and $\mathcal U_{\alpha j} = \delta_{\alpha j}$) we recover the SM tree-level prediction for $R_P$ (e.g. for $\ell_\alpha = e$ and $\ell_\beta = \mu$) as
\begin{equation}
    R_P^{\text{SM}} = \frac{m_e^2(m_P-m_e^2)^2}{m_\mu^2(m_P^2 - m_\mu^2)^2}\,,
\end{equation}
to which (small) electromagnetic corrections, that account for effects such as internal bremsstrahlung and structure-dependence, need to be added~\cite{Cirigliano:2007xi}.
As can be seen, there is a strong helicity suppression induced by the charged lepton masses ($R_P^\text{SM}\propto \frac{m_e^2}{m_\mu^2}$) rendering this ratio very sensitive to the presence of New Physics.
The general expression of $\Delta r_P$ is then given by
\begin{equation}
    \Delta r_P = \frac{m_\beta^2(m_P-m_\beta^2)^2}{m_\alpha^2(m_P^2 - m_\alpha^2)^2}\frac{\sum_{j=1}^{N_{\text{max}}^{(\ell_\alpha)}} F^{\alpha j} G^{\alpha j}}{\sum_{k=1}^{N_{\text{max}}^{(\ell_\beta)}} F^{\beta k} G^{\beta k}}\,.
\end{equation}
As can be seen, $\Delta r_P$ can significantly deviate from $0$, either due to the presence of additional neutrinos in the final state or due to a deviation from unitarity of the SM-like would-be PMNS mixing matrix.
The latter is the case if additional sterile states have non-negligible mixings with the mostly active neutrinos but are too heavy to be kinematically allowed as a final state. 

\medskip
Another (simple) tree-level process sensitive to LFU is the $\tau\to\ell_{\alpha}\nu\bar\nu$ decay.
Also here, we can define a ratio of decay widths,
\begin{equation}
    R_\tau \equiv \frac{\Gamma(\tau^-\to \mu^-\nu\bar\nu)}{\Gamma (\tau^-\to e^-\nu\bar\nu)}\,. 
    \label{eqn:Rtau2}
\end{equation}
The individual decay widths, in the SM extended by sterile fermions, and under the assumption of Majorana neutrinos, can be cast as~\cite{Abada:2013aba}
\begin{eqnarray}
    \Gamma(\ell_\beta\to \ell_\alpha \nu\bar\nu) = \sum_{i = 1}^{N_\text{max}^{(\ell_\alpha)}}\sum_{j = 1}^i \Gamma_{ij}\,,
\end{eqnarray}
with
\begin{eqnarray}
    \Gamma_{ij} &=& \frac{G_F^2 (2 - \delta_{ij})}{m_{\ell_\beta}^3(2 \pi)^3}\int_{(m_{\ell_\alpha} + m_{\nu_i})^2}^{(m_{\ell_\beta} + m_{\nu_j})^2} \mathrm d s_{\alpha i} \left[\frac{1}{4}|\mathcal U_{\beta i}|^2 |\mathcal U_{\alpha j}|^2 (s_{\alpha i} - m_{\ell_\alpha}^2 - m_{\nu_i}^2)(m_{\ell_\beta}^2 + m_{\nu_j}^2 - s_{\alpha i})\right.\nonumber\\
    &+&\left. \frac{1}{2} \mathrm{Re}(\mathcal U_{\beta i}^\ast\,\mathcal U_{\alpha j} \,\mathcal U_{\beta j}\,\mathcal U_{\alpha i}^\ast) m_{\nu_i} m_{\nu_j}\left(s_{\alpha i} - \frac{m_{\nu_i}^2 + m_{\nu_j}^2}{2} \right) \right]\nonumber\\
    &\times& \frac{1}{s_{\alpha i}}\sqrt{(s_{\alpha i} - m_{\ell_\alpha}^2 - m_{\nu_i}^2)^2 - 4 m_{\nu_i}^2m_{\ell_\alpha}^2}\sqrt{(m_{\ell_\beta}^2 + m_{\nu_j}^2 - s_{\alpha i})^2 - 4 m_{\nu_j}^2 m_{\ell_\alpha}^2}\nonumber\\
    &+& i \leftrightarrow j\,,
    \label{eqn:llnunu}
\end{eqnarray}
in which the Dalitz variable is defined as $s_{\alpha i} = (p_{\ell_\alpha} + p_{\nu_i})^2$ and $p_{\ell_\alpha}$, $p_{\nu_i}$ are the corresponding momenta of the charged lepton $\ell_\alpha$ and neutrino $\nu_i$.
In addition to the total decay width, several other observables in the charged lepton decays, as is the case of angular distributions, can be constructed. For a comprehensive review see~\cite{Pich:2013lsa}.
Here, we only focus on the LFU ratio defined in Eq.~\eqref{eqn:Rtau2}.

\medskip
Heavy neutal leptons also contribute to semi-leptonic decays of the $\tau$-lepton and further ratios sensitive to New Physics effect can be used to constrain HNL.
For a comprehensive overview and phenomenological impact see for instance~\cite{Abada:2013aba}.

\subsection{Electroweak precision observables}
The addition of (fermion) singlets to the SM with a sizeable active-sterile mixing can affect electroweak precision observables at tree-level (charged currents) and at higher order.
In particular, the non-unitarity of the would be PMNS matrix, $\tilde U_\text{PMNS}$, implies that the couplings to the $W$- and $Z$-bosons are suppressed with respect to their SM values. 
This has drastic implications on several precision observables; in particular on the invisible $Z$-decay width, the value of the Fermi constant $G_F$, the mass of the $W$ boson and the weak mixing angle $\sin^2\theta_w$.

The comparison of the SM prediction of the invisible $Z$ decay width to the LEP measurement~\cite{ParticleDataGroup:2020ssz},
\begin{eqnarray}
    \Gamma_\text{SM}(Z\to \nu\bar\nu) &=& (501.69 \pm 0.06)\:\mathrm{MeV}\,,\\
    \Gamma_\text{exp}(Z\to \nu\bar\nu) &=& (499.0 \pm 1.5)\:\mathrm{MeV}\,,
\end{eqnarray}
suggests that the experimental value is $\sim2\,\sigma$ below the theoretical expectation of the SM.
In the presence of massive Majorana neutrinos the modified decay width is given by~\cite{Abada:2013aba}
\begin{eqnarray}
    \Gamma(Z\to \nu\bar\nu) &=& \sum_{i,j = 1}^{N_\text{max}} (1-\delta_{ij})\sqrt{2}G_F\frac{\lambda^{\frac{1}{2}}(M_Z, m_{\nu_i}, m_{\nu_j})}{48\pi M_Z}\Bigg[ 2 |C_{ij}|^2\left(2M_Z^2 - m_{\nu_i}^2 - m_{\nu_j}^2 - \frac{(m_{\nu_i}^2 - m_{\nu_j}^2)^2}{M_Z^2}\right)\nonumber\\
    &\phantom{=}&  -12m_{\nu_i}m_{\nu_j} \mathrm{Re}(C_{ij}^2)\Bigg]\,,
\end{eqnarray}
in which $N_\text{max}$ denotes the heaviest neutrino state that is kinematically allowed as a final state. 
If now the masses of the sterile states are sufficiently low, they might contribute positively to the width;
however, for masses above the electroweak scale, the contribution is generically negative leading to a reduction of the $Z$-boson width.
In either case, the invisible $Z$ width provides a very important constraint.

Furthermore, the presence of (heavy) sterile states indirectly alters the ``usual'' muon decay $\mu\to e \bar\nu_e \nu_\mu$.
The width is given by~\cite{Fernandez-Martinez:2016lgt}
\begin{equation}
    \Gamma_\mu = \frac{m_\mu^5 G_F^2}{192\pi^3}\sum_{i} |\mathcal U_{\mu i}|^2 \sum_j |\mathcal U_{e j}|^2 \equiv \frac{m_\mu^5G_\mu^2}{192\pi^3}\,,
\end{equation}
so that the Fermi constant $G_F$, as determined from the muon decay ($G_\mu$) acquires a correction which will propagate to most electroweak observables.
For masses of the sterile states that are kinematically allowed to be produced as (on-shell) final states in muon decays, the modified decay width is given in Eq.~\eqref{eqn:llnunu}.
In the case of heavy sterile states, the modified Fermi constant is then given by
\begin{equation}
    G_F = G_\mu \left(\sum_{i} |\mathcal U_{\mu i}|^2 \sum_j |\mathcal U_{e j}|^2\right)^{-\frac{1}{2}}\,.
\end{equation}
In particular, the relation between $G_\mu$ and $M_W$ leads to a secondary constraint through kinematic measurements of $M_W$
\begin{equation}
    G_\mu \simeq \frac{\pi \alpha_e \sqrt{\sum_{i} |\mathcal U_{\mu i}|^2 \sum_j |\mathcal U_{e j}|^2}}{\sqrt{2}M_W^2(1 - M_W^2/M_Z^2)}\,.
\end{equation}
Consequently, the corrections will also propagate to the weak mixing angle $\sin^2\theta_w$.

\subsection{Deviation from unitarity of the PMNS}
\label{sec:unitaritynu}
As previously discussed, the introduction  of fermionic sterile states gives rise to many corrections to precision observables.
For masses above the electroweak scale, these can be conveniently encoded in the matrix $\eta$, parametrising the deviation from unitarity of the would-be PMNS matrix $\tilde U_\text{PMNS}$ (cf. Eq.~\eqref{eqn:Utildeeta}).
Inverting and expanding Eq.~\eqref{eqn:Utildeeta} yields an approximate expression for $\eta$ given by
\begin{equation}
\eta \simeq \frac{1}{2}\left(\mathbb{1} - \tilde{U}_{\mathrm{PMNS}} \, \tilde{U}_{\mathrm{PMNS}}^\dagger\right) \, .
\end{equation}
In addition to the precision observables previously mentioned, which can be re-expressed via the diagonal elements of $\eta$, the deviation from unitarity also alters the determination of CKM elements which rely on (semi-)~leptonic meson and $\tau$-lepton decays~\cite{Fernandez-Martinez:2016lgt,Coutinho:2019aiy}.
In particular, the (inclusive) determination of $V_{us}$ from super-allowed nuclear $\beta$ decays (using the unitarity of $V_\text{CKM}$) and the exclusive measurements of Kaon and $\tau$-lepton decays currently exhibits a $\sim 3-4\,\sigma$ tension, the so-called Cabibbo angle anomaly.
This tension can be (indirectly) alleviated in the presence of heavy sterile states, due to small deviations from unitarity of $\tilde U_\text{PMNS}$ leading to small non-vanishing entries in the matrix $\eta$~\cite{Coutinho:2019aiy}.

\smallskip
A combined fit~\cite{Fernandez-Martinez:2016lgt} to precision data then allows to establish model-independent upper bounds on the diagonal entries of $\eta$, independent of the mass and number of the sterile states (provided the mass is above the electroweak scale).
Non-vanishing off-diagonal entries will lead to cLFV interactions and can be constrained using data on cLFV observables.
Since contributions to cLFV observables in SM extensions via heavy sterile fermions first appear at the one-loop level, they are sensitive to the mass and number of the sterile states and model-independent constraints are not directly possible.
However, a robust constraint of theoretical nature can be derived. 
Given that $\eta$ is a Hermitian matrix, its entries must fulfill the Schwarz inequality
\begin{equation}
    |\eta_{\alpha\beta}| \leq \sqrt{\eta_{\alpha\alpha}\eta_{\beta\beta}}\,,
    \label{eqn:schwarz}
\end{equation}
which is only an exact equality in the case of $3$ additional sterile fermions.
In~\cite{Fernandez-Martinez:2016lgt} a combined fit to precision data was performed, which allowed to establish the following conservative bounds at $95\%\,\text{C.L.}$:
\begin{equation}
    \sqrt{2\eta_{ee}} < 0.050\,,\quad \sqrt{2\eta_{\mu\mu}} < 0.021\,,\quad \sqrt{\eta_{\tau\tau}} < 0.075\,,
\end{equation}
where the bounds on the off-diagonal entries can be derived via Eq.~\eqref{eqn:schwarz}.

The non-unitarity can be further probed in neutrino scattering experiments, since these lead to neutrino non-standard interactions with matter.

\subsection{Other constraints}

If the considered heavy neutral leptons are Majorana fermions, they will generically lead to LNV interactions.
On the one hand, HNL are known to lead to modifications of the predictions for the effective mass to which the amplitude of neutrinoless double beta decay is proportional to, $m_{ee}$. In the presence of $n_S$ heavy states, the contributions to $m_{ee}$ (first introduced in Chapter~\ref{chap:lepflav} Eq.~\eqref{eqn:0nubb_chap2}) can be written as~\cite{Blennow:2010th,Abada:2014nwa}
\begin{equation}
\label{eq:def:0nubb_nS}
m_{ee} \simeq \sum_{i=1}^{3+n_s} \, \mathcal U_{e i}^2 \, p^2 \, \frac{m_{ i}}{p^2-m_{i}^2} \simeq \sum_{i=1}^3 \, \mathcal U^2_{e i} \, m_i + \sum_{k=4}^{3+n_s} \mathcal U_{e k}^2 \, p^2 \, \frac{m_{k}}{p^2-m_{k}^2} \; ,
\end{equation}
in which, as before, $p^2$ corresponds to the neutrino virtual momentum, with $p^2 \simeq -(100 \, \mathrm{MeV})^2$. 
In the case of light sterile fermions, the latter can contribute positively to $0\nu\beta\beta$ decays while for heavy masses they have an indirect impact due to the non-unitarity of the PMNS.

One can also consider LNV $\tau$-lepton and meson decays of the form $\tau^- \to \ell^+ h_1^- h_2^-$ and $h_1^+\to \ell_\alpha^+\ell_\beta^+ h_2^-$. 
However, contributions of sterile states to these decays are usually suppressed, except if their masses are within the kinematical momentum transfer such that a resonant enhancement is possible. 
For further details see for instance~\cite{Abada:2017jjx}.

At colliders, one can perform direct searches for sterile neutrinos being produced either in decays of $Z$, $W$ and Higgs bosons or from LNV interactions in same-sign di-lepton signatures arising from the process $pp\to W^\ast\to \ell^\pm \nu_s\to \ell^\pm\ell^\pm + 2\:\text{jets}$~\cite{Antusch:2016ejd}.

Finally, sterile neutrinos with masses below the TeV-scale are subject to strong constraints from a number of cosmological observations~\cite{Kusenko:2009up,Smirnov:2006bu,Hernandez:2013lza,Hernandez:2014fha,Vincent:2014rja}, such as Big Bang Nucleosynthesis and Large Scale Structure formation.
These constraints severely restrict the the spectra of sterile states with masses in the range $1\:\mathrm{eV} - 100\:\mathrm{MeV}$.

\section{Heavy neutral leptons and charged lepton flavour violation}
\label{sec:cLFVobs}
The role of heavy neutral leptons in what concerns cLFV (see for instance~\cite{Riemann:1982rq,Riemann:1999ab,Illana:1999ww,Mann:1983dv,Illana:2000ic,Alonso:2012ji,Ilakovac:1994kj,Ma:1979px,Gronau:1984ct,Deppisch:2004fa,Deppisch:2005zm,Dinh:2012bp,Abada:2014kba,Abada:2015oba,Abada:2015zea,Abada:2016vzu,Abada:2018nio,Arganda:2014dta}) and lepton number violation - see for instance~\cite{Ali:2001gsa,Atre:2005eb,Atre:2009rg,Chrzaszcz:2013uz,Deppisch:2015qwa,Cai:2017mow,Abada:2017jjx,Drewes:2019byd,Maiezza:2015lza,Helo:2013dla,Blaksley:2011ey,Ibarra:2011wi} -
has been extensively explored in recent years.
Several studies revealed a promising potential of SM extensions via HNL in what concerns cLFV: depending on the mass regime and mixings with the active states, one could expect significant contributions to several observables, well within the future experimental sensitivity (with particularly interesting prospects in the $\mu-e$ sector). 
Moreover, in given scenarios, distinctive patterns and correlations of observables were identified, which in turn could be explored to probe and test these SM extensions, see for example~\cite{Hambye:2013jsa,Calibbi:2017uvl}. 

In this section we present the cLFV processes under the assumption of $3+n_S$ massive neutrinos, providing the expressions for the decay rates and other relevant amplitudes.
The cLFV observables addressed here receive contributions at the one-loop level arising from dipoles ($\gamma$- and $Z$-penguins) and/or from boxes. The vertex diagrams contributing to the cLFV decays are depicted in Fig.~\ref{Fig:penguins} and the box diagrams (adapted for the $\mu-e$ conversion rate for the sake of illustration) are presented in Fig.~\ref{Fig:boxes}.

\begin{figure}[h!]
\hspace*{0,5cm}\begin{equation*}
\vcenter{\hbox{
    \begin{tikzpicture}
        \begin{feynman}
            \vertex (m) [blob] at (0,0) {};
            \vertex (B) at (-1.5, 0);
            \vertex (f1) at (1.5, 1.5) {\(\ell_\alpha\)};
            \vertex (f2) at (1.5, -1.5) {\(\ell_\beta\)};
            \diagram* {
            (B) -- [boson, edge label={\small\(\gamma, Z\)}] (m) -- [fermion] (f2),
            (f1) -- [fermion] (m),
            };
        \end{feynman}
    \end{tikzpicture}}}
     =\quad
    \vcenter{\hbox{
    \begin{tikzpicture}
        \begin{feynman}
            \vertex (m) at (0,0);
            \vertex (B) at (-1.5, 0);
            \vertex (f1) at (1, 1);
            \vertex (f2) at (1, -1);
            \vertex (f12) at (2.5, 1) {\(\ell_\alpha\)};
            \vertex (f22) at (2.5, -1) {\(\ell_\beta\)};
            \diagram* {
            (B) -- [boson, edge label={\small\(\gamma, Z\)}] (m) -- [boson, edge label={\small\(W\)}] (f1),
            (m) -- [boson, edge label'={\small\(W\)}] (f2),
            (f1) -- [fermion, edge label={\small\(\nu_i\)}] (f2),
            (f1) -- [anti fermion] (f12),
            (f2) -- [fermion] (f22),
            };
        \end{feynman}
    \end{tikzpicture}
    }}
 +\quad
    \vcenter{\hbox{
    \begin{tikzpicture}
        \begin{feynman}
            \vertex (m) at (0,0);
            \vertex (B) at (-1.5, 0);
            \vertex (f1) at (1, 1);
            \vertex (f2) at (1, -1);
            \vertex (f12) at (2.5, 1) {\(\ell_\alpha\)};
            \vertex (f22) at (2.5, -1) {\(\ell_\beta\)};
            \diagram* {
            (B) -- [boson, edge label={\small\(Z\)}] (m) -- [anti fermion, edge label={\small\(\nu_i\)}] (f1),
            (m) -- [fermion, edge label'={\small\(\nu_j\)}] (f2),
            (f1) -- [boson, edge label={\small\(W\)}] (f2),
            (f1) -- [anti fermion] (f12),
            (f2) -- [fermion] (f22),
            };
        \end{feynman}
    \end{tikzpicture}
    }}
     +\hdots
\end{equation*}

\caption{
Vertex diagrams  contributing the cLFV decays. The flavour of the charged leptons is denoted by $\alpha, \beta, ... = e, \mu,\tau$; in the neutral fermion internal lines, $i,j=1,..., 3+n_S$ denote the neutral fermion mass eigenstates.
 }
 \label{Fig:penguins}
\end{figure}
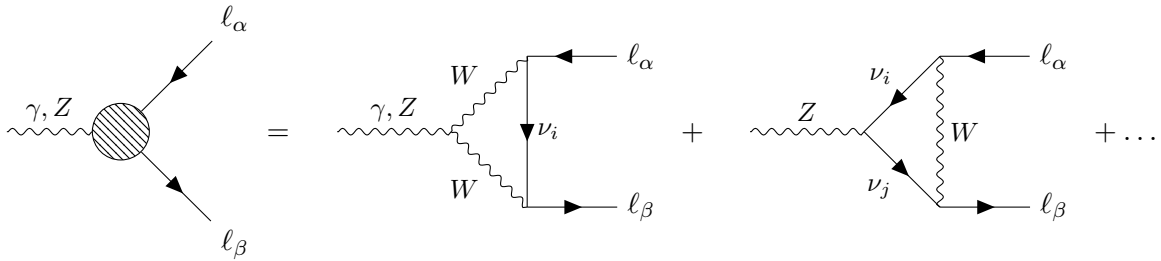

\begin{figure}[h!]
\begin{equation*}
\vcenter{\hbox{
\begin{tikzpicture}
    \begin{feynman}
        \vertex (f1) at (0,1.0) {\(\mu\)};
        \vertex (f11) at (2.0,1.0);
        \vertex (f12) at (4.0, 1.0);
        \vertex (f2) at (6.0, 1.0) {\(e\)};
        \vertex (f3) at (0, -1.0) {\(u\)};
        \vertex (f31) at (2.0, -1.0);
        \vertex (f32) at (4.0, -1.0);
        \vertex (f4) at (6.0, -1.0) {\(u\)};
        \diagram*{
        (f1) -- [fermion] (f11) -- [fermion, edge label={\small\(\nu_i\)}] (f12) -- [fermion] (f2),
        (f3) -- [fermion] (f31) -- [fermion, edge label'={\small\(d_j\)}] (f32) -- [fermion] (f4),
        (f11) -- [boson, edge label'={\small \(W\)}] (f31),
        (f12) -- [boson, edge label={\small \(W\)}] (f32),
        };
    \end{feynman}
\end{tikzpicture}
}}
\quad\quad
\vcenter{\hbox{
\begin{tikzpicture}
    \begin{feynman}
        \vertex (f1) at (0,1.0) {\(\mu\)};
        \vertex (f11) at (2.0,1.0);
        \vertex (f12) at (4.0, 1.0);
        \vertex (f2) at (6.0, 1.0) {\(e\)};
        \vertex (f3) at (0, -1.0) {\(u\)};
        \vertex (f31) at (2.0, -1.0);
        \vertex (f32) at (4.0, -1.0);
        \vertex (f4) at (6.0, -1.0) {\(u\)};
        \diagram*{
        (f1) -- [fermion] (f11) -- [fermion, edge label={\small\(\nu_i\)}] (f12) -- [fermion] (f2),
        (f3) -- [fermion] (f31) -- [fermion, edge label'={\small\(d_j\)}] (f32) -- [fermion] (f4),
        (f11) -- [boson, edge label'={\small \(W\)},near start] (f32),
        (f12) -- [boson, edge label={\small \(W\)},near start] (f31),
        };
    \end{feynman}
\end{tikzpicture}
}}
\end{equation*}
\caption{Example of  box diagrams (depicted for the box contributions to neutrinoless $\mu-e$ conversion). In the quark internal lines, $j=1,...,3$ 
  runs over the quark families; 
  in the neutral fermion ones, $i=1,...,3+n_S$.}  
\label{Fig:boxes}
\end{figure}
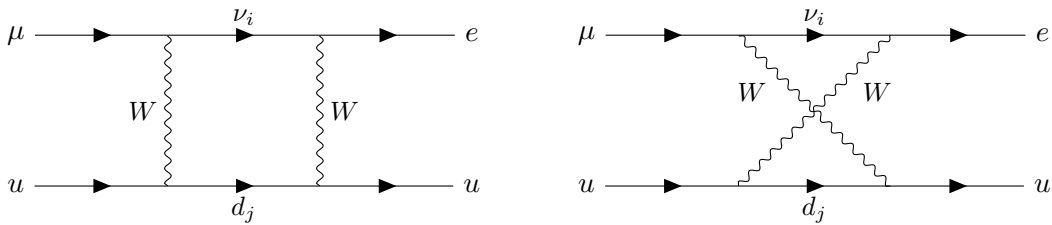

\mathversion{bold}
\subsection{Leptonic decays: $\ell_\beta \to \ell_\alpha\gamma$ and $\ell_\beta \to \ell_\alpha \ell_\gamma \ell_\gamma^\prime$}
\mathversion{normal}
We first provide the expressions for the branching ratios of the ``pure'' leptonic cLFV decays, i.e. the radiative 
decays $\ell_\beta \to \ell_\alpha \gamma$ and the 3-body decays\footnote{In the most general 3-body cLFV decay, $\ell_\beta \to \ell_\alpha \ell_\gamma \ell_\gamma^\prime$ the primed final state denotes the possibility of having equal or opposite charges for both $\ell_\gamma$ and $\ell_\gamma^{(\prime)}$.}
$\ell_\beta \to \ell_\alpha \ell_\gamma \ell_\gamma^\prime$.

In SM extensions via
$n_s$ heavy sterile fermions, the rates for the radiative and three-body decays are given by~\cite{Alonso:2012ji}  
\begin{equation}
    \mathrm{BR}(\ell_\beta\to \ell_\alpha \gamma) \,=
    \frac{\alpha_w^3\,
      s_w^2}{256\,\pi^2}\,\frac{m_{\beta}^4}{M_W^4}\,
\frac{m_{\beta}}{\Gamma_{\beta}}\, 
    \left|G_\gamma^{\beta \alpha} \right|^2\:, 
\end{equation}
\begin{eqnarray}
    \mathrm{BR}(\ell_\beta\to 3\ell_\alpha) &=&
    \frac{\alpha_w^4}{24576\,\pi^3}\,\frac{m_{\beta}^4}{M_W^4}\,
\frac{m_{\beta}}{\Gamma_{\beta}}\times\left\{2\left|\frac{1}{2}F_\text{box}^{\beta
      \alpha\alpha\alpha} +F_Z^{\beta\alpha} - 2 s_w^2\,(F_Z^{\beta\alpha} -
    F_\gamma^{\beta\alpha})\right|^2 \right.  \nonumber\\ 
     &+& \left. 4 s_w^4\, |F_Z^{\beta\alpha} -
    F_\gamma^{\beta\alpha}|^2 + 16
    s_w^2\,\mathrm{Re}\left[(F_Z^{\beta\alpha} - \frac{1}{2}F_\text{box}^{\beta
        \alpha\alpha\alpha})\,G_\gamma^{\beta \alpha
        \ast}\right]\right.\nonumber\\ 
     &-&\left. 48 s_w^4\,\mathrm{Re}\left[(F_Z^{\beta\alpha} -
      F_\gamma^{\beta\alpha})\,G_\gamma^{\beta\alpha \ast}\right] + 32
    s_w^4\,|G_\gamma^{\beta\alpha}|^2\left[\log\frac{m_{\beta}^2}{m_{\alpha}^2}
      - \frac{11}{4}\right] \right\}\,.  
\end{eqnarray}
Here, $m_{\beta}$ ($\Gamma_\beta$) denotes the mass (total width)
of the decaying charged lepton of flavour $\beta$. 

The more general 3-body cLFV decays, relevant only for the $\tau$-lepton, are given by~\cite{Ilakovac:1994kj}
\begin{eqnarray}
    \mathrm{BR}(\ell_\beta^-\to \ell_\alpha^-\ell_\gamma^+\ell_\gamma^-) &=& \frac{\alpha_w^4}{24576\pi^3}\frac{m_{\beta}}{M_W^4}{m_\beta}{\Gamma_\beta}\times\left\{\left|F_\text{box}^{\beta\alpha\gamma\gamma} - F_Z^{\beta\alpha} - 2s_w^2(F_Z^{\beta\alpha} - F_\gamma^{\beta\alpha})\right|^2 \right.\nonumber\\
    &+&\left. 4s_w^4 \left|F_Z^{\beta\alpha} - F_\gamma^{\beta\alpha} \right|^2 + 8 s_w^2\mathrm{Re}\left[(F_Z^{\beta\alpha} + F_\text{Box}^{\beta\alpha\gamma\gamma})G_\gamma^{\beta\alpha\ast}\right]\right.\nonumber\\
    &-&\left.32 s_w^4 \mathrm{Re}\left[(F_Z^{\beta\alpha} + F_\gamma^{\beta\alpha})G_\gamma^{\beta\alpha\ast}\right] + 32 s_w^4 \left|G_\gamma^{\beta\alpha}\right|^2\left[\log\frac{m_\beta^2}{m_\alpha^2} - 3\right]\right\}\,,
\end{eqnarray}
and the doubly flavour violating decay, exclusively mediated by box-diagrams, can be cast as
\begin{equation}
    \mathrm{BR}(\ell_\beta^-\to\ell_\alpha^-\ell_\gamma^+\ell_\alpha^-) = \frac{\alpha_w^4}{49152\pi^3}\frac{m_\beta^4}{M_W^4}\frac{m_\beta}{\Gamma_\beta}\left|F_\text{box}^{\beta\alpha\gamma\alpha}\right|^2\,.
\end{equation}

The form factors present in the above equations are given by~\cite{Alonso:2012ji, Ilakovac:1994kj} 
\begin{eqnarray}
    G_\gamma^{\beta \alpha} &=& \sum_{i =1}^{3 + n_s}
    \mathcal{U}_{\alpha i}^{\phantom{\ast}}\,\mathcal{U}_{\beta i}^\ast\,
    G_\gamma(x_i)\:,\label{eq:cLFV:FF:Ggamma} \\
     F_\gamma^{\beta \alpha} &=& \sum_{i =1}^{3 + n_s}
    \mathcal{U}_{\alpha i}^{\phantom{\ast}}\,\mathcal{U}_{\beta i}^\ast
    \,F_\gamma(x_i)\:,
   \\ 
    F_Z^{\beta \alpha} &=& \sum_{i,j =1}^{3 + n_s}
    \mathcal{U}_{\alpha i}^{\phantom{\ast}}\,\mathcal{U}_{\beta j}^\ast
    \left[\delta_{ij} \,F_Z(x_j) + 
    C_{ij}\, G_Z(x_i, x_j) + C_{ij}^\ast \,H_Z(x_i,
    x_j)\right]\:, 
    \label{eq:cLFV:FF:FZ}
    \\  
    F_\text{box}^{\beta\alpha\gamma\delta} &=& \sum_{i,j=1}^{3+n_S} \mathcal U_{\beta j}^\ast\,\mathcal U_{\delta j}^\ast\,\mathcal U_{\alpha i}\,\mathcal U_{\gamma i} G_\text{box}(x_i, x_j) - \mathcal U_{\beta j}^\ast\,\mathcal U_{\delta i}^\ast\,(\mathcal U_{\alpha j}\,\mathcal U_{\gamma i} + \mathcal U_{\delta j} \,\mathcal U_{\alpha i}) F_\text{Xbox}(x_i, x_j)\,,\\
    F_\text{box}^{\beta \alpha\alpha\alpha} &=&\sum_{i,j = 1}^{3+n_s}
    \mathcal{U}_{\alpha i}^{\phantom{\ast}}\,\mathcal{U}_{\beta
      j}^\ast\left[\mathcal{U}_{\alpha i}^{\phantom{\ast}} \,\mathcal{U}_{\alpha
        j}^\ast\, G_\text{box}(x_i, x_j) - 2 \,\mathcal{U}_{\alpha
        i}^\ast \,\mathcal{U}_{\alpha j}^{\phantom{\ast}}\, F_\text{Xbox}(x_i, x_j)
      \right]\:,\label{eq:cLFV:FF:Fbox}
    \end{eqnarray}
in which the sums run over the neutral mass eigenstates ($i,j=1,...,3+n_S$). 
The loop functions are given in Appendix~\ref{app:loopfunctions_neutrinos}, with the corresponding arguments defined as $x_i ={m_{i}^2}/{M_W^2}$,
and we recall that $C_{ij}$ was introduced in Eq.~\eqref{eqn:Cij}

\mathversion{bold}
\subsection{LFV $Z$-boson decays}
\mathversion{normal}
\label{sec:LFVZ}
The topology of cLFV $Z$ decays is closely related to contributions at the origin of several cLFV leptonic decays and transitions ($Z$-penguins)\footnote{For a recent discussion of LFV Higgs decays, see~\cite{Arganda:2014dta,Arganda:2015naa,Arganda:2015uca,Arganda:2017vdb}.}.

For convenience, we summarise here the analytical expressions needed for the LFV $Z$-decays in the Feynman-t'Hooft gauge, given in Refs.~\cite{DeRomeri:2016gum, Abada:2014cca, Abada:2015zea, Illana:1999ww}, in the convention of \texttt{LoopTools}~\cite{Hahn:1998yk}.
The decay width (for $\alpha\neq\beta$) is given by
\begin{equation}
    \Gamma(Z\to \ell_\alpha\bar\ell_\beta) = \frac{\alpha_w^3}{192\pi^2 c_w^2}M_Z\left|\mathcal F_Z^{\beta\alpha}\right|^2\,,
\end{equation}
in which the form factor $\mathcal F_Z^{\beta\alpha} = \sum_{i=1}^{10} \mathcal F_{Z,\beta\alpha}^{(i)}$ receives contributions from 10 different diagrams, as given in~\cite{DeRomeri:2016gum, Abada:2014cca, Illana:1999ww}.
The contributions of the different diagrams (neglecting the charged lepton masses) are given by
\begin{eqnarray}
    \mathcal F_{Z,\beta\alpha}^{(1)} &=& \frac{1}{2}\sum_{i,j = 1}^{3 + n_s}\mathcal U_{\alpha i}\,\mathcal U_{\beta j}^\ast\left[-C_{ij} x_i x_j M_W^2 C_0 + C_{ij}^\ast \sqrt{x_i x_j}\left(M_Z^2 C_{12} - 2C_{00} + \frac{1}{2}\right)\right]\ ,\\
    \mathcal F_{Z,\beta\alpha}^{(2)} &=& \sum_{i,j = 1}^{3 + n_s}\mathcal U_{\alpha i}\,\mathcal U_{\beta j}^\ast\left[-C_{ij}\left((C_0 + C_1 + C_2 + C_{12}) - 2C_{00} + 1\right) +  C_{ij}^\ast \sqrt{x_i x_j}M_W^2 C_0 \right]\,,
\end{eqnarray}
where $C_{0, 1, 2, 12, 00}\equiv C_{0, 1, 2, 12, 00}(0, M_Z^2, 0, M_W^2, m_{\nu_i}^2, m_{\nu_j}^2)$ are the Passarino-Veltman functions~\cite{Passarino:1978jh} in \texttt{LoopTools}~\cite{Hahn:1998yk} notation\footnote{We evaluate all Passarino-Veltman functions with the public Fortran code \texttt{LoopTools}~\cite{Hahn:1998yk} wrapped into our dedicated python code.}.
We further have
\begin{eqnarray}
    \mathcal F_{Z,\beta\alpha}^{(3)} &=& 2\, c_w^2 \sum_{i=1}^{3+n_s}\mathcal U_{\alpha i}\,\mathcal U_{\beta i}^\ast \left[ M_Z^2 (C_1 + C_2 + C_{12}) - 6C_{00} + 1\right]\,,\\
    \mathcal F_{Z,\beta\alpha}^{(4)} + \mathcal F_{Z,\beta\alpha}^{(5)} &=& -2\, s_w^2 \sum_{i=1}^{3+n_s}\mathcal U_{\alpha i}\,\mathcal U_{\beta i}^\ast\, x_i M_W^2 C_0\,,\\
    \mathcal F_{Z,\beta\alpha}^{(6)} &=& -(1 - 2 s_w^2)\sum_{i=1}^{3+n_s}\mathcal U_{\alpha i}\,\mathcal U_{\beta i}^\ast\, x_i C_{00}\,,
\end{eqnarray}
with $C_{0, 1, 2, 12, 00}\equiv C_{0, 1, 2, 12, 00}(0, M_Z^2, 0, m_{\nu_i}^2, M_W^2, M_W^2)$, and
\begin{equation}
    \mathcal F_{Z,\beta\alpha}^{(7)} + \mathcal F_{Z,\beta\alpha}^{(8)} + \mathcal F_{Z,\beta\alpha}^{(9)} + \mathcal F_{Z,\beta\alpha}^{(10)} = \frac{1}{2}(1 - 2 c_w^2)\sum_{i=1}^{3+n_s}\mathcal U_{\alpha i}\,\mathcal U_{\beta i}^\ast \left[(2 + x_i)B_1 + 1\right]\,,
\end{equation}
in which we have $B_1 \equiv B_1(0,m_{\nu_i}^2, M_W^2)$.

Since current searches do not distinguish the charges of the final state leptons, for numerical purposes one should thus consider the averaged decay rate, that is
\begin{equation}
    \Gamma (Z\to \ell_\alpha^\pm \ell_\beta^\mp) = \frac{1}{2}\left[\Gamma(Z\to \ell_\alpha^+\ell_\beta^-) + \Gamma(Z\to \ell_\alpha^-\ell_\beta^+)\right]\,.
\end{equation}

\section{cLFV in muonic atoms and heavy neutral leptons}
\label{sec:muonicatoms}
Several cLFV observables concern processes occurring in the presence of a (short-lived) muonic atom. 
As mentioned before, these include muonium oscillations and decay ($\text{Mu}-\overline{\text{Mu}}$, $\text{Mu}\to e e $), neutrinoless muon-electron conversion in nuclei ($\mu-e$, N), and the Coulomb enhanced decay
$\mu e \to e e$.

\mathversion{bold}
\subsection{Neutrinoless $\mu-e$ conversion in heavy nuclei}
\mathversion{normal}
The expression for the coherent conversion rate\footnote{In the present discussion we only consider the coherent conversion; for a general discussion of spin-dependent contributions to the process, we refer to Refs.~\cite{Cirigliano:2017azj,Davidson:2017nrp}.} ocurring in the presence of a nucleus (N) can be cast as~\cite{Alonso:2012ji}   
\begin{equation}\label{eq:def:CRfull}
    \mathrm{CR}(\mu - e,\,\mathrm{N}) = \frac{2 G_F^2\,\alpha_w^2\,
      m_\mu^5}{(4\pi)^2\,\Gamma_\text{capt.}}\left|4 V^{(p)}\left(2
    \widetilde F_u^{\mu e} + \widetilde F_d^{\mu e}\right) + 4 V^{(n)}\left(
    \widetilde F_u^{\mu e} + 2\widetilde F_d^{\mu e}\right)  + s_w^2
    \frac{G_\gamma^{\mu e}D}{2 e}\right|^2\,.  
\end{equation}
In the above expression, $\Gamma_\text{capt.}$ denotes the capture rate for the nucleus N; 
$D$, $V^{(p)}$ and $V^{(n)}$ correspond to nuclear form factors whose
values are given in~\cite{Kitano:2002mt}, with $e$ 
being the unit electric charge. 
For three nuclei of interest, we reproduce the relevant nuclear data in Table~\ref{TabTiAuData}.

\renewcommand{\arraystretch}{1.3}
\begin{table}[h!]
  \begin{center}
  \begin{tabular}{|l|c|c|c|c|} \hline
    Nucleus & $D [m_\mu^{5/2}]$
    & $V^{(p)} [m_\mu^{5/2}]$ & $V^{(n)} [m_\mu^{5/2}] $ & $\Gamma_{\text{capture}}[10^6\:\mathrm{s}^{-1}]$\\ \hline
    $\phantom{\text{Ti}}_{22}^{48}\text{Ti}$    & 0.0864 & 0.0396 & 0.0468  & 2.59\\
    $\phantom{\text{Au}}^{197}_{79}\text{Au}$ &  0.189 & 0.0974 & 0.146 & 13.07\\
    $\phantom{\text{Al}}_{13}^{27}\text{Al}$ & 0.0362 & 0.0161 & 0.0173 & 0.7054\\
   \hline
\end{tabular}
\end{center}

\caption{Overlap integrals $D, V$ and $\Gamma_\text{capture}$
  for Titanium, Gold and Aluminium nuclei, as reported in~\cite{Kitano:2002mt}
  (Tables I and VIII).}
\label{TabTiAuData}
\end{table}
\renewcommand{\arraystretch}{1.}

The form factors present in the above equation are given by~\cite{Alonso:2012ji, Ilakovac:1994kj} 
\begin{eqnarray}
    \widetilde F^{\mu e}_d &=& -\frac{1}{3}s_w^2 F_\gamma^{\mu e} - F_Z^{\mu e}\left(\frac{1}{4} - \frac{1}{3}s_w^2 \right) + \frac{1}{4}F^{\mu e dd}_\text{box}\ ,\\
    \widetilde F^{\mu e}_u &=& \frac{2}{3}s_w^2 F_\gamma^{\mu e} + F_Z^{\mu e}\left(\frac{1}{4} - \frac{2}{3}s_w^2 \right) + \frac{1}{4}F^{\mu e uu}_\text{box}\,.
\end{eqnarray}
In addition to the form factors previously defined in Eqs.~(\ref{eq:cLFV:FF:Ggamma} - \ref{eq:cLFV:FF:Fbox}), $\widetilde{F}_{u,d}^{\mu e}$ call upon the following box contributions, 
\begin{eqnarray}
     F_\text{box}^{\mu e uu} &=& \sum_{i = 1}^{3 + n_s}\sum_{q_d = d, s,
      b} \mathcal{U}_{e i}^{\phantom{\ast}}\,\mathcal{U}_{\mu i}^\ast\, V_{u q_d}^{\phantom{\ast}}\,V_{u
      q_d}^\ast \:F_\text{box}(x_i, x_{q_d})\,, 
    \label{eq:cLFV:FF:mueuu}\\
    F_\text{box}^{\mu e dd} &=& \sum_{i = 1}^{3 + n_s}\sum_{q_u = u, c,
      t} \mathcal{U}_{e i}^{\phantom{\ast}}\,\mathcal{U}_{\mu i}^\ast\, V_{q_u
      d}^{\phantom{\ast}}\,V_{q_u d}^\ast \:F_\text{Xbox}(x_i, x_{q_u})\,, 
    \label{eq:cLFV:FF:muedd}    
\end{eqnarray}
in which $x_{q} ={m_{q}^2}/{M_W^2}$ and $V$ is the CKM quark mixing matrix.  

\subsection{Muonium oscillations and decay}
\label{sec:intromuonium}
As mentioned in Chapter~\ref{chap:lepflav}, the spontaneous conversion of a Muonium atom to its anti-atom 
($\overline{\text{Mu}} = e^+\mu^-$)~\cite{Feinberg:1961zza}, as well as its decay to a pair of electrons, have been identified as promising cLFV observables. 

\begin{figure}[h!]
\begin{equation*}
\vcenter{\hbox{
\begin{tikzpicture}
    \begin{feynman}
        \vertex (f1) at (0,1.0) {\(\mu^-\)};
        \vertex (f11) at (2.0,1.0);
        \vertex (f12) at (4.0, 1.0);
        \vertex (f2) at (6.0, 1.0) {\(\mu^+\)};
        \vertex (f3) at (0, -1.0) {\(e^+\)};
        \vertex (f31) at (2.0, -1.0);
        \vertex (f32) at (4.0, -1.0);
        \vertex (f4) at (6.0, -1.0) {\(e^-\)};
        \diagram*{
        (f1) -- [fermion] (f11) -- [edge label={\small\(\nu_i\)}, insertion=0.5] (f12) -- [anti fermion] (f2),
        (f3) -- [anti fermion] (f31) -- [insertion=0.5, edge label'={\small\(\nu_j\)}] (f32) -- [fermion] (f4),
        (f11) -- [boson, edge label'={\small \(W\)}] (f31),
        (f12) -- [boson, edge label={\small \(W\)}] (f32),
        };
    \end{feynman}
\end{tikzpicture}
}}
\quad\quad
\vcenter{\hbox{
\begin{tikzpicture}
    \begin{feynman}
        \vertex (f1) at (0,1.0) {\(\mu^-\)};
        \vertex (f11) at (2.0,1.0);
        \vertex (f12) at (4.0, 1.0);
        \vertex (f2) at (6.0, 1.0) {\(\mu^+\)};
        \vertex (f3) at (0, -1.0) {\(e^+\)};
        \vertex (f31) at (2.0, -1.0);
        \vertex (f32) at (4.0, -1.0);
        \vertex (f4) at (6.0, -1.0) {\(e^-\)};
        \diagram*{
        (f1) -- [fermion] (f11) -- [insertion=0.5, edge label={\small\(\nu_i\)}] (f12) -- [anti fermion] (f2),
        (f3) -- [anti fermion] (f31) -- [insertion=0.5, edge label'={\small\(\nu_j\)}] (f32) -- [fermion] (f4),
        (f11) -- [boson, edge label'={\small \(W\)},near start] (f32),
        (f12) -- [boson, edge label={\small \(W\)},near start] (f31),
        };
    \end{feynman}
\end{tikzpicture}
}}
\end{equation*}
\caption{Box diagrams with Majorana mass insertions contributing to $\mathrm{Mu}-\overline{\mathrm{Mu}}$ oscillations.}  
\label{fig:boxes_maj}
\end{figure}
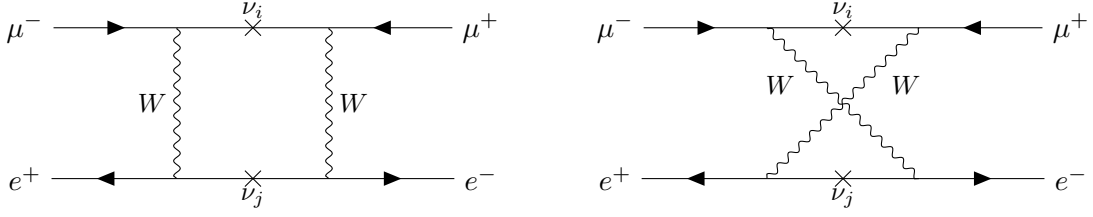

In the presence of $(V -A) \times (V - A)$ interactions, Muonium anti-Muonium oscillations can be described by the following effective
four-fermion interaction, in terms of an effective coupling $G_{M\overline{M}}$, 
\begin{equation}
 \mathcal{L}_\text{eff}^{M\overline{M}} \,= \, \frac{G_{
     M\overline{M}}}{ \sqrt{2} } 
\left[\, {\overline \mu}\,  \gamma^{\alpha} (1 - \gamma_5) \,e
\,\right] \left[ \,{\overline \mu}\, \gamma_{\alpha} (1 - \gamma_5)\,
e \,\right] \, .
\label{eq:def:Leff_anti-muonium}
\end{equation}
In extensions of the SM with sterile 
neutrinos
Mu-$\overline{\text{Mu}}$ conversion occurs at the loop level, being exclusively mediated by a set of 4 independent box diagrams; while a first set is common to both Dirac and Majorana
neutrinos 
a second one is only present if neutrinos are Majorana particles,
 effectively amounting to two Majorana mass insertions (see Refs.~\cite{Clark:2003tv,Cvetic:2005gx}). 
The corresponding diagrams with Majorana mass insertions are shown in Fig.~\ref{fig:boxes_maj}.
Working in the unitary gauge, the computation of the different box diagrams allows to write the effective coupling of Eq.~(\ref{eq:def:Leff_anti-muonium}) as~\cite{Clark:2003tv,Cvetic:2005gx}:
\begin{equation}
\frac{G_{M\overline{M}}}{\sqrt{2}}\,=\,
-\frac{G_F^2 M_W^2}{16\pi^2}\left[\sum_{i,j =1}^{3+n_s}
2\, \mathcal U_{\mu i}^\ast\,\mathcal U_{\mu j}^\ast\, \mathcal U_{e i}\, \mathcal U_{e j} F_\text{Xbox}(x_i, x_j)+ (\mathcal U_{e i})^2\,(\mathcal U_{\mu j}^\ast)^2 G_\text{box}(x_i, x_j)
\right]
\label{eq:Gmumu}\,,
\end{equation}
where $F_\text{Xbox}(x_i, x_j)$ and $G_\text{box}(x_i, x_j)$ are the relevant loop functions\footnote{We note a sign difference between the function $F_\text{box}$ in Ref.~\cite{Cvetic:2005gx} and the function $F_\text{Xbox}$ in our convention.} given in Appendix~\ref{app:loopfunctions_neutrinos}, with $x_i =\frac{m_{\nu_i}^2}{M_W^2}, i=1,...,3+n_s$
(further details can be found in \cite{Abada:2015oba}).

\bigskip
In the presence  of New Physics, Muonium can also undergo the cLFV decay  $\text{Mu} \, \to \, e^+ \, e^-$. In the SM extended by $n_s$ heavy neutral leptons,
the cLFV Muonium decay rate is given by~\cite{Cvetic:2006yg}  
\begin{equation}\label{eq:Mudecay:BR}
\text{BR}(\text{Mu} \to e^+ e^-) \, =\, 
\frac{\alpha_e^3}{\Gamma_\mu \, 32 \pi^2}\, 
\frac{m_e^2 m_\mu^2}{(m_e + m_\mu)^3}\,
\sqrt{1 -4\, \frac{m_e^2}{(m_e + m_\mu)^2}}\,
|\mathcal{M}_\text{tot}|^2\ , 
\end{equation}
in which $\Gamma_\mu = G_F^2 m_\mu^5 /(192 \pi^3)$ denotes the muon decay width, with 
$|\mathcal{M}_\text{tot}|$ the full amplitude (summed
(averaged) over final (initial) spins)~\cite{Cvetic:2006yg}, 
\begin{align}
|\mathcal{M}_\text{tot}|^2\,  =\, & 
\frac{\alpha_w^4}{16 M_W^4}
\bigg\{
\left( m_e\,m_\mu^{3}+2\,{m_e}^{2}m_\mu^{2}
+{m_e}^{3}m_\mu \right) {\left| 2F_Z^{\mu e} + F_\mathrm{Box}^{\mu e e e}\right|}^{2}
\nonumber\\
&+ 4 \sin^2 \theta_w\left( 2\,m_e\,m_\mu^{3}
+3\,{m_e}^{2}m_\mu^{2}+3\,{m_e}^{3}m_\mu \right)\
\mathrm{Re} \left[(2F_Z^{\mu e}+F_\mathrm{Box}^{\mu e e e})
(F_\gamma^{\mu e} -F_Z^{\mu e})^*\right]\nonumber\\
&+ 12 \sin^2 \theta_w
\left( m_e\,m_\mu^{3}+ 2\,{m_e}^{2}m_\mu^{2}+{m_e}^{3}m_\mu \right)
\ \mathrm{Re} \left[(2F_Z^{\mu e}+F_\mathrm{Box}^{\mu e e e})
G_\gamma^{\mu e *}\right] \nonumber\\
 &+ 4 \sin^4\theta_w \left( 7\,m_e\,{m_\mu}^{3}+12\,{m_e}^{2}m_\mu^{2}
 +9\,{m_e}^{3}m_\mu \right) {\left|F_\gamma^{\mu e}-F_Z^{\mu
     e}\right|}^{2}\nonumber\\ 
&+ 4 \sin^4\theta_w \left( -2\,m_\mu^{4}
+12\,m_e\,m_\mu^{3}+36\,{m_e}^{2}m_\mu^{2}+18\,{m_e}^{3}m_\mu \right)
\ \mathrm{Re} \left[(F_\gamma^{\mu e} - F_Z^{\mu e}) G_\gamma^{\mu e*}
\right]\nonumber\\
&+ 4 \sin^4\theta_w \left( {\frac {m_\mu^{5}}{m_e}}+2\,m_\mu^{4}
+8\,m_e\,{m_\mu}^{3}+24\,{m_e}^{2}m_\mu^{2}
+9\,{m_e}^{3}m_\mu \right) \left|{G_\gamma^{\mu e}}\right|^{2}\bigg\}
\ ,
\label{eq:amplMudecay}
\end{align}
where the corresponding form factors have been previously introduced (see Eqs.~(\ref{eq:cLFV:FF:Ggamma} - \ref{eq:cLFV:FF:Fbox}), setting $\beta=\mu$ and $\alpha=e$). 

\mathversion{bold}
\subsection{Coulomb-enhanced decay $\mu e \to e e$}
\mathversion{normal}
As also mentioned in Chapter~\ref{chap:lepflav}, several recent studies~\cite{Koike:2010xr,Uesaka:2016vfy,Uesaka:2017yin} have identified the complementary role of the cLFV decay of a bound $\mu^-$ 
in a muonic atom into a pair of electrons, 
\begin{equation}\label{eq:me2ee}
  \mu^{-} \, {e^{-}} \to\, {e^{-}}\,{e^{-}}\,.
\end{equation}
Notice that in the above transition, the initial states are a $\mu^-$ and a 1s atomic $e^-$, which are 
bound in the Coulomb field of a nucleus~\cite{Koike:2010xr}. 
The rate of the $\mu^- e^- \to e^- e^-$ process can be written as~\cite{Koike:2010xr}
\begin{eqnarray}\label{eq:cross.section:vrel.psi}
\Gamma(\mu^- e^- \to e^- e^- , \text{ N})\, &=& \, \sigma_{\mu e \to
  ee} v_\text{rel} \, 
|\psi_\text{1s}^{(e)} (0; Z-1)|^2\, , \quad \nonumber\\
 \text{with } 
\quad\psi_\text{1s}^{(e)} (0; Z-1)\, &=&\, 
\frac{\left[ (Z-1)\, \alpha_e\, m_e \right]^{3/2} }{\sqrt \pi} ,
\end{eqnarray}
where $\,\sigma_{\mu e \to ee} v_\text{rel}$ denotes the electroweak cross-section (dependent on the underlying NP model), and $Z$ the atomic number of the nucleus. 
Although the NP sources of flavour-violation are formally the same as those contributing to other $\mu-e$ transitions (specifically $\mu \to 3e$ or Muonium decay), there are several important differences. 
As discussed in~\cite{Uesaka:2015qaa}, the associated rate can be significantly enhanced in large $Z$ atoms\footnote{For small atoms, the enhancement can be well described
by a dependence $\approx (Z-1)^3$; however, for large $Z$ atoms - and as shown in Fig.~1 of~\cite{Uesaka:2015qaa} - the rate can be enhanced by as much as an additional order of magnitude.} (especially the contributions from contact
interactions~\cite{Uesaka:2015qaa}). Moreover, and when compared to the apparently similar $\mu^+ \to e^+ e^- e^-$ process, the Coulomb-enhanced has a larger phase space and from an experimental point of view, also a cleaner signature). It leads to a two-body final state, 
with nearly back-to-back emitted electrons, each with a well-defined
energy ($E_{e^-}\sim m_\mu/2$)~\cite{Koike:2010xr}. 

Neglecting (long-range) photonic interactions, which are typically subdominant for SM extensions via heavy neutral leptons~\cite{Uesaka:2016vfy,Uesaka:2017yin,Kuno:2019ttl}, the dominant contributions are due to (contact) interactions 
arising from photon 
and $Z$ penguin diagrams, as well as from box diagrams.
Considering a muonic atom, with an atomic number $Z$,
the branching ratio of the process can be cast as
\begin{align}
&\text{BR}(\mu^+ e^- \to e^+ e^- , \text{ N})\, \equiv 
\, \tilde{\tau}_{\mu}\,
  \Gamma(\mu^+ e^- \to e^+ e^- , \text {N}) \nonumber \\ 
& = 24\pi \, 
f_\text{Coul.}(Z)\, 
\alpha_w  \left( \frac{m_e}{m_{\mu}} \right)^{3}\,
  \frac{\tilde{\tau}_{\mu}}{\tau_{\mu}}\,
  \left(16 \, \left | \frac{1}{2} \left (\frac{g_w}{4 \pi} \right)^2
  \left (\frac{1}{2}F^{\mu eee}_\text{Box} + F_Z^{\mu e} - 
  2 \sin^2\theta_w \left (F_Z^{\mu e} - F_\gamma^{\mu e} \right)
  \right) \right|^2 + \right . \nonumber\\ 
& \left . +  4  \, \left| \frac{1}{2} \left(\frac{g_w}{4 \pi} \right)^2 
  2 \sin^2\theta_w \left (F_Z^{\mu e} - F_\gamma^{\mu e} \right)
 \right|^2 \right) \, , 
\label{eq:br-4fermi}
\end{align}
in which the cLFV form factors have already been defined. 
In the above,  $\tau_{\mu}$ denotes the lifetime of a free muon ($\tau_{\mu}= 2.197 \times 10^{-6}$~s~\cite{Agashe:2014kda}) and 
$\tilde{\tau}_{\mu}$ corresponds to the lifetime of a muonic atom (for the case of Aluminium, one has $\tilde{\tau}_{\mu} =  8.64 \times 10^{-7}$~s~\cite{Suzuki:1987jf}); 
moreover, for small atoms, one approximates $f_\text{Coul.}(Z) \approx(Z - 1)^{3}$.

\chapter{Charged lepton flavour violation and leptonic CP violation}
\label{chap:CPV}
\minitoc

\noindent
As extensively discussed in the previous chapters, lepton flavour violating observables constitute powerful probes of SM extensions featuring heavy neutral leptons.
Irrespective of the actual mechanism of neutrino mass generation under consideration, the mixings of the new states with the active left-handed neutrinos will lead to modifications in both leptonic charged and neutral currents, with a deep phenomenological impact. 
Via their mixings with the light (mostly active) states, and as a  consequence of a departure from unitarity of the would-be PMNS mixing matrix, the new states open the door to contributions to numerous observables.
In addition to the masses of the new states
and their mixings to the active neutrinos, constructions relying on heavy sterile states also open the door to new sources of CP violation: other than new Dirac CP violating (CPV) phases, should the massive states be of Majorana nature, further phases can be present.

For SM extensions featuring $n_S$ additional neutral fermions the mixing matrix $\mathcal U$ contains a total of $(3+n_S)(2 + n_S)/2$ rotation angles, $(2 + n_S)(1 + n_S)/2$ Dirac phases and $2+n_S$ Majorana phases.
The role of these additional CPV phases has been explored in analyses dedicated to CP violating observables, as is the case of electric dipole moments (EDM) of charged leptons~\cite{Abada:2015trh,Abada:2016awd,Novales-Sanchez:2016sng,deGouvea:2005jj}. 
New CPV phases have been also recently shown to play a crucial role in what concerns interference effects in LNV (and cLFV) semi-leptonic meson and tau decays~\cite{Abada:2019bac}, when more than one HNL is involved. 
Noticeably, while branching fractions of semi-leptonic meson or tau decays  into 
same-sign and opposite-sign di-leptons are expected to be of the same order in the case of SM extensions by a single heavy  Majorana fermion, this is no longer the case 
when the SM is extended by at least two HNLs, due to the possible
interferences that might arise in the presence of multiple states. 
Depending on the CPV phases, one can have a modification 
(enhancement/suppression) of the rates of LNV  
modes and of the lepton number conserving ones. 
As a consequence, the non-observation of a given mode need not be
interpreted in terms of reduced active-sterile couplings, but it could be instead understood in terms of interference effects due to the presence of several sterile states. 
(This effect is particularly amplified for processes 
with different charged leptons in the  final state.) 
Likewise, an experimental signal of a lepton number conserving process and the non-observation of the corresponding 
lepton number violating one do 
not necessarily rule out that the mediators are Majorana fermions~\cite{Abada:2019bac}. 

Similar studies have explored the role of a second heavy neutrino concerning the possibility of resonant CP violation~\cite{Bray:2007ru}, the effect of CP violation in high-scale seesaw scenarios in the context of renormalisation group running and leptogenesis~\cite{Petcov:2005yh,Petcov:2006pc}, the impact for forward-backward asymmetries at an electron-positron collider~\cite{Dev:2019rxh}, while others have compared the expected number of events associated with same-sign and opposite-sign dileptons at colliders in the framework of Left-Right symmetric models~\cite{Anamiati:2016uxp,Das:2017hmg,Dev:2019rxh}. 

\bigskip
In this chapter we focus exclusively on the role of CPV  phases (Dirac and Majorana) concerning an extensive array of cLFV observables. 
We work under the assumption of a unique source of lepton flavour violation, a generalised leptonic mixing matrix which now incorporates the active-sterile mixings.
As we have pointed out in~\cite{Abada:2021zcm}, the results of a thorough analysis suggest that the presence of leptonic CP violating phases can strongly affect the predictions (either suppressing or enhancing the otherwise expected rates), and possibly  
lead to a loss of correlation between observables (typically present in simple SM extensions via heavy sterile fermions). As it is subsequently argued in~\cite{Abada:2021zcm}, the confrontation of unexpected cLFV patterns upon observation of certain channels could be suggestive of the presence of non-vanishing phases.
As an example, one could have sizeable rates 
for $\mu\to eee$, and comparatively suppressed rates for $\mu \to e\gamma$.
The results presented in this chapter can be readily generalised for more complete NP models relying on the inclusion of heavy neutral fermions, provided that all complex degrees of freedom are systematically included (e.g. in Casas-Ibarra parametrisation~\cite{Casas:2001sr}), and predictions for cLFV observables revisited.

\mathversion{bold}
\section{``$3 + n_S$'' effective model}
\mathversion{normal}
As mentioned in the previous chapters, heavy neutral leptons, with masses ranging from the GeV to the tens of TeV, are among the most interesting minimal extensions of the SM, as they can be at the source of significant contributions  to numerous observables, both at high-intensities and at colliders. 
Interestingly, the most minimal tree-level mechanism for neutrino mass generation - the type I seesaw - calls upon the introduction of at least two such states to account for oscillation data.  
The type I seesaw~\cite{Minkowski:1977sc,Yanagida:1979as,Glashow:1979nm,Gell-Mann:1979vob,Mohapatra:1979ia}  
and its low-scale variants,  
such as the Inverse 
Seesaw (ISS)~\cite{Schechter:1980gr,Gronau:1984ct,Mohapatra:1986bd},  
the Linear Seesaw (LSS)~\cite{Barr:2003nn,Malinsky:2005bi}  and the 
$\nu$-MSM~\cite{Asaka:2005an,Asaka:2005pn,Shaposhnikov:2008pf}, all call upon extending the SM via additional sterile fermions, allowing for Dirac and Majorana mass terms for the neutral lepton sector. 
Irrespective of the actual mechanism of neutrino mass generation under consideration, the mixings of the new states with the active left-handed neutrinos will lead to modifications in both leptonic charged and neutral currents, with a deep phenomenological impact. 
To study and numerically assess the impact of the heavy states, 
it is often convenient to consider simplified ``ad-hoc'' models, in which one adds $n_s$ sterile fermions to the SM field content. Such an approach allows to identify the most relevant effects and the consequences for the observables under scrutiny, and paves the way to a subsequent thorough study of complete models of neutrino mass generation via sterile fermions. 

In the case of an extension via 2 heavy neutral leptons, and following for instance~\cite{Abada:2015trh}, the (enlarged) neutrino mixing matrix $\mathcal{U}$
can be parametrised through five subsequent rotations $R_{ij}$ (with $i\neq j$), and a diagonal matrix including the four physical Majorana phases, $\varphi_i$ 
\begin{eqnarray}
    \mathcal{U} \,= \,R_{45}\,R_{35}\,R_{25}\,R_{15}\,
    R_{34}\,R_{24}\,R_{14}\,R_{23}\,R_{13}\,R_{12}\times\mathrm{diag}(1, e^{i\varphi_2}, e^{i\varphi_3}, e^{i\varphi_4}, e^{i\varphi_5})\,.
    \label{eqn:allrot}
\end{eqnarray}
The above rotations are of the form (illustrated by $R_{45}$):
\begin{equation}\label{eq:R45}
    R_{45} = \begin{pmatrix}
                1 & 0 & 0 & 0 & 0\\
                0 & 1 & 0 & 0 & 0\\
                0 & 0 & 1 & 0 & 0\\
                0 & 0 & 0 & \cos\theta_{45} & \sin \theta_{45} e^{-i\delta_{45}}\\
                0 & 0 & 0 & -\sin\theta_{45} e^{i\delta_{45}} & \cos\theta_{45}
            \end{pmatrix}\,.
\end{equation}

As already noticed (and clear from the $W$ vertex in Eq.~(\ref{eqn:wlnu})), the mixing in charged current interactions is parametrised via a
rectangular $3 \times (3 +n_S)$ (i.e. $3\times 5$) mixing matrix, of which the $3 \times 3$ sub-block encodes the mixing between the left-handed leptons, $\tilde U_\text{PMNS}$. 

\section{Phases do matter}
\label{sec:phases.matter}
In what follows we offer a first insight into the role of CP violating phases regarding a subset of (representative) observables.
All other observables considered  in the full phenomenological analysis of Section~\ref{sec:num:analysis} can be understood from a generalisation of the discussion here carried. 
As mentioned before, we work under the hypothesis of having $n_s=2$  massive sterile states; the neutral spectrum thus comprises 5 states, with masses $m_i$ (with $i=1,...,5$), including the 3 light (mostly active) neutrinos and two heavier states, with masses $m_{4,5}$. The leptonic mixings (whose precise origin is not specified) are parametrised by a $5\times5$ unitary mixing matrix, $\mathcal{U}$.
As previously described, the full mixing matrix $\mathcal{U}$ 
can be cast in terms\footnote{For concreteness, we have chosen the six Dirac CP phases $\delta_{\alpha 4}$ and $\delta_{\alpha 5}$ to be non-vanishing.} of 10 real mixing angles $\theta_{ij}$, 6 Dirac phases  $\delta_{ij}$ and 4 Majorana phases, $\varphi_i$. 

\bigskip
The study carried in this section is accompanied by a brief analytical discussion, relying on a simplified approach to this ``3+2 toy model''. In particular, and for the purpose of deriving clear (and compact) analytical expressions, in this section we work under the following assumptions:
firstly, the mixings between the active and the sterile states (i.e., $\theta_{\alpha i}$ with $\alpha = e, \mu, \tau$ and $i=4,5$) are assumed to be sufficiently small so that $\cos \theta_{\alpha i} \approx 1$ to a very good approximation. The $3\times 2$ rectangular matrix encoding the active-sterile mixings can then be parametrised as \begin{equation}
\mathcal{U}_{\alpha (4,5)} \approx 
\left (\begin{array}{cc}
s_{14} e^{-i(\delta_{14}-\varphi_4)} &
s_{15} e^{-i(\delta_{15}-\varphi_5)} \\
s_{24} e^{-i(\delta_{24}-\varphi_4)} &
s_{25} e^{-i(\delta_{25}-\varphi_5)} \\
s_{34} e^{-i(\delta_{34}-\varphi_4)} &
s_{35} e^{-i(\delta_{35}-\varphi_5)} 
\end{array}
\right)\,,
\end{equation}
with $s_{\alpha i} = \sin \theta_{\alpha i}$, and where 
$\delta_{\alpha i}$ ($\varphi_i$) denote Dirac (Majorana) phases. We further assume $\theta_{\alpha 4} \approx \theta_{\alpha 5}$, and take the heavy states to be nearly mass-degenerate\footnote{This offers the advantage of further simplifying the expressions, allowing to factor out the loop functions from the sums over the heavy states.}, $m_4 \approx m_5 \gg \Lambda_\text{EW}$ (of the order of a few~TeV); a regime that typically allows for significant contributions to cLFV observables, see e.g.~\cite{Abada:2018nio}. 

We emphasise that all numerical results (displayed in the present section and throughout the present chapter) are obtained without relying on any approximation, taking into account all the contributions present in the most general setup. 

As this first discussion is dedicated to understanding and rendering visible the role of phases, no experimental constraints will be applied (certain observables might thus reach values already in disagreement with current experimental bounds). 

\subsection{cLFV decay rates: sensitivity to CPV 
phases}
In what follows, we focus on $\mu-e$ sector flavour violation, and consider
the following subset of observables:
BR($\mu \to e \gamma$), BR($\mu \to 3 e $) and BR($Z \to e \mu$). We then devote a brief dedicated discussion to $\mu-e$ conversion in nuclei. The corresponding expressions in the context of HNL can be found in Chapter~\ref{sec:numassgen}.

\begin{figure}
    \centering
\mbox{\hspace*{-5mm}    \includegraphics[width=0.51\textwidth]{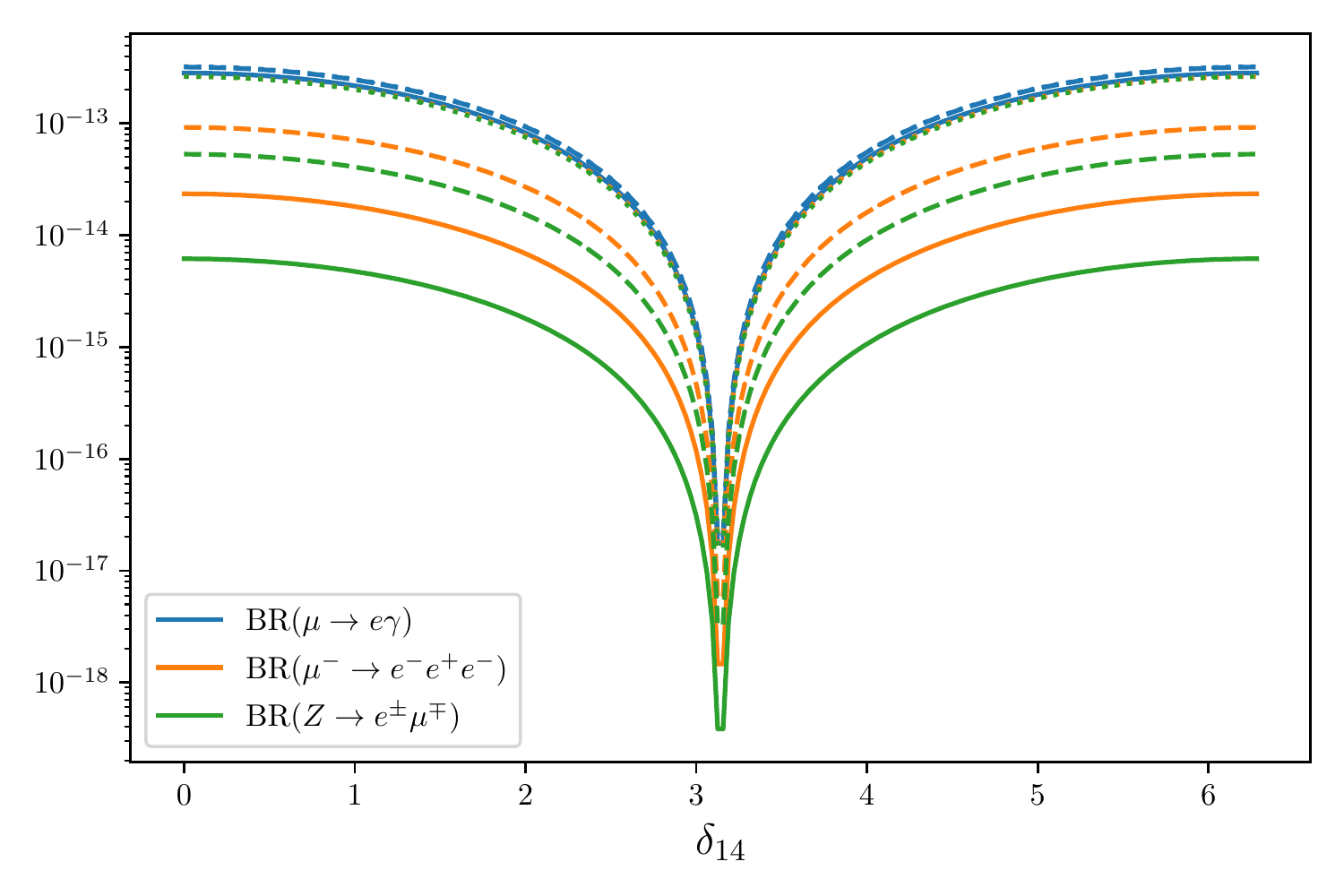} \hspace*{2mm}   
    \includegraphics[width=0.51\textwidth]{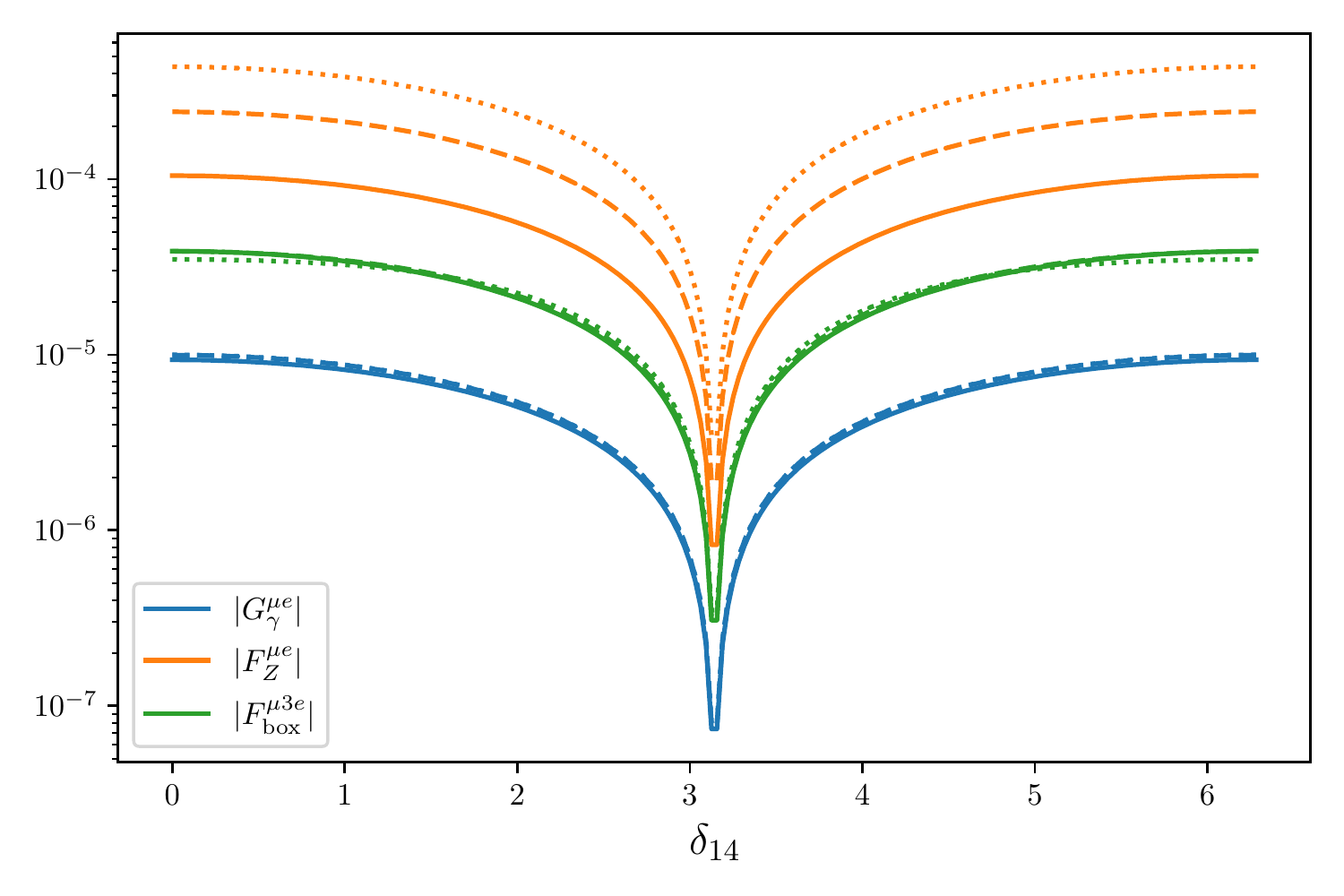}}
    \caption{Dependence of cLFV observables and several form factors (contributing to the different cLFV decay rates) on the CP violating Dirac phase $\delta_{14}$  (all other phases set to  zero).
    On the left panel we present BR($\mu \to e \gamma$) (blue), BR($\mu \to 3 e $) (orange) and BR($Z \to e \mu$) (green); on the right one finds $|G_\gamma^{\beta \alpha}|$ (blue), $|F_Z^{ \beta \alpha}|$ (orange) and $|F_\text{box}^{\beta3\alpha}|$ (green), choosing for illustrative purposes $\alpha=e$ and $\beta=\mu$. 
    In both panels, solid, dashed and dotted lines respectively correspond to the following heavy fermion masses: $m_4=m_5=1, 5, 10~\text{TeV}$. Figures from~\cite{Abada:2021zcm}.}
\label{fig:cLFV.FF.delta14}
\end{figure}

\paragraph{The role of Dirac phases} 
In Fig.~\ref{fig:cLFV.FF.delta14} we display the dependence of the above mentioned cLFV rates (and their form factors)
on the Dirac phases.
We set as an illustrative (benchmark) choice the following 
values for the mixing angles,
$\theta_{14}=\theta_{15}= 10^{-3}$, 
$\theta_{24}=\theta_{25}= 0.01$ and 
$\theta_{34}=\theta_{35}= 0$. 
Moreover, all phases are set to zero except the Dirac phase $\delta_{14}$. 
We also consider three representative values of the heavy fermion masses $m_4 = m_5 =1, 5, 10$~TeV (associated with solid, dashed and dotted lines).
As can be seen in the left panel, all considered observables have a clear dependence on $\delta_{14}$ (the only non-vanishing phase considered), with the associated rates exhibiting a strong cancellation (typically amounting to around four orders of magnitude) for $\delta_{14} = \pi$, for all considered masses of the heavy sterile states. This behaviour can be understood by considering the pattern shown by the form factors contributing to cLFV radiative and 3-body muon decays, all displaying an (analogous) suppression for $\delta_{14} = \pi$. 

Working in the limits above referred to, 
in Appendix~\ref{app:analytic.phase.observables} we present analytical expressions for the form factors contributing to the purely leptonic decays, including the full dependence on all phases. Regarding the dipole contributions, and in the case in which only 
$\delta_{14}\neq 0$, one has 
\begin{equation}\label{eq:Gmue:delta14}
    G_\gamma^{\mu e} \approx s_{1 4}s_{2 4} e^{-\frac{i}{2}(\delta_{14})} 2 \cos\left(\frac{\delta_{14}}{2}\right) G_\gamma (x_{4,5})\,,
\end{equation}
thus implying that in the simplest case of $\mu \to e \gamma$ decays, the corresponding branching fraction for the radiative decays is given by
\begin{equation}
    \mathrm{BR}(\mu \to e \gamma)\propto |G_\gamma^{\mu e}|^2 \approx 4 s_{14}^2 s_{24}^2 \cos^2\left(\frac{\delta_{14}}{2}\right)  G_\gamma^2(x_{4,5})\,,
\end{equation}
with $x_{4,5}=m_4^2/M_W^2=m_5^2/M_W^2$, thus indeed approximately vanishing for $\delta_{14}=\pi$. Similar results can be obtained for the photon penguin form factor $F_\gamma^{\mu e}$,  
as well as for one of the terms in the form factor
$F_Z^{\mu e}$ (i.e. $F_Z^{(1)}$, see Appendix~\ref{app:analytic.phase.observables}), all contributing to the rate of $\mu \to 3 e$. 
For $F_Z^{(2)}$, after carrying the sum over $\rho=e, \mu, \tau$, one finds\footnote{While one can in general neglect the contribution 
of the light (mostly active) neutrinos to the form factors here considered, that is not the case for $F_Z^{(2)}$, since the associated loop function $G_Z(x,y)$ does not vanish in the limit $x\sim 0, y \gg 1$.
As can be seen in Appendix~\ref{app:analytic.phase.observables}, despite being more complex, the term corresponding to the ``light-heavy'' contribution exhibits a similar dependence on the Dirac phases; here we only consider the dominant ``heavy-heavy'' contribution.}
\begin{equation}
F_Z^{(2)} \approx 4\,   s_{14} s_{24}  e^{-\frac{i}{2}(\delta_{14})} \left(s_{14}^2 + s_{24}^2 
\right)\, \cos\left(\frac{\delta_{14}}{2}\right) \widetilde G_Z (x_{4,5})\,,
\end{equation}
also exhibiting a suppression for $\delta_{14}=\pi$.
In the above equation we introduced $\widetilde G_Z(x_{4,5})\equiv  G_Z(x_{4,5}, x_{4,5})$, which we also use in the following for loop functions that depend on 2 parameters, in the limit of degenerate masses (cf. Appendix~\ref{app:loopfunctions_neutrinos}).
Finally, the remaining contributing term can be approximately given by 
\begin{equation}
F_Z^{(3)} \approx 4\,   s_{14} s_{24}  e^{-\frac{i}{2}(\delta_{14})} \cos\left(\frac{\delta_{14}}{2}\right)
\left[s_{14}^2  \cos(\delta_{14})
+ s_{24}^2  
\right] \widetilde H_Z (x_{4,5})\,,
\end{equation}
again revealing the same behaviour\footnote{Due to the  contributions associated with the combination $C_{ij}$ (sum over all flavours), mixings involving the tau sector also contribute; taking them into account would lead to a similar dependence on 
$\cos \delta_{14}/2$.}, which is also 
present in 
the box diagram contributions to $\mu \to 3 e$. 
The latter lead to 
\begin{align}
& F_\text{box}^{(1)} \approx  4\,   s_{14}^3 s_{24} 
\cos(\delta_{14}) \cos\left(\frac{\delta_{14}}{2}\right) 
\widetilde G_\text{box} (x_{4,5})\,, \nonumber \\
& F_\text{box}^{(2)} \approx  -8\,   s_{14}^3 s_{24} 
\cos\left(\frac{\delta_{14}}{2}\right) 
\widetilde F_\text{Xbox} (x_{4,5})\,.
\end{align}
This brief discussion explains the behaviour of the different form factors, as depicted in the right panel of 
Fig.~\ref{fig:cLFV.FF.delta14}.
Albeit carried in a 
limiting case (degenerate heavy states, identical mixings, etc.), this 
discussion is helpful in understanding the  more complex behaviours which will emerge upon the numerical analysis 
presented in Section~\ref{sec:num:analysis}. 

\bigskip
Concerning cLFV $Z$ decays\footnote{In view of the very distinct topological contributions, we do not include cLFV Higgs decays in the present work (for a recent discussion see~\cite{Arganda:2014dta,Arganda:2015naa,Arganda:2015uca,Arganda:2017vdb}).}, we do not explicitly discuss their phase dependence here; let us just notice that the form factors parameterising  $Z$ decays (cf. Section~\ref{sec:LFVZ}) can be understood as $Z$-penguins at non-vanishing momentum transfer, i.e. $q^2 = M_Z^2$, and are therefore expected to have a very similar behaviour in what concerns the impact of the CPV phases. 
This can be quantitatively confirmed in Fig.~\ref{fig:cLFV.FF.delta14} upon comparison of the dependence of the $Z$-penguin form factor $F_Z^{\mu e}$ and the branching fraction of the $Z \to e \mu$ decay on $\delta_{14}$.

\begin{figure}[h!]
    \centering
\includegraphics[width=0.48\textwidth]{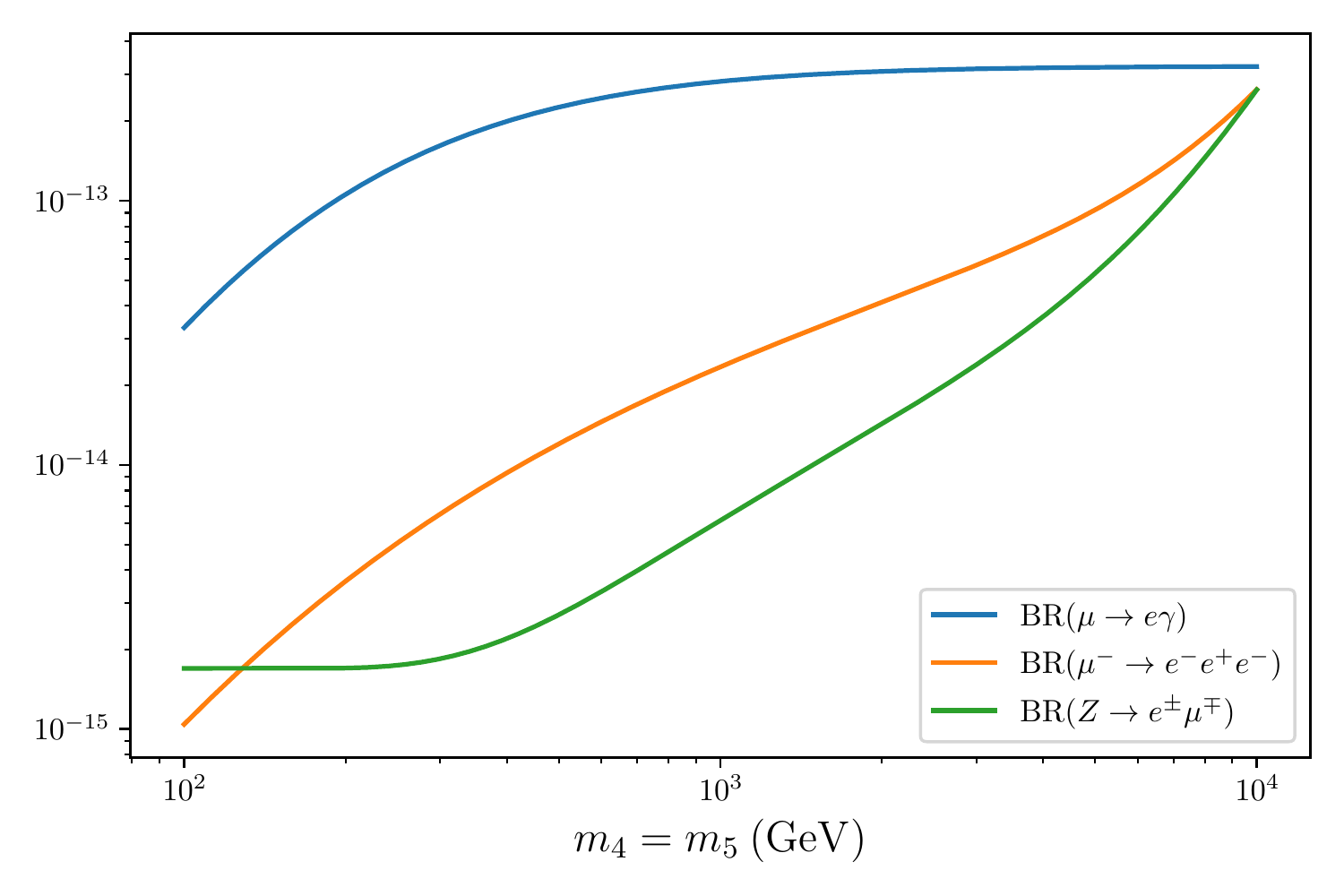}
\includegraphics[width=0.48\textwidth]{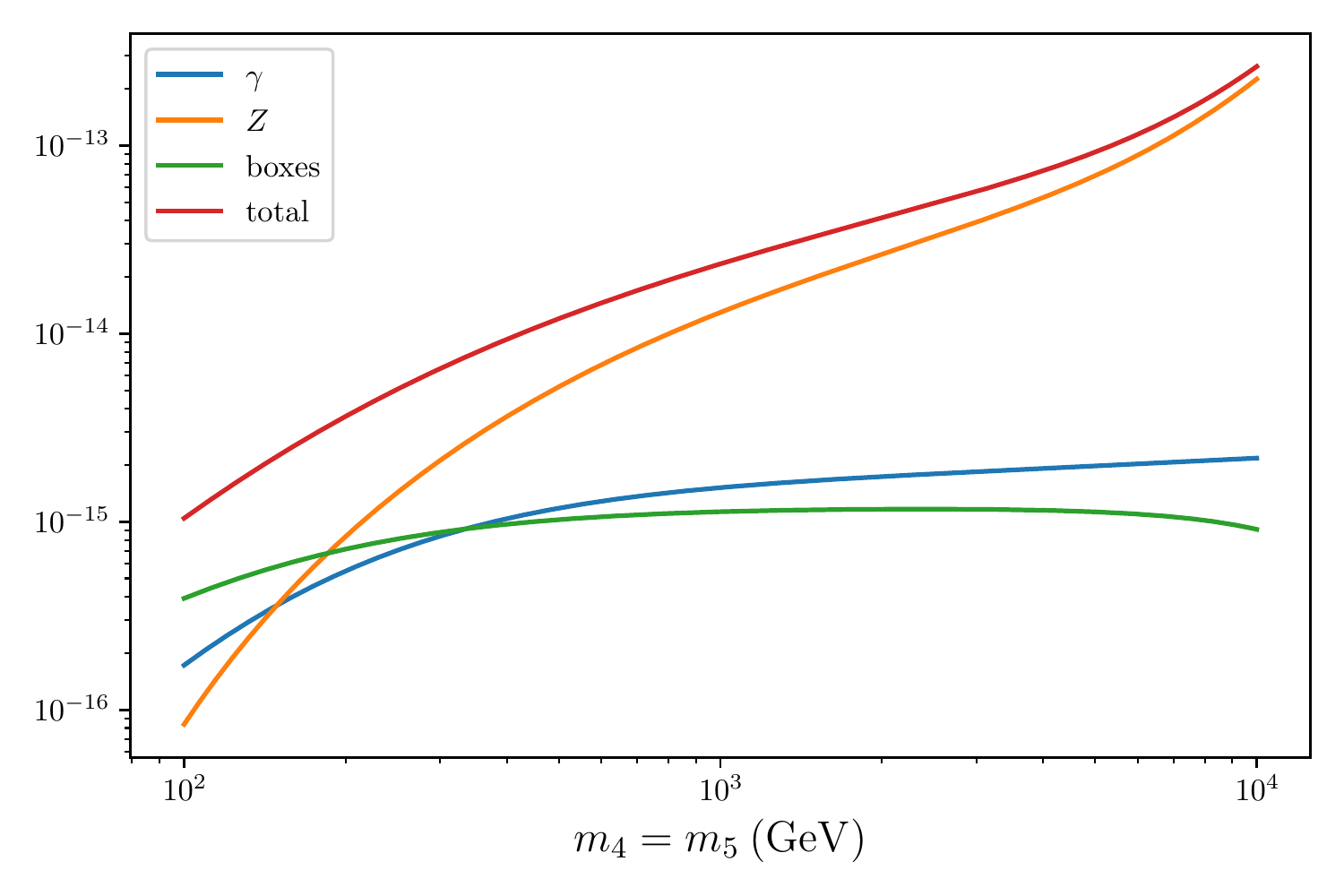}
\caption{cLFV observables (left panel) and choice of contributing form factors to the different rates (right panel), as a function of the degenerate heavy sterile mass, $m_4=m_5$ (in~GeV), for vanishing CPV phases. 
On the left panel we present
BR($\mu \to e \gamma$) (blue), BR($\mu \to 3 e $) (orange) and BR($Z \to e \mu$) (green); on the right, one finds the contributions of the $\gamma$-penguin form factors
$F_\gamma^{\beta \alpha}$ and $G_\gamma^{\beta \alpha}$ (blue), the $Z$-penguin form factor $F_Z^{ \beta \alpha}$ (orange) and the box form factor $F_\text{box}^{\beta3\alpha}$ (green) to the total branching ratio of decays of the form $\ell_\beta\to3\ell_\alpha$ (red), 
choosing for illustrative purposes $\alpha=e$ and $\beta=\mu$. 
Figures from~\cite{Abada:2021zcm}.
}
\label{fig:clfvFF_M}
\end{figure}

\bigskip
For completeness, and before moving to the possible impact of the Majorana phases, notice that the individual form factors - and hence the full rates - depend on the mass of the new fermions (already manifest from the behaviour shown in Fig.~\ref{fig:cLFV.FF.delta14}), as shown in both panels of Fig.~\ref{fig:clfvFF_M} (which were obtained considering vanishing values of all phases).

\paragraph{Impact of  Majorana phases on cLFV decay rates}

\begin{figure}[h!]
    \centering
\mbox{\hspace*{-5mm}   
    \includegraphics[width=0.51\textwidth]{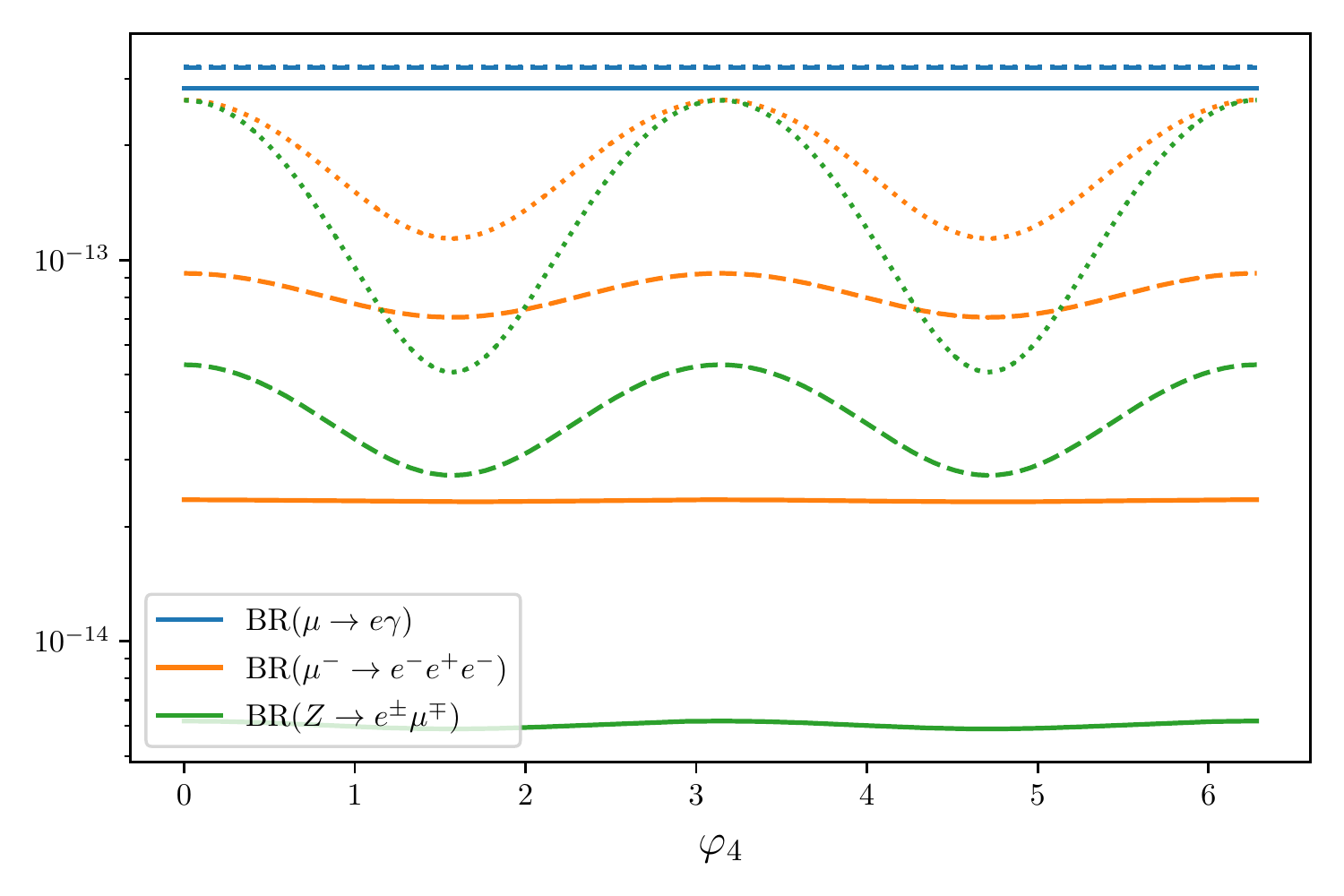} \hspace*{2mm}   
    \includegraphics[width=0.51\textwidth]{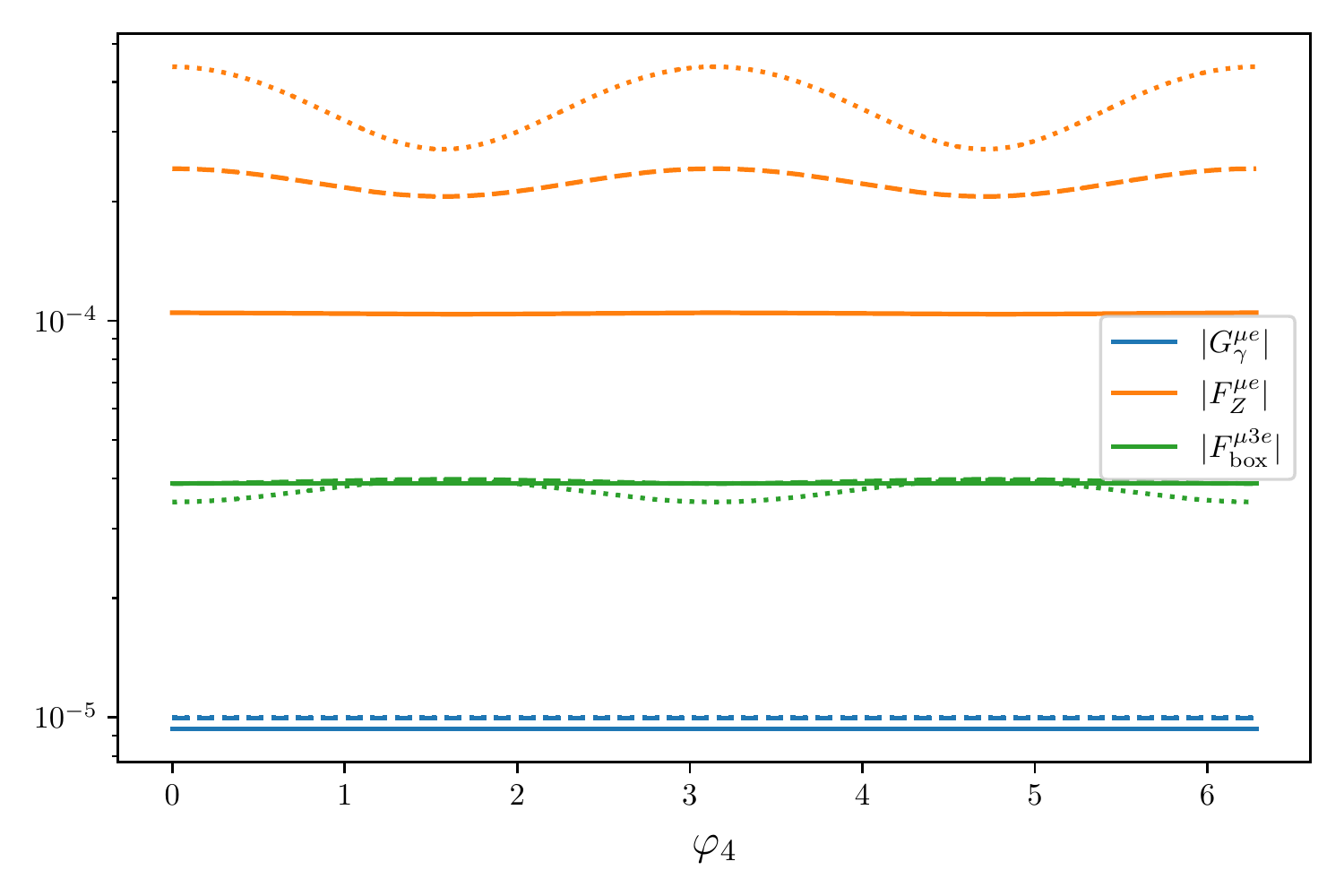}
    }
\caption{
Dependence of cLFV observables and several contributing form factors on the CP violating Majorana phase $\varphi_{4}$  (with all other phases set to  zero).
On the left panel we present
BR($\mu \to e \gamma$) (blue), BR($\mu \to 3 e $) (orange) and BR($Z \to e \mu$) (green); on the right, one has 
$|G_\gamma^{\mu e}|$ (blue), $|F_Z^{ \mu e}|$ (orange) and $|F_\text{box}^{\mu 3e}|$ (green).
In both panels, solid, dashed and dotted lines respectively correspond to $m_4=m_5=1, 5, 10~\text{TeV}$.
Figures from~\cite{Abada:2021zcm}.}
\label{fig:cLFV.FF.phi4}
\end{figure}

In what follows, we set the mixing angles to the same values as before, and choose the same three values of the heavy fermion masses. All phases are set to zero except $\varphi_{4}$.  The results for the observables (the same as studied concerning the Dirac phase) and the contributing form factors are displayed in Fig.~\ref{fig:cLFV.FF.phi4}, as a function of the Majorana phase $\varphi_{4}$. 

As expected, there  is no dependence of the radiative decays on the Majorana phase (cf. the full expression for BR($\ell_\beta \to \ell_\alpha \gamma$) given in Appendix~\ref{app:analytic.phase.observables}). This is also true 
for all dipole and dipole-like contributions.    
In contrast, the three-body decays (and the cLFV $Z$ decays) do exhibit a non-negligible dependence on the Majorana phase, as can be verified from both panels of Fig. \ref{fig:cLFV.FF.phi4}.
This is especially true for heavier mass regimes, in which the relative contribution of the form factors sensitive to $\varphi_4$ 
($Z$-penguins and to a lesser extent box-contributions)
become more important (cf. right-handed panel of Fig.~\ref{fig:clfvFF_M}).
Indeed, in the simplified limits of the form factors (see 
Appendix~\ref{app:analytic.phase.observables}), one verifies that only two contributions in the form factors depend on the Majorana phase, $F_Z^{(3)}$ and $F_\text{box}^{(1)}$. 
In the presence of a single non-vanishing Majorana phase, their expressions are:
\begin{align}\label{eq:Majorana:FF}
    & F_Z^{(3)} \approx 4s_{14}s_{24} \left(s_{14}^2 +
    s_{24}^2 \right) \cos^2(\varphi_4) \widetilde H_Z(x_{4,5}) \,, \nonumber \\
    & F_\text{box}^{(1)} \approx 4 s_{14}^3 s_{24} \cos^2(\varphi_4) \widetilde G_\text{box}(x_{4,5})\,.
\end{align}
The impact of the Majorana phase on the cLFV $Z$ decays can be also understood in analogy from the dependence of the corresponding $Z$ penguin form factor. This is readily visible from inspection of Fig.~\ref{fig:cLFV.FF.phi4}, which reveals a very similar dependence on $\varphi_4$.

\begin{figure}
    \centering
\mbox{\hspace*{-5mm}
\includegraphics[width=0.51\textwidth]{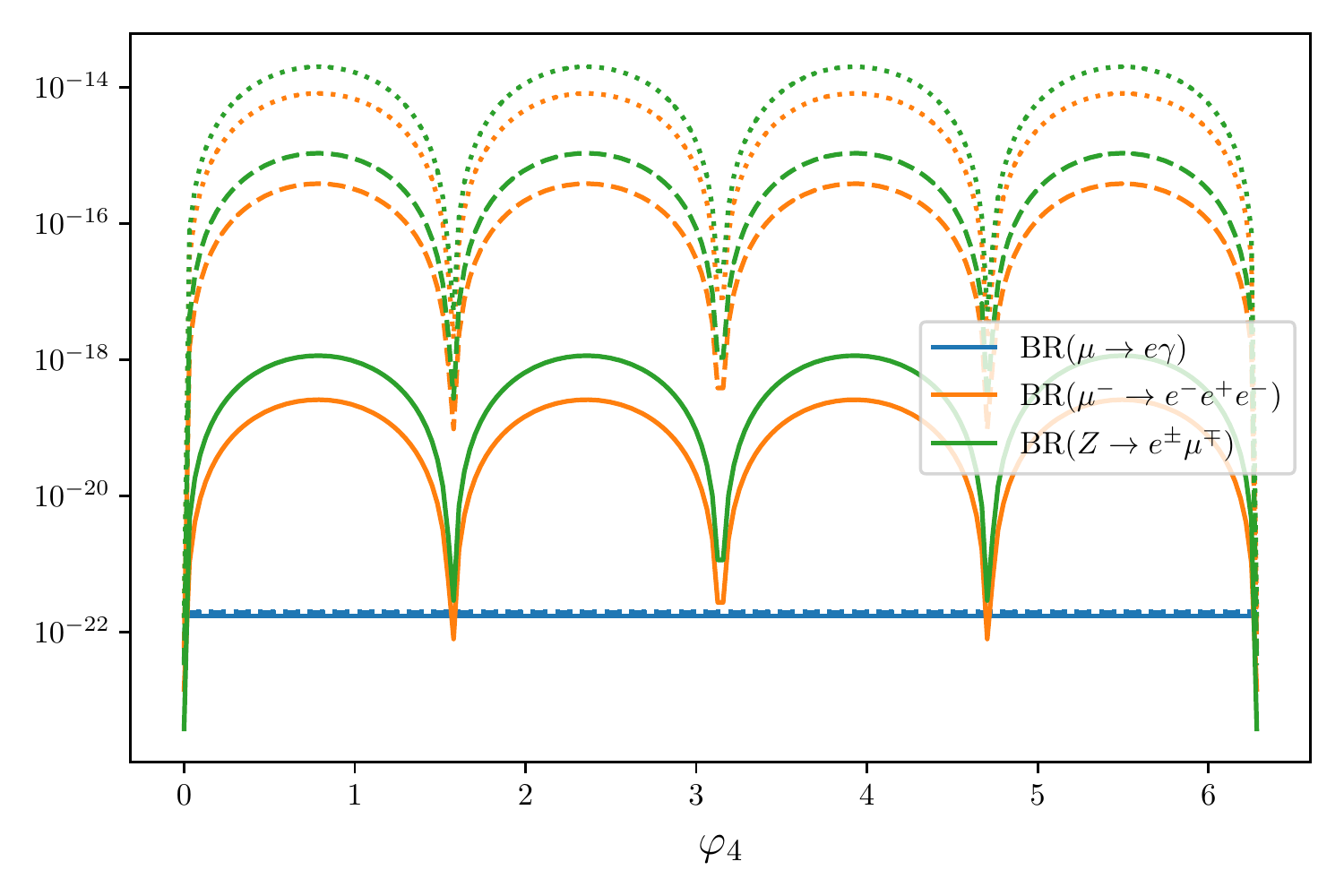} \hspace*{2mm}
    \includegraphics[width=0.51\textwidth]{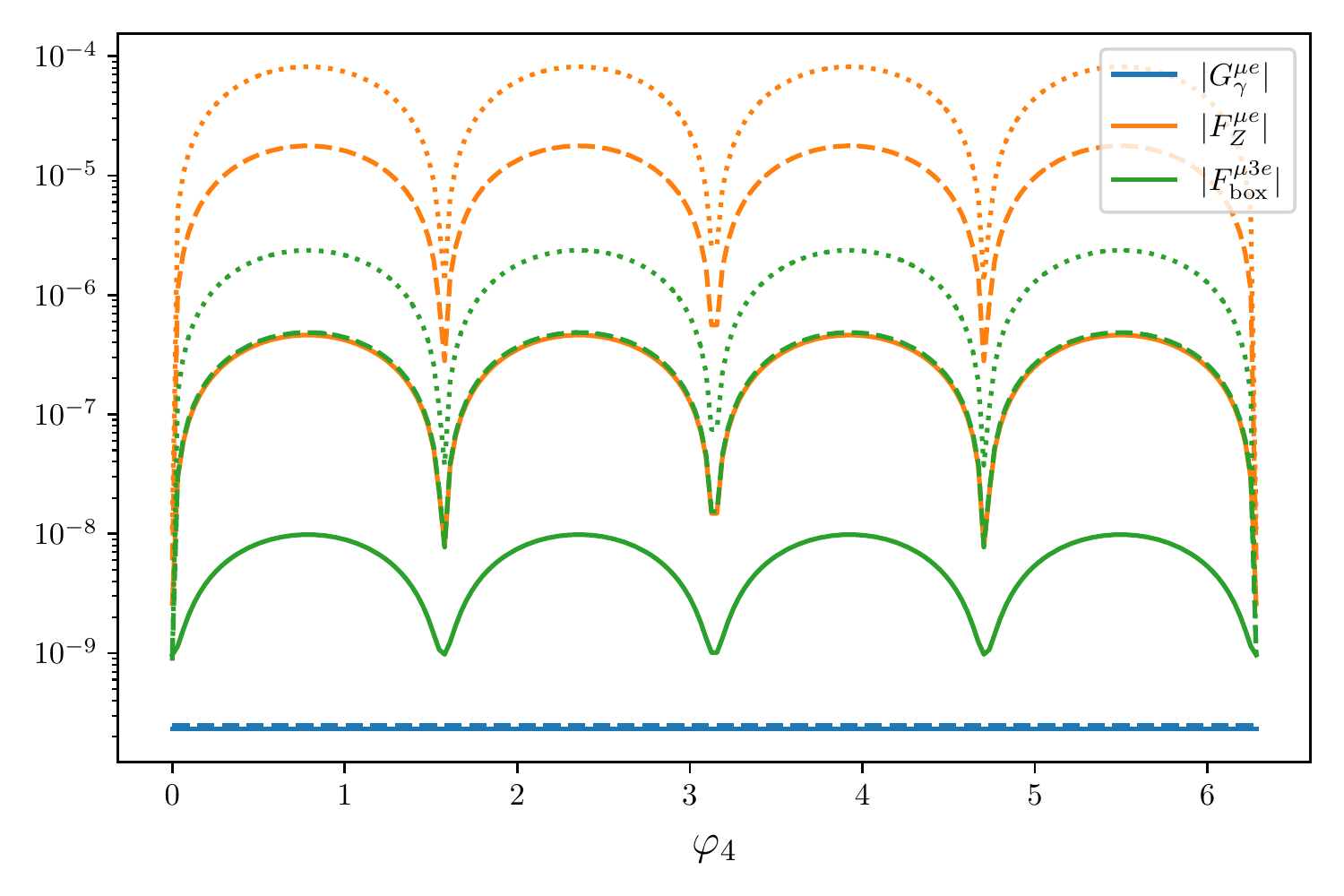}}
\caption{Dependence of cLFV observables (left) and several contributing form factors contributing to the decay rates (right) on the
Majorana CPV phase $\varphi_{4}$, for non-vanishing Dirac CPV phase, $\delta_{14}=\pi$.
We again present on the left
BR($\mu \to e \gamma$) (blue), BR($\mu \to 3 e $) (orange) and BR($Z \to e \mu$) (green), while on the right one finds $|G_\gamma^{\mu e}|$ (blue), $|F_Z^{ \mu e}|$ (orange) and $|F_\text{box}^{\mu 3e}|$ (green). We consider fixed values of the mass of heavy sterile states: 
solid, dashed and dotted lines respectively correspond to $m_4=m_5=1, 5, 10~\text{TeV}$.
Figures from~\cite{Abada:2021zcm}.}
\label{fig:clfvFF_phi_delta_pi}
\end{figure}

\paragraph{Joint Dirac-Majorana phase effects}
A first view of the joint effect of Majorana and Dirac phases can be obtained by setting one to a fixed non-vanishing value, while the other is varied over its full range (i.e. $\in [0, 2\pi ]$). This is shown in Fig.~\ref{fig:clfvFF_phi_delta_pi}, where we re-evaluate the dependence of the cLFV rates, and of a subset of form factors, on the Majorana phase $\varphi_4$ (similar to what was presented in Fig.~\ref{fig:cLFV.FF.phi4}), but now taking $\delta_{14}=\pi$. 

The effects arising from the presence of both phases 
are clearly manifest, especially when compared with the plots of Figs.~\ref{fig:cLFV.FF.delta14} and~\ref{fig:cLFV.FF.phi4}. Recall that $\delta_{14}=\pi$
was found to lead to a strong cancellation of the form factors (see discussion of Fig.~\ref{fig:cLFV.FF.delta14}); thus, the non-vanishing contributions to the observables (3-body and $Z$ decays) are associated with the form factors exhibiting a non-trivial dependence on $\varphi_4$ - as mentioned before.
Relying again on the simple analytical estimates (see Appendix~\ref{app:analytic.phase.observables}), one now verifies that taking $\delta_{14}=\pi$ leads to the following modification of the form factors sensitive to the Majorana phases (and hence of the observables): 
\begin{align}\label{eq:Majorana_fixedDirac:FF}
    & F_Z^{(3)} \propto
    s_{14}s_{24} \left(s_{14}^2 -
    s_{24}^2 \right) \sin(2\varphi_4)\,  \widetilde H_Z(x_{4,5}) \,, \nonumber \\
    & F_\text{box}^{(1)} \propto
    s_{14}^3 s_{24} \sin(2\varphi_4) \, \widetilde G_\text{box}(x_{4,5})\,,
\end{align}
explaining the modified pattern (shift of $\pi/2$ and dependence on $2\varphi_4$) visible in 
Fig.~\ref{fig:clfvFF_phi_delta_pi}.

\mathversion{bold}
\subsection{Neutrinoless $\mu-e$ conversion in nuclei and CP violating phases}
\begin{figure}[h!]
    \centering
\mbox{\hspace*{-5mm}
\includegraphics[width=0.51\textwidth]{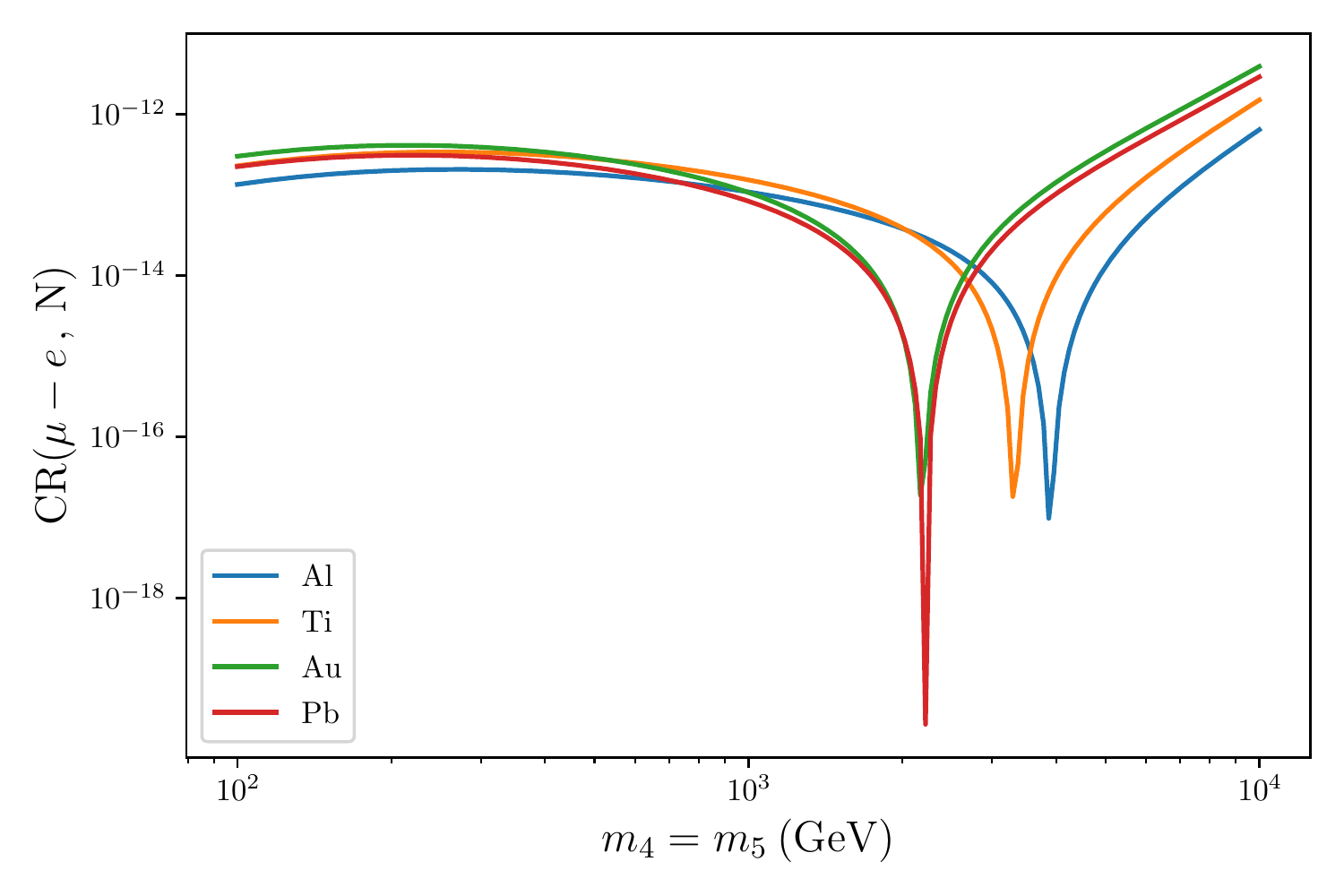} \hspace*{2mm}
\includegraphics[width=0.51\textwidth]{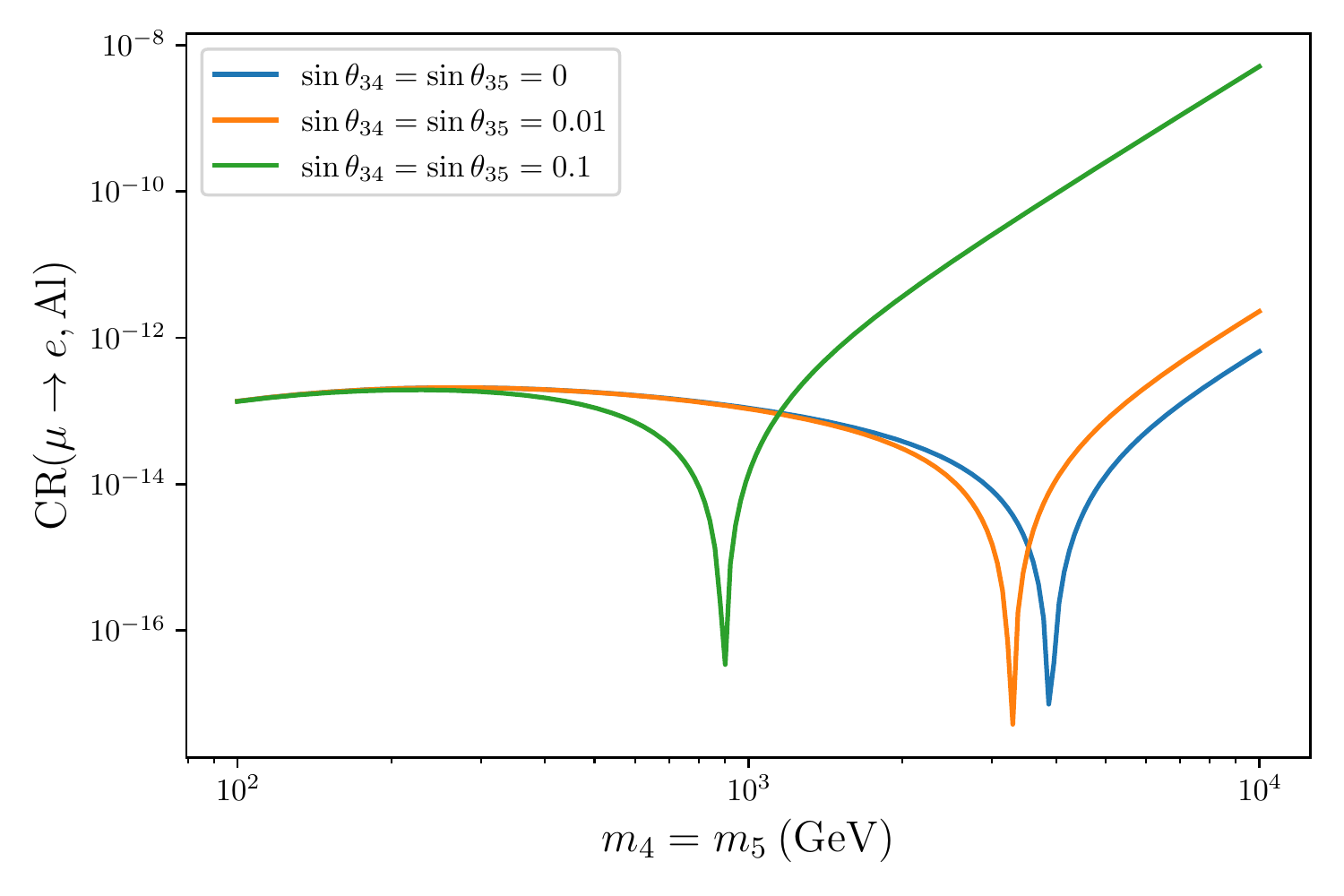}}
\caption{Neutrinoless $\mu-e$ conversion in nuclei as 
a function of the degenerate heavy sterile mass, $m_4=m_5$ (in~GeV). On the left, we set 
$\theta_{1j}=10^{-3}$, $\theta_{2j}=0.01$ and $\theta_{3j}=0$ ($j=4,5$), and consider different muonic atoms: Aluminium (blue), Titanium (orange), Gold (green) and Lead (red). On the right, CR($\mu-e$, Al) for different values of the 
tau-sterile mixing angles: $\theta_{3j}=0$ (blue), $\theta_{3j}=0.01$ (orange) and  $\theta_{3j}=0.1$ (green), again with $\theta_{1j}=10^{-3}$, $\theta_{2j}=0.01$ (with $j=4,5$).
Figures from~\cite{Abada:2021zcm}.}
\label{fig:CR_M:nuclei_tau}
\end{figure}

\mathversion{normal}
\label{sec:muecon_simple}
We begin by illustrating how the predictions for the conversion rate (in particular the dependence on the heavy fermion mass) reflect the nature of the muonic atom, i.e. 
the chosen target nucleus (see Eq.~(\ref{eq:def:CRfull})). This is shown on the left panel of Fig.~\ref{fig:CR_M:nuclei_tau}, where we display CR($\mu-e$, N) as a function of the (degenerate) heavy masses, for Aluminium, Titanium, Gold and Lead nuclei, which have been chosen for past and future experimental searches. We fix the active-sterile mixing angles as before ($\theta_{14}=\theta_{15}=10^{-3}$, $\theta_{24}=\theta_{25}=0.01$ and $\theta_{34}=\theta_{35}=0$), and set all phases to zero. We recover the behaviour originally pointed out in~\cite{Alonso:2012ji}, with the distinct rates all exhibiting a sharp cancellation for a given value  of the heavy fermion mass  - which we denote $m_{4,5}^c$ (the actual value of $m_{4,5}^c$, and the ``depth'' observed in the conversion rate, both depend on the considered nucleus).    

Until now, and for simplicity, we have not considered mixings of the sterile states to the third generation of leptons; however, and even though we are studying cLFV in the $\mu-e$ sector, certain observables are sensitive to tau mixings through the $C_{ij}$ coupling (sum over all flavours) which arises from the $Z$-penguin contribution, cf. Eq.~(\ref{eq:cLFV:FF:FZ}). On the right panel of Fig.~\ref{fig:CR_M:nuclei_tau}, we consider Aluminium nuclei, and again depict CR($\mu-e$, Al) vs. the heavy sterile masses, for different choices of $\theta_{3j}$ ($j=4,5$), keeping the other mixing angles and phases as before.  
As expected, and despite seemingly indirect, the impact is significant, both to the rate itself, and in what concerns the value of $m_{4,5}^c$ associated with the cancellation in the conversion rate. This is all the most important when $\theta_{3j} \gg \theta_{1j},\theta_{2j}$ 
($j=4,5$), as can be inferred from the green line. 

\begin{figure}
    \centering
\mbox{\hspace*{-5mm}\includegraphics[width=0.51\textwidth]{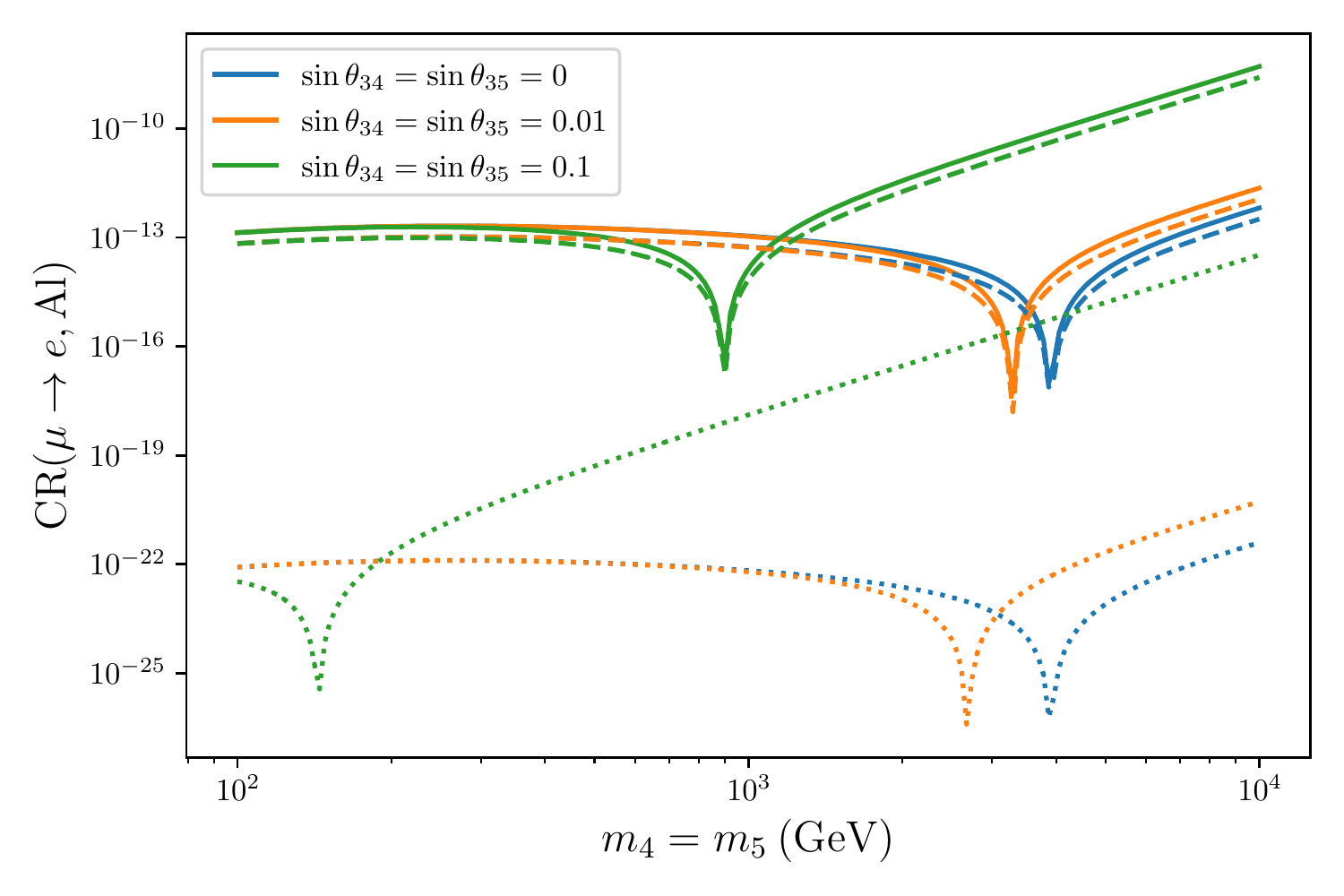}
\hspace*{2mm}
\includegraphics[width=0.51\textwidth]{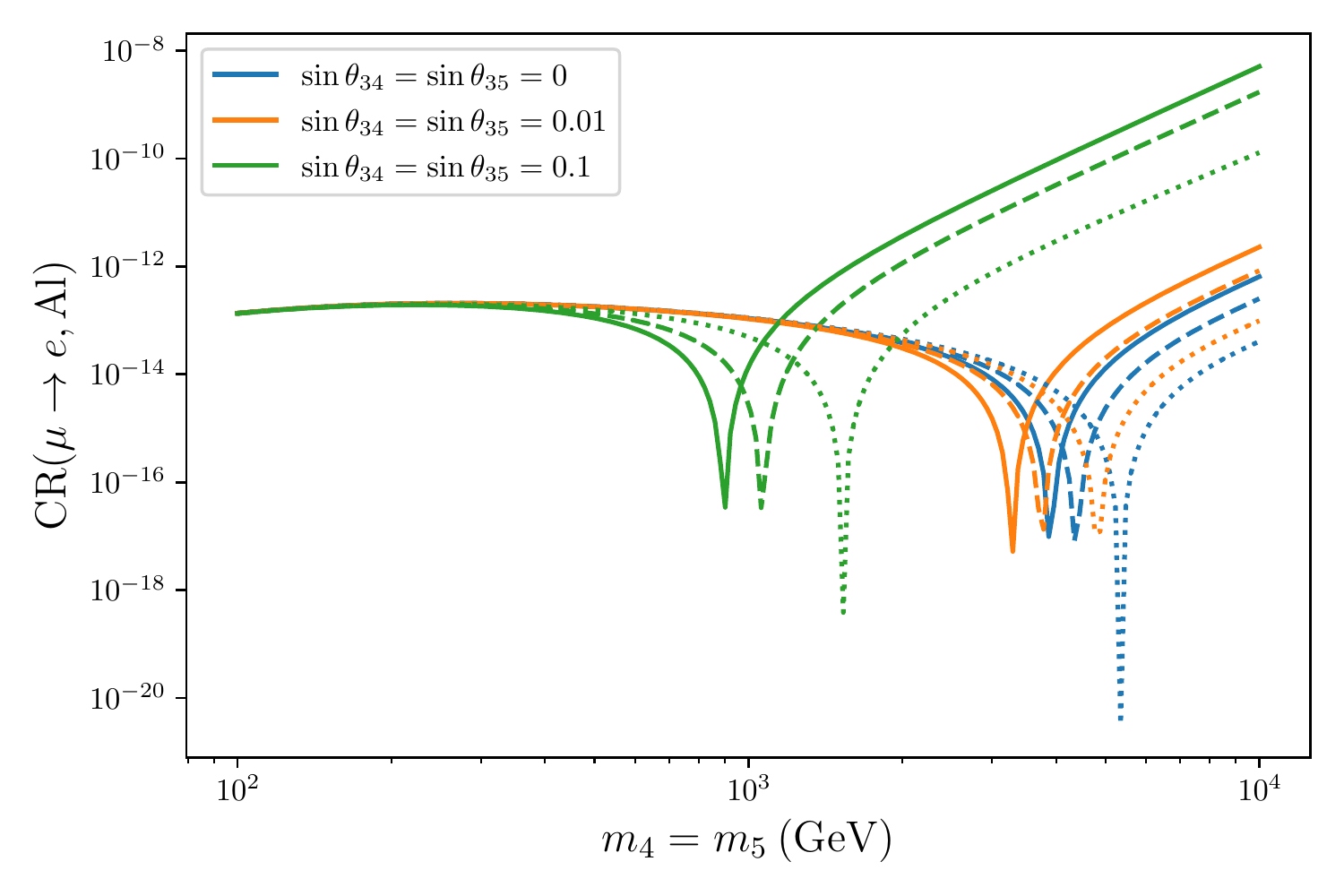}}
\caption{Neutrinoless $\mu - e$ conversion in Aluminium as 
a function of the degenerate heavy sterile mass, $m_4=m_5$ (in~GeV).  We set $\theta_{1j}=10^{-3}$, $\theta_{2j}=0.01$, for different values of the tau-sterile mixing angles: $\theta_{3j}=0$
 (blue), $\theta_{3j}=0.01$ (orange) and  $\theta_{3j}=0.1$ (green), with $j=4,5$.
 On the left, we set all Majorana phases to zero and consider three choices of the Dirac phase, $\delta_{14}=0, \pi/2$ and $\pi$, respectively corresponding to solid, dashed and dotted lines. Conversely, on the right panel
  all Dirac phases are set to zero, and we consider three choices of the Majorana phase, $\varphi_4=0, \pi/4$ and $\pi/2$, corresponding to solid, dashed and dotted lines.
  Figures from~\cite{Abada:2021zcm}.
  }
\label{fig:CR_Al_M:cancel_phases}
\end{figure}

The impact of the CPV phases on the conversion rate is studied in both panels of Fig.~\ref{fig:CR_Al_M:cancel_phases}: for three choices of $\theta_{3j}=0, 0.01$ and $0.1$, we consider the impact of Dirac and Majorana phases on 
CR($\mu-e$,~Al), displayed as a function of the masses of the heavy states. 
For the case of vanishing $\varphi_{4}$, the choice of Dirac phases leads to effects which could already be expected from the discussion of the previous subsection\footnote{We notice that although the full expression is considerably more involved, the form factors contributing to the conversion rate include those already presented for the radiative and 3-body cLFV decays. Additional ones (i.e. boxes with an internal quark line) only depend on a single combination of $\mathcal{U}_{e i}^{\phantom{\ast}}\,\mathcal{U}_{\mu i}^\ast$, and exhibit a behaviour similar to the dipole contributions, being only sensitive to Dirac phases.}. 
In particular, and as can be seen on the left panel of Fig.~\ref{fig:CR_Al_M:cancel_phases}, notice the very important suppression for $\delta_{14}=\pi$; for the latter case, and for sizeable 
$\theta_{3j}$ one also observes a significant displacement of $m_{4,5}^c$, lighter by almost an order of magnitude. This is the result of a cancellation of (numerically) very small terms. 

The effect of the Majorana phases (for vanishing Dirac phases) is more interesting: while having already a visible effect on the overall scale of CR($\mu-e$, Al) - a reduction by a factor $\sim 100$ between $\varphi_4=0$ and  
$\varphi_4=\pi/2$ -, they also modify the value of $m_{4,5}^c$
($m_{4,5}^c (\varphi_4=0)\approx 850~\text{GeV}$ while $m_{4,5}^c (\varphi_4=\pi/2)\approx 1.5~\text{TeV}$). 

\begin{figure}[h!]
\mbox{ \hspace*{-5mm}   \centering
\includegraphics[width=0.51\textwidth]{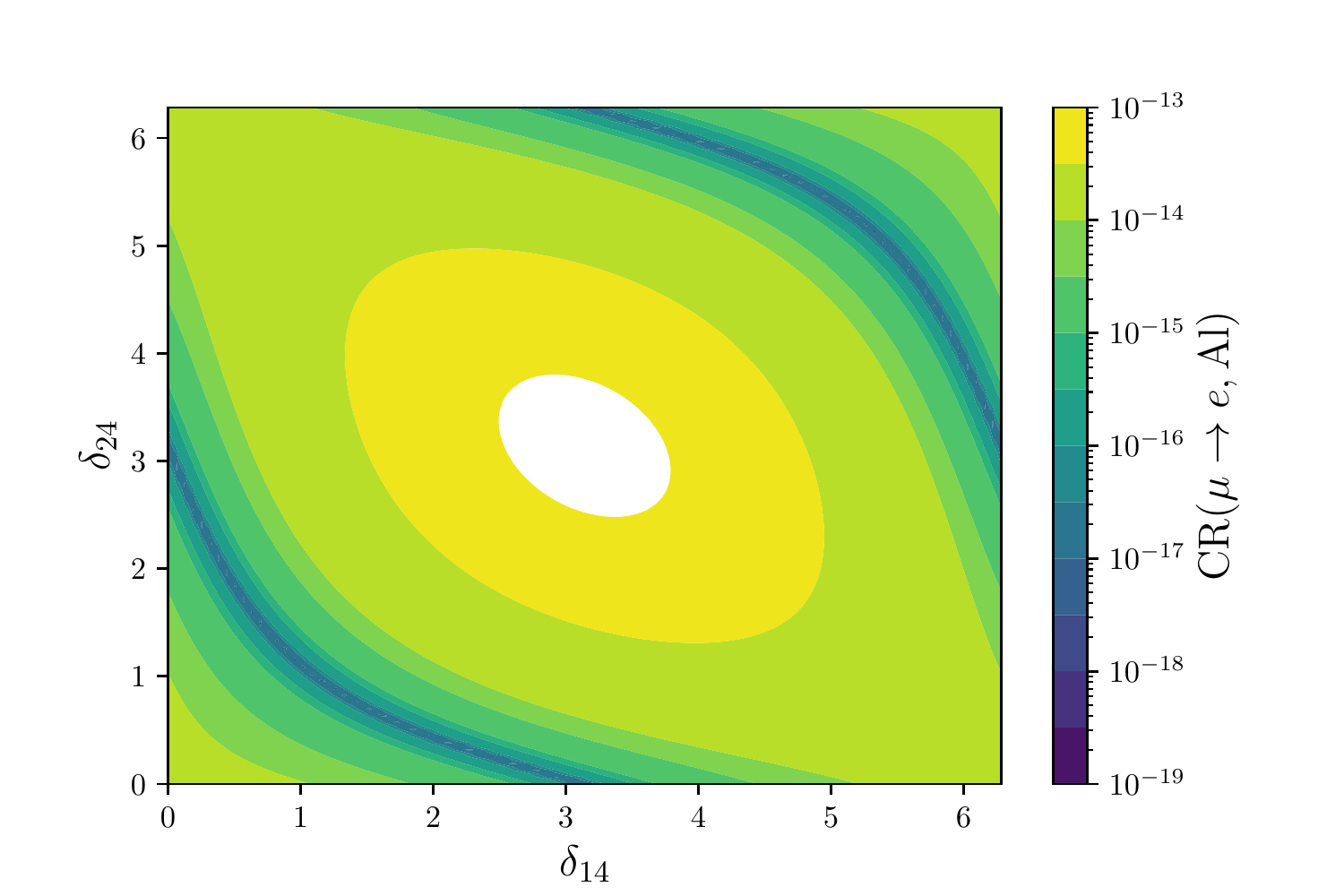}\hspace*{2mm}
\includegraphics[width=0.51\textwidth]{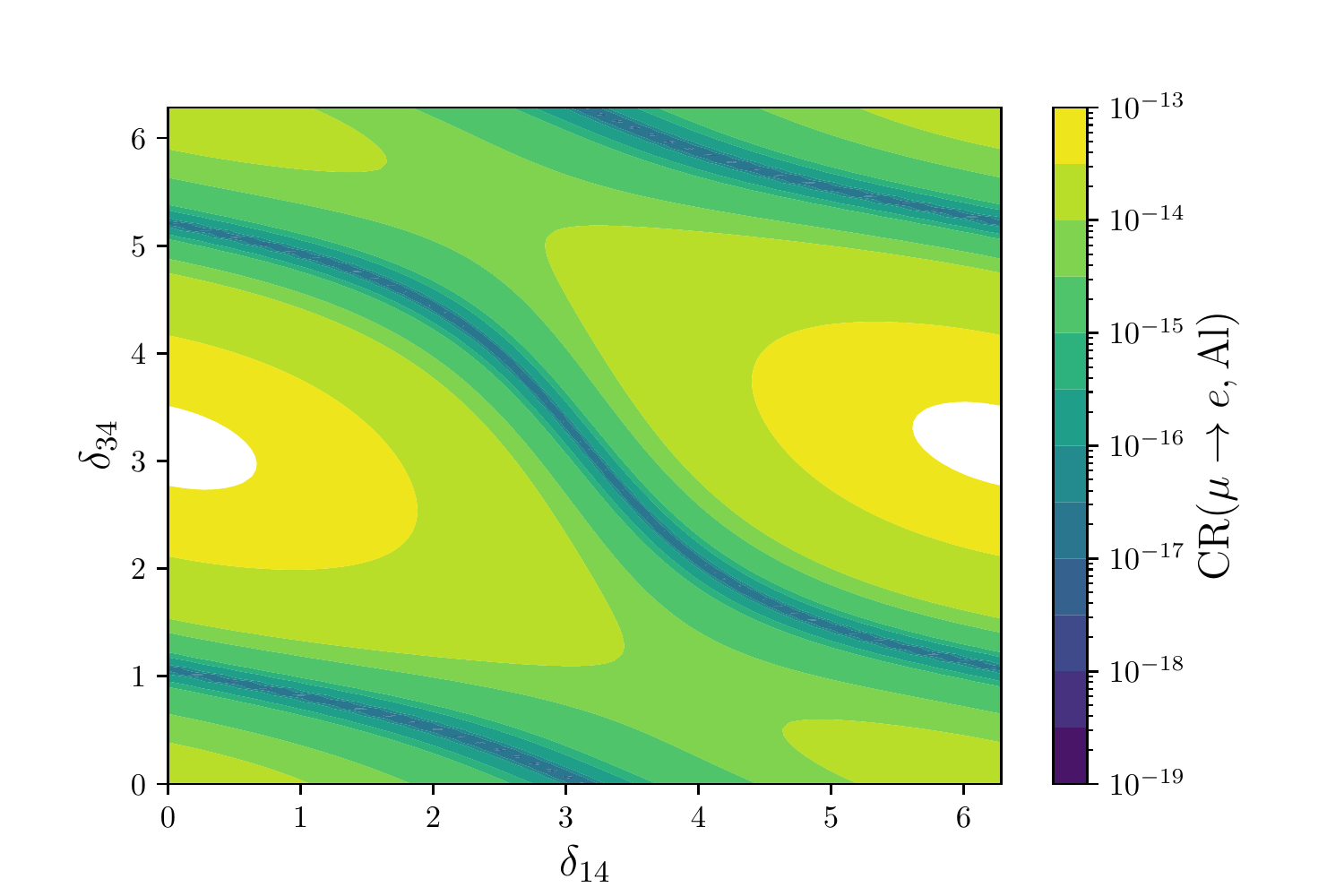}}\\
\mbox{\hspace*{-5mm}
\includegraphics[width=0.51\textwidth]{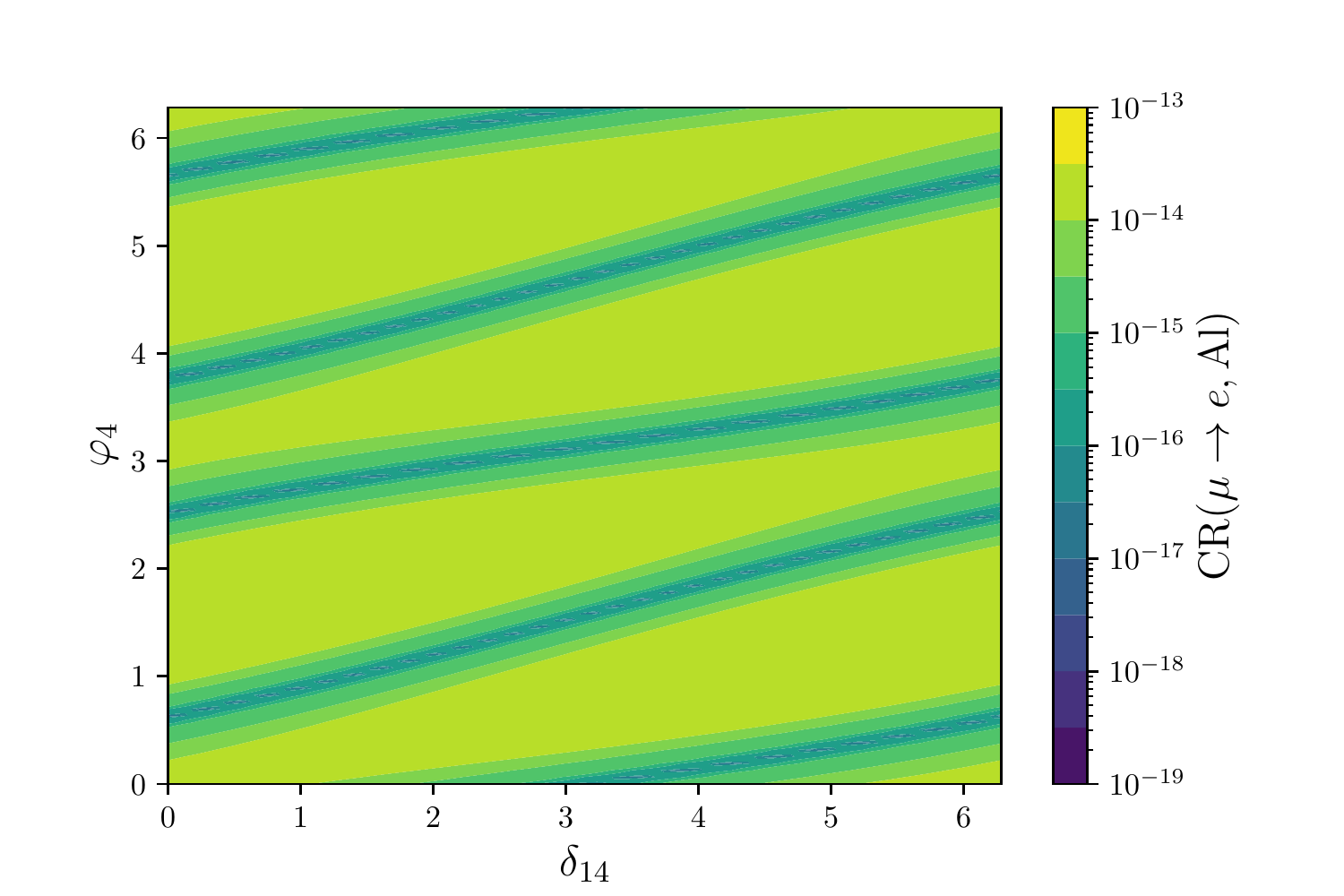}\hspace*{2mm}
\includegraphics[width=0.51\textwidth]{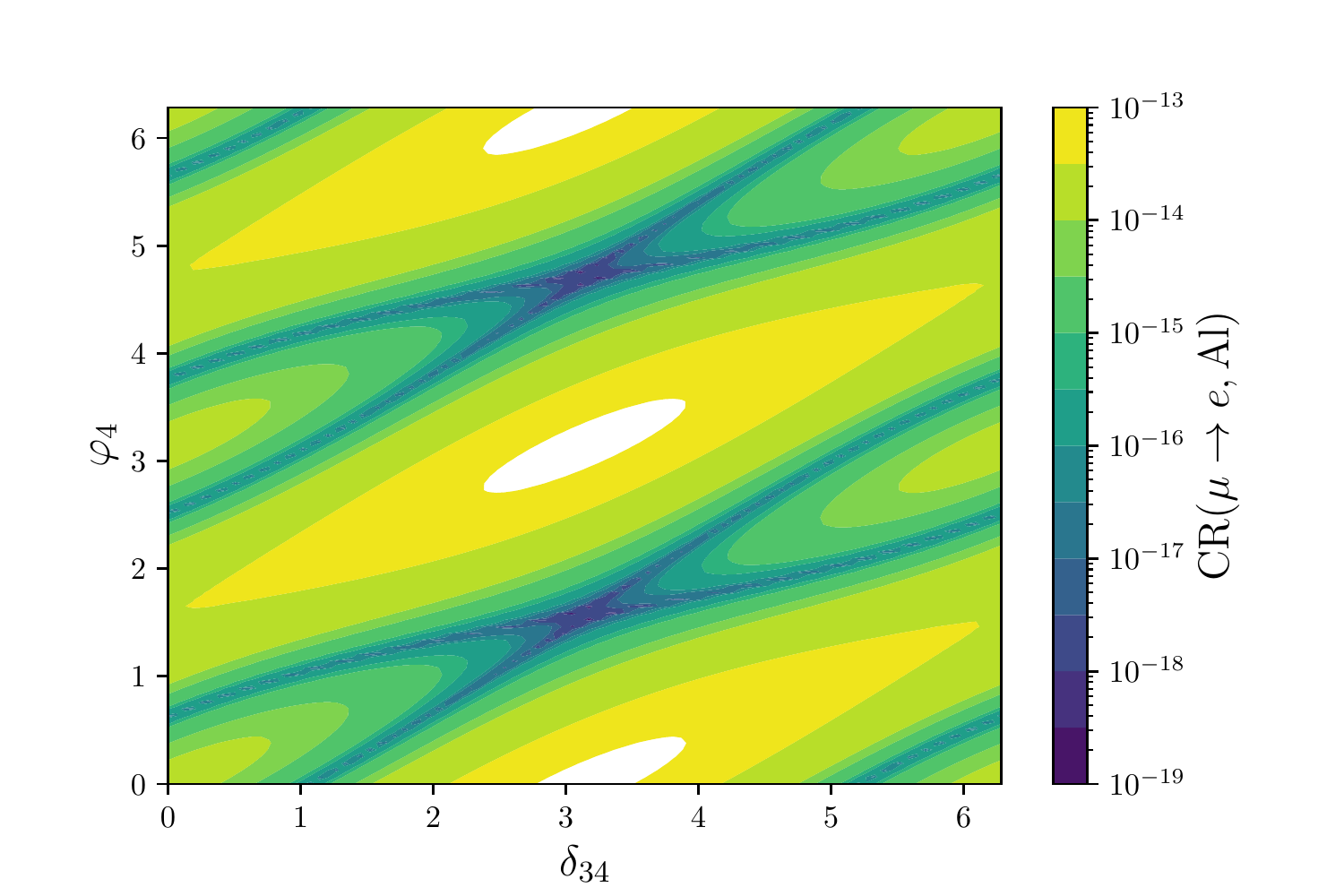}}
\caption{Contour plots for cLFV $\mu-e$ conversion in Aluminium, for fixed values of the degenerate heavy sterile mass, $m_4=m_5=1$~TeV, 
for $\theta_{1j}=10^{-3}$, $\theta_{2j}=0.01$ and $\theta_{3j}=0.1$ ($j=4,5$) and varying CPV phases: 
on the top row, spanned by pairs of Dirac phases, $(\delta_{14}-\delta_{24})$ and 
$(\delta_{14}-\delta_{34})$, respectively left and right panels; bottom row, spanned by Dirac-Majorana phases, 
$(\delta_{14}-\varphi_{4})$ and 
$(\delta_{34}-\varphi_{4})$, respectively left and right panels. 
The colour scheme denotes the associated value of CR($\mu-e$, Al) as indicated by the colour bar to the right of each plot (white regions denote CR($\mu-e$, Al)$> 10^{-13}$).
Figures from~\cite{Abada:2021zcm}.}
\label{fig:CR_Al:contour_phases}
\end{figure}

Finally, we consider for illustrative purposes the simultaneous effects of two phases (Dirac, or combined Dirac and Majorana). This is shown in the contour plots of Fig.~\ref{fig:CR_Al:contour_phases}, for which we again consider $m_4=m_5=1$~TeV, and fix the active-sterile mixings to 
$\theta_{1j}=10^{-3}$, $\theta_{2j}=0.01$ and $\theta_{3j}=0.1$ ($j=4,5$).

The different plots in Fig.~\ref{fig:CR_Al:contour_phases} summarise (and generalise) the previous findings of Figs.~\ref{fig:CR_M:nuclei_tau} and~\ref{fig:CR_Al_M:cancel_phases}: for fixed values of the ``standard'' input parameters (i.e. the mass of the heavy states and the active-sterile mixing angles), a variation of the phases - individually or as a joint effect - can lead to significant
changes in the predictions for the conversion rate.
This is seen in the panels of Fig.~\ref{fig:CR_Al:contour_phases}, with the rates ranging from as low as $10^{-18}$ (dark blue), to values above $10^{-14}$ (bright yellow), or even beyond 
$10^{-13}$ (white).
Here, one can also observe regions of constructive interference, which are of different origin.
For example, in the upper left plot of Fig.~\ref{fig:CR_Al:contour_phases} we show the conversion rate as a function of $\delta_{14}$ and $\delta_{24}$; if both phases are close to $\pi$, the suppression vanishes (as it depends on  $\cos((\delta_{14}-\delta_{24})/2)$), while the complex exponential multiplying the contributions depends on $(\delta_{14}+\delta_{24})/2$, thus leading to a ``sign-flip'' for values of $\delta_{14}\simeq\delta_{24}\simeq\pi$ (cf. App.~\ref{app:analytic.phase.observables}).
This affects the signs of the individual contributions
to the different form factors, and can lead to an overall constructive interference, as visible in the plots.
In the remaining panels in which one observes an enhancement of the rate with respect to vanishing CPV phases (conversion rate as a function of $\delta_{14}$ and $\delta_{34}$, and $\delta_{34}$ and $\varphi_{4}$ respectively), the source of constructive interference solely lies in the $Z$-penguin form factor. 
In this case, the interference occurs between terms that depend on the tau-sterile mixing angles $\theta_{3j}\,,\:j=4,5$ and the remaining terms, in conjunction with the effects of other phases.

\subsection{Muonium anti-Muonium oscillations}
Other cLFV observables rely on combinations of the already discussed form factors, so we will not address them individually. However, a few remarks concerning Muonium oscillations\footnote{The cLFV Muonium decays are also expected to be impacted in the same way, as can be understood from the corresponding form factors collected in Appendix~\ref{app:analytic.phase.observables}.} are in order, as $\text{Mu}-\overline{\text{Mu}}$ is unique in the sense that it only receives contributions from box diagrams, and thus offers a direct access to this topology (and associated form factors).
Notice that the oscillation probability is proportional to a single effective four-fermion coupling $G_{M\overline{M}}$~\cite{Clark:2003tv,Cvetic:2005gx}.
Moreover, as can be seen from the corresponding expressions for $G_{M\overline{M}}$ (see Eq.~(\ref{eq:Gmumu}) in Appendix~\ref{sec:intromuonium}), this observable is only sensitive to $\mu-e$ flavour violation. 

\begin{figure}
    \centering
    \mbox{ \hspace*{-5mm}   \centering
\includegraphics[width=0.51\textwidth]{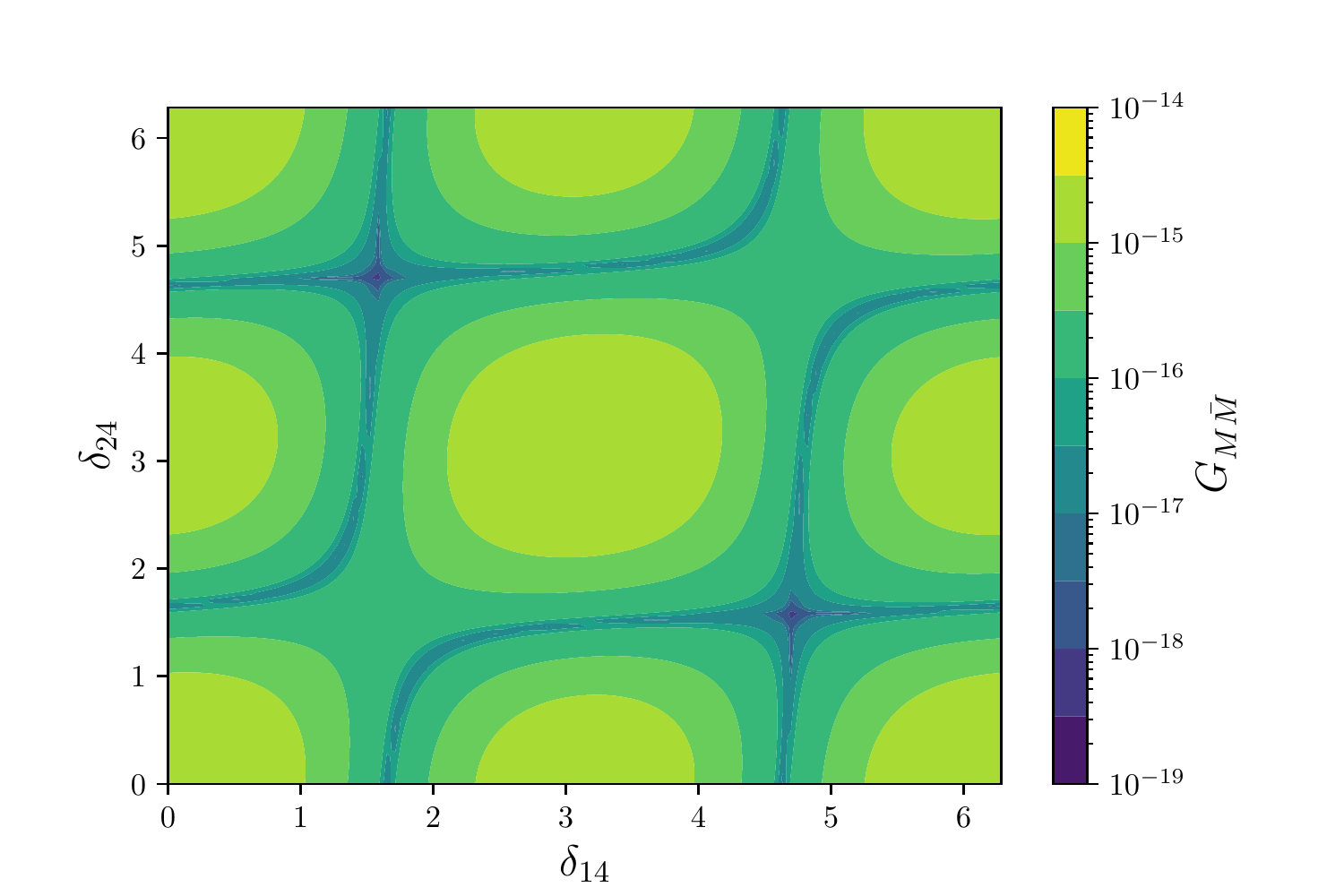}\hspace*{2mm}
\includegraphics[width=0.51\textwidth]{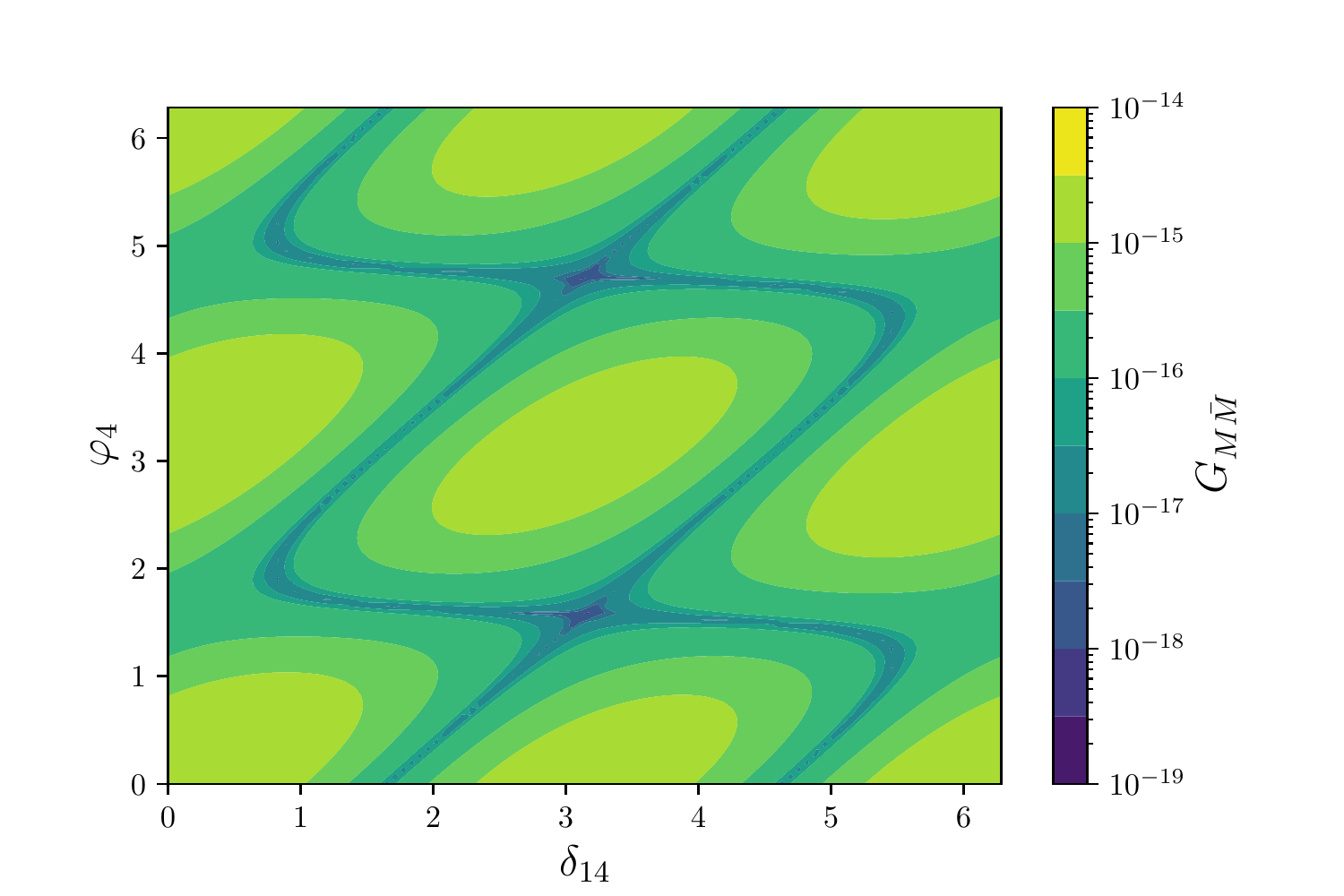}}
    \caption{Contour plots for the effective coupling $G_{M\overline{M}}$ of $\mathrm{Mu}-\overline{\mathrm{Mu}}$ oscillations, for fixed values of the degenerate heavy sterile mass, $m_4=m_5=1$~TeV, with 
$\theta_{1j}=10^{-3}$, $\theta_{2j}=0.01$ 
and varying pairs of CPV phases: 
$(\delta_{14}-\delta_{24})$ and 
$(\delta_{14}-\varphi_{4})$, respectively in the left and right panels.
Colour code as in Fig.~\ref{fig:CR_Al:contour_phases}.
Figures from~\cite{Abada:2021zcm}.}
    \label{fig:GMM:contour_phase}
\end{figure}

In Fig.~\ref{fig:GMM:contour_phase}, we present contour plots for the effective coupling $G_{M\overline{M}}$, spanned by varying pairs of Dirac CP violating phases ($\delta_{14}$ and $\delta_{24}$ on the left panel), and Dirac-Majorana CPV phases 
($\delta_{14}$ and $\varphi_{4})$ on the right panel). As previously done, we set $m_4=m_5=1$~TeV, and 
$\theta_{1j}=10^{-3}$, $\theta_{2j}=0.01$  ($j=4,5$). 
(Since they are solely generated by box diagrams, $\text{Mu}-\overline{\text{Mu}}$ oscillations are not sensitive to $\theta_{3j}$.)

\mathversion{bold}
\subsection{cLFV and CP violating phases in the 
$\tau-\ell$ sectors}
\mathversion{normal}

Leptonic cLFV tau decays (i.e. $\tau \to \ell_\alpha \gamma$ or  $\tau \to 3\ell_\alpha$
with $\alpha=e, \mu$) receive contributions from form factors whose structure is analogous to that of the corresponding muon decays (allowing for the different flavour composition of the final state leptons); likewise, tau leptons can be present as final states of cLFV $Z$ decays. Since the dependence of the observables on the CPV phases is in all similar to what has been discussed  for the $\mu-e$ sector, we refrain from a dedicated analytical study (these observables will be included in the numerical study of Section \ref{sec:num:analysis}).
Notice that due to the large tau mass, one can also have semi-leptonic cLFV tau decays, with the final state composed of a light lepton and (light) mesons. However, we will not address them in the present study. 

\subsection{Other possible enhancements}
\label{sec:enh_others}
\begin{figure}[t!]
    \centering
\mbox{   \hspace*{-5mm}  \includegraphics[width=0.51\textwidth]{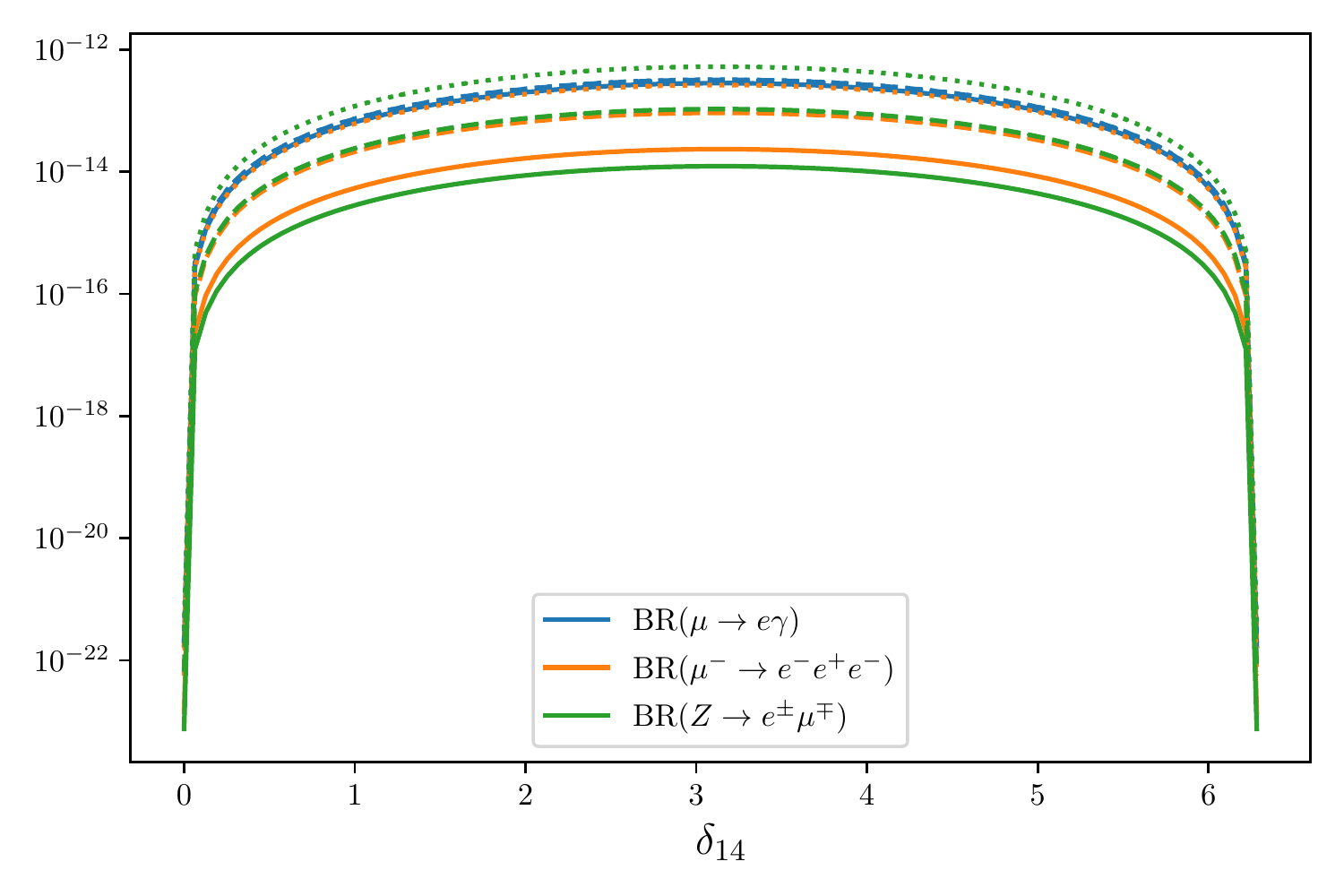}\hspace*{2mm} 
    \includegraphics[width=0.51\textwidth]{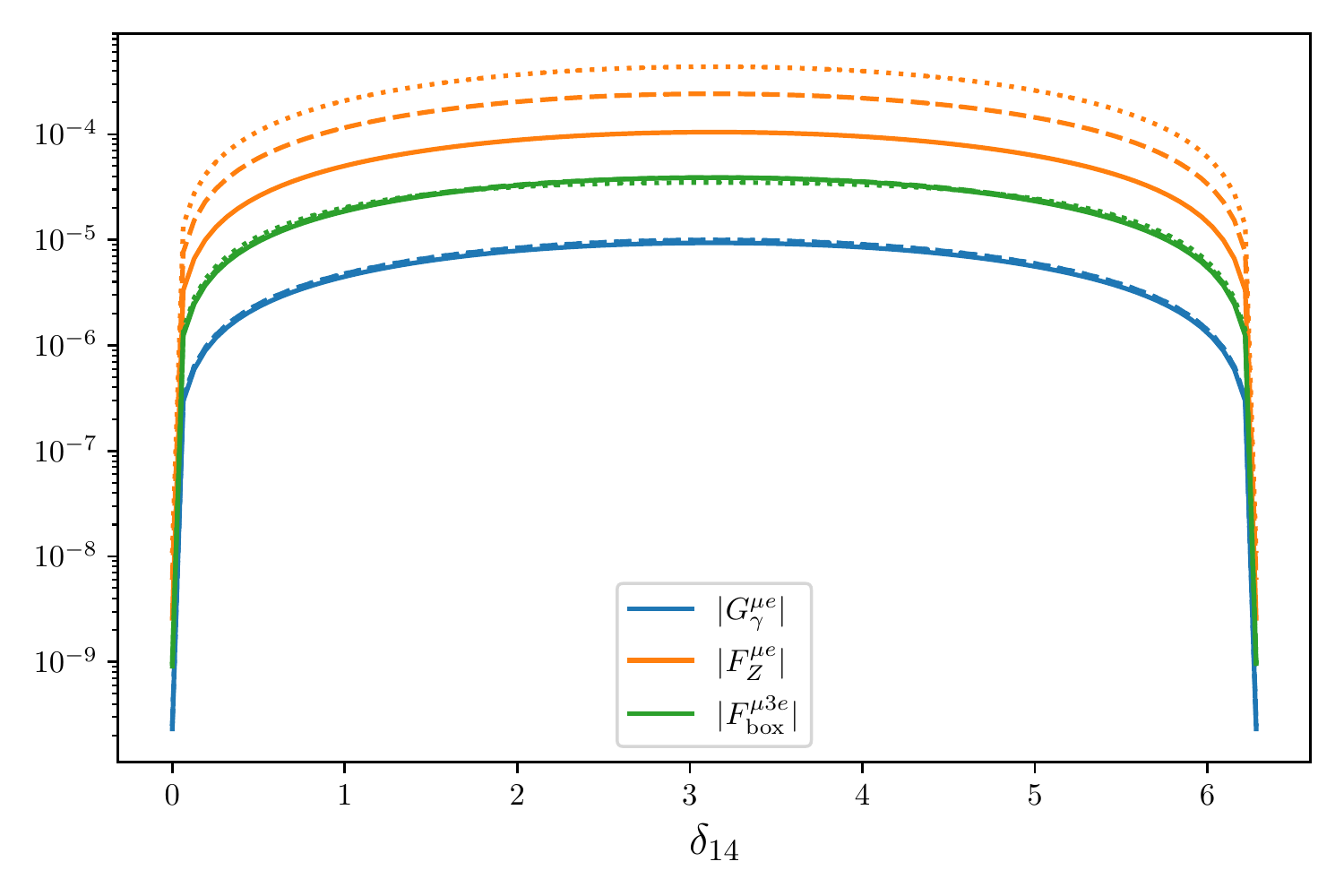}}\\
\mbox{  \hspace*{-5mm}   \includegraphics[width=0.51\textwidth]{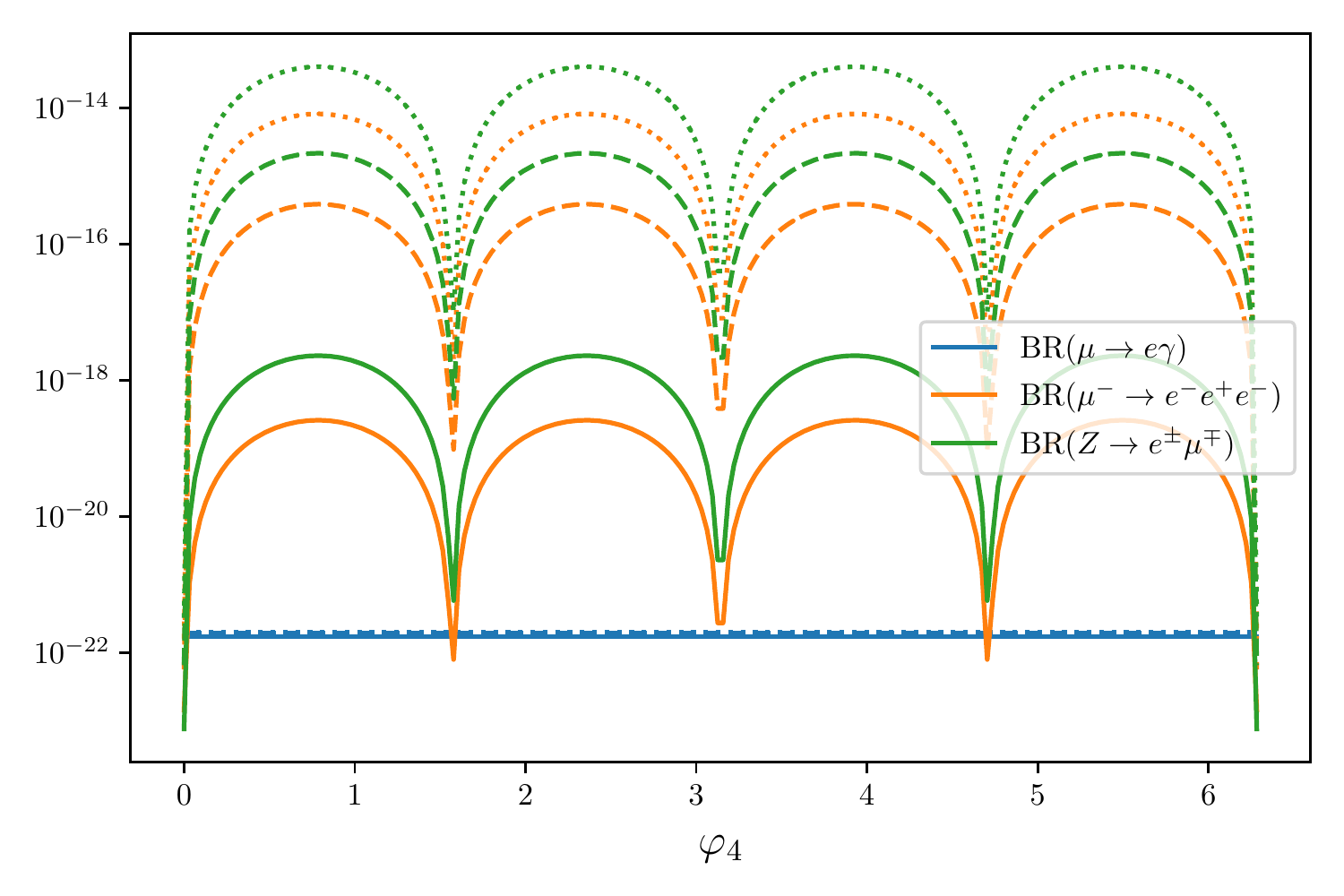}\hspace*{2mm}
    \includegraphics[width=0.51\textwidth]{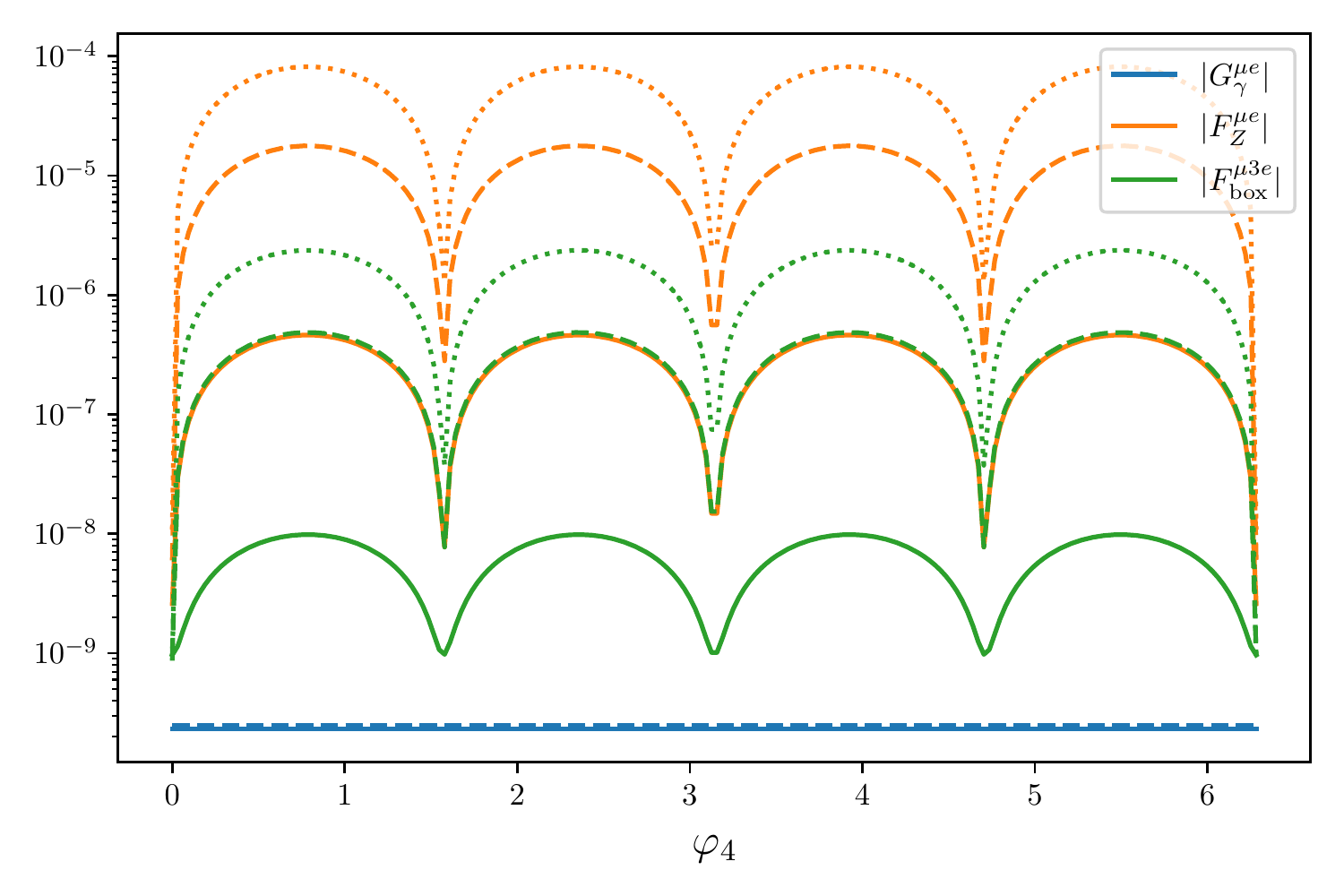}}
    \caption{Dependence of cLFV observables and several form factors (contributing to the different decay rates) on the CPV phases $\delta_{14}$ and $\varphi_4$, 
    as done in Figs.~\ref{fig:cLFV.FF.delta14} and~\ref{fig:cLFV.FF.phi4}, but with $\sin\theta_{15} = - \sin\theta_{14}$. 
    Figures from~\cite{Abada:2021zcm}.
    }
    \label{fig:cLFV.FF.delta14.signs}
\end{figure}

So far, and relying on the simplifying approximation $\theta_{\alpha 4} \approx \theta_{\alpha 5}$, we have mostly addressed effects of destructive interference leading to a strong suppression of the cLFV
observables (due to a cosine dependence of the corresponding form factors on the CPV phases, see Eqs. (\ref{eq:Gmue:delta14}-\ref{eq:Majorana_fixedDirac:FF})). 
However, in the most general case, 
different behaviours (in particular generic enhancements)
can be encountered upon 
relaxation of $\theta_{\alpha 4} \approx \theta_{\alpha 5}$. 
For example, by considering a simple sign difference in one of the flavours ($\theta_{14} = - \theta_{15}$), we are led to a generic cancellation as in 
this case, one finds a sinus-like dependence of the observables on the phases. 
For instance the photon dipole form factor is now given by 
\begin{equation}\label{eq:Gmue:delta14:minus}
    G_\gamma^{\mu e} \approx - i s_{1 4}s_{2 4} e^{-\frac{i}{2}(\delta_{14})} 2 \sin\left(\frac{\delta_{14}}{2}\right)  G_\gamma (x_{4,5})\,,
\end{equation}
to be compared with Eq.~(\ref{eq:Gmue:delta14}). 
Thus, non-vanishing phases now generically lead to enhancements, which can be quite important.
This is illustrated in Fig.~\ref{fig:cLFV.FF.delta14.signs}, where we show results analogous to those displayed in  Figs.~\ref{fig:cLFV.FF.delta14} and~\ref{fig:cLFV.FF.phi4},  but now taking $\theta_{14} = - \theta_{15}$.
It can be seen that the opposite sign of $\theta_{14}$ and $\theta_{15}$ effectively leads to the same behaviour as a shift in $\delta_{14} \to \delta_{14} + \pi$ (see Figs.~\ref{fig:cLFV.FF.delta14} and~\ref{fig:clfvFF_phi_delta_pi}). 

\section {Phenomenological study: interference effects of CPV phases}\label{sec:num:analysis}
Following the simple analysis of the previous section, which allowed a clear view of the effect of the CPV Dirac and Majorana phases, we now proceed to a more complete numerical study. In what follows, we will survey in a comprehensive way the simple ``3+2 toy model'', carrying a phenomenological analysis of a larger set of cLFV observables, now taking into account the available experimental constraints that were discussed in Chapters~\ref{chap:lepflav} and~\ref{sec:numassgen}. The latter include limits on the active-sterile mixings, results of direct and indirect searches for the heavy states, and EW precision tests, among many others. Finally, current bounds on searches for cLFV transitions are taken into account (cf. Table~\ref{tab:cLFVdata}).

In order to further explore the effects of the new non-vanishing CPV phases in a realistic way, we perform a random scan of the active-sterile mixing angles\footnote{Here we relax the assumption of $\theta_{\alpha 4} = \pm \theta_{\alpha 5}$ by means of adding gaussian noise with a relative $1\,\sigma$ deviation of $10\,\%$ to the samples, i.e.
$\theta_{\alpha 5} = \pm\theta_{\alpha 4} \pm 10\,\%\,$.}
for $m_4 = m_5 = 1,\,5\:\mathrm{TeV}$.  
Leading to the numerical results displayed in this section, we first perform a scan with the phases set to zero and (randomly) select $2000$ points consistent with all experimental data at the $3\,\sigma$ level (corresponding to the blue points in the plots presented in this section).
For each of the selected points we then randomly vary the phases $\delta_{14}$, $\delta_{24}$, $\delta_{34}$, and $\varphi_4$ in the interval $(0, \,2\pi)$, drawing 100 samples from a uniform distribution (shown in orange).
Finally, we further add to the data set 
additional points which  correspond to having systematically 
varied the phases on a grid for the ``special'' values $\{0, \frac{\pi}{4}, \frac{\pi}{2}, \frac{3\pi}{4}, \pi\}$ (shown in green).
This procedure allows to exhaustively study the impact of having non-vanishing CPV phases, especially regarding correlations between observables.
Due to computational limitations, in this section we still do not take into account $\varphi_5$ nor $\delta_{i5}$; in the limit of degenerate masses, and nearly degenerate angles, non-vanishing values of the latter can be understood as leading to a phase shift 
(e.g. $\propto \cos(\varphi_4 - \varphi_5)$, see Appendix~\ref{app:analytic.phase.observables}). 

\mathversion{bold}
\subsection{Correlation of $\mu-e$ observables} 
\begin{figure}[h!]
    \centering
\mbox{ \hspace*{-5mm}     \includegraphics[width=0.51\textwidth]{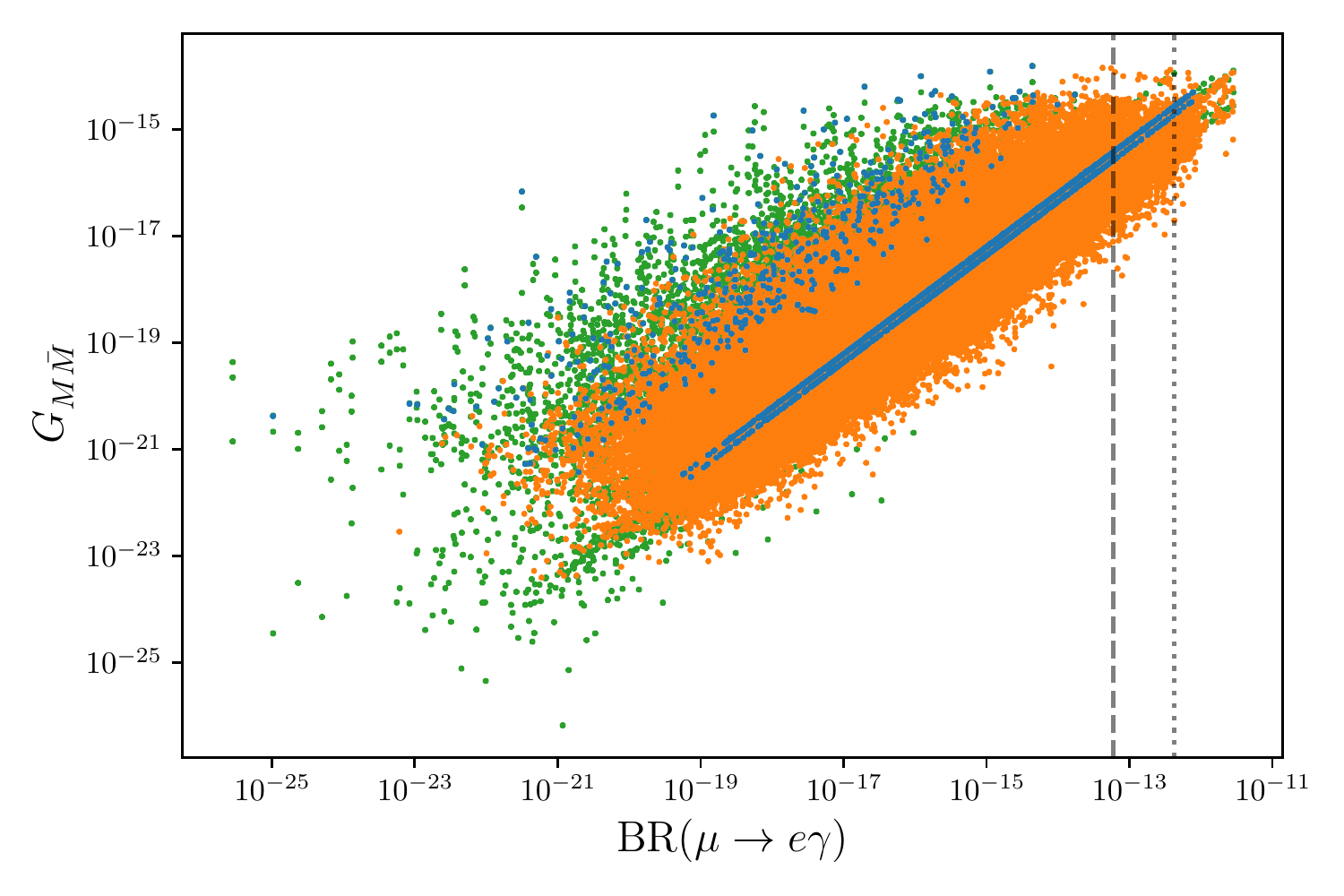}\hspace*{2mm}
    \includegraphics[width=0.51\textwidth]{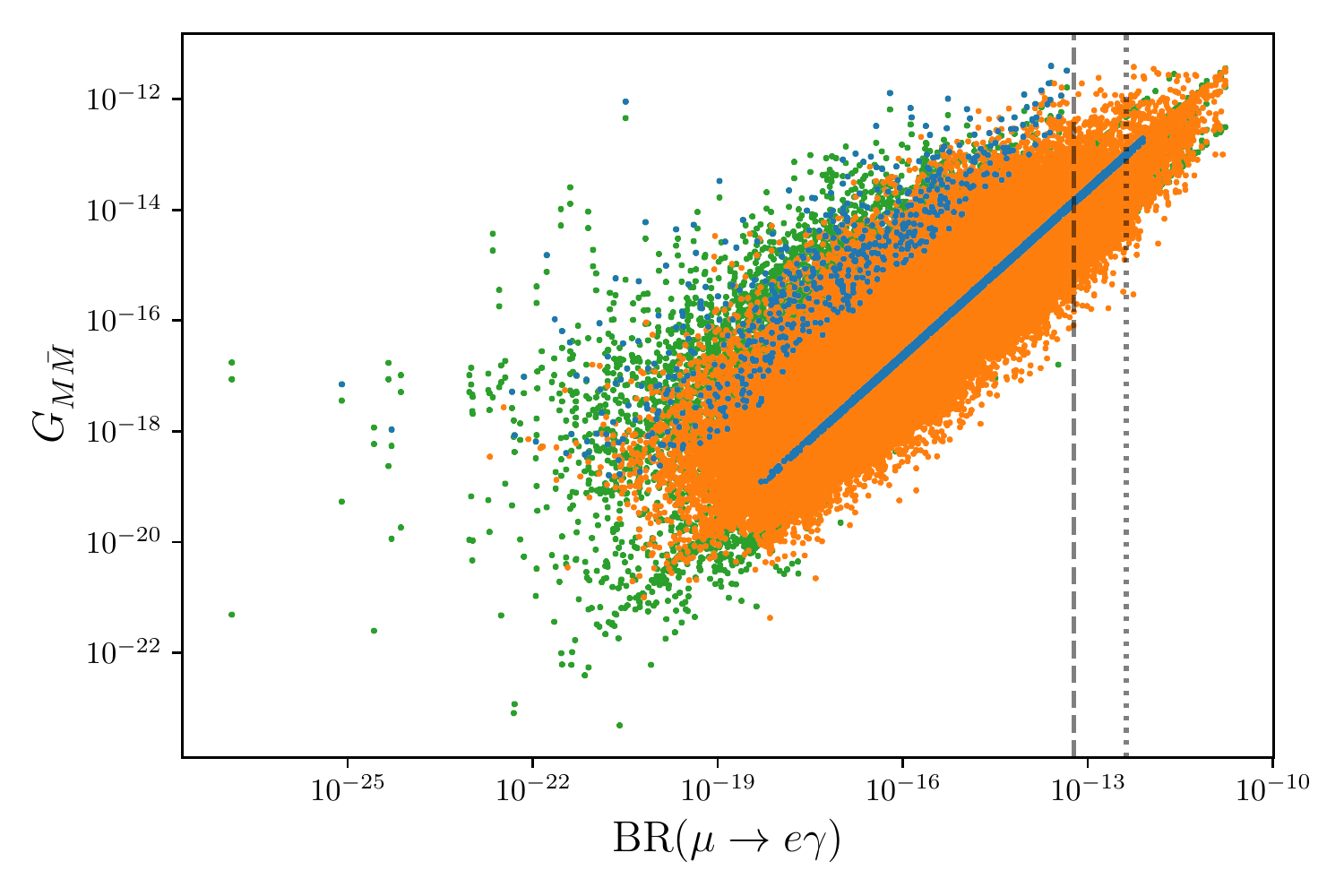}}\\
\mbox{ \hspace*{-5mm}     \includegraphics[width=0.51\textwidth]{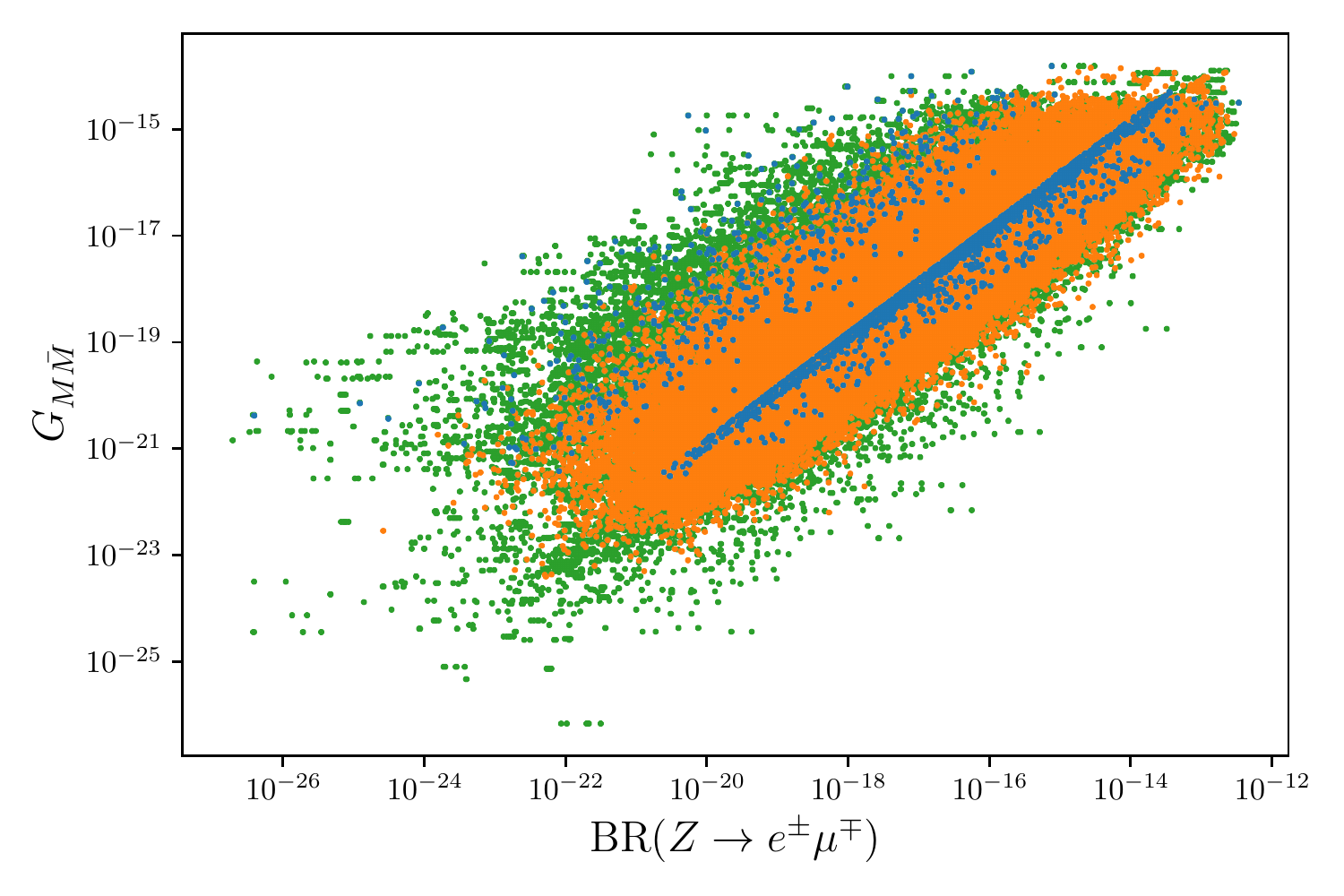}\hspace*{2mm}
    \includegraphics[width=0.51\textwidth]{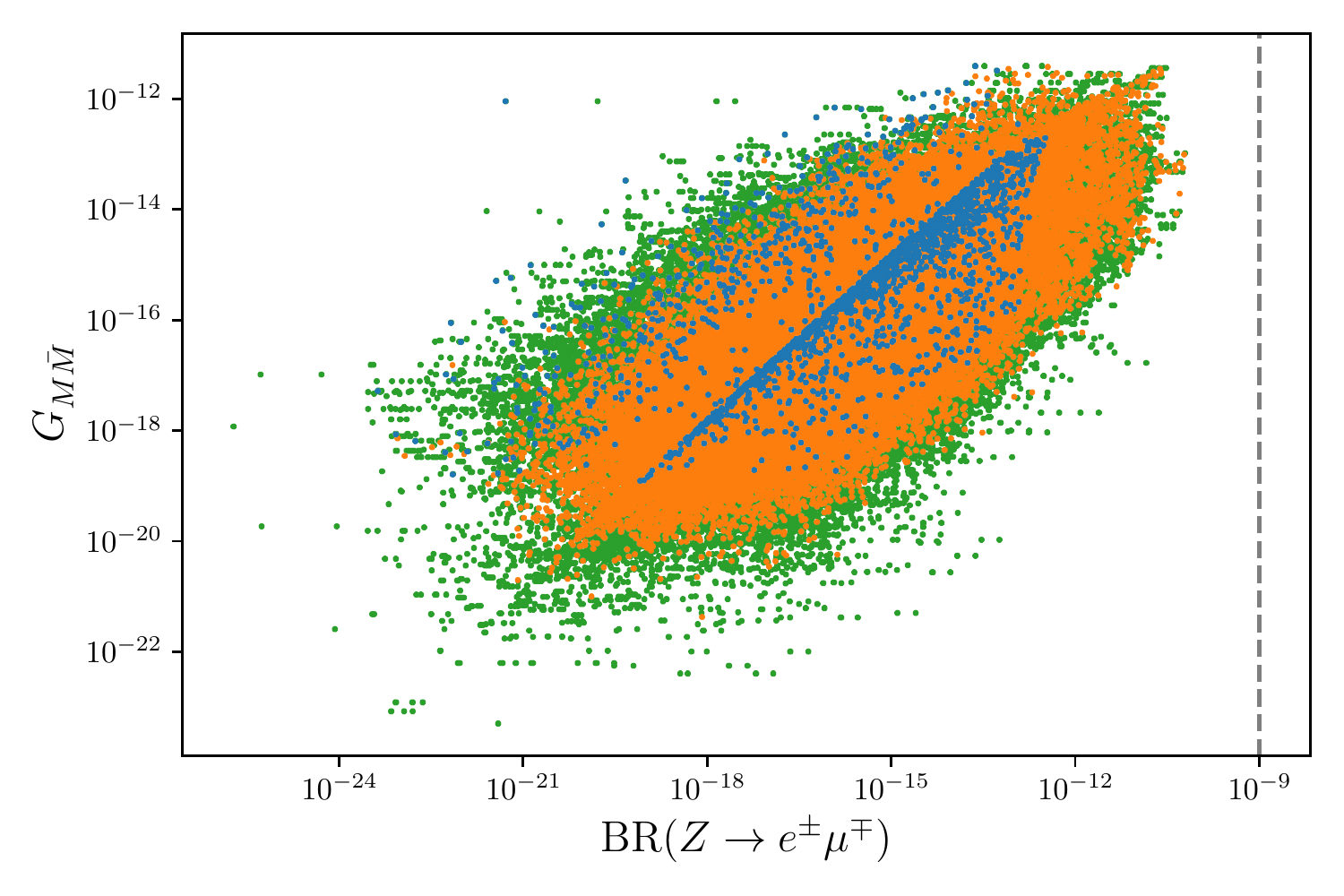}}\\
\mbox{ \hspace*{-5mm}     \includegraphics[width=0.51\textwidth]{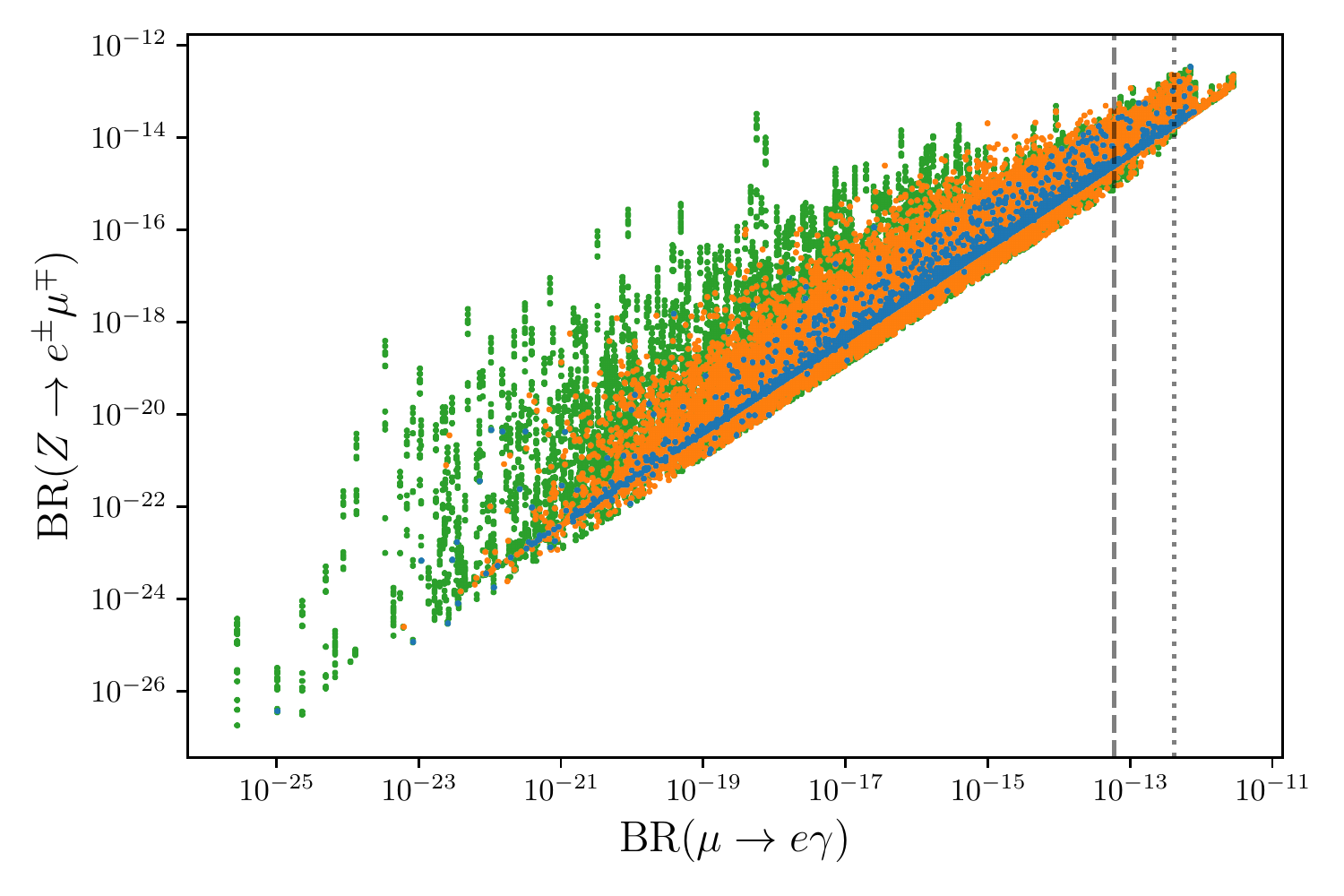}\hspace*{2mm} 
    \includegraphics[width=0.51\textwidth]{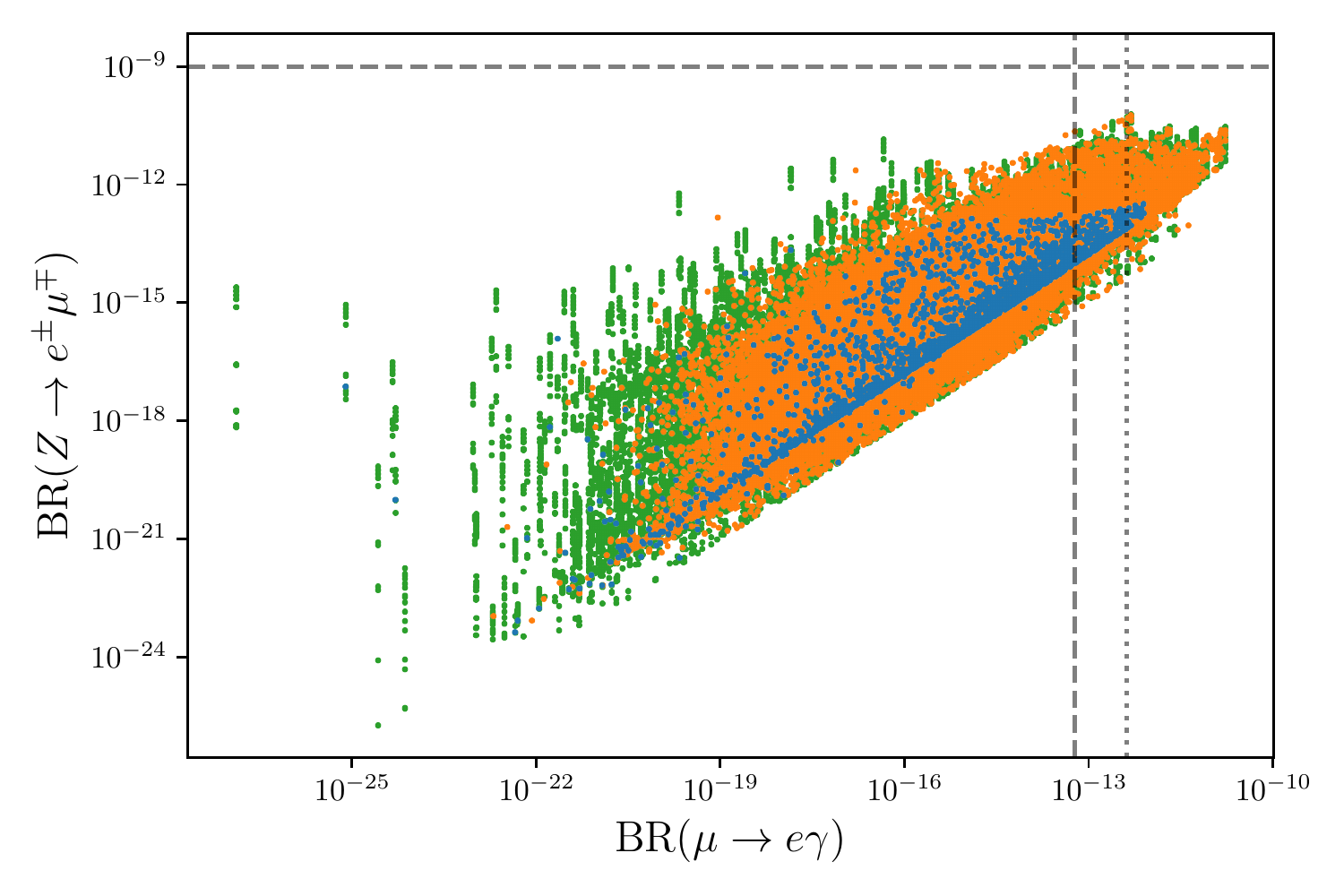}}
    \caption{Correlation of $\mu-e$ flavour violating observables (depending on unique topologies), for varying values of the CPV Dirac and Majorana phases: 
    blue points correspond to vanishing phases, orange denote random values of $\delta_{\alpha 4}$ and $\varphi_4$ in the interval $(0, 2\,\pi)$, and green points refer to $\delta_{\alpha 4}, \varphi_4 =\{0, \frac{\pi}{4}, \frac{\pi}{2}, \frac{3\pi}{4}, \pi\}$ (see text). Dotted  (dashed) lines denote current  bounds (future sensitivity).
    On the left panels, $m_4 = m_5 = 1\:\mathrm{TeV}$, while on the right we set  $m_4 = m_5 = 5\:\mathrm{TeV}$.
    Figures from~\cite{Abada:2021zcm}.}
    \label{fig:scatter_GMM_mueg_Zemu}
\end{figure}

\mathversion{normal}
We begin by considering cLFV observables in the $\mu-e$ sector which receive contributions from unique topologies (dipole, $Z$-penguins and boxes) at one-loop level, and address the impact of CPV phases on the expected correlations. More specifically, on Fig.~\ref{fig:scatter_GMM_mueg_Zemu}, we consider $\mu \to e \gamma$, $Z \to e \mu$ decays and the probability of muonium-antimuonium oscillations, $\propto G_{M\overline{M}}$, displaying the results for two heavy mass regimes, 1 and 5~TeV.

Focusing first on the blue sets of points\footnote{In several plots one observes that in some
extreme cases the predictions of certain observables already lie above the experimental limit; this merely reflects having taken points 
which are consistent with cLFV bounds at the $3\,\sigma$ level, while the denoted experimental limits correspond to 90\% C.L..} (corresponding to vanishing CP violating phases), one confirms that there is a strong correlation between the 
three considered $\mu -e$ flavour violating observables, as expected. This is particularly manifest in the upper row of 
Fig.~\ref{fig:scatter_GMM_mueg_Zemu}, since both 
BR($\mu \to e \gamma$) and $G_{M\overline{M}}$ do not depend on $\theta_{3j}$; on the other hand, non-vanishing values for $\theta_{3j}$ do contribute to $Z\to e\mu$ decays (through the $C_{ij}$ term), hence one observes a spread of the points along the central straight line. This well-known behaviour (correlated predictions for a given value of the propagator's mass) has been explored in the literature as a means to test the underlying BSM construction, see e.g.~\cite{Hambye:2013jsa,Calibbi:2017uvl}.  
Once CP violating phases are (randomly) taken into account, the correlation between the observables  is strongly affected - if not lost, as can be seen by the significant spread of the corresponding orange points. This is especially visible for the plots in the right  ($m_4 = m_5 = 5\:\mathrm{TeV}$). The situation becomes even more 
degraded once the ``special'' values of the phases are considered (those leading to vanishing values of certain contributions, as discussed in Section~\ref{sec:phases.matter}). As expected, the predictions for the observables are dramatically impacted, as visible from the green points. 
Finally, notice that the effect of phases 
can lead to either an increase or reduction of the cLFV rates with respect to the corresponding ones obtained in the vanishing phase 
limit. In some cases (like for BR($\mu \to e \gamma$), with $m_{4,5}=5$~TeV) the new predictions may even be in conflict with current experimental bounds.

\begin{figure}[h!]
    \centering
 \mbox{ \hspace*{-5mm}    \includegraphics[width=0.51\textwidth]{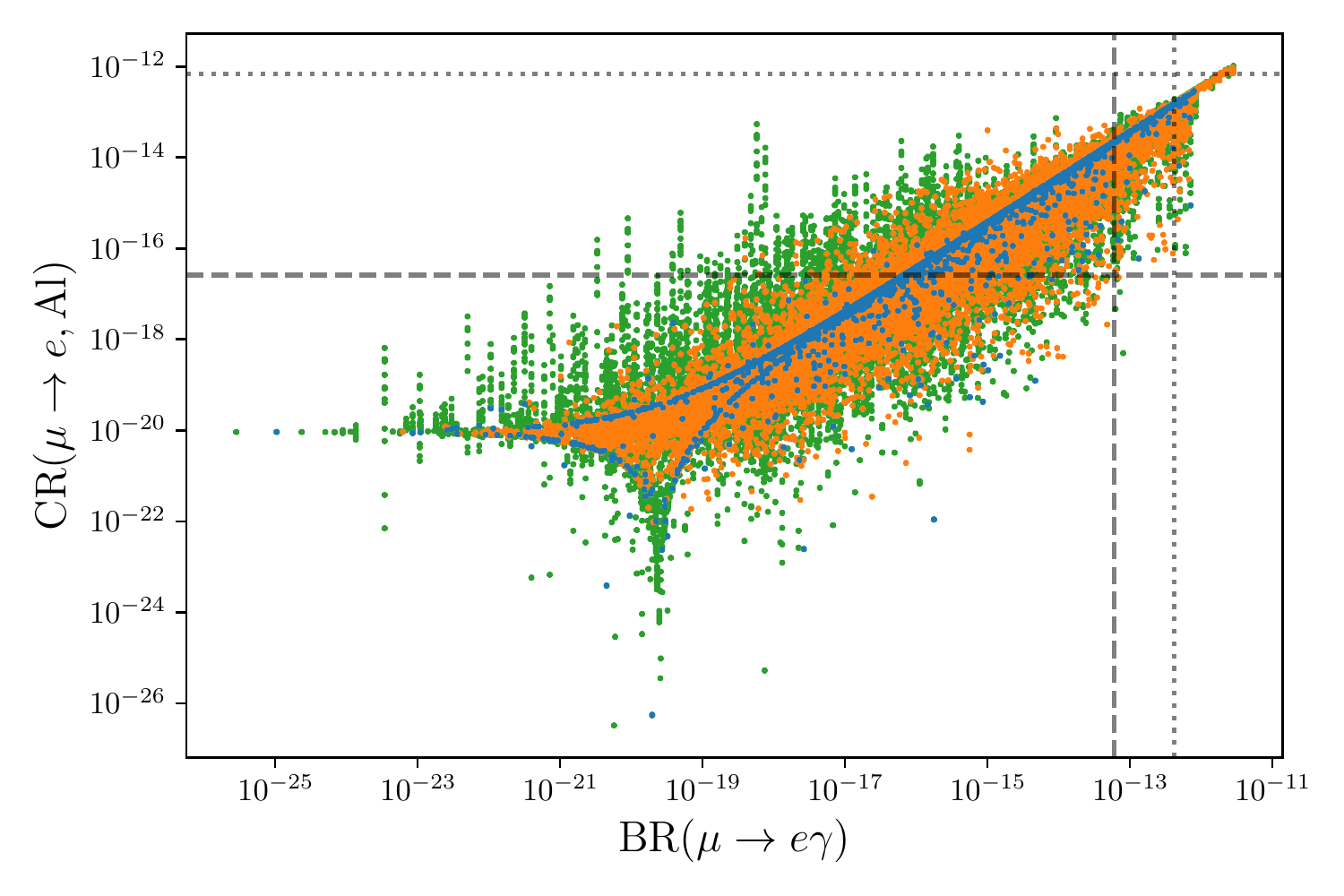}\hspace*{2mm}
    \includegraphics[width=0.51\textwidth]{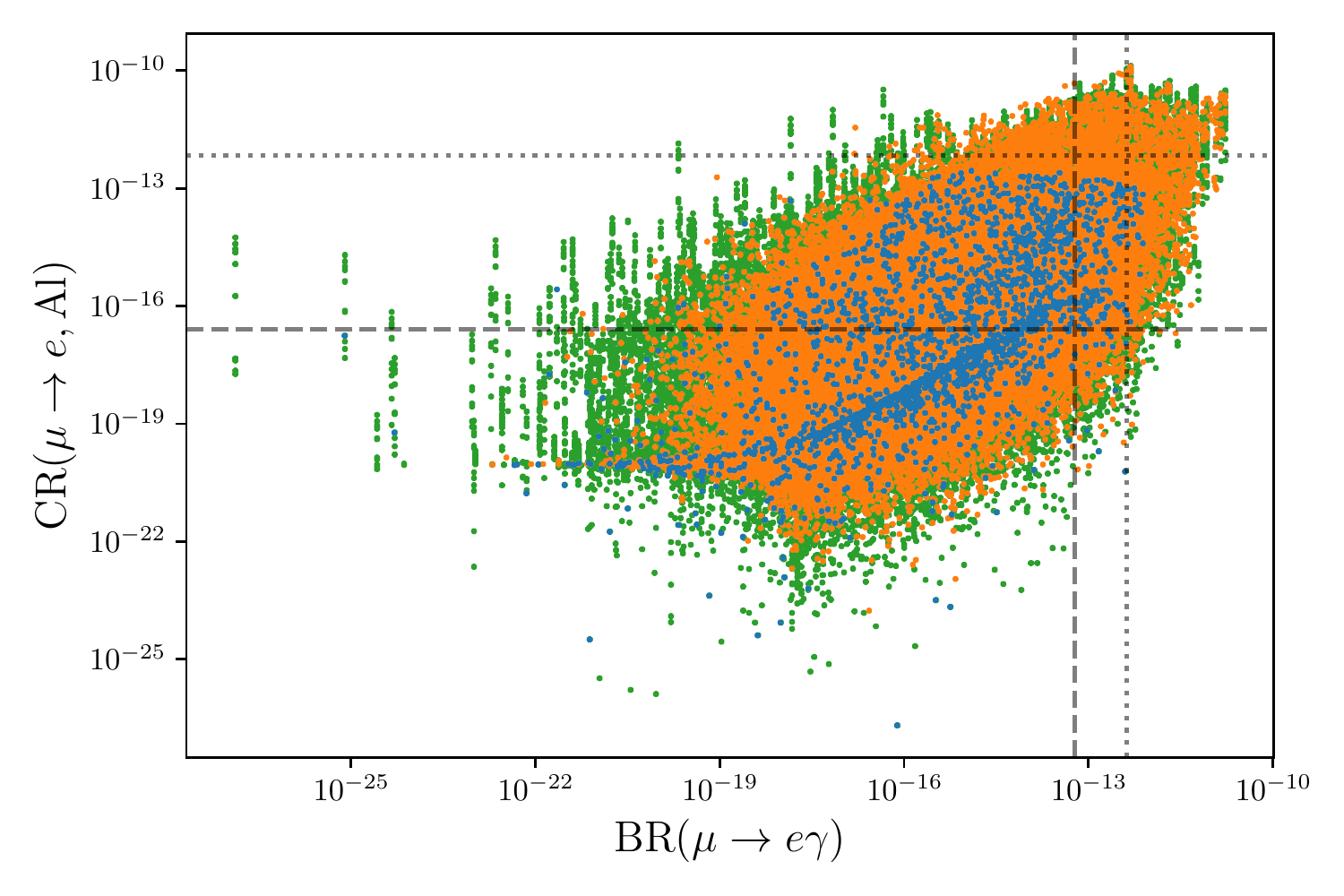}}\\
\mbox{ \hspace*{-5mm}     \includegraphics[width=0.51\textwidth]{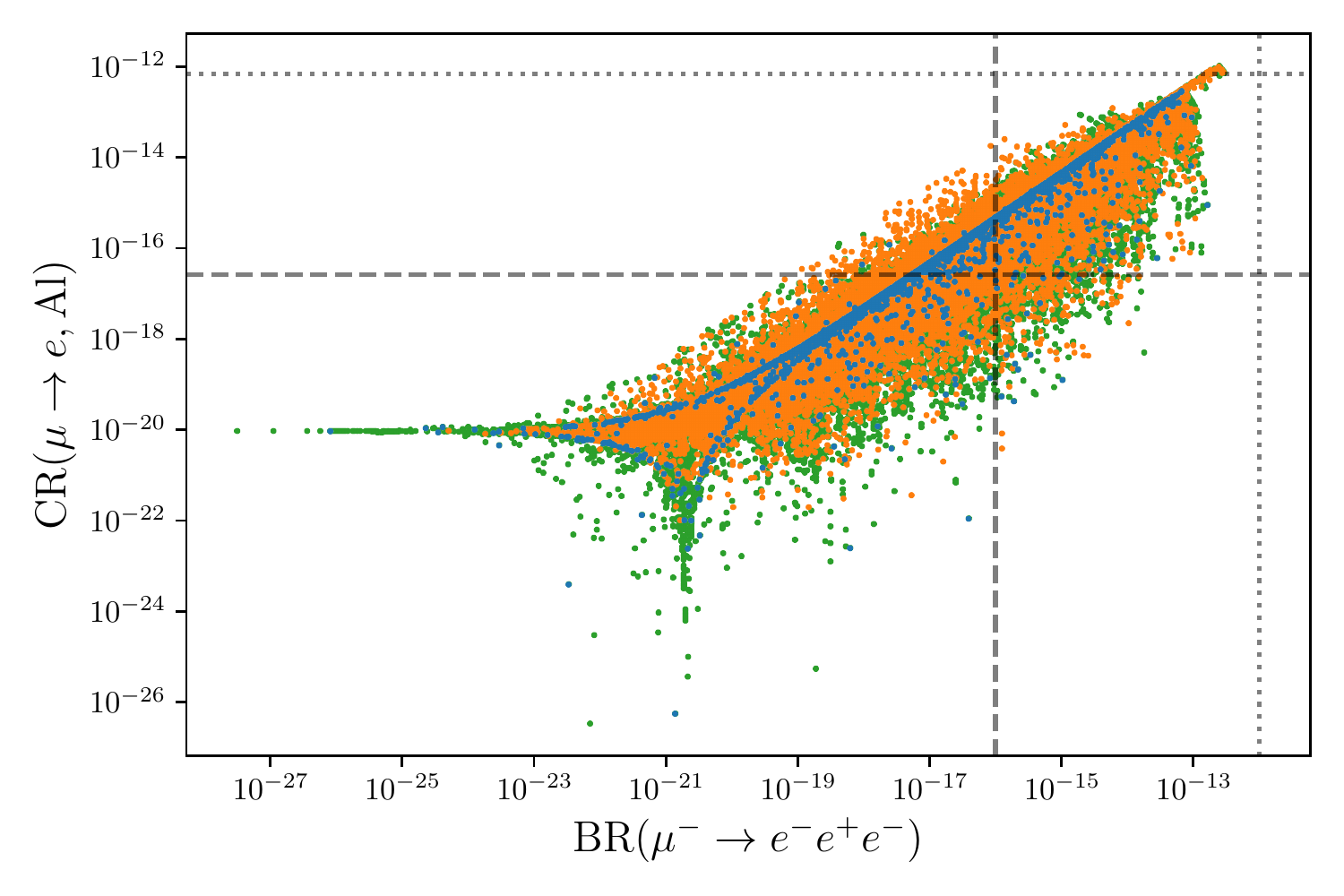}\hspace*{2mm}
    \includegraphics[width=0.51\textwidth]{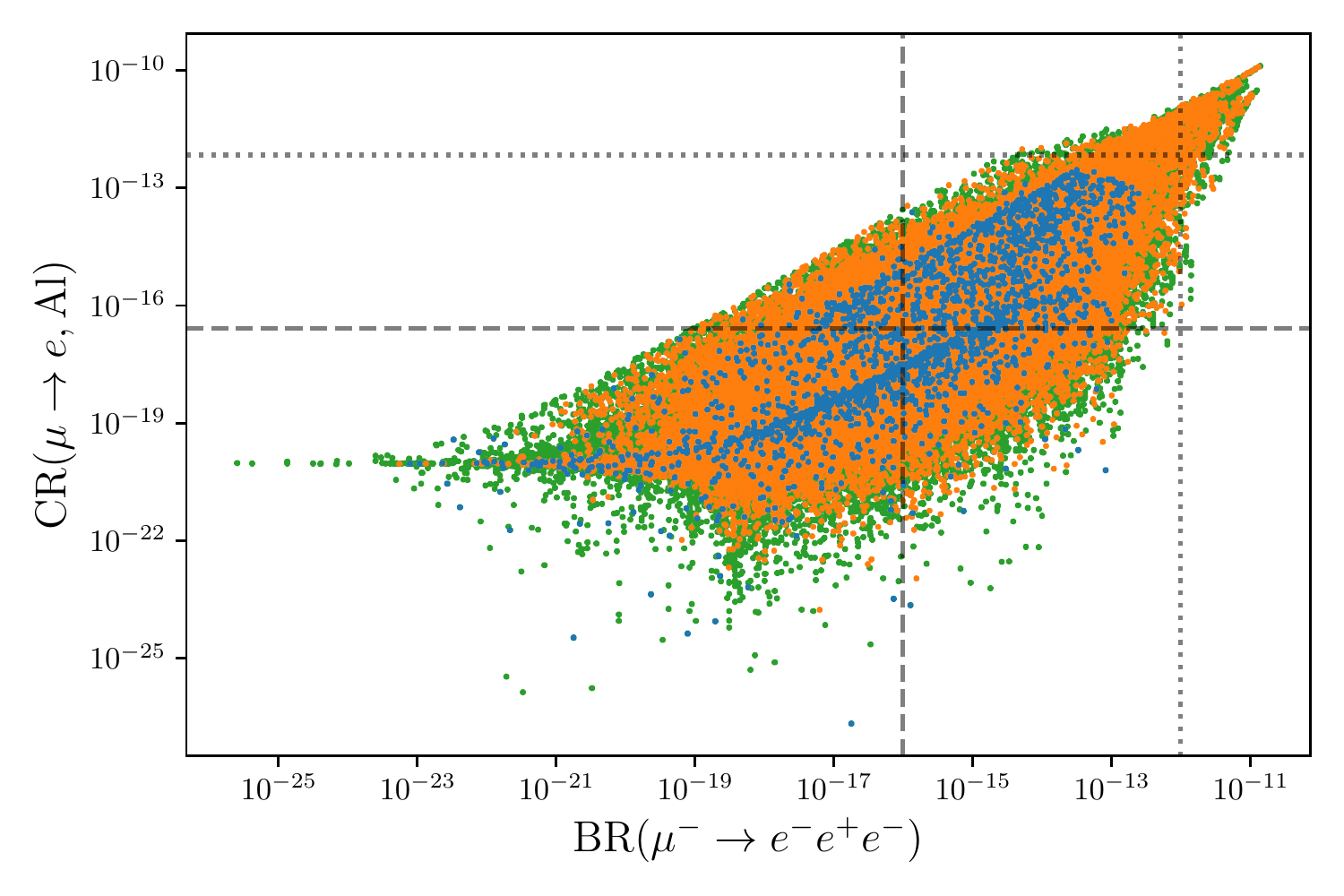}}
    \mbox{ \hspace*{-5mm}     \includegraphics[width=0.51\textwidth]{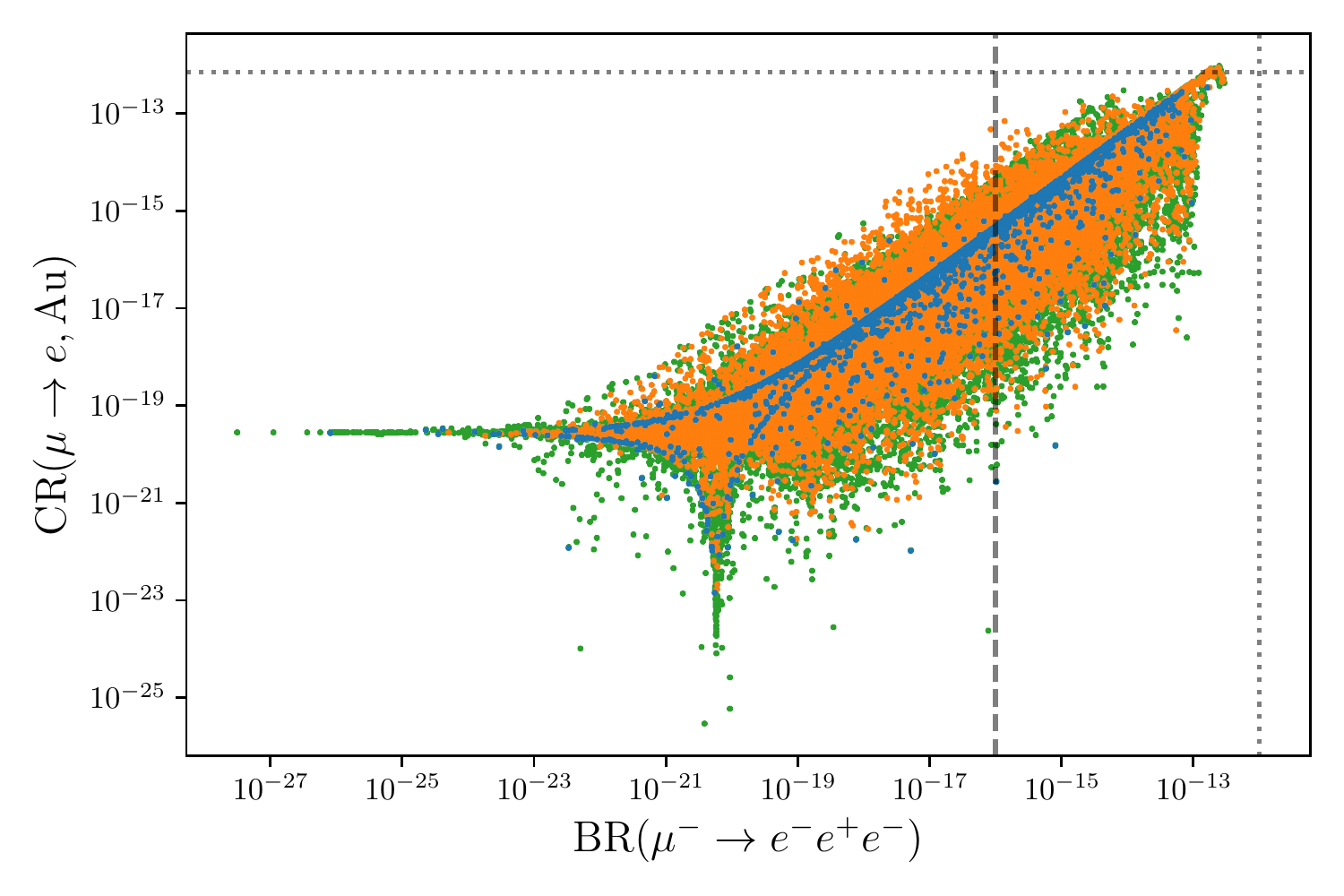}\hspace*{2mm}
    \includegraphics[width=0.51\textwidth]{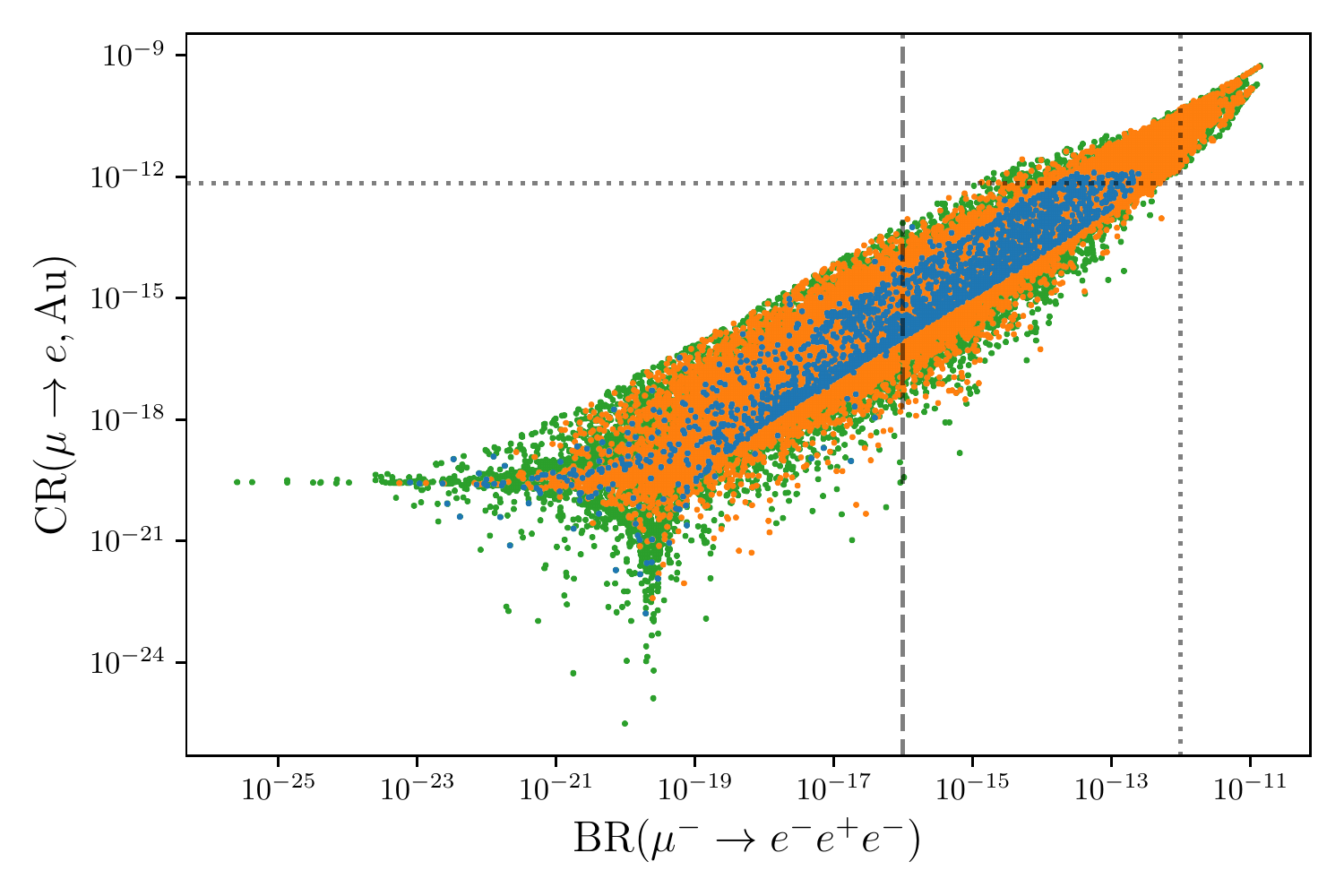}}
    \caption{
    Correlation of general $\mu-e$ flavour violating observables, for varying values of the CPV Dirac and Majorana phases. Line and colour code as in Fig.~\ref{fig:scatter_GMM_mueg_Zemu}.
    On the left panels, $m_4 = m_5 = 1\:\mathrm{TeV}$, while on the right we set  $m_4 = m_5 = 5\:\mathrm{TeV}$.
    Figures from~\cite{Abada:2021zcm}.}
    \label{fig:scatter_GMM_mu3e_CRmue}
\end{figure}

\medskip
In Fig.~\ref{fig:scatter_GMM_mu3e_CRmue} we consider the joint behaviour of general $\mu-e$ flavour violating observables, which now receive non-vanishing contributions from various  types of diagrams (dipole, penguins and boxes). In particular, on the first row we present CR($\mu-e$, Al) vs. BR($\mu \to e\gamma$), while on the second CR($\mu-e$,~Al) vs. BR($\mu \to 3e$), for 
$m_4 = m_5 = 1\text{ and }5\:\mathrm{TeV}$ (left and right columns, respectively). 
Despite the mixings between $\nu_\tau$ and the sterile states also leading to a spread in the case of vanishing phases
(for $\mu-e$ conversion and  3-body decays), one still observes a visible correlation between the different sets of observables (see blue points)\footnote{The spread and visible cancellations for vanishing CP-violating phases are a consequence of accidental cancellations in $\mu-e$ conversion as discussed in Section~\ref{sec:muecon_simple} and of accidental cancellations due to opposite-sign mixing angles (e.g. $\theta_{14}\approx - \theta_{15}$) leading to a suppression of dipole operators as discussed in Section~\ref{sec:enh_others}.}.
As already verified in Fig.~\ref{fig:scatter_GMM_mueg_Zemu}, the presence of CP violating phases - especially for certain values of the latter - leads to a strong loss of correlation, as visible from the dispersion of the orange and green points (again more important for $m_4 = m_5 = 5\:\mathrm{TeV}$). 
On the third and final row of Fig.~\ref{fig:scatter_GMM_mu3e_CRmue}, we present the prospects for the correlation between  $\mu \to 3 e$ decays and neutrinoless muon-electron conversion, but now for Gold (Au) nuclei.
Notice that there are significant differences between 
Al and Au, which become particularly manifest for  
$m_4 = m_5 = 5~\mathrm{TeV}$. In this case the correlation between the conversion rate in Gold and the three-body decays is more prominent than for Aluminium nuclei. This is a consequence of milder cancellations of the type discussed in Section~\ref{sec:muecon_simple}: as seen from 
Figs.~\ref{fig:CR_M:nuclei_tau} and~\ref{fig:CR_Al_M:cancel_phases}, in the limit of vanishing phases 
(and with $\theta_{3j}=0$), the values of the heavy propagator mass for which the accidental cancellation occurs ($m_{4,5}^c$) are considerably lower for Gold than Aluminium, and effects of CP phases and/or non-vanishing $\theta_{3j}$ tend to further shift $m_{4,5}^c$ to lower values. 

\bigskip
Consequently, the CP violating phases might thus have an important impact
regarding a future interpretation of data:
let us consider a hypothetical scenario in which collider searches strongly hint for the presence of sterile states with masses close to 1~TeV. Should BR($\mu \to 3 e$)$\approx 10^{-15}$ be measured in the future, one could expect an observation of CR($\mu-e$, Al)~$\approx \mathcal{O}(10^{-14})$, be it at COMET or Mu2e. However, in the presence of CP violating 
phases, the expected range for the muon-electron conversion is vast, with CR($\mu-e$, Al) potentially as low as 
$10^{-18}$.

\subsection{Prospects for other observables}
In view of the diversity of cLFV observables, and of the associated (future) experimental prospects, we have so far focused our phenomenological discussion on the most promising $\mu-e$ cLFV transitions and decays. 
Before concluding this section, we address the impact of the CP violating phases for the $\mu-\tau$ sector (i.e. for $\tau \to 3\mu$ and $Z \to \mu \tau$ decays), as well as for cLFV Muonium decays. 

\begin{figure}
    \centering
    \includegraphics[width=0.48\textwidth]{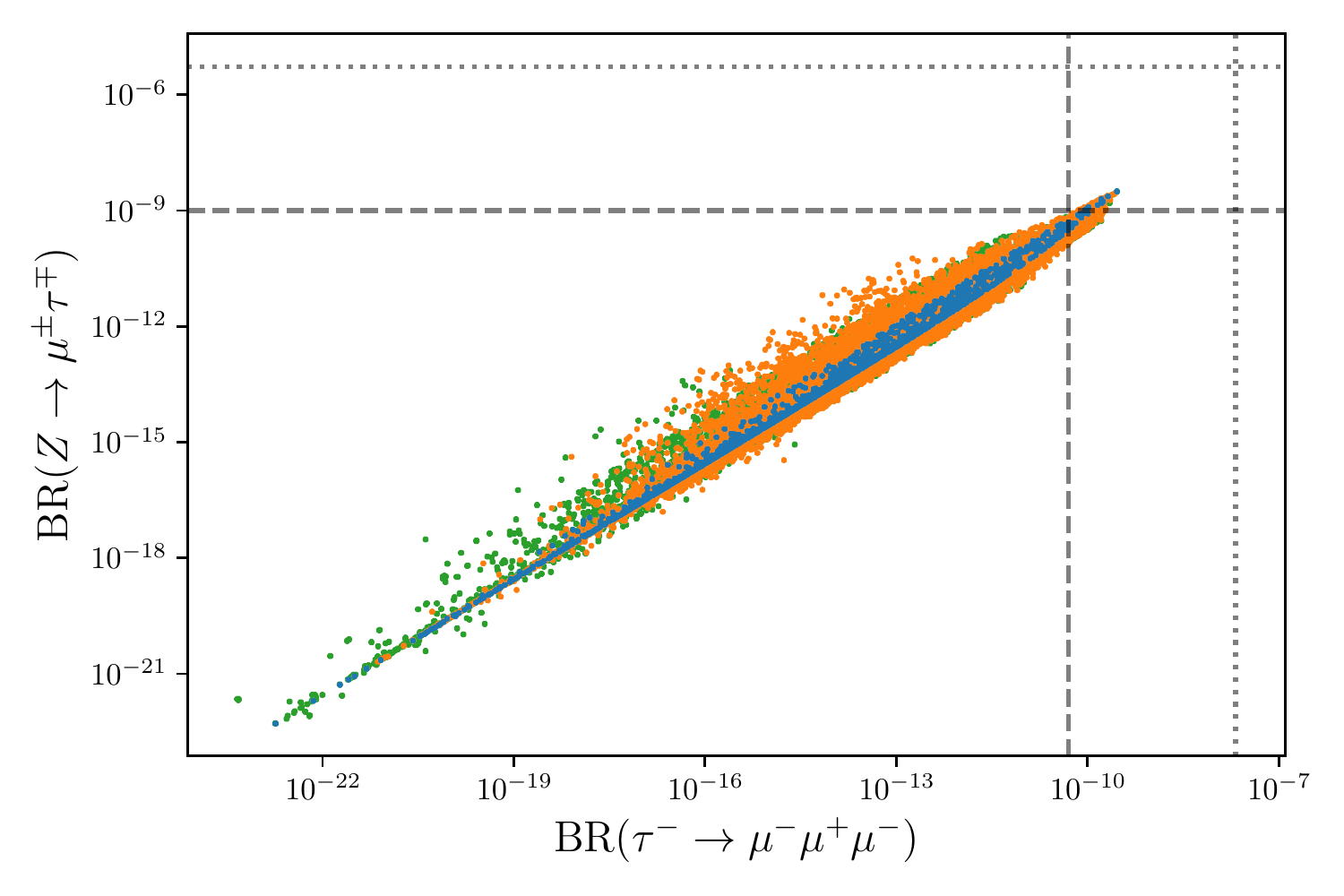}
    \includegraphics[width=0.48\textwidth]{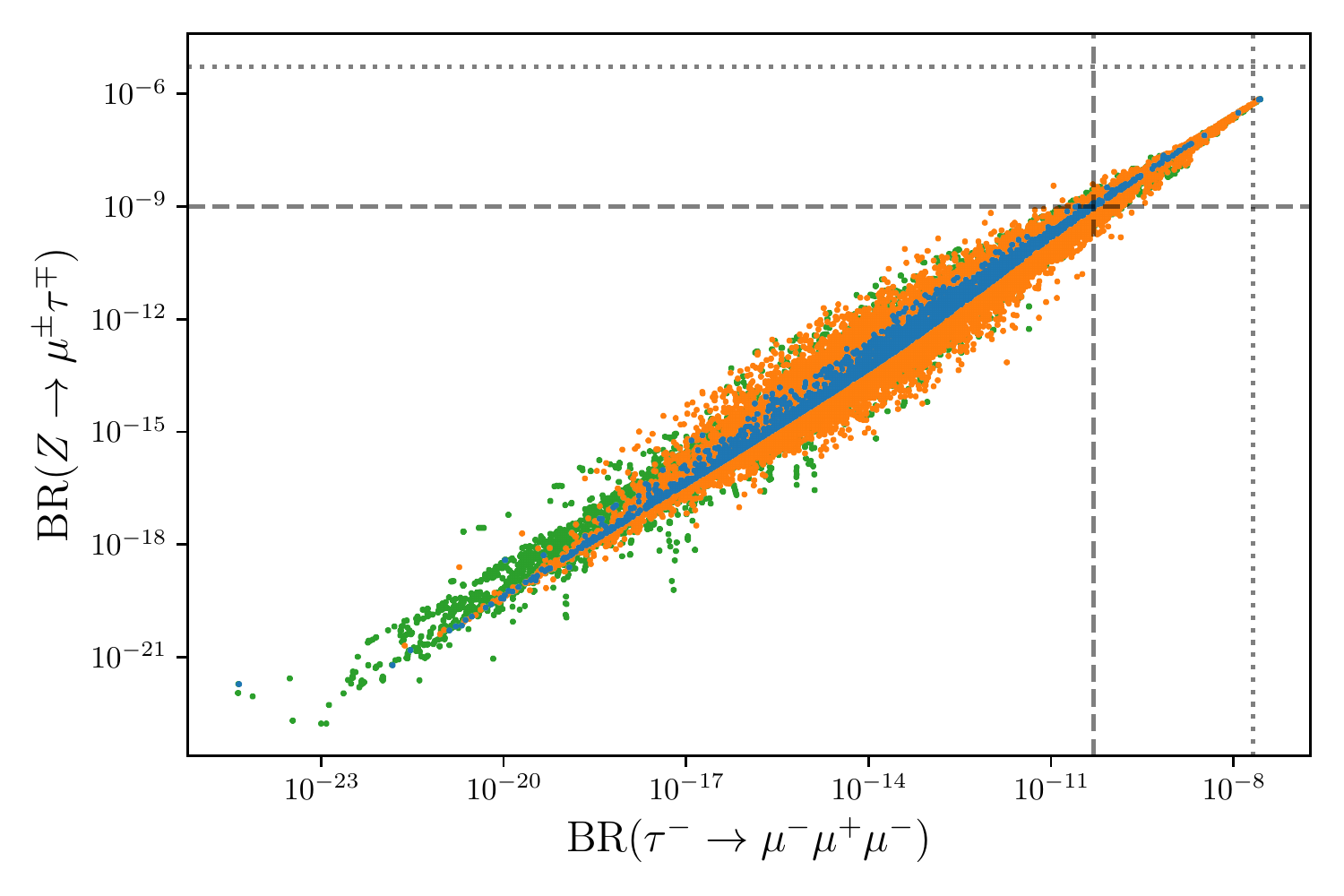}
    \caption{Correlation of $Z \to \mu \tau$ and $\tau \to 3\mu$ decays, for varying values of the CPV Dirac and Majorana phases. Line and colour code as in Fig.~\ref{fig:scatter_GMM_mueg_Zemu}.
    On the left panels, $m_4 = m_5 = 1\:\mathrm{TeV}$, while on the right we set  $m_4 = m_5 = 5\:\mathrm{TeV}$.
    Figures from~\cite{Abada:2021zcm}.}
    \label{fig:tau3mu:Ztaumu}
\end{figure}

In Fig.~\ref{fig:tau3mu:Ztaumu} we summarise the results of a study analogous to those presented in  Figs.~\ref{fig:scatter_GMM_mueg_Zemu} and~\ref{fig:scatter_GMM_mu3e_CRmue}, displaying the correlated behaviour of high- and low-energy $\mu-\tau$ sector cLFV observables for (non-) vanishing Dirac and Majorana CPV phases, and for two values of the degenerate heavy neutral leptons' masses. Although the general prospects for observation are comparatively less promising, one nevertheless encounters the same phase-induced distortion of the correlation between observables, which was present in the 
limit of vanishing phases.
Being also dominated by penguin transitions, the tau-lepton decay modes $\tau^-\to \mu^-e^+e^-$ and $\tau^-\to e^-\mu^+\mu^-$ (i.e. only one flavour violating vertex) do not offer any additional insight with respect to the $\tau\to 3\mu$ and $\tau \to 3e$ counterparts, and we find similar predictions for the associated rates.
On the other hand, tau-lepton decays with an additional flavour violating coupling, that is $\tau^-\to e^-\mu^+e^-$ and $\tau^-\to \mu^- e^+ \mu^-$, are transitions which are purely mediated by box diagrams. Thus, these are strongly suppressed when compared to other modes, with typically very small branching ratios, $\mathrm{BR} \lesssim 10^{-17}$, thus clearly beyond any future sensitivity reach.

\begin{figure}
    \centering
\mbox{\hspace*{-5mm}    \includegraphics[width=0.51\textwidth]{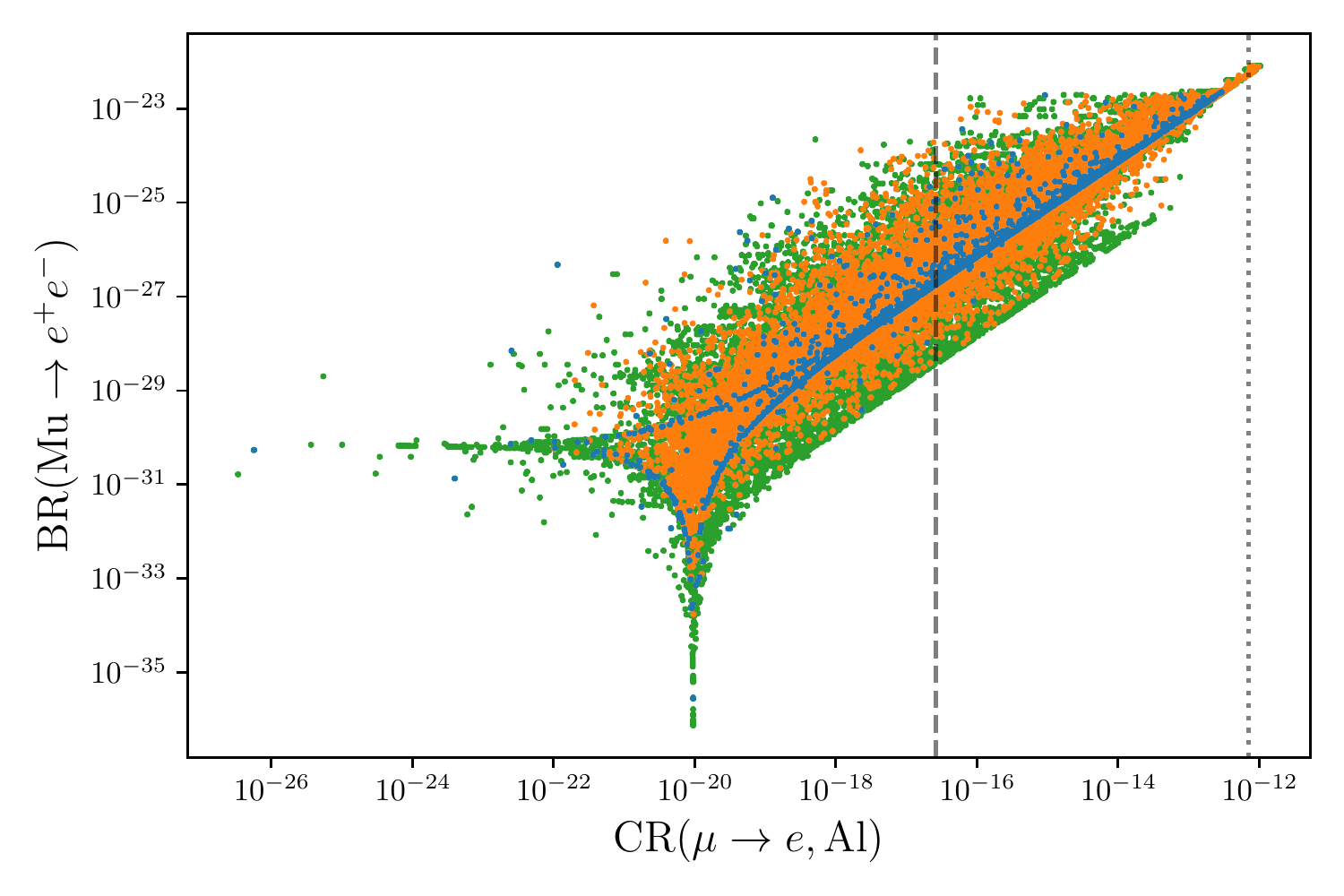}\hspace*{2mm}
    \includegraphics[width=0.51\textwidth]{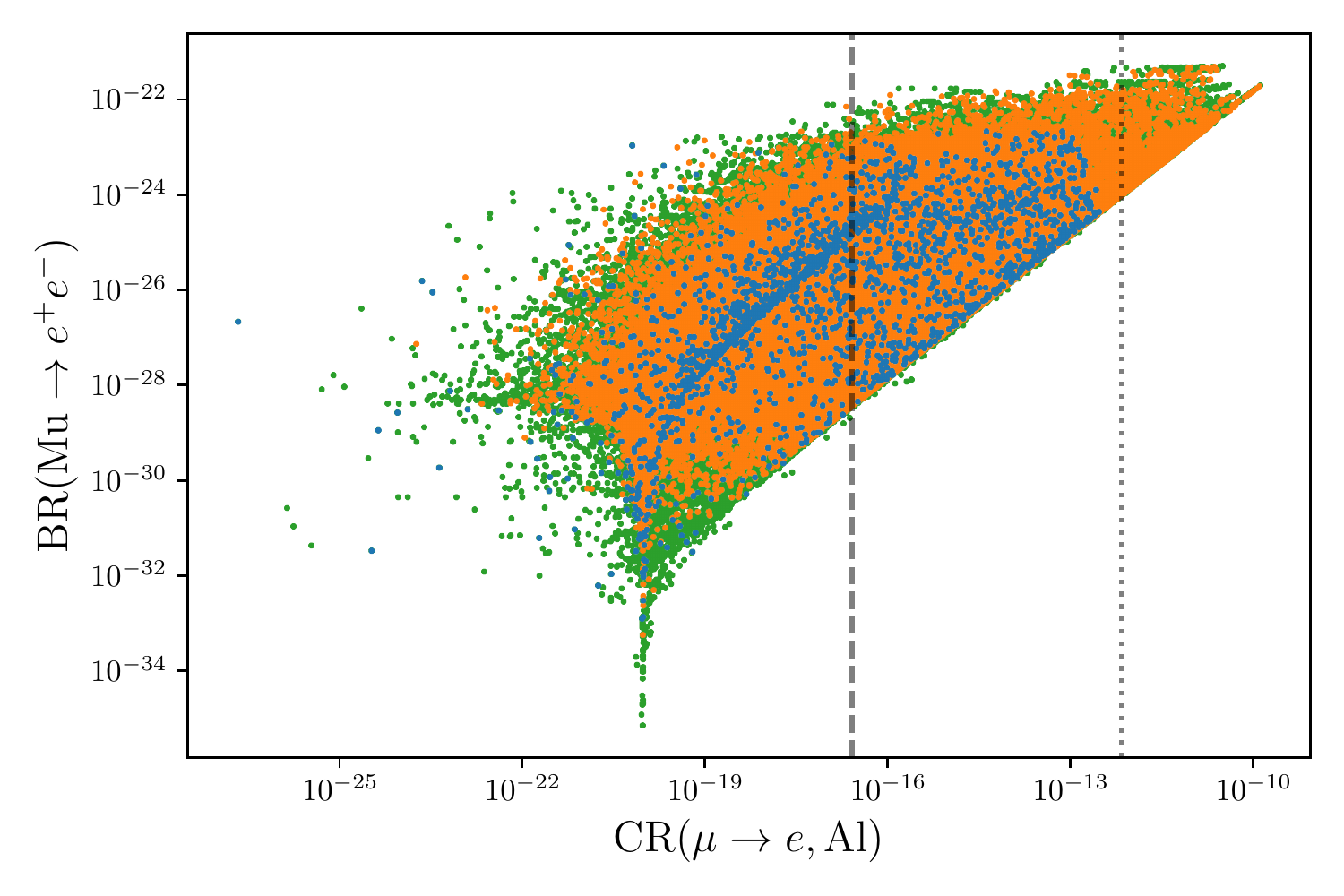}}
    \caption{Correlation of BR($\text{Mu}\to ee$) and CR($\mu-e$, Al), for varying values of the CPV Dirac and Majorana phases. Line and colour code as in Fig.~\ref{fig:scatter_GMM_mueg_Zemu}.
    On the left panel, $m_4 = m_5 = 1\:\mathrm{TeV}$, while on the right we set  $m_4 = m_5 = 5\:\mathrm{TeV}$.
    Figures from~\cite{Abada:2021zcm}.}
    \label{fig:Muoniumdecay}
\end{figure}

In Fig.~\ref{fig:Muoniumdecay}, we display for completeness\footnote{Other cLFV observables - such as the Coulomb enhanced decays $\mu e\to ee$ (see Refs.~\cite{Koike:2010xr,Uesaka:2016vfy,Uesaka:2017yin,Kuno:2019ttl,Abada:2015oba}), were also studied. Although the associated rate for $\mathrm{Al}$ typically lies some orders of magnitude below $\mu-e$ conversion, effects due to non-vanishing CPV phases can however lead to comparable rates, with a maximum of $\mathrm{BR}(\mu e\to ee,\,\mathrm{Al})\sim10^{-16}$.} the prospects for $\text{Mu}\to ee$ decays, depicting its correlation with CR($\mu-e$, Al).  
In the most optimal scenarios, one can expect BR($\text{Mu}\to ee)\sim \mathcal{O}(10^{-22})$.

\section{Overall view and further discussion of CPV phases}\label{sec:general-analysis}
So far, we have  analysed the implications of non-vanishing CPV phases in ``idealised and simplified'' scenarios, assuming certain relations between the model's parameters. 
We followed this approach to explore and maximise the effect of phases on the predictions for the cLFV observables.
Broader scenarios must be considered, and in the present section we relax several of the previous assumptions, aiming at a more comprehensive overview, and allowing for a better confrontation with (hypothetical) future data.

\subsection{Comprehensive overview of the parameter space}
The study in~\cite{Abada:2021zcm} is concluded by a final comprehensive overview of this very simple SM extension. The numerical data presented in this subsection is obtained as follows:
firstly, and in what concerns the masses of the two heavy  states\footnote{In more general scenarios in which the masses of the sterile states and their mixings with the active neutrinos are a priori unrelated to each other, effects of the CPV phases are expected to be less striking, but nevertheless important (and in general driven by the heaviest state).}, we no longer take them to be degenerate, but rather assume their masses to be sufficiently close to allow for interference effects\footnote{See Ref.~\cite{Abada:2019bac} for a related discussion regarding the mass splitting between $m_4$ and $m_5$.}; in practice, and for fixed 
$m_4$, random values of $m_5$ are obtained from 
half-normal distributions with the scale set to a value representative of the width of the sterile states (in this case $\sim50~\mathrm{GeV}$).
Concerning the active-sterile mixing angles, these are now independently varied: more specifically, we draw 
samples from log-uniform distributions, further randomly varying their signs.  
For $m_4=1$~TeV, the ranges of the parameters to be here explored 
are then
\begin{eqnarray}\label{eqn:6d_scan_ranges}
     && m_5 - m_4 \in\,[0.04, 210]\,\mathrm{GeV}\,,\nonumber \\
    && |\sin\theta_{14, 5}|\in\, [2.0\times 10^{-5}, 3\times10^{-3}]\,,\nonumber \\
    && |\sin\theta_{24,5}|\in\, [2.2\times 10^{-4}, 0.036]\,,\nonumber \\
    && |\sin\theta_{34,5}|\in\, [1.0\times 10^{-3}, 0.13]\,.
\end{eqnarray}
It is worth noticing that these ranges lead to scenarios complying with experimental bounds (see Chapter~\ref{chap:lepflav}).
In our analysis, we thus (randomly) select $10^4$ points consistent with all experimental data. 
For each tuple of mixing angles we then vary \textit{all} CPV phases associated with the sterile states, i.e. $\delta_{\alpha 4, 5}, \varphi_{4, 5}\,\in\,[0, 2\pi]$, drawing 100 values for each of the four from a uniform distribution.
The upper limits on the intervals for the mixing angles are inferred from requiring agreement with the most constraining  current bounds; clearly 
no lower limit for the mixing angles is phenomenologically relevant. 
However, we have limited ourselves to regimes that do not lead to cLFV predictions
excessively far away 
from the corresponding future experimental sensitivity. Thus, it is important to stress that the resulting predictions for the observables could in principle be extended to extremely tiny values of the rates, should we have explored all the allowed ranges for the mixing angles. In summary, no conclusions concerning lower limits for the cLFV observables should be drawn from this analysis.

\begin{figure}[t!]
    \centering
    \mbox{\hspace*{-8.5mm}
    \includegraphics[width = 0.53\textwidth]{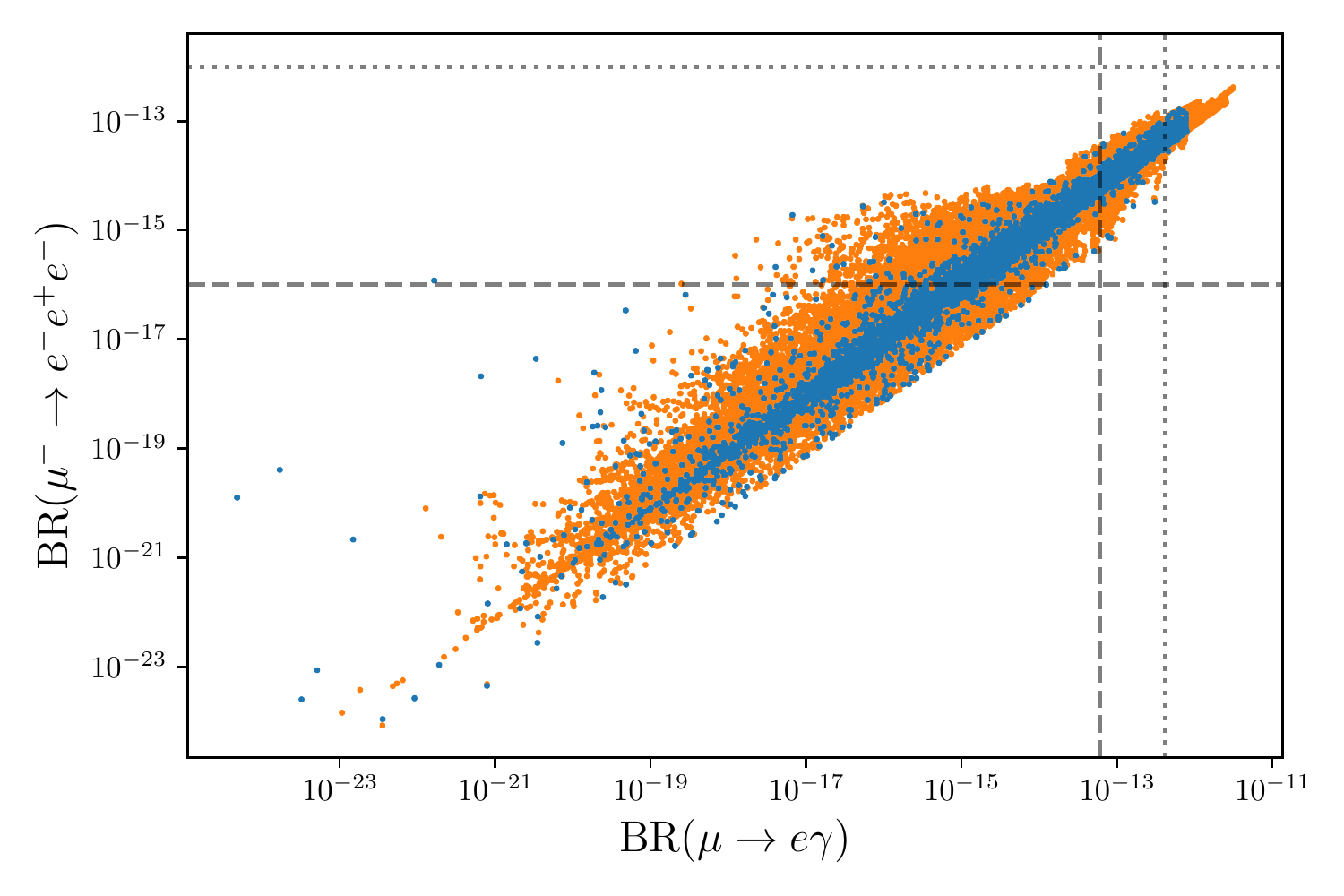}
    \hspace*{2mm} \includegraphics[width = 0.53\textwidth]{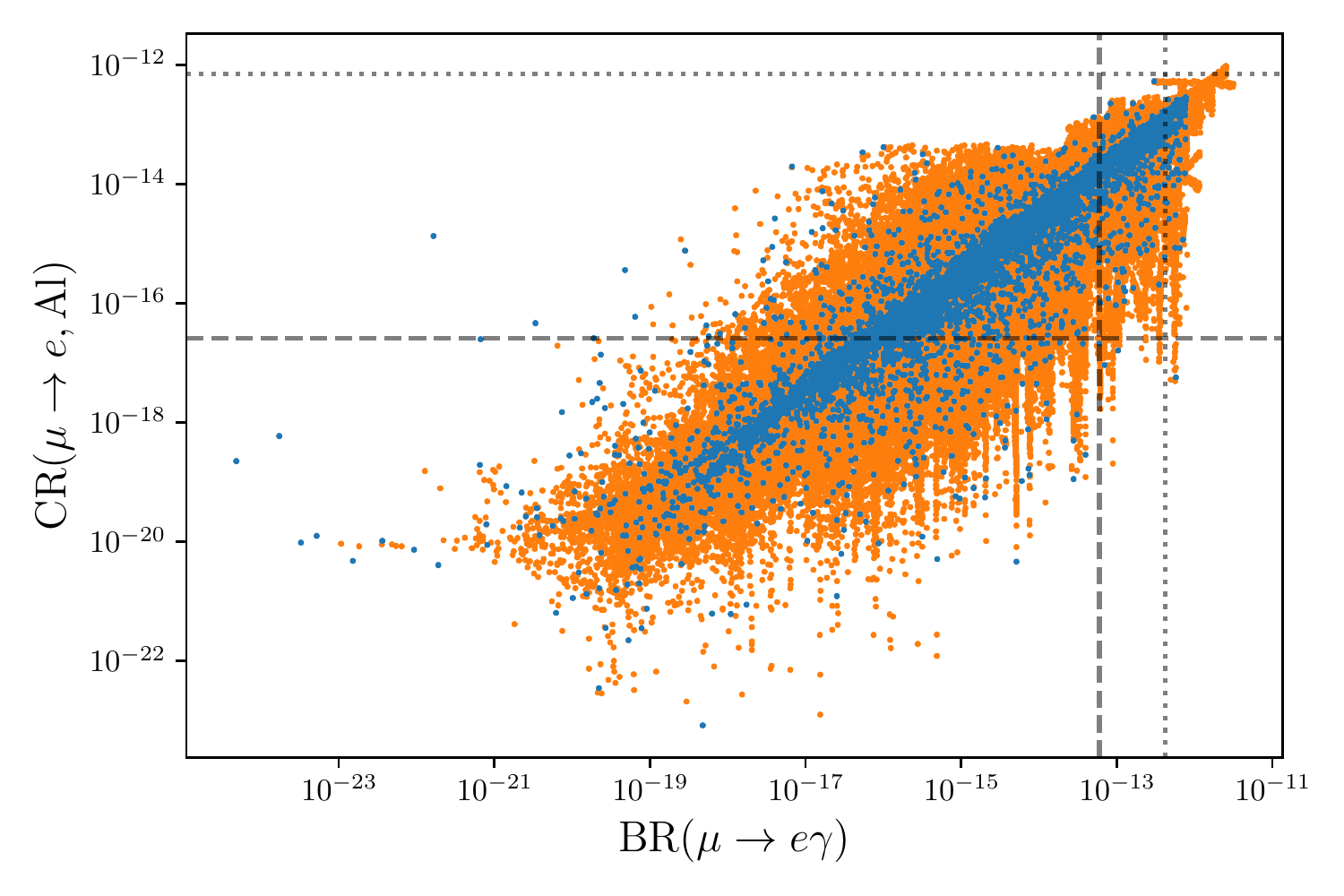}}\\
    \mbox{\hspace*{-8.5mm}\includegraphics[width = 0.53\textwidth]{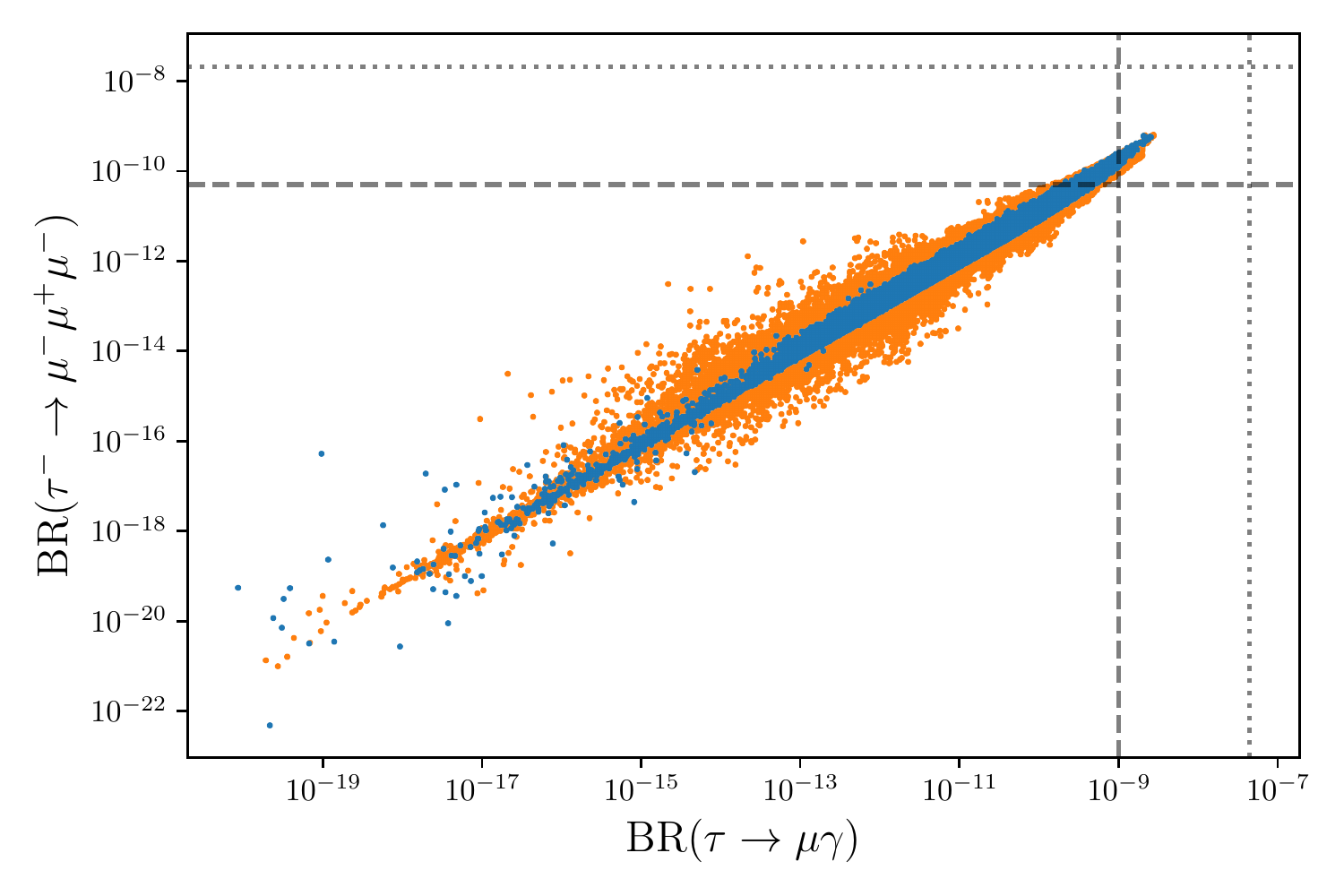}
    \hspace*{2mm} \includegraphics[width = 0.53\textwidth]{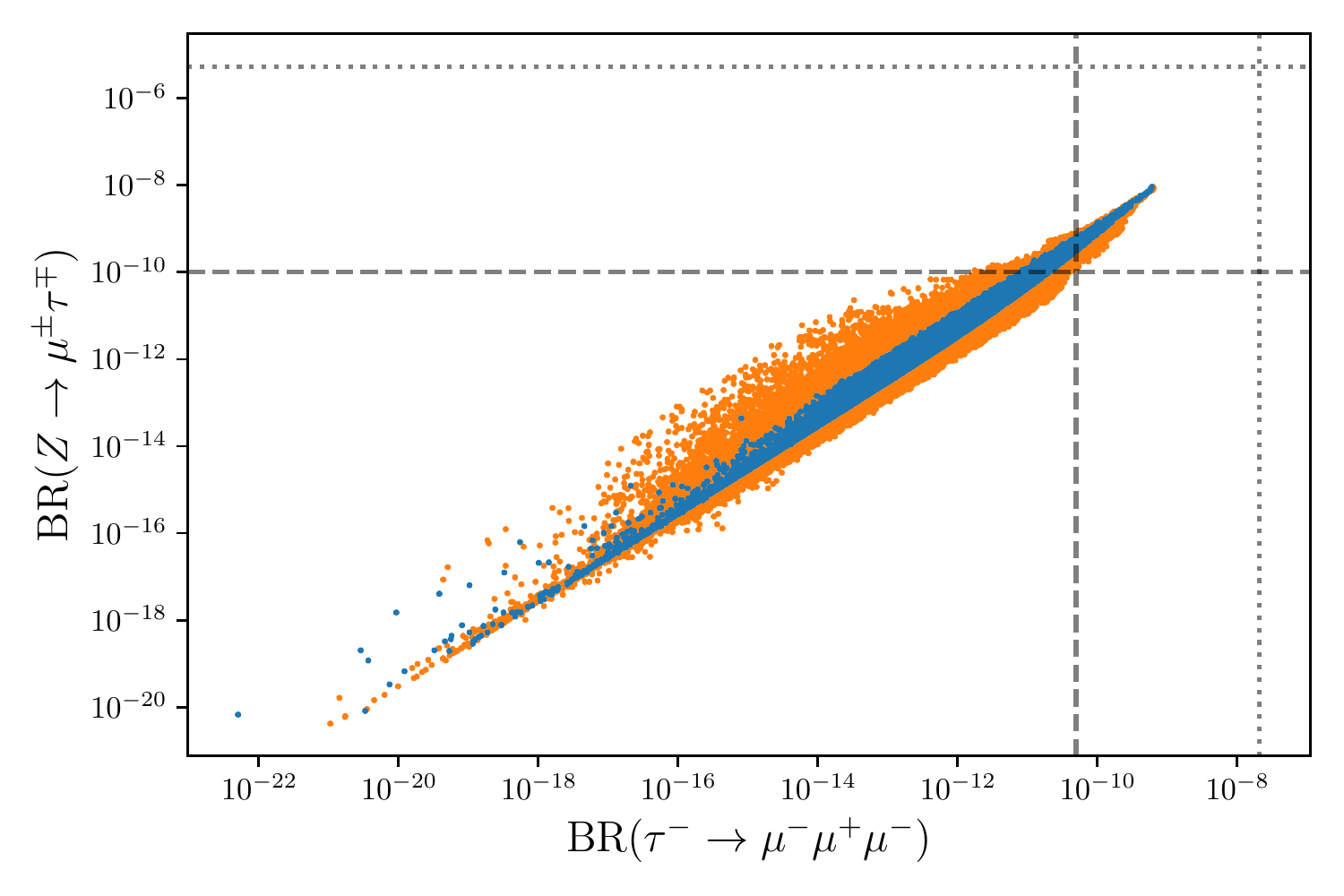}}
    \caption{General overview of cLFV observables (correlations) in the ``3+2 toy model'' parameter space. 
    All active-sterile mixing angles, as well as Dirac and Majorana CP phases, are randomly varied (see detailed description in the text). In all panels, $m_4=1$~TeV, with $m_5-m_4 \in [40~\text{MeV}, 210~\text{GeV}]$. Blue points correspond to vanishing phases, while orange denote random values of all phases ($\delta_{\alpha i}$ and $\varphi_i$, with $\alpha = e,\mu,\tau$ and $i=4,5$). Dotted (dashed) lines denote current bounds (future sensitivity) as given in Table~\ref{tab:cLFVdata}.
    Figures from~\cite{Abada:2021zcm}.}
    \label{fig:6dscan}
\end{figure}

The outcome of this comprehensive analysis is shown in Fig.~\ref{fig:6dscan}, where
we display the predictions for several cLFV rates (focusing on the $\mu-e$ and $\tau-\mu$ sectors), in particular in what concerns correlations between same-sector observables. As before, we present the predictions obtained in the 
case in which all CPV phases are set to zero (blue points), and then those corresponding to a random scan over \textit{all} Dirac and Majorana phases (orange points). 
For a fixed scale of the heavy propagators (with the latter sufficiently close in mass), and as would be expected, one finds correlated
same-sector observables.
In  addition to enlarging the range of the predictions for the different observables (possibly leading to a conflict with current bounds for certain $\mu-e$ observables), non-vanishing CP phases lead to a visible spread of the predicted rates. Should ``special'' values of the phases be imposed on the scan, the cancellation effects (and associated decrease of the rates) would have been even more striking. 

\medskip
Throughout the discussion we have focused on the $\mu-e$ and $\tau-\mu$ sectors, in view of the most promising prospects (both theoretical and experimental). In general, the associated predictions for $\tau-e$ cLFV transitions typically lie beyond future experimental sensitivity. However, and to fully explore this very minimal
SM extension, one would require a comprehensive probing of the associated cLFV predictions in all 3 flavour sectors (i.e. $\mu-e$, $\tau-\mu$ and $\tau-e$ transitions).

Profiting from the data collected leading to 
the results displayed in Fig.~\ref{fig:6dscan}, we have tried to infer which would be the required future sensitivity for the 
$\tau-e$ channels 
so that the regimes (mixing angles and CP phases)
leading to predictions for $\mu \to e\gamma$, $\mu \to 3e$, $\mu - e$ conversion in  Al, $\tau \to 3\mu$ and $Z \to \mu \tau$, all within future experimental sensitivities, would also be within reach of $\tau\to e\gamma$ and $\tau\to 3e$ dedicated searches.
Requiring that at least 68\% of the previously mentioned subset be within $\tau-e$ future reach would imply the following ideal experimental sensitivities\footnote{We have assumed the same ratio between the envisaged $\tau\to e\gamma$ and $\tau\to 3e$ sensitivities as the one of the future prospects of Belle II~\cite{Kou:2018nap}.}:  
\begin{equation}
    \text{BR}(\tau\to e\gamma) \geq 2\times 10^{-13}\,,
    \quad
    \text{BR}(\tau\to 3e) \geq 3\times 10^{-14}\,.
\end{equation}
In other words, should a signal of cLFV in $\mu-e$ and $\tau-\mu$ transitions be observed at the current and near-future facilities, an improvement of circa
$4$ orders of magnitude in the $\tau-e$ sensitivity is needed in order to obtain competitive constraints on 
these SM extensions via heavy neutral leptons from all flavour sectors.

\subsection{Reconciling cLFV predictions with future observations}
As discussed extensively in the previous (sub)sections, 
CPV phases can impact the predictions for the cLFV observables, 
enhancing or suppressing the distinct rates. 
To conclude the discussion, we have identified a small set of representative (benchmark) points, which reflect not only the effect on the rates, but also the impact that taking into account the CPV phases might have on 
the interpretation of 
experimental data (or negative search results). 

In Table~\ref{table:changingcLFVphases} we present the predictions for several cLFV observables for three configurations of active-sterile mixing angles, in the 
case of vanishing CPV phases (P$_i$), and for non-vanishing values of the phases (P$^\prime_i$):
\begin{eqnarray}
\label{eq:Pi:angles}
& \text{P}_1: &
s_{14} = 0.0023\,, \:s_{15} = -0.0024\,,\:
s_{24} = 0.0035\,, \:s_{25} = 0.0037\,, \:
s_{34} = 0.0670\,, \:s_{35} = -0.0654\,, \nonumber \\
& \text{P}_2: &
s_{14} = 0.0006\,, \: s_{15} = -0.0006\,, \: 
s_{24} = 0.008\,, \: s_{25} = 0.008\,, \: 
s_{34} = 0.038\,, \: s_{35} = 0.038\,, \nonumber \\
& \text{P}_3: &
s_{14} = 0.003\,, \: s_{15} = 0.003\,, \:  
s_{24} = 0.023\,, \:  s_{25} = 0.023\,, \:  
s_{34} = 0.068\,, \:  s_{35} = 0.068\,. 
\end{eqnarray}
The variants P$^\prime_i$ have identical mixing angles, but in association with the following phase configurations:
\begin{eqnarray}\label{eq:Pi:phases}
\text{P}^\prime_1:    
\delta_{14} = \frac{\pi}{2}\,, \:  
\varphi_4 = \frac{3\pi}{4}\,;\quad  
\text{P}^\prime_2:    
\delta_{24}=\frac{3\pi}{4}\,, \:  
\delta_{34} = \frac{\pi}{2}\,, \:  
\varphi_4 = \frac{\pi}{\sqrt{8}}\,; \quad  
\text{P}^\prime_3: 
\delta_{14}\approx \pi\,, \:  
\varphi_4\approx \frac{\pi}{2}\,.
\end{eqnarray}

\noindent We have chosen $m_4=m_5=5$ TeV for all three benchmark points.

\renewcommand{\arraystretch}{1.2}
\begin{table}[h!]
    \centering
    \begin{tabular}{|l|c|c|c|c|c|}
    \hline
 & BR($\mu \to e\gamma$) & BR($\mu \to 3e$) & CR($\mu - e$, Al) & BR($\tau \to 3\mu$)& BR($Z \to \mu \tau$)\\
 \hline\hline
$\text P_1$ & $ 3\times 10^{-16}$ \:\:$\circ$ & 
$ 1\times 10^{-15}$ \:\:$\checkmark$& 
$ 9\times 10^{-15}$ \:\:$\checkmark$& 
$ 2\times 10^{-13}$ \:\:$\circ$&
$ 3\times 10^{-12}$ \:\:$\circ$\\
$\text P_1^\prime$
& $ 1\times 10^{-13}$ \:\:$\checkmark$& 
$2\times 10^{-14}$ \:\:$\checkmark$& 
$1\times 10^{-16}$ \:\:$\checkmark$& 
$1\times 10^{-10}$ \:\:$\checkmark$& 
$2\times 10^{-9}$ \:\:$\checkmark$\\
\hline
\hline
$\text P_2$ 
& $2\times 10^{-23}$ \:\:$\circ$
& $2\times 10^{-20}$ \:\:$\circ$ 
& $2\times 10^{-19}$ \:\:$\circ$ 
&  $1\times 10^{-10}$ \:\:$\checkmark$
& $3\times 10^{-9}$ \:\:$\checkmark$\\
$\text P_2^\prime$
& $6\times 10^{-14}$ \:\:$\checkmark$
& $4\times 10^{-14}$ \:\:$\checkmark$
& $9\times 10^{-14}$ \:\:$\checkmark$
&  $8\times 10^{-11}$ \:\:$\checkmark$
& $1\times 10^{-9}$ \:\:$\checkmark$\\
 \hline
 \hline
 $\text{P}_3$ 
 & $2\times 10^{-11}$ \:\:{\footnotesize \XSolidBrush}
 & $3\times 10^{-10}$ \:\:{\footnotesize \XSolidBrush} 
 & $3\times 10^{-9}$ \:\:{\footnotesize \XSolidBrush}
 & $2\times 10^{-8}$ \:\:$\checkmark$
 & $8\times 10^{-7}$ \:\:$\checkmark$\\
$\text P_3^\prime$
& $8\times 10^{-15}$ \:\:$\circ$
  & $1\times 10^{-14}$ \:\:$\checkmark$
  & $6\times 10^{-14}$ \:\:$\checkmark$
  & $2\times 10^{-9}$ \:\:$\checkmark$
  & $1\times 10^{-8}$ \:\:$\checkmark$\\
 \hline
\end{tabular}
\caption{Predictions for several cLFV observables in association with three 
configurations with vanishing CPV phases,
P$_i$ ($i=1-3$) and associated variants with non-vanishing CP violating phases, P$_i'$, see Eqs.~(\ref{eq:Pi:angles}, \ref{eq:Pi:phases}). We have taken
$m_4 = m_5 = 5~\mathrm{TeV}$. The symbols 
({\small\XSolidBrush}, $\checkmark$, $\circ$) respectively denote 
rates already in conflict with current experimental bounds, predictions within future sensitivity and those beyond future experimental reach.
}\label{table:changingcLFVphases}
\end{table}
\renewcommand{\arraystretch}{1.}

\noindent
The first point (P$_1$) represents a case for which only two 
cLFV observables would be within future experimental reach, 
$\mu \to 3e$ and $\mu-e$ conversion in Aluminium; however, in the presence of CP phases (P$_1^\prime$), the predictions for the different considered observables are now \textit{all} within future sensitivity. 

\noindent
The points P$_2$ and P$_2^\prime$ correspond to a similar scenario, but for which only the two considered 
$\mu-\tau$ observables lie within future reach  in the case of vanishing phases.

\noindent
The third and final point (P$_3$) clearly illustrates the importance of taking into account the possibility of  CP violating phases upon interpretation of experimental data.
Negative search results for the different $\mu-e$ flavour violating transitions would lead to the exclusion of the associated mixing angles (for heavy masses $\sim5$~TeV); however, and should CPV phases be present, the considered active-sterile mixing regime can be readily reconciled with current bounds\footnote{A similar approach was pursued in Ref.~\cite{Heeck:2018ntp}, albeit for the $3\times3$ PMNS mixing matrix.} (with 
$\mu \to e \gamma$ now even lying beyond experimental reach). 
A similar exercise could be carried for other heavy mass regimes, leading to analogous conclusions.

\noindent
This demonstrates the crucial role of CPV phases in evaluating the viability of a given scenario in what regards conflict/agreement with the associated cLFV bounds.

\subsection{Concluding remarks}

In this chapter we have thoroughly addressed the impact of leptonic CP violating phases on the predictions of the rates of several cLFV observables, focusing on minimal SM extensions by
heavy Majorana sterile fermions. 
Despite their minimality and simplicity, these 
extensions can be interpreted as representative  of more complete constructions calling upon the addition  of heavy neutral fermions (as is the case of several low-scale seesaw realisations). 

In the study of~\cite{Abada:2021zcm}, we have considered a simple case with 2 heavy neutral fermions, taking them close in mass in order to explore the potential impact of the new CPV phases on cLFV observables.  
These states could very well be embedded in a seesaw, and the latter even incorporated in more complete BSM frameworks. The conclusions drawn in this work are thus always valid once one considers that the source of lepton flavour violation stems from the enlarged leptonic mixing. 

Building upon an analytical insight, the results of the numerical study in~\cite{Abada:2021zcm} reveal that the CP violating phases can indeed lead to important effects in  cLFV transitions and decays, with an 
impact for the rates (enhancement or suppression of the predictions obtained for vanishing phases).
Moreover, whenever correlations between observables would be typically expected (in association with the dominance of a given topology for certain regimes of the model), one also encounters a potential loss of correlation. 
Furthermore, the analysis suggests that  
the non-observation of a given observable (usually expected to be within experimental reach in view of the measurement of another one) should not be a conclusive reason to disfavour a given regime. 

Specific cLFV signatures have been extensively investigated and highlighted as powerful means to disentangle (and further learn about) certain mechanisms of neutrino mass generation; as an example, recall that while in type I seesaw constructions one typically finds BR$(\mu \to e\gamma)$/BR$(\mu \to 3 e) \sim 5-10$ (for masses of the propagators in the TeV-ballpark), for a type III seesaw one has BR$(\mu \to e\gamma)$/BR$(\mu \to 3 e) \sim 10^{-3}$, a consequence of having the cLFV 3-body decay occurring at the tree-level (see, e.g.~\cite{Hambye:2013jsa}). 
However, the presence of CP violating phases (Dirac and/or Majorana) in association to the new lepton mixings  can strongly impact such predictive scenarios. 

The conclusions drawn in~\cite{Abada:2021zcm} can be generalised for a given BSM construction, provided that all complex degrees of freedom are consistently taken into account, and predictions for cLFV observables re-evaluated in view of the potential presence of new CP violating phases.
This means for practical purposes that for instance in seesaw-type models relying on a Casas-Ibarra parametrisation, all possible complex values of $R$-matrix (cf. Eq.~\eqref{eqn:Rmatrix}) have to be sampled consistently, if $R$ is not predicted or constrained by a symmetry.

In the near future, should potentially new (unexpected) cLFV patterns emerge upon observation of certain processes, this could be interpreted as possibly hinting towards the presence of non-vanishing CP violating phases (under the working hypothesis of SM extensions via heavy neutral fermions).
\chapter{Flavour and CP symmetries in the inverse seesaw}
\label{chap:ISS}
\minitoc

\noindent
The Standard Model of particle physics can successfully explain a plethora of experimental observations. 
Yet, as extensively discussed in Chapters~\ref{chap:SM} and~\ref{chap:lepflav}, the existence of three generations of SM fermions,
the origin of neutrino masses, the features of lepton and quark mixing, as well as the striking differences between these remain open issues, in the absence of a full complete ``theory of everything''. 
Symmetries acting on flavour space can address the first and the third point~\cite{Ishimori:2010au,King:2013eh,Feruglio:2019ybq,Grimus:2011fk}\footnote{Other possible solutions include for instance geometric flavour constructions and GUT scenarios.}, while different types of new particles can be added to the SM
in order to generate at least two non-vanishing neutrino masses~\cite{Minkowski:1977sc,Yanagida:1979as,Glashow:1979nm,Gell-Mann:1979vob,Mohapatra:1979ia,Magg:1980ut,Schechter:1980gr,Cheng:1980qt,Lazarides:1980nt,Wetterich:1981bx,Mohapatra:1980yp,Foot:1988aq,Mohapatra:1986bd,GonzalezGarcia:1988rw,Mohapatra:1986aw,Bernabeu:1987gr,Cai:2017jrq}. 

Although a discussion of the flavour puzzle and its possible approaches via flavour symmetries are not the main focus of this thesis (which is motivated towards phenomenology) we nevertheless study an interesting possibility, that of a non-abelian discrete 
symmetry $G_f$ 
combined with a CP symmetry, both acting non-trivially on flavour space. This combination has proven to be highly constraining~\cite{Ecker:1983hz,Ecker:1987qp,Neufeld:1987wa,Grimus:1995zi,Harrison:2002kp,Grimus:2003yn,Feruglio:2012cw,Holthausen:2012dk,Chen:2014tpa} since, as long as $G_f$ and CP are broken to different residual symmetries $G_\ell$ among charged leptons and $G_\nu=Z_2 \times CP$ among the neutral states,
PMNS mixing matrix
depends on a single free parameter.
We select $G_f$ to be a member of the series of groups $\Delta (3 \, n^2)$~\cite{Luhn:2007uq} and $\Delta (6 \, n^2)$~\cite{Escobar:2008vc}, $n$ integer, because these have shown to lead to several interesting mixing patterns~\cite{Hagedorn:2014wha,Ding:2014ora,Ding:2015rwa,King:2014rwa,Ding:2013hpa,Feruglio:2013hia,Ding:2013bpa,Li:2013jya,Li:2014eia,Ding:2013nsa,Ding:2014hva,Ding:2014ssa}. Four of these, called Case 1), Case 2), Case 3 a) and Case 3 b.1), have been identified in~\cite{Hagedorn:2014wha}.
Flavour (and CP) symmetries have been studied in association with several scenarios of neutrino mass generation, see, e.g.,~\cite{Ma:2004zv,Grimus:2005mu,deMedeirosVarzielas:2005qg,Altarelli:2005yx,He:2006dk,Lin:2008aj,Ding:2013hpa,Feruglio:2013hia,Ding:2013bpa,Li:2013jya,Li:2014eia,Ding:2013nsa,Ding:2014hva,Ding:2014ssa,Hirsch:2009mx,Ibanez:2009du,Dorame:2012zv,CarcamoHernandez:2017kra,Borah:2017dmk,CarcamoHernandez:2019eme,Nomura:2019xsb,Nguyen:2020ehj,Camara:2020efq,Devi:2021ujp,Zhang:2021olk}.

As discussed in Chapter~\ref{sec:ISSintro}, in addition to being a theoretically well-motivated framework, the inverse seesaw mechanism can have an important phenomenological impact.

In this study, we thus endow an ISS framework with 
a flavour symmetry $G_f$ and a CP symmetry. We focus on the so-called $(3,3)$ ISS framework, in which the SM field content is extended by $3+3$ heavy sterile states, $N_i$ and $S_j$.
We note that different realisations of the ISS mechanism with flavour (and CP) symmetries have been considered in the literature, see, e.g.,~\cite{Hirsch:2009mx,Ibanez:2009du,Dorame:2012zv,CarcamoHernandez:2017kra,Borah:2017dmk,CarcamoHernandez:2019eme,Nomura:2019xsb,Nguyen:2020ehj,Camara:2020efq,Devi:2021ujp,Zhang:2021olk}.
In what concerns the flavour symmetries, the main features of the present ISS framework are the following: left-handed lepton doublets, and the sterile states $N_i$ and $S_j$ all transform as irreducible triplets of $G_f$,
while right-handed charged leptons are assigned to singlets, so that the three different charged lepton masses can be easily accommodated. While the source of
breaking of $G_f$ and CP to the residual symmetry $G_\ell$ 
is unique in the charged lepton sector (corresponding to the  charged lepton mass terms), the breaking to $G_\nu$ among the neutral states can be 
realised in different ways. 
Indeed, we can consider three minimal options, depending on which of the neutral fermion mass terms 
 encodes the symmetry breaking. 
Here, we will pursue
an option (henceforth called ``option 1''), in which only the Majorana mass matrix $\mu_S$ 
breaks $G_f$ and CP to $G_\nu$. In this way, $\mu_S$ is the unique source of lepton flavour
and lepton  number violation in the neutral sector. 
Similar to what is found for the charged lepton masses, light neutrino masses are not constrained
in this scenario, and 
their mass spectrum can follow either a normal ordering or an inverted ordering. 
The mass spectrum of the heavy sterile states is instead strongly restricted, since they combine to form three approximately degenerate pseudo-Dirac pairs (to a very high degree).

As we proceed to argue, analytical and numerical studies allow showing that the impact of these heavy sterile states on lepton mixing (i.e., results for lepton mixing angles, predictions for CP phases as well as (approximate) sum rules)
is always small, with relative deviations below $1\%$ from the results previously obtained in a model-independent scenario~\cite{Hagedorn:2014wha} (with the same symmetries $\Delta (3n^2)$ and $\Delta (6n^2)$).
This is a consequence of effects arising due to deviations from unitarity of the PMNS mixing matrix,
which are subject to stringent experimental limits.
The matrix encoding these effects turns out to be of a peculiar form in the considered scenario, being both flavour-diagonal and flavour-universal. 
Due to their pseudo-Dirac nature, the heavy states' contribution to $0\nu\beta\beta$ decay is always strongly suppressed. 
As we will discuss, and in stark contrast to typical ISS models, new contributions to cLFV are also negligible.
 
\section{Approach to lepton mixing}
\label{sec2}

We assume the existence of a flavour symmetry $G_f=\Delta (3 \, n^2)$ or $G_f=\Delta (6 \, n^2)$
and a $Z_3$ symmetry $Z_3^{(\mathrm{aux})}$, as well as a CP symmetry in the theory.\footnote{Since $\Delta (3 \, n^2)$ is a subgroup of $\Delta (6 \, n^2)$, it is sufficient to focus on the latter in the analysis.} These are broken (without specifying the breaking mechanism) to a residual $Z_3$ symmetry $G_\ell$, corresponding to the diagonal subgroup
of a $Z_3$ group contained in $G_f$ and $Z_3^{(\mathrm{aux})}$,\footnote{In the original study~\cite{Hagedorn:2014wha}, the residual symmetry $G_\ell$ was assumed to be fully contained in $G_f$. This was possible, since in~\cite{Hagedorn:2014wha}
the focus has been on the mass matrix combination $m_\ell^{\phantom{\dagger}} \, m_\ell^\dagger$ and not on the charged lepton mass matrix $m_\ell$ alone. Thus, only the transformation properties of LH lepton doublets were necessary.
However, when considering also $m_\ell$ and, consequently, RH charged leptons, a possibility to distinguish among these is needed. Nevertheless, the results for lepton mixing are not affected by this change.} in the charged lepton sector
and to $G_\nu= Z_2 \times CP$ (with $Z_2$ being a subgroup of $G_f$) among the neutral states. The $Z_2$ symmetry is given by the generator $Z$, denoted as $Z(\mathrm{\textbf{r}})$ in the representation $\mathrm{\textbf{r}}$.
 The CP symmetry is described by a CP transformation $X$ in flavour space. In the different representations $\mathrm{\textbf{r}}$ of $G_f$, $X (\mathrm{\textbf{r}})$ corresponds to a unitary matrix fulfilling
\begin{equation}
\label{eq:Xrcond}
X (\mathrm{\textbf{r}}) \, X(\mathrm{\textbf{r}})^\ast = X(\mathrm{\textbf{r}})^\ast \, X (\mathrm{\textbf{r}}) = \mathbb{1}\,,
\end{equation}
so that $X$ is always represented as a symmetric matrix.\footnote{For more details on this choice, see~\cite{Feruglio:2012cw}.} A consistent definition of a theory with $G_f$ and CP necessitates the fulfilment of the consistency 
condition
\begin{equation}
\label{eq:GfCPconsist}
X (\mathrm{\textbf{r}}) \, g (\mathrm{\textbf{r}})^\ast \, X (\mathrm{\textbf{r}})^\ast = g^\prime (\mathrm{\textbf{r}})\,,
\end{equation}
with $g$ and $g^\prime$ being elements of $G_f$ and $g^{(\prime)} (\mathrm{\textbf{r}})$
their representation matrices in the representation $\mathrm{\textbf{r}}$.
This condition must be fulfilled for all representations $\mathrm{\textbf{r}}$, or at least for the representations used for charged leptons and the neutral states.
Since the product $Z_2 \times CP$ is direct, $Z(\mathrm{\textbf{r}})$ and 
$X(\mathrm{\textbf{r}})$ commute
\begin{equation}
\label{eq:XZcond}
 X (\mathrm{\textbf{r}}) \, Z (\mathrm{\textbf{r}})^\ast - Z (\mathrm{\textbf{r}}) \, X (\mathrm{\textbf{r}}) =\mathbb{0}\,,
\end{equation}
for all representations $\mathrm{\textbf{r}}$.
The flavour and CP symmetries, together with their residuals, determine the lepton mixing pattern.
Since we follow the approach to lepton mixing presented in~\cite{Hagedorn:2014wha}, 
we further assume
that the index of $G_f$ is not divisible by three.
All choices of CP symmetries and residual $Z_2$ groups in the sector of the neutral states fulfil the conditions in Eqs.~(\ref{eq:Xrcond},\ref{eq:GfCPconsist},\ref{eq:XZcond}).

For convenience, we summarise below the relevant group theory aspects of $G_f$, i.e.~the generators and their form in the chosen irreducible representations $\mathrm{\textbf{r}}$ of $G_f$, as well as
details about the form of the CP transformation $X (\mathrm{\textbf{r}})$.

\medskip
The groups $\Delta (3 \, n^2)$ and $\Delta (6 \, n^2)$ are series of discrete symmetries for integer $n$. 
For $n \geq 2$, 
$\Delta (3 \, n^2)$ is non-abelian, while all groups $\Delta (6 \, n^2)$ have this property. The groups $\Delta (3 \, n^2)$ are isomorphic to the semi-direct product $(Z_n \times Z_n) \rtimes Z_3$ 
and can be described in terms of three generators $a$, $c$ and $d$ that fulfil the following relations
\begin{equation}
\label{eq:gen1}
a^3=e \; , \;\; c^n=e \; , \;\; d^n=e \; , \;\;
a \, c \, a^{-1} = c^{-1} \, d^{-1} \; , \;\; a \, d \, a^{-1} = c \; , \;\; c \, d= d \, c\,,
\end{equation} 
with $e$ being the neutral element of the group. For the groups $\Delta (6 \, n^2)$ that are isomorphic to $(Z_n \times Z_n) \rtimes S_3$, one adds the fourth generator $b$
to the set $\{ a, c, d\}$ which fulfils the relations
\begin{equation}
\label{eq:gen2}
b^2=e \; , \;\; (a \, b)^2=e \; , \;\; b \, c \, b^{-1} = d^{-1} \;\; \mbox{and} \;\; b \, d \, b^{-1} = c^{-1} \; .
\end{equation}
We note that all elements of the groups can be written in terms of these generators as
\begin{equation}
\label{eq:eleGf}
g= a^\alpha \, c^\gamma \, d^\delta \;\; \mbox{and} \;\; g= a^\alpha \, b^\beta \, c^\gamma \, d^\delta \;\; \mbox{with} \;\; \alpha=0,1,2, \; \beta=0, 1, \; 0 \leq \gamma, \delta \leq n-1 \; ,
\end{equation}
respectively.
For the analysis of lepton mixing we are interested in the generators in the irreducible faithful (complex) three-dimensional representation ${\mathbf 3}$ and in the (trivial) singlet ${\mathbf 1}$. 
For ${\mathbf 3}$
 we have 
\begin{eqnarray}
\label{eq:genrep3}
&&a ({\mathbf 3})  = \left( \begin{array}{ccc}
1 & 0 & 0\\
0 & \omega & 0\\
0 & 0 & \omega^2
\end{array}
\right) \; , \;\; b ({\mathbf 3}) = \left( \begin{array}{ccc}
1 & 0 & 0\\
0 & 0 & \omega^2 \\
0 & \omega & 0
\end{array}
\right) \; ,
\\ \nonumber
&&c ({\mathbf 3}) =\frac 13 \, \left( \begin{array}{ccc}
1 + 2 \, \cos \phi_n & 1- \cos \phi_n-\sqrt{3} \, \sin \phi_n & 1- \cos \phi_n+\sqrt{3} \, \sin \phi_n \\
1- \cos \phi_n+\sqrt{3} \, \sin \phi_n & 1 + 2 \, \cos \phi_n & 1- \cos \phi_n-\sqrt{3} \, \sin \phi_n\\
1- \cos \phi_n-\sqrt{3} \, \sin \phi_n & 1- \cos \phi_n+\sqrt{3} \, \sin \phi_n & 1 + 2 \, \cos \phi_n
\end{array}
\right)\,,
\end{eqnarray}
with $\omega= e^{\frac{2 \, \pi \, i}{3}}$ and $\phi_n= \frac{2 \, \pi}{n}$, while for ${\mathbf 1}$ we have
\begin{equation}
\label{eq:genrep1}
a ({\mathbf 1}) = b ({\mathbf 1}) = c ({\mathbf 1}) =1 \; .
\end{equation}
We note that the generator $d$ can be obtained from the generators $a$ and $c$, since we find $d= a^2 \, c \, a$ from Eq.~(\ref{eq:gen1}).

\vspace{0.1in}
\noindent For completeness, we list the set of used CP symmetries.
CP symmetries are associated with the automorphisms of the flavour group $G_f$. In particular, the automorphism
\begin{equation}
\label{eq:X0auto}
a \; \rightarrow \; a \; , \;\; c \; \rightarrow \; c^{-1} \; , \;\; d \; \rightarrow \; d^{-1} \;\; \mbox{and} \;\; b \; \rightarrow \; b\,,
\end{equation}
for $G_f =\Delta (3 \, n^2)$ and $G_f=\Delta (6 \, n^2)$, respectively, corresponds to the CP transformation $X_0$ that is of the following form in the representations ${\mathbf 1}$ and ${\mathbf 3}$ 
\begin{equation}
\label{eq:X0}
X_0 ({\mathbf 1}) = 1 \;\; \mbox{and} \;\; X_0 ({\mathbf 3}) = \left( \begin{array}{ccc}
1 & 0 & 0\\
0 & 0 & 1\\
0 & 1 & 0
\end{array}
\right) 
\; .
\end{equation}
All other CP transformations $X$ of interest correspond to the composition of the automorphism in Eq.~(\ref{eq:X0auto}) and a group transformation $g$. The CP transformation $X (\mathrm{{\mathbf r}})$ in the representation $\mathrm{{\mathbf r}}$ is of the form
\begin{equation}
\label{eq:Xr}
X (\mathrm{{\mathbf r}}) = g (\mathrm{{\mathbf r}}) \, X_0 (\mathrm{{\mathbf r}}) \;\; \mbox{with} \;\; g (\mathrm{{\mathbf r}}) = a (\mathrm{{\mathbf r}})^\alpha \, c(\mathrm{{\mathbf r}}) ^\gamma \, d(\mathrm{{\mathbf r}}) ^\delta \;\; \mbox{and} \;\; g (\mathrm{{\mathbf r}}) = a(\mathrm{{\mathbf r}})^\alpha \, b(\mathrm{{\mathbf r}})^\beta \, c(\mathrm{{\mathbf r}})^\gamma \, d(\mathrm{{\mathbf r}})^\delta
\end{equation}
for $G_f=\Delta (3 \, n^2)$ and $G_f=\Delta (6 \, n^2)$, respectively, as long as $X (\mathrm{{\mathbf r}})$ represents a symmetric matrix in flavour space, see Eq.~(\ref{eq:Xrcond}). 

In the following, we first review the implementation of these symmetries and their residuals in the model-independent scenario that has been considered in~\cite{Hagedorn:2014wha}, and then turn to the $(3,3)$ ISS framework, focusing on one particular implementation, called option 1. We further comment on two other minimal options at the end of this section.

\subsection{Description of a model-independent scenario}
\label{sec21}
In order to establish a baseline of the symmetry-based predictions for lepton mixing, we briefly review how the symmetries are acting on the Weinberg operator, as discussed in~\cite{Hagedorn:2014wha}, thus offering a model-independent way to study the implications of the considered flavour symmetries.
In the model-independent scenario, we consider the mass terms
\begin{equation}
\label{eq:masseseff}
-\bar{\ell}_{\alpha L} \, (m_\ell)_{\alpha\beta} \, \ell_{\beta R} - \frac 12 \, \overline{\nu}^c_{\alpha L} \, (m_\nu)_{\alpha\beta} \, \nu_{\beta L} + \mathrm{H.c.}\,,
\end{equation}
for charged leptons, $m_\ell$, and for neutrinos, $m_\nu$, and with indices $\alpha, \beta = e, \mu, \tau$. 
While charged leptons acquire their (Dirac) masses from the Yukawa couplings to the Higgs,  
the LNV neutrino mass term can be effectively generated by means of the Weinberg operator, 
\begin{equation}
\label{eq:YukawaWeinberg}
-(y_\ell)_{\alpha\beta} \, \overline{L}_\alpha \, H \, \ell_{\beta R} + 
\frac{1}{\Lambda_{\mathrm{LN}}} \, (y_\nu)_{\alpha\beta} \, \Big( \overline{L}^c_\alpha \, H \Big) \, \Big( L_\beta \, H \Big)  + \mathrm{H.c.}
\end{equation}
with LH lepton doublets defined as $L_\alpha = \Big( \begin{array}{c} \nu_{\alpha L} \\ \ell_{\alpha L} \end{array} \Big) \sim (\mathbf{2}, - \frac 12)$, RH charged leptons
$\ell_{\alpha R} \sim (\mathbf{1}, -1)$ and the Higgs doublet $H \sim (\mathbf{2}, \frac 12)$ under $SU(2)_L \times U(1)_Y$. 
$\Lambda_\mathrm{LN}$ defines the scale at which lepton  number is broken and Majorana neutrino masses are generated.
After electroweak symmetry breaking, the mass matrices $m_\ell$
and $m_\nu$ are given by  
\begin{equation}
\label{eq:massesEWth}
m_\ell= y_\ell \, \frac{v}{\sqrt{2}} \;\; \mbox{and} \;\; m_\nu= y_\nu \, \frac{v^2}{\Lambda_{\mathrm{LN}}} \; ,
\end{equation}
(with the Higgs vev $v=246\:\mathrm{GeV}$).
The physical (mass) basis, denoted by $\hat{\phantom{x}}$, is related to the interaction basis by the unitary transformations
\begin{equation}
\label{eq:Umassbasis}
\ell_{L} = U_\ell \, \hat{\ell}_{ L} \; , \;\; \ell_{R} = U_R \, \hat{\ell}_{R} \;\; \mbox{and} \;\; \nu_{ L} = U_\nu \, \hat{\nu}_{ L} \; .
\end{equation}
The mass matrices $m_\ell$ and $m_\nu$ are then diagonalised
 as follows\begin{equation}
\label{eq:massesdiag}
U_\ell^\dagger \, m_\ell \, U_R = m_\ell^\mathrm{diag} = \mathrm{diag}\left(m_e, m_\mu, m_\tau \right)
\;\; \mbox{and} \;\;
U_\nu^T \, m_\nu \, U_\nu = m_\nu^\mathrm{diag} = \mathrm{diag}\left(m_1, m_2, m_3 \right)\,,
\end{equation}
and the (unitary) PMNS mixing matrix $U_{\mathrm{PMNS}}$ appears in the charged current interactions\footnote{Notice that a few conventions concerning notation of bases and fermion fields differ from the rest of thesis, in order to avoid confusion and match the conventions of the corresponding literature.}
\begin{equation}
\label{eq:LCCdefPMNS}
- \frac{g}{\sqrt{2}} \, \overline{\hat{\ell}}_{L} \, \slash\!\!\!\!W^{-} \, U_{\mathrm{PMNS}} \, \hat{\nu}_L \;\; \mbox{with} \;\; U_{\mathrm{PMNS}} = U_\ell^\dagger \, U_\nu \; .
\end{equation}
When it comes to the implementation of $G_f$ and CP, and of the residual symmetries $G_\ell$ and $G_\nu$, we first specify the assignment of LH lepton doublets $L_\alpha$ and RH charged leptons $\ell_{\alpha R}$. In order to
constrain as much as possible the resulting lepton mixing pattern, we assign $L_\alpha$ to an irreducible, faithful (complex)\footnote{Only for the choice $n=2$ of the index of $G_f$ this representation is real.} three-dimensional representation $\mathbf{3}$ of $G_f$. This representation can be chosen without loss of generality (see \cite{King:2013vna} for details) as the representation $\mathbf{3_{(n-1, 1)}}$ and $\mathbf{3_{1 \, (1)}}$ in the convention of~\cite{Luhn:2007uq} and~\cite{Escobar:2008vc}, respectively. Right-handed charged leptons $\ell_{\alpha R}$ transform as the trivial singlet $\mathbf{1}$ of $G_f$.
In order to distinguish the different flavours, we employ the $Z_3$ symmetry $Z_3^{(\mathrm{aux})}$ and assign $\ell_{eR} \sim 1$, $\ell_{\mu
R} \sim \omega$ and $\ell_{\tau R} \sim \omega^2$ with $\omega= e^{\frac{2 \, \pi \, i}{3}}$. Left-handed lepton doublets $L_\alpha$
do not carry a non-trivial charge under $Z_3^{(\mathrm{aux})}$. 

The residual symmetry $G_\ell$ is fixed to the diagonal subgroup of the $Z_3$ group, arising from the generator $a$ of $G_f$, see Eqs.~(\ref{eq:genrep3}, \ref{eq:genrep1}),  and $Z_3^{(\mathrm{aux})}$. Since $a (\mathbf{3})$ is diagonal, see Eq.~(\ref{eq:genrep3}), the mass matrix $m_\ell$ of charged leptons is diagonal. In our analysis, we assume that charged lepton masses are canonically ordered\footnote{For results arising in the case of non-canonically ordered charged lepton masses, see~\cite{Hagedorn:2014wha}.} so that the contribution to lepton mixing from the charged lepton sector is trivial, i.e.
\begin{equation}
\label{eq:Ul}
U_\ell = \left( \begin{array}{ccc}
1 & 0 & 0\\
0 & 1 & 0\\
0 & 0 & 1
\end{array}
\right) \; .
\end{equation}
The lepton mixing pattern thus only depends on the choice of $G_f$, the CP symmetry and the residual $Z_2$ symmetry among the neutral states. In general, the light neutrino mass matrix $m_\nu$ is constrained by the conditions~\cite{Feruglio:2012cw} 
\begin{equation}
\label{eq:ZXmnu}
Z (\mathbf{3})^T \, m_\nu \, Z (\mathbf{3}) = m_\nu \;\; \mbox{and} \;\; X (\mathbf{3}) \, m_\nu \, X (\mathbf{3}) = m_\nu^\ast \; .
\end{equation}
The CP transformation $X (\mathbf{3})$ can be written as
\begin{equation}
\label{eq:XOmega3}
X (\mathbf{3}) = \Omega (\mathbf{3}) \, \Omega (\mathbf{3})^T\,,
\end{equation}
with $\Omega (\mathbf{3})$ being unitary; furthermore $\Omega (\mathbf{3})$ can be chosen such that $\Omega (\mathbf{3})^\dagger \, Z (\mathbf{3}) \, \Omega (\mathbf{3})$ is diagonal.
In this basis, rotated by $\Omega (\mathbf{3})$, the light neutrino mass matrix is block-diagonal and real.
Since $Z (\mathbf{3})$ generates a $Z_2$ symmetry, two of its eigenvalues are equal. This explains why the resulting matrix is block-diagonal and why a rotation around a free angle $\theta$, encoded in the rotation matrix $R_{fh} (\theta)$
(with the indices $f$ and $h$ determined by the pair of degenerate eigenvalues of $Z (\mathbf{3})$), is necessary in order
to arrive at a basis in which $m_\nu$ is diagonal. 
Furthermore, positive semi-definiteness of the light neutrino masses is ensured by a diagonal matrix $K_\nu$, with entries taking values $\pm 1$ and $\pm i$. 
Hence, $U_\nu$ is given by
\begin{equation}
\label{eq:Unumodind}
U_\nu = \Omega (\mathbf{3}) \, R_{fh} (\theta) \, K_\nu \; .
\end{equation}
The explicit form of $\Omega (\mathbf{3})$ and the value of the indices $f$ and $h$ in the different cases, Case 1) through Case 3 b.1), can be found in Appendix~\ref{app:leptonmixing}.
Here we just give the general forms of the generator $Z$ of the residual $Z_2$ symmetry and the CP transformation $X$ for the different cases.
For Case 1) we have
\begin{equation}
    Z = c^{n/2}\,,\quad X = abc^s d^{2s} X_0\,,
\end{equation}
with $s$ restricted to $0\leq s\leq n-1$.
Case 2) is defined as
\begin{equation}
    Z = c^{n/2}\,,\quad X = c^s d^t X_0\,,
\end{equation}
with $s,t$ being restricted to $0\leq s,t\leq n-1$.
Finally, the Cases 3 a) and 3 b.1) rely on
\begin{equation}
    Z = b c^m d^m\,,\quad X = b c^s d^{m-s} X_0\,,
\end{equation}
with $m,s$ being restricted to $0\leq m,s\leq n-1$.
The difference between Case 3 a) and Case 3 b.1) is in their definition of the PMNS: the PMNS of Case 3 b.1) is a permutation of the PMNS in Case 3 a). For more details see Appendix~\ref{app:leptonmixing}.

Since the charged leptons' physical basis coincides with the interaction basis, see Eq.~\eqref{eq:Ul}, we have $U_\ell = \mathbb{1}$ and thus $U_{\mathrm{PMNS}}=U_\nu$.
The angle $\theta$ can take values between $0$ and $\pi$ and is fixed by accommodating the measured lepton mixing angles as well as possible.

\mathversion{bold}
\subsection{$(3,3)$ ISS framework}
\mathversion{normal}
\label{sec22}
In the $(3,3)$ ISS framework six neutral states, singlets under the SM gauge group, are added to the SM field content. In the following, these are denoted by $N_i$ and $S_j$ with $i,j = 1, 2, 3$. 
The Lagrangian giving rise to masses for the neutral particles (i.e. light neutrinos and heavy sterile states) reads
\begin{equation}
\label{eq:LISS}
- (y_D)_{\alpha i} \, \overline{L}^c_\alpha \, H \, N^c_i - (M_{NS})_{ij} \, \overline{N}_i \, S_j - \frac 12 \,  (\mu_S)_{kl} \, \overline{S}^c_k \, S_l + \mathrm{H.c.}
\end{equation}
with $\alpha=e, \mu, \tau$ and $i,j,k,l=1,2,3$. 
In the basis $\left(\nu_{\alpha L}, N^c_i, S_j \right)$,\footnote{In the following, we neglect possible contributions to the masses of the neutral particles arising from radiative corrections.} the mass matrix is of the form
\begin{equation}
\label{eq:MMaj}
\mathcal{M}_{\mathrm{Maj}} = 
\left( \begin{array}{ccc}
 \mathbb{0} & m_D & \mathbb{0} \\
 m_D^T & \mathbb{0} & M_{NS} \\
 \mathbb{0} & M_{NS}^T & \mu_S
\end{array}
\right) \;\; \mbox{with} \;\; m_D = y_D \, \frac{v}{\sqrt{2}}\,\text.
\end{equation}

In the limit $|\mu_S| \ll |m_D| \ll |M_{NS}|$ the light neutrino mass matrix is given at leading order 
in $(|m_D|/|M_{NS}|)^2$ by
\begin{equation}
\label{eq:mnuLO}
m_\nu = m_D \, \Big( M_{NS}^{-1} \Big)^{T} \, \mu_S \, M_{NS}^{-1} \, m_D^T\,\text.
\end{equation}
The contribution at subleading order reads~\cite{Hettmansperger:2011bt}\footnote{Note the different choice of basis in~\cite{Hettmansperger:2011bt}.}
\begin{equation}
\label{eq:mnuNLO}
\!\!m_\nu^1= - \frac 12 \, m_D \, \Big( M_{NS}^{-1} \Big)^T \, \Big[ \mu_S \, M_{NS}^{-1} \, m_D^T \, m_D^\ast \, \Big( M_{NS}^{-1} \Big)^\dagger + \Big( M_{NS}^{-1} \Big)^\ast \, m_D^\dagger \, m_D \, \Big( M_{NS}^{-1} \Big)^T \, \mu_S \Big] \, M_{NS}^{-1} \, m_D^T \, .
\end{equation}
The source of lepton number breaking in the ISS framework is $\mu_S$ and light neutrino masses vanish in the limit $\mu_S \, \rightarrow \, 0$, upon which lepton number conservation is restored. 

The matrix $\mathcal{M}_{\mathrm{Maj}}$ is diagonalised as
\begin{equation}
\label{eq:MMajdiag}
\mathcal{U}^ T \, \mathcal{M}_{\mathrm{Maj}} \, \mathcal{U} = \mathcal{M}_{\mathrm{Maj}}^{\mathrm{diag}} \,,
\end{equation}
with
\begin{equation}
\label{eq:formBigU}
\mathcal{U} = \left( \begin{array}{cc}
\tilde{U}_\nu & S \\ T & V
\end{array}
\right)\,,
\end{equation}
in which $\tilde{U}_\nu$ is a three-by-three, $S$ a three-by-six, $T$ a six-by-three and $V$ a six-by-six matrix. 
The mass spectrum contains the three light (mostly active) neutrinos and six heavy (mostly sterile) states; their masses are denoted by $m_i$, with $i = 1, 2, 3$ corresponding to the light neutrinos, and $i = 4, ..., 9$ regarding the heavy neutral mass eigenstates.
For $|\mu_S| \ll |M_{NS}|$, the heavy masses are given to good approximation by $M_{NS}$, with $\mu_S$ determining the
mass splitting between the states forming pseudo-Dirac pairs.

We note that at leading order $\tilde{U}_\nu$ approximately diagonalises the light neutrino mass matrix (c.f. Eq.~\eqref{eq:mnuLO}) as
\begin{equation}
    \label{eq:Unutildemu}
    \tilde{U}_\nu^T m_\nu\tilde{U}_\nu\approx \:\mathrm{diag} (m_1, m_2, m_3)\,.
\end{equation}
While $\mathcal{U}$ is unitary, $\mathcal{U} \, \mathcal{U}^\dagger=\mathcal{U}^\dagger \, \mathcal{U} = \mathbb{1}$, none of the matrices $\tilde{U}_\nu$, $S$, $T$ and $V$ has a priori this property. We can define the (in general non-unitary) PMNS mixing matrix as
\begin{equation}
\label{eq:defUPMNStilde}
\tilde{U}_{\mathrm{PMNS}} = U_\ell^\dagger \, \tilde{U}_\nu  \; .
\end{equation}
The non-unitarity of $\tilde{U}_{\mathrm{PMNS}}$, induced by the mixing of the active neutrinos with the (heavy) sterile states, can be conveniently captured in the matrix $\eta$,
with flavour indices $\alpha, \beta=e, \mu, \tau$, and it is defined as in Eq.~\ref{eqn:Utildeeta}, with $\eta$ hermitian, as discussed in Chapter~\ref{sec:massivenu}.
Note that
\begin{equation}
\label{eq:UPMNStildeeta2}
\tilde{U}_{\mathrm{PMNS}} \, \tilde{U}_{\mathrm{PMNS}}^\dagger \approx \mathbb{1} - 2 \, \eta \; .
\end{equation}
For $U_\ell=\mathbb{1}$, which is always the case in our analysis, the following equality also holds 
\begin{equation}
\label{eq:UnuUl1}
\tilde{U}_{\nu} = \Big( \mathbb{1} - \eta \Big) \, U_0 \; .
\end{equation}
The size of $\eta$ and its form in flavour space are given at leading order by
\begin{equation}
\label{eq:eta}
\eta = \frac 12 \, m_D^\ast \, \Big( M_{NS}^{-1} \Big)^\dagger \, M_{NS}^{-1} \, m_D^T\,.
\end{equation}
We can estimate the form of the matrix $T$ as 
\begin{equation}
\label{eq:matT}
T = \left( \begin{array}{c}
\mathbb{0} \\ - M_{NS}^{-1} \, m_D^T \, \tilde{U}_\nu
\end{array}
\right) \approx \left( \begin{array}{c}
\mathbb{0} \\ - M_{NS}^{-1} \, m_D^T \, U_0
\end{array}
\right) \; ,
\end{equation}
while for $S$ one has  
\begin{equation}
\label{eq:matS}
S = \left(\mathbb{0} \; , \;\; m_D^\ast \, \Big( M_{NS}^{-1} \Big)^\dagger  
\right) \, V \; ,
\end{equation}
and $V$ approximately diagonalises the lower six-by-six matrix of $\mathcal{M}_{\mathrm{Maj}}$, i.e.~
\begin{equation}
\label{eq:defV}
V^T \, \left( \begin{array}{cc}
\mathbb{0} & M_{NS} \\
M_{NS}^T & \mu_S
\end{array}
\right) \, V \approx \:\mathrm{diag}\left(m_4, ..., m_9\right)\,.
\end{equation}
The matrix $\mu_S$, a complex symmetric matrix, is itself diagonalised by
\begin{equation}
 \label{eq:muS}
U_S^T \, \mu_S \, U_S = \left(
\begin{array}{ccc}
\mu_1 & 0 & 0 \\
0 & \mu_2 & 0 \\
0 & 0 & \mu_3
\end{array}
\right)\,,
\end{equation}
with $\mu_i$ real and positive semi-definite, and $U_S$ unitary.

\vspace{0.1in}
\noindent Like in the model-independent scenario, the charged lepton sector leaves the residual symmetry $G_\ell$ invariant. For this reason, we assign the three generations of LH lepton doublets $L_\alpha$ and of 
RH charged leptons $\ell_{\alpha R}$ to the same representations under $G_f$, the $Z_3$ group $Z_3^{(\mathrm{aux})}$ and the CP symmetry as in the model-independent scenario. As a consequence, also in the $(3,3)$ ISS framework
the charged lepton mass matrix $m_\ell$ is diagonal and the contribution to the lepton mixing matrix is $U_\ell=\mathbb{1}$.  The group $G_\nu=Z_2 \times CP$ is the residual symmetry among the neutral states. In the $(3,3)$ 
ISS framework, we also have to assign the heavy sterile states, $N_i$ and $S_j$ with $i,j= 1,2,3$, to representations of $G_f$, $Z_3^{(\mathrm{aux})}$ and CP.
In the following, we identify three minimal options to choose these representations, allowing to locate the source of flavour violation in different terms of the lagrangian.

\subsubsection*{Option 1}

For option 1, we assume that $N_i$ and $S_j$ each transform like the LH lepton doublets $L_\alpha$, namely as the triplet $\mathbf{3}$ under $G_f$.  
Furthermore, the heavy sterile states are neutral under $Z_3^{(\mathrm{aux})}$. As a consequence of this assignment, the Dirac neutrino Yukawa matrix $y_D$, and 
consequently the mass matrix $m_D$ as well as the matrix $M_{NS}$, are non-vanishing in the limit of unbroken $G_f$, $Z_3^{(\mathrm{aux})}$ and CP. They take a particularly simple form 
 \begin{equation} 
 \label{eq:mDopt1}
 m_D = y_0 \, \left(
 \begin{array}{ccc}
 1 & 0 & 0\\
 0 & 1 & 0\\
 0 & 0 & 1
 \end{array}
 \right) \, \frac{v}{\sqrt{2}} \;\; \mbox{with} \;\; y_0 > 0\,,
 \end{equation}
 and
 \begin{equation}
 \label{eq:MNSopt1}
 M_{NS} = M_0 \, \left(
 \begin{array}{ccc}
  1 & 0 & 0\\
 0 & 1 & 0\\
 0 & 0 & 1
 \end{array}
 \right) \;\; \mbox{with} \;\; M_0 > 0 \, .
 \end{equation}
Thus, the only source of $G_f$ and CP breaking in the sector of the neutral states is the matrix $\mu_S$. In order to preserve the residual symmetry $G_\nu$, the matrix $\mu_S$ is constrained by the following equations
\begin{equation}
\label{eq:muScond}
Z (\mathbf{3})^T \, \mu_S \, Z (\mathbf{3}) = \mu_S \;\; \mbox{and} \;\; X (\mathbf{3}) \, \mu_S \, X (\mathbf{3}) = \mu_S^\ast \; ,
\end{equation}
implying that $\mu_S$ has to fulfil the same relations as $m_\nu$ (cf. Eq.~(\ref{eq:ZXmnu})). Hence, the matrix $U_S$, which diagonalises $\mu_S$, is of the same form as $U_\nu$, see Eq.~(\ref{eq:Unumodind}), 
\begin{equation}
 \label{eq:USopt1}
U_S = \Omega (\mathbf{3}) \, R_{fh} (\theta_S) \; .
\end{equation}
Note that we do not mention explicitly a matrix equivalent to $K_\nu$ in Eq.~(\ref{eq:Unumodind}), as we assume for concreteness in our analysis that it is the identity matrix. 

Thus, for option 1, $\mu_S$ is the unique source of lepton  number violation and lepton flavour violation. Nevertheless, lepton number, $G_f$ and CP can be broken in different ways, explicitly or spontaneously, and at vastly
different scales in concrete models.

Plugging $m_D$, $M_{NS}$ and $\mu_S$
 from Eqs.~(\ref{eq:mDopt1},\ref{eq:MNSopt1},\ref{eq:muS},\ref{eq:USopt1}) into the form of $m_\nu$ in Eq.~(\ref{eq:mnuLO}), we find at leading order
 \begin{equation}
 \label{eq:mnuLOopt1}
m_\nu = \frac{y_0^2 \, v^2}{2 \, M_0^2} \, \mu_S = \frac{y_0^2 \, v^2}{2 \, M_0^2} \, U_S^\ast \, \left( \begin{array}{ccc}
\mu_1 & 0 & 0 \\
0 & \mu_2 & 0 \\
0 & 0 & \mu_3
\end{array}
\right) \, U_S^\dagger \, .
\end{equation}
Consequently, the matrix $\tilde{U}_\nu$, which diagonalises $m_\nu$ at leading order (neglecting the correction $\eta$ that encodes the deviation from unitarity of $\tilde{U}_\nu$), is given by
\begin{equation}
\label{eq:UnuU0opt1}
\tilde{U}_\nu \approx U_0 = U_S = \Omega (\mathbf{3}) \, R_{fh} (\theta_S) \; , 
\end{equation}
and the light neutrino masses read
\begin{equation}
\label{eq:m123LOopt1}
m_i = \frac{y_0^2 \, v^2}{2 \, M_0^2} \, \mu_i \;\; \mbox{for} \;\; i=1,2,3 \; .
\end{equation}
Assuming $y_0 \sim 1$ and $M_0 \sim 1000 \, \mbox{GeV}$, we can estimate the size of $\mu_i$ to be of the order of $\mathrm{eV}$. The ratio between $m_D$ and $M_{NS}$, evaluating the impact 
of the heavy sterile states, is then $\frac{y_0 \, v}{\sqrt{2} \, M_0} \sim 0.17$.
Since the mass squared differences of neutrinos have been determined from neutrino oscillation data and the sum of neutrino masses is constrained by cosmological measurements (see Section~\ref{sec:osci}), 
the values of $\mu_i$ are further restricted. 
Since $U_\ell=\mathbb{1}$ and $\tilde{U}_\nu$ is at leading order of the form given in Eq.~(\ref{eq:UnuU0opt1}), we have for the PMNS mixing matrix
\begin{equation}
\label{eq:UPMNSLOopt1}
\tilde{U}_{\mathrm{PMNS}}\approx\Omega (\mathbf{3}) \, R_{fh} (\theta_S)\,,
\end{equation}
with $\theta_S$ being constrained by the measured values of the lepton mixing angles, like $\theta$ in Eq.~(\ref{eq:Unumodind}). We note that we consider the free angle $\theta_S$ to vary in the range $0$ and $\pi$.
The results in the $(3,3)$~ISS framework (at leading order) are thus identical to those obtained in the model-independent scenario. However, they can be altered by two effects: the inclusion of the subleading contribution $m_\nu^1$ to the light
neutrino mass matrix in Eq.~(\ref{eq:mnuNLO}) and effects of non-unitarity of $\tilde{U}_\nu$, see Eq.~(\ref{eq:eta}). This is studied in detail analytically in Section~\ref{sec4}
and numerically in Appendix~\ref{app:numericalISS}.
The experimental constraints on $\eta$ are discussed in Chapter~\ref{sec:unitaritynu} and their implications on the considered scenario in Section~\ref{sec:unitarityISS}. 

We briefly discuss the form of the matrices $S$, $T$ and $V$, as well as the mass spectrum of the heavy sterile states analytically. With Eqs.~(\ref{eq:matT},\ref{eq:mDopt1},\ref{eq:MNSopt1},\ref{eq:UnuU0opt1}) the matrix $T$ reads at leading order
\begin{equation}
\label{eq:matTopt1}
T  = \left( \begin{array}{c}
\mathbb{0} \\ - \frac{y_0 \, v}{\sqrt{2} \, M_0} \, U_S
\end{array}
\right) \; .
\end{equation}
From the definition of $V$ in Eq.~(\ref{eq:defV}) and with the form of $M_{NS}$ in Eq.~(\ref{eq:MNSopt1}) and $\mu_S$ in Eqs.~(\ref{eq:muS},\ref{eq:USopt1}), we find at leading order for $V$
 \begin{equation}
\label{eq:matVopt1}
V = \frac{1}{\sqrt{2}} \, \left(
\begin{array}{cc}
i \, U_S^\ast & U_S^\ast\\
-i \, U_S & U_S
\end{array}
\right) \; ,
\end{equation}
while the matrix $S$ in Eq.~(\ref{eq:matS}) reads 
\begin{equation}
\label{eq:matSopt1}
S = \frac{y_0 \, v}{2 \, M_0} \, \left(- i \, U_S \; , \;\; U_S \right) \, . 
\end{equation}
We note that the approximate analytical results for $\tilde{U}_\nu$, $S$, $T$ and $V$ have been compared to the numerical ones for one choice of parameters for Case 1) and we find good agreement in form and magnitude of their entries.
The mass spectrum of the heavy sterile states (arising from the diagonalisation through $V$ in Eq.~(\ref{eq:matVopt1})) is at leading order
\begin{equation}
\label{eq:heavyMopt1}
m_{3 + i} = M_0 - \frac{\mu_i}{2} \;\; \mbox{and} \;\; m_{6+i}= M_0 + \frac{\mu_i}{2} \;\; \mbox{with} \;\; i=1,2,3 \,.
\end{equation}
All heavy sterile states are thus degenerate in mass to a very high degree for typical choices of $M_0$ and $\mu_S$, e.g.~$M_0 \sim 1000 \, \mathrm{GeV}$ and $\mu_S \lesssim 1 \, \mathrm{keV}$.

\vspace{0.2in}
Beyond option 1, there are two further minimal options, option 2 and option 3, in which only one of the mass matrices $m_D$, $M_{NS}$ and $\mu_S$ carries non-trivial flavour information.
These options share a common feature: in both the matrix $\mu_S$ has a trivial flavour structure.
Thus, for these options the sources of lepton flavour and lepton  number violation are decoupled. 
For option 2,
$m_D$ contains all flavour information, while $M_{NS}$ is flavour-diagonal and flavour-universal, 
so that the mass spectrum of the heavy sterile
states will be degenerate to a high degree, like for option 1. 
Instead, for option 3 the entire flavour structure is encoded in the matrix $M_{NS}$, while 
$m_D$ is flavour-diagonal
and flavour-universal. In this way, the heavy sterile states have in general different masses. We note that the realisation of option 2 and option 3 requires in general that (at least) the assignment of the three sterile
states $S_i$, $i=1,2,3$, under the flavour symmetry $G_f$ be altered compared to option 1, in order to ensure that the matrix $\mu_S$ is non-vanishing in the limit of unbroken $G_f$, $Z_3^{(\mathrm{aux})}$ and CP.
However, this can always be achieved by an appropriate choice of $G_f$. Obviously, one can also consider less minimal options in which two of the three mass matrices $m_D$, $M_{NS}$ and $\mu_S$,
if not all three of them, have a non-trivial flavour structure.

\mathversion{bold}
\section{Impact of heavy sterile states of the $(3,3)$ ISS on lepton mixing}
\mathversion{normal}
\label{sec4}

As already mentioned in Section~\ref{sec22}, there are two possible effects that can have an impact on lepton mixing: the inclusion of the subleading
contribution $m_\nu^1$ to the light neutrino mass matrix $m_\nu$ and effects of non-unitarity of $\tilde{U}_\nu$, which are encoded in  $\eta_{\alpha\beta}$.
A numerical analysis of examples for each case, Case 1) through Case 3 b.1), can be found in
Appendix~\ref{app:numericalISS} and confirms the analytical results, which we proceed to discuss.

\subsection{Subleading contribution to the light neutrino mass matrix}
\label{sec41}

When plugging in the form of the matrices
$m_D$, $M_{NS}$ and $\mu_S$ for option 1, see Eqs.~(\ref{eq:mDopt1},\ref{eq:MNSopt1},\ref{eq:muS},\ref{eq:USopt1}), the subleading contribution to the light neutrino mass matrix, shown in Eq.~(\ref{eq:mnuNLO}), takes a simple form:
\begin{equation}
\label{eq:mnuNLOopt1}
m_\nu^1 = - \frac{y_0^4 \, v^4}{4 \, M_0^4}  \, \mu_S = - \frac{y_0^4 \, v^4}{4 \, M_0^4}  \,  \, U_S^\ast \, \left( \begin{array}{ccc}
\mu_1 & 0 & 0 \\
0 & \mu_2 & 0 \\
0 & 0 & \mu_3
\end{array}
\right) \, U_S^\dagger \; .
\end{equation}
Comparing with the leading order contribution $m_\nu$, found in Eq.~(\ref{eq:mnuLOopt1}), we see that $m_\nu^1$ has exactly the same form in flavour space and is suppressed by
a factor $\frac{y_0^2 \, v^2}{2 \, M_0^2}$. Thus, this subleading contribution does not introduce any change in the lepton mixing parameters and only slightly corrects the values of the 
light neutrino masses, e.g.~for $y_0 \sim 1$ and $M_0 \sim 1000 \, \mathrm{GeV}$ the correction is around $0.03$ with respect to the leading order result, see Eq.~(\ref{eq:m123LOopt1}).
Such a correction can be compensated by re-adjusting the values of the parameters $\mu_i$.

\mathversion{bold}
\subsection{Effects of non-unitarity of $\tilde{U}_\nu$}
\label{sec42}
\mathversion{normal}

The deviation from unitarity of $\tilde{U}_\nu$ is encoded in $\eta$, see Eq.~(\ref{eq:eta}). For option 1, the form of $\eta$ turns out to be flavour-diagonal and
flavour-universal, since both $m_D$ and $M_{NS}$ have this property, see Eqs.~(\ref{eq:mDopt1},\ref{eq:MNSopt1})
\begin{equation}
\label{eq:etaopt1}
\eta = \frac{y_0^2 \, v^2}{4 \, M_0^2} \, \mathbb{1} \equiv \eta_0 \, \mathbb{1} \; .
\end{equation}
Furthermore, it is independent of the particular case, Case 1) through Case 3 b.1), which we confirm numerically.

For $y_0 \sim 1$ and $M_0 \sim 1000 \, \mathrm{GeV}$ we have $\eta_0 \sim 0.015$, while for $y_0 \sim 0.1$ it is suppressed by further two orders of magnitude (at constant $M_0$).
The features of being flavour-diagonal and flavour-universal are numerically confirmed. 
The size of $\eta$ and its dependence on $y_0^2$ as well as $\frac{1}{M_0^2}$ are also very well fulfilled.

Since $\eta$ is flavour-diagonal as well as flavour-universal and $\eta_0$ is positive, the presence of $\eta$ effectively leads to a suppression of all elements of $U_0 = U_S$, see Eq.~(\ref{eq:UnuU0opt1}).
We can thus easily estimate the deviations expected in the results for the lepton mixing parameters (mixing angles and CP invariants/CP phases) extracting them in the same
way as for the unitary case, i.e.~the $(3,3)$ ISS framework at leading order and the model-independent scenario (MIS).\footnote{We extract the lepton mixing parameters using Eqs.~(\ref{eq:sin2thij},\ref{eq:Jarlskog:1985ht},\ref{eq:I1I2}) in Appendix~\ref{app:leptonmixing} (with $U_\text{PMNS}$ replaced by $\tilde{U}_\nu$).} We consider relative deviations between the non-unitary results, $(\sin^2\theta_{ij})_{\mathrm{ISS}}$, $(J_{\mathrm{CP}})_{\mathrm{ISS}}$ and 
$(I_i)_{\mathrm{ISS}}$, and the unitary ones, $(\sin^2\theta_{ij})_{\mathrm{MIS}}$, $(J_{\mathrm{CP}})_{\mathrm{MIS}}$ and $(I_i)_{\mathrm{MIS}}$,\footnote{When considering these relative deviations, we always assume that
$(\sin^2\theta_{ij})_{\mathrm{MIS}}$, $(J_{\mathrm{CP}})_{\mathrm{MIS}}$ and $(I_i)_{\mathrm{MIS}}$ do not vanish.}
\begin{equation}
\Delta \sin^2\theta_{ij}=\frac{(\sin^2\theta_{ij})_{\mathrm{ISS}} - (\sin^2\theta_{ij})_{\mathrm{MIS}}}{(\sin^2\theta_{ij})_{\mathrm{MIS}}} \; , \;\;
\Delta J_{\mathrm{CP}}=\frac{(J_{\mathrm{CP}})_{\mathrm{ISS}}-(J_{\mathrm{CP}})_{\mathrm{MIS}}}{(J_{\mathrm{CP}})_{\mathrm{MIS}}} \;\; \mbox{and} \;\;
\Delta I_i=\frac{(I_i)_{\mathrm{ISS}}-(I_i)_{\mathrm{MIS}}}{(I_i)_{\mathrm{MIS}}} 
\end{equation}
and alike for the sines of the CP phases $\delta$, $\alpha$ and $\beta$. In doing so, we can find formulae for the relative deviations that are valid for all cases, Case 1) through Case 3 b.1).
The exact numerical values of these deviations can in general (slightly) depend on the chosen case and other parameters, such as the index of $G_f$, the choice of the residual $Z_2$ symmetry in the sector of the neutral states, and the value of the free angle 
$\theta_S$. We comment on this in the detailed numerical analysis whose results are summarised in Appendix~\ref{app:numericalISS}.

For $\Delta \sin^2 \theta_{ij}$ we have
\begin{equation}
\label{eq:Deltasin2thetaij_gen}
\!\!\!\!\!\!\Delta \sin^2 \theta_{13} \approx -2 \, \eta_0 \; , \;\; \Delta \sin^2 \theta_{12} \approx - \frac{2 \, \eta_0}{1-|U_{e3}|^2} \approx -2.04 \, \eta_0 \; , \;\; \Delta \sin^2 \theta_{23} \approx - \frac{2 \, \eta_0}{1-|U_{e3}|^2} \approx -2.04 \, \eta_0
\end{equation}
for $|U_{e3}|^2 \approx 0.022$~\cite{Esteban:2020cvm}. For the CP invariants $J_{\mathrm{CP}}$, $I_1$ and $I_2$ we find
\begin{equation}
\label{eq:DeltaJarlskog:1985htI1I2_gen}
\Delta J_{\mathrm{CP}} \approx - 4 \, \eta_0 \; , \;\; \Delta I_1 \approx  - 4 \, \eta_0 \; , \;\; \Delta I_2 \approx  - 4 \, \eta_0 \; .
\end{equation}
With this information we can also extract $\Delta \sin\delta$, $\Delta \sin\alpha$ and $\Delta \sin\beta$, arriving at
\begin{equation}
\label{eq:DeltasinCPphases_gen}
\Delta\sin\delta \approx -2.82 \, \eta_0 \; , \;\; \Delta\sin\alpha \approx -2.95 \, \eta_0 \; , \;\; \Delta\sin\beta \approx -2.95 \, \eta_0
\end{equation}
for $|U_{e2}|^2 \approx 0.30$, $|U_{e3}|^2 \approx 0.022$ and $|U_{\mu3}|^2 \approx 0.56$~\cite{Esteban:2020cvm}.
For $y_0 \sim 1$ and $M_0 \sim 1000 \, \mathrm{GeV}$ we expect 
\begin{equation}
\label{eq:estimateDsy01}
\Delta \sin^2 \theta_{ij} \approx -0.03 \; , \;\; \Delta J_{\mathrm{CP}} \approx \Delta I_i \approx -0.06 \; , \;\; \Delta\sin\delta\approx -0.042, \; \Delta\sin\alpha\approx \Delta\sin\beta \approx -0.044  \; .
\end{equation}
Due to the suppression of all elements of $U_0 = U_S$, all relative deviations are expected to be negative. Furthermore, their size slightly depends on 
the considered quantity and is generally not expected to exceed values of a few percent.
These estimates are confirmed numerically (see Appendix~\ref{app:numericalISS}).
It is important to note that certain features, like the vanishing of the sine
and the periodicity of some of the CP phases in terms of the group theory parameters, remain preserved exactly, since the flavour structure of 
the light neutrino mass matrix is not changed and the deviation from unitarity only amounts to a common rescaling of all elements of the PMNS mixing matrix.

\mathversion{bold}
\subsection{Symmetry endowed (3,3) ISS: unitarity constraints on $\tilde{U}_\nu$}
\mathversion{normal}
\label{sec:unitarityISS}

In what follows, we discuss the constraints arising from the
violation of unitarity of the PMNS mixing matrix $\tilde{U}_\nu$, 
as encoded in the  
matrix $\eta$, which turn out to be of paramount importance for the phenomenological study of the $(3,3)$ ISS endowed with  flavour symmetries.  
As can be seen from Eq.~(\ref{eq:etaopt1}), $\eta$ is determined by 
the chosen regimes for $y_0$ and $M_0$, which 
characterise the impact of the heavy sterile states on the lepton
mixing parameters. Thus, the experimental limits on the quantities
$\eta_{\alpha\beta}$, $\alpha,\beta=e,\mu,\tau$, are at the source of
the most important constraints on the present (3,3) ISS framework.

Before discussing how the limits on $\eta_{\alpha\beta}$ crucially 
constrain $y_0$ and hence the combination of $y_0$ and $M_0$, let us
first emphasise two points: 
we have checked numerically that the form of the 
quantities $\eta_{\alpha\beta}$ does not depend on the specific case, 
Case 1) through Case 3 b.1), as expected from the analytical estimate
in Eq.~(\ref{eq:etaopt1}); 
furthermore, we also confirm numerically that the matrix 
$\eta$ is flavour-diagonal and flavour-universal, and that
$\eta_0$ is proportional to $y_0^2$ (and inversely proportional to
$M_0^2$), as can be also seen from Eq.~(\ref{eq:etaopt1}). 
The (indirect) experimental constraints on $\eta_{\alpha\beta}$ are taken
from~\cite{Fernandez-Martinez:2016lgt} and  
are given by\footnote{
We use the bounds obtained in~\cite{Fernandez-Martinez:2016lgt}, although
the form of $\eta$ is flavour-diagonal and flavour-universal in the case at hand.}
\begin{equation}
\label{eq:etaexp}
\left|\eta_{\alpha\beta}\right| \leq \left(
\begin{array}{ccc}
1.3 \times 10^{-3} & 1.2 \times 10^{-5} & 1.4 \times 10^{-3}\\
1.2 \times 10^{-5} & 2.2 \times 10^{-4} & 6.0 \times 10^{-4}\\
1.4 \times 10^{-3} & 6.0 \times 10^{-4} & 2.8 \times 10^{-3}
\end{array}
\right) \;\; \mbox{at the $1\,\sigma$ level.}
\end{equation}
As can be verified, the diagonal element subject to the
strongest experimental bounds is $\eta_{\mu\mu}$, 
$\left|\eta_{\mu\mu}\right| \leq 2.2 \, (4.4) \, [6.6] \times 10^{-4}$ at
the $1 (2) [3]\, \sigma$ level. 
We thus use this limit in the subsequent analysis.

\begin{figure}[t!]
\begin{center}
\parbox{3in}{\hspace{-0.3in}\includegraphics*[scale=0.55]{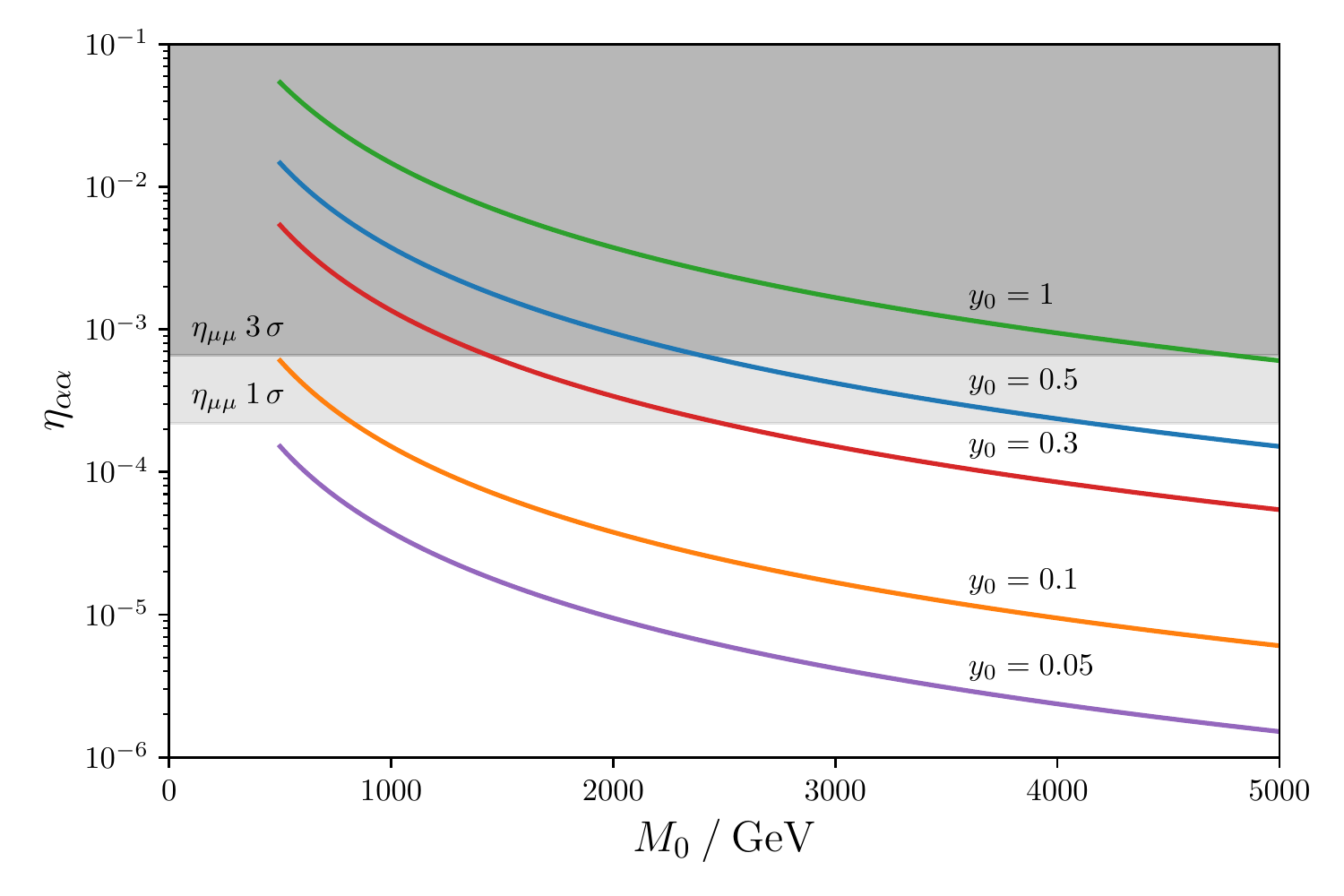}} 
\hspace{-0.1in}
\parbox{3in}{\includegraphics*[scale=0.55]{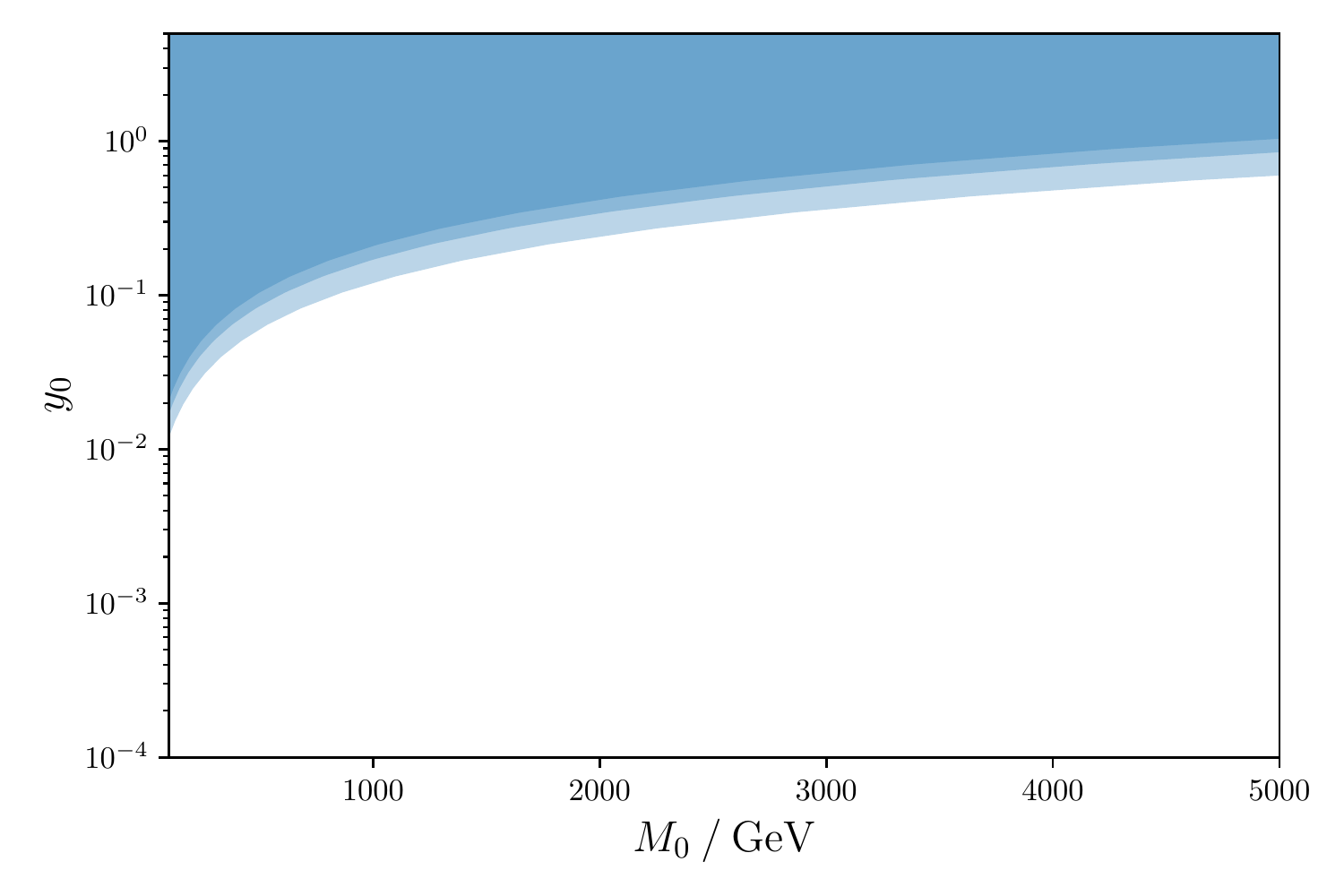}} 
\end{center}
\caption{{\small \textbf{Constraints from unitarity of 
\mathversion{bold}{$\tilde{U}_\nu$}}. \textbf{Left plot}:
    $\eta_{\alpha\alpha}$  with respect to the mass scale $M_0$ (in GeV) 
for different values of the Yukawa coupling: $y_0=0.05$ (purple line), $y_0=0.1$ (orange line), $y_0=0.3$ (red line), $y_0=0.5$ (blue line)  
and $y_0=1$ (green line). The grey-shaded regions denote the
areas excluded by the strongest constraint on the flavour-diagonal
entries of $\eta$ (arising from $\eta_{\mu\mu}$)
at $1 \, \sigma$ level (light grey)~\cite{Fernandez-Martinez:2016lgt} 
and $3 \, \sigma$ level (dark grey). 
\textbf{Right plot}: Disfavoured regions of the ($M_0-y_0$) plane, with $M_0$
given in GeV, due to 
conflict with experimental bounds on $\eta_{\alpha\alpha}$, at $1 \, \sigma$, $2 \, \sigma$ and 
$3 \, \sigma$ (respectively denoted by light, medium and dark blue). 
Figures from~\cite{Hagedorn:2021ldq}.
\label{fig:Constraints_eta}}}
\end{figure}

The maximal size of the Yukawa coupling $y_0$, compatible with the
experimental constraints on $\eta_{\alpha\beta}$, can be read from
the left plot in Fig.~\ref{fig:Constraints_eta}: for
$y_0$ as small as $y_0=0.1$, the unitarity constraints can be evaded 
for values of $M_0$ as low as $M_0 \gtrsim 500$~GeV (at the $3 \,
\sigma$ level); larger values,  
$y_0=0.5$, already require $M_0 \gtrsim 2400$~GeV, and for 
$y_0=1$ one must have $M_0 \gtrsim 4800$~GeV in order to be in
agreement with the bounds of Eq.~(\ref{eq:etaexp}) at the $3 \, \sigma$ level, i.e. $|\eta_{\mu\mu}|\lesssim 6.6\times10^{-4}$. 
This is illustrated in the right panel of
Fig.~\ref{fig:Constraints_eta} by an exclusion plot in the
($M_0-y_0$) plane. Our numerical analyses rely on 
regimes of $y_0$ and $M_0$ compatible with experimental data at the $3\, \sigma$ 
level,\footnote{As will be discussed in detail 
in Section~\ref{sec:clfv}, the predictions for
cLFV observables will not lead to any additional
constraints on the parameter space of the $(3,3)$ ISS framework 
with $G_f$ and CP in the case of option 1.
Thus, the only relevant constraints are those arising from the effects
of non-unitarity of $\tilde{U}_\nu$.} and regimes in conflict with experimental bounds on $\eta_{\alpha\beta}$ are clearly indicated in the discussion in Appendix~\ref{app:numericalISS}.
  
%

\vspace{0.3in}
In summary, we find that the effects of non-unitarity (of the PMNS mixing matrix,  $\tilde{U}_\nu$) on the lepton mixing parameters and on the (approximate) sum rules relating them, turn out to be below the $1\%$ level, once experimental limits on the quantities $\eta_{\alpha\beta}$
are taken into account. 
Consequently, the results obtained for option 1 of the $(3,3)$ ISS framework are very similar to those obtained in the model-independent scenario~\cite{Hagedorn:2014wha}. 
In particular, the dependence of the CP phases on the group theory parameters (especially those determining the CP transformation $X$) and the 
vanishing of a CP phase for certain choices of group theory parameters, are not affected.

\mathversion{bold}
\section{Results for neutrinoless double beta decay}
\mathversion{normal}
\label{sec6}

In the following, we briefly comment on $0\nu\beta\beta$ decay prospects for option 1 of the $(3,3)$ ISS framework.
First, we recall that  in the presence of light neutrinos and of heavy sterile states, the effective mass $m_{ee}$, accessible in $0\nu\beta\beta$ decay experiments, is given by~\cite{Blennow:2010th}
\begin{equation}
\label{eq:0nubb_HSS}
m_{ee} \simeq \sum_{i=1}^{3+n_s} \, \mathcal U_{e i}^2 \, p^2 \, \frac{m_{ i}}{p^2-m_{i}^2} \simeq \sum_{i=1}^3 \, \mathcal U^2_{e i} \, m_i + \sum_{k=4}^{3+n_s} \mathcal U_{e k}^2 \, p^2 \, \frac{m_{k}}{p^2-m_{k}^2} \; ,
\end{equation}
where $n_s$ denotes the number of heavy sterile states, in our case $n_s=6$, and the virtual momentum $p^2$ is estimated as $p^2 \simeq -(100 \, \mathrm{MeV})^2$. 
For $i=1,2,3$, the mixing matrix elements 
$\mathcal U_{e i}$ coincide with the elements of the first row of the matrix $\tilde{U}_\nu$ (and hence $\tilde{U}_{\mathrm{PMNS}}$); for $k=4,..., 3+n_s=4, ..., 9$,  
in our case $\mathcal U_{e k}$ are approximately given by 
\begin{equation}
\label{eq:matchingUek}
\mathcal U_{e k} \approx -i \, \left( \frac{y_0 \, v}{2 \, M_0} \right) \, (U_S)_{1 \, k-3} \;\; \mbox{for} \;\; k=4,..., 6 \;\; \mbox{and} \;\; \mathcal U_{e k} \approx \left( \frac{y_0 \, v}{2 \, M_0} \right) \, (U_S)_{1 \, k-6} \;\; \mbox{for} \;\; k=7,..., 9
\end{equation}
according to the expression for $S$ presented in Eq.~(\ref{eq:matSopt1}). 
For $i=1,2,3$ $m_{i}$  correspond to the light neutrino masses; we recall that, according to Eq.~\eqref{eq:heavyMopt1} for option 1 of the $(3,3)$ ISS framework, the masses of the heavy sterile states
$m_{k}$ (with $k=4,..., 3+n_s=4, ..., 9$) are approximately degenerate
\begin{equation}
\label{eq:mnkapprox}
m_{k} \approx M_0 \,.
\end{equation}
Thus, we have 
\begin{equation}
\label{eq:0nubb_HSS_alone}
m_{ee} \simeq \sum_{i=1}^3 \, \mathcal U^2_{e i} \, m_i + \left( \frac{p^2 \, M_0}{p^2-M_0^2} \right) \, \left( \frac{y_0^2 \, v^2}{4 \, M_0^2} \right) \, \left( - \sum_{k=1}^3 (U_S)_{1 k}^2 + \sum_{k=1}^3 (U_S)_{1 k}^2 \right) = \sum_{i=1}^3 \, 
\mathcal U^2_{e i} \, m_i \; ,
\end{equation}
implying that the contribution of the heavy sterile states to $m_{ee}$ is very suppressed due to their pseudo-Dirac nature. Consequently, we expect that the results for $m_{ee}$
be very similar to those obtained in the model-independent scenario, as studied for example in~\cite{Hagedorn:2016lva}.
 \begin{figure}[t!]
\begin{center}
\parbox{3in}{\hspace{-0.3in}\includegraphics*[scale=0.55]{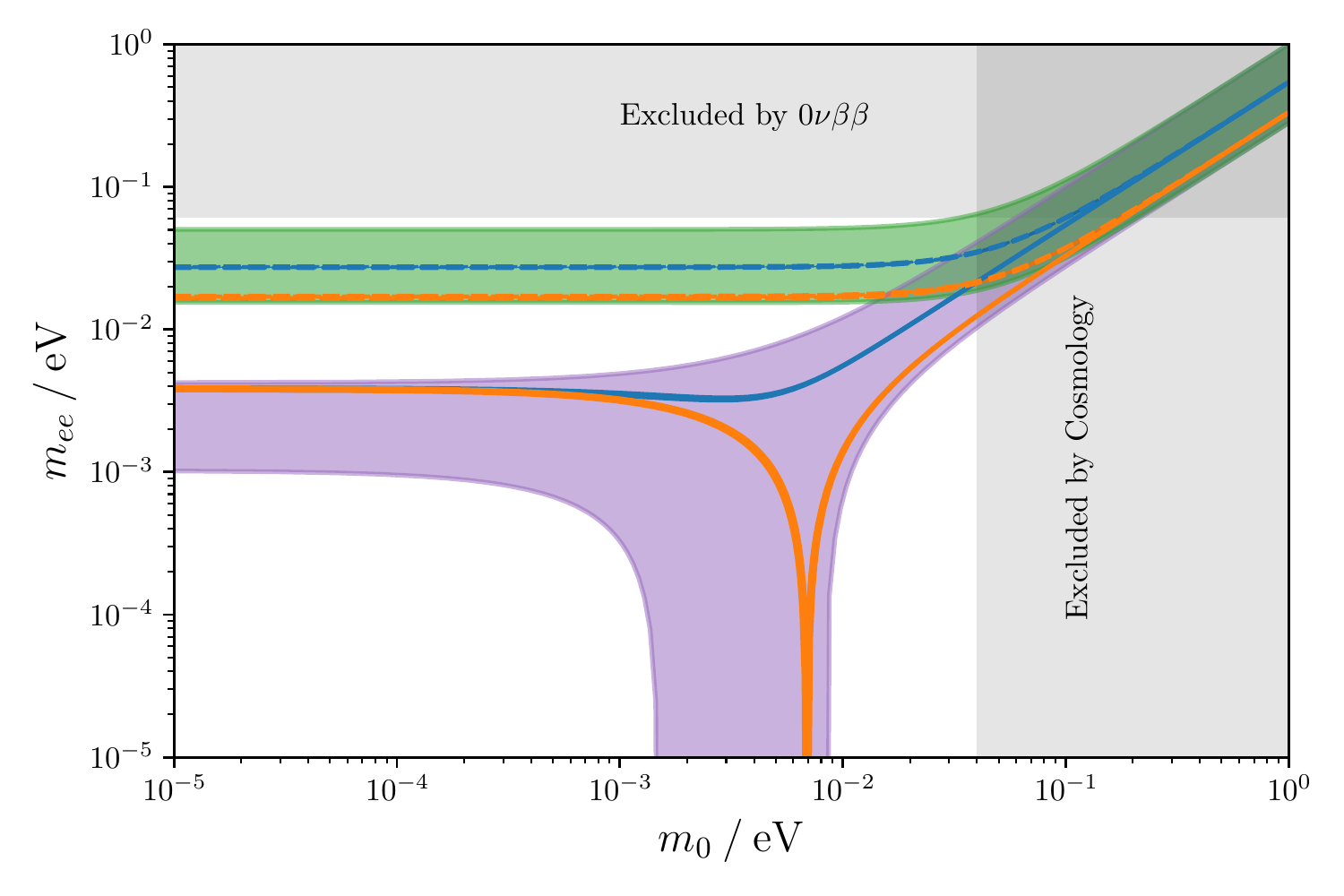}}
\hspace{-0.1in}
\parbox{3in}{\includegraphics*[scale=0.55]{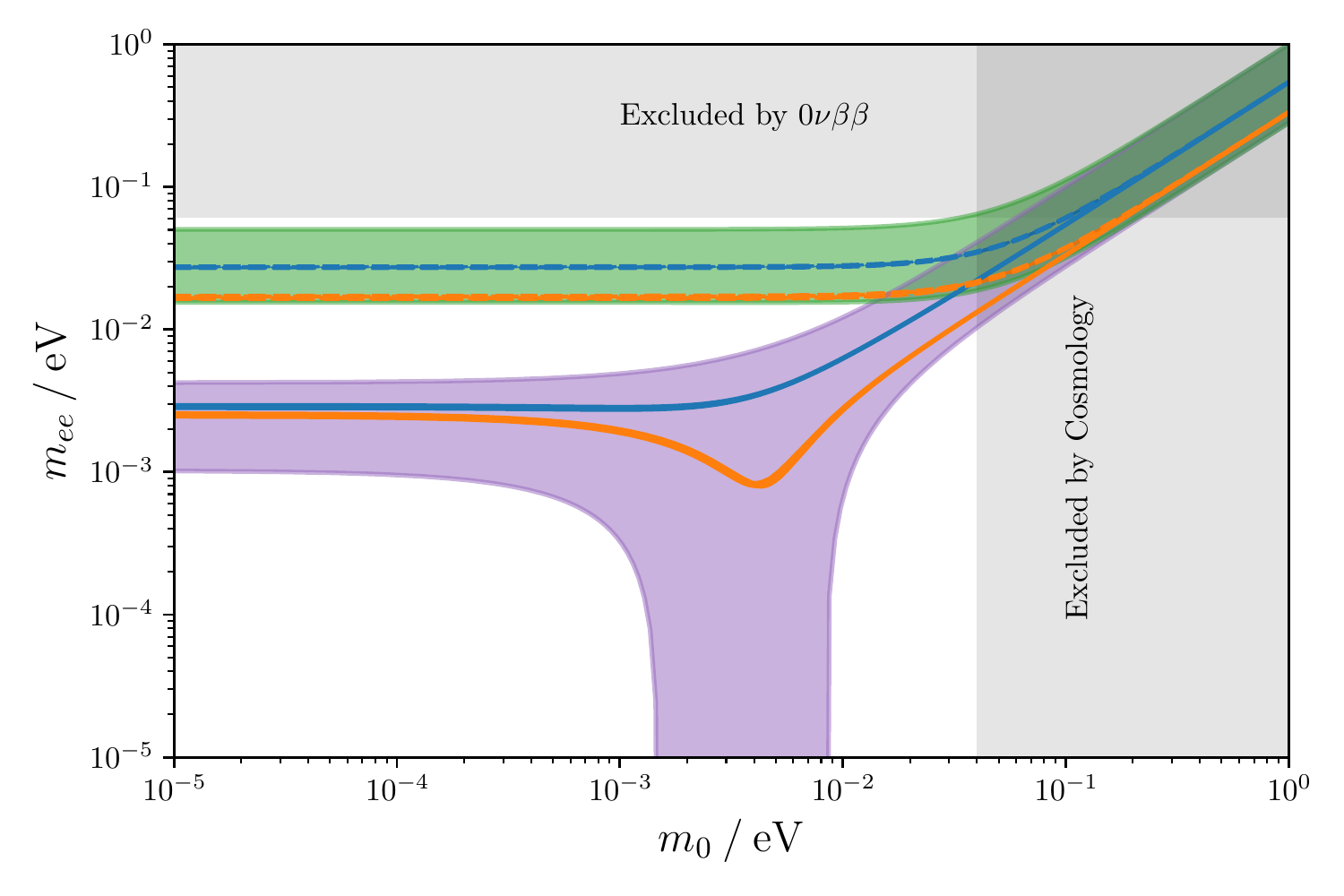}}
\end{center}
\caption{{\small \textbf{Results for \mathversion{bold}{$0\nu\beta\beta$} decay for Case 3 b.1)}. Effective mass $m_{ee}$ as a function of the lightest neutrino mass $m_0$ (both in eV)
obtained for option 1 of the $(3,3)$ ISS framework. 
Leading to the results in the left plot, we take $n=20$, $m=10$ and $s=9$ (blue curves) and $s=10$ (orange curves), while the right plot is for $n=20$, $m=11$ and $s=9$ (blue curves) and $s=10$ (orange curves). In both plots  we fix $M_0=1000$~GeV and $y_0=0.1$.
Solid (dashed) curves correspond to a NO (IO)  light neutrino mass spectrum.
The purple (green) shaded area arises upon variation of the lepton mixing parameters and mass squared differences within the experimentally preferred
$3 \, \sigma$ ranges for NO (IO)~\cite{Esteban:2020cvm}. 
Values of $m_0$ disfavoured by the cosmological bound on the sum of the light neutrino masses~\cite{Planck:2018vyg} are identified by a vertical grey band, while regimes of
$m_{ee}$ already disfavoured by searches
for $0\nu\beta\beta$ decay~\cite{KamLAND-Zen:2016pfg} are indicated by a horizontal grey band.
Figures from~\cite{Hagedorn:2021ldq}.
\label{fig:Case3b1_0nubb}}}
\end{figure}

For completeness, we show two plots for $m_{ee}$ in Fig.~\ref{fig:Case3b1_0nubb}, where we have set $y_0=0.1$ and $M_0=1000$~GeV.
These plots were obtained for Case 3 b.1), and  we have chosen $n=20$.
In the left plot of Fig.~\ref{fig:Case3b1_0nubb} we fix $m=10$, while in the right one $m=11$.
In both plots we display results for two values of $s$, $s=9$ (blue) and $s=10$ (orange). 
Solid (dashed) curves correspond to a NO (IO) light neutrino mass spectrum.
We remind that for $s=10$ both Majorana phases turn out to be trivial, thus allowing for the strong 
cancellation observed in association with the orange solid curve in the left plot. The thickness of the curves is determined by the variation
of the mass squared differences in their experimentally preferred $3 \, \sigma$ ranges~\cite{Esteban:2020cvm}, Table~\ref{tab:nufit} in Section~\ref{sec:osci}.
The purple (green) shaded area arises upon variation of the lepton mixing parameters  and of the mass squared differences within the experimentally preferred
$3 \, \sigma$ ranges for NO (IO). 
The upper bound on the lightest neutrino mass 
$m_0$ arises from the cosmological bound on the sum of the light neutrino masses~\cite{Planck:2018vyg},
see Eq.~\eqref{eq:sumnuexp} in Section~\ref{sec:osci}. 
In Fig.~\ref{fig:Case3b1_0nubb} we have depicted the experimental limit on $m_{ee}$ obtained by the KamLAND-Zen Collaboration (using the isotope $^{136} \mathrm{Xe}$)~\cite{KamLAND-Zen:2016pfg}, 
\begin{equation}
\label{eq:meeKamLAND-Zen:2016pfg}
m_{ee} < (61 \div 165) \, \mathrm{meV} \; ,
\end{equation}
with the above range resulting from different 
theoretical estimates of the nuclear matrix elements. 
With further improvement of the experimental limits, certain combinations of group theory parameters in the different cases could be disfavoured, at least, if the light neutrino mass spectrum is assumed to follow IO.

\section{Impact for charged lepton flavour violation}\label{sec:clfv} 
We now proceed to discuss the impact of endowing the (3,3) ISS
realisation with flavour and CP symmetries concerning cLFV
observables, such as radiative and three-body lepton decays,
and neutrinoless $\mu-e$ conversion in matter. 
  
Before addressing the cLFV rates,
it is important to recall that in a regime of sufficiently small
$\mu_i$, the heavy Majorana states are approximately mass-degenerate
in pairs, and have opposite CP-parity, thus effectively leading to the
formation of pseudo-Dirac pairs, whose phases are closely related by (see Eq.~\eqref{eq:matSopt1})
\begin{eqnarray}
    \mathcal{U}_{\alpha j } &=& \mathcal{U}_{\alpha j+3}\, e^{i
      (\varphi_{\alpha j} - \varphi_{\alpha j+3})}\:, \quad \text{with} \quad  
    \varphi_{\alpha j} - \varphi_{\alpha j+3} = -\pi/2\,, \label{eq:cLFV:PD:phase}
\end{eqnarray}
in which $\mathcal{U}_{\alpha j }$ are elements of the
unitary nine-by-nine matrix (cf. Eq.~\eqref{eq:formBigU}), with $j = 4, 5, 6$ and $\alpha = e, \mu, \tau$. 
For option 1, not only is the mass splitting  extremely
small, typically $\mathcal{O}(1-100\:\mathrm{eV})$ but, as can be
seen from Eq.~(\ref{eq:heavyMopt1}), the pseudo-Dirac pairs are
themselves degenerate in mass up to a very good approximation.
In view of the above, the loop functions entering the distinct
observables (see Appendix~\ref{app:loopfunctions_neutrinos}) can be taken
universal for the heavy states,  
$f(x_i) = f(x_0), \, \forall i = 4,5, ..., 9$ with $x_0 \simeq M_0^2/M_W^2$, where $M_W$ is the mass of the $W$-boson.

The full expressions for the cLFV rates arising in SM extensions via
$n_s$ heavy sterile states for the radiative and three-body decays are given in Section~\ref{sec:cLFVobs}.

For simplicity, here we only
  focus on identical flavour final states for the three-body cLFV decays,
  although one expects similar results for $\ell_\beta \to \ell_\alpha \ell_\gamma
  \ell_\gamma$ decays.
  
It is worth noticing that the combination $\sum_{i=4}^9
\mathcal{U}_{\alpha i}^{\phantom{\ast}} \,\mathcal{U}_{\beta i}^\ast$, present in several of the cLFV form factors,  can be recast in
terms of the unitarity violation of the PMNS mixing matrix, $\tilde
U_\nu$. As usually done~\cite{Xing:2007zj},  
one can write  
\begin{equation}\label{eq:cLFV:UtildeAU}
\tilde{U}_\nu \, = \, A\,U_0\,,
\end{equation}
in which $U_0$ is a unitary three-by-three matrix
and $A$ is a triangular matrix, 
\begin{equation}
A\, =\, 
\begin{pmatrix}
\alpha_{11} & 0& 0\\
\alpha_{21} & \alpha_{22} & 0\\
\alpha_{31} & \alpha_{32} & \alpha_{33}
\end{pmatrix}\,,
\end{equation}
and we define
\begin{equation}
    \mathcal A \equiv \, A \,A^\dagger = \tilde{U}_\nu^{\phantom{\dagger}} \, \tilde{U}_\nu^\dagger\,.
\end{equation}
Recalling the definition of the quantity $\eta$ (see
Eqs~(\ref{eq:UPMNStildeeta2})), it is manifest
that one has  
\begin{equation}
\mathcal{A}_{\alpha \beta} \, =\, \delta_{\alpha \beta} - 2 \eta_{\alpha \beta}\,.
\end{equation}
Unitarity of the full nine-by-nine matrix $\mathcal{U}$ implies that 
\begin{equation}
\sum_{i=4}^9 \mathcal{U}_{\alpha i}^{\phantom{\ast}}\,\mathcal{U}_{\beta i}^\ast
=\delta_{\alpha\beta} - \sum_{i=1}^3 \mathcal{U}_{\alpha
  i}^{\phantom{\ast}}\,\mathcal{U}_{\beta i}^\ast =\delta_{\alpha\beta} - (\tilde{U}_\nu^{\phantom{\dagger}}\, \tilde{U}_\nu^\dagger)_{\alpha \beta} =\delta_{\alpha \beta} -
\mathcal{A}_{\alpha \beta} = 2\, \eta_{\alpha \beta}\,.
\end{equation}
For option 1 one has $\eta_{\alpha \beta} =0$, $\forall
\alpha\neq \beta$, so that  
$\eta$ and thus $\mathcal{A}$ (and also $A$) are diagonal. This is of
paramount importance for the cLFV observables, since - and as
discussed below - any contribution proportional to $\eta_{\alpha
  \beta}$ will vanish (for $\alpha\neq \beta$).

\mathversion{bold}
\subsection{Dipole terms - radiative decays $\mathbf{\ell_\beta \to
    \ell_\alpha \gamma}$} 
\mathversion{normal}
Since the contribution of the light (mostly active) neutrinos to the
dipole form factor can be neglected 
(the relevant limits of the loop functions can be found in
Appendix~\ref{app:loopfunctions_neutrinos}) one has
\begin{equation}
G_\gamma^{\beta \alpha} \, \simeq\, G_\gamma (x_0) 
\sum_{i=4}^9 \mathcal{U}_{\alpha i}\,\mathcal{U}_{\beta i}^\ast\,,
\end{equation}
or, and in view of the above discussion, 
\begin{equation}
G_\gamma^{\beta \alpha} \, \simeq\, 
G_\gamma (x_0)\, \left(\delta_{\alpha \beta } -
\mathcal{A}_{\alpha \beta} \right) \,=\, 2\, G_\gamma (x_0)\,  
\eta_{\alpha \beta}\,.
\end{equation}
As an illustrative example, for radiative cLFV muon decays one has 
$G_\gamma^{\mu e} = -G_\gamma (x_0)\, \alpha_{11}\,\alpha_{21}^\ast$.
For the present scenario, in which $\mathcal{A}$ and $\eta$
are diagonal, one thus finds $G_\gamma^{\beta \alpha} \, \simeq\,0\,$.

In line with the analytical discussion on lepton mixing carried in
Section~\ref{sec22}, let us also emphasise that similar results can be
obtained relying on the approximate analytical expression for $S$, which for
option 1 is given at leading order in $\mu_S/M_{NS}$ 
by
$    S \,\simeq \,\frac{y_0\, v}{2\,M_0}\,(- i U_S,\, U_S)$ (see Eq.~(\ref{eq:matSopt1})).
The form factor can be recast as 
\begin{align}
    G_\gamma^{\mu e} &\simeq G_\gamma(x_0)\left\{\sum_{i = 4}^6
    \mathcal{U}_{ei} \,\mathcal{U}_{\mu i}^\ast +  \sum_{i = 7}^9
    \mathcal{U}_{ei}\, \mathcal{U}_{\mu i}^\ast\right\} \nonumber\\ 
    &\propto G_\gamma(x_0)\left\{\sum_{i = 1}^3 (i U_S)_{ei}
    \,(iU_S)_{\mu i}^\ast +  \sum_{i = 1}^3 (U_S)_{ei}\, (U_S)_{\mu
      i}^\ast\right\} \,= \,2 G_\gamma(x_0) \left\{\sum_{i = 1}^3
    (U_S)_{ei} \,(U_S)_{\mu i}^\ast \right\}  = 0\:,
\end{align}
due to the orthogonality of the ($\mu - e$) rows of $U_S$ (which we
recall to be a unitary matrix, determined by the group theory parameters and
the free angle $\theta_S$). 

\mathversion{bold}
\subsection{Photon and $Z$ penguin form factors} 
\mathversion{normal}

Relevant for both ${\ell_\beta \to 3 \ell_\alpha}$
and $\mu-e$ conversion, these
include several contributions (reflecting the fact that two neutral
fermions can propagate in the loop).

A reasoning analogous to the one conducted for 
$G_\gamma^{\beta \alpha}$ leads to 
$F_\gamma^{\beta \alpha} \, \simeq\, F_\gamma (x_0)\, 
\left( \delta_{\alpha \beta} -\mathcal{A}_{\alpha \beta} \right)\,=\,
2\,F_\gamma (x_0)\, \eta_{\alpha\beta}$, which thus vanishes for the
flavour violating decays.
Likewise, the first term on the right-hand side of Eq.~(\ref{eq:cLFV:FF:FZ})
leads to the same result, 
\begin{equation}
\sum_{i,j =1}^{9}
\mathcal{U}_{\alpha i}\,\mathcal{U}_{\beta j}^\ast \left[\delta_{ij}
  F_Z(x_i) \right] \simeq 
F_Z(x_0)\left( \delta_{\alpha \beta} -\mathcal{A}_{\alpha \beta}
\right)\,=\, 2F_Z(x_0)\, \eta_{\alpha\beta}\,=\, 0\, .  
\end{equation}

Both terms associated with $G_Z(x,y)$ and $H_Z(x,y)$ correspond
to two neutral leptons
propagating in the loop. Although the loop functions do tend to zero
for the case of very light internal fermions, the same does not occur for
$G_Z$ if at least one of the states is heavy, i.e. $G_Z (0,x_i)$ (see
Appendix~\ref{app:loopfunctions_neutrinos}).
Introducing the following limits for the loop functions,
\begin{equation}
{\bar G}_Z(x) \, =\, \lim_{x_i \gg 1} G_Z (0,x_i)\, ,
\quad \quad 
{\bar{\bar G}}_Z(x) \, =\, \lim_{x_i \approx x_j \gg 1} G_Z (x_i,x_j)
\,,
\end{equation}
respectively corresponding to ``heavy-light" and
``heavy-heavy" (combinations of) fermion propagators, one thus has
\begin{equation}
\sum_{i,j = 1}^{9}
\mathcal{U}_{\alpha i}\,\mathcal{U}_{\beta j}^\ast 
\,C_{ij}\, G_Z(x_i, x_j)\simeq
{\bar G_Z(x_0)} \left[ 2 \mathcal{A}
  \,(\mathbb{1}-\mathcal{A})\right]_{\alpha \beta} 
+
{\bar{\bar G}}_Z(x_0) \left[ (\mathbb{1}-\mathcal{A})^2\right]_{\alpha \beta},
\end{equation}
or in terms of $\eta$,
${\bar G_Z(x_0)} \left[ 4(\mathbb{1}-2\eta)\,\eta\right]_{\alpha
      \beta} + 4 \,{\bar{\bar G}}_Z(x_0) \,
[\eta^2]_{\alpha\beta}\,=\,0$,
as previously argued.

For the $H_Z(x,y)$-associated terms, only the ``heavy-heavy" case (two
heavy sterile states in 
the loop) can potentially contribute in a non-negligible way.
However, the corresponding contribution also vanishes, 
as a consequence of the nature of the (degenerate) heavy states, which
as mentioned
form pseudo-Dirac pairs. 
Defining (see Appendix~\ref{app:loopfunctions_neutrinos})   
\begin{equation}
{\bar{\bar H}}_Z(x) \, =\, \lim_{x_i \approx x_j \gg 1} H_Z (x_i,x_j)\,,
\end{equation}
one then finds (taking into account Eq.~(\ref{eq:cLFV:PD:phase}))
\begin{align}
&\sum_{i,j =1}^{9}
\mathcal{U}_{\alpha i}\,\mathcal{U}_{\beta j}^\ast 
\,C_{ij}^\ast\, H_Z(x_i, x_j) \simeq
{\bar{\bar H}}_Z(x_0) \sum_{i,j =4}^{9} 
\sum_{\rho =1}^{3} 
\mathcal{U}_{\alpha i}\,\mathcal{U}_{\rho i}\,
\mathcal{U}_{\beta j}^\ast\,\mathcal{U}_{\rho j}^\ast  \nonumber \\
& \simeq
{\bar{\bar H}}_Z(x_0) \sum_{\rho =1}^{3}
\left\{\left[
\sum_{i =4}^{6}  \left(\mathcal{U}_{\alpha i}\,
\mathcal{U}_{\rho i} +
e^{i \pi/2} \mathcal{U}_{\alpha i}\,
e^{i \pi/2} \mathcal{U}_{\rho i}
\right)\right] \Bigg[ 
\sum_{j =4}^{6} \left(\mathcal{U}_{\beta j}^\ast\,
\mathcal{U}_{\rho j}^\ast +
e^{-i \pi/2}\mathcal{U}_{\beta j}^\ast\,
e^{- i \pi/2}\mathcal{U}_{\rho j}^\ast
\right)
\Bigg]\right\}\,=\, 0\,,
\end{align}
which is a direct consequence of the pseudo-Dirac nature of the heavy states.

\subsection{Box diagrams}
Several form factors contribute to both the three-body decays
$\ell_\beta \to 3 \ell_\alpha$, and neutrinoless $\mu-e$
conversion. The first ($F^{\beta3\alpha}_\text{box}$) can be decomposed in two terms,
``box" and cross-box ``Xbox", respectively associated with the loop
functions $G_\text{box}$ and $F_\text{Xbox}$. Similar contributions
(single internal neutral lepton) are present for the latter ($F^{\mu e
qq}_\text{box}$).

Only diagrams with two heavy neutrinos are at the source of 
non-vanishing contributions to $G_\text{box}$; however, and
analogously to what occurred for the previously discussed
$H_Z(x,y)$-associated terms, the contributions vanish, 
due to having the heavy states forming, to an excellent
approximation, pseudo-Dirac pairs. 

A priori, one can have contributions to  the $F_\text{Xbox}$ form factors
from ``light-light" and ``heavy-heavy" fermion propagators in
the box. However, both turn out to be proportional to $\mathcal{A}_{\alpha \beta}$ and thus to $\eta_{\alpha \beta}$, and
are hence vanishing.

The additional form factors relevant for 
${\mu-e}$ conversion, $F^{\mu eqq}_\text{box}$ lead to contributions again
proportional to $\eta_{e \mu}$, thus also vanishing in the present scenario.

\mathversion{bold}
\subsection{cLFV for option 1 of the $(3,3)$ ISS with flavour and CP symmetry}
\mathversion{normal}
In the present scenario, no new
contributions to the different cLFV observables due to the exchange of
heavy states are expected.\footnote{Numerical evaluations confirm that
  the rates are typically $\mathcal{O}(10^{-50})$.} Such a
``stealth" realisation of the ISS - which in general can account for
significant contributions to the observables~\cite{Abada:2014vea,Abada:2014zra,Abada:2015rta,Abada:2017ieq}, well within    
experimental sensitivity - is due to two peculiar features of 
option 1. First and most importantly, recall that here $\mu_S$ is the unique source of
flavour violation in the sector of neutral states; this is in contrast with other
ISS realisations in which the Dirac neutrino Yukawa couplings (and possibly
$M_{NS}$) are non-trivial in flavour space.  
Moreover, notice that for option 1 of the flavour symmetry-endowed ISS the heavy mass spectrum is composed of three degenerate
pseudo-Dirac pairs (to an excellent approximation), which further
suppresses any new contribution. 

Thus, cLFV processes will not offer any additional source of insight
in what concerns the underlying discrete flavour symmetries nor the
mass scale of the heavy states; however 
the observation of at least one cLFV transition would
strongly disfavour the 
flavour symmetry-endowed ISS in its option 1, with strictly
diagonal and universal $M_{NS}$ and $m_D$ in flavour space.

\begin{figure}
    \centering
    \mbox{\includegraphics[width=0.48\textwidth]{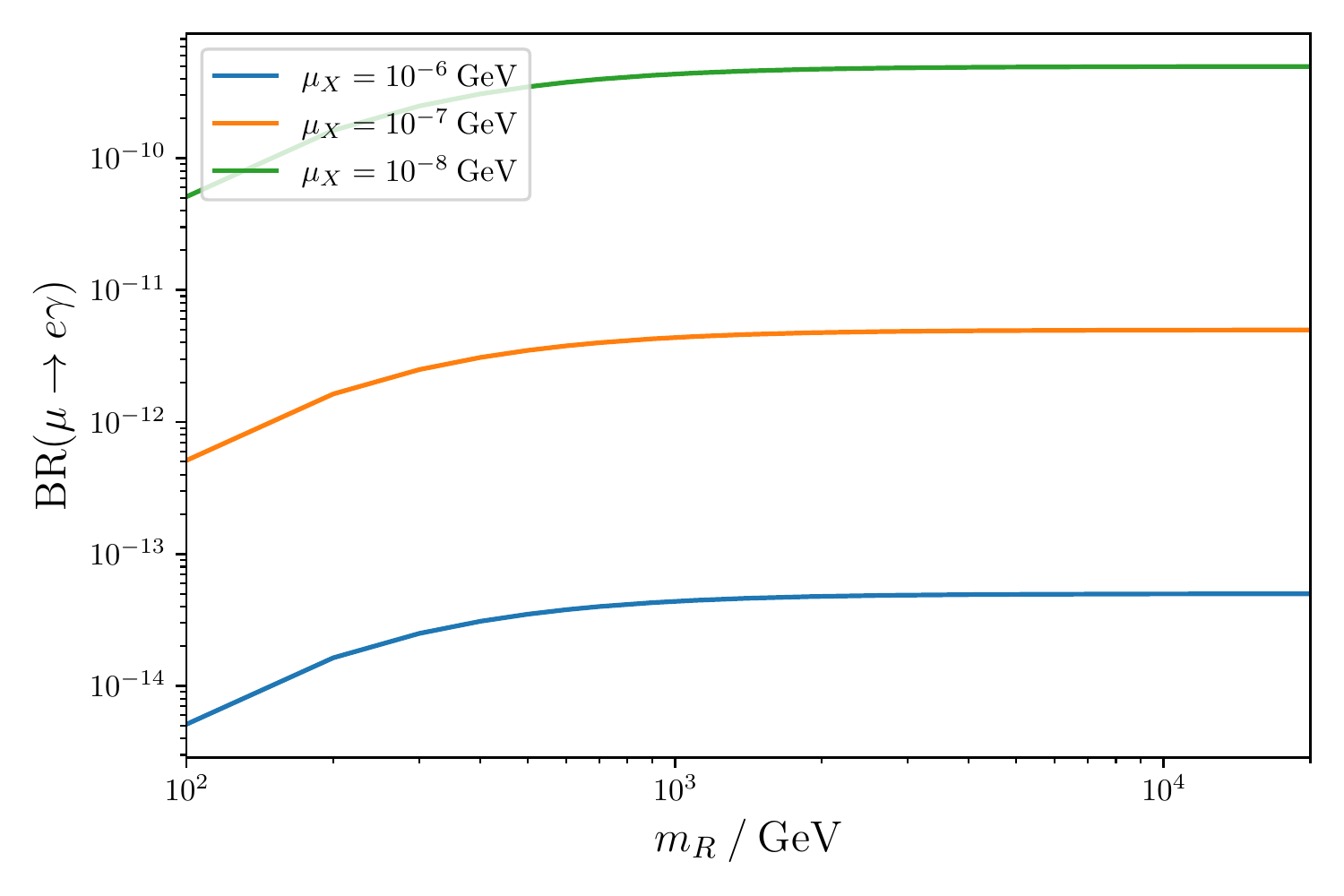}\includegraphics[width=0.48\textwidth]{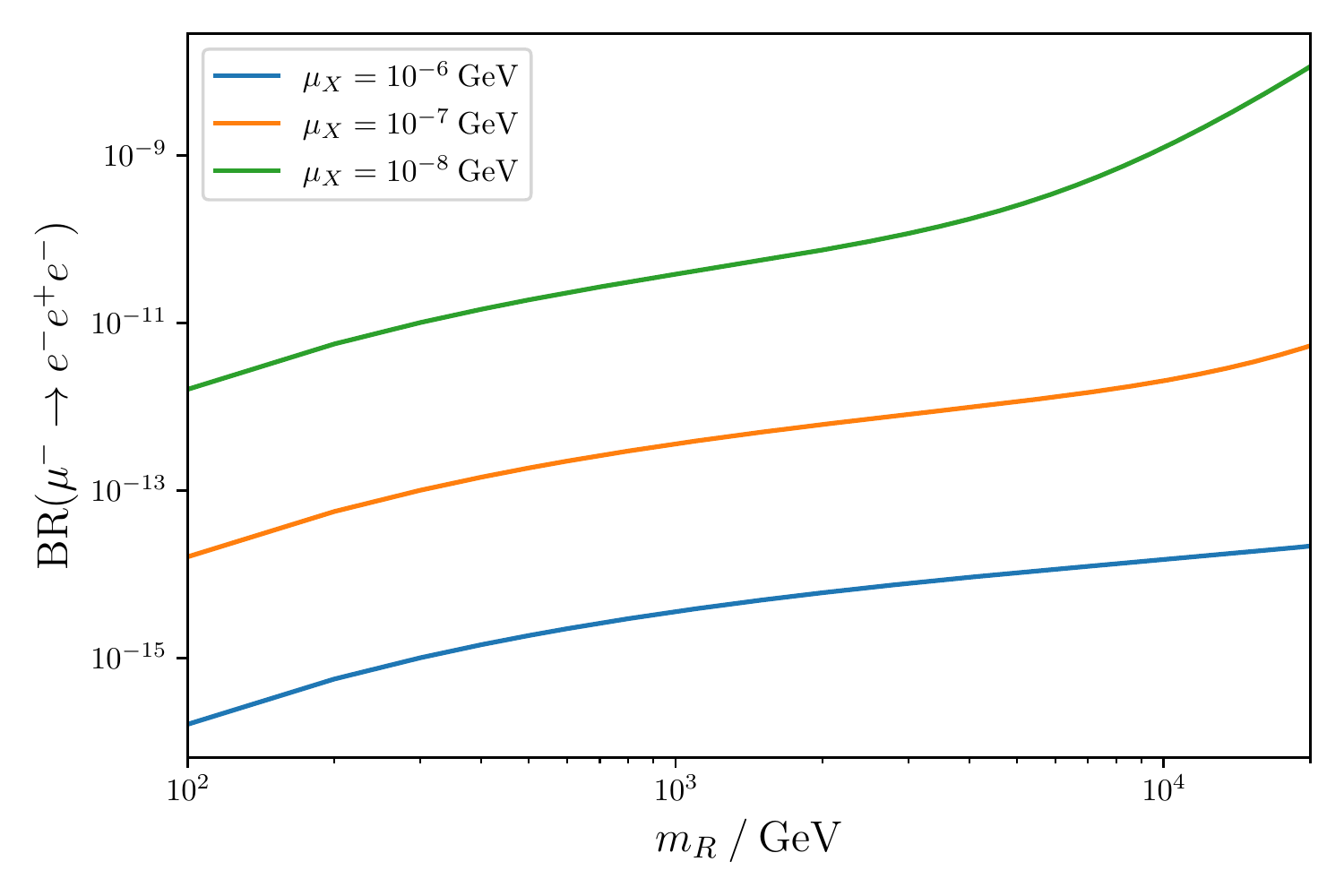}}
    \caption{Predictions of a ``normal'' ISS framework in which the flavour structure has been encoded in $m_D$, making use of the Casas-Ibarrra parametrisation in Eq.~\eqref{eqn:ISSCI}. The mass terms $M_{NS} = m_R\:\mathbb{1}$ and $\mu_S = \mu_X\:\mathbb{1}$ are chosen to be diagonal and universal.}
    \label{fig:ISSnormal}
\end{figure}

For illustrative purposes, let us however consider in comparison a ``standard'' $(3,3)$ ISS (without a symmetry at work) with the flavour structure encoded in $m_D$.
Following Eq.~\eqref{eqn:ISSCI}, we use a Casas-Ibarra parametrisation in order to accommodate oscillation data, and we take $R = \mathbb{1}$, $M_{NS} = m_R \:\mathbb{1}$ and $\mu_S = \mu_X\:\mathbb{1}$ (in the present convention).
A non-trivial structure in $m_D$ then leads to a very rich flavour phenomenology, as can be seen in Fig.~\ref{fig:ISSnormal}, in which we show the predicted rates for $\mu\to e\gamma$ and $\mu\to eee$ vs. the heavy mass $m_R$, for three different choices of $\mu_X$ (see legend).

\section{Further discussion}
\label{summ}
We have considered an inverse seesaw mechanism with $3+3$ heavy
sterile states, endowed with a flavour symmetry $G_f=\Delta (3 \,
n^2)$ or $G_f=\Delta (6 \, n^2)$ 
and a CP symmetry. The
peculiar breaking of the flavour and CP symmetry to different residual
symmetries $G_\ell$ in the charged lepton 
sector and $G_\nu$ in the sector of the neutral states, is 
the key to rendering this scenario predictive (and possibly testable). 
In the inverse seesaw mechanism, several terms in the Lagrangian 
determine the mass spectrum of the neutral states, in association with  
three matrices, $m_D$, $M_{NS}$ and $\mu_S$. 
Several realisations of the residual symmetry $G_\nu$ are possible, and here we have focused on one of the three minimal options, which we have
called ``option 1''. 
In this option only the Majorana mass matrix $\mu_S$ breaks $G_f$ and
CP to $G_\nu$, while $m_D$ and $M_{NS}$ preserve $G_f$ and CP.
In the sector of the neutral states, lepton number and lepton flavour violation are thus
both encoded in $\mu_S$.  
Left-handed lepton doublets and the $3+3$ heavy sterile states are assigned to the same triplet 
$\mathbf{3}$ of $G_f$, whereas right-handed
charged leptons are in singlets. 

In~\cite{Hagedorn:2014wha} mixing patterns arising from the breaking of $G_f$ and CP to $G_\ell$
and $G_\nu$ have been analysed in a model-independent framework, and four of them have been identified as particularly interesting for leptons.
In~\cite{Hagedorn:2021ldq}, we have studied examples of lepton mixing for each of the different mixing patterns, Case 1) through Case 3 b.1), both analytically and numerically, within a class of possible realisations of the ISS, here called option 1.
For option 1, a significant consequence of the presence of the heavy
sterile states is that for certain regimes there is a sizeable
deviation from unitarity of the PMNS mixing matrix, and thus potential
conflict with the associated experimental bounds. 
This leads to stringent constraints on the Yukawa coupling $y_0$ and on the mass scale $M_0$, 
so that regimes of large $y_0$ and small $M_0$ are disfavoured.
In the viable regimes, the impact of the heavy sterile states on lepton mixing turns out to be
 small:
deviations typically below $1\%$ are found upon comparison of the results of the $(3,3)$ ISS framework to those derived in the model-independent scenario.
We have also discussed the potential impact of this ISS framework for
several observables. An interesting implication of option 1 here
discussed is that the heavy sterile states are degenerate to a very
good approximation, and combine to form three pseudo-Dirac pairs. 
As a consequence, the results for neutrinoless double
beta decay are hardly modified, compared to results obtained in the model-independent scenario.
We have also addressed in detail charged lepton flavour violating
processes: in sharp contrast to what generally occurs for 
inverse seesaw models (see, e.g.~\cite{Abada:2015oba,Abada:2014kba}), 
the cLFV rates are highly
suppressed, similar to what occurs in the
Standard Model with three light (Dirac) neutrinos. 
This is a consequence of having strictly flavour-diagonal and flavour-universal deviations from unitarity of the PMNS mixing matrix (and also due to a very high degree of 
degeneracy in the heavy mass spectrum).

Throughout this discussion we have assumed that the desired breaking of the flavour and CP symmetries can be realised,  and that the appropriate residual symmetries are preserved by the different mass matrices.
As has been shown in the literature,
it is possible to achieve the breaking of flavour (and CP) 
in different ways, e.g.~spontaneously, if flavour (and CP) symmetry breaking fields acquire non-vanishing vacuum expectation values, in supersymmetric theories (see for instance~\cite{Hagedorn:2018bzo}), or explicitly via boundary conditions
in a model with an extra dimension (see e.g.~\cite{Hagedorn:2011un,Hagedorn:2011pw}). The predictive power of concrete models 
is usually higher than the one of the model-independent approach: for example, by choosing a certain set of flavour (and CP) symmetry breaking fields, the ordering of the light neutrino mass  spectrum
can be predicted, and by extending the flavour (and CP) symmetry to the flavour sector of the new particles, as for instance supersymmetric particles or Kaluza-Klein states, many flavour observables can be constrained and correlated. 
It could thus be interesting to consider the construction of 
such models. 

It is well-known that in concrete models
corrections to the desired breaking of flavour (and CP) can arise. This can for instance be the case if flavour (and CP) symmetry breaking fields, whose vacuum expectation values preserve the residual symmetry $G_\ell$, couple 
at a higher order to the neutral states as well. We have not discussed such corrections in our analysis, but we can briefly comment on their expected impact on lepton mixing as well as
predictions for branching ratios of different charged lepton flavour violating processes. Considering, for example, that corrections invariant under $G_\ell$ contribute to the mass matrices $m_D$ and $M_{NS}$,
we expect that lepton mixing can still be correctly explained for corrections not larger than a few percent\footnote{See also~\cite{Hagedorn:2011un, Hagedorn:2011pw} for a similar analysis in the context of a type-I seesaw 
mechanism, implemented in a model with a warped extra dimension and a flavour symmetry $G_f$.} and possibly by re-fitting the value of the free angle $\theta_S$. 
At the same time, the branching ratios of charged lepton flavour violating processes would still remain strongly suppressed, beyond the reach of current and future experiments.\footnote{Notice that corrections that are invariant under $G_\ell$ only contribute to the diagonal entries of $m_D$ and $M_{NS}$, so that even in the presence of the latter the matrix $\eta$ will still be diagonal (see Eq.~\eqref{eq:eta}). Moreover, 
in the considered mass regime, the dominant loop functions have an asymptotic logarithmic behaviour (or are even constant, cf. Appendix~\ref{app:loopfunctions_neutrinos}), thus being insensitive to percent 
level changes in the mass splitting of the heavy states; 
this thus still leads to a strong GIM cancellation in the cLFV rates.}

As mentioned, here we have focused on one of the three minimal options to
realise the residual symmetry $G_\nu$ in the sector of the neutral
states. It could be interesting to analyse 
 lepton mixing, as well as neutrinoless double beta decay, effects of
 non-unitarity of the PMNS mixing matrix $\tilde{U}_\nu$, and charged lepton flavour violating processes for the
 other two options, called option 2 and option 3. 
Both these options
could potentially lead to larger effects in charged lepton flavour violating
processes. For option 2 the non-trivial flavour structure is encoded in the
Dirac neutrino mass matrix $m_D$ and thus strongly resembles ISS
constructions typically associated with sizeable predictions to
numerous leptonic observables.  
Furthermore, a non-trivial flavour structure in $M_{NS}$ for option 3 also leads to off-diagonal terms in $\eta$, thus potentially having a strong impact on cLFV processes.
Should this be the case, a study of possible correlations among the lepton mixing parameters and the different charged lepton flavour violating processes for the distinct cases (Case 1) through Case 3 b.1)) could be valuable and may even help testing the hypotheses of $G_f$, CP and the residual symmetries $G_\ell$ and
$G_\nu$.   
Going beyond the three minimal options, 
we can also consider options,
in which at least two of the three mass matrices $m_D$, $M_{NS}$ and
$\mu_S$ carry non-trivial flavour information.  

Further variants could be also envisaged, possibly including versions of the inverse
seesaw mechanism, for instance with two right-handed neutrinos $N_i$ and two (three)
neutral states $S_j$~\cite{Abada:2014vea}, or even a minimal radiative inverse seesaw mechanism~\cite{Dev:2012sg}.
\chapter{Anomalies in nuclear transitions: hints for light flavoured New Physics?}
\label{sec:g-2paper}
\minitoc

\begin{figure}
    \centering
    \mbox{\includegraphics[width=0.48\textwidth]{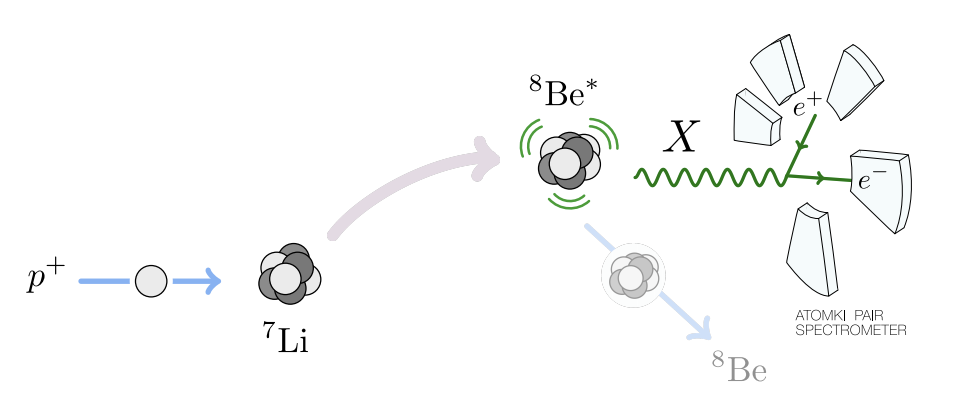}\hspace{2mm}\includegraphics[width=0.48\textwidth]{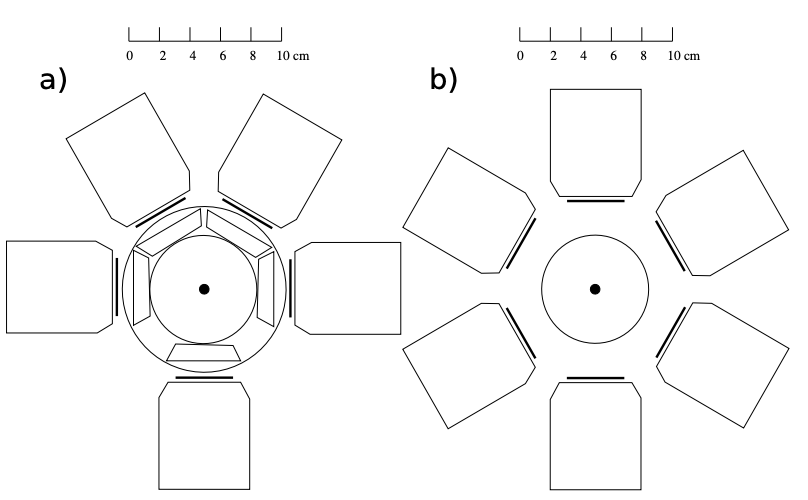}}
    \caption{Experimental setup of the Atomki experiment. \textbf{Left}: Schematics of the production and de-excitation of the $^8$Be$^\ast$ nuclei. \textbf{Right}: Original (a) and improved (b) Atomki electron pair spectrometers respectively relying  on 5 and 6 telescopes. Figures taken from~\cite{Feng:2016ysn, Gulyas:2015mia}.}
    \label{fig:atomkiexp}
\end{figure}

\noindent
A few years ago, the Atomki Collaboration 
reported their results~\cite{Krasznahorkay:2015iga} 
on the measurement of the angular
correlation of electron-positron internal pair creation (IPC) for two
nuclear transitions of Beryllium atoms
($^8$Be$^\ast \to $ $^8$Be), with a significant excess being observed  at large angles
for one of them. The experimental setup is shown schematically in Fig.~\ref{fig:atomkiexp}.

The magnetic dipole ($M1$) transitions under study
concerned the decays of the excited isotriplet and isosinglet states, 
respectively denoted $^8$Be$^{*'}$ and $^8$Be$^{*}$, into the
fundamental state ($^8$Be$^{0}$). The transitions are summarised
below, together with the associated energies:
\begin{align}
^8\text{Be}^{*'} (j^\pi = 1^+,T=1)\, 
\rightarrow \, 
^8\text{Be}^{0} (j^\pi = 0^+,T=0)\,, \quad 
E\,=\, 17.64\text{ MeV}\,; \nonumber \\
^8\text{Be}^{*} (j^\pi = 1^+,T=0)
\,\rightarrow\, 
^8\text{Be}^{0} (j^\pi = 0^+,T=0)\,, \quad 
E\,=\, 18.15\text{ MeV}\,,
\end{align}
in which $j^\pi$ and $T$ correspond
to the spin-parity and isospin of the nuclear states, respectively. A significant enhancement of the IPC was observed at large angles in the
angular correlation of the 18.15~MeV transition; it was subsequently
pointed out that such an anomalous result could be potentially interpreted
as the creation and decay of an unknown intermediate light particle with
mass $m_{X}$=16.70$\pm0.35 $(stat)$\pm 0.5 $(sys)~MeV~\cite{Krasznahorkay:2015iga}.  

Recently, a re-investigation of the $^8$Be anomaly with an
improved set-up corroborated the earlier 
results for the 18.15~MeV transition~\cite{Krasznahorkay:2017qfd,Krasznahorkay:2017bwh,Krasznahorkay:2018snd,Krasznahorkay:2019lgi};
moreover, it allowed constraining the mass of the hypothetical mediator
to $m_X = 17.01(16)$~MeV and its branching ratio 
(normalised to its $\gamma$-decay) to $\Gamma_X/\Gamma_\gamma = 6(1)\times 10^{-6}$. The $e^+e^-$ pair correlation in the 17.64~MeV
transition of $^8$Be was also revisited, and again no significant
anomalies were found~\cite{Krasznahorkay:2017gwn,Gulyas:2015mia}. A
combined interpretation of the data of $^8\text{Be}^*$ decays 
(only one set exhibiting an anomalous behaviour)
in terms of a new light particle, in association with
the possibility of mixing between the two different excited $^8$Be
isospin states ($^8\text{Be}^{*'}$ and $^8\text{Be}^{*}$)
might suggest a larger preferred mass
for the new mediator; this would lead to a large phase space suppression,
therefore potentially explaining the null results for
$^8\text{Be}^{*'}$ decay. In turn, it can further entail 
significant changes in the preferred quark (nucleon)
couplings to the new particle mediating the anomalous IPC, 
corresponding to significantly smaller normalised branching fractions 
than those of the preferred fit of~\cite{Krasznahorkay:2019lyl}. 

\begin{figure}
    \centering
    \hspace*{-5mm}\mbox{\includegraphics[width=0.35\textwidth]{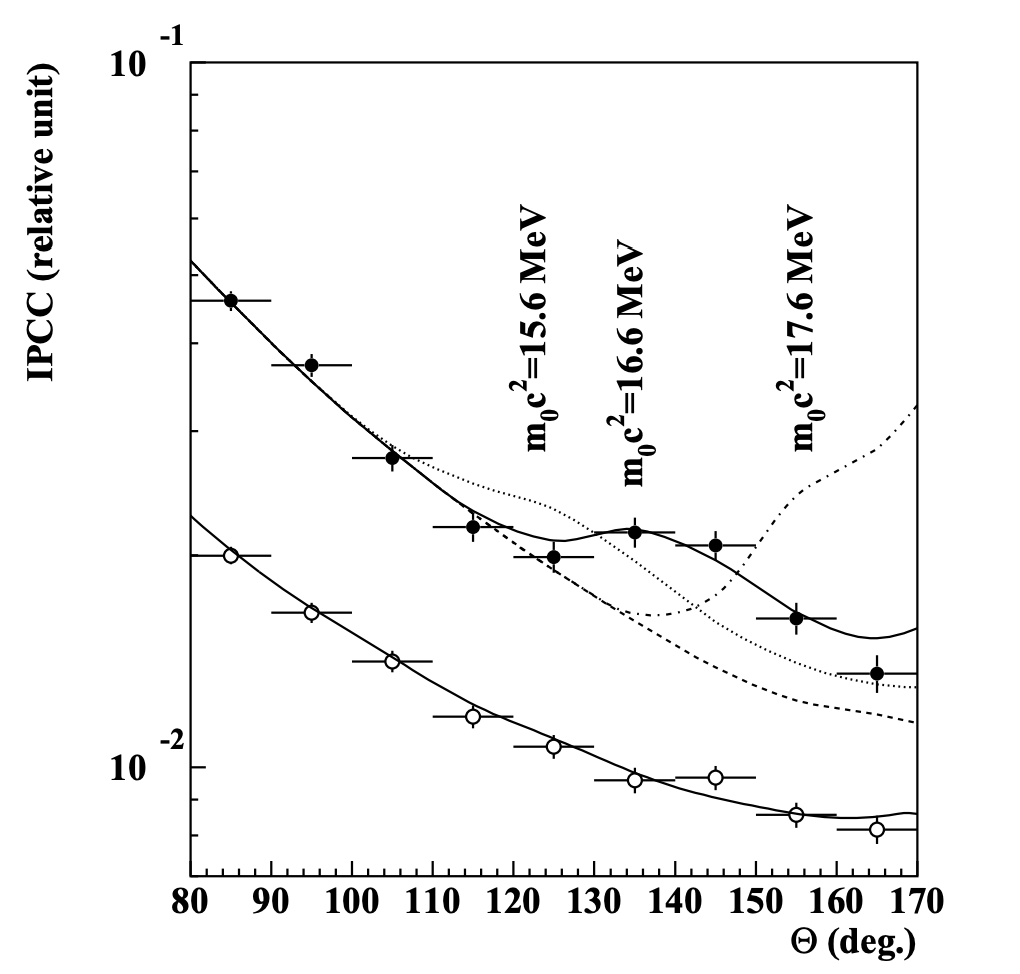} \includegraphics[width=0.34\textwidth]{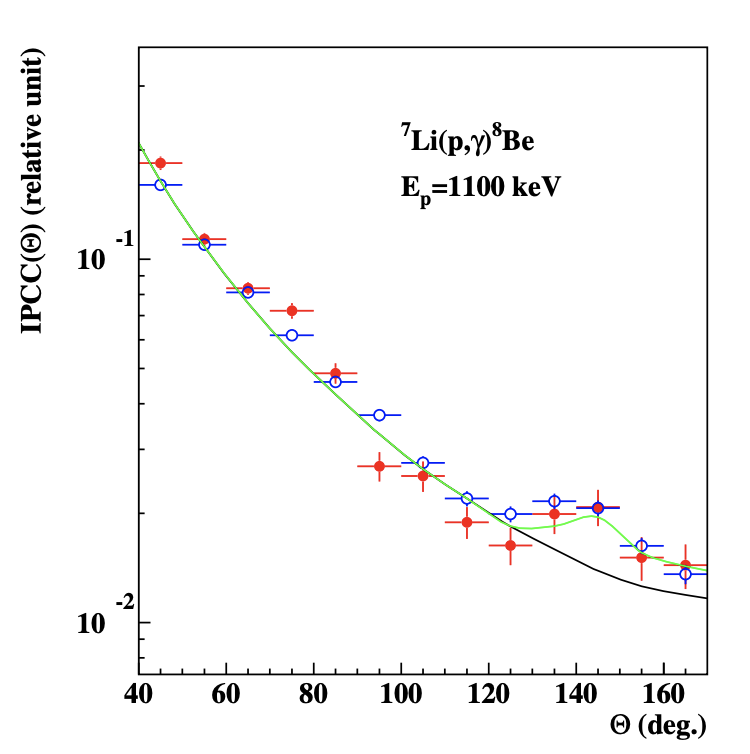}
    \includegraphics[width=0.33\textwidth]{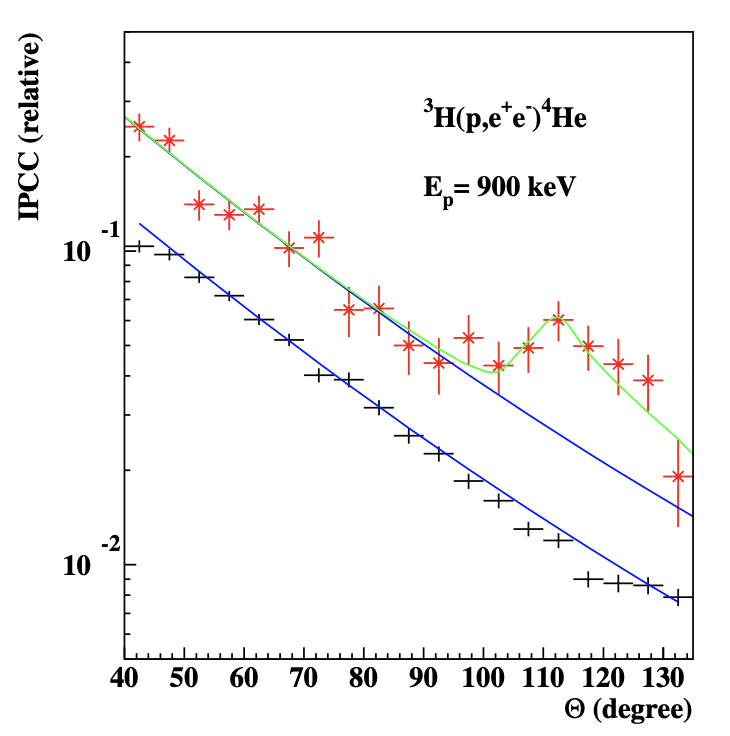}}
    \caption{Experimental data of the IPC correlation obtained by Atomki, plotted versus the opening angle $\Theta$. \textbf{Left}: Results of the 2015 run~\cite{Krasznahorkay:2015iga}. Closed circles correspond to the $18.15~\mathrm{MeV}$ transition, open circles to the $17.64~\mathrm{MeV}$ control channel.  \textbf{Middle}:  Comparison between the old (blue) and improved data (red)~\cite{Krasznahorkay:2019lgi}. \textbf{Right}: Data of the  $^4$He transitions~\cite{Krasznahorkay:2019lyl,Firak:2020eil}. Figures taken from~\cite{Krasznahorkay:2015iga,Krasznahorkay:2019lgi,Krasznahorkay:2019lyl,Firak:2020eil}.}
    \label{fig:atomkidata}
\end{figure}

Further anomalies in nuclear transitions have been observed, in
particular concerning the 21.01~MeV  $0^-\rightarrow 0^+$  transition
of $^4$He~\cite{Krasznahorkay:2019lyl,Firak:2020eil}, resulting in 
another anomalous IPC corresponding to the angular correlation of
electron-positron pairs at 115$^\circ$, with 7.2$\sigma$ significance. 
The result can also be potentially interpreted as the creation and
subsequent decay of a light particle: the corresponding mass and
width, $m_{X}$=16.98$\pm0.16\text{(stat)}\pm0.20 (\text{syst})$~MeV, 
and $\Gamma_{X}$= $3.9\times
10^{-5}$~eV, lie in a range similar to that suggested by the anomalous 
$^8\mathrm{Be}$ transition.
An overview of the $^8$Be and $^4$He data is shown in Fig.~\ref{fig:atomkidata}.

If the anomalous IPC observations are to be interpreted as being
mediated by a light new state, 
the latter can a priori be a scalar,
pseudo-scalar, vector, axial vector, or even a spin-2 particle,
provided it decays into electron-positron pairs. 
The possible existence of new interactions, in addition to those
associated with the SM gauge group, has been a
longstanding source of interest, both for particle and astroparticle physics.
Numerous experimental searches have been dedicated to look for
theoretically well-motivated light mediators, such as 
axions (spin-zero), dark photons\footnote{A dark photon is a massive vector boson with a ``dark'' gauge coupling; none of the SM fermion are charged under the associated gauge group. If there is however additional fermion content that couples (possibly at higher order) to both the dark photon and the SM photon, the additional $U(1)$ gauge group will be kinetically mixed with the SM $U(1)_Y$ group, thus leading to a small coupling of the dark photon to SM fermions at higher order. Dark photons are thus often invoked as a mediator between SM and dark matter, or can be a dark matter candidate themselves.} (spin-1) or light $Z^\prime$ (spin-1)~\cite{Ni:1999di,Heckel:2006ww,Baessler:2006vm,Hammond:2007jm,Heckel:2008hw,Vasilakis:2008yn,Serebrov:2009ej,Ignatovich:2009zz,Serebrov:2009pa,Karshenboim:2010cg,Karshenboim:2010cj,Petukhov:2010dn,Karshenboim:2011dx,Hoedl:2011zz,Raffelt:2012sp,Yan:2012wk,Tullney:2013wqa,Chu:2012cf,Bulatowicz:2013hf,Mantry:2014zsa,Hunter:2013gba,Salumbides:2013dua,Leslie:2014mua,Arvanitaki:2014dfa,Stadnik:2014xja,Afach:2014bir,Terrano:2015sna,Leefer:2016xfu,Ficek:2016qwp,Ji:2016tmv,Crescini:2017uxs,Dzuba:2017puc,Delaunay:2017dku,Stadnik:2017hpa,Rong:2017wzk,Safronova:2017xyt,Ficek:2018mck,Rong:2018yos,Dzuba:2018anu,Kim:2017yen}. 
For a parity conserving scenario,
the hypothesis of an intermediate scalar boson has already been
dismissed~\cite{Feng:2016jff}, due conservation of angular momentum in
the $1^+ \rightarrow 0^+$ $^8$Be transition. 
Having a pseudo-scalar mediator has been also severely constrained (and disfavoured) by experiments -- for an axion-like particle $a$ with a mass of $m_a\approx 17~\text{MeV}$ and an interaction term 
$g_{a} a F^{\mu\nu}\tilde{F}_{\mu\nu}$, all couplings in the range 
$ 1 /(10^{18}~\text{GeV}) < g_{a} < 1/ (10~\text{GeV})$ are
excluded~\cite{Hewett:2012ns,Dobrich:2015jyk} (although this can 
be partially circumvented in the presence of additional non-photonic
couplings~\cite{Ellwanger:2016wfe}). 
A potential first explanation of the anomalous IPC in $^8$Be 
in the context of simple $U(1)$ extensions of the
SM was discussed in~\cite{Feng:2016jff,Feng:2016ysn}, relying on the 
exchange of a 16.7~MeV, $j^\pi$ = 1$^+$
vector gauge boson. 
In~\cite{Ellwanger:2016wfe} the possibility of a light pseudo-scalar
particle with $j^\pi = 0^-$ was examined, while a potential
explanation based on an axial vector particle (including an ab-initio
computation for the relevant form factors) was carried
in~\cite{Kozaczuk:2016nma}. 
Further ideas have been put forward and discussed (see, for 
instance,~\cite{Gu:2016ege,Seto:2016pks,DelleRose:2017xil,DelleRose:2018eic,BORDES:2019wcp,Nam:2019osu,Pulice:2019xel,Wong:2020hjc,Tursunov:2020wfy,Kirpichnikov:2020tcf,Koch:2020ouk,Jentschura:2020zlr}). 

The favoured scenario of a new light vector boson is nevertheless
heavily challenged by numerous experimental constraints: 
the dark photon hypothesis is strongly disfavoured in view of negative
searches for associate production in rare light meson decays 
(e.g., $\pi^0 \rightarrow \gamma A'$ at NA48/2 which, for a dark
photon mass $\mathcal{O}(17~\text{MeV})$, constrains its
couplings to be strictly ``protophobic'', in stark contrast with the
requirements to explain the anomalous IPC in $^8$Be); the
generalisation towards a protophobic vector boson arising from 
a gauged $U(1)$ extension of the SM (potentially with
both vector and axial couplings to the SM fields) is also subject 
to stringent constraints from the measurements of atomic parity
violation in Caesium and neutrino-electron scattering (as well as
non-observation of coherent neutrino–nucleus scattering), which force 
the leptonic couplings of the gauge boson to be too small to account
for the anomalous IPC in $^8$Be. However, this problem can be
circumvented in the presence of additional vector-like leptons as
noted in~\cite{Feng:2016ysn}.

Interestingly, extensions of the SM which include light new physics states coupled to
the standard charged leptons are a priori expected to have 
implications for precision tests of leptonic observables, and even have the potential to address (solving, or at least rendering less severe)
other anomalies, as is the case of those concerning the anomalous magnetic
moment of light charged leptons, usually expressed
in terms of $a_\ell \equiv (g-2)_{\ell}/2$ ($\ell = e, \mu$), see Section~\ref{sec:g-2section}.
Several attempts have been recently conducted to
simultaneously explain the tensions in both $\Delta a_{e,\mu}$
(see for example~\cite{Crivellin:2018qmi,Liu:2018xkx,Dutta:2018fge,Han:2018znu,Endo:2019bcj,Kawamura:2019rth,Abdullah:2019ofw,Badziak:2019gaf,CarcamoHernandez:2019ydc,Hiller:2019mou,Cornella:2019uxs,CarcamoHernandez:2020pxw,Haba:2020gkr,Bigaran:2020jil,Jana:2020pxx,Calibbi:2020emz,Chen:2020jvl,Yang:2020bmh}): in particular, certain 
scenarios have explored a chiral enhancement, 
due to heavy vector-like leptons in the
one-loop dipole operator, which can potentially lead to sizeable
contributions for the leptonic magnetic moments; 
however, this can open the door to charged lepton flavour violating
interactions (new physics fields with non-trivial
couplings to both muons and electrons can potentially lead to sizeable
rates for $\mu\rightarrow e\gamma, \mu\rightarrow 3e$ and $\mu-e$
conversion), already in conflict with current data~\cite{Tanabashi:2018oca}.
Controlled couplings of
electrons and muons BSM fields in the loop (subject to
``generation-wise'' mixing between SM and heavy vector-like fields)
can be achieved, 
and this further allows to evade the potentially 
very stringent constraints from cLFV $\mu-e$ transitions.

\bigskip
In this chapter, we explore a simple New Physics model, based on an 
extended gauge group 
$SU(3)\times SU(2)_L \times U(1)_Y \times U(1)_{B-L}$, with the SM
particle content extended by heavy vector-like 
fermion fields, in addition to the light 
$Z^\prime$ associated with a low-scale breaking of $U(1)_{B-L}$ by an
extra scalar field. This ``prototype model'' offers a minimal
scenario to successfully explain the anomalous 
internal pair creation in $^8$Be 
while being consistent with various experimental bounds. 
However, the couplings of the light $Z^\prime$ to fermions are strongly
constrained by experimental searches: the measurement of the atomic parity
violation in Caesium proves to be one of the most stringent constraints in
what concerns couplings to the electrons. 
Likewise, neutrino-electron
scattering and the non-observation of 
coherent neutrino–nucleus scattering impose equally stringent constraints on $Z^\prime$-neutrino couplings (the tightest bounds being due to the TEXONO~\cite{Deniz:2009mu} and CHARM-II~\cite{Vilain:1993kd} experiments). 
Consequently, we fit the $Z'$ model to the data of TEXONO and CHARM-II in order to find upper bounds on the involved couplings.

In what follows, we begin with a description of the model and its construction in~Section~\ref{sec:model}, 
in which we detail the couplings of the exotic states to SM fields,
and their impact for the new neutral current interactions.
After a brief description of the new contributions to charged lepton
anomalous magnetic moments (in a generic way) in 
Section~\ref{sec:g-2}, Section~\ref{sec:IPCcon}
is devoted to discussing
how a the light $Z^\prime$ can successfully explain the several
reported results on the anomalous IPC in $^8$Be atoms, including a 
discussion of potentially relevant isospin-breaking effects.
We revisit the available experimental constraints in 
Section~\ref{sec:phenocon},
and subsequently investigate how these impact the model's parameter
space in Section~\ref{sec:combined:explanation}, in particular 
the viable regimes allowing for a combined explanation of 
$^8$Be anomaly, as well as the tensions in $(g-2)_{e,\mu}$\footnote{Notice that the content of this chapter is based on the original publication~\cite{Hati:2020fzp}. Since then a new experimental measurement and an improved SM prediction became available (cf. Section~\ref{sec:g-2section}), superseding the measurement and SM prediction we adhere to here, with only very minor impact on the presented results.}. 

\mathversion{bold}
\section{A light vector boson from a $U(1)_{B-L}$: a prototype model}
\mathversion{normal}
\label{sec:model}
In order to establish a framework for a light vector boson, we consider a minimal gauge extension of the SM gauge group,
$SU(3)\times SU(2)_L \times U(1)_Y \times U(1)_{B-L}$. 

Extensions with a locally gauged $U(1)_{B-L}$ give rise
to new gauge and gauge-gravitational anomalies in the theory, which
need to be cancelled.
In particular, the gauged $U(1)_{B-L}$ gives rise to the triangular
gauge anomalies -  
$\mathcal{A}\left[U(1)_{B-L}\left(SU(2)_{L}\right)^2\right]$,
$\mathcal{A}\left[\left(U(1)_{B-L}\right)^3\right]$,
$\mathcal{A}\left[U(1)_{B-L} \left(  U(1)_{Y}\right)^2\right]$,  and
$\mathcal{A}\left[\mbox{Gravity}^2 \times U(1)_{B-L}\right]$. 
While the first two automatically vanish for the SM field content,
the other two require a (positive) contribution from additional fields. 
One of the most conventional and economical ways to achieve this
relies on the introduction of a SM singlet neutral fermion, right-handed neutrinos $N_R$,
with a charge $B-L=-1$, for each fermion generation. 
In the present model, 
the $U(1)_{B-L}$ is broken at a low scale by a SM singlet scalar, $h_X$,
which acquires a vacuum expectation value (VEV) $v_X$,
responsible for a light vector boson, with a mass $M_{Z^\prime} 
\sim \mathcal{O}(17~\text{MeV})$.

A successful explanation of the $^8$Be anomaly
through a light $Z^\prime$~\cite{Feng:2016ysn} further requires the
presence of additional fields vector-like fields. In particular, constraints arising from
neutrino-electron scattering experiments 
require the addition of this exotic particle content as
discussed in more detail in Section~\ref{sec:phenocon}.
Thus, the model also
includes three generations of isodoublet vector-like
leptons, denoted by $L$.
The modification of only the left-handed $Z'-\nu\nu$ couplings naturally implies a strong parity violation.
The associated electron couplings are then in turn highly constrained by measurements of atomic parity violation, so that the addition of further vector-like isosinglet leptons is necessary, to counteract the unwanted effects.
Therefore, we add another three generations of vector-like isosinglet leptons (denoted by $E$) to the field content.

The field content of the model and the transformation properties under
the extended gauge group $SU(3)\times SU(2)_L \times U(1)_Y \times U(1)_{B-L}$
are summarised in Table~\ref{tab:fields}. 
\begin{table}[ht!]
\begin{center}
  \begin{tabular}{|c|c|c|c|c|}
  \hline
      Field & $SU(3)_c$ & $SU(2)_L$ & $U(1)_Y$ & $U(1)_{B-L}$\\
  \hline
  \hline
  $Q = \left(u_L, \, d_L\right)^T$ & $ \mathbf{3} $ & $ \mathbf{2} $ &
  $ \frac{1}{6} $ & $ \frac{1}{3} $\\ 
  $ \ell = \left(\nu_L, \, e_L\right)^T $ & $ \mathbf{1} $ & $
  \mathbf{2} $ & $ -\frac{1}{2} $ & $ -1 $\\ 
  $ u_R $ & $ \mathbf{3} $ & $ \mathbf{1} $ & $ \frac{2}{3} $ & $
  \frac{1}{3} $\\ 
  $ d_R $ & $ \mathbf{3} $ & $ \mathbf{1} $ & $ -\frac{1}{3} $ & $
  \frac{1}{3} $\\ 
   $ e_R $ & $ \mathbf{1} $ & $ \mathbf{1} $ & $ -1 $ & $ -1 $\\
  \hline
  $ h_\text{SM} $ & $ \mathbf{1} $ & $ \mathbf{2} $ & $ \frac{1}{2} $ & $ 0 $ \\
  \hline
  \hline
  $ N_R$ & $ \mathbf{1} $ & $ \mathbf{1} $ & $ 0 $ & $ -1 $\\
  $ L_{L,R} = \left( L_{L,R}^0 , \, L_{L,R}^-\right)^T $ & $
  \mathbf{1} $ & $ \mathbf{2} $ & $ -\frac{1}{2} $ & $ 1 $ \\ 
  $ E_{L,R} $ & $ \mathbf{1} $ & $ \mathbf{1} $ & $ -1 $ & $ 1 $ \\
  \hline
  $ h_{X} $ & $ \mathbf{1} $ & $ \mathbf{1} $ & $ 0 $ & $ 2 $\\
  \hline
  \end{tabular}
  \end{center}
\caption{Field content of the model and
  transformation properties under the gauge group $SU(3)\times SU(2)_L
  \times U(1)_Y \times U(1)_{B-L}$.} 
  \label{tab:fields}
\end{table}
%


\subsection{Gauge sector}\label{sec:mixing}
In the unbroken phase, the relevant part of the kinetic Lagrangian,
including mixing~\footnote{We recall that
  kinetic mixing always appears at least at the one-loop level
  in models with fermions which are charged under
  both $U(1)$s. Here we parametrise these corrections through an effective
  coefficient, $\epsilon_k$.} between the hypercharge boson $B$ and
the $U(1)_{B-L}$ boson $B^\prime$, is given by 
\begin{equation}
  \mathcal L^\text{gauge}_\text{kin.} \supseteq
  -\frac{1}{4}\tilde{F}_{\mu\nu} \tilde{F}^{\mu\nu}
  - \frac{1}{4} \tilde{F^\prime}_{\mu\nu}\tilde{F^\prime}^{\mu\nu}
  + \frac{\epsilon_k}{2}\,\tilde{F}_{\mu\nu}\tilde{F^\prime}^{\mu\nu} +
  \sum_{f} i \bar f \,\tilde{\slashed{D}}\, f\,.
\end{equation}
In the above, $\tilde{F}_{\mu\nu}$ and $\tilde{F^\prime}_{\mu\nu}$
correspond to the field strengths of the (kinetically mixed)
hypercharge boson $\tilde B$ and the $U(1)_{B-L}$ boson $\tilde
B^\prime$;
$\epsilon_k$ denotes the kinetic mixing parameter.
The relevant part of the gauge covariant derivative is given by
\begin{equation}
  \tilde{D}_\mu \,=  \,\partial_\mu + \dots + i \,g^\prime \, Y_f \,
  \tilde{B}_\mu
  + i \,g_{B-L}  \,Q_f^{B-L} \, \tilde{B}^\prime_\mu\,, 
\end{equation}
with the hypercharge and $B-L$ charge written as
$Y_f = Q_f-T_{3\,f}$ and $Q^{B-L}_f$, 
respectively;
the corresponding gauge couplings are denoted by $g^\prime$ and $g_{B-L}$.
The gauge kinetic terms can be cast in matrix form as
\begin{equation}
	-\frac{1}{4} \tilde{F}_{\mu\nu}\tilde{F}^{\mu\nu} -\frac{1}{4}
        \tilde{F^\prime}_{\mu\nu}\tilde{F^\prime}^{\mu\nu} + \frac{\epsilon_k}{2}
        \tilde{F}_{\mu\nu}\tilde{F^\prime}^{\mu\nu} =
        -\frac{1}{4}\begin{pmatrix} \tilde{F}_{\mu\nu} &
          \tilde{F^\prime}_{\mu\nu}\end{pmatrix} 
	\begin{pmatrix} 1 & -\epsilon_k\\
          -\epsilon_k & 1\end{pmatrix}
	  \begin{pmatrix}\tilde{F}^{\mu\nu} \\
            \tilde{F^\prime}^{\mu\nu}\end{pmatrix}\,,
\end{equation}
which can then be brought to the diagonal canonical form
\begin{equation}
  \mathcal{L}^\text{gauge}_\text{kin.} =
  -\frac{1}{4} F_{\mu\nu}F^{\mu\nu} -\frac{1}{4} F^\prime_{\mu\nu}F^{\prime\mu\nu}
  + \sum_f i \bar f \,{\slashed{D}} \,f
\end{equation}
by a linear transformation of the fields,
\begin{equation}
  \tilde{B}_\mu = B_\mu + \frac{\epsilon_k}{\sqrt{1 -
      \epsilon_k^2}}B^\prime_\mu\:\text,\quad\quad
  \tilde{B^\prime}_\mu = \frac{1}{\sqrt{1-\epsilon_k^2}} B^\prime_\mu
  \, . 
\end{equation}
This transformation is obtained by a Cholesky decomposition, allowing
the resulting triangular matrices to be absorbed into a redefinition
of the gauge fields. 
The neutral part of the gauge covariant derivative can then be written as
\begin{equation}
  D_\mu \,=  \,\partial_\mu + \dots + i  \,g_w \, T_{3\,f}  \,W_{3\,\mu}
  + i  \,g^\prime  \,Y_f \,B_\mu + i\left(\varepsilon^\prime  
\,g^\prime  \,Y_f
  + \varepsilon^\prime_{B-L} \,  Q^{B-L}_f\right) B^\prime_\mu \,, 
  \label{eqn:covdiv}
\end{equation}
in which we have introduced the following coupling strengths
\begin{equation}\label{eq:def:epsilonprime}
  \varepsilon^\prime_{B-L} = \frac{g_{B-L}}{\sqrt{1 - \epsilon_k^2}} \,,
  \quad\quad 
\varepsilon^\prime = \frac{\epsilon_k}{\sqrt{1-\epsilon_k^2}}\:\text.
\end{equation}
Note that due to the above transformation, the mixing now appears
in the couplings of the physical fields.
In the broken phase 
(following electroweak symmetry breaking, and the
subsequent $U(1)_{B-L}$ breaking), the Lagrangian includes the following mass terms 
\begin{equation}
  \mathcal{L}^\text{gauge}_\text{mass}
  \supseteq (D_\mu\braket{h_\text{SM}})^\dagger \,
  (D^\mu\braket{h_\text{SM}}) + (D_\mu\braket{h_{X}})^\dagger\,
  (D^\mu\braket{h_{X}})\,,
\end{equation}
with the covariant derivative $D_\mu$ defined in Eq.~(\ref{eqn:covdiv}).
The resulting mass matrix, in which the neutral bosons mix amongst
themselves, can be diagonalised, leading to the following relations
between physical and interaction gauge boson states
 \begin{equation}
  \begin{pmatrix}A^\mu\\ Z^\mu\\Z^{\prime\,\mu}\end{pmatrix}
  = \begin{pmatrix}
  \cos\theta_w & \sin\theta_w & 0\\
  - \sin\theta_w \cos\theta^\prime & \cos\theta_w\cos\theta^\prime &
  \sin\theta^\prime\\ 
  \sin\theta_w\sin\theta^\prime & -\cos\theta_w\sin\theta^\prime &
  \cos\theta^\prime 
  \end{pmatrix}
  \begin{pmatrix}B^\mu\\
    W_3^\mu\\B^{\prime\,\mu}\end{pmatrix}\text,
 \end{equation}
with the mixing angle $\theta^\prime$ defined as
\begin{equation}
  \tan2\theta^\prime \,=\,
  \frac{2\,\varepsilon^\prime \,g^\prime\, \sqrt{g_w^2 + {g^\prime}^2}}{
    {\varepsilon^\prime}^2\,{g^\prime}^2 + 4\, m_{B^\prime}^2/v^2 - g_w^2
    - {g^\prime}^2} \, ,
\end{equation}
in which $m_{B^\prime}^2 = 4 \,{\varepsilon^\prime}_{B-L}^2 \,v_X^2$ is the mass
term for the $B^\prime$-boson induced by $v_X$ (the VEV of the scalar singlet $h_X$ responsible for $U(1)_{B-L}$
breaking), and $\theta_w$ is the standard weak
mixing angle. The mass eigenvalues of the neutral vector bosons are given by 
\begin{equation}
M_{A} = 0\,, \quad \quad
M_{Z,\,Z^\prime} = \frac{g_w}{\cos\theta_w}
\frac{v}{2}\left[\frac{1}{2}\left(\frac{{\varepsilon^\prime}^2 +
    {4\,m_{B^\prime}^2}/{v^2}}{g_w^2 + {g^\prime}^2} + 1\right) \mp
  \frac{g^\prime\,\cos\theta_w\,
  \varepsilon^\prime}{g_w\,\sin2\theta^\prime}
  \right]^{\frac{1}{2}}\text.  
\end{equation}
In the limit of small $\varepsilon^\prime$ (corresponding to small
kinetic mixing, cf. Eq.~(\ref{eq:def:epsilonprime}),
one finds the following approximate expressions for
the mixing angle and the masses of the $Z$ and $Z^\prime$ bosons,
\begin{equation}\label{eq:mix2}
  M_Z^2 \simeq \frac{g_w^2 + {g^\prime}^2}{4} \,v^2\,,\quad M_{Z^\prime}^2\simeq
  m_{B^\prime}^2\,,\quad \tan2\theta^\prime \simeq -2
  \varepsilon^\prime \,\sin\theta_w\,.
\end{equation}
The relevant terms of the gauge covariant derivative
can now be expressed as~\footnote{Corrections in the $Z$ coupling due
  to mixing with the $Z^\prime$ only appear at order
  ${\varepsilon^\prime}^2\:\text{or}\:\varepsilon^\prime\varepsilon^\prime_{B-L}$
  and will henceforth be neglected.} 
\begin{equation}\label{eq:mix3}
  D_\mu \,\simeq \,\partial_\mu + \dots + i \frac{g}{\cos\theta_w}\,
  (T_{3\,f} - \sin^2\theta_W \,Q_f)\,Z_\mu + ie\, Q_f \,A_\mu +
  ie\,(\varepsilon\,
  Q_f + \varepsilon_{B-L}\,Q_f^{B-L})\,Z^\prime_\mu\, , 
\end{equation}
in which the kinetic mixing parameter and the $B-L$ gauge
coupling have been redefined as 
\begin{equation}\label{eq:epsilon:redefine}
\varepsilon \,= \, {\varepsilon^\prime
  \cos\theta_w}\,, \quad \quad
\varepsilon_{B-L} \,=\,
{\varepsilon^\prime_{B-L}}/{e}\,.
\end{equation}

\subsection{Lepton sector: masses and mixings}
The lepton masses (both for SM leptons and the additional vector-like
leptons) arise from the following generic terms in the Lagrangian
\begin{eqnarray}
  \mathcal{L}^\text{lepton}_{\text{mass}} &=&
  -y_\ell^{ij} \,h_\text{SM} \,\bar\ell_L^i \,e_R^j
  + y_\nu^{ij} \,\tilde{h}_\text{SM} \,\bar\ell_L^i \,N_R^j -\frac{1}{2}\,y_M^{ij}
  \,h_{X} \,\bar N_R^{i\,c} \,N_R^j -
  \lambda_L^{ij} \,h_{X}\, \bar\ell_L^i\,L_R^j -
  \lambda_E^{ij} \,h_{X}\,\bar E_L^{i} \,e_R^j\nonumber\\ 
  &\phantom{k}& - M_L^{ij} \,\bar L_L^i \,L_R^j - M_E^{ij} \,\bar E_L^i\,
  E_R^j - h^{ij}\, h_\text{SM} \,\bar L_L^i \,E_R^j + k^{ij} \,
  \tilde{h}_\text{SM} \,\bar
  E_L^i \,L_R^j + \mathrm{H.c.}\,,
  \label{eqn:yuk}
\end{eqnarray}
in which $y$, $\lambda$, $k$ and $h$ denote Yukawa-like interactions
involving the SM leptons, heavy right-handed neutrinos and the vector-like 
neutral and charged leptons; as mentioned in the beginning of this
section, and in addition to the three SM generations of neutral and
charged leptons (i.e., 3 flavours), the model includes three
generations of isodoublet and isosinglet vector-like leptons.
In Eq.~(\ref{eqn:yuk}), each coupling or mass term thus runs over
$i,j=1...3$, i.e. $i$ and $j$ denote the three generations intrinsic
to each lepton species.

As emphasised in Ref.~\cite{Crivellin:2018qmi}, 
intergenerational couplings between the SM charged leptons and
the vector-like fermions should be very small, in order to avoid
the otherwise unacceptably large rates
for cLFV processes,
as for instance $\mu\to e\gamma$. In what follows, and to further
circumvent excessive
flavour changing neutral current interactions mediated by the
light $Z^\prime$,
we assume the couplings $h$, $k$, $\lambda_L$ and $\lambda_E$, 
as well as the masses $M_{L,E}$, to be diagonal,
implying that the SM fields (neutrinos and charged leptons)
of a given generation can only mix with vector-like fields
of the same generation. 

After electroweak and $U(1)_{B-L}$ breaking,
the mass matrices for the charged leptons and neutrinos can be cast
for simplicity in a ``chiral basis'' spanned, for each generation, by
the following vectors: $(e_L,L_L^-,E_L)^T$, $(e_R,L_R^-,E_R)^T$ and
$(\nu, N^c,L^{0},L^{0c})_L^T$. 
The charged lepton mass matrix is thus given by
\begin{equation}
  \mathcal L^{\ell}_{\text{mass}} =\begin{pmatrix}\bar e_L& \bar
  L_L^-& \bar E_L\end{pmatrix} 
  \cdot M_\ell \cdot
  \begin{pmatrix}e_R\\L_R^-\\E_R\end{pmatrix}
  = \begin{pmatrix}\bar e_L& \bar L_L^-& \bar E_L\end{pmatrix}
  \begin{pmatrix}
    y \frac{v}{\sqrt{2}} & \lambda_L \frac{v_{X}}{\sqrt{2}} & \mathbb{0}\\
    \mathbb{0}& M_L & h \frac{v}{\sqrt{2}}\\
    \lambda_E \frac{v_{X}}{\sqrt{2}} & k \frac{v}{\sqrt{2}} & M_E\\
  \end{pmatrix}
  \begin{pmatrix}e_R\\L_R^-\\E_R\end{pmatrix}\,\text,
\end{equation}
in which every entry should be understood as a $3\times3$ block (in
generation space).
The full charged lepton mass matrix can be (block-) diagonalised by a
bi-unitary rotation 
\begin{equation}\label{eq:mix4}
  M_\ell^\text{diag} = U_L^\dagger \,M_\ell\, U_R\,,
\end{equation}
where the rotation matrices $U_{L,R}$ can be obtained by a
perturbative expansion, justified in view of relative size of the SM
lepton masses and the much heavier ones 
of the vector-like leptons ($M_{L,E}$). In this study, we used
$\frac{(yv, \, hv_X,\, kv_X)}{M_{L,E}}\ll1$ as the (small) expansion
parameters, and followed the algorithm prescribed in~\cite{Grimus:2000vj}. 
Up to third order in the perturbation series, we thus obtain
\begin{equation}\label{eq:def:UellL}
  U_L = \begin{pmatrix}
          \mathbb{1} - \frac{\lambda_L^2 v_X^2}{4 M_L^2} & \frac{\lambda_L
            v_X}{\sqrt{2} M_L} - \frac{\lambda_L^3
            v_X^3}{4\sqrt{2}M_L^3} & \frac{(k \lambda_L M_E + h
            \lambda_L M_L + \lambda_E M_E y)v v_X}{2 M_E^3}\\ 
          \frac{\lambda_L^3 v_X^3}{4\sqrt{2}M_L^3} - \frac{\lambda_L
            v_X}{\sqrt{2}M_L} & \mathbb{1} - \frac{\lambda_L^2 v_X^2}{4 M_L^2}
          & \frac{(k M_E M_L + h(M_E^2 + M_L^2))v}{\sqrt{2}M_E^3} \\ 
          \frac{(h \lambda_L M_E - \lambda_E M_L y)v v_X}{4 M_E^3} & -
          \frac{(k M_E M_L + h(M_E^2 + M_L^2))v}{\sqrt{2}M_E^3} & \mathbb{1} 
        \end{pmatrix}
\end{equation}
and
\begin{equation}\label{eq:def:UellR}
  U_R = \begin{pmatrix}
          \mathbb{1} - \frac{\lambda_E^2 v_X^2}{4 M_E^2} & \frac{\lambda_L v
            v_X}{2M_L^2} - \frac{\lambda_E(k M_E M_L + h(M_E^2 +
            M_L^2))v v_X}{2M_E^3 M_L} & \frac{\lambda_E
            v_X}{\sqrt{2}M_E} - \frac{\lambda_E^3 v_X^3}{4\sqrt{2}
            M_E^3}\\ 
          \frac{(h \lambda_E M_L - \lambda_L M_E y)v v_X}{2 M_E M_L^2}
          & \mathbb{1} & \frac{(h M_E M_L + k(M_E^2 +
            M_L^2))v}{\sqrt{2}M_E^3}\\ 
          \frac{\lambda_E^3 v_X^3}{4\sqrt{2} M_E^3} - \frac{\lambda_E
            v_X}{\sqrt{2} M_E} & -\frac{(h M_E M_L + k(M_E^2 +
            M_L^2))v}{\sqrt{2}M_E^3} & \mathbb{1} - \frac{\lambda_E^2 v_X^2}{4
            M_E^2} 
        \end{pmatrix}\,\text.
\end{equation}

\bigskip
Concerning the neutral leptons, the symmetric (Majorana) mass
matrix can be written as 
\begin{eqnarray}
   \mathcal L^{\nu}_{\text{mass}}&=& \begin{pmatrix}\nu^T&  N^{c\:T}&
     L^{0\,T}&  L^{0\:c\:T}\end{pmatrix}_L C^{-1} 
  \cdot M_{\nu}\cdot
  \begin{pmatrix}\nu\\N^c\\  L^{0} \\L^{0\:c}\end{pmatrix}_L\nonumber\\
  &=& \begin{pmatrix}\nu^T&  N^{c\:T}& L^{0\,T}&  L^{0\:c\:T}\end{pmatrix}_L C^{-1}
  \begin{pmatrix}
    \mathbb{0} & y_\nu \frac{v}{\sqrt{2}} & \mathbb{0} &\lambda_L \frac{v_{X}}{\sqrt{2}}\\
    y_\nu \frac{v}{\sqrt{2}} & y_M \frac{v_{X}}{\sqrt{2}} & \mathbb{0} & \mathbb{0}\\
    \mathbb{0} & \mathbb{0} & \mathbb{0} & M_L\\
    \lambda_L \frac{v_{X}}{\sqrt{2}} & \mathbb{0} & M_L & \mathbb{0}\\
  \end{pmatrix}
  \begin{pmatrix}\nu\\N^c\\  L^{0} \\L^{0c}\end{pmatrix}_L\,\text,
  \label{eqn:numass}
\end{eqnarray}
in which each entry again corresponds to a $3\times3$ block
matrix. Following the same perturbative approach, and in this case up
to second order in perturbations of $\frac{y_\nu v}{y_M v_X},
\:\frac{y_\nu v}{M_L},\: \frac{y_M v_X}{M_L}\ll 1$, 
the mass matrix of Eq.~\eqref{eqn:numass}
can be block-diagonalised via a single unitary rotation
\begin{equation}
  M_\nu^\text{diag} = \tilde U_\nu^T\, M_\nu \,\tilde U_\nu\,,
\end{equation}
with 
\begin{equation}
  \tilde U_\nu = \begin{pmatrix}
           \mathbb{1} - \frac{\lambda_L^2 v_X^2}{4 M_L^2} - \frac{v^2
             y_\nu^2}{2 v_X^2 y_M^2} & \frac{v y_\nu}{v_X y_M} &
           \frac{\lambda_L v_X}{2 M_L} & \frac{\lambda_L v_X}{2
             M_L}\\ 
           -\frac{v y_\nu}{v_X y_M} & \mathbb{1} -\frac{v^2 y_\nu^2}{2 v_X^2
             y_M^2} & \mathbb{0} & \mathbb{0}\\ 
           -\frac{\lambda_L v_X}{\sqrt{2}M_L} & -\frac{\lambda_L v
             y_\nu}{\sqrt{2} M_L y_M} & \frac{1}{\sqrt{2}} -
           \frac{\lambda_L^2 v_X^2}{4\sqrt{2}M_L^2} &
           \frac{1}{\sqrt{2}} - \frac{\lambda_L^2
             v_X^2}{4\sqrt{2}M_L^2}\\ 
           \mathbb{0}&\mathbb{0}&-\frac{1}{\sqrt{2}} & \frac{1}{\sqrt{2}}
          \end{pmatrix}
          \,\text.
\end{equation}
We notice that the light (active) neutrino masses are generated via a
type-I seesaw mechanism 
\cite{Minkowski:1977sc,GellMann:1980vs,Yanagida:1979as,Mohapatra:1979ia,Schechter:1980gr,Mohapatra:1980yp,Schechter:1981cv},
relying on the Majorana mass term of the singlet right-handed 
neutrinos, $\sim v_X y_M/\sqrt 2$, which is dynamically generated upon
the breaking of $U(1)_{B-L}$; contributions from the vector-like
neutrinos arise only at higher orders and can therefore be safely neglected.
Up to second order in the relevant expansion parameters,
one then finds for the light (active) neutrino masses
\begin{equation}
	m_{\nu} \simeq -\frac{y_\nu^2 v^2}{v_X y_M}\,.
\end{equation}
As already mentioned, we work under the assumption that with the
exception of the neutrino Yukawa couplings $y_\nu$, all other couplings are
diagonal in generation space. 
Thus, the entire flavour structure
at the origin of leptonic mixing is encoded in the Dirac mass matrix
$(\propto v y_\nu)$, which can be itself diagonalised by a unitary matrix
$U_P$ as  
\begin{equation}
	\hat y_\nu \,= \,U_P^T\, y_\nu\, U_P\,.
\end{equation}
By carefully choosing the entries of $y_\nu$, e.g. via a Casas-Ibarra parametrisation (cf. Eq.~\ref{eqn:ynuci}) neutrino mixing data can easily be accommodated.
In view of the comparatively low scale of $B-L$ breaking, the neutrino Yukawa couplings must be very small, which also leads to small mixings with the (mostly) sterile states.
Furthermore, the masses of the (mostly) sterile states are expected to be comparatively small as well.
In view of the small mixings and the small masses, the effects of the sterile states in cLFV transitions can be safely neglected.

The full diagonalisation of the $12\times12$ neutral lepton mass matrix is
then given by 
\begin{equation}
	U_\nu \,= \,\tilde U_\nu \,\mathrm{diag}(U_P, \mathbb{1}, \mathbb{1}, \mathbb{1})\,,
\end{equation}
in which $\mathbb{1}$ denotes a $3\times3$ unity matrix.
In turn, this allows generalising the lepton currents (see Chapters~\ref{chap:lepflav} and~\ref{sec:numassgen}) as
\begin{equation}
  \mathcal L_{W^\pm} \,=\, -\frac{g}{\sqrt{2}} \,W_\mu^- \,
  \sum_{\alpha=e,\,\mu,\,\tau}\sum_{i = 1}^{9}\sum_{j = 1}^{12}
  \bar\ell_i \,(U_L^\dagger)_{i\,\alpha}\,\gamma^\mu \,P_L\,
  (U_\nu)_{\alpha\,j}\,\nu_j + \mathrm{H.c.}\,, 
\end{equation}
in which we have explicitly written the sums over flavour and mass
eigenstates (9 charged lepton mass eigenstates, and 12 neutral
states). 
Once again, one is led to a not necessarily unitary~\cite{Xing:2007zj,Blennow:2016jkn,Fernandez-Martinez:2016lgt,Escrihuela:2015wra,Hati:2019ufv}
PMNS matrix corresponding to
the upper $3\times 3$ block of $\sum_\alpha
(U_L^\dagger)_{i\,\alpha}(U_\nu)_{\alpha\,j}$ (i.e. $i,j=1,2,3$,
corresponding to the lightest, mostly SM-like states of both charged
and neutral lepton sectors). 
The deviation from unitarity of the would-be PMNS matrix is further constrained from electroweak precision data as discussed in Chapters~\ref{chap:lepflav} and~\ref{sec:numassgen}.
These constraints can be easily satisfied by respectively adjusting the entries of $y_M$ and $y_\nu$, without further phenomenological impact on the observables discussed in this chapter.

\mathversion{bold}
\subsection{New neutral current interactions: $Z^\prime$ and $h_X$}
\mathversion{normal}
\label{section:newneutralcurrent}

Having obtained all the relevant elements to characterise the lepton
and gauge sectors, we now address the impact of the additional fields
and modified couplings on the new neutral currents, in particular
those mediated by the light $Z^\prime$, which will be the key
ingredients to address the distinct anomalies here considered. 
The new neutral currents mediated by the $Z^\prime$ boson, $ i
Z^\prime_\mu J_{Z^\prime}^\mu $
can be expressed as
\begin{align}
  J_{Z^\prime}^\mu \,=  \,e  \,\bar f_i  \,\gamma^\mu \,
  \left(\varepsilon^V_{ij} + \gamma^5  \,\varepsilon^A_{ij}\right) \,  f_j \, ,
	\label{eq:vector-current}
\end{align}
in which $f$ denotes a SM fermion (up- and down-type quarks, charged
leptons, and neutrinos) and the coefficients $\varepsilon^{V,A}_i$ are
the effective couplings in units of $e$. For the up- and down-type
quarks ($f=u,d$) the axial part of the $Z^\prime$ coupling formally
vanishes, $\varepsilon^{A}_q=0$, while the vector part is given by 
\begin{eqnarray}\label{eq:epsq}
  \varepsilon^{V}_{qq} &=&
  \varepsilon \,Q_q + \varepsilon_{B-L} \,Q_q^{B-L}\,\text.
\end{eqnarray}
On the other hand, and due to the mixings with the vector-like fermions,
the situation for the lepton sector is different: the
modified left- and right-handed couplings for the charged leptons
now lead to mixings between different species, as cast below
(for a given generation)
\begin{eqnarray}
  g_{Z^\prime, \,L}^{\ell_a \ell_b} &=&  \sum_{c=1,2,3}
  \left(\varepsilon \,Q_c + \varepsilon_{B-L} \,Q_c^{B-L}
  \right)(U_L^\dagger)^{ac}\,U_L^{cb}\, , 
  \label{eq:epslL}\\ 
  g_{Z^\prime,\, R}^{\ell_a \ell_b} &=&  \sum_{c=1,2,3}
  \left(\varepsilon \,Q_c + \varepsilon_{B-L} \,Q_c^{B-L}
  \right)(U_R^\dagger)^{ac}\,U_R^{cb}\,\text.
  \label{eq:epslR}
\end{eqnarray}
In the above, the indices $a,\,b,\,c$ refer to the 
mass eigenstates of different species:
the lightest one ($a,b,c=1$) corresponds to the (mostly)
SM charged lepton, while the two heavier ones (i.e. $a,b,c=2,3$)
correspond to the isodoublet and isosinglet heavy vector-like
leptons.
This leads to the following vectorial and axial couplings
\begin{eqnarray}\label{eq:epsl}
  g^{V}_{\ell_a \ell_b} = \frac{1}{2} \left({g_{Z^\prime, \,L}^{\ell_a
      \ell_b}+ g_{Z^\prime,\, R}^{\ell_a \ell_b}}\right)\, , \quad
  g^{A}_{\ell_a \ell_b} = \frac{1}{2} \left({g_{Z^\prime, \,R}^{\ell_a
      \ell_b}- g_{Z^\prime,\, L}^{\ell_a \ell_b}}\right)\,\text. 
\end{eqnarray} 
Similarly, the new couplings to the (Majorana) neutrinos are given by
\begin{eqnarray}\label{eq:epsnu}
   g^{V}_{\nu_a \nu_b}  &=&\sum_{c} \varepsilon_{B-L}\,
   \mathrm{Im}\left(Q_c^{B-L} \,(U_\nu^\ast)^{ca}\, U_\nu^{c
     b}\right)\,\text,\\ 
   g^{A}_{\nu_a \nu_b}  &=&-\sum_{c} \varepsilon_{B-L}\,
   \mathrm{Re}\left(Q_c^{B-L} \,(U_\nu^\ast)^{ca} \,U_\nu^{c
     b}\right)\,\text.\label{eqn:gAnu}
\end{eqnarray}
Note that the vector part of the couplings vanishes for
$\nu_a = \nu_b$ (with $a,b=1,2$),
which corresponds to the Majorana 
mass eigenstates with purely Majorana masses (cf. Eq.~\ref{eqn:numass}).  
For the lightest (mostly SM-like) physical states ($a,b=1$) one has
\begin{eqnarray}
	\varepsilon^A_{\nu_\alpha\nu_\alpha} =
        - g_{Z^\prime,\,L}^{\nu_\alpha\nu_\alpha} &\simeq& \,
        \varepsilon_{B-L} \left(1 - \frac{\lambda_{L\:\alpha}^2
          v_{X}^2}{M_{L\:\alpha}^2} \right)\label{eqn:nunu}\,\text,\\ 
  g_{Z^\prime,\,L}^{\ell_\alpha\ell_\alpha} & \simeq&  - \varepsilon
  + \left( \frac{\lambda_{L\:\alpha}^2 v_{X}^2}{M_{L\:\alpha}^2} -
  1\right)\,\varepsilon_{B-L}\label{eqn:eeL}\,\text,\\ 
  g_{Z^\prime,\,R}^{\ell_\alpha\ell_\alpha} & \simeq& - \varepsilon +
  \left(\frac{\lambda_{E\:\alpha}^2 v_{X}^2}{M_{E\:\alpha}^2} -
  1\right)\varepsilon_{B-L}\,\text,\label{eqn:eeR}\\ 
	\varepsilon^{V}_{\ell_\alpha\ell_\alpha} &\simeq&  -
        \varepsilon +\frac{1}{2}\left(\frac{\lambda_{L\,\alpha}^2
          v_X^2}{M_{L\,\alpha}^2} + \frac{\lambda_{E\,\alpha}^2
          v_X^2}{M_{E\,\alpha}^2} - 2 \right)\varepsilon_{B-L}\,, \\ 
	\varepsilon^{A}_{\ell_\alpha\ell_\alpha} &\simeq&
        \frac{1}{2}\left(\frac{\lambda_{E\,\alpha}^2
          v_X^2}{M_{E\,\alpha}^2} - \frac{\lambda_{L\,\alpha}^2
          v_X^2}{M_{L\,\alpha}^2}  \right)\varepsilon_{B-L}\,, 
\end{eqnarray}
in which the subscript $\alpha\in\{e, \,\mu,\,\tau\}$ now
explicitly denotes the SM lepton flavour.
Note that flavour changing (tree-level) couplings are absent by
construction, as a consequence of having imposed strictly diagonal
couplings and masses ($\lambda_{L,\,E}$, $M_{L,\,E}$) for the
vector-like states. The ``cross-species couplings'' are defined in
Eqs.~\eqref{eq:epslL}, \eqref{eq:epslR} and \eqref{eqn:gAnu}. 

\bigskip
Finally, the scalar and pseudo-scalar couplings of $h_X$ to the
charged leptons (SM- and vector-like species) can be conveniently
expressed as 
\begin{equation}
	\frac{v_X}{\sqrt{2}}\,g_S \,=\, m^\ell_\text{diag} -
        \frac{1}{2}\left(C_{LR} + C_{RL}\right) 
\end{equation}
and
\begin{equation}
	\frac{v_X}{\sqrt{2}}\,g_P \,=\,  \frac{1}{2}\left(C_{LR} -
        C_{RL}\right)\,\text, 
\end{equation}
where
\begin{equation}
	C_{LR} = (C_{RL})^\dagger =
	U_L^\dagger
  	\begin{pmatrix}
  	\frac{y v}{\sqrt{2}} & 0 & 0\\
  	0&M_L&\frac{h v}{\sqrt{2}}\\
  	0&\frac{k v}{\sqrt{2}} & M_E
  	\end{pmatrix}
 	 U_R\,\text,
\end{equation}
with $U_{L,R}$ as defined in Eqs.~(\ref{eq:def:UellL}, \ref{eq:def:UellR}).
Further notice that corrections to the tree-level couplings of the SM
Higgs and $Z$-boson appear only at higher orders in the
perturbation series of the mixing matrices, and are expected to be of
little effect.

\section{New physics
  contributions to the anomalous magnetic moments}\label{sec:g-2} 
Having laid down the ingredients of this minimal prototype construction, we no begin by investigating how the new fields and interaction allow to explain the anomalous magnetic lepton moments.

\begin{figure}[h!]
\begin{center}
\feynmandiagram [layered layout, horizontal=b to c] {
i1 [particle=\(\ell_{R,L}\)]
a -- [fermion] b
-- [fermion, edge label=\(f\)] c1
-- [fermion, edge label=\(f^\prime\)] c
-- [fermion] f1 [particle=\(\ell_{L,R}\)] d,
b -- [insertion={[style=black]0.5}] c,
c-- [photon, half left, edge label=\({Z^\prime, Z, W}\)] b,
}; \quad \quad 
\feynmandiagram [layered layout, horizontal=b to c] {
i1 [particle=\(\ell_{R,L}\)]
a -- [fermion] b
-- [fermion, edge label=\(f\)] c1
-- [fermion, edge label=\(f^\prime\)] c
-- [fermion] f1 [particle=\(\ell_{L,R}\)] d,
b -- [insertion={[style=black]0.5}] c,
c-- [scalar, half left, edge label=\({h_X, H}\)] b,
};
\caption{Illustrative Feynman diagrams for the one-loop contributions to
  $(g-2)_{e,\mu}$ induced by the new states and couplings (with a
  possible mass insertion inside the 
  loop or at an external leg). The internal states ($f,\,f^\prime$)
  are leptons; the photon can be
  attached to any of the charged fields.}
\label{fig:feyn}
\end{center}
\end{figure}
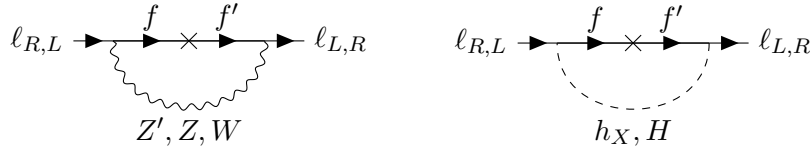

The field content of the model here proposed gives rise to new contributions to the
anomalous magnetic moments of the light charged leptons, in the
form of several one-loop diagrams mediated by the extra $Z^\prime$ and $h_X$
bosons, as well as the new heavy vector-like fermions, which can also
propagate in the loop. The new contributions are schematically
illustrated in Fig.~\ref{fig:feyn}. Notice that due to the potentially
large couplings, the contributions induced by the $Z^\prime$ or even $h_X$
can be dominant when compared to the SM ones.
%

Generic one-loop contributions generated by the exchange of a 
neutral vector boson (NV) and a negatively charged internal fermion,
$\Delta a_\ell^\text{NV}$,
can be expressed as~\cite{Jegerlehner:2009ry}
\begin{equation}
  \Delta a_\ell^\text{NV} \,=  \,
  \sum_{i}\left[\frac{g_V^{\ell\,i} \,{g_V^{\ell\,i}}^\ast}{4\pi^2} \,
    \frac{m_\ell^2}{m_B^2} \,F(\lambda, \rho_i) +
    \frac{g_A^{\ell\,i} \,{g_A^{\ell\,i}}^\ast}{4\pi^2} \,
    \frac{m_\ell^2}{m_B^2} \,F(\lambda, -\rho_i)\right]\,\text,
\end{equation}
with $\Delta a_\ell$ as defined in Eqs.~(\ref{eq:amu:delta},\ref{Delta_aeCS});  
$g_{V(A)}$ is the vector (axial-vector) coupling\footnote{The sum in
  Eq.~\eqref{mdmvcl} runs over 
  all fermions which have non-vanishing couplings to the external
  leptons, so that in general $i = 1,\,2,\,\dots 9$; however note that
  only fermions belonging to the same generation (but possibly of
  different species e.g., SM leptons and isosinglet or isodoublet
  vector-like leptons) have a non-vanishing entry.}  and $m_B$ is the mass
of the exchanged vector boson. The function $F(\lambda, \rho)$ is
defined as follows
\begin{equation}{\label{mdmvcl}}
  F(\lambda, \rho)  \,= \,
  \frac{1}{2}\int_0^1 \frac{2x \,(1-x) \,\left[x-2(1-\rho)\right] + \lambda^2
    \,x^2 \,(1-\rho)^2 \,(1+\rho -x)}{(1-x)(1-\lambda^2 x) +
    (\rho\lambda)^2  \,x}\, dx \,\text,
\end{equation}
in which $\rho_i = M_{f_i} / m_\ell$ with $M_{f_i}$ denoting the
internal fermion mass and with $\lambda = m_\ell / M_B$.

\noindent
Generic new contributions due to a neutral scalar mediator (NS),
$\Delta a_\ell^\text{NS}$, are given by
\begin{equation}
  \Delta a_\ell^\text{NS}\, =\,
  \sum_{i}\left(\frac{g_S^{\ell\,i}\,{g_S^{\ell\,i}}^\ast}{4\pi^2}\,
  \frac{m_\ell^2}{m_S^2}\,G(\lambda, \rho_i) +
  \frac{g_P^{\ell\,i}\,{g_P^{\ell\,i}}^\ast}{4\pi^2}\,
  \frac{m_\ell^2}{m_S^2}\,G(\lambda, -\rho_i)\right)\,\text,
\end{equation}
with
\begin{equation}
  G(\lambda,\rho) \,=\,
  \frac{1}{2}\int_{0}^{1} dx \frac{x^2\,(1 + \rho - x)}{(1-x)\,
    (1-\lambda^2 \,x) + (\rho\,\lambda)^2 \,x}\,\text,
\end{equation}
in which $g_{S(P)}$ denotes the scalar (pseudo-scalar) coupling
and $m_S$ is the mass of the neutral scalar $S$. Note that the loop
functions of a vector or a scalar mediator have an overall positive
sign, whereas the contributions of axial and pseudo-scalar mediators
are negative. This allows for a partial cancellation between
vector and axial-vector contributions, as well as
between scalar and pseudo-scalar
ones, which are, as will be subsequently discussed, crucial
to explain the relative (opposite) signs of
$\Delta a_e$ and $\Delta a_\mu$. As expected, such cancellations
naturally rely on the interplay of the $Z^\prime$
and $h_X$ couplings. 

\mathversion{bold}
\section{Explaining the anomalous IPC in $^8$Be}
\mathversion{normal}
\label{sec:IPCcon} 
We proceed to discuss how the presence of a light $Z^\prime$ boson and
the modified neutral currents can successfully address the  
internal pair creation anomaly in $^8$Be atoms.
As already
  mentioned in the Introduction, in~\cite{Krasznahorkay:2019lyl} it
  has been reported that a peak in the electron-positron pair angular
  correlation was observed in the electromagnetically forbidden
  $M0$ transition depopulating the $21.01$ MeV $0^{-}$ state in
  $^4$He, which could be explained by the creation and subsequent decay of a
  light particle in analogy to $^8$Be. However, in the absence of any fit
  for normalised branching fractions we will not include this in our
  analysis.

Firstly, let us consider one of the quantities which is extremely
relevant for the IPC excess - the width of the $Z^\prime$ decay into
a pair of electrons.   
At tree level, the latter is given by 
\begin{align}
\label{eqn:VffZp}
\Gamma(Z^\prime\to e^{+} e^{-}) \,= \,\left(| \varepsilon_{ee}^V|^2 + |
\varepsilon_{ee}^A|^2 \right) \frac{\lambda^{1/2}(M_{Z^\prime}, m_e ,
  m_{e})}{24 \,\pi \,M_{Z^\prime}} \,,
\end{align}
where the K\"all\'en function is defined as before:
$\lambda(a,b,c) = \left(a^{2} - \left(b-c\right)^{2} \right)\left(a^{2} -
\left(b+c\right)^{2} \right)$.

In what follows we discuss the bounds on the $Z^\prime$ which are directly
connected with an explanation of the $^8$Be anomaly. 
A first bound on the couplings of the $Z^\prime$ can be obtained from
the requirement that the $Z^\prime$ be sufficiently short lived for its
decay to occur inside the Atomki spectrometer, whose length is 
$\cal{O}$(cm). This gives rise to 
a lower bound on the couplings of the $Z^\prime$ to electrons 
\begin{align}\label{eq:4.2}
|\varepsilon^V_{ee}|
\gtrsim
{1.3 \times 10^{-5}} \sqrt{\text{BR}(Z^\prime\to e^+e^-)} \, .
\end{align}

The most important bounds clearly arise from the requirement that
$Z^\prime$ production (and decay) complies with the (anomalous)
data on the electron-positron angular correlations for the $^8$Be
transitions. 
We begin by recalling that the relevant quark (nucleon) couplings
necessary to explain the anomalous IPC in $^8$Be can be determined
from a combination of
the best fit value for the normalised branching fractions
experimentally measured. In what follows, this is done here for both the cases of isospin conservation and breaking.

\begin{figure}[hbt!]
\begin{center}
\mbox{    \includegraphics[width = 0.48 \textwidth]{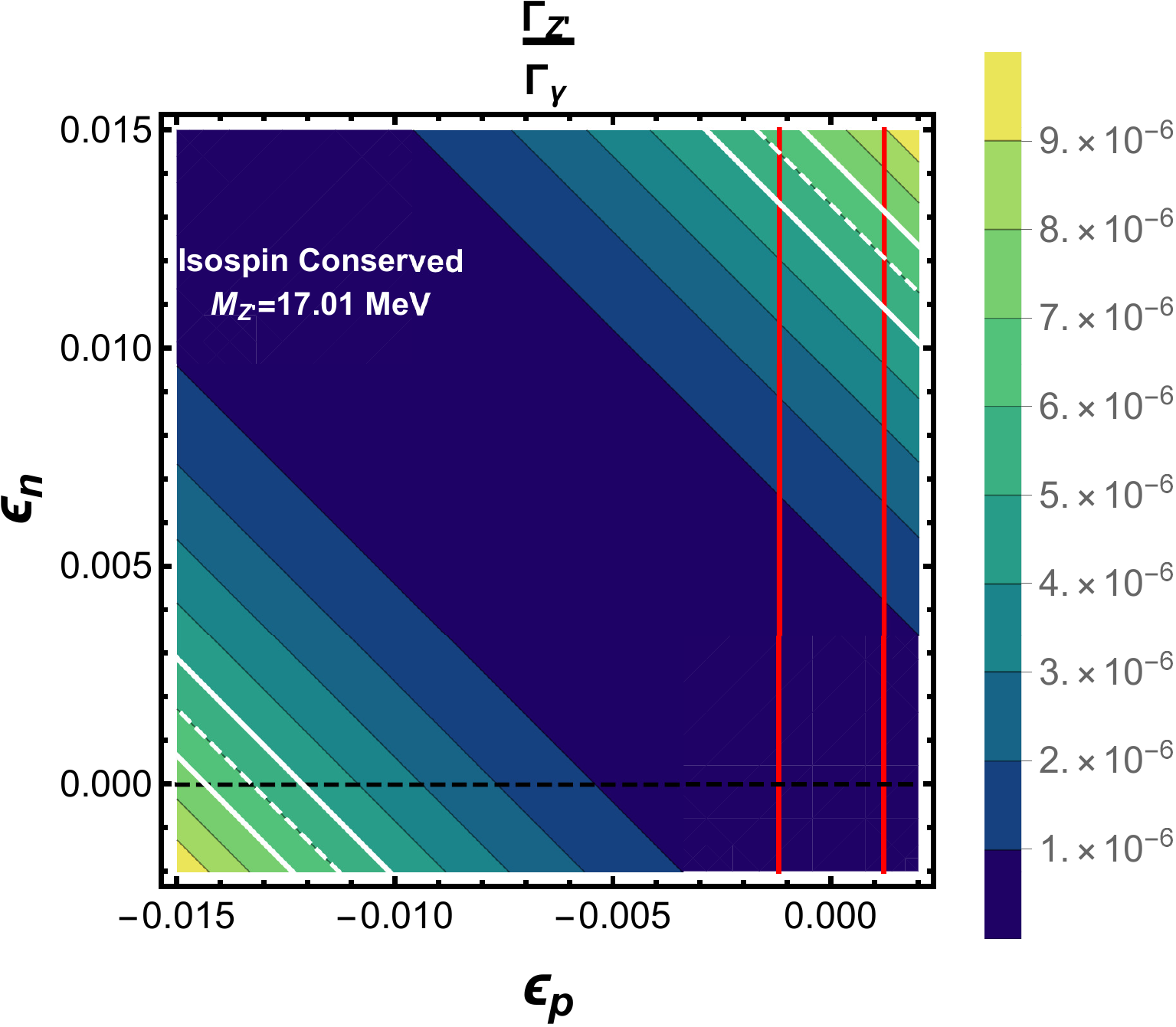} \quad\quad 
     \includegraphics[width = 0.48 \textwidth]{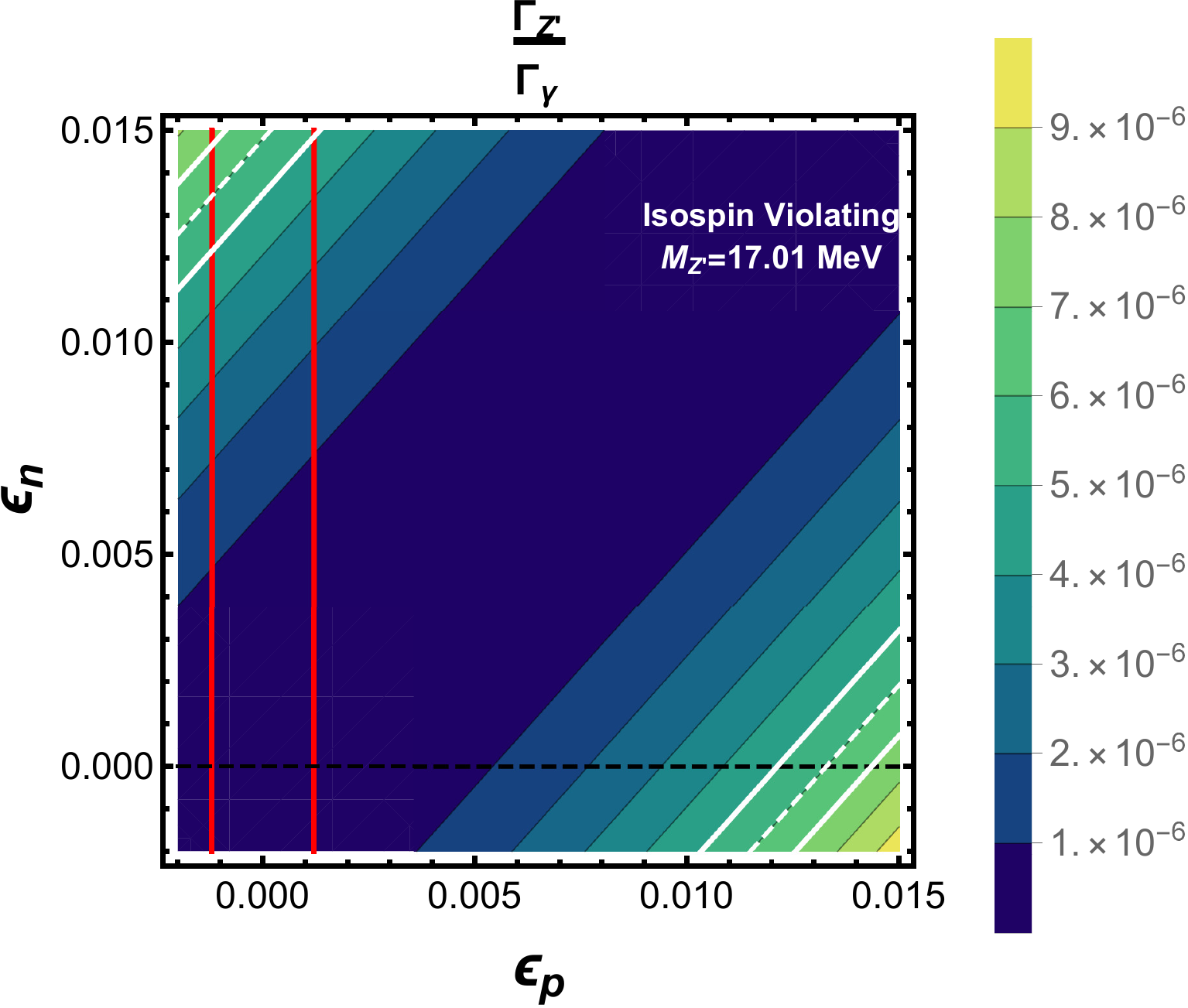}}
     \vspace*{3mm}\\
\mbox{		  \includegraphics[width = 0.48 \textwidth]{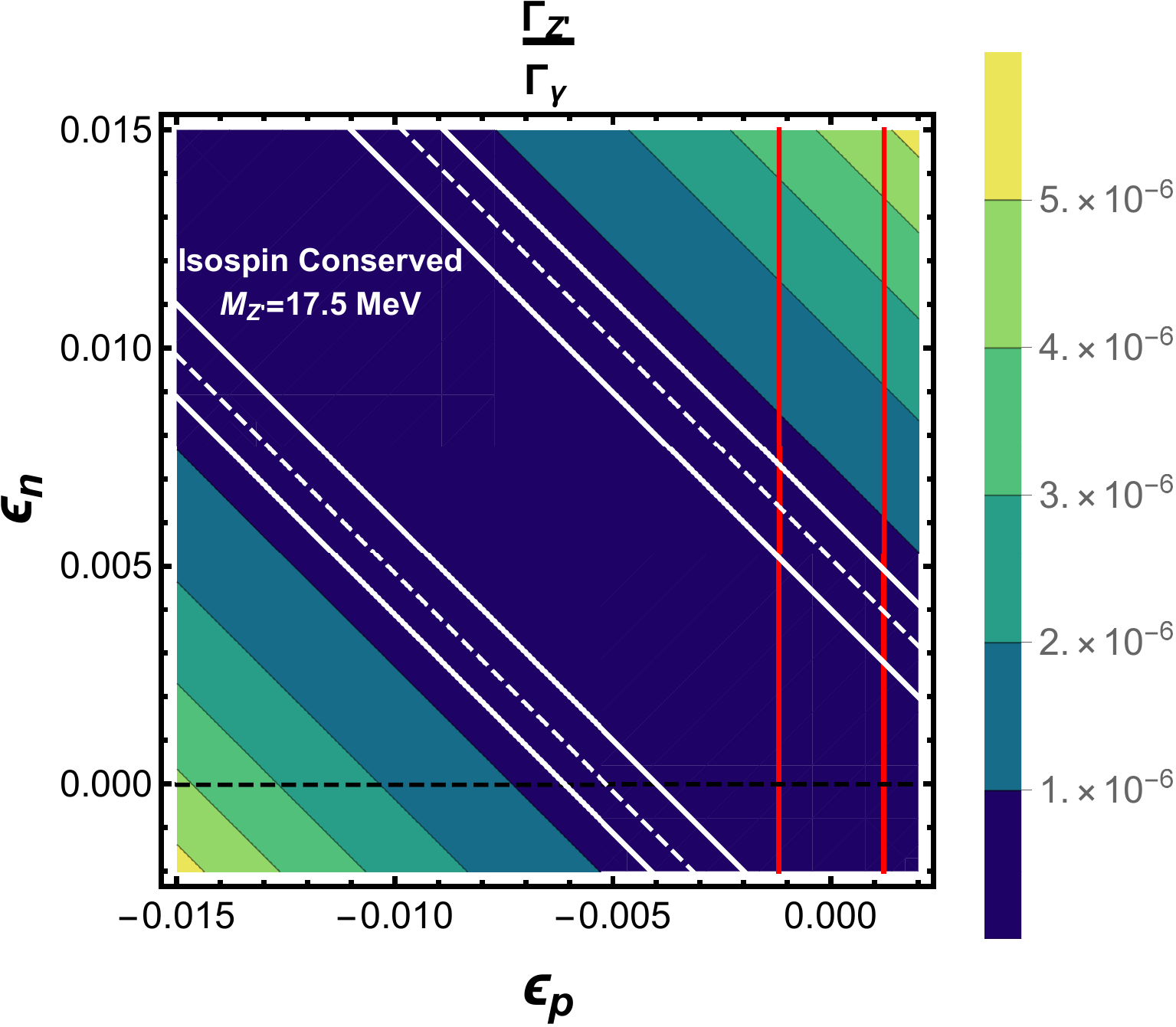} \quad \quad 
       \includegraphics[width = 0.48 \textwidth]{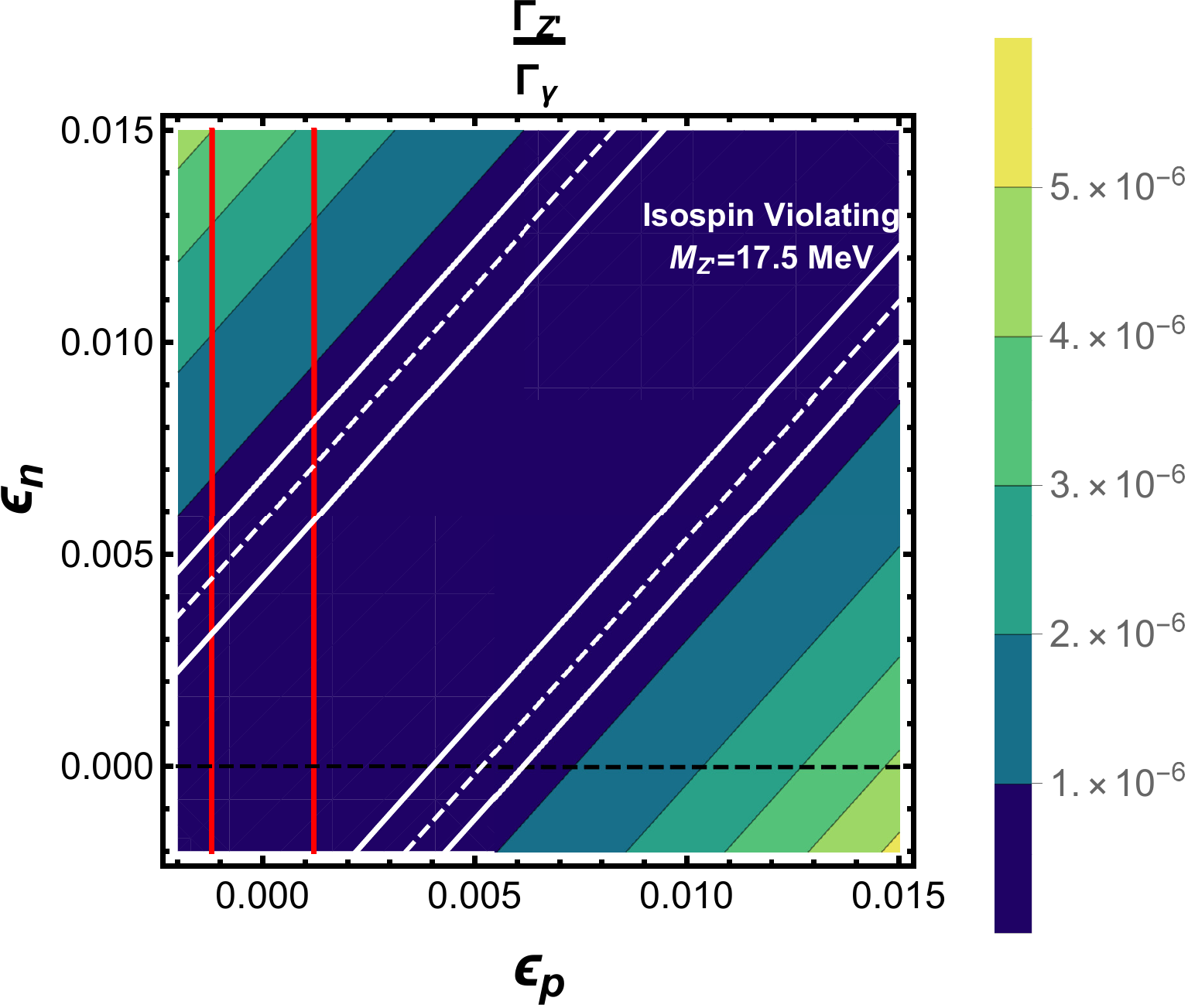}}
       \caption{On the left (right) panels,
         contour plots of the ratio $\Gamma_{Z^\prime}/\Gamma_{\gamma}$
         (see Eq.~(\ref{eq:def:rationGammaZpgamma})
         for the isospin conserving (violating) limit. The white dashed
and solid lines correspond to the best fit and to the 1$\sigma$ interval for
the experimental best fit values for $\Gamma_{Z^\prime}/\Gamma_{\gamma}$,
under the assumption BR$(Z^\prime\to e^{+} e^{-})=1$.
The region between the two red
vertical lines corresponds to the viable protophobic region of the
parameter space, as allowed by NA48/2 constraints, while the horizontal
dashed line corresponds to the pure dark photon limit.
On both upper panels we have taken $M_{Z^\prime}$=17.01~MeV, as well as
the experimental best fit value 
$\Gamma_{Z^\prime}/\Gamma_{\gamma}=6 (1) \times 10^{-6}$ (following the fit
values reported in~\cite{Krasznahorkay:2018snd}). The lower
panels illustrate the case in which $M_{Z^\prime}$=17.5~MeV, for an
experimental best fit value $\Gamma_{Z^\prime}/\Gamma_{\gamma}=0.5 (0.2)
\times 10^{-6}$, in agreement with the values quoted in~\cite{Feng:2016ysn}
(for which we have taken a conservative estimate of the error in 
$\Gamma_{Z^\prime}/\Gamma_{\gamma} \sim 0.2 \times 10^{-6}$, following the
uncertainties of~\cite{Krasznahorkay:2018snd}). Figures taken from~\cite{Hati:2020fzp}.
}
\label{fig:lim}
\end{center}
\end{figure}

\paragraph{Isospin conservation}
In the isospin conserving
limit, the normalised branching fraction
\begin{equation}\label{eq:def:rationGammaZpgamma}
\frac{\Gamma({^8\text{Be}^*} \to
  {^8\text{Be}}\, Z^\prime)}{
  \Gamma({^8\text{Be}^*} \to {^8\text{Be}}\, \gamma)}
\equiv
\frac{\Gamma_{Z^\prime}}{\Gamma_{\gamma}}
\end{equation}  
is a particularly convenient observable because the relevant nuclear
matrix elements cancel in the ratio, giving 
\begin{align}\label{eq:4.3}
	\frac{
	\Gamma({^8\text{Be}^*} \rightarrow {^8\text{Be}}+Z^\prime)
	}{
	\Gamma({^8\text{Be}^*} \rightarrow {^8\text{Be}}+\gamma)
	}
	&=
	(\varepsilon^V_p+\varepsilon^V_n)^2
	\frac{|\mathbf{k}_{Z^\prime}|^3}{|\mathbf{k}_\gamma|^3}
	=
	(\varepsilon^V_p+\varepsilon^V_n)^2
	\left[1 - \left(\frac{M_{Z^\prime}}{18.15\text{ MeV}}
          \right)^2\right]^{3/2}
	\, ,
\end{align}
in which $\varepsilon^V_p = 2\,\varepsilon^V_{uu} + \varepsilon^V_{dd}$
and $\varepsilon^V_n = \varepsilon^V_{uu} + 2 \,\varepsilon^V_{dd}$.
The purely vector quark currents (see Eq.~\eqref{eq:vector-current}) are
expressed as 
\begin{align}
	J_{Z^\prime}^{\mu \, (\mathrm{q})}
	= \sum_{i=u,d} \varepsilon^V_{ii} e J_i^\mu\,  \quad 
	(J_i^\mu = \bar q_i \gamma^\mu q_i) \, .
	\label{eq:quark-vector-current}
\end{align}
The cancellation of the nuclear matrix elements in the ratio of
Eq.~\eqref{eq:4.3} can be understood as described below. Following the
prescription of Ref.~\cite{Feng:2016ysn}, it is convenient to parametrise
the matrix element for nucleons in terms of the Dirac and Pauli form
factors $F^{Z^\prime}_{1,p} (q^2)$ and $F^{Z^\prime}_{2,p}
(q^2)$~\cite{Perdrisat:2006hj}, so that the proton matrix element can be
written as 
\begin{align}
J_p^\mu \equiv
\langle p(k') | J_{Z^\prime}^{\mu \, (\mathrm{q})} | p(k) \rangle  & =
e \,\overline{u}_p(k')  \left\{ F^{Z^\prime}_{1,p} (q^2)
\,\gamma^\mu + F^{Z^\prime}_{2,p} (q^2)
\,\sigma^{\mu \nu} \,\frac{q_\nu} {2 M_p }\right\} u_p (k) \,.
\end{align}
Here  $| p(k) \rangle$ corresponds to a proton state and $u_p (k)$
denotes the spinor corresponding to a free proton. The nuclear
magnetic form factor is then given by $G_{M,p}^{Z^\prime}(q^2) =
F^{Z^\prime}_{1,p} (q^2) + F^{Z^\prime}_{2,p}
(q^2)$~\cite{Perdrisat:2006hj,Yennie:1957,Ernst:1960zza,Hand:1963zz}. The
nucleon currents can be combined to obtain the isospin currents as 
\begin{equation}
	J_{0}^\mu \,=\, J_p^\mu \,+\, J_n^\mu\,, \quad \quad
	J_{1}^\mu \,=\, J_p^\mu \,-\, J_n^\mu
	\,.
	\label{eq:isospin-current}
\end{equation}
In the isospin conserving limit,
$\langle {^8\text{Be}} | J_1^\mu |{^8\text{Be}^*}\rangle = 0$, since
both the exited and the ground state of $^8\text{Be}$ are
isospin singlets. Defining the $Z^\prime$ hadronic current as
\begin{align}
	J_{Z^\prime}^{\mu\, h} \,=\,
 \sum_{i=u,d} \varepsilon^V_{ii} \,e \,J_i^\mu
	= (2\,\varepsilon^V_{uu}+\varepsilon^V_{dd})\,e\, J_p^\mu
	+ (\varepsilon^V_{uu}+2\,\varepsilon^V_{dd})\,e\, J_n^\mu  \ ,
\end{align}
with $p, n$ denoting protons and neutrons, one obtains
\begin{align}
	\langle {^8\text{Be}} | J_{Z^\prime}^{\mu\, h} |{^8\text{Be}^*}\rangle
	&= \frac{e}{2} \,(\varepsilon_p + \varepsilon_n)
	\langle {^8\text{Be}} | J_0^\mu |{^8\text{Be}^*}\rangle\,,
	\label{eq:isospin:matrix:elements1}
	\\
	\langle {^8\text{Be}} | J_\text{EM}^\mu |{^8\text{Be}^*}\rangle
	&= \frac{e}{2}\,
	\langle {^8\text{Be}} | J_0^\mu |{^8\text{Be}^*}\rangle \, ,
	\label{eq:isospin-matrix-elements}
\end{align}
in which $\varepsilon_p = 2\,\varepsilon^V_{uu} + \varepsilon^V_{dd}$
and $\varepsilon_n = \varepsilon^V_{uu} + 2 \,\varepsilon^V_{dd}$. From
Eq.~\eqref{eq:isospin-matrix-elements} it follows that the relevant
nuclear matrix elements cancel in the normalised branching fraction
of Eq.~\eqref{eq:4.3} (in the isospin conserving
limit). Therefore,
using the best fit values for
the mass  $M_{Z^\prime}$=17.01~(16)~MeV~\cite{Krasznahorkay:2019lyl}, and 
the normalised branching
fraction $\Gamma_{Z^\prime}/\Gamma_{\gamma}=6 (1) \times 10^{-6}$,  
Eq.~\eqref{eq:4.3} leads to the following constraint
\begin{align}\label{eq:4.4.1}
  |\varepsilon^V_p + \varepsilon^V_n| \approx
	\frac{1.2 \times 10^{-2}}{\sqrt{\text{BR}(Z^\prime \to e^+e^-)}}
		\,.
\end{align}
On the top left panel of Fig.~\ref{fig:lim} we display the
 plane spanned by $\varepsilon_p$ vs. $\varepsilon_n$, for a hypothetical
$Z^\prime$ mass of $M_{Z^\prime}$=17.01~MeV, and for
the experimental best fit value
$\Gamma_{Z^\prime}/\Gamma_{\gamma}=6 (1) \times 10^{-6}$ (following the
most recent best fit values reported in~\cite{Krasznahorkay:2018snd}). 
Notice that a large departure of $|\varepsilon_p|$ from the
protophobic limit is excluded by NA48/2 constraints~\cite{Raggi:2015noa}, which are
depicted by the two red vertical lines. The region between the latter 
corresponds to the viable protophobic regime still currently
allowed. The horizontal dashed line denotes the limiting case of
a pure dark photon. 

\paragraph{Isospin breaking}
In the above discussion it has been implicitly assumed that the $^8$Be
states have a well-defined isospin; however, as extensively noted in the
literature~\cite{Barker:1966zza,Wiringa:2000gb,Pieper:2004qw,Wiringa:2013fia,Pastore:2014oda},
the $^8$Be states are in fact isospin-mixed. In order to take the latter
effects into account, isospin breaking in the electromagnetic
transition operators arising from the neutron--proton mass difference
was studied in detail in Ref.~\cite{Feng:2016ysn}, and found to have
potentially serious implications for the quark-level couplings
required to explain the $^8$Be signal. In what follows we summarise
the most relevant points, which will be included in the present discussion.

\noindent
For a doublet of spin $J$, the
physical states (with superscripts $s_1$ and $s_2$) can be defined
as~\cite{Pastore:2014oda} 
\begin{equation}
  \Psi_J^{s_1} \,= \,\alpha_J \,\Psi_{J, T=0} \,+\, \beta_J\, \Psi_{J, T=1}\,,
  \quad\quad
  \Psi_J^{s_2} \,=\, \beta_J \,\Psi_{J, T=0} \,- \,\alpha_J\, \Psi_{J, T=1} \,,
  \label{mixing}
\end{equation}
in which the relevant mixing parameters $\alpha_J$ and $\beta_J$ can
be obtained by computing the widths of the isospin-pure states using
the  	
Quantum Monte Carlo (QMC) approach~\cite{Pastore:2014oda}. As pointed out
in~\cite{Feng:2016ysn}, this procedure may be used for the
electromagnetic transitions of isospin-mixed states as well. However,
the discrepancies with respect to the experimental results are
substantial,
even after including the meson-exchange currents in the relevant matrix
element~\cite{Pastore:2014oda}. To address this deficiency,
an isospin breaking effect was introduced
in the hadronic form factor of the electromagnetic transition operators
themselves in Ref.~\cite{Feng:2016ysn}
(following~\cite{Gardner:1995uq,Gardner:1995ya}). This has led to 
changes in the relative strength of the isoscalar and isovector
transition operators which appear as a result of isospin-breaking in
the masses of isospin multiplet states, e.g. the nonzero
neutron-proton mass difference. The isospin-breaking contributions
have been 
incorporated through the introduction of a spurion, which regulates
the isospin-breaking effects within an isospin-invariant framework
through a ``leakage'' parameter (controlled by non-perturbative
effects). The ``leakage'' parameter
is subsequently determined by matching the resulting
$M1$
transition rate of the 17.64~MeV decay of $^8$Be with its experimental
value, using the matrix elements of Ref.~\cite{Pastore:2014oda}. This
prescription leads to the corrected ratio of partial
widths~\cite{Feng:2016ysn},  
\begin{equation}
\frac{\Gamma(^8\text{Be}^* \rightarrow {^8\text{Be}}+Z^\prime)}{
\Gamma(^8\text{Be}^* \rightarrow {^8\text{Be}}+\gamma)}
=| 0.05 \, (\varepsilon^V_p + \varepsilon^V_n)   + 0.95 \,
(\varepsilon^V_p - \varepsilon^V_n) |^2 
\left[1 - \left(\frac{M_{Z^\prime}}{18.15\text{ MeV}}\right)^2\right]^{3/2} \, ,
\label{eq:IVH}
\end{equation}
and consequently, to new bounds on the relevant quark (nucleon)
couplings necessary to explain the anomalous IPC in $^8$Be.  
On the upper right panel of Fig.~\ref{fig:lim}, we display the
isospin-violating scenario of Eq.~\eqref{eq:IVH}, in the
$\varepsilon_p$ vs. $\varepsilon_n$ plane, for $M_{Z^\prime}$=17.01~MeV and
for the experimental best fit value 
$\Gamma_{Z^\prime}/\Gamma_{\gamma}=6 (1) \times
10^{-6}$~\cite{Krasznahorkay:2018snd}.
A comparison with the case of isospin conservation (upper left plot)
reveals a~$15\%$
modification with respect to 
the allowed protophobic range of $\varepsilon_n$ in
the isospin violating case.

\bigskip
Other than the best fit values for the mass of the mediator
and normalised branching fraction for the
predominantly isosinglet $^8$Be excited state with an excitation energy
18.15~MeV (here denoted as $^8\text{Be}^*$), it is 
important to take into account the IPC null results
for the predominantly isotriplet excited
state ($^8\text{Be}^{*'}$), as emphasised in~\cite{Gulyas:2015mia}. 
In particular, in the presence of a finite
isospin mixing, the latter IPC null result would call for 
a kinematic suppression, thus implying a larger preferred mass
for the $Z^\prime$, in turn leading to a large phase space suppression.
This may translate into (further) significant changes for
the preferred quark (nucleon) couplings
to the $Z^\prime$ (corresponding to a heavier $Z^\prime$, 
and to significantly smaller
normalised branching fractions when compared to the preferred fit
reported in~\cite{Krasznahorkay:2019lyl}).
Considering the benchmark value\footnote{Since no public results are
  available to the best of our knowledge, we use the values quoted
  from a private communication in~\cite{Feng:2016ysn}.} 
$\Gamma_{Z^\prime}/\Gamma_{\gamma}=0.5 \times 10^{-6}$~\cite{Feng:2016ysn}, 
we obtain the following constraint in the isospin conserving limit,
\begin{align}\label{eq:4.4}
  |\varepsilon^V_p + \varepsilon^V_n| \approx
 \frac{(3-6) \times 10^{-3}}{\sqrt{\text{BR}(Z^\prime\to e^+e^-)}}
   \, . 
\end{align}
Leading to the above limits, we have used a conservative estimate for
the error in $\Gamma_{Z^\prime}/\Gamma_{\gamma}$ ($\sim 0.2 \times 10^{-6}$)
following the quoted uncertainties in~\cite{Krasznahorkay:2018snd}.
In Fig.~\ref{fig:lim}, the bottom panels illustrate
the relevant parameter space for the isospin conserving and isospin
violating limits (respectively left and right plots).

To summarise, it is clearly important to further improve
the estimation of nuclear isospin violation, and perform
more accurate fits for the null result of IPC in $^8 \text{Be}^{*'}$
(in addition to the currently available fits for the
predominantly isosinglet $^8$Be excited state). This will allow
determining the ranges for the bounds on the relevant quark (nucleon)
couplings of the $Z^\prime$ necessary to explain the anomalous IPC in
$^8$Be.
However, in view of the guesstimates here mentioned, in the subsequent
numerical analysis we will adopt conservative ranges for
different couplings (always under the assumption $\text{BR}(Z^\prime \to e^+
e^-)\simeq 1$),
\begin{eqnarray}
| \varepsilon^V_n | &\;=\;& (2-15) \times 10^{-3}\, ,
\label{eqn:epsn}
\\
| \varepsilon^V_p | &\;\lesssim \; & 1.2 \times 10^{-3}\, .
\label{eqn:epsp}
\end{eqnarray}

\section{Phenomenological constraints on neutral (vector and axial) couplings}
\label{sec:phenocon}

If, and as discussed in the previous section,
the new couplings of fermions to the light $Z^\prime$ must satisfy
several requirements to explain the anomalous IPC in $^8$Be, there are
extensive constraints arising from various experiments, both regarding its
leptonic and hadronic couplings. In this section, we collect the
most important ones, casting them in a model-independent way,  
and subsequently summarising the results of
the new fit carried for the case of light Majorana 
neutrinos (which is the case in the model under consideration).

\mathversion{bold}
\subsection{Experimental constraints on a light $Z^\prime$ boson}
\mathversion{normal}
\label{sec:phenocon:Zprime} 
The most relevant constraints arise
from negative $Z^\prime$ searches in beam dump
experiments, dark photon bremsstrahlung and production,
experiments measuring atomic parity violation, and neutrino-electron scattering.

\paragraph{Searches for $\pmb{Z^\prime}$ in electron beam dump experiments}

\noindent
The non-observation of a $Z^\prime$ in experiments such
as SLAC E141, Orsay and NA64~\cite{Banerjee:2018vgk, Banerjee:2019hmi}, as well as 
searches for dark photon bremsstrahlung from electron and nuclei
scattering, can be interpreted in a two-fold way:
(i) absence of $Z^\prime$ production due to excessively feeble couplings;
(ii) excessively rapid $Z^\prime$ decay, occurring even prior to the dump.
Under assumption (i) (i.e. negligible production), one finds the following bounds
\begin{equation}
	{\varepsilon^V_{ee}}^2 + {\varepsilon^A_{ee}}^2 < 1.1\times10^{-16}\,,
\end{equation}
while (ii) (corresponding to fast decay) leads to 
\begin{equation}
	\sqrt{{|\varepsilon^V_{ee}|}^2 + {|\varepsilon^A_{ee}|}^2}
	 \gtrsim 6.8\times10^{-4}\: \sqrt{\text{BR}(Z^\prime\to e^+e^-)}\,.
	 \label{eqn:dump}
\end{equation}

\paragraph{Searches for dark photon production}

\noindent
A bound can also be obtained from the KLOE-2 experiment, which has
searched for $e^+ e^- \to X \gamma$, followed by the decay $X \to e^+
e^-$~\cite{Anastasi:2015qla}, leading to 
\begin{align}\label{eq:4.6}
 {\varepsilon^V_{ee}}^2+{\varepsilon^A_{ee}}^2   < \frac{4 \times
   10^{-6}}{{\text{BR}(Z^\prime\to e^+e^-)}}\,. 
\end{align}
Similar searches were also performed at BaBar, although the latter
were only sensitive to slightly heavier candidates, with
masses $m_X > 20$~MeV~\cite{Lees:2014xha}.

\paragraph{Light meson decays}

\noindent
In addition to the (direct) requirements that an explanation of the $^8$Be anomaly
imposes on the couplings of the $Z^\prime$ to quarks - already discussed in
Section~\ref{sec:IPCcon}-, important constraints on the latter arise
from several light meson decay experiments. 
For instance, this is the case of 
searches for $\pi^0 \to \gamma Z^\prime(Z^\prime\to ee)$ and $K^+ \to \pi^+ Z^\prime(Z^\prime\to ee)$
at the NA48/2~\cite{Raggi:2015noa} experiment,
as well as searches for 
$\phi^+ \to \eta^+ Z^\prime(Z^\prime\to
ee)$ at KLOE-2~\cite{Anastasi:2015qla}.
Currently, the most stringent constraint does arise
from the rare pion decays searches which lead, for
$M_{Z^\prime} \simeq 17$ MeV~\cite{Raggi:2015noa}, to the following bound 
\begin{align}
\vert 2 \varepsilon^V_{uu}  + \varepsilon^V_{dd}  \vert \,=\,\vert
\varepsilon^V_{p}  \vert \,\lesssim \,\frac{1.2 \times
  10^{-3}}{\sqrt{\textrm{BR}(Z^\prime \to e^+ e^-)}}\, . 
\end{align}
If one confronts the range for $|\varepsilon^V_p + \varepsilon^V_n|$
required to explain the anomalous IPC in $^8$Be (see
Eq.~\eqref{eq:4.4}), with the comparatively small allowed regime for
$\vert \varepsilon^V_{p}  \vert$  from the above equation, it is
clear that in order to explain the anomaly in $^8$Be the neutron
coupling $\varepsilon^V_n$ must be sizeable 
(This enhancement of neutron couplings (or suppression of the proton
ones) is also often referred to as
a ``protophobic scenario'' in the literature). Further (subdominant)
bounds can also be obtained from 
neutron-lead scattering, proton fixed target experiments and other
hadron decays, but  we will not take them into account in the present study

\paragraph{Constraints arising from parity-violating experiments}

\noindent
Very important constraints on the product of vector and axial
couplings of the $Z^\prime$ to electrons arise from the parity-violating
M{\o}ller scattering,
measured at the SLAC E158 experiment~\cite{Anthony:2005pm}.
For $M_{Z^\prime} \simeq 17$ MeV, it yields~\cite{Kahn:2016vjr}
\begin{align}
\vert \varepsilon^V_{ee}  \varepsilon^A_{ee} \vert \lesssim 1.1 \times 10^{-7}.
\end{align}
Further useful constraints on a light $Z^\prime$ couplings can be
inferred from atomic parity violation in Caesium, in particular 
from the measurement of the effective weak charge of the
Cs atom~\cite{Davoudiasl:2012ag,Bouchiat:2004sp, Roberts:2014bka,Dzuba:2012kx}. At the $2\sigma$
level~\cite{Porsev:2009pr}, these yield
\begin{align}\label{eqn:parity}
|\Delta Q_w|\, =\, |\frac{2\sqrt{2}}{G_F}\, 4\pi\alpha\,
\varepsilon^A_{ee} \left[\varepsilon^V_{uu} (2 Z + N) +
  \varepsilon^V_{dd} (Z + 2 N) \right] \left( \frac{\mathcal
  K(M_{Z^\prime})}{M_{Z^\prime}^2}\right)| \,\lesssim \,0.71\,, 
\end{align}
in which $\mathcal K$ is an atomic form factor, with
$\mathcal K(17 \,\mathrm{MeV})\simeq 0.8$~\cite{Bouchiat:2004sp}. 
For the anomalous IPC favoured values of $\varepsilon^V_{uu(dd)}$, 
the effective weak charge of the
Cs atom measurement\footnote{There are also measurements of the effective weak charge of other fermions, notably of the proton which was performed by the Qweak experiment~\cite{Androic:2018kni}. The bound which can be inferred from the result obtained by Qweak is however an order of magnitude weaker than the one from the Caesium measurement. For a new measurement of the effective weak charge of the electron, the MOLLER experiment~\cite{Benesch:2014bas} was proposed with an anticipated relative uncertainty of $2.4\%$, which would lead to a bound on the axial coupling to electrons comparable to the one from the Caesium measurement.} provides a very strong constraint on
$|\varepsilon^A_{ee}|$, $|\varepsilon^A_{ee}| \lesssim 2.6 \times
10^{-9}$,
which is particularly relevant for the present scenario, as it renders a
combined explanation of $(g-2)_e$ and the anomalous IPC particularly
challenging. As we will subsequently discuss, the constraints on
$|\varepsilon^A_{ee}|$ exclude a large region of the parameter space,
leading to a ``predictive'' scenario for the $Z^\prime$ couplings.

\noindent
Finally, neutrino--electron scattering provides stringent constraints
on the $Z^\prime$  neutrino
couplings~\cite{Bilmis:2015lja,Khan:2016uon,Lindner:2018kjo}, with the
tightest bounds arising from the TEXONO and CHARM-II experiments.
In particular, for the mass range $M_{Z^\prime} \simeq 17\:\mathrm{MeV}$,
the most stringent
bounds are in general due to the TEXONO
experiment~\cite{Deniz:2009mu}. While for some simple $Z^\prime$
constructions the couplings are flavour-universal, the extra fermion
content in our model leads to a decoupling of the lepton families in
such a way that only the couplings to electron neutrinos can be
constrained with the TEXONO data. 
For muon neutrinos, slightly weaker but nevertheless very
relevant bounds can be
obtained from the CHARM-II experiment~\cite{Vilain:1993kd}.

\subsection{Neutrino-electron scattering}
\label{sec:subsec:MajoranaFit} 
In the present model, neutrinos are Majorana particles, which implies
that the corresponding flavour conserving pure vector part of the
$Z^\prime$-couplings vanishes. The fits performed in
Refs.~\cite{Bilmis:2015lja,Khan:2016uon,Lindner:2018kjo} are thus not
directly applicable to the model under consideration; consequently we have performed
new two-dimensional fits to simultaneously constrain the axial
couplings to electron and muon neutrinos, and the vector coupling to
electrons,  following the prescription of Ref.~\cite{Lindner:2018kjo}.
The results of the fits first appeared in~\cite{Hati:2020fzp}.

In general, the differential cross section for neutrino and
antineutrino scattering can be computed as~\cite{Lindner:2018kjo}
\begin{eqnarray}
\frac{d\sigma}{dT}(\bar\nu e^- \to \bar\nu e^-) &=
\frac{m_e}{4\,\pi}\left[G_+^2 + G_-^2\left(1 - \frac{T}{E_\nu} \right)^2
  - G_+ G_- \frac{m_e \,T}{E_\nu^2} \right]\,\text,\\ 
\frac{d\sigma}{dT}(\nu e^- \to \nu e^-) &= \frac{m_e}{4\,\pi}\left[G_-^2
  + G_+^2\left(1 - \frac{T}{E_\nu} \right)^2 - G_+ G_- \frac{m_e\,
    T}{E_\nu^2} \right]\,\text, 
\end{eqnarray}
where $T$ is the recoil energy of the electron and $E_\nu$ the energy
of the (anti)neutrino. 
The coefficients $G_{\pm}$ are defined as
\begin{eqnarray}
G_{\pm} \,=\, \sum_{i = W, Z, Z^\prime} \frac{1}{P_i}\,
(g_{V_i}^{\nu\nu} - g_{A_i}^{\nu\nu})\,(g_{V_i}^{ee} \pm g_{A_i}^{ee})\,\text.
\end{eqnarray}
In the above, the sum runs over all relevant vector bosons (i.e.
$W$, $Z$ and $Z^\prime$), with $P_i$ denoting the denominator 
of the corresponding propagators;  $g_{V_i}$ and $g_{A_i}$ correspond
to the vector and axial couplings of the involved vector bosons to
(anti)neutrinos and electrons.
Since the energy of the neutrinos is well below the masses 
of the relevant gauge bosons, we will work with the following approximations
\begin{eqnarray}
P_W \approx -\frac{\sqrt{2}\,g^2}{8\,G_F}\,\text,\quad 
P_Z \approx -\frac{\sqrt{2}\,g^2}{8 \,G_F \,c_w^2}\,\text,\quad 
P_{Z^\prime} \sim -(2 \,m_e \,T +M_{Z^\prime}^2)\,\text. 
\end{eqnarray}
For the case of the model under study, 
the vector and axial coefficients are given by
\begin{eqnarray}
	g_{V_W} &=& - g_{A_W} = \frac{g}{2\,\sqrt{2}}\quad\text{(for
          both $\nu$ and $e$),}\\ 
	g_{A_Z}^{\nu\nu} &=& -\frac{g}{2\,c_w}\,\text,\\
	g_{V_Z}^{ee} &=& - \frac{g\,(1 - 4\,s_w^2)}{4\,c_w}\,\text,\\
	g_{A_Z}^{ee} &=& \frac{g}{4\,c_w}\,\text,\\
	g_{V_{Z^\prime}}^{ee} &=& e \varepsilon_{ee}^V\,\text,\\
	g_{A_{Z^\prime}}^{\nu\nu} &=&2\, e \,\varepsilon_{\nu\nu}^A\,\text,
\end{eqnarray}
with all other remaining coefficients vanishing.
In order to take into account the fact that for Majorana neutrinos
the $\nu$ and $\bar\nu$ final  states are indistinguishable, 
a factor of 2 is present in the (axial) neutrino coefficients
(effectively allowing to double the contributions from amplitudes
involving two neutrino operators~\cite{Rosen:1982pj}). 
As argued earlier, the axial coupling to electrons has to be
negligibly
small in order to comply with constraints from atomic parity violation
and, for practical purposes, these will henceforth be set to zero in our analyses.

\medskip
\paragraph{Data from the CHARM-II experiment}

\noindent
To fit the data from the CHARM-II experiment (extracted from 
Table 2 of~\cite{Vilain:1993kd}), one can directly compare the
differential cross-section, averaged over the binned recoil energy $T$,
with the data. 
For neutrinos and antineutrinos, the average energies are 
$\braket{E_{\nu_\mu}} = 23.7\,\mathrm{GeV}$ and
$\braket{E_{\bar\nu_\mu}} = 19.1\,\mathrm{GeV}$, respectively.
Since no data correlation from the CHARM-II samples is available, 
we assume all data to follow a gaussian distribution, 
and accordingly define the $\chi^2$ function 
\begin{equation}
\chi^2_\text{CHARM-II} \,=  \,\sum_{i}\left(\frac{\sigma_i - \sigma_{i,
    \text{exp}}}{\Delta \sigma_{i, \text{exp}}} \right)^2\,\text, 
\end{equation}
where $i$ runs over the different bins.
The $\chi^2$-function is minimised, and its $1\sigma$ and $2 \sigma$ contours around
the minimum are computed.

\begin{figure}[h!]
    \hspace*{-10mm}\mbox{\includegraphics[width=0.55\textwidth]{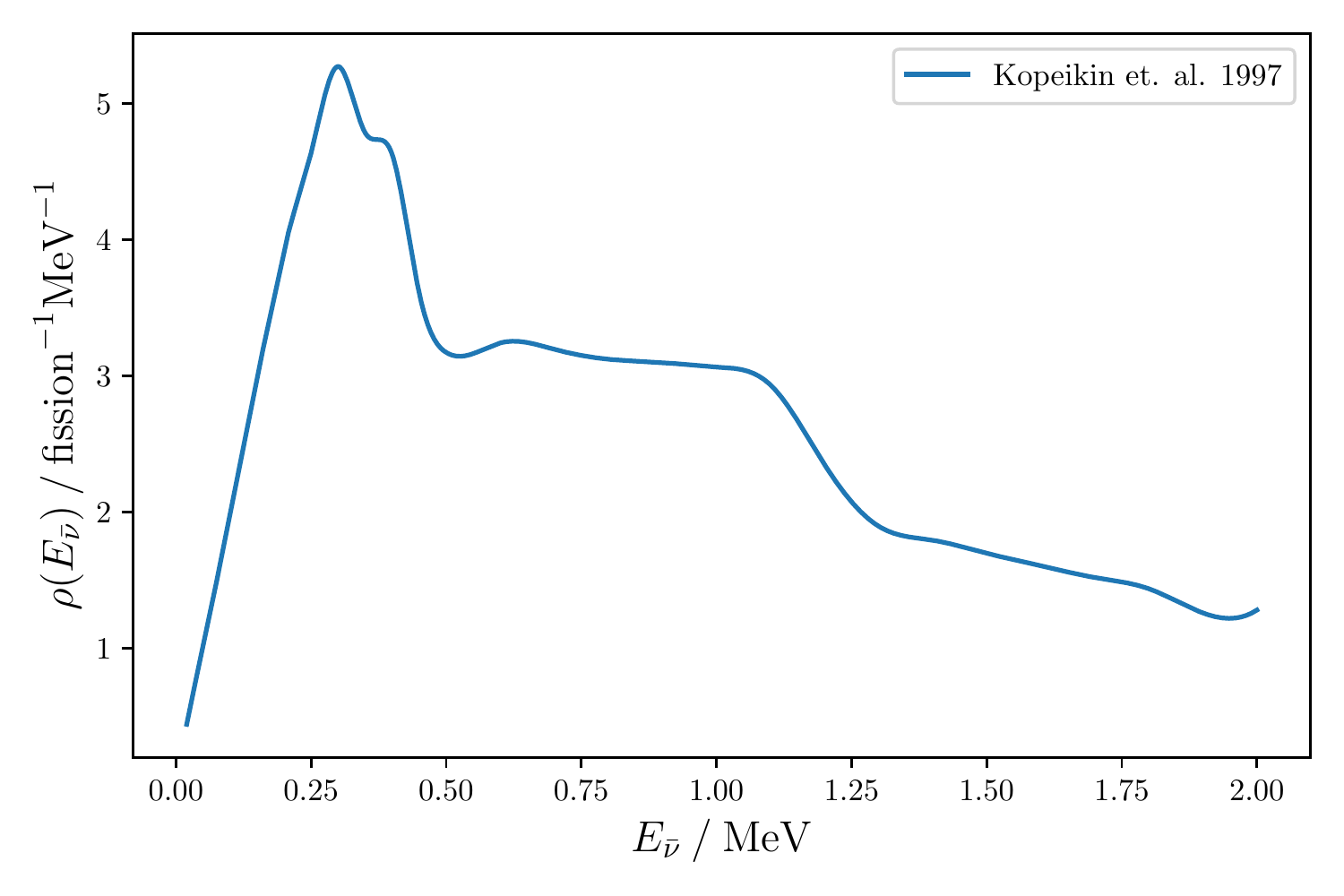}
    \includegraphics[width=0.55\textwidth]{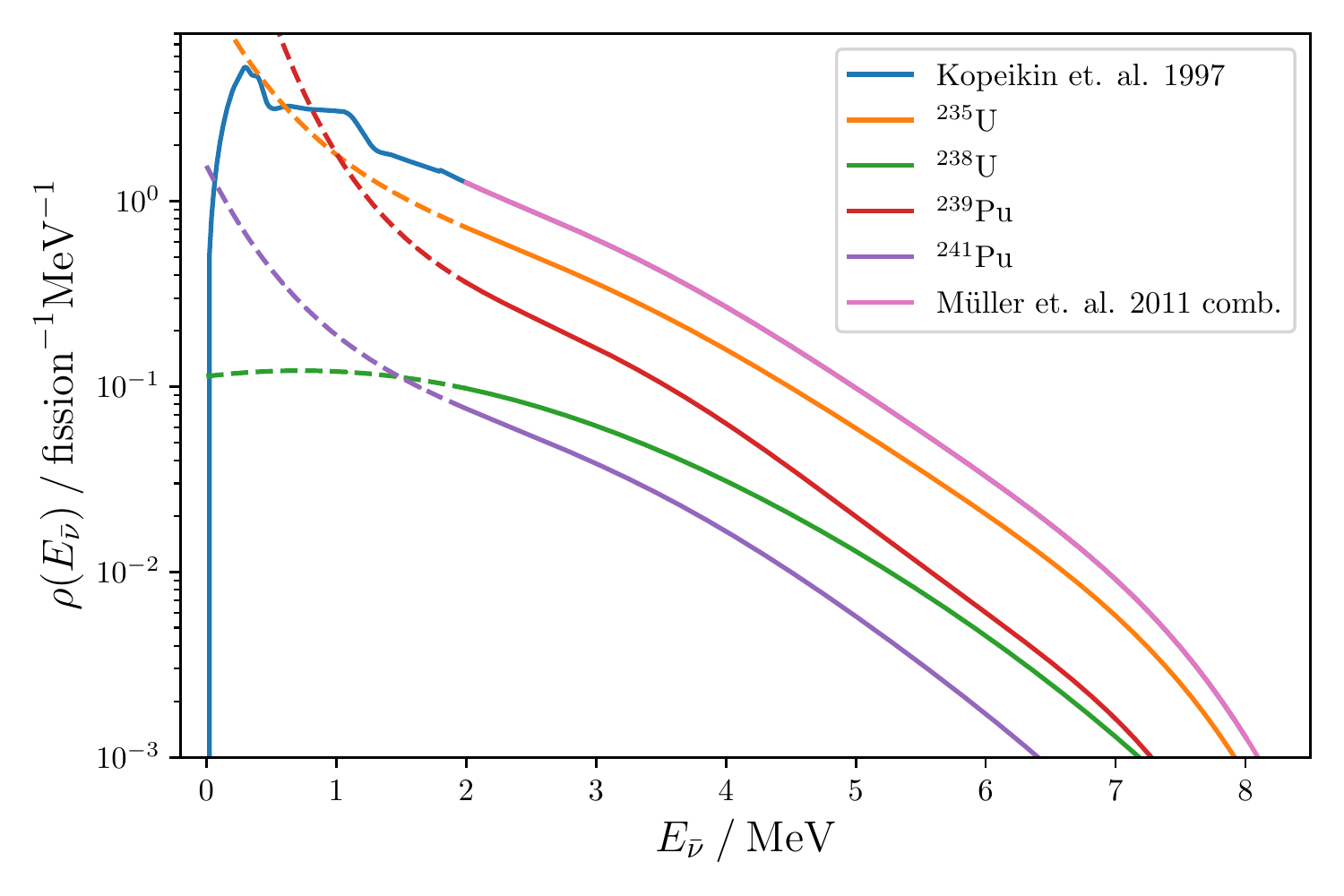}}
    \caption{Neutrino spectra at a power reactor with the standard fuel composition. \textbf{Left}: Low energy part, interpolated from the results obtained in~\cite{Kopeikin:1997ve}.
    \textbf{Right}: High energy part from the fit of~\cite{Mueller:2011nm}. The dashed lines indicate the range in which the parametrisation of~\cite{Mueller:2011nm} breaks down and the low energy part of slow neutron capture is valid.}
    \label{fig:reactorspectra}
\end{figure}

\medskip
\paragraph{Data from the TEXONO experiment}
The analysis of the TEXONO data~\cite{Deniz:2009mu} is comparatively more
involved than that of CHARM-II. 
Since TEXONO is a reactor experiment, the computation of the binned
event rate requires knowledge of the reactor anti-neutrino flux.
Following the approach of~\cite{Lindner:2018kjo}, 
the event rate can be computed as
\begin{equation}
R(T_1, T_2) \,= \,
\frac{\rho_e}{T_2 - T_1}\int\phi(E_{\bar\nu})\left[\int_{\bar
    T_1}^{\bar T_2}\frac{d\sigma}{dT} dT \right] d E_{\bar\nu}\,\text, 
\end{equation}
in which $T_{1,2}$ are the bin edges for the electron's recoil energy, 
$\phi(E_{\bar\nu})$ is the neutrino flux, 
$\rho_e$ the electron density of the target material and 
$\bar T_{1,2} = \mathrm{min}(T_{1,2}, T_\text{max})$;
the maximum recoil energy $T_\text{max}$ can be defined as
\begin{equation}
T_\text{max} \,= \,\frac{2E_{\bar\nu}^2}{M + 2 \,E_{\bar\nu}}\,\text.
\end{equation}
The (anti)neutrino flux is given by~\cite{Kopeikin:2004cn}
\begin{equation}
\phi(E_{\bar\nu}) = \frac{1}{4\,\pi\, R^2}\frac{W_{\text{th}}}{\sum_i f_i
  E_{f,i}}\left(\sum_i f_i \rho_i(E_{\bar\nu}) \right)\,\text, 
\end{equation}
in which the sums run over the reactor fuel constituents $i$; for each
of the latter, $f_i$ is the fission rate, 
$E_{f,i}$ the fission energy and $\rho_{i}(E_{\bar\nu})$ the neutrino
spectrum. The remaining intrinsic parameters are 
$W_\text{th}$ - the total thermal energy of the reactor, 
and $R$ which corresponds to the distance between reactor and detector
(details of the reactor and general experimental set-up can be found
in~\cite{Wong:2006nx}). 
In what concerns the neutrino spectra, and 
depending on the different reactor fuel constituents, we use 
the fit of~\cite{Mueller:2011nm}, in which 
spectra between $2-8\,\mathrm{MeV}$ are parametrised by 
the exponential of a fifth degree polynomial.
The lower energy part of
the spectrum, which is governed by slow neutron capture, 
has been obtained in~\cite{Kopeikin:1997ve}, 
and is given in the form of numerical results 
for the approximate standard fuel composition of pressurised water
reactors ($\sim 55\%\: {^{235}\mathrm{U}}\,,\:\sim 7\%\: {^{238}\mathrm{U}}\,,\:\sim 32\%\: {^{239}\mathrm{Pu}}\,,\:\sim 6\%\: {^{241}\mathrm{Pu}}$).
Although the low energy part is not immediately relevant for our study (since the 
TEXONO data consists of 10 equidistant bins between $3-8\,\mathrm{MeV}$), both parts of the spectrum are shown in Fig.~\ref{fig:reactorspectra}.

\noindent
We have thus obtained the electron density of the detector material 
$\rho_e$ of the TEXONO experiment by fitting the SM expectation of the 
binned event rate to the SM curve given in Fig.~16 of~\cite{Deniz:2009mu}. Our result is as follows
\begin{equation}
\rho_e \,\simeq \,2.77\times10^{26}\:\mathrm{kg}^{-1}\,\text.
\end{equation}
Finally, and to define the $\chi^2$ function for the TEXONO experiment
data, we again rely on the experimental data Fig.~16 of 
Ref.~\cite{Deniz:2009mu}, leading to
\begin{equation}
\chi^2_\text{TEXONO} \,= \,\sum_{i}\left(\frac{R_i - R_{i,
    \text{exp}}}{\Delta R_{i, \text{exp}}} \right)^2\,\text, 
\end{equation}
where $i$ counts the different bins in the recoil energy.

\begin{figure}
\centering
\includegraphics[width=0.7\textwidth]{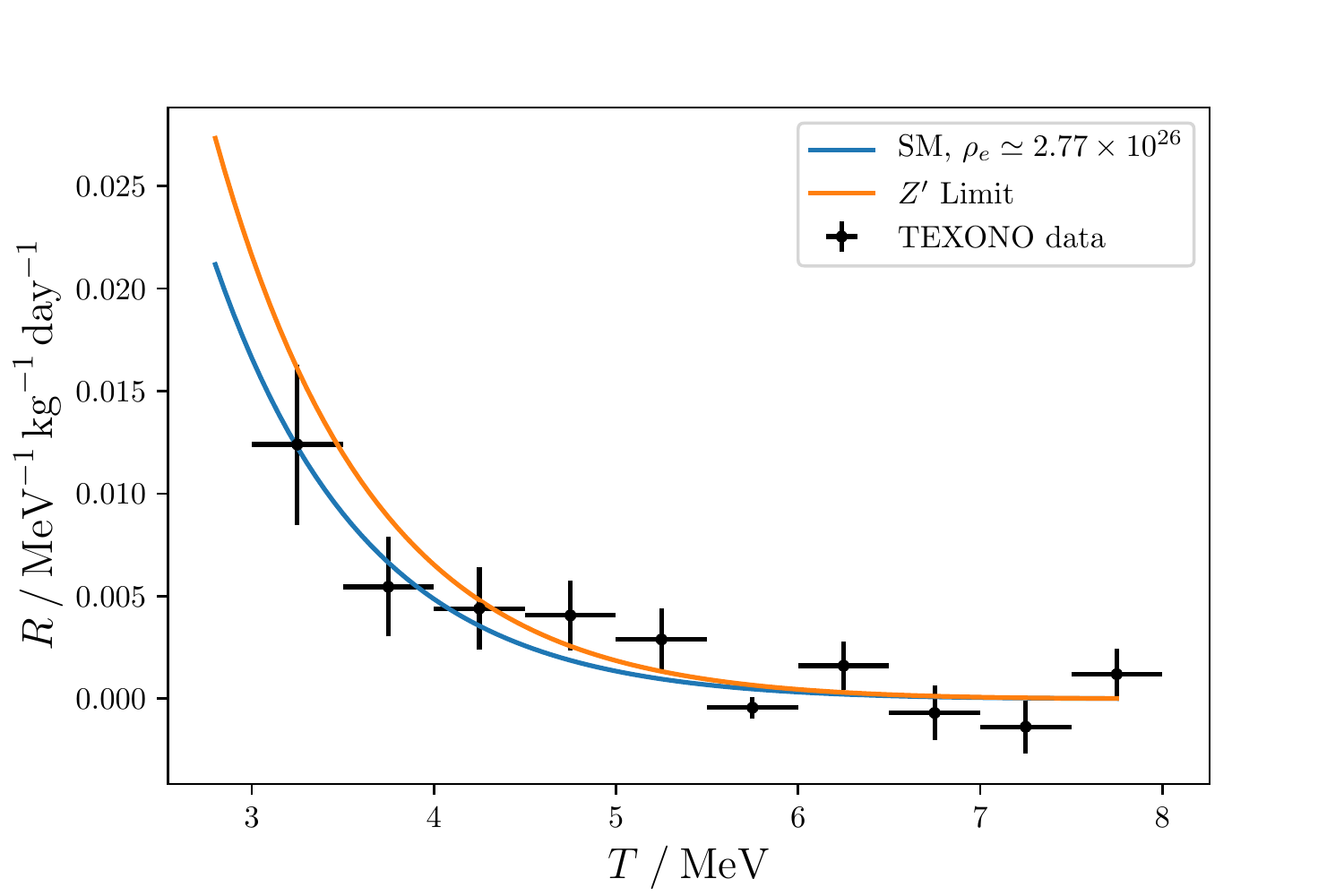}
\caption{Data of the TEXONO experiment (neutrino rate $R$ in units of 
$\mathrm{MeV}^{-1}\,\mathrm{kg}^{-1}\,\mathrm{day}^{-1}$ as a function
  of the binned recoil energy $T$)~\cite{Deniz:2009mu}, to which we superimpose 
our SM and $Z^\prime$ predictions, respectively corresponding to blue and
orange lines. Figure taken from~\cite{Hati:2020fzp}.
}
\label{fig:TEXONO_data}
\end{figure}

In Fig.~\ref{fig:TEXONO_data} we display 
the experimental data obtained by TEXONO, together with the (fitted) 
SM curve as well as the $Z^\prime$ prediction. Leading to the 
$Z^\prime$ curve we have taken the minimum couplings to electrons allowed 
by NA64~\cite{Banerjee:2018vgk,Banerjee:2019hmi}, and the maximum values of the 
couplings to neutrinos as derived from the TEXONO data~\cite{Deniz:2009mu}.

\begin{figure}
\centering
\includegraphics[width=0.7\textwidth]{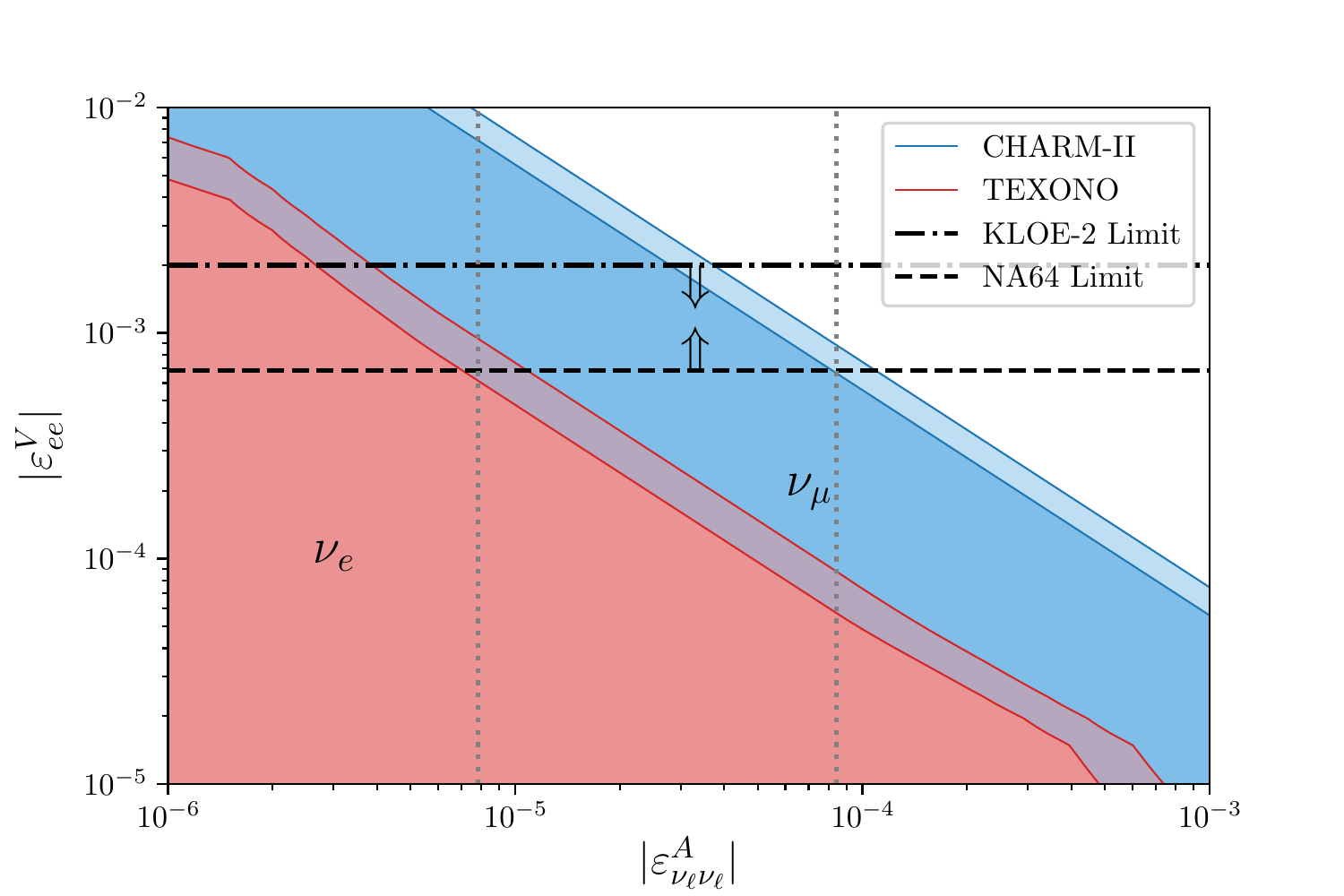}
\caption{New $\chi^2$-fit of the $\bar\nu_e\:e$ scattering data of TEXONO
  (red) and the $\bar\nu_\mu\:e$ scattering data of CHARM-II
  (blue), displaying the 1- and 2-$\sigma$ allowed regions around the best fit
  point (respectively darker and lighter colours).
 The lower bound of NA64 (dashed line) and
  the upper bound by KLOE-2 (dash-dotted line) are also shown, with
  the arrows identifying the viable allowed regions. The obtained upper
  limits on the axial coupling to neutrinos,
  cf. Eq~\eqref{eqn:nu_scat_lim}, are marked by dotted lines:
  the TEXONO data mostly constrains the couplings to electron neutrinos while
  the CHARM-II data is responsible for the constraints on 
  the couplings to muon neutrinos. Figure taken from~\cite{Hati:2020fzp}.
}
\label{fig:nu_scat}
\end{figure}

The particular likelihood contours
deviating $1$ and $2\,\sigma$ from the best fit point for the
neutrino--electron scattering data (which is found to lie very close
to the SM prediction), are shown in Fig.~\ref{fig:nu_scat}. 
Applying the constraints on the electron vector coupling
$\varepsilon^V_{ee}$ obtained from NA64~\cite{Banerjee:2018vgk,Banerjee:2019hmi} and KLOE-2~\cite{Anastasi:2015qla} leads to the
limits
\begin{eqnarray}\label{eq:nuelim}
	|\varepsilon_{\nu_e\nu_e}^A| &\lesssim& 7.8\times
	10^{-6}\,\text,\nonumber\\ 
	|\varepsilon_{\nu_\mu\nu_\mu}^A| &\lesssim& 8.4\times 10^{-5}\,\text,
	\label{eqn:nu_scat_lim}
\end{eqnarray}
leading to which we have assumed
the smallest allowed electron coupling $|\varepsilon_{ee}^V| \sim
6.8\times 10^{-4}$.
Note that interference effects between the charged and neutral
currents (as discussed in~\cite{Feng:2016ysn,Bilmis:2015lja,Khan:2016uon,Lindner:2018kjo})
do not play an important role in this scenario, due to vanishing
neutrino vector couplings. 

\bigskip
To conclude the discussion, we list below a summary of the relevant
constraints so far inferred on the couplings of the $Z^\prime$ to matter: combining the required ranges of couplings needed to explain the
anomalous IPC signal with the relevant bounds from other experiments,
we have established the following ranges for the couplings (assuming
$\text{BR}(Z^\prime \to e^+ e^-) = 1$),
\begin{eqnarray}
2\times 10^{-3} \;\lesssim\;| \varepsilon^V_n | &\;\lesssim\;& 15
\times 10^{-3}\, , 
\\
| \varepsilon^V_p | &\;\lesssim \; & 1.2 \times 10^{-3}\, ,
\\
0.68\times 10^{-3} \;\lesssim\;| \varepsilon^V_{ee} | &\;\lesssim\;&  2
\times 10^{-3}\, , 
\label{eq:lepton:allowed:couplings1}
\\
| \varepsilon^A_{ee} | &\;\lesssim\;& 2.6 \times 10^{-9}\, ,\\
|\varepsilon^A_{\nu_e\nu_e}| &\lesssim& 7.8\times 10^{-6}\, ,\\
|\varepsilon^A_{\nu_\mu\nu_\mu}| &\lesssim& 8.4\times 10^{-5}\, .
\end{eqnarray}

\mathversion{bold}
\section{Addressing the anomalous IPC in $^8$Be: impact for a combined
explanation of $(g-2)_{e,\mu}$}
\mathversion{normal}
\label{sec:combined:explanation}

As a first step, we apply the previously obtained model-independent
constraints on the $Z^\prime$ couplings to the specific structure of
the present model. After taking the results of (negative) collider
searches for the exotic matter fields into account, we will be able to infer
an extremely tight range for $\varepsilon$ (which we recall to
correspond to a redefinition of the effective kinetic mixing
parameter, cf. Eq.~(\ref{eq:epsilon:redefine})). In turn, this will
imply that very little freedom is left to explain the experimental
discrepancies in the light charged lepton anomalous
magnetic moments, the latter requiring an interplay of the $h_{\ell \ell}$
and $k_{\ell \ell}$ couplings.

\subsection{Constraining the model's 
parameters}\label{subsec:model-constraint}
The primary requirements  to 
explain the anomalous IPC in $^8$Be concern 
the physical mass of the $Z^\prime$, which should approximately be  
\begin{equation}  \label{eq:7.1}
  M_{Z^\prime} \,\approx 17\,\mathrm{MeV}\,,
\end{equation}
and the strength of its couplings to nucleons (protons and neutrons),
as given in Eqs.~(\ref{eqn:epsn}, \ref{eqn:epsp}). With 
$\varepsilon^V_{qq}$ as defined in Eq.~\eqref{eq:epsq},
and recalling that
$\varepsilon_p = 2\,\varepsilon^V_{uu} + \varepsilon^V_{dd}$ and
$\varepsilon_n = \varepsilon^V_{uu} + 2 \,\varepsilon^V_{dd}$, one
obtains the following constraints on $\varepsilon_{B-L}$ and
$\varepsilon$
\begin{eqnarray}
  |\varepsilon_n^V| &=& |\varepsilon_{B-L}| \,=\,
  (2 - 15)\times 10^{-3}\,
  , \label{eq:7.2}\\ 
  |\varepsilon_p^V| &=& |\varepsilon + \varepsilon_{B-L}| \lesssim 1.2
  \times 10^{-3}\,. 
  \label{eq:7.3}
\end{eqnarray}
Furthermore, this implies an upper bound for the VEV of $h_X$,
$v_X \lesssim 14\:\mathrm{GeV}$, since
\begin{equation}
  M_{Z^\prime} \approx m_{B^\prime} \,=\, 2\, e\, \vert
  \varepsilon_{B-L}\vert \,
  v_X\,.
\end{equation}

In the absence of heavy vector-like leptons, there are no other
sources of mixing in the lepton section in addition to the PMNS.
This would imply that the effective couplings of the $Z^\prime$ to
neutrinos are identical to that of the neutron (up to a global sign),
that is
\begin{equation}  
\label{eq:7.4}
  \varepsilon_{\nu\nu}^A \, = \, \varepsilon_{B-L}\,, 
\end{equation}
which, in view of Eq.~\eqref{eq:7.2}, leads to
$ \varepsilon_{\nu\nu}^A = (2 - 15)\times 10^{-3}$. However,
the bounds of the TEXONO experiment~\cite{Deniz:2009mu} for 
neutrino-electron scattering (cf. Eq.~\eqref{eq:nuelim})
imply that for the minimal allowed electron coupling
$|\varepsilon_{ee}^V| \gtrsim 6.8\times 10^{-4}$ one requires 
\begin{equation}  \label{eq:7.5}
  |\varepsilon_{\nu\nu}^A|\,\lesssim\, 7.8 \,(84)\,\times\, 10^{-6}\,,
\end{equation}
for electron (muon) neutrinos. As can be inferred, this is 
in clear conflict with the values of $ \varepsilon_{\nu\nu}^A$
required to explain the $^8$Be anomalous IPC, which are $\mathcal{O}(10^{-3})$. 

\noindent
In order to circumvent this problem,
the effective $Z^\prime$ coupling to the SM-like neutrinos must be
suppressed. 
Here is where the role of the exotic fermions becomes crucial: the additional vector-like leptons open the possibility
of having new sources of mixing between the distinct species of
neutral leptons; the effective neutrino coupling derived in 
Section~\ref{section:newneutralcurrent}
allows to suppress the couplings by a factor
$\sim (1 - {\lambda_{L\,\alpha}^2 v_{X}^2/M_{L\,\alpha}^2})$ (see
Eq.~(\ref{eqn:nunu}), with $\alpha$ denoting SM flavours), hence implying
\begin{equation}  \label{eq:7.6}
  |1 - \frac{\lambda_L^2 v_X^2}{M_L^2}| \,\lesssim 0.01\,.
\end{equation}
Thus, up to a very good approximation, we can assume 
$\lambda_L v_X\simeq M_L$ for each lepton generation $\alpha$.   
On the other hand, from Eqs.~\eqref{eq:epsl} and~\eqref{eqn:parity} it
follows that the bound from atomic parity violation in Caesium tightly
constrains the isosinglet vector-like lepton coupling
$\lambda_E$ (for the first lepton generation)\footnote{In what
  follows, we will not explicitly include the flavour indices, as it
  would render the notation too cumbersome, but rather describe it in
  the text.}, leading to 
\begin{eqnarray}
  |\varepsilon_{ee}^A| \,=  \,\left|\frac{1}{2}\left(\frac{\lambda_E^2 \,
    v_X^2}{M_E^2} - \frac{\lambda_L^2 \,
    v_X^2}{M_L^2}\right)\varepsilon_{B-L}\right| \,\lesssim \,
  2.6\times10^{-9}\,\text, 
  \label{eqn:limit_axial_0}
\end{eqnarray}
which in turn implies 
\begin{eqnarray}
  \left|\frac{\lambda_E^2  \,v_X^2}{M_E^2}
  - \frac{\lambda_L^2  \,v_X^2}{M_L^2}\right|
   \,\lesssim 2.6\times 10^{-6}\,\text.
  \label{eqn:limit_axial}
\end{eqnarray}
Notice that this leads to a tight correlation between the
isosinglet and isodoublet vector-like lepton couplings,
$\lambda_E$ and $\lambda_L$, respectively.
More importantly, the above discussion renders manifest the
necessity of having the additional field content (a minimum of two
generations of heavy vector-like leptons).

Together with Eqs.~\eqref{eqn:eeL} and~\eqref{eqn:eeR},
Eqs.~\eqref{eq:7.6} and~\eqref{eqn:limit_axial}
suggest that the $Z^\prime$ coupling to electrons is now almost solely
determined by $\varepsilon$.
In particular, the KLOE-2~\cite{Anastasi:2015qla} limit of Eq.~\eqref{eq:4.6} for
$\varepsilon_{ee}$ now implies
\begin{eqnarray}
  |\varepsilon|\,< \,0.002\,.
  \label{eqn:limit_eps}
\end{eqnarray}

Further important constraints on the model's parameters arise from the 
masses of the vector-like leptons, which are bounded from both below
and above. On the one hand, 
the perturbativity limit of the couplings $\lambda_L$ and $\lambda_E$
implies an upper bound on the vector-like lepton masses.
On the other hand, direct searches for vector-like leptons exclude
vector-like lepton masses below
$\sim 100\,\mathrm{GeV}$~\cite{Achard:2001qw}
(under the assumption these dominantly decay into $W\nu$ pairs).
This bound can be relaxed if other decay modes exist, for instance
involving the $Z^\prime$ and $h_X$ as is the case in our scenario.
However, and given the similar decay and production mechanisms,
a more interesting possibility is to recast the results of LHC
dedicated searches for SUSY searches, in particular for sleptons (superpartners of leptons decaying into a neutralino and a charged SM lepton) for the case of
vector-like leptons decaying into $h_X$ and a charged SM lepton. 
Taking into account
the fact that the vector-like lepton's cross section is
a few times larger than the selectron's or
smuon's~\cite{Feng:2016ysn,Khachatryan:2014qwa}, one can roughly
estimate that vector-like leptons with a mass $\sim 100\,\mathrm{GeV}$
can decay into a charged lepton and an $h_X$ with mass
$\sim~(50-70)\,\mathrm{GeV}$. 
Therefore, as a benchmark choice we fix the tree-level mass of the
vector-like leptons of all generations to $M_L = M_E = 90\,\mathrm{GeV}$
(which yields a physical mass $\sim110\,\mathrm{GeV}$, once the
corrections due to mixing effects are taken into account). In turn,
this implies that the couplings $\lambda_{L,E}$ should be sizeable 
$\lambda_E^e \approx \lambda_L^e \sim 6.4$ (for the first
generation, due to the very stringent parity violation
constraints)\footnote{Couplings so close to the perturbativity limit of
  $\mathcal{O}(4\pi)$ can potentially lead to Landau poles at
  high-energies, as a consequence of running effects. To avoid this,
  the low-scale model here proposed should be embedded into an
  ultra-violet complete framework.}, while for the second generation
one only has $\lambda_L^\mu \sim 6.4$. 
(We notice that
smaller couplings, complying
with all imposed constraints can still be accommodated, at the price
of extending the particle content to include additional exotic fermion
states.)
In agreement with the the above discussion, we further 
choose $m_{h_X} = 70 \,\mathrm{GeV}$ as a benchmark value. Since
$h_X$ can also decay into two right handed neutrinos (modulo a
substantially large Majorana coupling $y_M$), leading to a signature
strongly resembling that of slepton pair production, current negative
search results then lead to constraints on $\varepsilon_{B-L}$.
For the choice $m_{h_X} = 70 \,\mathrm{GeV}$, $\varepsilon_{B-L}$
should be close to its smallest allowed value $\varepsilon_{B-L} =
0.002$ ~\cite{Feng:2016ysn}, which in turn implies the following range
for $\varepsilon$
 \begin{eqnarray}
  -0.0032 \,\lesssim \,\varepsilon\, \lesssim\, -0.0008\,.
   \label{eqn:limit_eps_coll}
 \end{eqnarray}
The combination of the previous constraint with the one inferred from
the KLOE-2 limit on the couplings of the $Z^\prime$ to electrons,
see Eq.~\eqref{eqn:limit_eps}, allows to derive the viability range for
$\varepsilon$,  
\begin{eqnarray}
 -0.002 \,\lesssim \,\varepsilon \,\lesssim \,-0.0008\,.
  \label{eqn:limit_eps_coll2}
\end{eqnarray}

\bigskip

Before finally addressing the feasibility of a combined explanation to
the atomic $^8$Be and $(g-2)_{e,\mu}$ anomalies, let us notice
that in the study of
Ref.~\cite{Dror:2017nsg} the authors have derived significantly stronger
new constraints on the parameter space of new (light) vector states,
$X$, arising in $U(1)_X$ extensions of the SM, such as  $U(1)_{B-L}$
models. The new bounds can potentially disfavour some well-motivated
constructions, among which some aiming at addressing the $^8$Be
anomalies, and arise in general from an energy-enhanced emission
(production) of the longitudinal component ($X_L$) via anomalous
couplings\footnote{As discussed in~\cite{Dror:2017nsg}, 
such an enhancement can occur if the model's content is
such that a new set of heavy fermions with vector-like couplings to
the SM gauge bosons, but chiral couplings to $X$, is introduced to
cancel potentially dangerous chiral anomalies. Explicit Wess-Zumino
terms must be introduced to reinforce the SM gauge symmetry, which in
turn breaks the $U(1)_X$, leading to an energy-enhanced emission of
$X_L$. Moreover, the SM current that $X$ couples to may also be broken
at tree level, due to weak-isospin violation ($W \bar{u} d$ or 
$W\ell\bar{\nu}$ vertices may break $U (1)_X$, if $X$ has different
couplings to fermions belonging to a given $SU(2)_L$ doublet and
lacks the compensating coupling to the $W$).
In such a situation the
longitudinal $X$ radiation from charged current processes can be again
enhanced, leading to very tight constraints from 
$\pi\rightarrow e\nu_e +X$, or $W \to \ell \nu_\ell + X$. }.

\mathversion{bold}
\subsection{A combined explanation of $(g-2)_{e,\mu}$}
\mathversion{normal}
In view of the stringent constraints on the parameter space of the
model, imposed both from phenomenological arguments and from a
satisfactory explanation of the anomalous IPC in $^8$Be, one must now
consider whether it is still possible to account for the observed
tensions in the electron and muon anomalous magnetic moments. As
discussed in Section~\ref{sec:g-2}, the
discrepancies between SM prediction and experimental observation have
an opposite sign for electrons and muons, and exhibit a scaling
behaviour very different from the na\"ive expectation (powers of the
lepton mass ratio). 

Given the necessarily small mass of the $Z^\prime$ and the large couplings
between SM leptons and the heavier vector-like states
($\lambda_{L,E}$), in most of the parameter space
the new contributions to $(g-2)_{e,\mu}$ are
considerably larger than what is suggested from experimental
data. Firstly, recall that due to the opposite sign of the loop
functions for (axial) vector and 
(pseudo)scalar contributions, a cancellation between the latter
contributions allows for an overall suppression of each $(g-2)_{e,\mu}$. 
Moreover, a partial cancellation between the distinct diagrams can
lead to $\Delta a_\mu$ and $\Delta a_e$ with opposite signs; this
requires nevertheless a large axial coupling to electrons, which is 
experimentally excluded. However, an asymmetry in the
couplings of the SM charged leptons to the vector-like states
belonging to the same generation can overcome the problem,
generating a sizeable ``effective'' axial coefficient
$g_A^{\ell\ell}$: while for electrons Eq.~\eqref{eqn:limit_axial}
implies a strong relation between $\lambda_L$ and $\lambda_E$, the
(small) couplings $h_{\ell}$ and $k_{\ell}$ remain essentially 
unconstrained\footnote{Being diagonal in generation space, we
  henceforth denote the couplings via a single index, i.e.
  $h_{\ell} = h_{\ell\ell}$, etc., for simplicity.} and can induce
such an asymmetry, indeed leading to the desired ranges for the
anomalous magnetic moments. 

\begin{figure}
    \begin{center}
    \mbox{\hspace{-1cm}
      \includegraphics[width=0.55\textwidth]{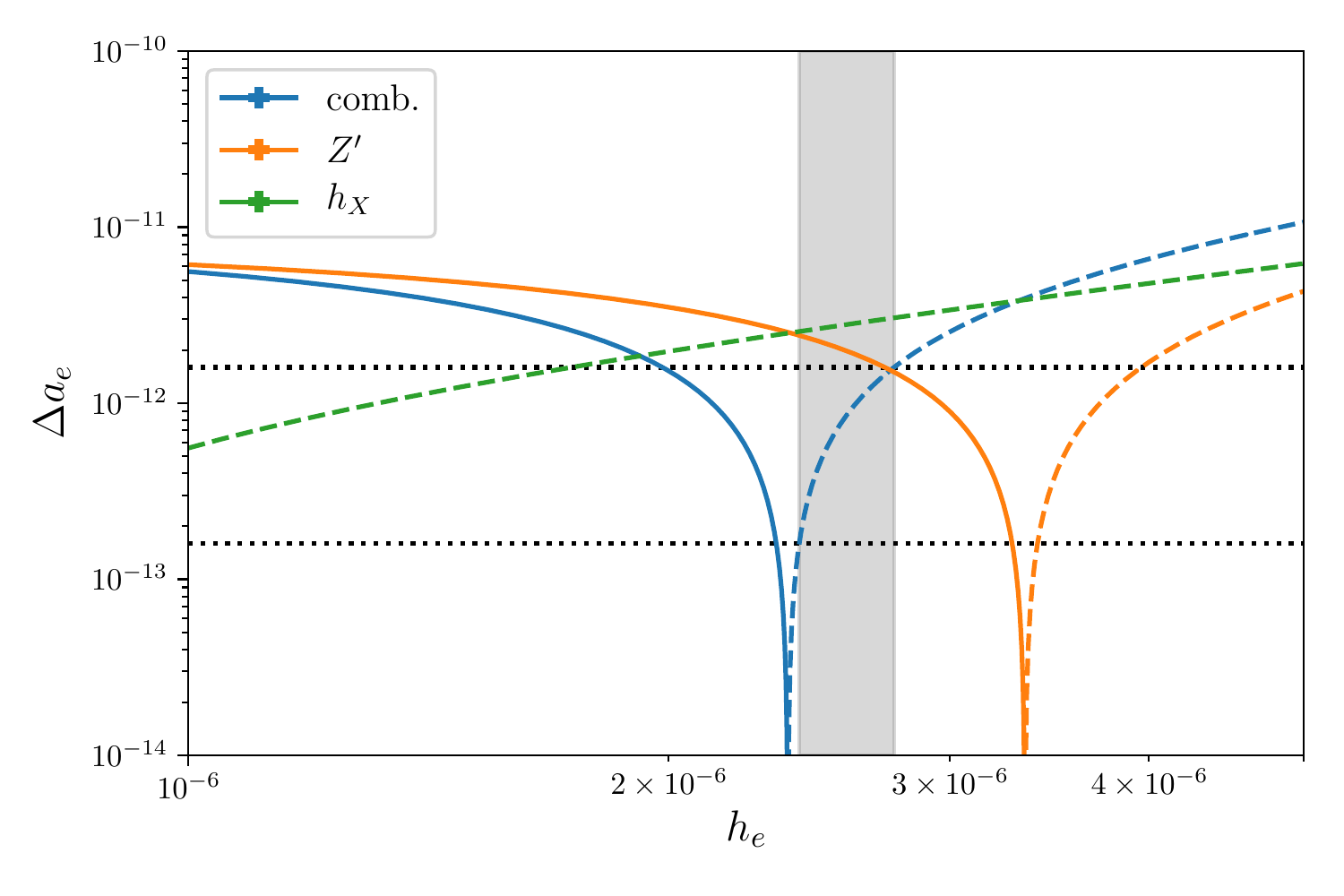} 
   	\includegraphics[ width=0.55\textwidth]{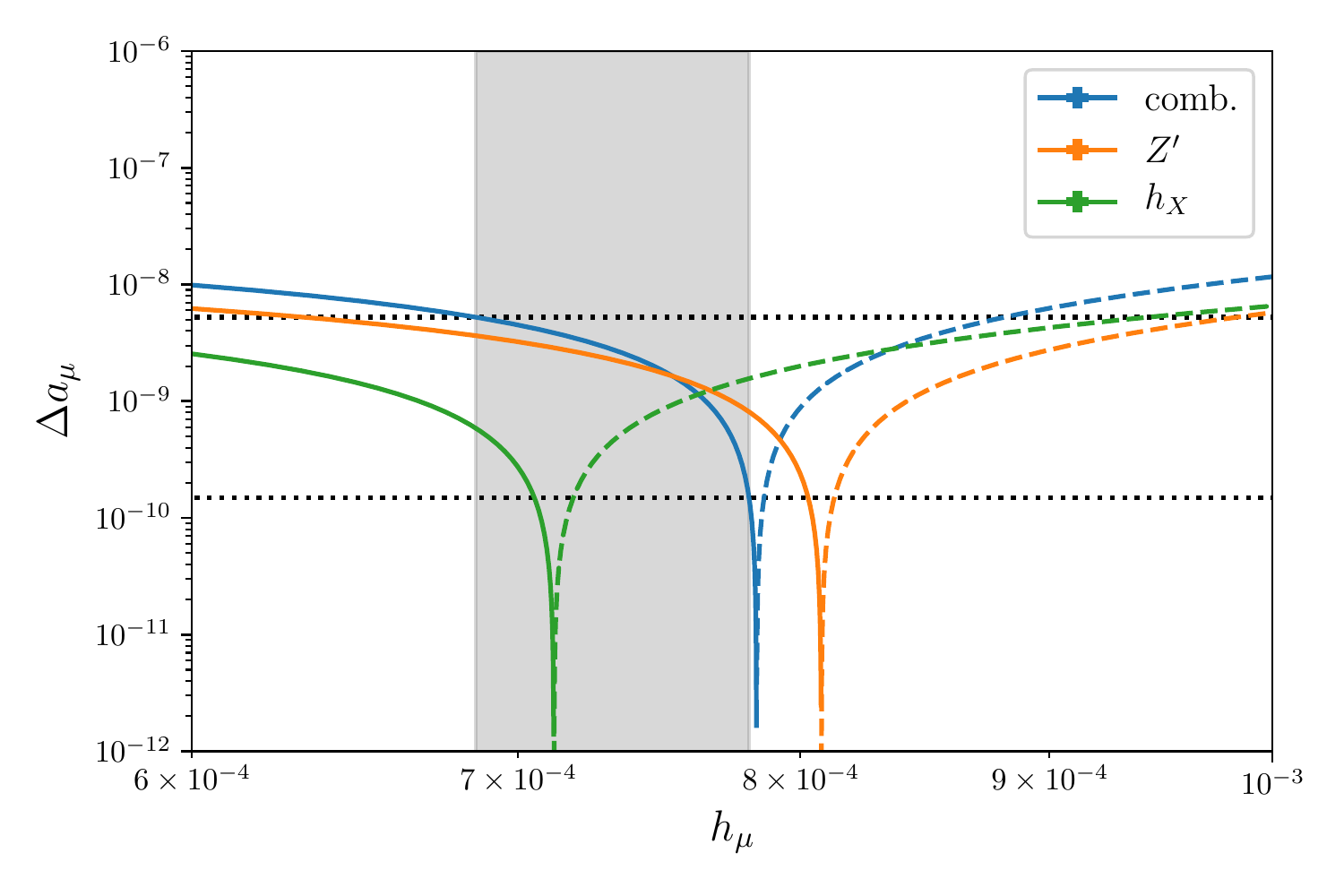}}
    \end{center}
    \caption{Contributions to the anomalous magnetic moment of charged
      leptons, $|\Delta a_\ell|$, as a function of the $h_{\ell}$
      coupling for $\ell=e$ (left) and  $\ell=\mu$ (right).
      Solid (dashed) lines correspond to positive (negative) values of 
      $\Delta a_\ell$; the colour code denotes contributions from the
      $Z^\prime$ (orange) and from $h_X$ (green), as well as the
      combined one (blue). Horizontal (dotted) lines denote the
      $2\sigma$ and $3\sigma$ regions of the electron and muon
      $\Delta a_\ell$. A vertical opaque region corresponds to the
      $h_{\ell}$ interval for which the
      combined contributions to $\Delta a_{e (\mu)}$
      lie within the $2\sigma$ ($3\sigma$) region.  
      Leading to this figure, we have selected a benchmark choice of
      parameters complying with all the constraints mentioned in
      Section~\ref{sec:phenocon}: $M_L = M_E = 90\,\mathrm{GeV}$,
      $\lambda_E = \lambda_L = M_L / v_X$, $m_{h_X} = 70\,\mathrm{GeV}$,
      $\varepsilon = -8\times10^{-4}$, $\varepsilon_{B-L} = 0.002$ and 
      $k_\ell = 10^{-7}$. Figures taken from~\cite{Hati:2020fzp}.}
\label{fig:cancellation}
\end{figure}

\begin{figure}[t!]
  \begin{center}
    \mbox{
 \hspace{-1.5cm}\includegraphics[width = 0.65 \textwidth]{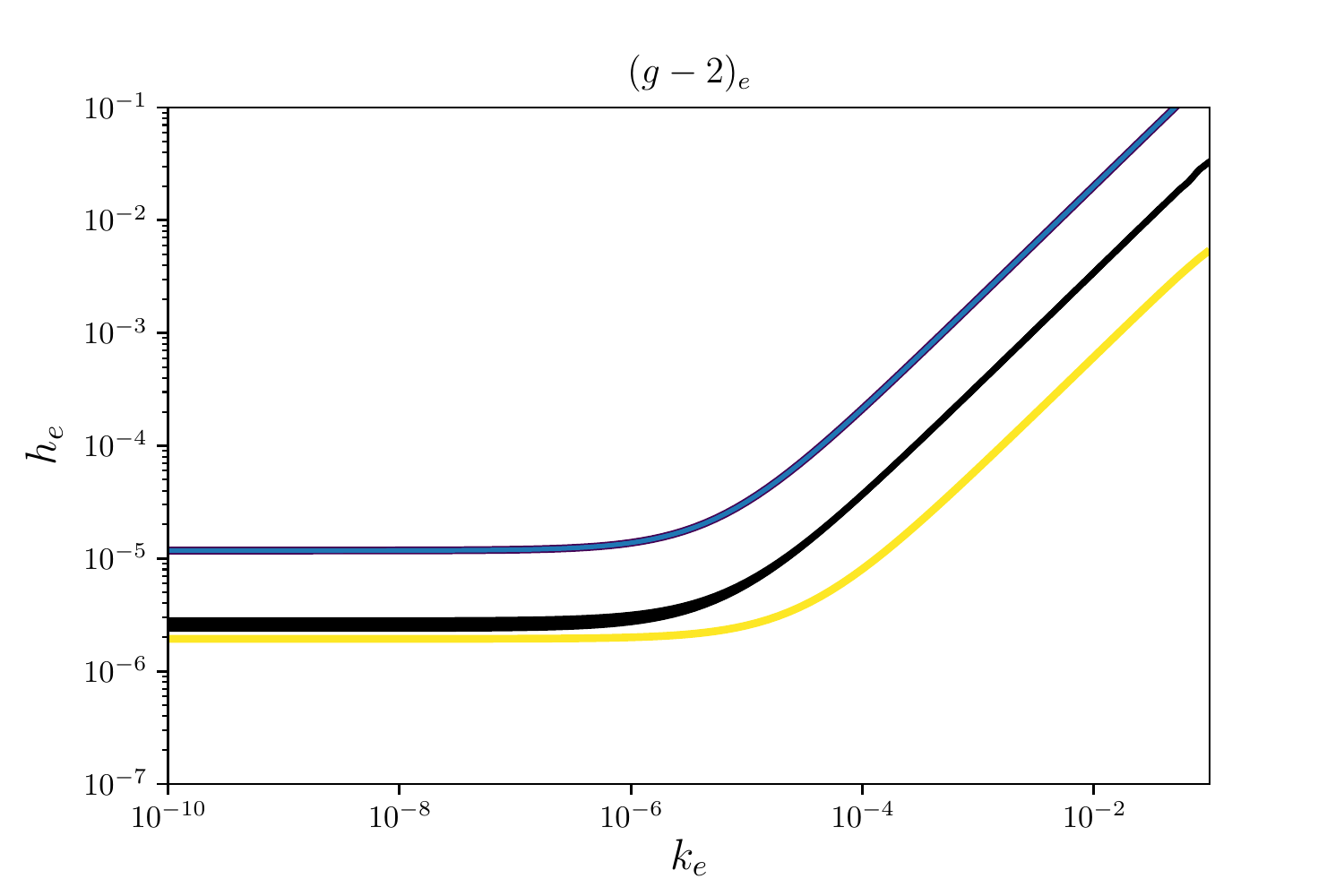}
 \hspace{-1.2cm}\includegraphics[width = 0.65 \textwidth]{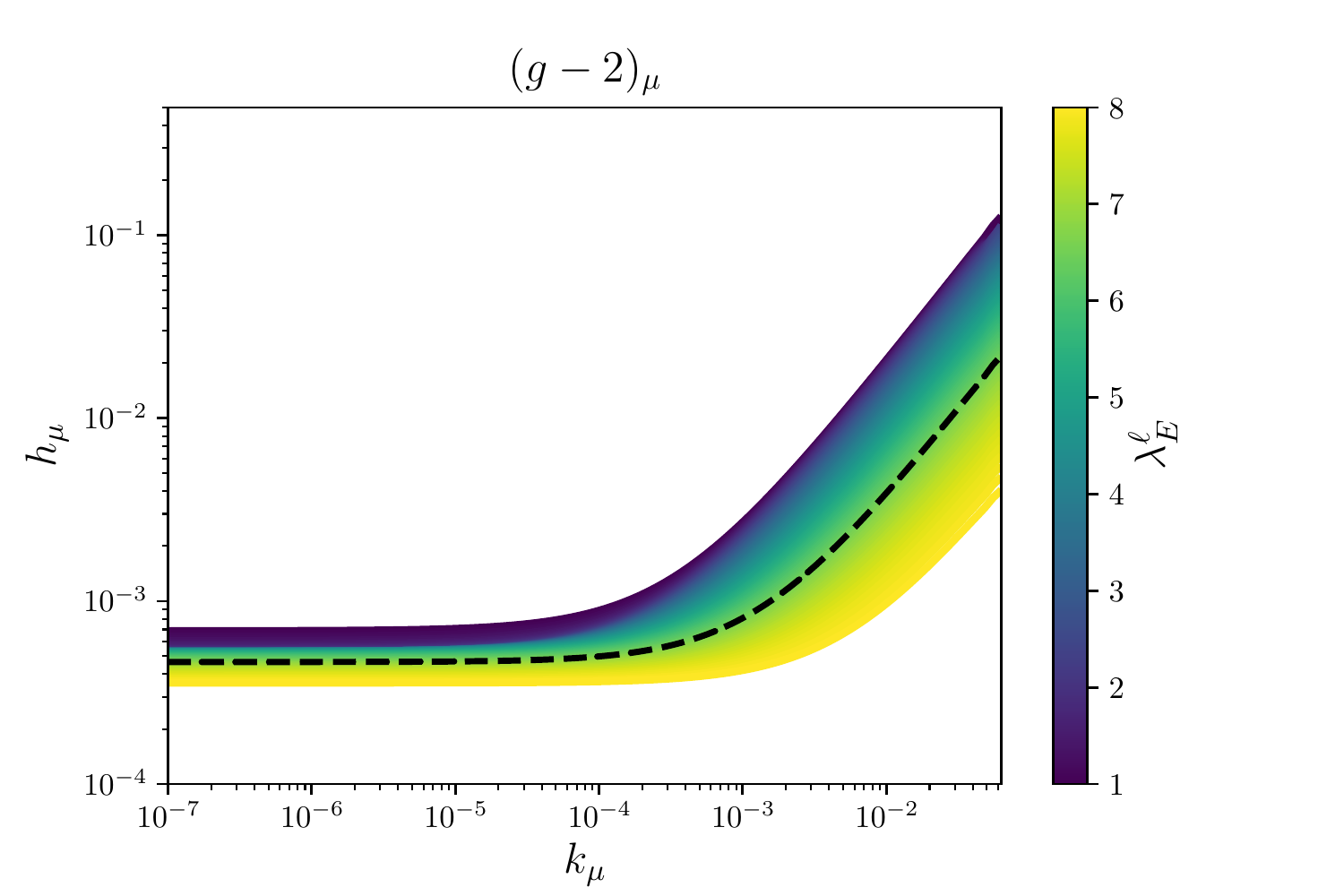}} 
\caption{Viable regions in $h_\ell$ vs. $k_\ell$ parameter space: on the left
  (right) $\ell=e\, (\mu)$). In both panels the colour code denotes
  the value of $\lambda_E^\ell$ ($\lambda_E = 1- 8$, from dark violet
  to yellow). 
  On the left panel, only the central black line complies with
  $(g-2)_e$ at the $2 \sigma$ level (i.e. $\lambda_E^e\sim  6.4$);
  for the right panel, all the coloured region allows to satisfy
   $(g-2)_\mu$ at $2 \sigma$ (the dashed black line illustrates the
  value $\lambda_E^\mu\sim  6.4$). All other relevant parameters
  fixed as leading to Fig.~\ref{fig:cancellation}.
  Figures taken from~\cite{Hati:2020fzp}.
    }
    \label{fig:g-2comb}
  \end{center}
\end{figure}

This interplay of the different (new) contributions can be understood
from Fig.~\ref{fig:cancellation}, which illustrates the $h_X$ and the
$Z^\prime$ contributions to the electron and muon $|\Delta a_\ell|$, as a
function of the $h_{\ell}$ coupling for $\ell=e$ (left) and 
$\ell=\mu$ (right).
The $h_X$-induced contribution to $(g-2)_\ell$
changes sign when the pseudo-scalar dominates over the scalar
contribution (for the choices of the relevant Yukawa couplings
$h_\ell$ and $k_\ell$). Likewise, 
a similar effect occurs for the $Z^\prime$ contribution
when the axial-vector contribution dominates over the vector one.
The transition
between positive (solid line) and negative (dashed line) contributions
- from $Z^\prime$ (orange), $h_X$ (green) and combined (blue) -
is illustrated by the sharp kinks visible in the logarithmic plots
of Fig.~\ref{fig:cancellation}.
In particular, notice that the negative electron $\Delta a_e$ is
successfully induced by the flip of the sign of the $h_X$
contribution, while a small positive muon $\Delta a_\mu$ arises from
the cancellation of the scalar  and the $Z^\prime$ contributions.
Leading to the numerical results of  Fig.~\ref{fig:cancellation}
(and in the remaining of our numerical analysis), we have taken as 
benchmark values $\varepsilon_{B-L} = 2\times 10^{-3}$ and
$\varepsilon = -8\times 10^{-4}$
(which are consistent with the criterion for explaining
the anomalous IPC in $^8$Be and respect all other imposed constraints). 
We emphasise that as a consequence of their already extremely
constrained ranges, both the $B-L$ gauge coupling and the kinetic
mixing parameter have a very minor influence on the contributions
to the anomalous magnetic lepton moments (when varied in the allowed
ranges). 
Furthermore, the masses $M_{L,E}$ and $m_{h_X}$ can be slightly varied 
with respect to the proposed benchmark values, with only a minor impact 
on the results; a mass-splitting between $M_E$ and $M_L$ (for each generation) slightly modifies the slope of the curves presented in Fig.~\ref{fig:g-2comb}, while an overall scaling to increase $M_{L,E}$ 
would imply taking (even) larger values for most of the couplings in their allowed regions. (Notice however that the model's parameter space is severely constrained, so that any departure from the benchmark values is only viable for a comparatively narrow band in the parameter space.)

To conclude the discussion, and provide a final illustration of how
constrained the parameter space of this simple model becomes, we
display in Fig.~\ref{fig:g-2comb}
the regions complying at the $2 \sigma$ level 
with the observation of $(g-2)_\ell$
in the planes spanned by $h_\ell$ and $k_\ell$ 
(for $\ell=e,\mu$). The colour code reflects the size of the
corresponding entry of $\lambda_E^\ell$, which
is varied in the interval $[1,8]$ (recall that for the electron
anomalous magnetic moment, $\lambda_L^e=\lambda_E^e\sim  6.4$). 
All remaining parameters are fixed to the same values used for the numerical 
analysis leading to Fig.~\ref{fig:cancellation}.

Notice that, as mentioned in the discussion at the beginning of the
section (cf. \ref{subsec:model-constraint}), the extremely stringent
constraints on the $Z^\prime$ couplings arising from  atomic parity
violation and electron neutrino scatterings render the model
essentially predictive in what concerns $(g-2)_e$: only the narrow
black band of the $(h_e-k_e)$ space succeeds in complying with all
available constraints, while both addressing the IPC $^8$Be anomaly, and
saturating the current discrepancy between SM and observation on  
$(g-2)_e$. For the muons, and although $h_\mu$ remains strongly
correlated with $k_\mu$, the comparatively larger freedom associated
with $\lambda_E^\mu$ (recall that no particular relation between
$\lambda_L$ and $\lambda_E$ is required by experimental data) 
allows to identify a wider band in $(h_\mu-k_\mu)$
space for which $\Delta a_\mu$ is satisfied at  $2 \sigma$. 

Finally, notice that the $h_\ell$ and  $k_\ell$ are forced into
a strongly hierarchical pattern, at least in what concerns the first
two generations.

\section{Further discussion}\label{sec:concs}
As previously discussed, the discrepancy between the SM prediction and experimental observation
regarding the anomalous magnetic moment of the muon is perhaps the
most longstanding anomaly, currently exhibiting a tension around 
$4.2\,\sigma$; more recently, the electron $(g-2)$ also started to
display tensions between theory and observation (around $2.5\,\sigma$ with $\alpha_e$ from Caesium, $1.7$ with $\alpha_e$ from Rubidium), 
all the most intriguing since instead of following a na\"ive scaling
proportional to powers of the light lepton masses, the comparison of 
$\Delta a_{e,\mu}$ suggests the presence of a New Physics which
would couple in a very distinct way to electrons and muons.

In recent years, an anomalous
angular correlation was observed for the 18.15~MeV nuclear transition of 
$^8$Be atoms, in particular an enhancement of the IPC at large angles,
with a similar anomaly having been observed in  $^4$He transitions. 

An interesting possibility is to interpret the atomic
anomalies as being due to the presence of a light vector boson, with a
mass close to $17$~MeV. Should such a state have non-vanishing
electroweak couplings to the standard fields, it could also have an
impact on $\Delta a_{e,\mu}$. 
We have thus investigated the phenomenological implications
of a BSM construction in which the light vector boson arises from a
minimal extension of the gauge group via an additional
$U(1)_{B-L}$. Other than the scalar field (whose VEV is responsible
for breaking the new $U(1)$), three generations of  Majorana right-handed
neutrinos, as well as of heavy vector-like leptons are added to the SM
field content. As discussed here, the new matter fields play an
instrumental role both in providing additional sources of leptonic
mixing, and in circumventing the very stringent experimental constraints. 

As we have discussed, the interplay of the (one-loop)
contributions of the $Z^\prime$ and the $U(1)_{B-L}$ breaking Higgs
scalar can further saturate the discrepancies in both 
$(g-2)_{e,\mu}$ anomalies. In particular, 
a cancellation between the new contributions 
is crucial to reproduce the observed pattern of
opposite signs of $\Delta a_e$ and $\Delta a_\mu$.
In view of the extensive limits on the $Z^\prime$ couplings,
arising from experimental searches, and which are 
further constrained by the requirements to explain  
the anomalous IPC in $^8$Be atoms, 
a combined explanation of the different anomalies renders the model
ultimately predictive in what concerns the electron 
$(g-2)$. 
We emphasise that even though we have considered a particular
$U(1)_{B-L}$ extension here - a minimal working
``prototype model'' - the general idea can be straightforwardly 
adopted and incorporated into other possible protophobic $U(1)$ extensions of the SM.

\medskip
Future measurements of 
right-handed neutral couplings, or axial couplings, for the second
generation charged leptons could further constrain the new muon
couplings. Although this clearly goes beyond the scope of the study carried out in~\cite{Hati:2020fzp},
one could possibly envisage parity-violation experiments carried in
association with muonic atoms.
As an example, in experiments designed to test parity non-conservation (PNC) with atomic radiative capture (ARC), the 
measurement of the forward-backward asymmetry of the photon 
radiated by muons ($2s \to 1s$ transition) is sensitive to 
(neutral) muon axial couplings~\cite{McKeen:2012sh}. Further 
possibilities include scattering experiments, such as MUSE at 
PSI~\cite{Gilman:2017hdr}, or  
studying the muon polarisation in $\eta$ decays (REDTOP 
experiment proposal~\cite{Gatto:2019dhj}), which could allow a measurement of the 
axial couplings of muons. 
 
\chapter{Testing the standard model with (semi-) leptonic heavy flavour decays}
\label{chap:bphysics}
\minitoc

\noindent
Due to their heavy mass ($\gtrsim 5\:\mathrm{GeV}$) and associated large phase space, hadrons containing a $b$ quark can decay into a plethora of possible final states.
In particular, charged and neutral $B$-mesons ($|B^0\rangle\sim |\bar b d\rangle\,,\:|B^+\rangle \sim |\bar b u\rangle$) can be copiously produced at $B$-factories such as Belle and BaBar (cf. related discussion in Section~\ref{sec:clfv_intro}); together with their relatively long lifetime ($\sim \mathcal O(\mathrm{ps})$), this renders $B$-mesons powerful laboratories to measure and test the properties and symmetries of the SM, and possibly search for New Physics.

\medskip
In particular, $\mathcal O(100)$ hadronic decay modes (and $B_{(s)} -\bar B_{(s)}$ mixing) are used to measure CP properties of the (heavy) quark sector, while measurements of inclusive and exclusive (semi-) leptonic tree-level decays offer opportunities to determine the CKM elements $V_{ub}$ and $V_{cb}$.

\begin{figure}[h!]
    \centering
    \includegraphics[width=\textwidth]{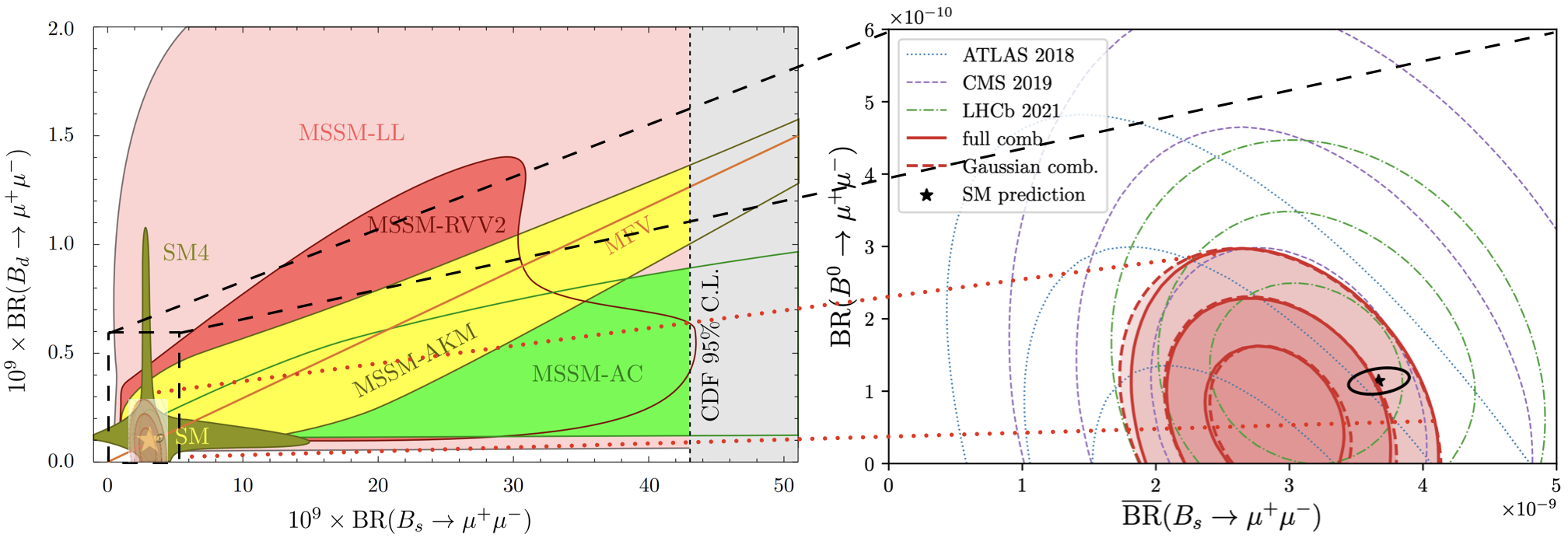}
    \caption{Evolution of $B_{(s)}\to \mu^+\mu^-$ data since 2010, together with predictions of several well-motivated New Physics frameworks and the SM (see text for further details). Figure adapted from~\cite{Straub:2010ih,Altmannshofer:2021qrr}.}
    \label{fig:bsmumu_overview}
\end{figure}

Furthermore, so-called rare (FCNC) decays offer invaluable opportunities to test the (accidental) symmetries of the SM and search for New Physics effects virtually present at low-energy.
Since in the SM flavour changing neutral currents are absent at tree-level, these are necessarily loop level decays, which are furthermore GIM-suppressed, thus leading to extremely small associated branching fractions ($\mathcal O (10^{-10} - 10^{-7})$).
Thus, they consist of very sensitive probes for New Physics; for instance, any new contribution at tree-level will lead to sizeable deviations with respect to the SM prediction.
This is for instance the case of $B_{(s)}\to \mu^+\mu^-$ decays, recently measured by the LHC collaborations ATLAS~\cite{Aaboud:2018mst}, CMS~\cite{Chatrchyan:2013bka,Sirunyan:2019xdu} and LHCb~\cite{LHCb:2021qbv,LHCb:2021awg,Aaij:2017vad}.
The SM prediction is given by~\cite{Bobeth:2013uxa,Straub:2018kue,Altmannshofer:2021qrr}
\begin{equation}
    \mathrm{BR}(B_s\to \mu^+\mu^-)^\text{SM} = (3.67\pm 0.15)\times 10^{-9}\,,\quad\quad\mathrm{BR}(B^0\to \mu^+\mu^-)^\text{SM} = (1.14\pm 0.12)\times 10^{-10}\,,
\end{equation}
consistent with most recent LHCb measurement~\cite{Altmannshofer:2021qrr} (for ``unofficial'' combinations see also~\cite{Geng:2021nhg,Angelescu:2021lln,Altmannshofer:2021qrr})
\begin{eqnarray}
    \mathrm{BR}(B_s\to \mu^+\mu^-)^\text{exp} &=& (3.09_{-0.43 -0.11}^{+0.46+0.15})\times 10^{-9}\,,\nonumber\\
    \quad\quad\mathrm{BR}(B^0\to \mu^+\mu^-)^\text{exp} &=& (1.20_{-0.74}^{+0.83}\pm 0.14)\times 10^{-10} \:\:(< 2.6\times 10^{-10}\:\: 95\%\: \text{C.L.})\,.
\end{eqnarray}
Prior to the first observation of these decays, many well-motivated New Physics frameworks such as a ``four family SM''~\cite{Buras:2010pi} and several supersymmetric flavour models~\cite{Antusch:2007re,Ross:2004qn,Agashe:2003rj,Hall:1995es}, predicted significant enhancements of the $B_{(s)}\to \mu^+\mu^-$ decay widths.
An overview of the status (experimental and theoretical) in 2010~\cite{Straub:2010ih} and of a contemporary one~\cite{Altmannshofer:2021qrr} of these decays are shown in Fig.~\ref{fig:bsmumu_overview}, in which one can see that most of the aforementioned models' parameter space are now excluded. If observed, a sizeable enhancement of these decays would have been a strong hint of the presence of the associated New Physics states.

Furthermore, as already discussed in Section~\ref{sec:LFUVnu}, semi-leptonic meson decays constitute an invaluable laboratory to search for the violation of lepton flavour universality, which will be discussed in the next section.

\bigskip
In addition to their excellent experimental accessibility, on the theoretical side, certain approximations such as QCD factorisation (QCDF) and (in many cases) heavy quark effective theory (HQET) are applicable to the $B$-meson sector, allowing for precise predictions.
Leptonic and semi-leptonic meson decays are in general mediated by weak interactions. 
For instance, the matrix element of a charged pseudo-scalar meson leptonic decay can be written as
\begin{equation}
    \mathcal M(P\to \ell\nu) \sim \langle\ell\nu|\mathcal O| P\rangle\,,
\end{equation}
in which $\mathcal O$ is some short-distance operator mediated by a $W$ boson, which can be ``integrated out'' (cf. Section~\ref{sec:eft}).
For the case of a $B$-meson, whose valence quark content is $|B^-\rangle \sim |b \bar u\rangle$, the short-distance operator can be parametrised as
\begin{equation}
    \mathcal O \simeq (\bar \ell\, \Gamma_\mu \nu) \frac{1}{M_W^2}(\bar b\, \Gamma^\mu u)\,, 
\end{equation}
in which $\Gamma_\mu$ is some combination of Dirac matrices.
We note here, that the operator $\mathcal O$ factorises into a leptonic and a hadronic part.
Thus, the matrix element can also be factorised, leading to
\begin{equation}
    \langle \ell\nu |\mathcal O | B^-\rangle = \langle \ell\nu | \bar \ell \Gamma_\mu \nu|0\rangle \frac{1}{M_W^2}\langle 0| \bar b \Gamma^\mu u| B^-\rangle\,,
\end{equation}
in which ``$0$'' denotes the vacuum. 
The hadronic part of the calculation, that is the ``annihilation'' of the $B$-meson, is then purely encoded in the second part of the full matrix element.
Since the $B$-meson is a pseudo-scalar ($J^P = 0^-$) and QCD conserves parity, the Lorentz structure is necessarily $\Gamma_\mu = \gamma_\mu\gamma_5$.
The hadronic matrix element is thus given by
\begin{equation}
    \langle 0| \bar b \gamma_\mu\gamma_5 u |B^-(p)\rangle = i p_\mu f_B\,,
\end{equation}
in which $f_B$ is the $B$-meson decay constant.
Using the axial Ward identity, it is further defined via
\begin{equation}
    \langle 0 | \bar b \gamma_5 u | B^-(p)\rangle = - i\frac{m_B^2 f_B}{m_u + m_b}\,.
\end{equation}
The decay constant $f_B$ encodes the non-perturbative QCD dynamic of the $B$-meson and has to be calculated via Lattice QCD, or measured in experiment.

For semi-leptonic meson decays one also assumes that QCD factorisation is valid, separating the short-distance dynamics from QCD; for instance for a $B\to D$ transition (see more in the following sections)
\begin{equation}
    \langle D(k)|\bar c\gamma_\mu b|\bar B(p)\rangle = \left[(p + k)_\mu - \frac{m_B^2 - m_D^2}{q^2}q_\mu\right]F_1(q^2) + q_\mu \frac{m_B^2 - m_D^2}{q^2}F_0(q^2)\,,
\end{equation}
in which $F_{0,1}$ denote the corresponding form factors.
Furthermore, if one of the valence quarks of a given meson-to-meson transition is significantly heavier than the other one (as is the case for both $B$ and $D$ mesons), the form factors can be approximated by expanding the ``internal dynamics'' around the background of a non-relativistic heavy quark. This is the key idea of HQET, which allows for precise predictions of the decays.

\mathversion{bold}
\section{Anomalies in $B$-meson decays}
\mathversion{normal}
As extensively discussed in Section~\ref{sec:LFUVnu}, one of the key predictions of the SM is the universality of interactions for the charged leptons of different generations. Extensive experimental observations confirm that this is indeed the case for several electroweak precision observables, as for example for $Z\to \ell \ell$ decays~\cite{Tanabashi:2018oca, ALEPH:2005ab}. However, certain recent experimental measurements suggest that hints for the violation of lepton flavour universality might be present in a number of observables, which would thus unambiguously point towards the presence of New Physics. The LFUV observables under current intense scrutiny are the FCNC quark transitions $b\to s \ell^+\ell^-$, 
and the charged current quark transitions $b\to c \ell^- \nu$: the former are loop-suppressed within the SM, thus providing a high sensitivity to probe New Physics effects; the latter can occur at the tree-level and are only subject to CKM suppression within the SM. 
Among these observables, ratios of potentially LFU violating $B$-meson decays are of particular interest, since they are free of the theoretical hadronic uncertainties arising from the form factors, as these cancel out in the ratios. The most relevant LFUV ratios for our study are  $R_{D^{(*)}}$ (corresponding to the charged current transition $b\to c \ell^- \nu$) and  $R_{K^{(\ast)}}$ (corresponding to the neutral current transition $b\to s \ell^+\ell^-$), respectively defined as
\begin{equation}\label{eq:RDRK:def}
  R_{D^{(*)}} \,= \,\frac{\text{BR}(B \to D^{(*)} \,\tau^- \,\bar\nu)}{
    \text{BR}(B \to  D^{(*)}\, \ell^- \,\bar\nu)}\, , \quad R_{K^{(\ast)}} \,= \,
  \frac{\text{BR}(B \to K^{(*)}\, \mu^+\,\mu^-)}{\text{BR}(B \to
  K^{(*)}\, e^+\,e^-)}\,,
\end{equation}
where $\ell=e,\,\mu$. 
A number of experimental measurements~\cite{Lees:2012xj,Lees:2013uzd,Amhis:2019ckw,Huschle:2015rga,Adachi:2009qg, Bozek:2010xy,Aaij:2015yra,Hirose:2016wfn, Abdesselam:2019dgh, Aaij:2019wad, Aaij:2017vbb, Abdesselam:2019wac, Aaij:2015esa, Wehle:2016yoi} shows deviations from the theoretical SM predictions~\cite{Ligeti:2016npd,Crivellin:2016ejn, Amhis:2019ckw,Bigi:2016mdz,Bigi:2017jbd,Bordone:2016gaq,Capdevila:2017bsm,Iguro:2020cpg}. In particular, the current measurements of $R_D$~\cite{Amhis:2019ckw, Abdesselam:2019dgh} and
$R_{D^\ast}$~\cite{Aaij:2015yra,Hirose:2016wfn,Amhis:2019ckw,Abdesselam:2019dgh}
respectively reveal $1.4\sigma$ and $2.5\sigma$ deviations with respect to their SM predictions~\cite{Bigi:2016mdz,Bigi:2017jbd,Iguro:2020cpg} and, when combined, this amounts to a deviation of $3.1\sigma$ from the SM expectation~\cite{Ligeti:2016npd,Crivellin:2016ejn,Amhis:2019ckw}. In the neutral current $b\to s \ell^+\ell^-$ transitions,
the measurements of $R_K$~\cite{Aaij:2019wad,Aaij:2014ora} in the di-lepton invariant mass squared bin $[1.1,6]~\text{GeV}^{2}$ show a deviation from the corresponding SM prediction~\cite{Bordone:2016gaq,Capdevila:2017bsm} at the level of $2.5\sigma$; for $R_{K^\ast}$, the measurements 
in the di-lepton invariant mass squared bins $q^2\in[1.1,6]~\text{GeV}^{2}$ and $q^2\in[0.045,1.1]~\text{GeV}^{2}$~\cite{Aaij:2017vbb} reveal tensions with the SM expectations~\cite{Bordone:2016gaq,Capdevila:2017bsm}
with significances of $2.5\sigma$ and $2.4\sigma$,
respectively.
The recent Belle collaboration results for $R_{K^{\ast}}$ in the analogous bins~\cite{Abdesselam:2019wac} are consistent with both the SM and the LHCb measurements~\cite{Aaij:2017vbb}; these results suffer from large statistical uncertainties.
Furthermore, the LHCb measurement of $R_K$ in the bin $[1.1,6]~\text{GeV}^{2}$ has been recently updated~\cite{LHCb:2021trn}, now exhibiting a $3.1\,\sigma$ tension with respect to the SM prediction.

In addition to the LFUV ratios, further discrepancies with respect to the SM have also been identified in a small number of lepton flavour specific observables in $b\to s \ell^+\ell^-$ neutral current transitions - this is the case of several angular observables in both charged and neutral $B^{0,+}\to K^\ast \mu^+\mu^-$ decays (as recently reported by the LHCb collaboration~\cite{Aaij:2020nrf, Aaij:2020ruw}), for which tensions between observation and SM expectations lie around the $3\sigma$ level.
Very recent measurements~\cite{LHCb:2021zwz} of the differential branching fraction of $B_s\to\phi\mu^+\mu^-$ decays further corroborate the picture.
Moreover, LHCb recently updated~\cite{LHCb:2021qbv,LHCb:2021awg} their analysis of $B_{(s)}\to \mu^+\mu^-$ decays leading to an improved measurement of the $B_{(s)}\to \mu^+\mu^-$ branching fractions.

In what follows, we first briefly review the current experimental status and the theoretical computation of the anomalous observables.
We then provide fits of New Physics Wilson coefficients to the data separately for $b\to c\ell\nu$ and $b\to s\ell\ell$ transitions. 
In the end, we attempt at putting forward a combined interpretation of the data using SMEFT, and discuss possible implications and further ways to corroborate these intriguing hints of New Physics.
\mathversion{bold}
\section{New Physics in $b\to c\ell\nu$}
\mathversion{normal}

\begin{figure}[h!]
    \centering
    \includegraphics[width=0.8\textwidth]{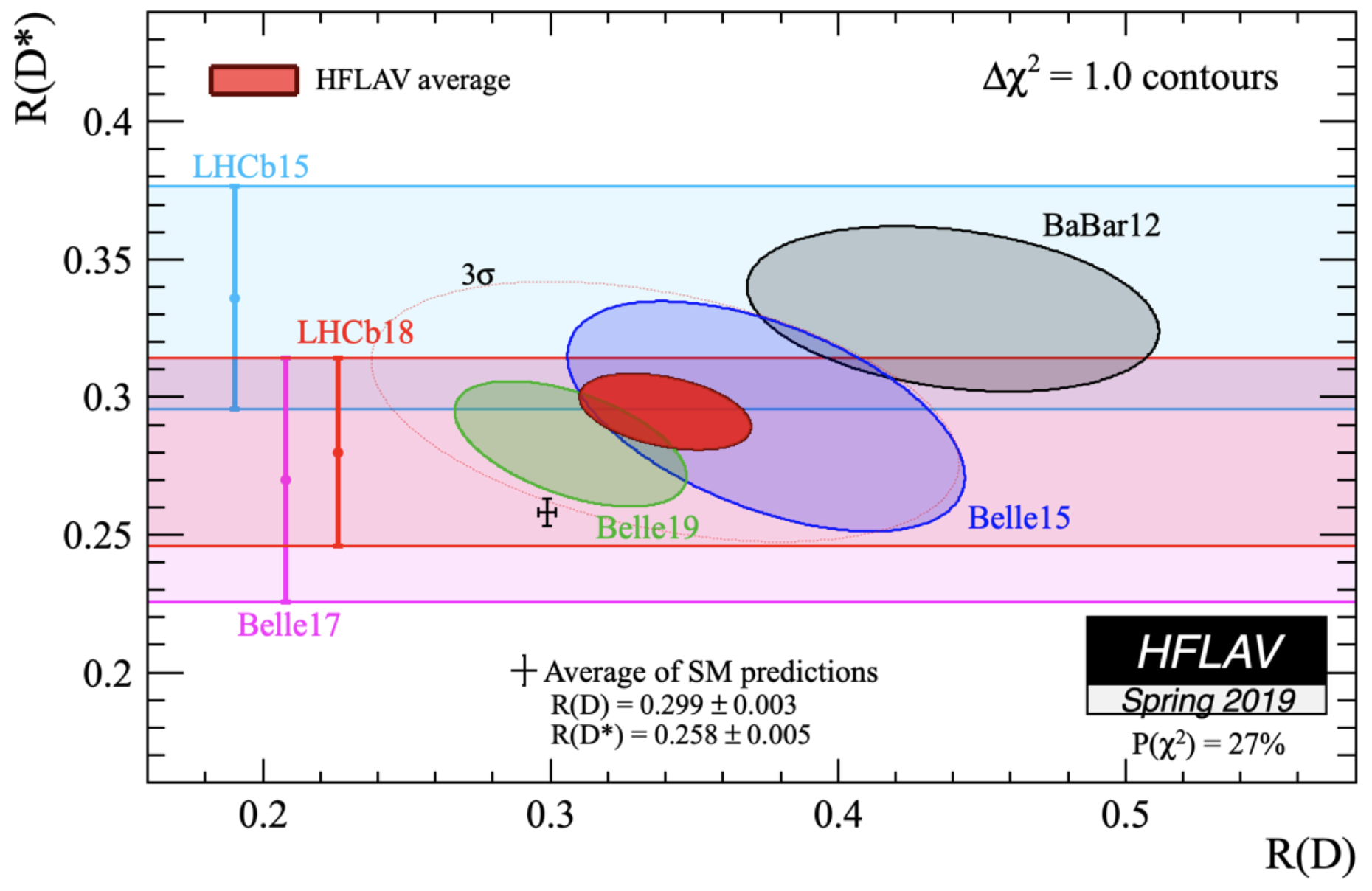}
    \caption{Current combination of experimental $R_{D^{(\ast)}}$ data together with an average of the SM predictions, as obtained by the HFLAV collaboration~\cite{Amhis:2019ckw}. Figure taken from~\cite{Amhis:2019ckw}.}
    \label{fig:rdrdshflav}
\end{figure}

As previously stated, a number of reported results from several experimental collaborations have
suggested a possible violation of lepton flavour
universality in the charged current decay mode $B \to D^{(*)} \ell \nu$, parametrised by the $R_{D}$ and $R_{D^{(*)}}$ ratios (see Eq.~(\ref{eq:RDRK:def})). 
The latest average values of these observables, given by the HFLAV collaboration~\cite{Amhis:2019ckw}, are 
\begin{eqnarray}\label{eq:RD*:expSMsigma}
 R_{D}\, =\, 0.340\,\pm\, 0.027\,\pm\,0.013\,,\quad
& R_{D}^{\text{SM}}\, =\, 0.299\, \pm\, 0.003 \, \quad
& (1.4 \sigma) \,;
 \nonumber\\
R_{D^\ast}\, =\, 0.295\,\pm\, 0.011\,\pm\,0.008\,,\quad
& R_{D^\ast}^{\text{SM}}\, =\, 0.258\, \pm\, 0.005 \, \quad
& (2.5 \sigma) \,.
 \end{eqnarray}
An overview of the current experimental data (individually and combined) and an average of the SM predictions by the HFLAV collaboration~\cite{Amhis:2019ckw} is shown in Fig.~\ref{fig:rdrdshflav}.
The HFLAV collaboration estimates the global significance of the SM deviation at $3.08\,\sigma$, but with improvements of the hadronic part of the computation, leading to an improved SM prediction, it can be slightly larger, reaching $\sim 4\,\sigma$ (see e.g.~\cite{Bordone:2019vic,Bordone:2019guc}).

\noindent
The relevant effective Lagrangian for the charged current
transitions $d_k\to u_j\bar{\nu}\ell^{-}$ can be expressed as
\begin{eqnarray}\label{eq:Heff:RD}
\mathcal L_\text{eff} = -{4\,G_F \over \sqrt2} \,V_{jk} &\times&\left[
	(\delta_{\ell i}+C_{V_L}^{jk;\ell i}) (\bar{u}_j \,\gamma_\mu\, P_L\, d_k) (\bar{\ell}\, \gamma^\mu\, P_L \,\nu^i) +
	C_{V_R}^{jk;\ell i} (\bar{u}_j \,\gamma_\mu \,P_R \,d_k) (\bar{\ell} \,\gamma^\mu\, P_L \,\nu^i) \right. \nonumber\\
	&+& C_{S_L}^{jk;\ell i} (\bar{u}_j \,P_L \,d_k) (\bar{\ell} \,P_L\, \nu^i)
	  + C_{S_R}^{jk;\ell i} (\bar{u}_j \,P_R \,d_k) (\bar{\ell}\, P_L\, \nu^i) \nonumber\\
	&+& \left. C_{T_L}^{jk;\ell i} (\bar{u}_j \sigma_{\mu\nu} \,P_L\, d_k) (\bar{\ell} \sigma^{\mu\nu}\, P_L\, \nu^i) \right] + \mathrm{H.c.}\,,
\end{eqnarray}
in which we have assumed the neutrinos to be left-handed and where, for the SM, we have $C_i=0$, $\forall i\in \{S_L,S_R,V_L,V_R,T_L\}$.
For the convenient
double ratios $R_{D}/R_{D}^\text{SM}$ and $R_{D^\ast}/R_{D^\ast}^\text{SM}$ (which combine the current experimental averages with the SM predictions), the current data can be summarised as 
\begin{equation}
R_{D}/R_D^\text{SM} \,= \,1.14 \pm 0.10\, , \quad
R_{D^{\ast}}/R_{D^{\ast}}^\text{SM} 
\,= \,1.14 \pm 0.06\,,
\end{equation}
where the statistical and systematical errors have been added in quadrature. 

In order to analyse the impact of New Physics on these observables, one has to calculate the individual decay rates using the effective Lagrangian in Eq.~\eqref{eq:Heff:RD} for $b\to c\ell\nu$ transitions~\cite{Fajfer:2012vx,Tanaka:2012nw,Sakaki:2013bfa}.
Fixing the quark indices in Eq.~\eqref{eq:Heff:RD} to $j = c$ and $k=b$, in the following we only keep the $\ell$ and $i$ indices for simplicity.

The following differential decay rates (calculated using helicity amplitudes~\cite{Tanaka:2012nw}) obtained in~\cite{Sakaki:2013bfa} are given by
\begin{eqnarray}
    \frac{d\Gamma(\bar{B} \to D\ell \bar \nu_i)}{dq^2} &=& \frac{G_F^2 |V_{cb}|^2}{192\pi^3 m_B^2}q^2\sqrt{\lambda_D(q^2)}\left(1 - \frac{m_\ell^2}{q^2}\right)^2\times\nonumber\\
    &\phantom{=}& \times \Bigg\{\left|\delta_{\ell i} + C_{V_L}^{\ell i} + C_{V_R}^{\ell i}\right|^2\left[\left(1 + \frac{m_\ell^2}{2q^2}\right){H_{V,0}^{s}}^2 + \frac{3}{2}\frac{m_\ell^2}{q^2}{H_{V,t}^s}^2\right] + \frac{3}{2}\left|C_{S_R}^{\ell i} + C_{S_L}^{\ell i}\right|^2 {H_{S}^s}^2\nonumber\\
    &\phantom{=}& + 8\left|C_{T_L}^{\ell i}\right|^2\left(1 + \frac{2m_\ell^2}{q^2}\right) {H_T^s}^2 + 3\mathrm{Re}\left[(\delta_{\ell i} + C_{V_L}^{\ell i} + C_{V_R}^{\ell i})(C_{S_R}^{\ell i \ast} + C_{S_L}^{\ell i \ast})\right]\frac{m_\ell}{\sqrt{q^2}}H_S^s H_{V,t}^s\nonumber\\
    &\phantom{=}& -12 \mathrm{Re}\left[(\delta_{\ell i} + C_{V_L}^{\ell i} + C_{V_R}^{\ell i})C_{T_L}^{\ell i\ast}\right]\frac{m_\ell}{\sqrt{q^2}} H_T^s H_{V,0}^s\Bigg\}\,,
\end{eqnarray}
and
\begin{eqnarray}
    \frac{d\Gamma(\bar{B} \to D^\ast\ell \bar \nu_i)}{dq^2} &=& \frac{G_F^2 |V_{cb}|^2}{192\pi^3 m_B^2}q^2\sqrt{\lambda_{D^\ast}(q^2)}\left(1 - \frac{m_\ell^2}{q^2}\right)^2\times\nonumber\\
    &\phantom{=}& \times \Bigg\{\left(\left|\delta_{\ell i} + C_{V_L}^{\ell i}\right|^2 + \left|C_{V_R}^{\ell i}\right|^2\right)\left[\left(1 + \frac{m_\ell^2}{2 q^2}\right)\left(H_{V,+}^2 + H_{V,-}^2 + H_{V,0}^2\right) + \frac{3}{2}\frac{m_\ell^2}{q^2}H_{V,t}^2\right]\nonumber\\
    &\phantom{=}& -2\mathrm{Re}\left[(\delta_{\ell i} + C_{V_L}^{\ell i})C_{V_R}^{\ell \ast}\right]\left[\left(1 + \frac{m_\ell^2}{2 q^2}\right)\left(H_{V,0}^2 + 2 H_{V,+}H_{V,-}\right) + \frac{3}{2}\frac{m_\ell^2}{q^2} H_{V,t}^2\right]\nonumber\\
    &\phantom{=}&+\frac{3}{2}\left|C_{S_R}^{\ell i} - C_{S_L}^{\ell i}\right|^2 + 8 \left|C_{T_L}^{\ell i}\right|^2\left(1 + \frac{2m_\ell^2}{q^2}\right)\left(H_{T,+}^2 + H_{T, -}^2 + H_{T,0}^2\right)\nonumber\\
    &\phantom{=}& + 3\mathrm{Re}\left[(\delta_{\ell i} + C_{V_L}^{\ell i} - C_{V_R}^{\ell i})(C_{S_R}^{\ell i \ast} - C_{S_L}^{\ell i\ast}\right]\frac{m_\ell}{\sqrt{q^2}} H_S H_{V,t}\nonumber\\
    &\phantom{=}& - 12 \mathrm{Re}\left[(\delta_{\ell i} + C_{V_L}^{\ell i})C_{T_L}^{\ell i\ast}\right]\frac{m_{\ell}}{\sqrt{q^2}}\left(H_{T,0} H_{V,0} + H_{T,+} H_{V,+} - H_{T,-}H_{V,-}\right)\nonumber\\
    &\phantom{=}& + 12\mathrm{Re}\left[C_{V_R}^{\ell i}C_{T_L}^{\ell i \ast}\right]\frac{m_\ell}{\sqrt{q^2}}\left(H_{T,0}H_{V,0} + H_{T,+}H_{V,-} - H_{T,-}H_{V,+}\right)\Bigg\}\,,
\end{eqnarray}
with $\lambda_{D^{(\ast)}}(q^2) = ((m_B - m_{D^{(\ast)}})^2-q^2)((m_B + m_{D^{(\ast)}})^2 - q^2)$.
In the above equations, the SM corresponds to $\delta_{\ell i}$.
Following QCDF, the meson transitions are encoded in ($q^2$ dependent) hadronic amplitudes which for $\bar B\to M\ell\bar\nu_i \:\:(M = D, D^\ast)$ are defined as~\cite{Sakaki:2013bfa}
\begin{eqnarray}
    H_{V_{1,2},\lambda}^{\lambda_M}(q^2) &=& \varepsilon_\mu^\ast(
    \lambda) \langle M(\lambda_M)|\bar c\gamma^\mu(1 \mp \gamma_5)b|\bar B\rangle\,,\nonumber\\
    H_{S_{1,2},\lambda}^{\lambda_M}(q^2) &=& \langle M(\lambda_M)| \bar c (1 \pm \gamma_5)b|\bar B\rangle\,,\nonumber\\
    H_{T, \lambda\lambda'}^{\lambda_M}(q^2)&=& -H_{T,\lambda'\lambda}^{\lambda_M}(q^2) = \varepsilon_\mu^\ast(\lambda)\varepsilon_\nu^\ast(\lambda')\langle M(\lambda_M)|\bar c\sigma^{\mu\nu} (1 - \gamma_5)b|\bar B\rangle\,,
\end{eqnarray}
where $\lambda_M$ denotes the final state meson helicity ($\lambda_M = s$ for $D$ and $\lambda_M = 0, \pm 1$ for $D^\ast$), and $\varepsilon$ denotes the polarisation vector of intermediate virtual bosons, with the associated helicity $\lambda = 0,\pm, t$.
The hadronic matrix elements are given in terms of form factors, commonly parametrised as
\begin{eqnarray}
    \langle D(k)|\bar c\gamma_\mu b|\bar B(p)\rangle &=& \left[(p + k)_\mu - \frac{m_B^2 - m_D^2}{q^2}q_\mu\right]F_1(q^2) + q_\mu \frac{m_B^2 - m_D^2}{q^2}F_0(q^2)\,,\\
    \langle D(k)|\bar c b|\bar B(p)\rangle &=& \frac{m_B^2 - m_D^2}{m_b - m_c}F_0 (q^2)\,,\\
    \langle D(k) |\bar c \sigma_{\mu\nu} b|\bar B(p)\rangle &=& - i (p_\mu k_\nu - k_\mu p_\nu)\frac{2 F_T(q^2)}{m_B + m_D}\,,
\end{eqnarray}
in which $F_1$ is the vector form factor, $F_T$ the tensor form factor and $F_0$ the scalar form factor.
Similarly, for the $B\to D^\ast$ transition, one commonly defines
\begin{eqnarray}
    \langle D^\ast(k, \varepsilon)|\bar c \gamma_\mu b|\bar B(p)\rangle &=& - i \epsilon_{\mu\nu\rho\sigma} \varepsilon^{\nu\ast} p^\rho k^\sigma \frac{2 V(q^2)}{m_B + m_{D^\ast}}\,,\nonumber\\
    \langle D^\ast (k,\varepsilon)|\bar c \gamma_\mu\gamma_5 b|\bar B(p)\rangle &=& \varepsilon_\mu^{\ast}(m_B + m_{D^\ast}) A_1(q^2) - (p + k)_\mu (\varepsilon^\ast q)\frac{A_2(q^2)}{m_B + m_{D^\ast}}\nonumber\\
    &\phantom{=}& - q_\mu(\varepsilon^\ast q)\frac{2m_{D^\ast}}{q^2}[A_3 (q^2) - A_0(q^2)]\,,\nonumber\\
    \langle D^\ast (k, \varepsilon)| \bar c \gamma_5 b|\bar B(p)\rangle &=& - (\varepsilon^\ast q)\frac{2m_{D^\ast}}{m_b + m_c}A_0 (q^2)\,,\nonumber\\
    \langle D^\ast (k, \varepsilon)|\bar c \sigma_{\mu\nu} b|\bar B(p)\rangle &=& \epsilon_{\mu\nu\rho\sigma}\Bigg\{ - \varepsilon^{\rho\ast}(p + k)^\sigma T_1(q^2) + \varepsilon^{\rho\ast}q^\sigma \frac{m_B^2 - m_{D^\ast}^2}{q^2}[T_1(q^2) - T_2(q^2)]\nonumber\\
    &\phantom{=}& + 2\frac{(\varepsilon^\ast q)}{q^2}p^\rho k^\sigma\left[T_1(q^2) - T_2(q^2) - \frac{q^2}{m_B^2 - m_{D^\ast}^2}T_3(q^2)\right]\Bigg\}\,,\nonumber\\
    \langle D^\ast (k, \varepsilon)|\bar c \sigma_{\mu\nu} q^\nu b|\bar B(p)\rangle &=& 2\epsilon_{\mu\nu\rho\sigma}\varepsilon^{\nu\ast} p^\rho k^\sigma T_1(q^2)\label{eq:tensdef1}\,,\nonumber\\
    \langle D^\ast (k, \varepsilon)|\bar c \sigma_{\mu\nu} \gamma_5 q^\nu b|\bar B(p)\rangle &=& [(m_B^2 - m_{D^\ast}^2)\varepsilon_\mu^\ast - (\varepsilon^\ast q)(p+k)_\mu] T_2(q^2)\nonumber\\
    &\phantom{=}& - (\varepsilon^\ast q)\left[q_\mu - \frac{q^2}{m_B^2 - m_{D^\ast}^2} (p+k)_\mu\right] T_3(q^2)\label{eq:tensdef2}\,,
\end{eqnarray}
in which $V$ is the vector form factor, $T_i$ the tensor form factors (defined via Eqs.\eqref{eq:tensdef1} and \eqref{eq:tensdef2}) and $A_i$ the axial vector form factors that fulfil
\begin{eqnarray}
    A_3(q^2) &=& \frac{m_B + m_{D^\ast}}{2 m_{D^\ast}} A_1(q^2) - \frac{m_B - m_{D^\ast}}{2 m_{D^\ast}} A_2(q^2)\,.
\end{eqnarray}
With the hadronic matrix elements the hadronic amplitudes can be calculated, see e.g.~\cite{Tanaka:2012nw,Sakaki:2013bfa}.
The form factors $F_i$, $V$, $A_i$ and $T_i$ can be parametrised in terms of Isgur-Wise functions using heavy quark effective theory (HQET)~\cite{Caprini:1997mu,Bernlochner:2017jka}, further improved with input from Lattice QCD~\cite{MILC:2015uhg,Na:2015kha,Harrison:2017fmw,FermilabLattice:2014ysv,Aoki:2016frl}, QCD sum rule calculations~\cite{Neubert:1992wq,Neubert:1992pn,Ligeti:1993hw} and Lightcone sum rule (LCSR) calculations~\cite{Faller:2008tr,Gubernari:2018wyi,Braun:2017liq}.

As a first step, under the simplifying assumption of a non-vanishing single type of New Physics operator at a time - i.e. $C_i\neq 0$, $i\in \{S_L,S_R,V_L,V_R,T_L\}$ -, it is possible to draw some qualitative conclusions from the approximate numerical forms for the double ratios using a HQET formalism~\cite{Iguro:2018vqb,Sakaki:2012ft,Tanaka:2012nw,Hati:2015awg,Neubert:1991td,Hagiwara:1989cu,Caprini:1997mu}.

In particular, and if one assumes that all the relevant Wilson coefficients are real, then the following qualitative observations can be readily made. The operator corresponding to $C_{V_L}$ contains the same Lorentz structure as the SM contribution and the New Physics amplitude adds to the SM one, thus leading to similar enhancements to both $R_{D}$ and $R_{D^\ast}$, which are proportional to $(1+C_{V_L})^2$. In turn, this leads to similar fractional enhancements to $R_{D}/R_{D}^\text{SM}$ and $R_{D^\ast}/R_{D^\ast}^\text{SM}$. Therefore, $C_{V_L}$ is one of the most favoured choices for explaining the anomalous $R_{D}$ and $R_{D^\ast}$ data. On the other hand, if the New Physics contribution is purely 
a right-handed vector current
($C_{V_R}$ type), then for a real $C_{V_R}$, $R_{D}$ is proportional to $(1+C_{V_R})^2$ while $R_{D^\ast}$ is roughly proportional to $(1-C_{V_R})^2$. Under such circumstances, it is then not possible to simultaneously explain both $R_{D}$ and $R_{D^\ast}$ data. 
However, and as discussed in~\cite{Iguro:2018vqb}, this conclusion is no longer valid for a complex $C_{V_R}$. The scalar operators corresponding to $C_{S_L}$ and $C_{S_R}$ contain the pseudo-scalar Dirac bilinear and therefore are not subject to helicity suppressions, leading to stringent constraints from the (relatively large) branching ratios of $B_c\rightarrow \tau \nu$. The tensor operator, corresponding to $C_{T_L}$, is subject to tensions from the recent measurement of the $D^{*}$ longitudinal polarisation  $f_L^{D^\ast}$, which is currently about $1.6\, \sigma$ higher than the SM prediction  and has a discriminatory power between the scalar and tensor solutions~\cite{Alok:2017qsi,Blanke:2019qrx,Shi:2019gxi}. Choices based on pure right-handed operators seem to be disfavoured by LHC data~\cite{Shi:2019gxi,Cornella:2019hct}. Finally, scenarios that only present scalar contributions are in conflict with both LHC and $B_c\rightarrow \tau \nu$ data. 

As an illustrative example of a two-dimensional New Physics hypothesis, we present in Fig.~\ref{fig:rdrdsfit} a fit\footnote{For more details on the fit and numerics see Appendix~\ref{app:stats}.} of New Physics Wilson coefficients to the data on $b\to c\ell\nu$ transitions (which are listed in appendix~\ref{app:BFCCC}). 
Here we assume New Physics to be only present in $C_{V_L}^{bc\tau\nu}$ and in a lepton flavour universal\footnote{The assumption of a lepton flavour universal $C_{V_R}$ can be justified by the SMEFT matching conditions under the assumption of a linear electroweak phase transition~\cite{Murgui:2019czp}.} $C_{V_R}^{bc\ell\nu}$.
The best fit point is given by $C_{V_L}^{bc\tau\nu} = 0.073 \pm 0.018\,,\:\:C_{V_R}^{bc\ell\nu} = -0.032 \pm0.020$.
For a more complete model-independent analysis see e.g.~\cite{Murgui:2019czp}.

The inherent New Physics scale which is implied by the best fit values is given by (cf. discussion in Section~\ref{sec:eft})
\begin{equation}
    \Lambda_\text{NP} = \left(\frac{4 G_F}{\sqrt{2}} |V_{cb}| |C_{V_L}^{bc\tau\nu}|\right)^{-1/2}\simeq 3.3\:\mathrm{TeV}\,,
    \label{eqn:npscale_rd}
\end{equation}
interestingly well within LHC reach.
\begin{figure}[h!]
    \centering
    \includegraphics[width=0.8\textwidth]{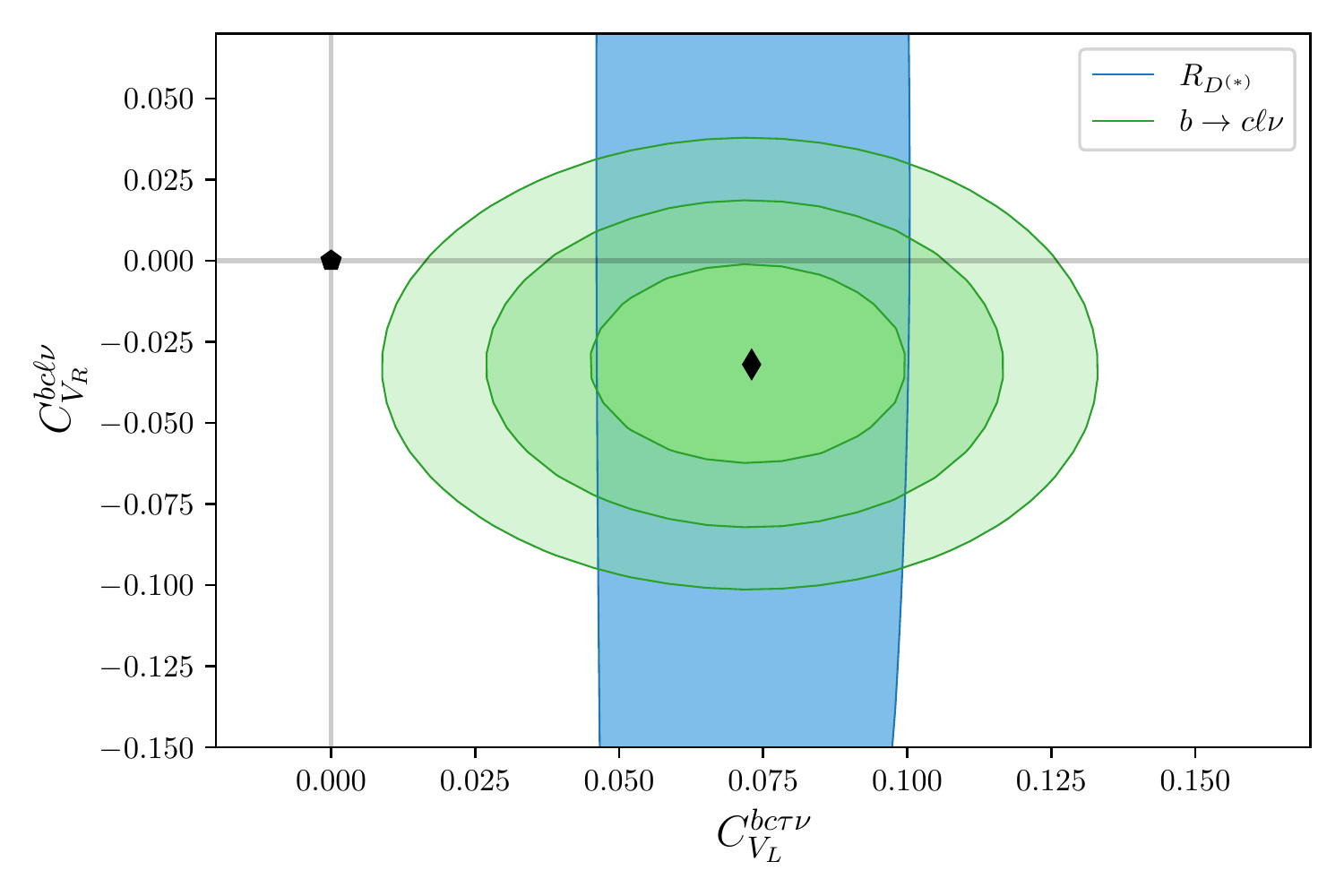}
    \caption{Illustrative (example) fit of New Physics contributions to left- and right-handed charged current vector operators on $R_{D^{(\ast)}}$ and $b\to c\ell\nu$ data. Notice, that the contribution to $C_{V_R}^{bs\ell\nu}$ is lepton flavour universal.
    The best fit point is given by $C_{V_L}^{bc\tau\nu} = 0.073 \pm 0.018\,,\:\:C_{V_R}^{bc\ell\nu} = -0.032 \pm0.020$, denoted by the diamond.}
    \label{fig:rdrdsfit}
\end{figure}

\mathversion{bold}
\section{New Physics in $b\to s\ell\ell$}
\begin{figure}[]
    \centering
    \mbox{\includegraphics[width=0.48\textwidth]{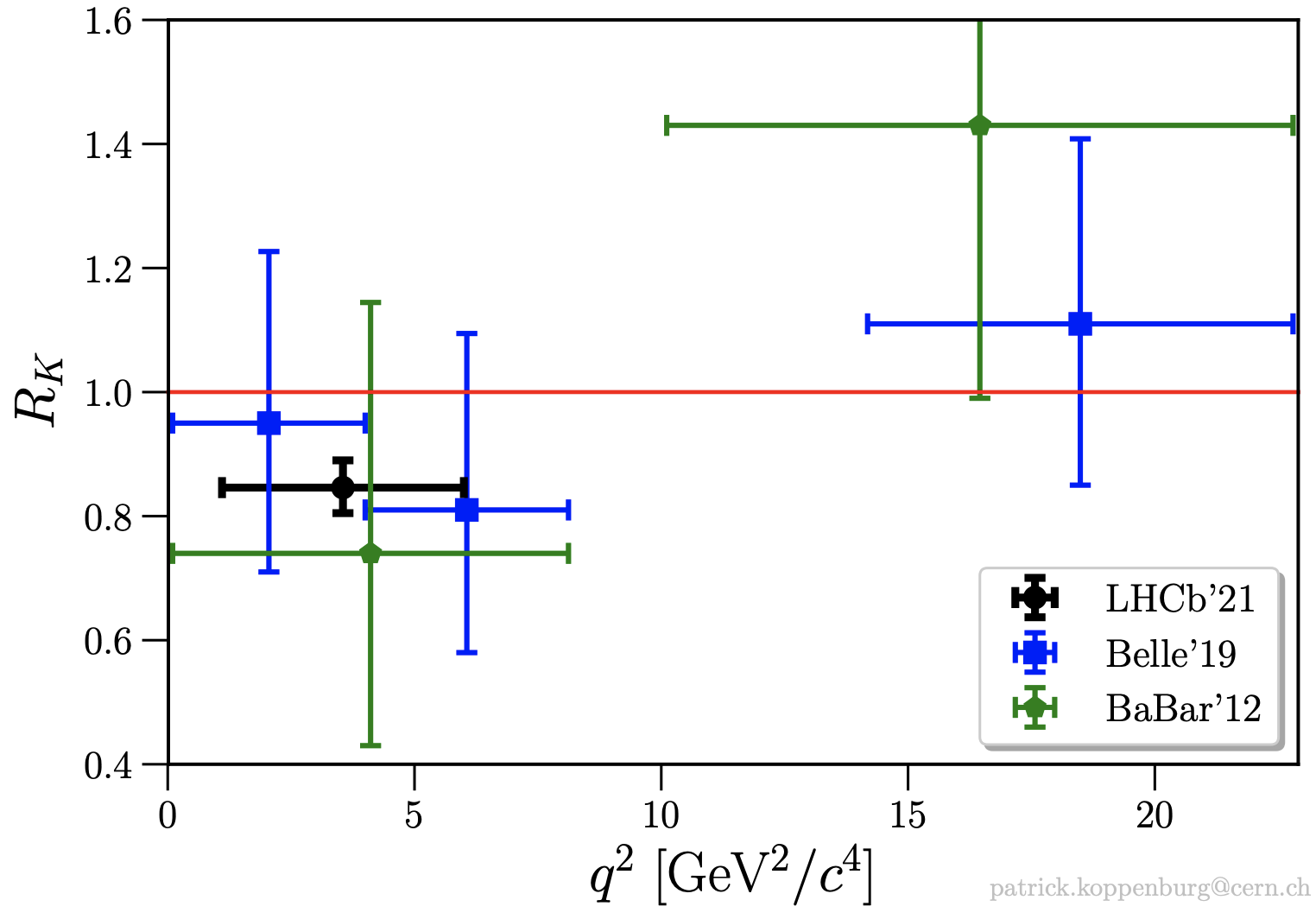}\includegraphics[width=0.48\textwidth]{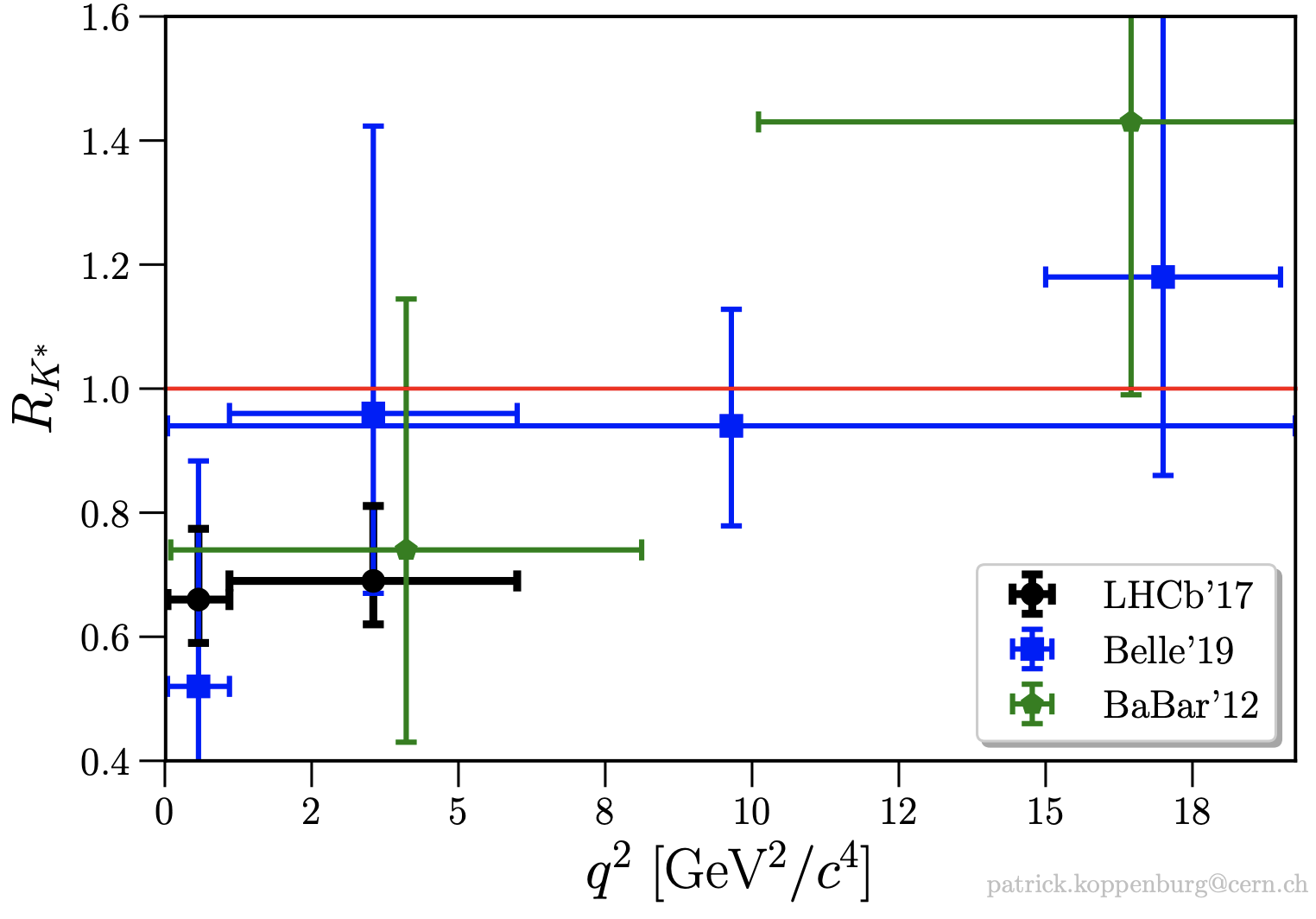}}
    \caption{Overview of $R_{K^{(\ast)}}$ measurements by BaBar~\cite{BaBar:2012mrf}, Belle~\cite{Belle:2019oag,BELLE:2019xld} and LHCb~\cite{LHCb:2017avl,LHCb:2021trn}. Figures taken from~\cite{koppenburgtalk}.}
    \label{fig:rkrks_overview}
\end{figure}

\mathversion{normal}
\label{sec:bsll}
A number of anomalies reported in $b\to s\ell\ell$ observables currently stand as promising hints of NP, among them those parametrised by the 
$R_{K^{(*)}}$ ratios, defined in Eq.~(\ref{eq:RDRK:def}).
The latest averages of the reported anomalous experimental data, together with the SM predictions can be expressed as~\cite{Aaij:2019wad,Aaij:2017vbb,Abdesselam:2019wac,LHCb:2021trn}
	\begin{eqnarray}\label{eq:RK*:expSMsigma}
	  R_{K [1.1,6]}^{\text{LHCb}}\, &=&\, 0.846\,\pm^{0.042}_{0.039}\,
	  \pm^{0.013}_{0.012}\,, \quad
	 R_{K}^{\text{SM}}\, =\, 1.0003\, \pm\, 0.0001
	  \,,
	 \nonumber\\
	 R_{K^\ast[0.045, 1.1]}^{\text{LHCb}}\, &=&\, 0.66 ^{+0.11}_{-0.07} \,\pm\,
	 0.03\,, \quad
	 R_{K^\ast[0.045, 1.1]}^{\text{Belle}}\, =\, 0.52 ^{+0.36}_{-0.26} \,\pm\,
	 0.05\,, \quad
	R_{K^\ast[0.045, 1.1]}^{\text{SM}}\,  \sim\,  0.93 \,,
	 \nonumber\\
	 R_{K^\ast [1.1, 6]}^{\text{LHCb}} \,&=& \,0.69 ^{+0.11}_{-0.07}\, \pm
	0.05\,, \quad
	 R_{K^\ast [1.1, 6]}^{\text{Belle}} \,= \,0.96 ^{+0.45}_{-0.29}\, \pm
	0.11\,, \quad
	 R_{K^\ast [1.1, 6]}^{\text{SM}} \,\sim \, 0.99\, ,
	\end{eqnarray}
where the di-lepton invariant mass squared bin ($[q^2_\text{min}, q^2_\text{max}]$ in $\text{GeV}^{2}$) is identified by the associated subscripts. 
An overview of the current measurements of $R_{K^{(\ast)}}$ is shown in Fig.~\ref{fig:rkrks_overview}.
Further anomalies have also been reported in the neutral current decay modes of $B$-mesons for semi-leptonic final states including muon pairs\footnote{Notice that here we refer to the neutral and \textit{charged} $B$-meson decays, i.e. 
$B^{0,+} \to K^{*} \mu \mu$ decays.}.
Among them, one concerns the observable $d \mathrm{BR}(B_s\to\phi\mu\mu)/ dq^2$ in the bin
$q^2\in[1,6]\,\mathrm{GeV}^2$~\cite{Aaij:2015esa}, presently exhibiting a tension with the SM prediction around $3\sigma$. Further discrepancies with respect to the SM, typically at the $3\sigma$ level, have also emerged in relation to the angular observables. In particular, this is the case of
$P_5^{\prime}$ in $B \to K^\ast \ell^+ \ell^-$	processes: 
the results from the LHCb collaboration for $P_5^{\prime}$ regarding muon final states ($B \to K^\ast	\mu^+ \mu^-$ decays) reveal a discrepancy with respect to the SM~\cite{Aaij:2013qta,Aaij:2015oid}.
The $P_5'$ results for electrons reported by the 
Belle collaboration~\cite{Abdesselam:2016llu,Wehle:2016yoi}  show a better agreement with theoretical SM expectations than those for muons. 
More recently, similar measurements have also been reported by the ATLAS~\cite{Aaboud:2018krd} and CMS~\cite{Sirunyan:2017dhj} collaborations. 
The 2015 LHCb results~\cite{Aaij:2015oid} and the ATLAS result~\cite{Aaboud:2018krd} for $P_5^{\prime}$ in the low dimuon invariant mass-squared
range, $q^2\in[4,6]\,\mathrm{GeV}^2$, indicate a $\approx 3.3\sigma$ discrepancy with respect to the
SM prediction~\cite{Aebischer:2018iyb}. Belle
results corroborate the latter findings, showing a deviation of $2.6\sigma$ from the SM expectation in the bin $q^2\in[4,8]\,\mathrm{GeV}^2$~\cite{Wehle:2016yoi}. 
The reported CMS measurement (possibly as a consequence of insufficient statistics) is still consistent with the SM expectation within $1\sigma$~\cite{Sirunyan:2017dhj}. 

\begin{figure}[]
    \centering
    \includegraphics[width=0.6\textwidth]{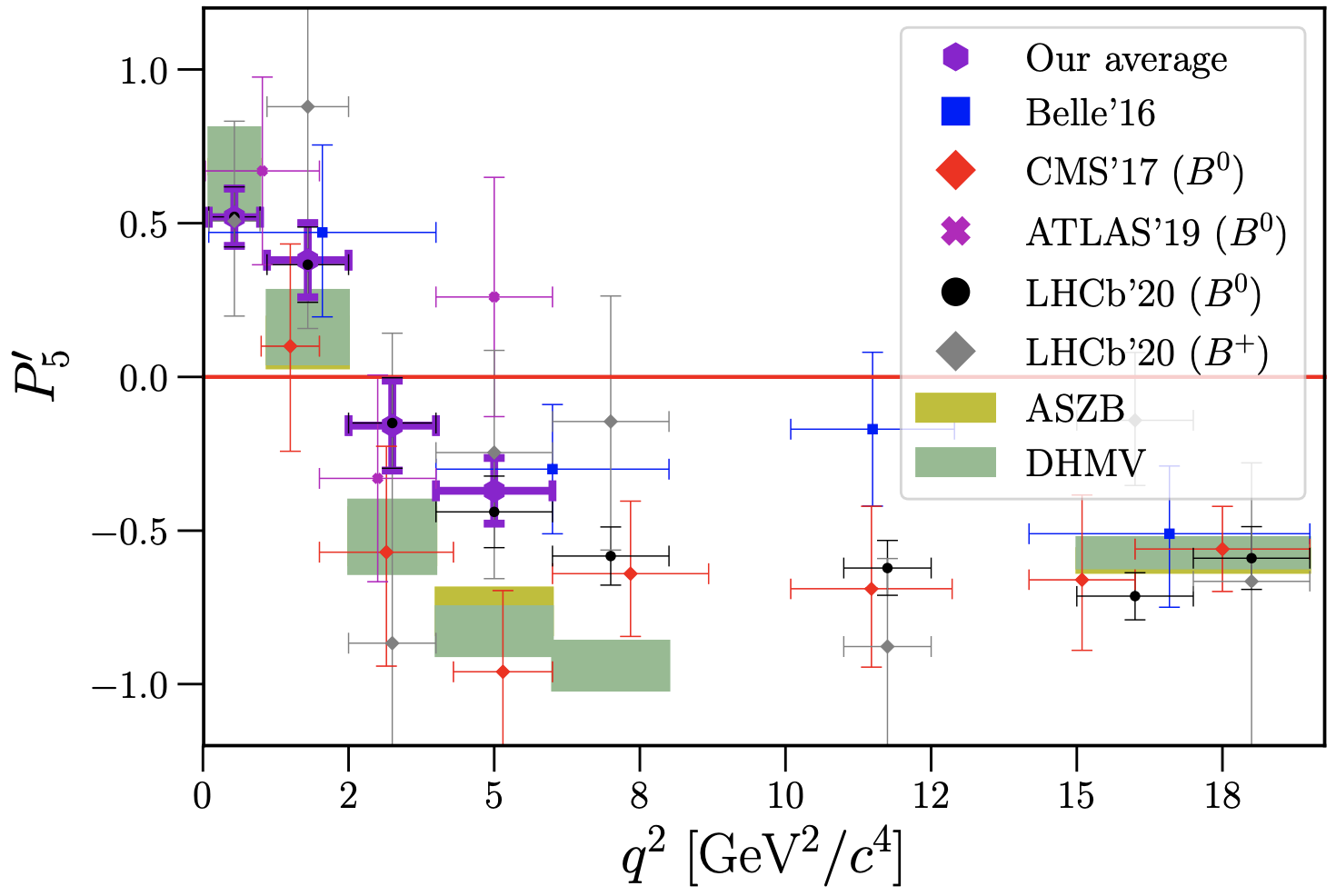}
    \caption{Overview of several measurements of the angular observable $P_5'$ in several bins, by Belle~\cite{Belle:2016fev}, CMS~\cite{CMS:2017rzx}, ATLAS~\cite{ATLAS:2018gqc}, and LHCb~\cite{LHCb:2020lmf,LHCb:2020gog}, together with the SM predictions~\cite{Bharucha:2015bzk,Descotes-Genon:2014uoa}. Figure taken from~\cite{koppenburgtalk}.}
    \label{fig:p5p_overview}
\end{figure}

Among the angular observables it is important to stress that $F_L$, $P_4^{\prime}$ , $P_5^{\prime}$ and $P_8^{\prime}$ have been a driving force in the evolution of the global fits.
Very recently, the LHCb collaboration has updated the results for the angular observables relying on 4.7 fb$^{-1}$ of data~\cite{Aaij:2020nrf, Aaij:2020ruw}: local discrepancies of $2.5\sigma$ and $2.9\sigma$, respectively in the bins $q^2\in[4,6]\,\mathrm{GeV}^2$ and $q^2\in[6,8]\,\mathrm{GeV}^2$ GeV$^2$, were reported.
An overview of measurements of $P_5'$ is shown in Fig.~\ref{fig:p5p_overview}, together with the SM predictions relying on two sets of form factors.
(The definition of this observable will be subsequently provided in Section~\ref{sec:bksll}.)
While these lepton flavour dependent observables are also sensitive to the presence of NP~\cite{Altmannshofer:2008dz,Bobeth:2011nj,Matias:2012xw,DescotesGenon:2012zf,Matias:2014jua}, they are nevertheless subject to hadronic uncertainties (for example form factors, power corrections and charm resonances~\cite{Khodjamirian:2010vf,Khodjamirian:2012rm,Lyon:2014hpa,Descotes-Genon:2014uoa,Capdevila:2017ert,Blake:2017fyh,Jager:2012uw,Jager:2014rwa,Ciuchini:2015qxb,Ciuchini:2016weo,Bobeth:2017vxj,Gubernari:2020eft}) contrary to the LFUV ratios, which are in general free of the latter sources of uncertainty.

\bigskip
Analogously to what was done for $b\to c\ell\nu$ transitions, a way to consistently analyse the aforementioned anomalous experimental data is to adopt the ``effective approach''.
Within WET, the effective Lagrangian for a general $d_j \rightarrow d_i \ell^- \ell^{\prime +}$ transition can be expressed as~\cite{Buchalla:1995vs,Bobeth:1999mk,Ali:2002jg,Hiller:2003js,Bobeth:2007dw,Bobeth:2010wg}
\begin{equation}\label{eqn:effL}
\mathcal L_\text{eff} =
 \frac{4 G_F}{\sqrt{2}} V_{3j}\,V_{3i}^{\ast}\Big[ \sum_{
    \begin{array}{c}
      k=7,8,9,\\
      10,S,P
    \end{array}
  } \hspace{-5mm}\left(C_k (\mu) \,\mathcal{O}_k(\mu) + 
  C_k^{'} (\mu)\,\mathcal{O}_k^{'} (\mu)\right) + 
  C_T (\mu) \,\mathcal{O}_T(\mu) +
  C_{T_5} (\mu) \,\mathcal{O}_{T_5}(\mu)\Big]\, ,
\end{equation}
with $V_{ij}$ denoting the CKM matrix and in which the relevant operators are defined as
\begin{align}\label{eqn:operators}
\mathcal O_7^{ij} &= \,\frac{e \,m_{d_j}}{(4\pi)^{2}}
(\bar d_i\,
\sigma_{\mu\nu}\,P_R\, d_j)\,F^{\mu\nu}\:\text,   &\mathcal O_8^{ij} &=\, \frac{g_s m_{d_j}}{(4\pi)^2}(\bar d_i \sigma_{\mu\nu} P_R d_j)\,G^{\mu\nu}
\:\text, \nonumber \\
\mathcal{O}_{9}^{ij;\ell \ell^{\prime}} &=\,
\frac{e^{2}}{(4\pi)^2}(\bar d_i \,\gamma^{\mu}\,P_L d_j)(\bar \ell\,
\gamma_\mu \,\ell^\prime)\:\text, 
&\mathcal{O}_{10}^{ij;\ell \ell^{\prime}} &=\, \frac{e^{2}}{(4\pi)^2}
(\bar
d_i \,\gamma^{\mu}\,P_L\, d_j)(\bar \ell \,\gamma_\mu\,\gamma_5 \,\ell^\prime)\:\text, \nonumber \\
\mathcal{O}_S^{ij;\ell \ell^{\prime}} &=\, \frac{e^{2}}{(4\pi)^2}(\bar
d_i \,P_R \,d_j)(\bar\ell\,\ell^\prime)\:\text,
&\mathcal{O}_P^{ij;\ell \ell^{\prime}} &=\, \frac{e^{2}}{(4\pi)^2}(\bar
d_i \,P_R \,d_j)(\bar\ell\,\gamma_5 \,\ell^\prime)\:\text,  \nonumber \\
\mathcal O_{T}^{ij;\ell \ell^{\prime}} &=\,
\frac{e^{2}}{(4\pi)^{2}}(\bar d_i \sigma_{\mu\nu}\, d_j)(\bar \ell
\sigma^{\mu\nu} \,\ell^\prime)\:\text, 
&\mathcal O_{T5}^{ij;\ell \ell^{\prime}} &=\, \frac{e^{2}}{(4\pi)^{2}}(\bar
d_i \sigma_{\mu\nu} \,d_j)(\bar \ell \sigma^{\mu\nu}\,\gamma_5\,
\ell^\prime)\:\text,
\end{align}
where the primed operators $\mathcal O_{7, 8, 9, 10, S, P}^{\prime}$
correspond to the exchange $P_L \leftrightarrow P_R$.  
In the SM, out of the above operators, only $O_{7,8,9,10}$ receive non-vanishing contributions at the $b$-quark (renormalisation) scale, usually set to $\mu_b \simeq 4.8\:\mathrm{GeV}$.
Additionally, in the SM\footnote{One could also consider New Physics contributions to the four-quark operators, but these are stringently constrained by meson mixing observable.}, there are also the charged current four-quark operators $\mathcal O_1$ and $\mathcal O_2$, and the four-quark penguin operators $\mathcal O_{3,...,6}$, which receive contributions at the electroweak scale $\mu_\text{EW}\simeq 2M_W$ and then mix (via RG evolution down to $\mu_b$) into $O_{7,8,9}$. The SM Wilson coefficients have been evaluated up to next-to-next-to-leading order (NNLO) and QCD corrections up to next-to-next-to-leading logarithm (NNLL)~\cite{Bobeth:1999mk,Bobeth:2001jm,Huber:2005ig,Gambino:2003zm,Gorbahn:2004my,Bobeth:2003at,Misiak:2006ab,Huber:2007vv,Gorbahn:2005sa,Greub:2008cy,Altmannshofer:2008dz}. 
The operator mixing furthermore leads to the definition of so-called ``effective'' Wilson coefficients in which the effects of the four-quark operators $\mathcal O_{3,...,6}$ have been absorbed into $\mathcal O_{7,...,10}$.
Following~\cite{Altmannshofer:2008dz}, the latter are defined as
\begin{eqnarray}
    C_7^\text{eff} &=& \frac{4\pi}{\alpha_s} C_7 - \frac{1}{3} C_3 - \frac{4}{9}C_4 - \frac{20}{3}C_5 - \frac{80}{9} C_6 \,,\nonumber\\
    C_8^\text{eff} &=& \frac{4\pi}{\alpha_s}C_8 + C_3 - \frac{1}{6} C_4 + 20  C_5 - \frac{10}{3}C_6\,,\nonumber\\
    C_9^\text{eff} &=& \frac{4\pi}{\alpha_s}C_9 + Y(q^2) \,,\nonumber\\
    C_{10}^\text{eff} &=& \frac{4\pi}{\alpha_s}  C_{10}\,,
\end{eqnarray}
with 
\begin{eqnarray}
    Y(q^2) &=& h(q^2, m_c)\left(\frac{4}{3}C_1 + C_2 + 6 C_3 + 60 C_5\right)\nonumber\\
    &\phantom{=}& - \frac{1}{2}h(q^2, m_b)\left(7 C_3 + \frac{4}{3} C_4 + 76 C_5 + \frac{64}{3} C_6 \right)\nonumber\\
    &\phantom{=}& -\frac{1}{2} h(q^2, 0)\left(C_3 + \frac{4}{3} C_4 + 16 C_5 + \frac{64}{3} C_6\right)\nonumber\\
    &\phantom{=}&  + \frac{4}{3} C_3 + \frac{64}{9} C_5 + \frac{64}{27}C_6 \,,
\end{eqnarray}
in which the function $h$ encodes the quark loop contributions~\cite{Altmannshofer:2008dz}.
For $C_7$ and $C_9$ further NNLL QCD corrections, leading to mixing of the charged current operators $\mathcal O_{1,2}$ and the gluon penguin $\mathcal O_8$ into $\mathcal O_{7,9}$, have to be taken into account, see e.g.~\cite{Beneke:2001at,Seidel:2004jh,Greub:2008cy,deBoer:2017way} for further details. 
The coefficient $C_9$ is further susceptible to additional long-distance corrections, see e.g.~\cite{Khodjamirian:2010vf,Khodjamirian:2012rm,Descotes-Genon:2015uva}.
At the $b$-quark scale $\mu_b = 4.8\:\mathrm{GeV}$ (and at $q^2=0$), the SM Wilson coefficients are given by~\cite{Altmannshofer:2008dz}
\begin{equation}
    C_7^\text{eff} = -0.304\,,\quad C_8^\text{eff} =  -0.167\,,\quad C_9^\text{eff} - Y(q^2) = 4.211\,,\quad C_{10}^\text{eff} = -4.103\,.
    \label{eqn:wcsm}
\end{equation}
Furthermore, apart from operator mixing due to RG running under NNLL QCD corrections, there are several hadronic corrections that need to be taken into account: these can be separated into factorisable corrections (to the form factors) and non-factorisable corrections (that cannot be absorbed into form factors). For a review and an assessment of their impact on the observables see e.g.~\cite{Capdevila:2017ert}.
For the extremely challenging SM calculation of $b\to s\ell\ell$ observables see~\cite{Buras:1993xp,Ali:1996bm,Ali:1999mm,Ali:2002jg,Hiller:2003js,Bobeth:2007dw,Bobeth:2008ij,Bobeth:2010wg,Bobeth:2011gi,Bobeth:2011nj,Bobeth:2012vn,Beneke:2001at,Khodjamirian:2010vf,Altmannshofer:2008dz,Descotes-Genon:2013vna,Bharucha:2015bzk,Bobeth:2013uxa} and references therein.
Thus, in the following formulae concerning $b\to s\ell\ell$ transitions, the Wilson coefficients are implicitly understood as
\begin{equation}
    C_i = C_i^{\text{eff, SM}} + C_i^\text{corrections} + \Delta C_i^{\text{NP}}\,.
    \label{eqn:wcsmnp}
\end{equation}

\medskip
Given the above WET parametrisation of New Physics,
the first question to address concerns the set(s) of Wilson coefficients seemingly
preferred by the anomalous experimental data, which then leads to the identification of possible phenomenological candidates, and ultimately to the construction of UV complete extensions of the SM. 

In what follows, we will first examine the dependence of the numerous observables in the $b\to s\ell\ell$ system on New Physics Wilson coefficients.
The observables in $B\to K\ell\ell$ and $B\to K^\ast\ell\ell$ transitions are very distinct; while $B\to K\ell\ell$ corresponds to a true three-body final state, $B\to K^\ast \ell\ell$ is actually measured as a four-body final state due to the decay of the $K^\ast$ meson, given by $B\to K^\ast (\to K\pi)\ell\ell$. Consequently, the phenomenology of this decay is a lot richer and more complciated to treat theoretically.
Thus, we will discuss them separately in the following.
However, before we proceed, we will first discuss
the rare $B_{(s)}\to \ell\ell$ decays. 
As previously discussed, the measurement of this decay in agreement with the SM prediction sets tight constraints on possible New Physics contributions.
Following e.g. Refs.~\cite{Becirevic:2012fy,Bobeth:2013uxa,Becirevic:2016zri}, the decay width of the pseudo-scalar $B_s$ meson can be written as
\begin{eqnarray}
    \Gamma_{B_s\to \ell^+\ell^-} &=& f_{B_s}^2 m_{B_s}^2 \frac{G_F^2\alpha_e^2}{64\pi^3}|V_{tb}V_{ts}^\ast|^2\sqrt{1-\frac{4m_\ell^2}{m_{B_s}^2}}\times \Bigg[\frac{m_{B_s}^2}{m_b^2}\left|C_S^{bs\ell\ell} - C_S^{\prime\,bs\ell\ell}\right|^2\left(1-\frac{4m_\ell^2}{m_{B_s}^2}\right)\nonumber\\
    &\phantom{=}&  + \left|\frac{m_{B_s}}{m_b}(C_P^{bs\ell\ell} - C_P^{\prime\,bs\ell\ell}) + 2\frac{m_\ell}{m_{B_s}}(C_{10}^{bs\ell\ell} - C_{10}^{\prime\,bs\ell\ell})\right|^2\Bigg]\,,
    \label{eqn:bstomumu}
\end{eqnarray}
in which the decay constant is defined via
\begin{eqnarray}
    \langle0|\bar s\gamma_\mu \gamma_\mu\gamma_5 b|B_s(p)\rangle = i p_\mu f_{B_s}\,.
\end{eqnarray}
Especially the scalar and pseudo-scalar coefficients are strongly constrained by data~\cite{LHCb:2021qbv,LHCb:2021awg} on $B_{(s)}\to \mu^+\mu^-$, since they do not suffer from a suppression of the lepton mass. For a New Physics analysis concerning the impact of scalar and pseudo-scalar operators see e.g.~\cite{Becirevic:2012fy,Altmannshofer:2017wqy}.

\mathversion{bold}
\subsection{New Physics in $B\to K\ell^+\ell^-$}
\mathversion{normal}
\label{sec:bkll}
As previously discussed, we will consider the observables in $B\to K\ell\ell$ and $B\to K^\ast\ell\ell$ decays separately, starting with $B\to K\ell\ell$.
The (integrated) LFU ratio $R_{K}$ are more precisely defined as
\begin{equation}
    R_{K} \equiv \frac{\int_{q^2_\text{min}}^{q^2_\text{max}} \frac{d\Gamma(B\to K\mu^+\mu^-)}{dq^2} dq^2}{\int_{q^2_\text{min}}^{q^2_\text{max}} \frac{d\Gamma(B\to Ke^+e^-)}{dq^2} dq^2}\,.
\end{equation}
In this and the next subsection we will summarise the expressions for the ($q^2$-dependent) differential decay widths, and other observables in the decays.

Following Ref.~\cite{Bobeth:2007dw}, the full distribution of the decay $B\to K\ell^+\ell^-$ can be written as
\begin{equation}
    \frac{d^2 \Gamma (B\to K\ell^+\ell^-)}{d q^2 d\cos \theta} = a_\ell(q^2) + b_\ell(q^2)\cos\theta + c_\ell(q^2)\cos^2\theta\,,
    \label{eqn:bkllq2cos}
\end{equation}
in which $\theta$ is the angle between the $B$-meson and $\ell^-$ in the rest frame of the lepton pair.
The angular coefficients are given by 
\begin{eqnarray}
    a_\ell(q^2) &=& \mathcal C(q^2)\Bigg[q^2\left( \beta_\ell^2(q^2) |F_S(q^2)|^2 + |F_P(q^2)|^2\right) + \frac{\lambda(q^2)}{4}\left(|F_A(q^2)|^2 + |F_V(q^2)|^2\right)\nonumber\\
    &\phantom{=}&  + 4m_\ell^2 m_B^2 |F_A(q^2)|^2 + 2m_\ell(m_B^2 - m_K^2 + q^2)\mathrm{Re}\left(F_P(q^2) F_A^\ast(q^2)\right)\Bigg]\,,\nonumber\\
    b_\ell(q^2) &=& 2 \mathcal C(q^2)\Big\{q^2\left[\beta_\ell^2(q^2)\mathrm{Re}\left(F_S(q^2)F_T^\ast(q^2)\right) + \mathrm{Re}\left(F_P(q^2)F_{T5}^\ast (q^2)\right)\right]\nonumber\\
    &\phantom{=}& + m_\ell\left[\sqrt{\lambda(q^2)}\beta_\ell(q^2)\mathrm{Re}\left(F_S(q^2) F_V^\ast(q^2)\right) + (m_B^2 - m_K^2 + q^2)\mathrm{Re}\left(F_{T5}(q^2)F_A^\ast(q^2)\right)\right]\Big\}\,,\nonumber\\
    c_\ell(q^2) &=& \mathcal C(q^2)\Bigg[q^2 \left(\beta_\ell^2(q^2) |F_T(q^2)|^2 + |F_{T5}(q^2)|^2\right) = \frac{\lambda(q^2)}{4}\beta_\ell^2(q^2)\left(|F_A(q^2)|^2 + |F_V(q^2)|^2\right)\nonumber\\
    &\phantom{=}& + 2m_\ell \sqrt{\lambda(q^2)}\beta_\ell(q^2)\mathrm{Re}\left(F_T(q^2)F_V^\ast(q^2)\right)\Bigg]\,,
\end{eqnarray}
in which
\begin{eqnarray}
    \mathcal C(q^2) &=& \frac{G_F^2 \alpha_e^2 |V_{tb}V_{ts}^\ast|^2}{512\pi^5 m_B^3}\beta_\ell(q^2)\sqrt{\lambda(q^2)}\,,\nonumber\\
    \beta_\ell(q^2) &=& \left(1-\frac{4m_\ell^2}{q^2}\right)\,,\nonumber\\
    \lambda(q^2) &=& q^4 + m_B^4 + m_K^4 -2\left(m_B^2 m_K^2 + m_B^2 q^2 + m_K^2 q^2\right)\,.
\end{eqnarray}
The functions $F_i$ are defined via a Lorentz invariant decomposition of the decay amplitude and are given by
\begin{eqnarray}
    F_V(q^2) &=& \left(C_9^{bs\ell\ell} + C_9^{\prime\,bs\ell\ell}\right)f_+(q^2) + \frac{2 m_b}{m_B + m_K}\left(C_7^{bs} + C_7^{\prime\,bs} + \frac{4 m_\ell}{m_b} C_T\right)f_T(q^2)\,,\nonumber\\
    F_A(q^2) &=& \left(C_{10}^{bs\ell\ell} + C_{10}^{\prime\,bs\ell\ell}\right)f_+(q^2)\,,\nonumber\\
    F_S(q^2) &=& \frac{m_B^2 - m_K^2}{2 m_b}\left(C_S^{bs\ell\ell} + C_S^{\prime\,bs\ell\ell}\right)f_0(q^2)\,,\nonumber\\
    F_P(q^2) &=& \frac{m_B^2 - m_K^2}{2 m_b}\left(C_P^{bs\ell\ell} + C_P^{\prime\,bs\ell\ell}\right)f_0(q^2) \nonumber\\
    &\phantom{=}& - m_\ell\left(C_{10}^{bs\ell\ell} + C_{10}^{\prime\,bs\ell\ell}\right)\left[f_+(q^2) - \frac{m_B^2 - m_K^2}{q^2}\left(f_0(q^2) - f_+ (q^2)\right)\right]\,,\nonumber\\
    F_T(q^2) &=& \frac{2\sqrt{\lambda(q^2)}\beta_\ell(q^2)}{m_B + m_K}C_T^{bs\ell\ell} f_T(q^2)\,,\nonumber\\
    F_{T5}(q^2) &=& \frac{2\sqrt{\lambda(q^2)}\beta_\ell(q^2)}{m_B + m_K}C_{T5}^{bs\ell\ell} f_T(q^2)\,,
    \label{eqn:bkllcoeff}
\end{eqnarray}
in which the Wilson coefficients were introduced in Eq.~\eqref{eqn:operators}.
The form factors $f_{0,+,T}(q^2)$ in the above expressions are defined via the following hadronic matrix  elements:
\begin{eqnarray}
    \langle K(k)| \bar s\gamma_\mu b|B(p)\rangle &=& \left[(p+k)_\mu - \frac{m_B^2 - m_K^2}{q^2} q_\mu\right]f_+(q^2) + \frac{m_B^2 - m_K^2}{q^2} q_\mu f_0(q^2)\,,\nonumber\\
    \langle K(k) | \bar s\sigma_{\mu\nu} b |B(p)\rangle &=& - i \left(p_\mu k_\nu - p_\nu k_\mu\right) \frac{2 f_T(q^2)}{m_B + m_K}\,.
\end{eqnarray}
The form factors have been evaluated by means of LQCD and LCSR calculations, see e.g.~\cite{Gubernari:2018wyi}.

The differential decay width, necessary for $R_K$, is then given by
\begin{equation}
    \frac{d\Gamma(B\to K\ell^+\ell^-)}{dq^2} = 2 a_\ell(q^2) + \frac{2}{3}c_\ell(q^2)\,,
\end{equation}
which is obtained by integrating Eq.\eqref{eqn:bkllq2cos} over $\cos\theta$.
One can further define the normalised ``forward-backward asymmetry'' $A_{FB}$ and a so-called ``flat term'' $F_H$, given by~\cite{Bobeth:2007dw}
\begin{equation}
   A_{FB} = \frac{1}{2} \frac{\int_{q^2_\text{min}}^{q^2_\text{max}} b_\ell(q^2) dq^2}{\int_{q^2_\text{min}}^{q^2_\text{max}} a_\ell(q^2) + \frac{1}{3} c_\ell(q^2)dq^2}\,,\quad\quad F_H = \frac{\int_{q^2_\text{min}}^{q^2_\text{max}}a_\ell(q^2) + c_\ell(q^2) dq^2 }{\int_{q^2_\text{min}}^{q^2_\text{max}} a_\ell(q^2) + \frac{1}{3} c_\ell(q^2)dq^2}\,.
\end{equation}
Since both of these observables are normalised via the integrated decay width (in the specific bin), one expects a reduction of hadronic uncertainties, similar to what occurs for $R_K$.

\medskip
A few comments are in order concerning the appropriate $q^2$ ranges.
Firstly, the semi-leptonic $b\to s\ell\ell$ processes are plagued by quarkonia resonances in their $q^2$ distribution, making it impossible to obtain precise predictions- these are the narrow $s\bar s$ resonance $\phi$ and the (broader) charmonium ($c\bar c$) resonances $J/\psi$  and $\psi(2S)$ (higher resonances are subdominant).
Due to the breakdown of QCDF, no theoretical predictions for the $q^2$ bins covering the resonances can be obtained.
Furthermore, for large $q^2$, the LCSR calculations are not directly applicable and LQCD results should be employed. However, one can extrapolate the LCSR results to the large $q^2$ regions via HQET~\cite{Bobeth:2010wg,Bobeth:2012vn} and cross-check with LQCD results.

For these reasons, phenomenological and experimental studies are usually restricted to $q^2 < 6\:\mathrm{GeV}^2$ and $q^2 > 14\:\mathrm{GeV}^2$. 
\mathversion{bold}
\subsection{New Physics in $B\to K^\ast \ell^+\ell^-$}
\begin{figure}[]
    \centering
    \includegraphics[width=0.6\textwidth]{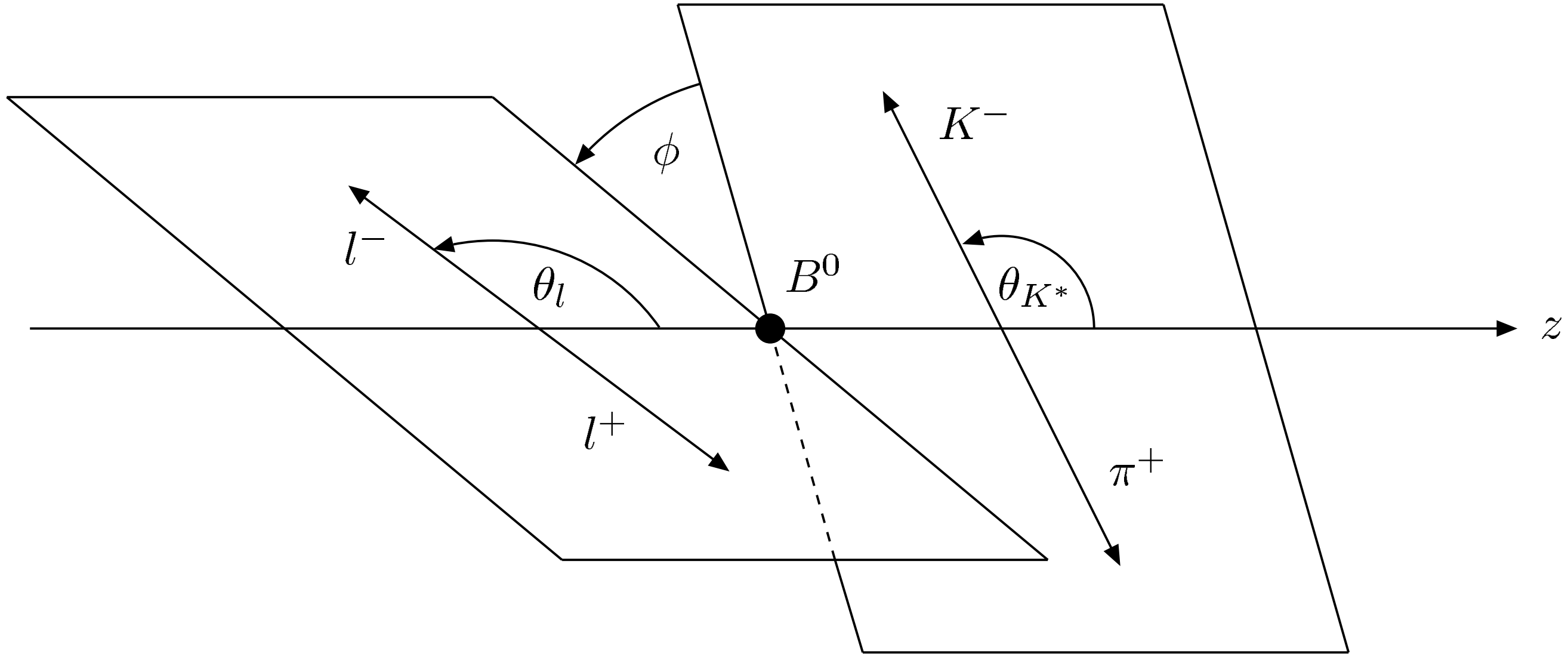}
    \caption{Geometry of the $B\to K^\ast (\to K\pi)\ell\ell$ decay. Figure taken from~\cite{Kruger:2005ep}.}
    \label{fig:angles}
\end{figure}

\mathversion{normal}
\label{sec:bksll}
Analogously to $R_K$, the $R_{K^\ast}$ ratio is defined as
\begin{equation}
    R_{K^\ast } \equiv \frac{\int_{q^2_\text{min}}^{q^2_\text{max}} \frac{d\Gamma(B\to K^\ast \mu^+\mu^-)}{dq^2} dq^2}{\int_{q^2_\text{min}}^{q^2_\text{max}} \frac{d\Gamma(B\to K^\ast e^+e^-)}{dq^2} dq^2}\,.
\end{equation}
Furthermore,  due to its kinematical structure, the decay $\bar B\to \bar K^\ast \ell^+\ell^-$ offers a very rich phenomenology, allowing to construct several independent observables.
Here, the actual decay that is observed in experiment is not $\bar B\to \bar K^\ast \ell^+\ell^-$, but $\bar B\to \bar K^\ast(\to K\pi) \ell^+\ell^-$, in which the angle between $K$ and $\pi$ is sensitive to the polarisation of $K^\ast$~\cite{Kruger:1999xa}.
Following~\cite{Kruger:1999xa,Altmannshofer:2008dz,Descotes-Genon:2013vna}, the full decay distribution is given by
\begin{eqnarray}
    \frac{d^4 \Gamma}{dq^2 d\cos\theta_\ell d\cos \theta_{K^\ast} d\phi} &=& \frac{9}{32\pi} I(q^2, \theta_\ell, \theta_{K^\ast}, \phi)\,,\nonumber\\
    I(q^2, \theta_\ell, \theta_{K^\ast}, \phi) &=& 
      I_1^s \sin^2\theta_{K^\ast} + I_1^c \cos^2\theta_{K^\ast}
      + (I_2^s \sin^2\theta_{K^\ast} + I_2^c \cos^2\theta_{K^\ast}) \cos 2\theta_\ell
\nonumber \\       
    &\phantom{=}& + I_3 \sin^2\theta_{K^\ast} \sin^2\theta_\ell \cos 2\phi 
      + I_4 \sin 2\theta_{K^\ast} \sin 2\theta_\ell \cos\phi 
\nonumber \\       
    &\phantom{=}& + I_5 \sin 2\theta_{K^\ast} \sin\theta_\ell \cos\phi
\nonumber \\      
    &\phantom{=}& + (I_6^s \sin^2\theta_{K^\ast} +
      {I_6^c \cos^2\theta_{K^\ast}})  \cos\theta_\ell 
      + I_7 \sin 2\theta_{K^\ast} \sin\theta_\ell \sin\phi
\nonumber \\ 
    &\phantom{=}& + I_8 \sin 2\theta_{K^\ast} \sin 2\theta_\ell \sin\phi
      + I_9 \sin^2\theta_{K^\ast} \sin^2\theta_\ell \sin 2\phi\,,
      \label{eqn:bksll_dist}
\end{eqnarray}
in which $\theta_\ell$ describes the angle between $K^\ast$ and $\ell^-$, $\theta_{K^\ast}$ the angle between $K^\ast$ and $K$, and $\phi$ corresponds to the angle between the di-lepton and di-meson planes. An overview of the geometry of the decay is shown geometrically in Fig.~\ref{fig:angles}.
The CP conjugated decay mode $B\to K^\ast \ell^+\ell^-$ is obtained by replacing $I^{(a)}_{1,2,3,4,7} \to \bar I^{(a)}_{1,2,3,4,7}$ and $I^{(a)}_{5,6,8,9}\to -\bar I^{(a)}_{5,6,8,9}$, in which the bar indicates that all weak phases have been conjugated.
We want to stress here that the angular coefficients $I_i^{(a)}$ are actual physical observables which can be measured in experiments. Furthermore, since the decay $B\to K^\ast \ell^+\ell^-$ is self-tagging, the coefficients $I_i$ and $\bar I_i$ can be extracted independently.
The angular coefficients can be written in terms of the so-called transversity amplitudes~\cite{Kruger:2005ep}, and are given by
\begin{eqnarray}
    I_1^s &=& \frac{(2+\beta_\ell^2)}{4} \left[|A_{\perp L}|^2 + |A_{\parallel L}|^2 + |A_{\perp R}|^2 + |A_{\parallel R}|^2 \right] 
            + \frac{4 m_\ell^2}{q^2} \mathrm{Re}\left(A_{\perp L}^{}A_{\perp R}^\ast + A_{\parallel L}^{}A_{\parallel R}^\ast\right)\,, \nonumber
\\
  I_1^c &=&  |A_{0 L}|^2 +|A_{0 R}|^2  + \frac{4m_\ell^2}{q^2} 
               \left[|A_t|^2 + 2\mathrm{Re}(A_{0 L}^{}A_{0 R}^\ast) \right] + \beta_\ell^2 |A_S|^2 \,,\nonumber
\\
  I_2^s &=& \frac{ \beta_\ell^2}{4}\left[ |A_{\perp L}|^2+ |A_{\parallel L}|^2 + |A_{\perp R}|^2+ |A_{\parallel R}|^2\right]\,,\nonumber
\\
  I_2^c &=& - \beta_\ell^2\left[|A_{0 L}|^2 + |A_{0 R}|^2\right]\,,\nonumber
\\
  I_3 &=& \frac{1}{2}\beta_\ell^2\left[ |A_{\perp L}|^2 - |A_{\parallel L}|^2  + |A_{\perp R}|^2 - |A_{\parallel R}|^2\right]\,,\nonumber
\\
  I_4 &=& \frac{1}{\sqrt{2}}\beta_\ell^2\left[\mathrm{Re} (A_{0 L}^{}A_{\parallel L}^\ast + A_{0 R}^{}A_{\parallel R}^\ast) \right]\,,\nonumber
\\
  I_5 &=& \sqrt{2}\beta_\ell\left[\mathrm{Re}(A_{0 L}^{}A_{\perp L}^\ast - A_{0 R}^{}A_{\perp R}^\ast)  
- \frac{m_\ell}{\sqrt{q^2}}\, \mathrm{Re}(A_{\parallel L} {A_S^\ast}+A_{\parallel R} {A_S^\ast})
\right]\,,\nonumber
\\
\label{eq:I6c}
  I_6^s  &=& 2\beta_\ell\left[\mathrm{Re} (A_{\parallel L}^{}A_{\perp L}^\ast - A_{\parallel R}^{}A_{\perp R}^\ast) \right]\,,\nonumber
\\
   I_6^c  & =&
 4 \beta_\ell  \frac{m_\ell}{\sqrt{q^2}}\, \mathrm{Re} \left[ A_{0 L} {A_S^\ast} + A_{0 R} {A_S^\ast} \right]\,,\nonumber
\\
  I_7 &=& \sqrt{2} \beta_\ell \left[\mathrm{Im} (A_{0 L}^{}A_{\parallel L}^\ast - A_{0 R}^{}A_{\parallel R}^\ast) 
+ \frac{m_\ell}{\sqrt{q^2}}\, \mathrm{Im}(A_{\perp L} {A_S^\ast}+A_{\perp R} {A_S^\ast})
\right]\,,\nonumber
\\
  I_8 &=& \frac{1}{\sqrt{2}}\beta_\ell^2\left[\mathrm{Im}(A_{0 L}^{}A_{\perp L}^\ast + A_{0 R}^{}A_{\perp R}^\ast) \right]\,,\nonumber
\\
\label{eq:AC-last}
  I_9 &=& \beta_\ell^2\left[\mathrm{Im} (A_{\parallel L}^{\ast}A_{\perp L} + A_{\parallel R}^{\ast}A_{\perp R})\right]\,.
\end{eqnarray}
In the above, the $s$ or $c$ superscripts refer to the appearance of the angular coefficients in association with a sine or cosine of the corresponding angle (see Eq.~\eqref{eqn:bksll_dist}.
The eight different transversity amplitudes appearing in the angular coefficients above are given by~\cite{Kruger:2005ep,Altmannshofer:2008dz}
\begin{eqnarray}
    A_{\perp L,R}  &=& \mathcal N \sqrt{2} \lambda^{1/2} \bigg[ \left[ (C_9 + C_9^{\prime}) \mp (C_{10} + C_{10}^{\prime}) \right] \frac{ V(q^2) }{ m_B + m_{K^\ast}} + \frac{2m_b}{q^2} (C_7 + C_7^{\prime}) T_1(q^2)\bigg]\,,\nonumber\\
A_{\parallel L,R}  &=& - \mathcal N \sqrt{2}(m_B^2 - m_{K^\ast}^2) \bigg[ \left[ (C_9 - C_9^{\prime}) \mp (C_{10} - C_{10}^{\prime}) \right] 
\frac{A_1(q^2)}{m_B-m_{K^\ast}}
\nonumber\\
&\phantom{=}& +\frac{2 m_b}{q^2} (C_7 - C_7^{\prime}) T_2(q^2)
\bigg]\,,\nonumber\\
A_{0L,R}  &=& - \frac{\mathcal N}{2 m_{K^\ast} \sqrt{q^2}}  \bigg\{ 
 \left[ (C_9 - C_9^{\prime}) \mp (C_{10} - C_{10}^{\prime}) \right]
\nonumber\\
 &\phantom{=}& \times 
\bigg[ (m_B^2 - m_{K^\ast}^2 - q^2) ( m_B + m_{K^\ast}) A_1(q^2) 
 -\lambda \frac{A_2(q^2)}{m_B + m_{K^\ast}}
\bigg] 
\nonumber\\
&\phantom{=}& + {2 m_b}(C_7 - C_7^{\prime}) \bigg[
 (m_B^2 + 3 m_{K^\ast}^2 - q^2) T_2(q^2)
-\frac{\lambda}{m_B^2 - m_{K^\ast}^2} T_3(q^2) \bigg]
\bigg\}\,,\nonumber \\
 A_t  &=&\frac{\mathcal N}{\sqrt{q^2}}\lambda^{1/2} \left[ 2 (C_{10} - C_{10}^{\prime}) + \frac{q^2}{m_\mu} (C_{P} - C_{P}^\prime)  \right] A_0(q^2) \,,\nonumber\\
 A_S  &=& - 2\mathcal N \lambda^{1/2} (C_{S} - C_{S}^\prime)  A_0(q^2)\,,
\end{eqnarray}
with the normalisation factor 
\begin{equation}
    \mathcal N = V_{tb}V_{ts}^\ast \left[\frac{G_F^2 \alpha_e^2}{3\times2^{10} \pi^5} q^2 \lambda^{1/2} \beta_\ell\right]^{1/2}\,,
\end{equation}
and $\lambda = m_B^4 + m_{K^\ast}^4 + q^4 - 2(m_B^2 m_{K^\ast}^2 +  m_{K^\ast}^2 q^2 + m_B^2 q^2)$.
The form factors are defined via the hadronic matrix elements as~\cite{Altmannshofer:2008dz,Khodjamirian:2010vf,Bharucha:2015bzk}
\begin{eqnarray}
    \langle \bar K^\ast (k, \varepsilon)|\bar s \gamma_\mu (1 - \gamma_5) b|\bar B(p)\rangle &=& -i \varepsilon_\mu^\ast(m_B + m_{K^\ast}) A_1(q^2) + i (2p-q)_\mu(\varepsilon^\ast q)\frac{A_2(q^2)}{m_B + m_{K^\ast}}\,,\nonumber\\
    &\phantom{=}& + i q_\mu(\varepsilon^\ast q)\frac{2 m_{K^\ast}}{q^2} \left[A_3 (q^2) - A_0(q^2)\right] + \epsilon_{\mu\nu\rho\sigma}\varepsilon^{\nu\ast} p^\rho k^\sigma \frac{2 V(q^2)}{m_B - m_{K^\ast}}\,,\nonumber\\
    A_3(q^2) &=& \frac{m_B + m_{K^\ast}}{2m_K^\ast} A_1(q^2) - \frac{m_B - m_{K^\ast}}{2 m_{K^\ast}} A_2(q^2)\,,\nonumber\\
    \langle\bar K^\ast(k,\varepsilon)|\bar s \sigma_{\mu\nu}q^\nu(1 + \gamma_5)b|\bar B(p)\rangle &=& 2i \epsilon_{\mu\nu\rho\sigma}\varepsilon^{\nu\ast} p^\rho k^\sigma T_1(q^2) + T_2(q^2)\left[\varepsilon_\mu^\ast(m_B^2 - m_{K^\ast}^2) - (\varepsilon^\ast q)(2p-q)_\mu\right] \nonumber\\
    &\phantom{=}& + T_3(q^2)(\varepsilon^\ast q)\left[q_\mu - \frac{q^2}{m_B^2 - m_{K^\ast}^2}(2p-q)_\mu\right]\,,\nonumber\\
    \langle \bar K^\ast | \bar s i\gamma_5 b|\bar B(p)\rangle &=& \frac{2m_{K^\ast}}{m_b + m_s}(\varepsilon^\ast q)A_0(q^2)\,,
\end{eqnarray}
and are calculated at low $q^2$ using LCSR~\cite{Khodjamirian:2010vf,Bharucha:2015bzk}.
The transversity amplitudes are further subject to non-factorisable corrections and charm-loop contributions~\cite{Altmannshofer:2008dz,Khodjamirian:2010vf,Bharucha:2015bzk,Capdevila:2017ert}.

\smallskip
Integrating Eq.~\eqref{eqn:bksll_dist} over the angles, one then obtains the differential decay widths and the CP-averaged branching fraction which, following~\cite{Descotes-Genon:2013vna}, is defined as
\begin{eqnarray}
    \langle \frac{d\Gamma}{dq^2}\rangle &=& \frac{1}{4}\int_{q^2_\text{min}}^{q^2_\text{max}} [3 I_1^c + 6 I_1^s - I_2^c - 2 I_2^s] dq^2\,,\nonumber\\
    \langle \frac{d\mathrm{BR}}{dq^2}\rangle &=& \frac{\langle d\Gamma/dq^2\rangle + \langle d\bar\Gamma/dq^2\rangle}{2 \Gamma_B}\,,
\end{eqnarray}
in which $\Gamma_B$ denotes the total width of the $B$-meson.
The angular coefficients can be in principle directly measured by experiments. 
However, due to the potentially large hadronic uncertainties, it is more desirable to construct ratios in which the uncertainties (at least partially) cancel.
So-called ``optimised observables'' were thus proposed and constructed in~\cite{Descotes-Genon:2013vna}, keeping experimental accessibility in mind as well.
The optimised observables are given by~\cite{Descotes-Genon:2013vna}
\begin{align}
    \langle P_1/dq^2\rangle &= \frac{1}{2}\frac{\int_{q^2_\text{min}}^{q^2_\text{max}}[I_3 + \bar I_3]dq^2}{\int_{q^2_\text{min}}^{q^2_\text{max}}[I_2^s + \bar I_2^s] dq^2}\,, &\langle P_2/dq^2\rangle &= \frac{1}{8}\frac{\int_{q^2_\text{min}}^{q^2_\text{max}}[I_6^s + \bar I_6^s]dq^2}{\int_{q^2_\text{min}}^{q^2_\text{max}}[I_2^s + \bar I_2^s] dq^2}\,,\nonumber\\
    \langle P_3/dq^2\rangle &= -\frac{1}{4}\frac{\int_{q^2_\text{min}}^{q^2_\text{max}}[I_9 + \bar I_9]dq^2}{\int_{q^2_\text{min}}^{q^2_\text{max}}[I_2^s + \bar I_2^s] dq^2} \,, &\langle P_4'/dq^2\rangle &= \frac{1}{2}\frac{\int_{q^2_\text{min}}^{q^2_\text{max}}[I_4 + \bar I_4]dq^2}{\mathcal N' |_{q^2_\text{min}}^{q^2_\text{max}}}\,,\nonumber\\
    \langle P_5'/dq^2\rangle &= \frac{1}{2}\frac{\int_{q^2_\text{min}}^{q^2_\text{max}}[I_5 + \bar I_5]dq^2}{\mathcal N' |_{q^2_\text{min}}^{q^2_\text{max}}}\,, &\langle P_6'/dq^2\rangle &= \frac{1}{2}\frac{\int_{q^2_\text{min}}^{q^2_\text{max}}[I_7 + \bar I_7]dq^2}{\mathcal N' |_{q^2_\text{min}}^{q^2_\text{max}}}\,,\nonumber\\
    \langle P_8'/dq^2\rangle &= \frac{1}{2}\frac{\int_{q^2_\text{min}}^{q^2_\text{max}}[I_8 + \bar I_8]dq^2}{\mathcal N' |_{q^2_\text{min}}^{q^2_\text{max}}}\,, & & \nonumber\\
    \langle A_{FB}\rangle &= -\frac{3}{4}\frac{\int_{q^2_\text{min}}^{q^2_\text{max}}[I_6^s + \bar I_6^s]dq^2}{\langle d\Gamma/dq^2\rangle + \langle d\bar\Gamma/dq^2\rangle}\,, &\langle F_{L}\rangle &= -\frac{\int_{q^2_\text{min}}^{q^2_\text{max}}[I_2^c + \bar I_2^c]dq^2}{\langle d\Gamma/dq^2\rangle + \langle d\bar\Gamma/dq^2\rangle}\,,
\end{align}
which also include the CP-averaged forward-backward asymmetry $A_{FB}$ and the longitudinal polarisation fraction $F_L$.
The normalisation $\mathcal N'$ is given by
\begin{equation}
    \mathcal N'|_{q^2_\text{min}}^{q^2_\text{max}} = \sqrt{-\int_{q^2_\text{min}}^{q^2_\text{max}}[I_2^s + \bar I_2^s]dq^2\int_{q^2_\text{min}}^{q^2_\text{max}}[I_2^c + \bar I_2^c]dq^2}\,.
\end{equation}
Similar observables can also be constructed for the $B_s\to \phi\mu^+\mu^-$ decay~\cite{Bharucha:2015bzk}.

\mathversion{bold}
\section{Global fits to $b\to s\ell\ell$ data}
\mathversion{normal}
\label{sec:bsllfits}
Let us then first proceed to obtain model-independent fits for different possible New Physics scenarios, in terms of non-vanishing contributions to one or several Wilson coefficients $C_i^{\text{NP}}$ in the FCNC $b\to s\ell\ell$ transitions (in addition to their SM values, cf. Eqs.\eqref{eqn:wcsm}  and~\eqref{eqn:wcsmnp}), at the $b$-quark scale, $\mu_b\simeq 4.8\:\mathrm{GeV}$.
In our fits we take into account the experimental data on the LFUV ratios $R_{K^{(\ast)}}$, the differential branching fractions and the CP-averaged angular observables in $B\to K^{(\ast)}\mu^+\mu^-$ and $B_s\to \phi\mu^+\mu^-$ decays\footnote{Since the decay $B_s\to \phi(\to K^+ K^-)\mu^+\mu^-$ is not self-tagging, i.e. the final states do not allow determining whether the decaying meson is $B_s$ or $\bar B_s$, not all of the associated angular observables defined in Section~\ref{sec:bksll} can be measured.}, the $B_{(s)}\to \mu^+\mu^-$ branching fractions and, additionally, in order to constrain the dipole contributions $C_7^{(\prime)}$, data on $b\to s\gamma$ and the data on $B\to  K^\ast e^+e^-$ at very large hadronic recoil (i.e. low $q^2$). All observables and measurements taken into account are listed in Appendix~\ref{app:bsll}.

Concerning the (New Physics) operators that must be taken into account, we firstly note that the operator $\mathcal O_8$ (cf. Eq.~\eqref{eqn:operators}) corresponds to a ``gluon dipole'' and does not directly contribute to the decay rates; after matching at the electroweak scale, its associated Wilson coefficient $C_8$ only mixes into $C_7$ and $C_9$ due to RGE effects.
Hence, we will neglect New Physics contributions to the latter at the $b$-quark scale.
Furthermore, we neglect the (pseudo-) tensor operators $\mathcal O_{T(5)}^{(\prime)}$, since they are not generated at dimension 6 in SMEFT.
Finally, the SMEFT tree-level matching conditions lead to $\Delta C_S = - \Delta C_P$ and $\Delta C_S^\prime = \Delta C_P^\prime$~\cite{Altmannshofer:2021qrr}.
Thus, the remaining operators to consider are $\mathcal O_{7,9,10}^{(\prime)}$ and $\mathcal O_{S,P}^{(\prime)}$.
Moreover, we will only consider the \textit{real} part of the Wilson coefficients, since we only take into account CP-averaged observables which are not sensitive to the imaginary part.
We begin by fitting all viable Wilson coefficients to $b\to s\ell\ell$ data and the dismiss, one by one, the coefficients of operators compatible with $0$, subsequently laying down the minimal New Physics hypotheses that are interesting for model building. 
Details about the fit can be found in Appendix~\ref{app:stats}.

Taking into account all viable Wilson coefficients leads to the fit results shown in Table~\ref{tab:8d}.
Albeit giving a good fit to $b\to s\ell\ell$ data with an improvement over the SM\footnote{We obtain the $p$-value of the SM to be $\sim 1\%$.} of $\sim 5.8\,\sigma$, it can be clearly seen that $C_S$, $C_P$, $C_7$ and $C_7'$ are preferred to be (almost) vanishing.

\begin{figure}[h!]
    \centering
    \mbox{\includegraphics[width=0.48\textwidth]{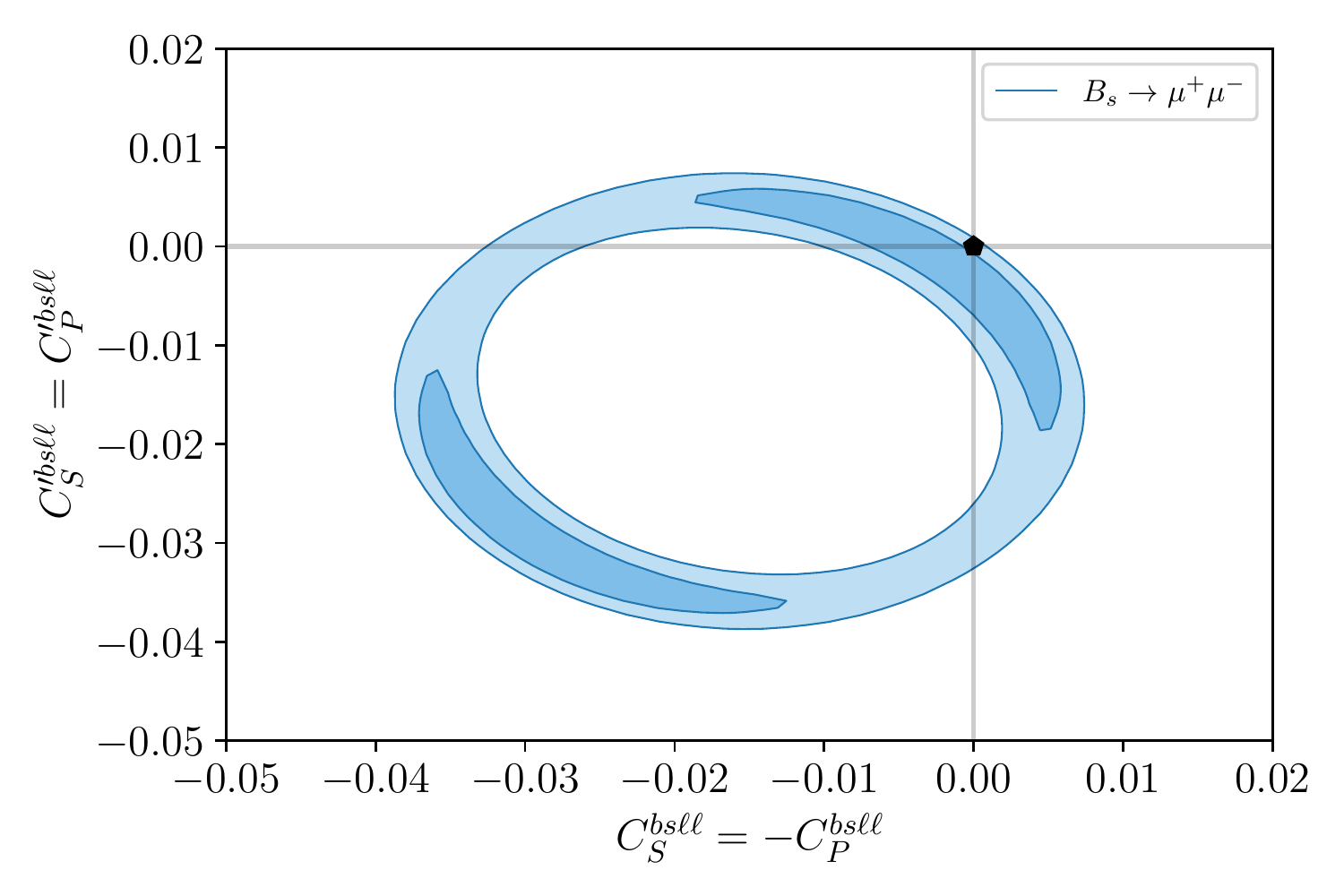}\includegraphics[width=0.48\textwidth]{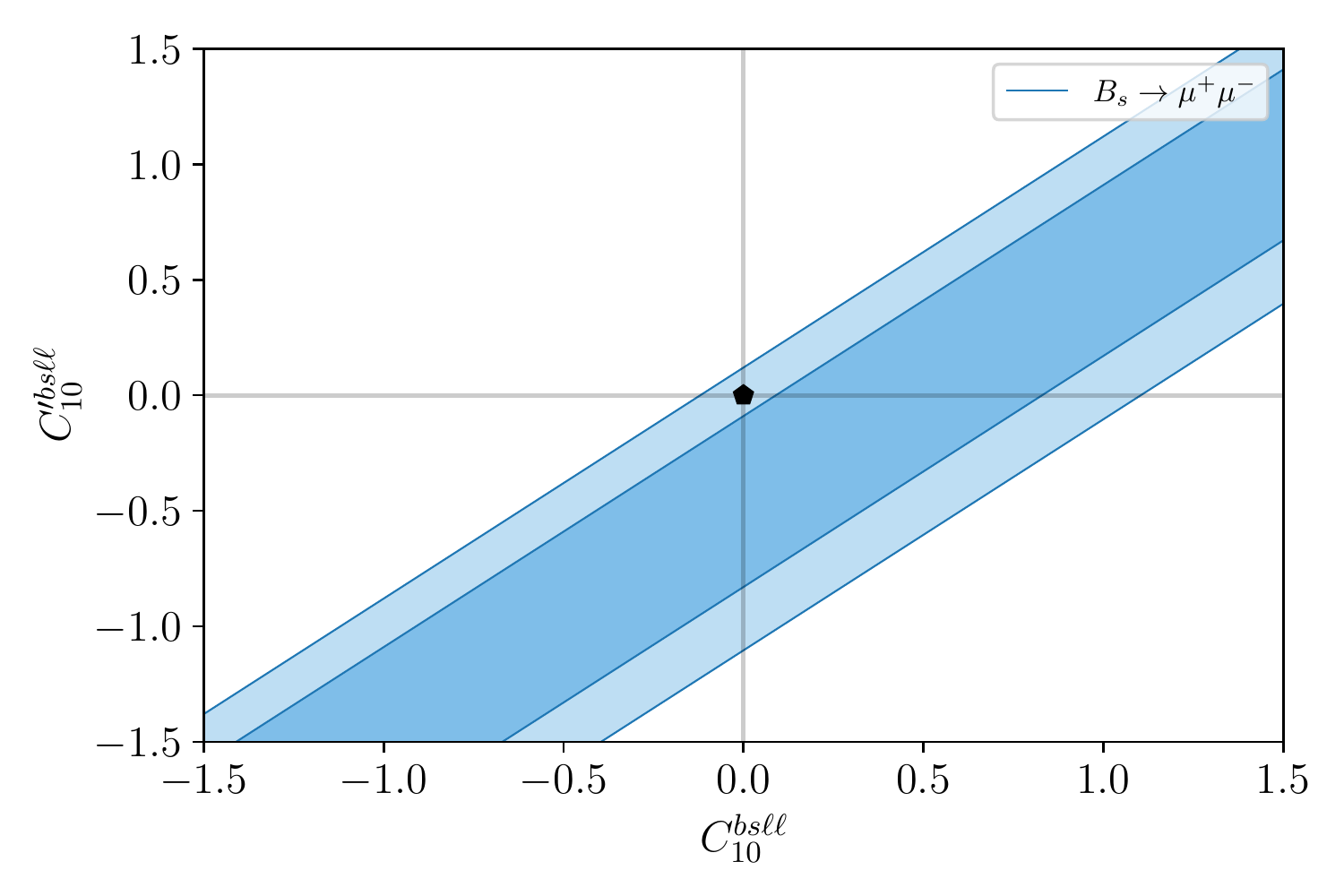}}
    \caption{Constraints derived from $B_s\to\mu^+\mu^-$ data on the relevant Wilson coefficients (cf. Eq.~\eqref{eqn:bstomumu}). The dark (light) shaded bands indicate the $1\,\sigma$ ($2\,\sigma$) confidence regions.
    \textbf{Left}: Constraints of $B_s\to \mu^+\mu^-$ on (pseudo-) scalar Wilson coefficients, with all others set to zero. \textbf{Right}: Constraints of $B_S\to \mu^+\mu^-$ on the left- and right-handed axial coefficients with all others set to zero.}
    \label{fig:CSCSp}
\end{figure}

\begin{figure}[h!]
    \centering
    \mbox{\hspace{-5mm}\includegraphics[width=0.55\textwidth]{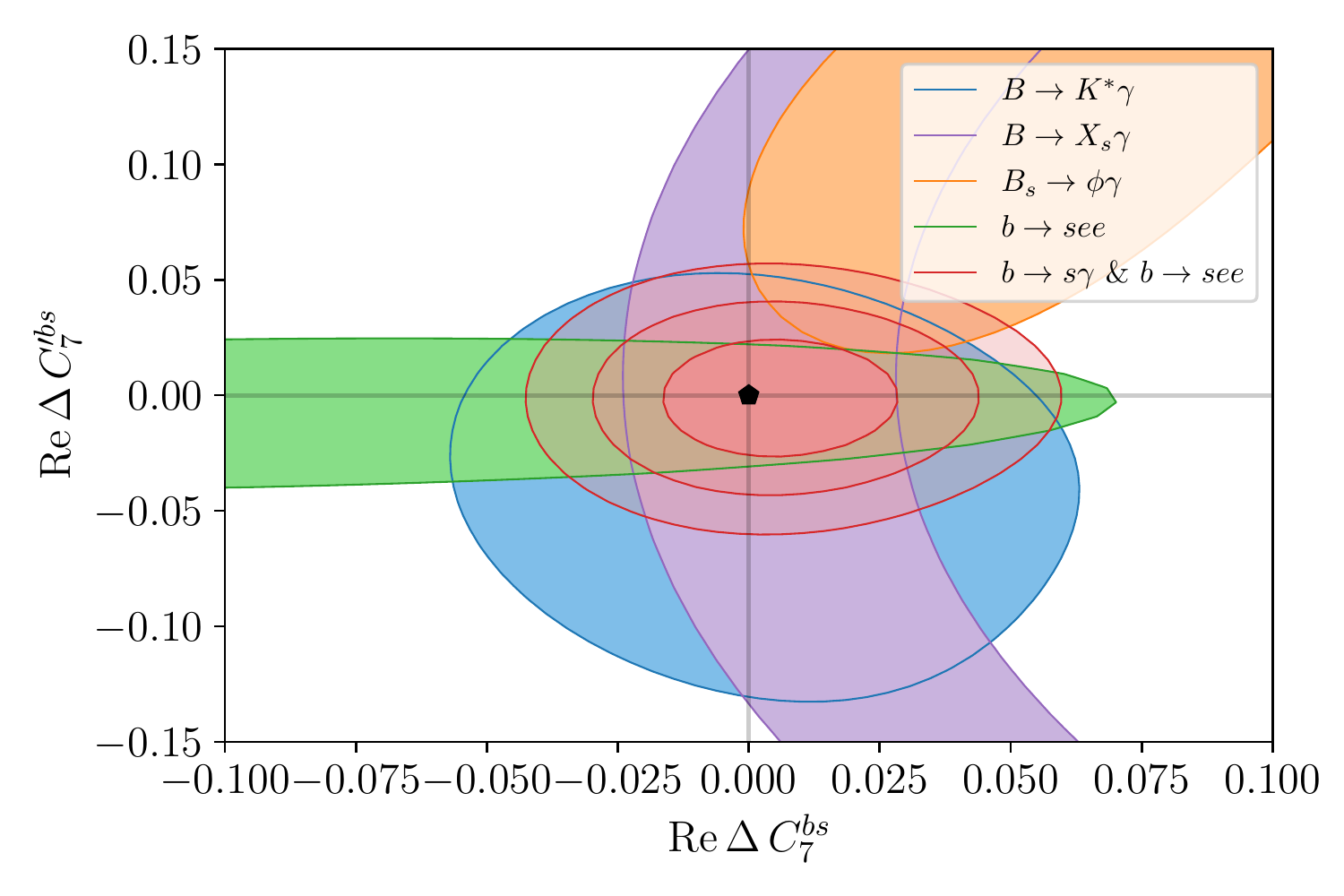}\includegraphics[width=0.54\textwidth]{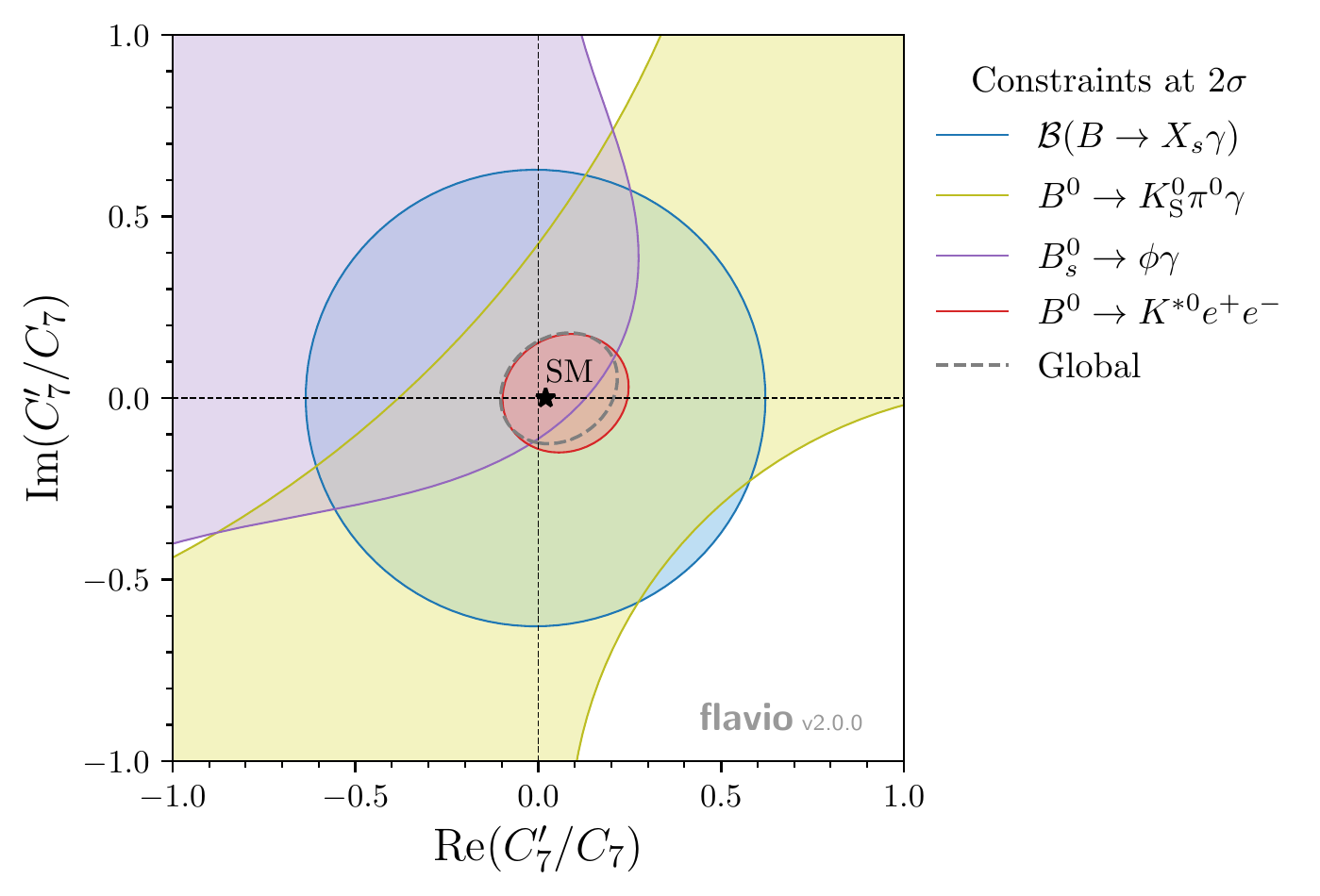}}
    \caption{Constraints on $\Delta C_7^{bs}$ and $\Delta C_7^{\prime bs}$ from data on $b\to s\gamma$ and $b\to see$ transitions at large hadronic recoil. \textbf{Left}: Constraints on the real parts of $\Delta C_7^{bs}$ and $\Delta C_7^{\prime bs}$, all likelihood contours at $1\,\sigma$ (global at $1 - 3\,\sigma$). \textbf{Right}: Constraints on the real and imaginary part of $C_7^{\prime bs}$ normalised by ${C_7^{bs}}_\text{SM} \simeq -0.3$, allowing to constrain the photon polarisation in $B^0\to K^{\ast 0}\gamma$, as obtained by LHCb~\cite{Aaij:2020umj}. Right figure taken from~\cite{Aaij:2020umj}.}
    \label{fig:C7C7p}
\end{figure}

\renewcommand{\arraystretch}{1.3}
\begin{table}[]
    \centering
    \begin{tabular}{|c|c|c|c|c|c|c|c|c|c|}
        \hline
         $\Delta C_9^{bs\mu\mu}$ & $\Delta C_{10}^{bs\mu\mu}$ & $\Delta C_9^{\prime bs\mu\mu}$ & $\Delta C_{10}^{\prime bs\mu\mu}$ & $\Delta C_7^{bs}$ \\ 
         \hline
         $-1.17\pm 0.16$ & $0.09\pm 0.14$ & $0.41\pm 0.34$ & $-0.19 \pm 0.20$ & $0.002\pm 0.014$ \\
         \hline
         \hline
         $\Delta C_7^{\prime bs}$ & $\Delta C_S^{bs\mu\mu} = -\Delta C_P^{bs\mu\mu}$ & $\Delta C_S^{\prime bs\mu\mu} = \Delta C_P^{\prime bs\mu\mu}$ & $\text{Pull}_{\text{SM}}$ & $p$-value\\
         \hline
         $0.006\pm 0.017$ & $-0.001 \pm 0.025$ & $-0.001\pm 0.025$ & $5.8$ & $49.7\%$\\
        \hline
    \end{tabular}
    \caption{Results of a fit of New Physics contributions to all viable Wilson coefficients to all available $b\to s\ell\ell$ data (as listed in Appendix~\ref{app:bsll}). The uncertainties correspond the gaussian uncertainties derived from the hessian matrix at the best fit point (central value).}
    \label{tab:8d}
\end{table}
\renewcommand{\arraystretch}{1.}
As previously discussed, New Physics contributions to scalar and pseudo-scalar operators $\mathcal O_{S,P}^{(\prime)}$ are tightly constrained by data on $B_s\to \mu\mu$ decays. This is shown in the left plot of Fig.~\ref{fig:CSCSp} (see also related discussion in~\cite{Altmannshofer:2021qrr}). Therefore, we will henceforth set these coefficients to zero.
Eliminating these coefficients from the fit leads to a slightly larger pull and $p$-value, as can be seen in the first part of Table~\ref{tab:6d4d}. This is purely due to the fact that the fit function depends on a smaller number of parameters (two less than for the msot general case). As can be seen, $C_7$ and $C_7'$ are still preferred to be close to zero. This scenario has also been considered in~\cite{Alguero:2019ptt,Alguero:2021anc} with which our results are in good agreement.
\renewcommand{\arraystretch}{1.3}
\begin{table}[]
    \centering
    \begin{tabular}{|c|c|c|c|c|c|c|c|}
        \hline
        $\Delta C_9^{bs\mu\mu}$ & $\Delta C_{10}^{bs\mu\mu}$ & $\Delta C_9^{\prime bs\mu\mu}$ & $\Delta C_{10}^{\prime bs\mu\mu}$ & $\Delta C_7^{bs}$ & $\Delta C_7^{\prime bs}$ & $\text{Pull}_{\text{SM}}$ & $p$-value\\
        \hline
        $-1.18^{+0.17}_{-0.16}$ & $0.11^{+0.15}_{-0.14}$ & $0.34^{+0.33}_{-0.33}$ & $-0.25^{+0.18}_{-0.17}$ & $0.001^{+0.014}_{-0.014}$ & $0.005^{+0.014}_{-0.014}$ & $6.1$ & $53.2\%$\\
        \hline
        \hline
        $-1.17_{-0.16}^{+0.17}$ & $0.11_{-0.13}^{+0.15}$ & $0.35_{-0.32}^{+0.32}$ & $-0.25_{-0.17}^{-0.18}$ & --- & --- &$6.5$   & $57.3\%$\\
        \hline
    \end{tabular}
    \caption{Results of fits to all $b\to s\ell\ell$ data considering six (four) independent Wilson coefficients in the top (bottom) row. The uncertainties correspond to the  frequentist intervals of the likelihood.}
    \label{tab:6d4d}
\end{table}
\renewcommand{\arraystretch}{1.}
New Physics contributions to $C_7$ and $C_7'$ are strongly constrained from data on $b\to s\gamma$ transitions. 
Moreover, semi-leptonic $b\to s\ell$ transitions at very large hadronic recoil are dominated by the photon penguins ($\mathcal O_7$ and $\mathcal O_7'$).
The LHCb collaboration recently investigated the angular distribution in $B\to K^\ast e^+e^-$ decays at very large hadronic recoil (low $q^2$) in the bin $[0.0008,0.257]\:\mathrm{GeV}^2$, in order to constrain the photon polarisation in $B^0\to K^{\ast 0}\gamma$ decays~\cite{Aaij:2020umj}.
In Fig.~\ref{fig:C7C7p} we present our results of a fit of $\Delta C_7$ and $\Delta C_7^\prime$ to $b\to s \gamma$ and $B\to K^\ast e^+e^-$ data (on the left), together with results of a fit performed by LHCb~\cite{Aaij:2020umj} (on the right).
It can be clearly seen that $C_7$ and $C_7^\prime$ are constrained to very small values.
We will therefore set them to zero in the following.
Consequently, we obtain for the (now) four-parameter fit the results which are shown in the second part of Table~\ref{tab:6d4d}, in good agreement with the results obtained in~\cite{Altmannshofer:2021qrr}. 
The changes of the best fit point (after neglecting the dipole coefficients) are basically negligible. Again, the pull and the $p$-value slightly increase, since the fit has less free parameters.

Upon inspection of the results shown in Tables~\ref{tab:8d} and~\ref{tab:6d4d} it is evident that all considered New Physics Wilson coefficients are compatible with zero at the $\sim 1-2\,\sigma$ level, with the remarkable exception of $\Delta C_9$, which is preferred at $\mathcal O(1)$.
The inherent New Physics scale which is implied by $\Delta C_9 \simeq \mathcal O(1)$ (cf. related discussion in Section~\ref{sec:eft}) is given by
\begin{equation}
    \Lambda_\text{NP} \simeq \left( \frac{4 G_F}{\sqrt{2}} |V_{tb}V_{ts}|\frac{ e^2}{(4\pi)^2} |\Delta C_9^{bs\mu\mu}|\right)^{-1/2} \simeq 35\:\mathrm{TeV}\,,
    \label{eqn:npscale_rk}
\end{equation}
one order of magnitude larger than the charged current anomalies.
In particular, $\Delta C_{10}$ and $\Delta C_{10}'$ are preferred to be rather small, since sizeable values lead to large contributions to $B_s\to \mu^+\mu^-$. Sizeable values for $\Delta C_{10}^{(\prime)}$ are only viable if $\Delta C_{10}\simeq \Delta C_{10}'$, which leads to a cancellation of the contributions in the $B_s\to \mu^+\mu^-$ decay rate (cf. Eq.~\eqref{eqn:bstomumu}).
This can lso be confirmed by the results previously displayed in Fig.~\ref{fig:CSCSp} (right plot).

\begin{figure}[h!]
    \centering
    \mbox{\includegraphics[width=0.48\textwidth]{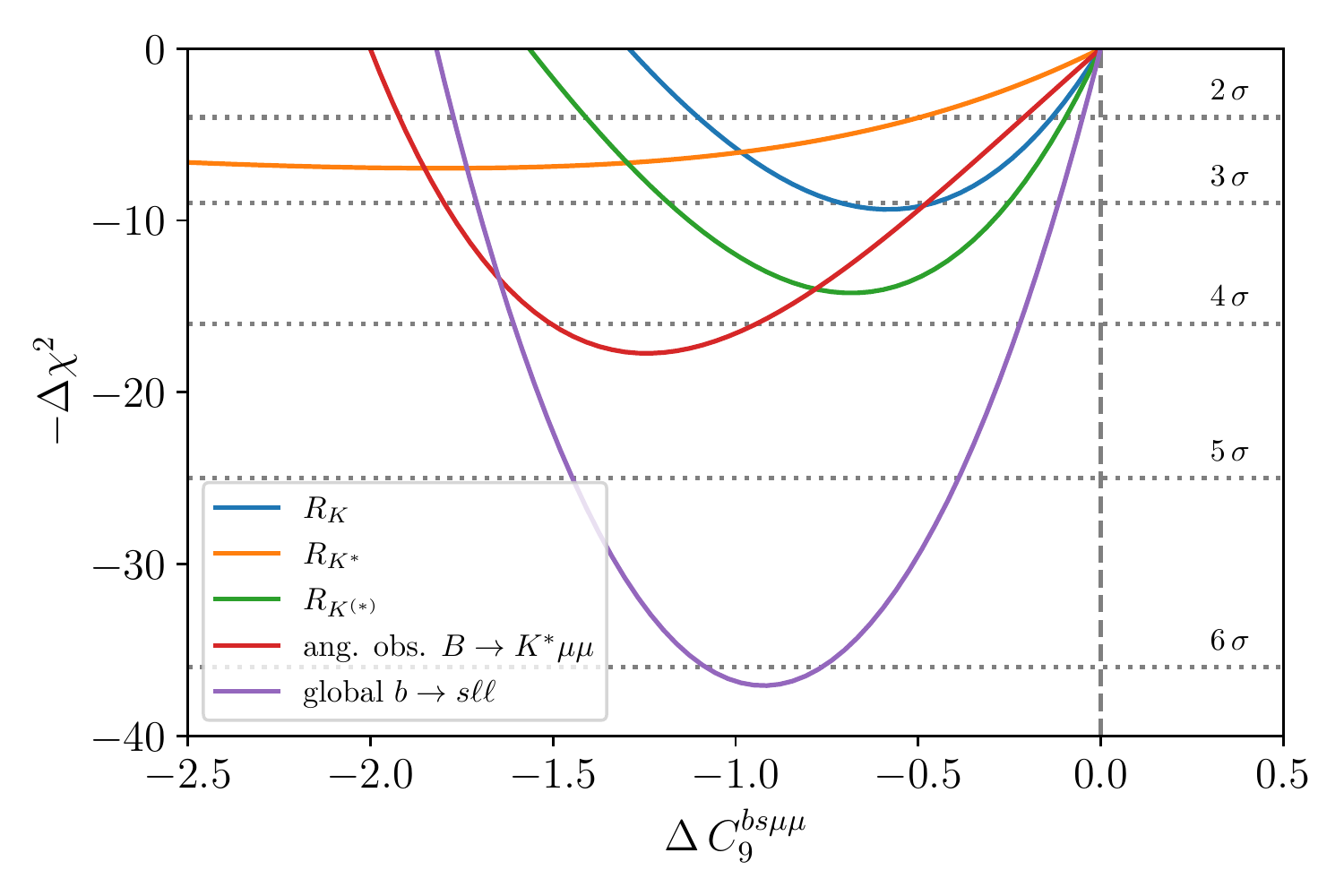}\includegraphics[width=0.48\textwidth]{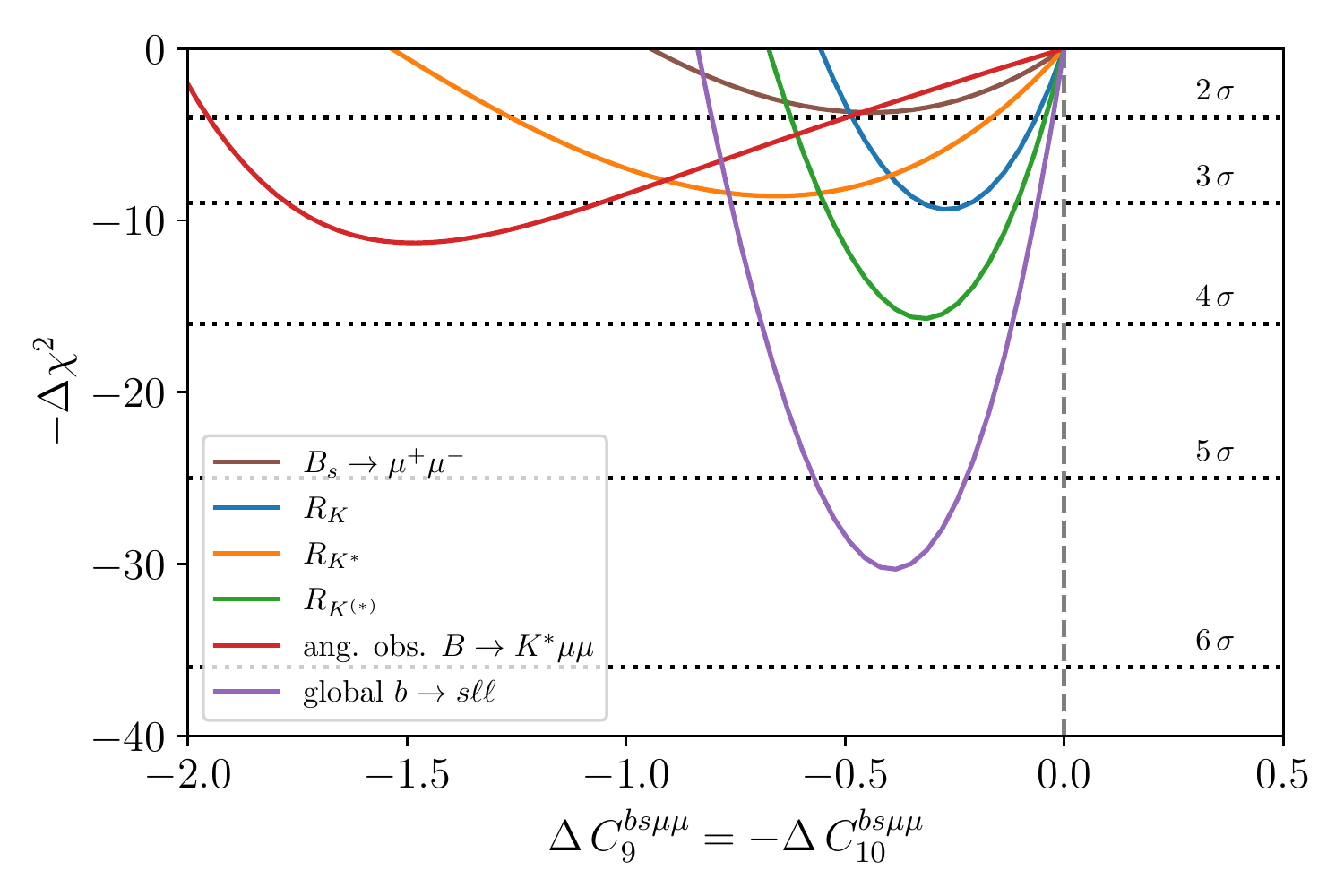}}
    \caption{Profile likelihoods for two different New Physics hypotheses, fitted to the data indicated by the plot legends.
    The profile likelihoods are cast as $-\Delta \chi^2 = -2\log(\mathcal L_\text{NP}/\mathcal L_\text{SM}$, thus indicating the improvement of the presence of New Physics with respect to the SM prediction.
    \textbf{Left}: Profile likelihoods assuming New Physics to be present in $C_9$. \textbf{Right}: Profile likelihoods assuming New Physics to be present as $\Delta C_9 = - \Delta C_{10}$.}
    \label{fig:1d}
\end{figure}

A sizeable New Physics contribution to $C_9$ is in fact preferred by both the LFUV observables $R_{K^{(\ast)}}$ and the angular observables.
However, accommodating $R_{K^{\ast}} < R_K$ is not possible and the small experimental value pushes towards more negative values of $C_9$.
This can be seen in the left plot of Fig.~\ref{fig:1d}, in which we present for non-vanishing values of $C_9^{bs\mu\mu}$ the improvement of agreement with data with respect to the SM prediction in terms of $\Delta \chi^2 = \chi^2_\text{SM} - \chi^2_\text{NP}$. The minimum of the purple line indicate the best fit point which is shown in Table~\ref{tab:1d2d}, while the other coloured lines correspond to subsets of the likelihood, as indicated by the plot legend.
The dotted lines correspond to the $n\,\sigma$ confidence region as derived from the ($d$-dimensional) cumulative distribution function of a $\chi^2$-distributed random variable.

\smallskip
Clearly, a New Physics contribution \textit{only} to $C_9$ is hard (if not impossible) to achieve in what concerns model building.
Naturally, the question then arises concerning which other coefficient should receive New Physics contributions (while having already ruled out contributions to $C_{S,P}^{(\prime)}$ and $C_7^{(\prime)}$).

\renewcommand{\arraystretch}{1.3}
\begin{table}[h!]
    \centering
    \begin{tabular}{|c|r|c|c|c|}
    \hline
    New Physics ``scenario'' & best-fit & $1\sigma$ range & $\text{pull}_\text{SM}$ & $p$-value\\
    \hline
    \hline
    $\Delta C_9^{bs\mu\mu}$ & $-0.92$ & $ [-1.07, -0.77] $ & $6.1$ & $29.2\%$\\
    \hline
    $\Delta C_9^{bs\mu\mu} = -\Delta C_{10}^{bs\mu\mu}$ & $-0.39$ & $[-0.47, -0.32]$ & $5.5$ & $18.3\%$\\
    \hline
    $\Delta C_9^{bs\mu\mu}$ & $-0.86$ & $ [-1.03, -0.66]$ & $ 5.8$ & $28.7\% $\\
    $\Delta C_{10}^{bs\mu\mu}$ & $0.10$ & $[-0.02, 0.22]$ & &\\
    \hline
    $\Delta C_9^{bs\mu\mu} = - \Delta C_{10}^{bs\mu\mu}$ & $-0.47$ & $[-0.55, -0.39]$ & $5.7$ & $25.5\%$\\
    $\Delta C_{9}^{\prime bs\mu\mu} = -\Delta C_{10}^{\prime bs\mu\mu}$ & $0.17$ & $[0.10, 0.24]$ & &  \\
    \hline
    $\Delta C_9^{bs\mu\mu} = - \Delta C_9^{\prime bs\mu\mu}$ & $-1.02$ & $[-1.18, -0.86]$ & $\mathbf{6.4}$ & $43.4\%$ \\
    $\Delta C_{10}^{bs\mu\mu} = \Delta C_{10}^{\prime bs\mu\mu}$ & $0.21$ & $[0.13, 0.29]$ & & \\
    \hline
    $\Delta C_9^{bs\mu\mu}$ & $-1.14$ & $[-1.28, -0.99]$ & $\mathbf{6.6}$ & $49.3\%$\\
    $\Delta C_9^{\prime bs\mu\mu}$ & $0.60$ & $[-0.78, -0.52]$ & &\\
    \hline
    \hline
    $\Delta C_9^{bs\mu\mu} = -\Delta C_{10}^{bs\mu\mu}$ & $-0.62$ & $[-0.79, -0.46]$ & $5.4$ & $20.6\%$\\
    $\Delta C_9^{bsee} = -\Delta C_{10}^{bsee}$ & $-0.30$ & $[-0.39, -0.12]$ & & \\
    \hline
    $\Delta C_9^{bs\mu\mu} = -\Delta C_{10}^{bs\mu\mu}$ & $-0.33$ & $[-0.41, -0.25]$ & $\mathbf{6.4}$ & $41.9\%$\\
    $\Delta C_9^\text{univ.}$ & $-0.86$ & $[-1.05, -0.66]$ & & \\
    \hline
    $\Delta C_9^{bs\mu\mu} = -\Delta C_{10}^{bs\mu\mu}$ & $-0.37$ & $[-0.55, -0.20]$ & $6.1$ & $40.0\%$\\
    $\Delta C_9^{bsee} = -\Delta C_{10}^{bsee}$ & $-0.04$ & $[-0.24, 0.15]$ & & \\
    $\Delta C_9^\text{univ.}$ & $-0.84$ & $[-1.06, -0.61]$ & & \\
    \hline
    \end{tabular}
    \caption{Fits of minimal New Physics hypotheses to $b\to s\ell\ell$ data inspired by model building. The most promising ``candidates'' are highlighted in boldface.}
    \label{tab:1d2d}
\end{table}
\renewcommand{\arraystretch}{1.}

If the underlying New Physics preserves $SU(2)_L$,
one is led to\footnote{In fact, the tree-level matching conditions of the SMEFT at the electroweak scale, up to higher order corrections, also lead to the preservation of $SU(2)_L$ and therefore to $\Delta C_9 = -\Delta C_{10}$. However, in practice this relation is spoiled due to RG running effects.} $\Delta C_9 = -\Delta C_{10}$, replicating the $V-A$ structure of the SM (cf. Eqs.~\eqref{eqn:operators} and ~\eqref{eqn:wcsm}).
As shown in Table~\ref{tab:1d2d}, this restriction gives a reasonable fit to the data, with an improvement of $5.5\,\sigma$ over the SM.
However, this turns out to be worse than only considering $C_9$, since $B_s\to \mu^+\mu^-$ data enforces small values for $C_{10}$, as previously discussed.
Furthermore, as can be seen in the right plot of Fig.~\ref{fig:1d}, the data on the angular observables pushes towards $\Delta C_{9}=-\Delta C_{10}\simeq -1.5$, while the LFUV observables are best accommodated for $C_9=-C_{10}\simeq -0.35$, consequently worsening the agreement with \textit{all} data simultaneously\footnote{One could for instance also consider a relation $C_9 = -C_9'$, that is New Physics coupling to a vector current of quarks and exclusively to right-handed leptons.
In this case one obtains a good fit to the angular observables~\cite{Descotes-Genon:2015uva}. 
However, since the decay width of $B\to K\mu^+\mu^-$ depends on the combination $C_9 + C_9'$ (cf. Eq.~\eqref{eqn:bkllcoeff}), $R_K$ cannot be accounted for in such a scenario. 
Other one-parameter combinations of Wilson coefficients also do not lead to good agreement with data~\cite{Descotes-Genon:2015uva}, and are therefore not discussed here.}.

\begin{figure}[h!]
    \centering
    \mbox{\hspace{-5mm}\includegraphics[width=0.52\textwidth]{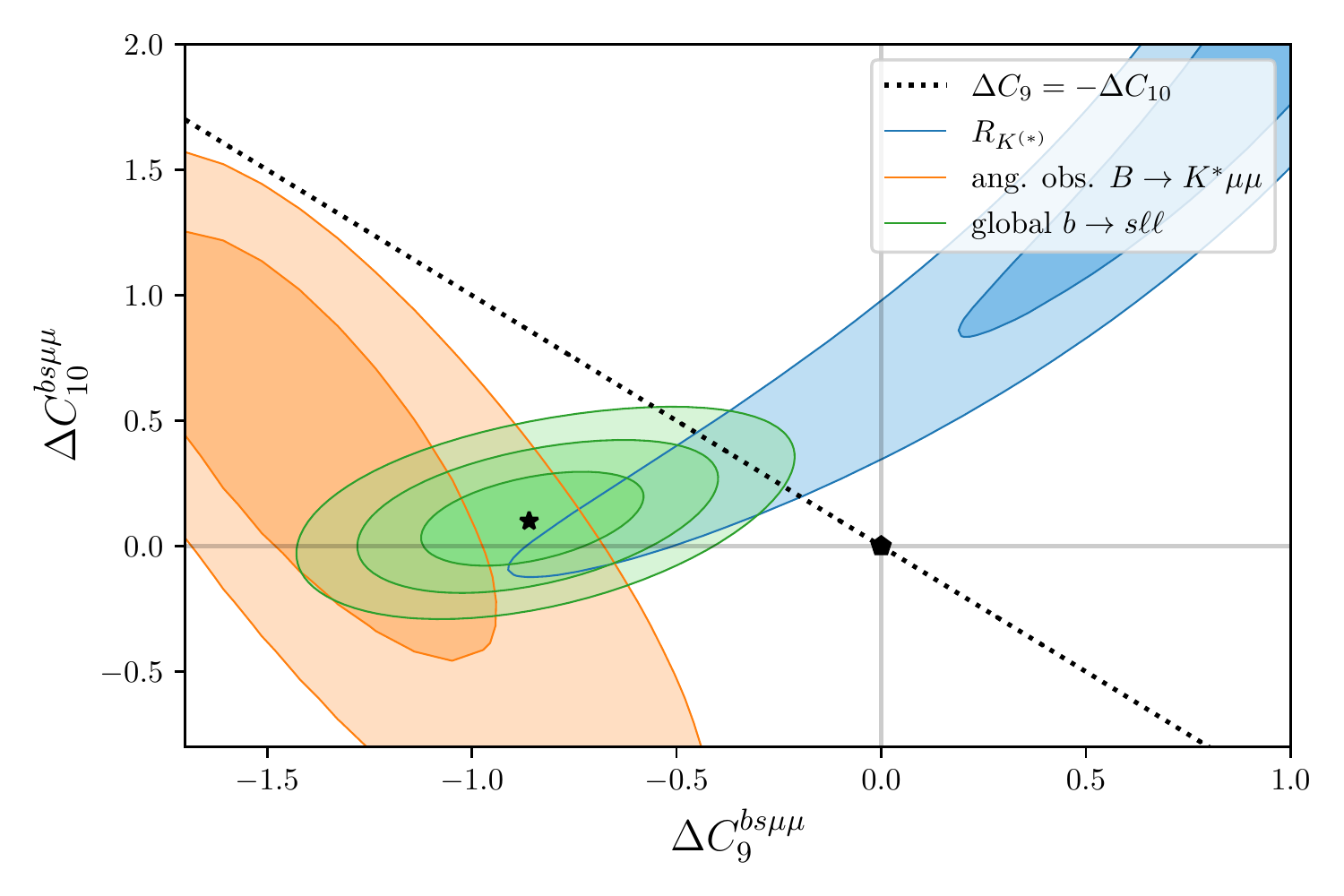}\includegraphics[width=0.52\textwidth]{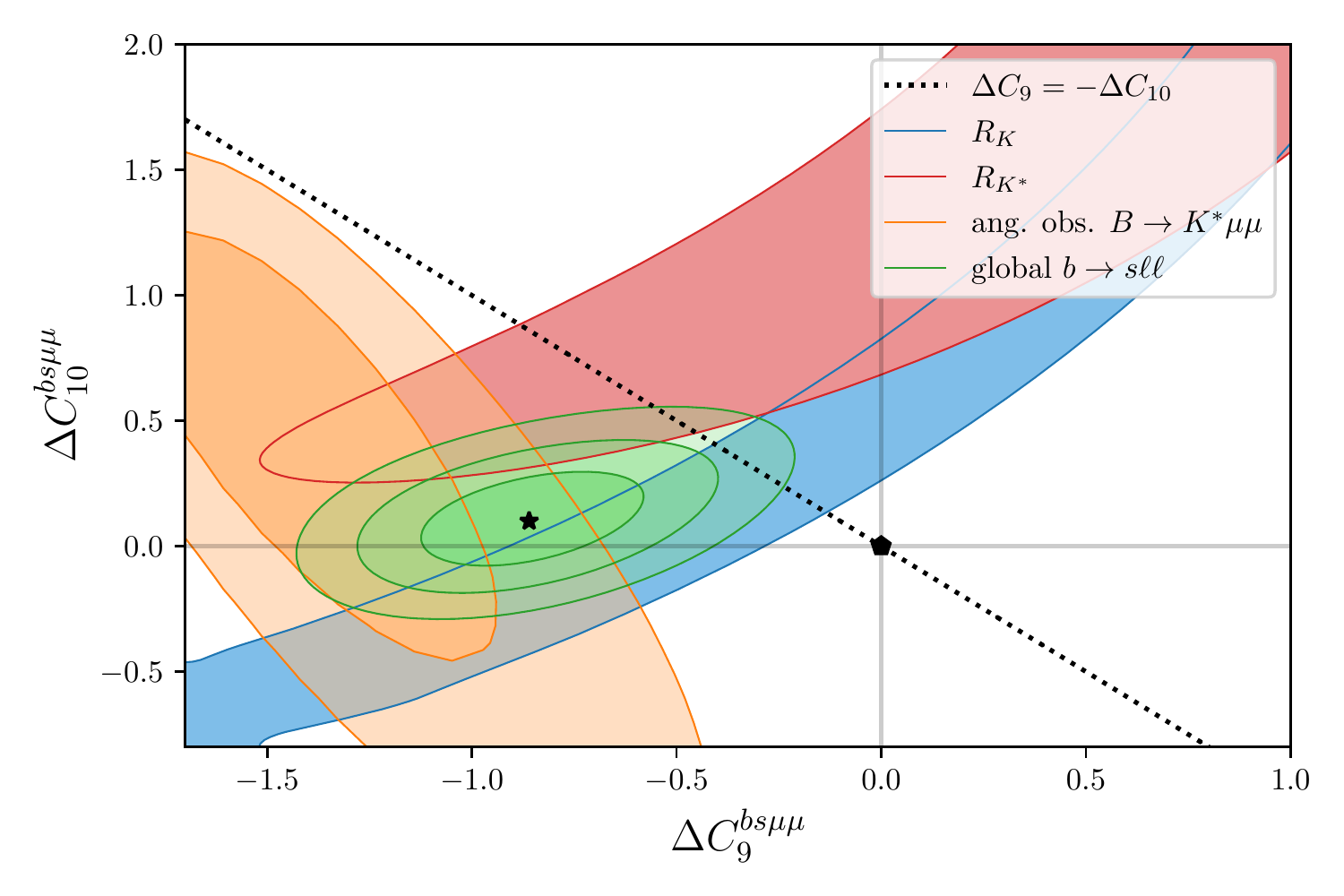}}
    \caption{Results of a fit to $b\to s\ell\ell$ data, assuming New Physics to be present in $C_9$ and $C_{10}$. The results are shown as two-dimensional likelihood contours. Figures updated from~\cite{Kriewald:2021hfc}.}
    \label{fig:C9mu_C10mu}
\end{figure}

Considering New Physics coupling only to left-handed quarks, but now relaxing the condition $\Delta C_9 = -\Delta C_{10}$, leads to a slight improvement of the fit. 
However, as can be seen in Table~\ref{tab:1d2d}, contributions to $C_{10}$ are still preferred to be small, due to the constraint from $B_s\to \mu^+\mu^-$ as discussed before.
Furthermore, there are two different tensions between the observables in this scenario:
this can be seen in the left plot of Fig.~\ref{fig:C9mu_C10mu}, in which we display the $1\,\sigma$ and $2\,\sigma$ likelihood contours around the respective best fit points of $R_{K^{(\ast)}}$ data (shades of blue) and the data of the angular observables (orange).
The global best fit point is indicated by the ``star'' symbol and the corresponding ($1,2,3\,\sigma$) likelihood contours in green.
Firstly, the $1\,\sigma$ likelihood contours of $R_{K^{(\ast)}}$ do not overlap with neither the $1\,\sigma$ contours of the angular observables nor with the global $b\to s\ell\ell$ data (see left plot in Fig.~\ref{fig:C9mu_C10mu}).
Secondly, $R_{K^\ast} < R_K$ cannot be satisfactorily accommodated in this scenario - as previously discussed (see also left plot of Fig.~\ref{fig:1d}) - as further visible in the right plot of Fig.~\ref{fig:C9mu_C10mu}.

\begin{figure}[h!]
    \centering
    \mbox{\hspace{-5mm}\includegraphics[width=0.52\textwidth]{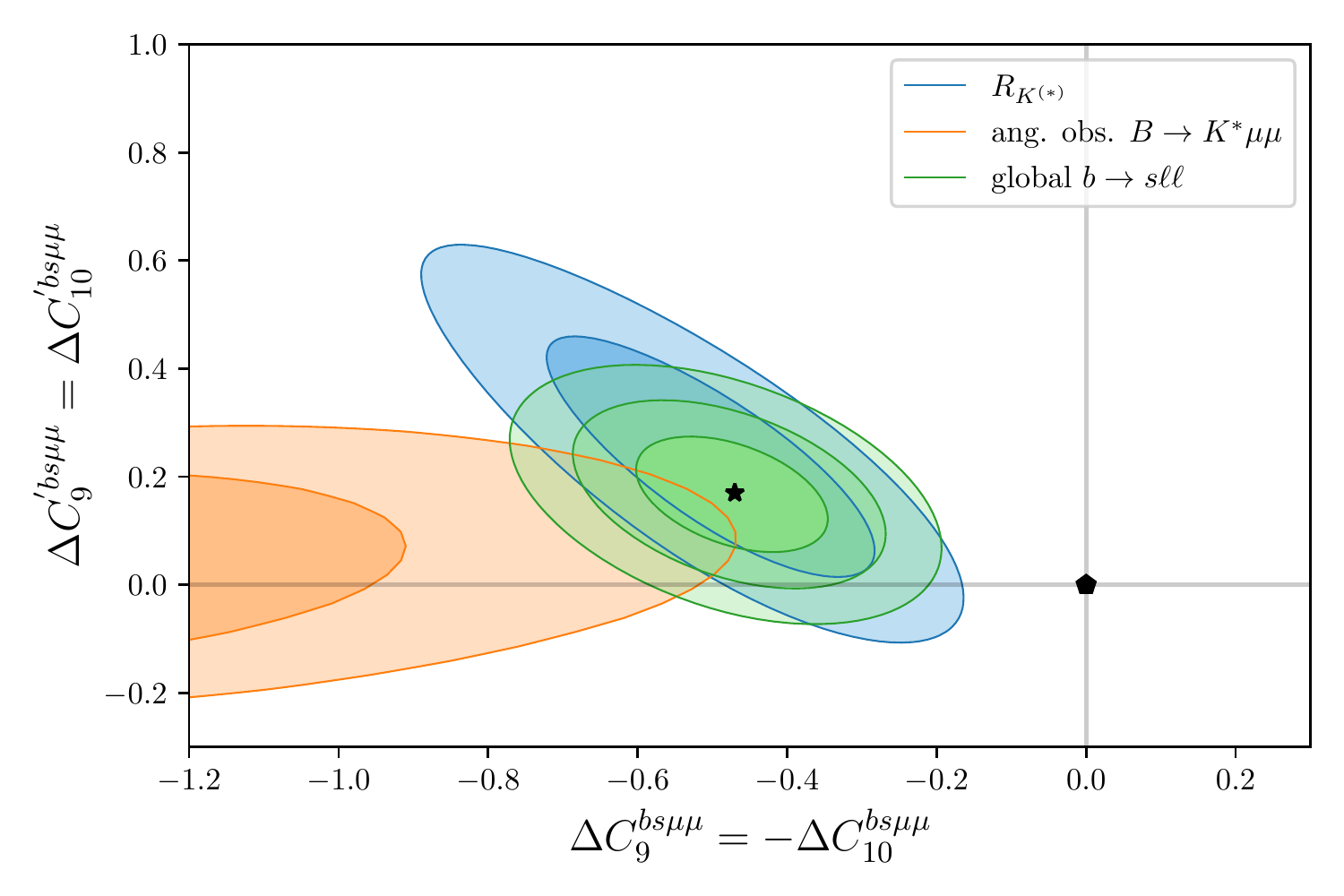}\includegraphics[width=0.52\textwidth]{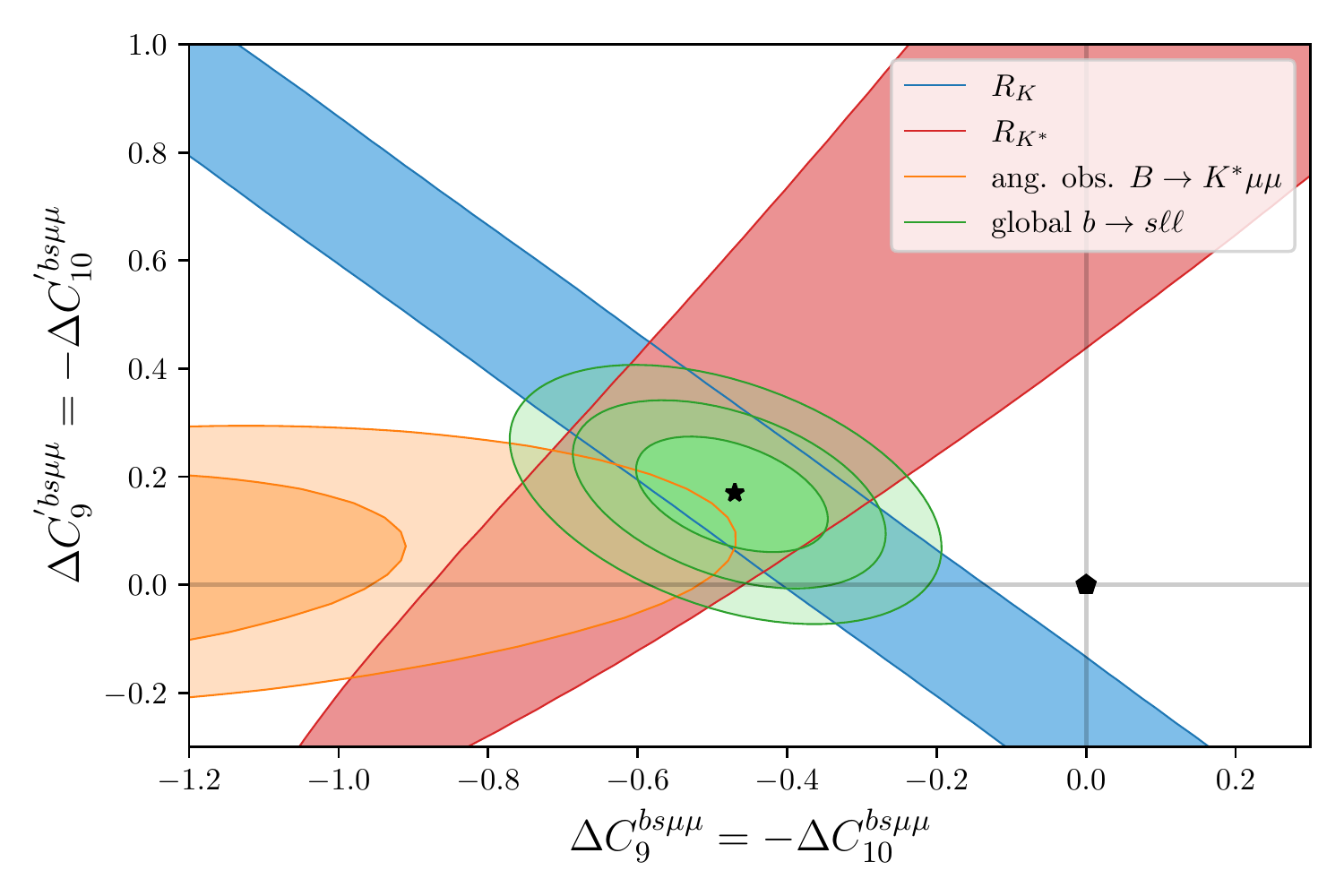}}
    \mbox{\hspace{-5mm}\includegraphics[width=0.52\textwidth]{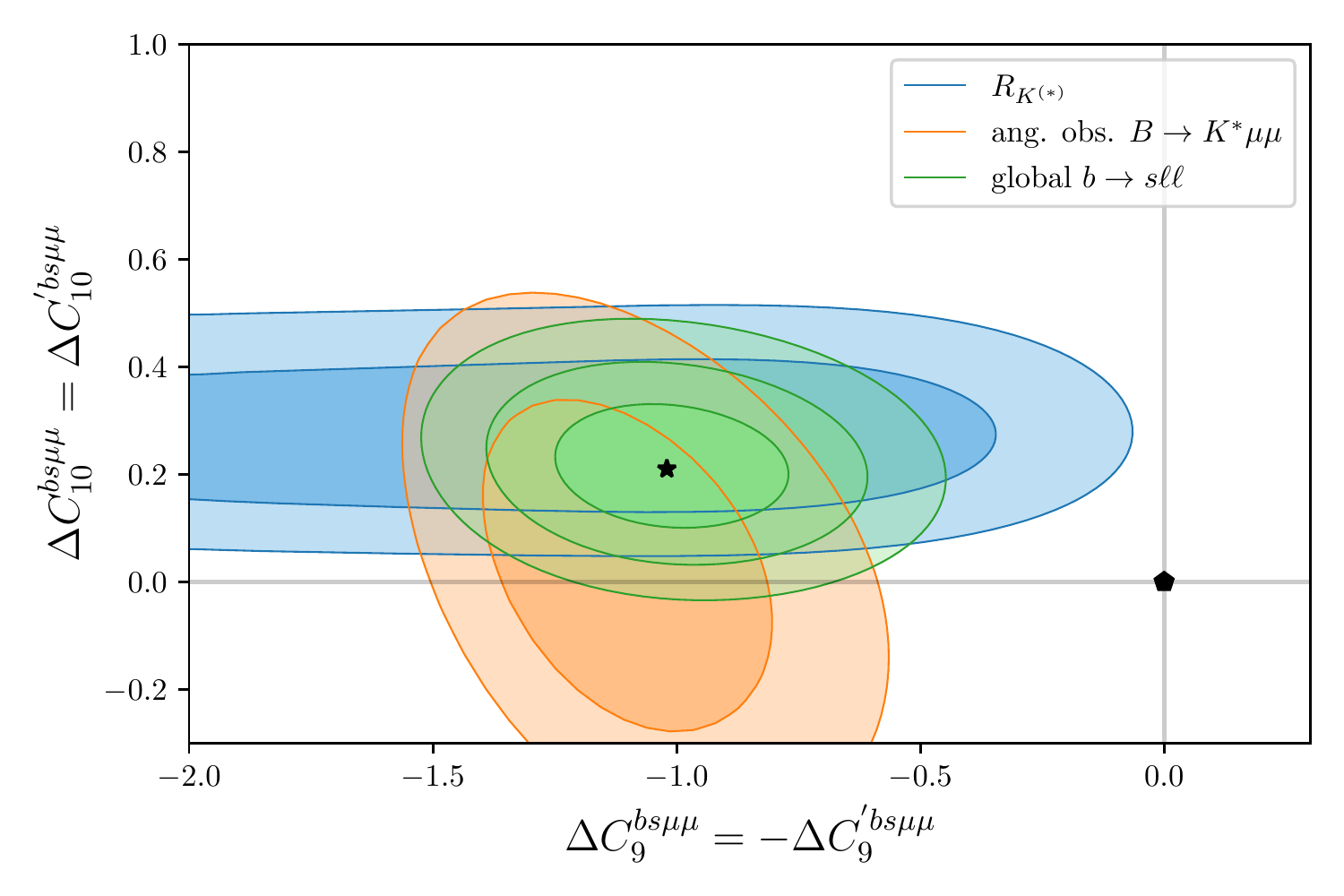}\includegraphics[width=0.52\textwidth]{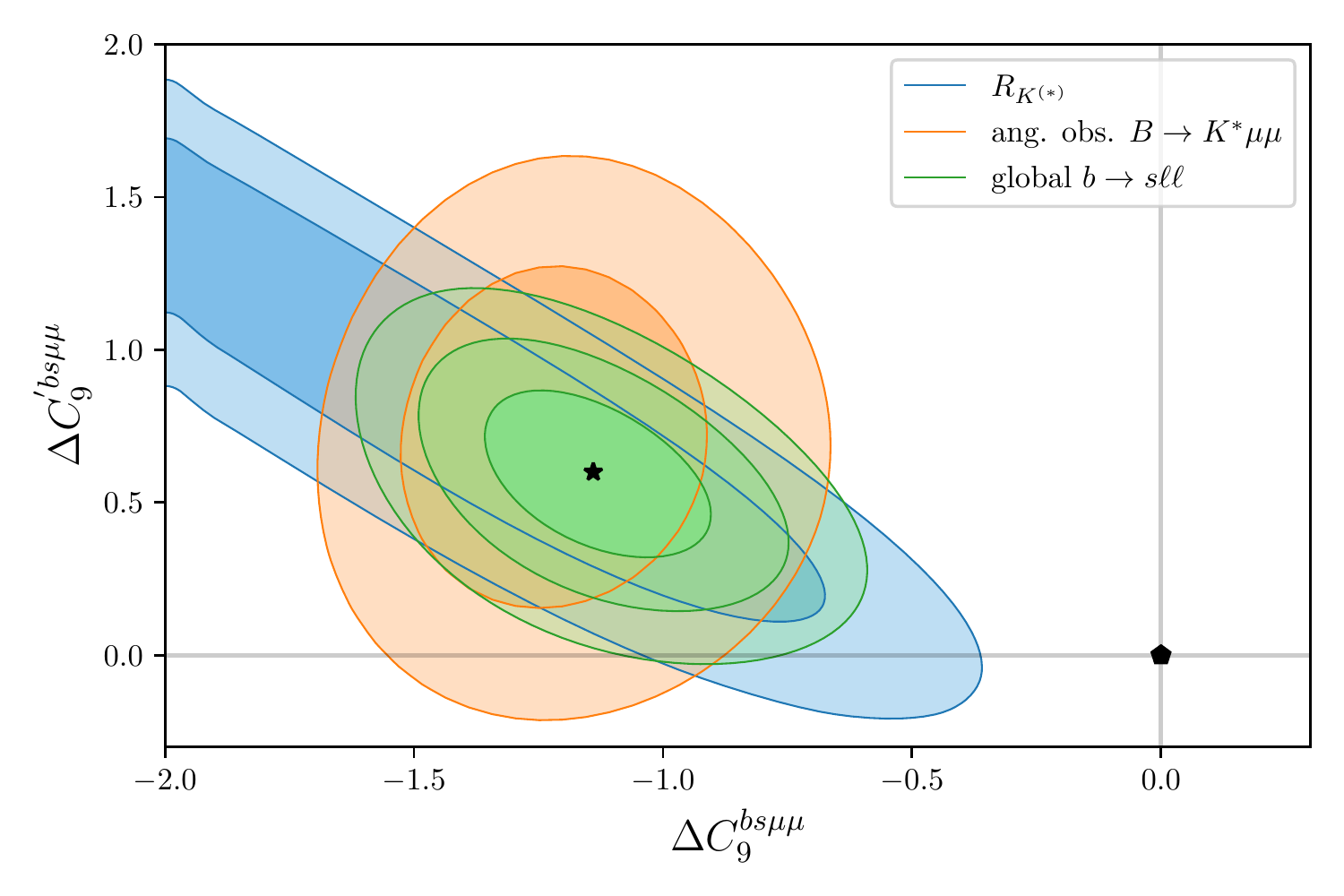}}
    \caption{Fits to $b\to s\ell\ell$ data considering New Physics also coupling to right-handed quarks, for different relations between Wilson coefficients. The best-fit points are shown in Table~\ref{tab:1d2d}.}
    \label{fig:rhfits}
\end{figure}

Furthermore, one can consider various combinations of right-handed quark currents in addition to New Physics contributions to $C_9$.
Firstly, leaving $SU(2)_L$ intact, we impose $\Delta C_9 = -\Delta C_{10}$ and $\Delta C_9' = - \Delta C_{10}'$, leading to a scenario in which New Physics is coupled to both left- and right-handed quarks, but exclusively couples to left-handed leptons.
The agreement between $R_{K^{(\ast)}}$ data and other observables is slightly improved, which can be seen in the upper two plots in Fig.~\ref{fig:rhfits}, however a tension with the angular data still remains.

Another interesting possibility is given by imposing the condition $\Delta C_9 = - \Delta C_9'$ and $\Delta C_{10} = \Delta C_{10}'$, which describes New Physics that couples an axial quark current to a vectorial lepton current and vice-versa, so that combinations of the currents are then linearly dependent.
This leads to a significantly better fit (cf. Table~\ref{tab:1d2d}) and lifts the tension between angular and LFUV observables, as can be seen in the bottom left plot of Fig.~\ref{fig:rhfits}.

Finally, we consider a scenario in which New Physics is only present in $C_9$ and $C_9'$. This corresponds to a case in which New Physics couples to left- and right-handed quarks differently, but only vectorially to charged leptons.
Here, the fit leads to an excellent agreement between the observables and the overall data (cf. Table~\ref{tab:1d2d}), which can be seen in the bottom right plot of Fig.~\ref{fig:rhfits}.

As an alternative, and as pointed out in~\cite{Kumar:2019qbv,Datta:2019zca}, one can consider New Physics contributions to $C_i^{bs ee}$ Wilson coefficients.
Again imposing $SU(2)_L$ invariance, we consider $\Delta C_9^{bs\mu\mu} = - \Delta C_{10}^{bs\mu\mu}$ vs. $\Delta C_9^{bsee} = - \Delta C_{10}^{bsee}$ which leads to a reasonable description of the data. 
However, as previously discussed, the angular observables cannot be well accommodated (cf. left plot in Fig.~\ref{fig:1d} and related discussion).

\begin{figure}[h!]
    \centering
    \mbox{\hspace{-5mm}\includegraphics[width=0.52\textwidth]{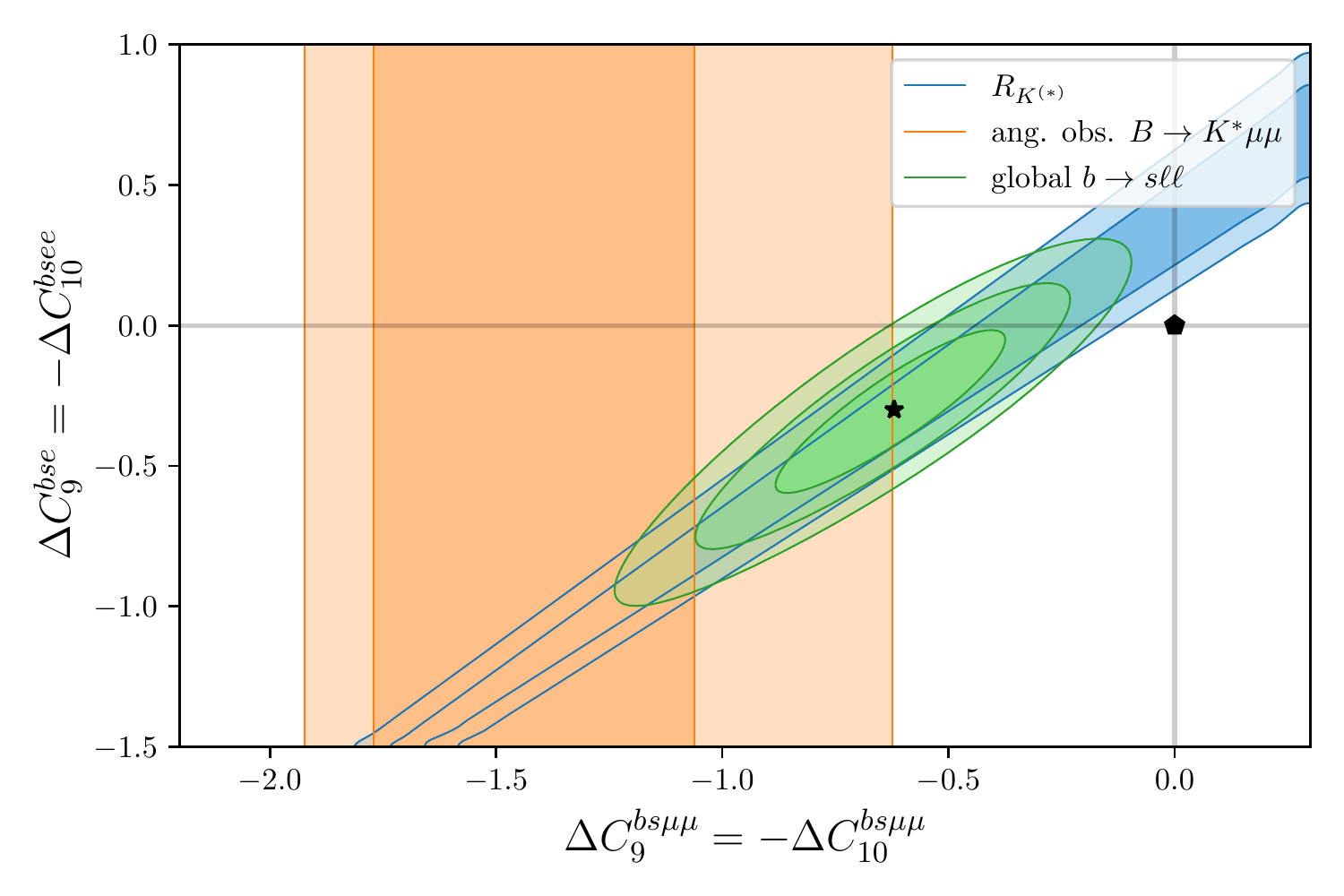}\includegraphics[width=0.52\textwidth]{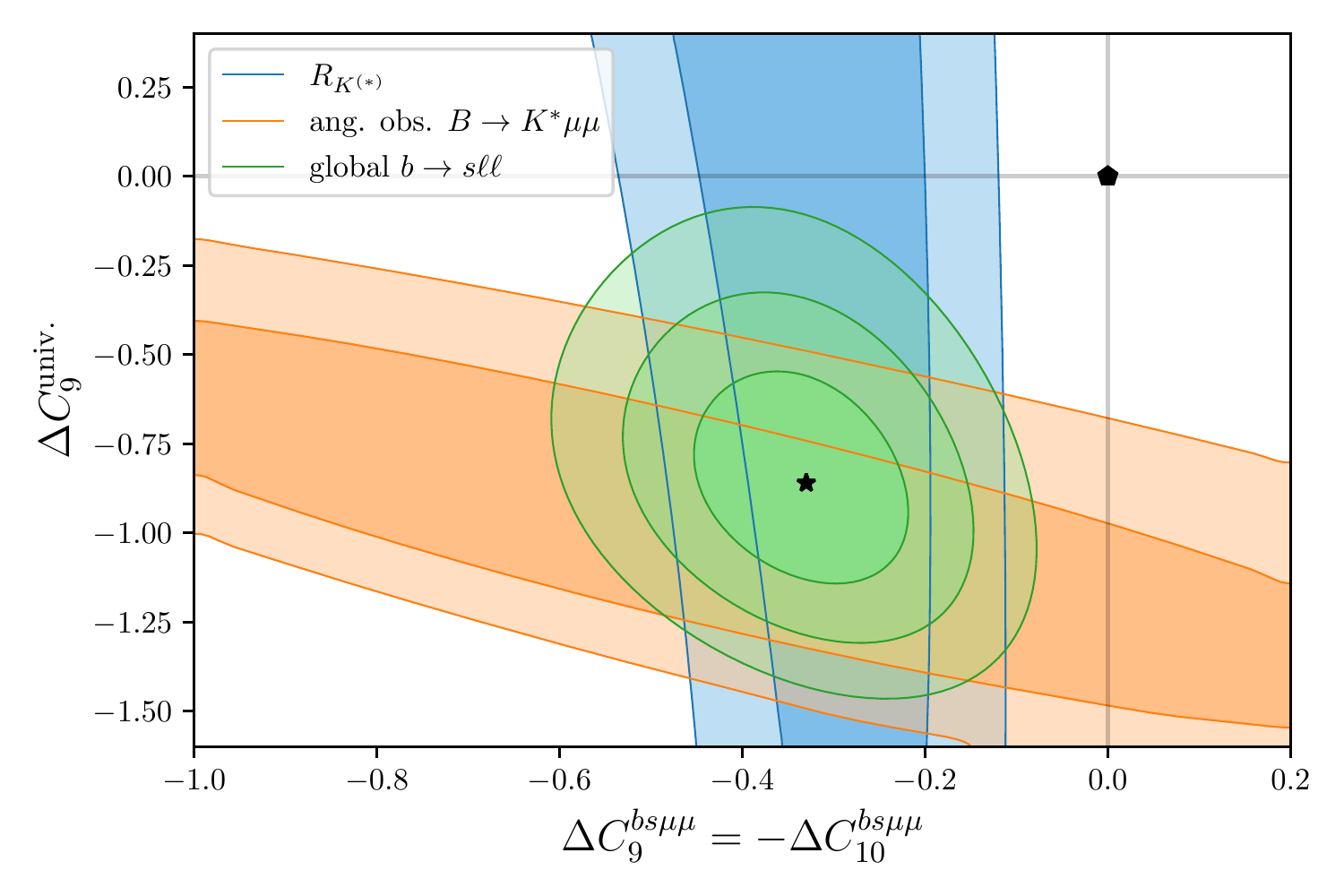}}
    \caption{Fits to $b\to s\ell\ell$ data considering New Physics also coupled to electrons. The best-fit points are shown in Table~\ref{tab:1d2d}. Figures updated from~\cite{Hati:2019ufv, Kriewald:2021hfc}.}
    \label{fig:ee_uni}
\end{figure}

A somewhat different idea, that was first proposed in~\cite{Alguero:2018nvb}, relies on introducing a lepton flavour universal contribution to $C_9$, in addition to $\Delta C_9^{bs\mu\mu} = - \Delta C_{10}^{bs\mu\mu}$.
Thus, the relevant New Physics contributions to the Wilson coefficients are given by
\begin{equation}
    C_9^{bs\mu\mu\text{NP}} = \Delta C_9^{bs\mu\mu} + \Delta C_9^\text{univ}\,,\quad \Delta C_{10}^{bs\mu\mu} = -\Delta C_9^{bs\mu\mu}\,,\quad \Delta C_9^{bsee} = \Delta C_9^\text{univ}\,.
\end{equation}
As can be seen in Table~\ref{tab:1d2d}, a fit of this scenario leads to an excellent description of the data, which is also shown in the right plot of Fig.~\ref{fig:ee_uni}.
A universal contribution to $C_9$ further dismisses the need for a $\Delta C_9^{bsee} = -\Delta C_{10}^{bsee}$ New Physics contribution (cf. Table~\ref{tab:1d2d}).

In principle, the universal contribution can also be envisaged in $C_9^{bs\tau\tau}$.
Should this be the case, it could be mimicked by underestimated or unknown long distance charm-loop effects, a possibility that was recently studied in~\cite{Isidori:2021vtc}, in which a lepton flavour universal contribution to $C_9$ is treated as a nuisance parameter in the fit.
Furthermore, the ``look elsewhere effect'' (LEE) is taken into account in the fit, leading to a global significance of the $b\to s\mu\mu$ anomalies (LFUV, angular observables and differential branching fractions) amounting to $4.3\,\sigma$~\cite{Isidori:2021vtc}.

However, a more interesting possibility was pointed out in~\cite{Crivellin:2018yvo}: the universal contribution might arise via RGE from a large $C_9^{bs\tau\tau}$ at the electroweak scale (or New Physics matching scale).
Since tree-level matching of SMEFT preserves $SU(2)_L$, and can thus naturally lead to $\Delta C_9 = - \Delta C_{10}$, a large $\Delta C_9^{bs\tau\tau}$ at the electroweak scale could then be connected to the charged current $R_{D^{(\ast)}}$ anomalies - a possibility we will explore in the next section.

\mathversion{bold}
\section{Combining $b\to c\ell\nu$ and $b\to s\ell\ell$ in the SMEFT}
\mathversion{normal}
\label{sec:smeft}
As previously discussed, the $b\to s\ell\ell$ anomalies can be well accommodated by having a LFU contribution to $C_9^{bs\ell\ell}$ in addition to a LFUV contribution of the form $\Delta C_9^{bs\mu\mu} = -\Delta C_{10}^{bs\mu\mu}$.
As pointed out in~\cite{Crivellin:2018yvo}, such a contribution might arise via RGE mixing.
In order to analyse this interesting possibility we will now consider semi-leptonic (dimension-6) SMEFT operators\footnote{Other operators, for instance four-quark operators or $Z$-penguins, can also generate contributions to $C_9$ and/or $C_{10}$ at the $b$-quark scale, see e.g.~\cite{Aebischer:2019mlg}.}, 
focusing on the following (left-handed) operators 
\begin{equation}
    \mathcal L_\text{SMEFT} \supseteq (C_1)_{\ell q}^{\alpha\beta i j} (\bar L_L^\alpha \gamma_\mu L_L^\beta)(\bar Q_L^{i}\gamma^\mu Q_L^j) + (C_3)_{\ell q}^{\alpha\beta i j} (\bar L_L^\alpha \gamma_\mu \tau^\alpha L_L^\beta)(\bar Q_L^{i}\gamma^\mu \tau^\alpha Q_L^j)\,,
    \label{eqn:smeftst}
\end{equation}
in which $Q_L$ and $L_L$ respectively denote the left-handed SM quark and lepton doublets, and $\tau^\alpha$ are the $SU(2)_L$ generators defined via the Pauli matrices as $\tau^\alpha = \sigma^\alpha/2$.
Matching the SMEFT operators at tree-level onto WET~\cite{DAmbrosio:2002vsn,Feruglio:2017rjo}, leads to neutral and charged current WET operators, which are related by $SU(2)_L$ invariance.
The matching conditions relevant for $b\to s\ell\ell$ and $b\to c\ell\nu$ transitions can be schematically written as~\cite{Feruglio:2017rjo} (up to the appropriate normalisations)
\begin{eqnarray}
    C_9^{bs\ell_\alpha\ell_\beta} = - C_{10}^{bs\ell_\alpha\ell_\beta} &\propto& (C_1)_{\ell q}^{\alpha\beta 2 3} + (C_3)_{\ell q}^{\alpha\beta 2 3}\,,\nonumber\\
    C_{V_L}^{bc \ell_\alpha\nu_\beta} &\propto& -(C_3)_{\ell q}^{\alpha\beta 2 3}\,.
    \label{eqn:smefttowet}
\end{eqnarray}
However, due to $SU(2)_L$ invariance, a sizeable contribution to $(C_1)_{\ell q}^{\alpha\beta 23}$ and $(C_3)_{\ell q}^{\alpha\beta 23}$ also leads to significant New Physics contributions in $B\to K^{(\ast)}\nu\bar\nu$.
It can be shown~\cite{Feruglio:2017rjo} that this decay is proportional to $B\to K^{(\ast)}\nu\bar\nu \propto (C_1)_{\ell q} - (C_3)_{\ell q}$, and so we impose the condition\footnote{This condition is however (mildly) spoiled due to RGE above the electroweak scale, and thus only holds approximately.} $(C_1)_{\ell q} = (C_3)_{\ell q}$ to evade this constraint.
A large LFU contribution to $C_9^{bs\ell\ell}$ is then generated via RGE above and below the electroweak scale by having a contribution in $(C_1)_{\ell q}^{3323} = (C_3)_{\ell q}^{3323}$, leading to a large $C_9^{bs\tau\tau}$ (below the electroweak scale), in addition to $C_{V_L}^{bc\tau\nu}$, which is necessary to accommodate the $b\to c\tau\nu$ data.

\begin{figure}
    \centering
    \includegraphics[width=0.5\textwidth]{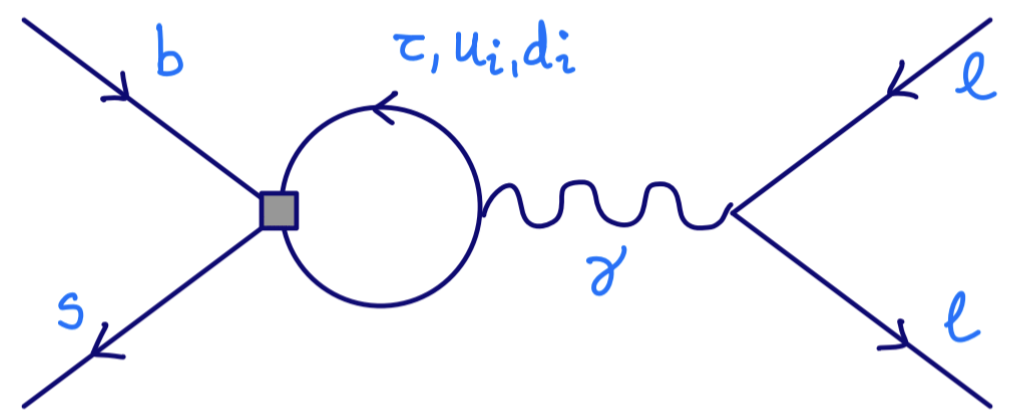}
    \caption{Example Feynman diagram leading to mixing of $C_9^{bs\tau\tau}$ into $C_9^{bs\mu\mu}$ and $C_9^{bsee}$ due to RGE. The small square denotes the insertion of a four-fermion operator (e.g. $\mathcal O_9^{bs\tau\tau}$) at the electroweak scale.}
    \label{fig:running_loop}
\end{figure}

As shown in the illsutrative example diagram in Fig.~\ref{fig:running_loop}, a large $C_9^{bs\tau\tau}$ then universally mixes into $C_9^{bs\mu\mu}$ and $C_9^{bsee}$, thus generating a potentially sizeable $C_9^\text{univ}$, as required by $b\to s\ell\ell$ data.

\begin{figure}
    \centering
    \includegraphics[width=0.6\textwidth]{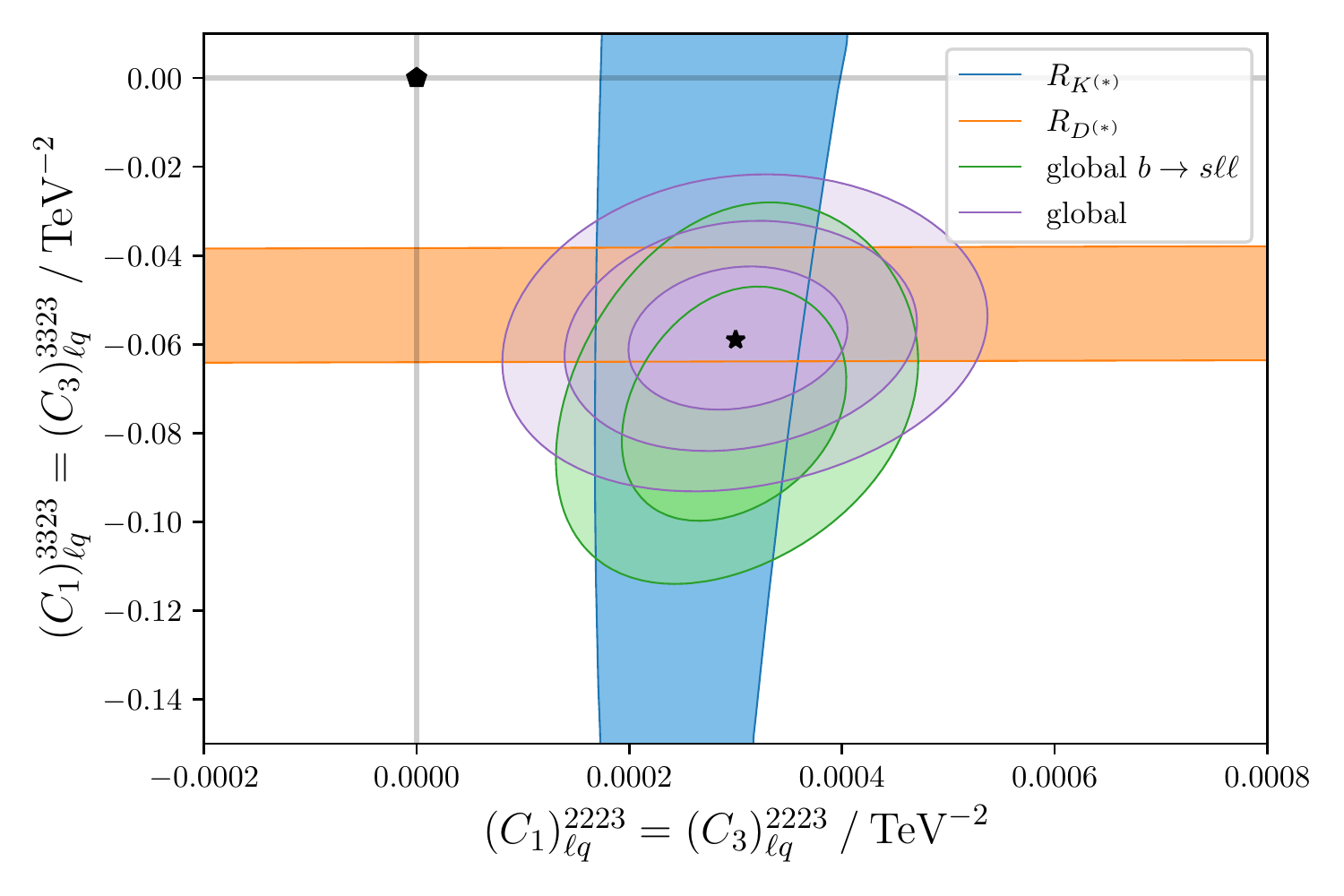}
    \caption{Results of a fit of SMEFT Wilson coefficients to $b\to s\ell\ell$ and $b\to c\tau\nu$ data assuming a minimal realistic New Physics hypothesis.}
    \label{fig:SMEFT}
\end{figure}

Upon comparison of the inherent New Physics scales implied by the $B$-decay anomalies (cf. Eqs.~\eqref{eqn:npscale_rd} and~\eqref{eqn:npscale_rk}), it is clear that, should the anomalies have a common origin, $(C_1)_{\ell q}^{3323} \sim 10^{2} \times (C_1)_{\ell q}^{2223}$ at a common matching scale.
Motivated by the reach of LHC, we set a moderate matching scale\footnote{Obviously, the matching scale can be different with only a minor impact on the fit results. However, motivated by model building efforts, a moderate matching scale of a few TeV seems appropriate in view of the near future direct collider reach.} at $2\:\mathrm{TeV}$ and carry out a fit combining $b\to c\ell\nu$ and $b\to s\ell\ell$ data (see Appendices~\ref{app:BFCCC} and~\ref{app:bsll} for the observables taken into account).
We obtain a best fit point given by
\begin{equation}
    (C_1)_{\ell q}^{2223} = (C_3)_{\ell q}^{2223} = (3.0_{-0.6}^{+0.7})\times 10^{-4}\:\mathrm{TeV}^{-2}\,,\quad(C_1)_{\ell q}^{3323} = (C_3)_{\ell q}^{3323} = -0.059 \pm 0.01\:\mathrm{TeV}^{-2}\,,
\end{equation}
amounting to a pull of $7.4\,\sigma$ with respect to the SM prediction, and a $p$-value of $59.4\%$.
Interestingly, as can be seen in Fig.~\ref{fig:SMEFT}, the agreement between the different observables is excellent.

Should the $B$-meson decay anomalies indeed be a low-energy manifestation of New Physics, this can be interpreted as a hint that both the charged and neutral current anomalies could be accommodated in a model with a single mediator that preserves $SU(2)_L$, and thus be taken as a guide for model building.
For other recent EFT analyses of the $B$-meson decay anomalies see e.g.~\cite{Ligeti:2016npd,Bigi:2017jbd,Capdevila:2017bsm,Alguero:2019ptt,Aebischer:2019mlg,Ciuchini:2019usw,Datta:2019zca,Arbey:2019duh,Shi:2019gxi,Bardhan:2019ljo,Alok:2019ufo,Bhattacharya:2019eji,Alok:2017qsi,Ghosh:2014awa,Glashow:2014iga,Bhattacharya:2014wla,Freytsis:2015qca,Ciuchini:2017mik,Jaiswal:2017rve,Jaiswal:2020wer,Bhattacharya:2019dot,Biswas:2020uaq,Bhattacharya:2018kig,Alguero:2021anc,Altmannshofer:2021qrr}.

Extensive efforts have also been devoted to explain the anomalies - either separately or combined - in terms of specific New Physics constructions: 
among the most minimal scenarios
studied, heavy $Z^\prime$ mediators were identified as possible solutions (see for example ~\cite{Altmannshofer:2014cfa,Crivellin:2015mga,Crivellin:2015lwa,Sierra:2015fma,Crivellin:2015era,Celis:2015ara,Bhatia:2017tgo,Kamenik:2017tnu,Chen:2017usq,Camargo-Molina:2018cwu,Darme:2018hqg,Baek:2018aru,Biswas:2019twf,Allanach:2019iiy,Crivellin:2020oup}); likewise, numerous studies addressed the scalar and the vector leptoquark hypotheses (e.g.~\cite{Hiller:2014yaa,Gripaios:2014tna,Sahoo:2015wya,Varzielas:2015iva,Alonso:2015sja,Bauer:2015knc,Hati:2015awg,Fajfer:2015ycq,Das:2016vkr,Becirevic:2016yqi,Sahoo:2016pet,Cox:2016epl,Crivellin:2017zlb,Becirevic:2017jtw,Cai:2017wry,Dorsner:2017ufx,Buttazzo:2017ixm,Greljo:2018tuh,Sahoo:2018ffv,Becirevic:2018afm,Hati:2018fzc,Fornal:2018dqn,deMedeirosVarzielas:2018bcy,Aebischer:2018acj,Aydemir:2019ynb,Mandal:2018kau,deMedeirosVarzielas:2019okf,Yan:2019hpm,Bigaran:2019bqv,Popov:2019tyc,Hati:2019ufv,Crivellin:2019dwb,Saad:2020ihm,Dev:2020qet,Saad:2020ucl,Balaji:2019kwe,Cornella:2019hct,Mandal:2019gff,Babu:2020hun,Martynov:2020cjd,Fuentes-Martin:2020bnh,Guadagnoli:2020tlx,Angelescu:2021lln,Greljo:2021xmg,Cornella:2021sby,Marzocca:2021azj,Du:2021zkq,Perez:2021ddi,Marzocca:2021miv,Marzocca:2021miv}); further examples include
$R-$parity violating supersymmetric models (see for instance~\cite{Deshpand:2016cpw,Altmannshofer:2017poe,Das:2017kfo,Earl:2018snx,Trifinopoulos:2018rna,Trifinopoulos:2019lyo,Cohen:2019cge,Earl:2019adq,Hu:2019ahp,Hu:2020yvs,Altmannshofer:2020axr}, as well as other interesting
constructions~\cite{Greljo:2015mma,Arnan:2017lxi,Geng:2017svp,Choudhury:2017qyt,Choudhury:2017ijp,Grinstein:2018fgb,Cerdeno:2019vpd,Crivellin:2019dun,Arnan:2019uhr,Gomez:2019xfw}). 

Despite the large number of alternatives, only a few select scenarios can successfully put forward a simultaneous explanation for both charged and neutral current $B$-meson decay anomalies. Standard Model extensions relying on a $V_1$ vector leptoquark transforming as $(\mathbf{3},\mathbf{1},2/3)$ under the SM gauge group have received considerable attention in the literature~\cite{Assad:2017iib,Buttazzo:2017ixm,Calibbi:2017qbu,Bordone:2017bld,Blanke:2018sro,Bordone:2018nbg,Kumar:2018kmr,Angelescu:2018tyl,Balaji:2018zna,Fornal:2018dqn,Baker:2019sli,Cornella:2019hct,DaRold:2019fiw,Hati:2019ufv, Fuentes-Martin:2019ign, Fuentes-Martin:2020luw, Fuentes-Martin:2020hvc}, being currently the only single-leptoquark solution capable of simultaneously addressing both charged and neutral current anomalies; at tree-level, $V_1$ generates $(C_1)_{\ell q} = (C_3)_{\ell q}$ and such constructions can be realised at sufficiently low scales.
This will be the topic of the next chapter.

\section{Outlook}
The numerous experimental hints concerning the presence of LFUV New Physics interactions in $B$-meson decays are certainly intriguing.
However, in the absence of a ``$5\,\sigma$ discovery'', they should be treated with caution.
In recent months (and years) many ideas to cross-check the hints on LFUV, and deepen our understanding of them, have been proposed; numerous efforts have been conducted, on the theoretical as well as on the experimental sides.

If the anomalies in $b\to s\ell\ell$ are indeed a genuine New Physics effect and not a statistical fluctuation, they should emerge in other observables as well, for instance in baryon decays such as $\Lambda_b$, $\Omega_b$ and $\Xi_b$ decays.
The LHCb collaboration has recently measured the LFU ratio~\cite{LHCb:2019efc}
\begin{equation}
    R_{pK} \equiv  \left.\frac{\mathrm{BR}(\Lambda_b \to p K^- \mu^+\mu^-)}{\mathrm{BR}(\Lambda_b \to p K^- e^+ e^-)}\right\vert_{0.1\leq q^2 \leq 6.0 \:\mathrm{GeV}^2} = 0.86_{-0.11}^{+0.14} \pm 0.05\,,
\end{equation}
which is another potential hint on LFUV in $b\to s\ell\ell$. Here, the $pK^-$ pair is the decay product of one of the $\Lambda$ resonances, similarly to $B\to K^\ast (\to \pi K)\ell\ell$ decays.
The LHCb collaboration has also conducted an analysis of the angular moments in the decay $\Lambda_b \to \Lambda \mu^+\mu^-$~\cite{LHCb:2018jna} at low hadronic recoil, resulting in a $2.6\,\sigma$ deviation from the SM prediction in one of them.
Further angular analysis of the excited decays are desirable~\cite{Boer:2014kda,Amhis:2020phx,Descotes-Genon:2019dbw}.

Furthermore, as it has been pointed out in~\cite{Alguero:2019pjc,Alguero:2021anc}, a precise measurement of the $B\to K^\ast\ell\ell$ angular observables
\begin{equation}
    \langle Q_{4,5}\rangle_\text{bin} \equiv \langle P_{4,5}^\prime\rangle_\text{bin}^\mu - \langle P_{4,5}^\prime\rangle_\text{bin}^e\,,
\end{equation}
could help disentangling long-distance effects via $c\bar c$-loops from New Physics effects.
Currently, there is only one set of measurements available, performed by Belle~\cite{Belle:2016fev}
\begin{equation}
    \langle Q_{4}\rangle_{1.0 < q^2 < 6.0\:\mathrm {GeV}^2} = 0.498 \pm 0.527 \pm 0.166\,,\quad \langle Q_{5}\rangle_{1.0 < q^2 < 6.0\:\mathrm {GeV}^2} = 0.656 \pm 0.485 \pm 0.103\,,
\end{equation}
which is in mild tension with the SM, which predicts (almost) vanishing values of these observables. Belle has performed this measurement in other $q^2$ bins as well, but with larger statistical uncertainties.
Recently, a phenomenological analysis disentangling $P$- and $S$-wave contributions in $B\to K^\ast \ell\ell$ was put out, suggesting that additional observables can be measured~\cite{Alguero:2021yus}.
Another interesting avenue to pursue is the interplay of di-neutrino modes (such as $B\to K^{(\ast)}\nu\bar\nu$) and $b\to s\ell\ell$~\cite{Bause:2021ply,Bause:2020xzj,Bause:2020auq,Descotes-Genon:2021doz}.  
Obviously, more data from LHCb on the $b\to s\ell\ell$ transitions is impatiently waited for, while Belle and Belle II will serve as crucial cross-checks.

\begin{figure}
    \centering
    \includegraphics[width=0.6\textwidth]{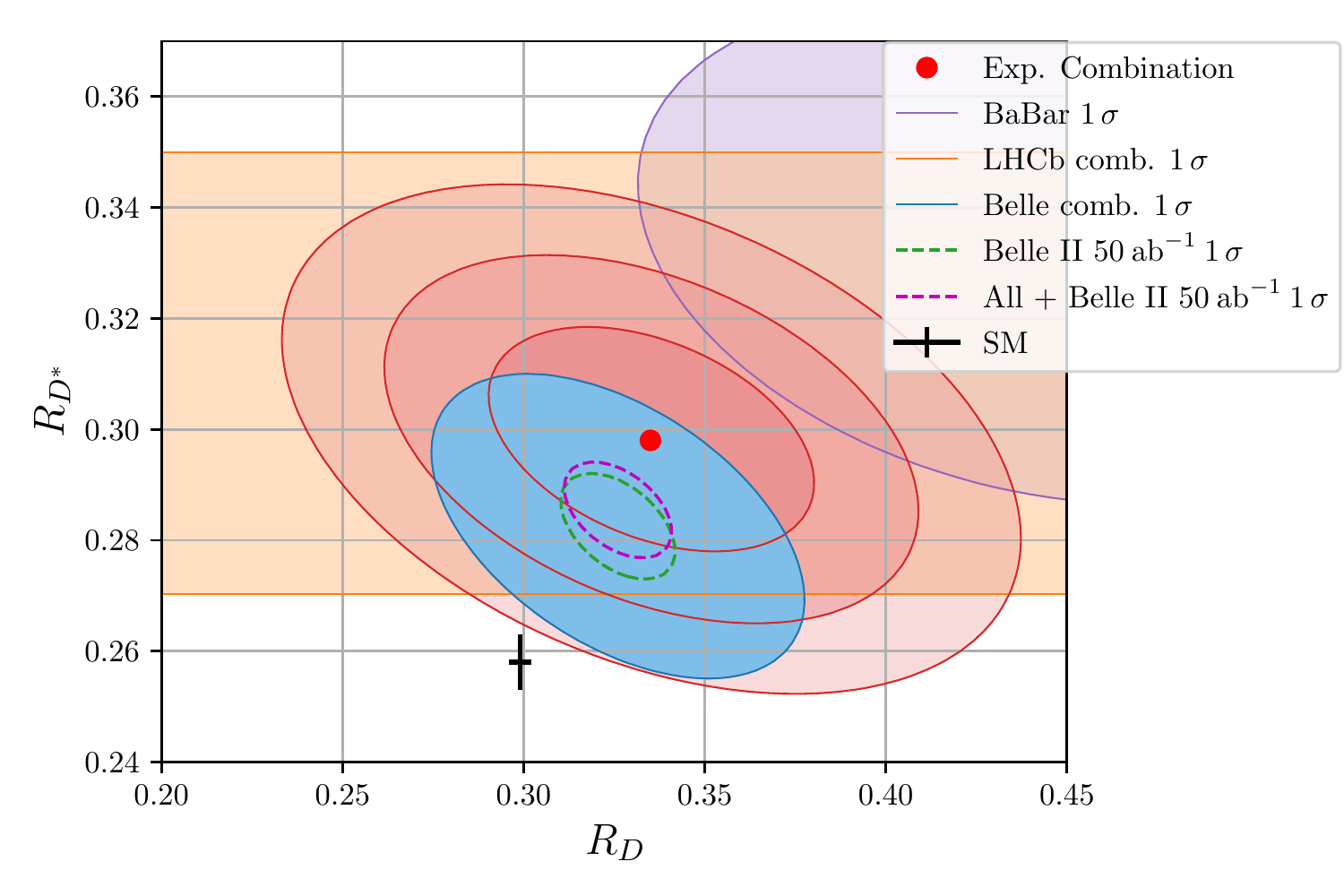}
    \caption{Projection of the future Belle II sensitivity to $R_{D^{(\ast)}}$, derived from a gaussian combination of current Belle data, extrapolated to the anticipated relative uncertainties.
    The projections are shown together with the current data obtained by BaBar, Belle and LHCb, and a combination of the projection with contemporary data. Colours are as indicated by the legend.}
    \label{fig:rdrds_proj}
\end{figure}

\medskip
For the charged current anomalies, whose measurements are currently dominated by Belle data, also several cross-checks have been proposed.
First and foremost, Belle II will greatly improve the precision of these measurements, anticipating to reach percent level accuracy with the full dataset of $50\:\mathrm{ab}^{-1}$~\cite{Belle-II:2018jsg}.
In Fig.~\ref{fig:rdrds_proj} we plot a na\"ive extrapolation of the current Belle average to this accuracy and combine it with the current data.
Obviously, one can also test the $b\to c\ell\nu$ LFUV anomalies with $b$-flavoured baryon decays, e.g. $\Lambda_b\to \Lambda_c\ell\nu$~\cite{Boer:2018vpx,Boer:2019zmp}.
Here, some theoretical progress has recently been made for the form factor calculations~\cite{Mannel:2015osa,Becirevic:2019xbn,Becirevic:2020nmb}, showing promising results.

Another way to test the anomalies was recently suggested in~\cite{Descotes-Genon:2021uez}, in which it is proposed to measure inclusive $\Upsilon(4S) \to \ell\ell' + \slashed{E_T} + \text{hadron(s)}$. It is shown that the dominant decays can be related to $b\to X\tau\nu/b\to X\ell\nu$, thus serving as an important test of LFUV in charged current $B$-meson decays.

Recent measurements by Babar~\cite{BaBar:2020nlq} of the ratio
\begin{equation}
    R_{\Upsilon(3S)}^{\tau\mu} \equiv \frac{\mathrm{BR}(\Upsilon(3S)\to \tau^+\tau^-)}{\mathrm{BR}(\Upsilon(3S)\to \mu^+\mu^-)} = 0.966 \pm 0.008_\text{stat} \pm 0.014_\text{syst}\,,
\end{equation}
exhibit a $2\,\sigma$ discrepancy with the SM prediction, which amounts to $R_{\Upsilon(3S)}^{\tau\mu} = 0.9948$~\cite{Aloni:2017eny}.
Although by far not as significant as the other tensions, interestingly it is in the same ballpark as predictions of New Physics models that accommodate $R_{D^{(\ast)}}$ data~\cite{Aloni:2017eny}.

Furthermore, it has been shown that the presence of New Physics in the SMEFT coefficients $\mathcal (C_{1,3})_{\ell q}^{3323}$, favoured in order to accommodate $R_{D^{(\ast)}}$, necessarily leads to a massive enhancement in $b\to s\tau\tau$ processes reaching predictions far larger than the SM one by as much as three orders of magnitude~\cite{Capdevila:2017iqn}.
Via sampling of the posterior distribution of our SMEFT fit (cf. Section~\ref{sec:smeft}), we obtain e.g.
\begin{eqnarray}
    \mathrm{BR}(B_s \to \tau^+ \tau^-)_\text{SMEFT} = (2.1\pm 0.67)\times 10^{-4}\,,
\end{eqnarray}
to be compared with the SM prediction $\mathrm{BR}(B_s \to \tau^+ \tau^-)_{\text{SM}} = (7.73 \pm 0.49) \times 10^{-7}$~\cite{Capdevila:2017iqn}.
Measurements of $b\to s\tau\tau$ processes are thus another crucial step towards fully understanding these anomalies, should they persist in the future.

\medskip
At high energies, there are interesting proposals to probe $\mathcal O_9^{bs\mu\mu}$ contact interactions directly at a future muon collider~\cite{Huang:2021biu}, leading to promising signatures in $\mu^+\mu^- \to b j$ events.
Current $R_{K^{(\ast)}}$ data then (model-independently) implies an excess over the SM of up to $3\,\sigma$~\cite{Huang:2021biu}.

Furthermore, one can explore SMEFT synergies between top-quark physics at the electroweak scale with the $B$-meson sector~\cite{Bissmann:2020mfi,Bissmann:2019gfc}.

\medskip
Finally, other LFUV tests in charm decays can be envisaged, due to a lot of recent theoretical progress~\cite{Fajfer:2015mia,Golz:2021imq,Hiller:2021zth,Bause:2020obd,Bause:2019vpr,DeBoer:2018pdx,Bharucha:2020eup}.

\mathversion{bold}
\chapter{$V_1$ vector leptoquarks for the $B$-meson decay anomalies}
\label{chap:lq}
\mathversion{normal}
\minitoc

\noindent
Motivated by the flavour puzzle, and in particular by the existence of three generations of fermions in both the quark and the lepton sectors, in the past many attempts have been made to understand this ``coincidence'' in the number of matter generations on a more fundamental level, relying on the use of symmetries.
In particular, Pati and Salam proposed in the 1970s to unify quarks and leptons into a single multiplet of a higher gauge symmetry (beyond the SM gauge group)~\cite{Pati:1973rp,Pati:1973uk,Pati:1974yy}.
It was suggested that lepton number could be a interpreted as a ``fourth colour'', allowing to unify quarks and leptons in common multiplets of a gauged $SU(4)$ group.
The associated gauge boson of the $SU(4)$ then necessarily couples  simultaneously to quarks and leptons, since the SM fermions carry the same charge under the new group; in other words the gauge boson can turn leptons into quarks and vice versa.
For this reason, bosons that have this property, are commonly called leptoquarks.
It has been subsequently shown that the original model by Pati and Salam~\cite{Pati:1974yy}, the so-called Pati-Salam model based on a gauge group given by $SU(4)_c\times SU(2)_L \times SU(2)_R$\footnote{The originally proposed gauge group is in fact $SU(4)_c\times SU(4)_L\times SU(4)_R$, but it has been argued that most of the New Physics features are retained in the simpler group $SU(4)_c\times SU(2)_L\times SU(2)_R$~\cite{Pati:1974yy}.}, can appear in the breaking pattern of the simple groups $SU(5)$ or $SO(10)$, which represent grand unified theory (GUT) frameworks~\cite{Georgi:1974sy,Georgi:1974my,Fritzsch:1974nn}.
The ambitious goal of matter unification into the same representation of a simple unified gauge group then also requires the existence of scalar leptoquarks, in order to achieve the desired breaking pattern of the unified gauge symmetry to the SM gauge group.
However, these scalar leptoquarks tend to lead to proton decay at tree-level and their masses must be extremely large in order to comply with the stringent bounds on the proton lifetime ($\tau_p \geq 10^{31}$ years~\cite{ParticleDataGroup:2020ssz}).

Furthermore, the introduction of supersymmetry has opened the door to a very rich leptoquark phenomenology, since squarks (scalar fields) carry coincidentally the correct quantum numbers to couple quarks to leptons.
This requires however explicit R-parity violation~\cite{Hall:1983id,Dawson:1985vr,Giudice:1997wb,Csaki:2011ge}, but can still be embedded into GUTs~\cite{Giudice:1997wb}.
Finally, leptoquarks can also appear in composite models~\cite{Abbott:1981re,Schrempp:1984nj,Wudka:1985ef}.

\bigskip
Although UV complete constructions (as GUTs) remain the most appealing and theoretically well-motivated New Physics constructions featuring leptoquarks, new states - as is the case of \textit{scalar} leptoquarks - can also be ``just so'' added to the SM field content, provided they respect all fundamental SM gauge symmetries.

The SM gauge symmetry allows in fact for twelve different leptoquark representations, leading to six scalar and six vector bosons.
Furthermore, one can define an additional quantum number, composed of the baryon number and lepton number carried by the leptoquark, which can be defined as~\cite{Buchmuller:1986zs,Dorsner:2016wpm}
\begin{equation}
    F = 3B + L\,,
\end{equation}
provided we assign $B = 1/3$ to quarks and $L=1$ to leptons.
Following the classification of~\cite{Buchmuller:1986zs,Dorsner:2016wpm}, we show in Table~\ref{tab:lqrep} all possible leptoquark representations that lead to renormalisable and SM gauge-invariant interactions with the SM fermions.
The leptoquarks with a non-vanishing fermion number $F$ can in addition have couplings to quark pairs, thus leading to proton decay at tree-level~\cite{Dorsner:2016wpm}.
If one aims at constructing a UV complete model of leptoquarks, these couplings then have to be suppressed (for instance via symmetry considerations) or the leptoquarks must be extremely heavy in order to comply with matter stability.
\renewcommand{\arraystretch}{1.3}
\begin{table}[]
    \centering
    \begin{tabular}{|c|c|c|c|}
        \hline
         $(SU(3)_c, SU(2)_L, U(1)_Y)$ & Spin & Name & $F$\\
         \hline
         $(\bar{\mathbf{3}}, \mathbf{1}, 1/3)$ & $0$ & $S_1$ & $-2$\\
         $(\bar{\mathbf{3}}, \mathbf{1}, 4/3)$ & $0$ & $\tilde S_1$ & $-2$\\
         $(\bar{\mathbf{3}}, \mathbf{1}, -2/3)$ & $0$ & $\bar S_1$ & $-2$\\
         $(\bar{\mathbf{3}}, \mathbf{3}, 1/3)$ & $0$ & $S_3$ & $-2$\\
         $(\mathbf{3}, \mathbf{2}, 7/6)$ & $0$ & $R_2$ & $0$\\
         $(\mathbf{3}, \mathbf{2}, 1/6)$ & $0$ & $\tilde R_2$ & $0$\\
         \hline
         \hline
         $(\mathbf{3}, \mathbf{1}, 2/3)$ & $1$ & $V_1$ & $0$\\
         $(\mathbf{3}, \mathbf{1}, 5/3)$ & $1$ & $\tilde V_1$ & $0$\\
         $(\mathbf{3}, \mathbf{1}, -1/3)$ & $1$ & $\bar V_1$ & $0$\\
         $(\mathbf{3}, \mathbf{3}, 2/3)$ & $1$ & $V_3$ & $0$\\
         $(\bar{\mathbf{3}}, \mathbf{2}, 5/6)$ & $1$ & $V_2$ & $-2$\\
         $(\bar{\mathbf{3}}, \mathbf{2}, -1/6)$ & $1$ & $\tilde V_2$ & $-2$\\
         \hline
    \end{tabular}
    \caption{List of all possible leptoquark representations that lead to renormalisable and SM gauge-invariant interactions with the SM fermions~\cite{Dorsner:2016wpm}.}
    \label{tab:lqrep}
\end{table}
\renewcommand{\arraystretch}{1.}

\medskip
From a phenomenological point of view it is very appealing to consider low-energy effects of leptoquarks; due to their tree-level couplings to leptons and quarks, the presence of leptoquarks leads to tree-level contributions to a plethora of semi-leptonic processes.
This can be particularly interesting for processes which are otherwise strongly suppressed (as is the case of $b\to s\ell\ell$ transitions) or absent in the SM (for instance semi-leptonic LFV transitions), which could then allow searching for leptoquark effects at the intensity frontier.

In addition, leptoquarks are often invoked to reconcile theory with observation; for instance, the $(g-2)_{e\,,\mu}$ anomalies can be addressed and accommodated in models featuring $\mathrm{TeV}$-scale scalar leptoquarks (see e.g.~\cite{Bigaran:2020jil,Dorsner:2020aaz}).
Furthermore, due to the leptoquark's ability to mediate semi-leptonic charged and neutral current transitions at tree-level, leptoquarks are often invoked to address the $B$-meson decay anomalies~\cite{Hiller:2014yaa,Gripaios:2014tna,Sahoo:2015wya,Varzielas:2015iva,Alonso:2015sja,Bauer:2015knc,Hati:2015awg,Fajfer:2015ycq,Das:2016vkr,Becirevic:2016yqi,Sahoo:2016pet,Cox:2016epl,Crivellin:2017zlb,Becirevic:2017jtw,Cai:2017wry,Dorsner:2017ufx,Buttazzo:2017ixm,Greljo:2018tuh,Sahoo:2018ffv,Becirevic:2018afm,Hati:2018fzc,Fornal:2018dqn,deMedeirosVarzielas:2018bcy,Aebischer:2018acj,Aydemir:2019ynb,Mandal:2018kau,deMedeirosVarzielas:2019okf,Yan:2019hpm,Bigaran:2019bqv,Popov:2019tyc,Hati:2019ufv,Crivellin:2019dwb,Saad:2020ihm,Dev:2020qet,Saad:2020ucl,Balaji:2019kwe,Cornella:2019hct,Mandal:2019gff,Babu:2020hun,Martynov:2020cjd,Fuentes-Martin:2020bnh,Guadagnoli:2020tlx,Angelescu:2021lln,Greljo:2021xmg,Cornella:2021sby,Marzocca:2021azj,Du:2021zkq,Perez:2021ddi,Marzocca:2021miv,Hati:2020cyn}, either in ad-hoc extensions of the SM, simplified models or in UV complete frameworks with or without gauge unification.

In order to find suitable New Physics candidates, in this case leptoquark representations, that can accommodate the $B$-meson decay anomalies, it is very useful to consider ``bottom-up'' model building, that is finding data-driven requirements a BSM mediator (or sets thereof) must fulfil to successfully reconcile theory with observation. 
After extensive studies of tree-level matching of leptoquark interactions to SMEFT, it has be shown that $V_1$ is the only leptoquark that leads to contributions in semi-leptonic SMEFT operators of the form $(C_1)_{\ell q} = (C_3)_{\ell q}$~\cite{Buttazzo:2017ixm}, allowing to 
accommodate both the charged and neutral current anomalies  (cf. Section~\ref{sec:smeft}).
The $V_1$ hypothesis has received increasing attention in the literature, as it is currently the only single leptoquark (and perhaps single BSM mediator) construction that successfully offers a simultaneous solution to the charged and neutral current $B$-meson decay anomalies~\cite{Assad:2017iib,Buttazzo:2017ixm,Calibbi:2017qbu,Bordone:2017bld,Blanke:2018sro,Bordone:2018nbg,Kumar:2018kmr,Angelescu:2018tyl,Balaji:2018zna,Fornal:2018dqn,Baker:2019sli,Cornella:2019hct,DaRold:2019fiw,Hati:2019ufv, Hati:2020cyn}.

If one aims at building a model with scalar leptoquarks, in order to accommodate the $B$-meson decay anomalies, one either has to consider combinations of leptoquarks (e.g. $S_1$ and $S_3$ or $R_2$ and $S_3$~\cite{Angelescu:2018tyl,Angelescu:2021lln}), or further additional field content. 
For example, in~\cite{Marzocca:2021azj} a model featuring the leptoquark $S_1$ and a singly-charged singlet scalar is considered.
The leptoquark generates sizeable contributions to $b\to c\tau\nu$ transitions at tree-level, while the interplay of the charged scalar and the leptoquark at loop-level leads to large contributions in $b\to s\ell\ell$ transitions.
Ideally, these leptoquark constructions should also allow addressing a viable mechanism for neutrino mass generation, the baryon asymmetry of the Universe, the dark matter problem, or be part of a UV complete model encompassing a solution to the latter issues.

Among the many possible SM extensions including leptoquarks, in what follows we focus on vector leptoquark
($V_1$) scenarios, using a simplified-model approach.

\mathversion{bold}
\section{Reconciling the $B$-meson decay anomalies with $V_1$ vector leptoquarks}
\mathversion{normal}
As previously highlighted, following the updated global fits carried out in the previous chapter, the vector leptoquark hypothesis belongs to the class of New Physics ``scenarios'' most favoured by current data.
Working under the hypothesis that $V_1$ is associated to some spontaneously broken gauge symmetry, its couplings to SM fermions are a priori universal.
However, and in order to account for experimental data, 
$V_1$ should have non-universal couplings to quarks and leptons, and the latter can be realised in a number of ways.
The most minimal possible scenario relies in the assumption that the vector leptoquark is an elementary gauge boson\footnote{There are also models in which the vector leptoquark appears as a composite field, see for instance~\cite{Barbieri:2016las,Cline:2017aed}; we will not consider them here.}, associated to a non-abelian gauge group extension of the SM, under which the SM fermion generations are universally charged; in the unbroken phase of the underlying extended gauge group, the leptoquark gauge couplings 
also remain universal. Despite its simplicity, 
this scenario is challenged by constraints from the
cLFV decays $K_L\rightarrow \mu e$ and $K\rightarrow \pi \mu e$: current limits (see Table~\ref{tab:semicLFV} in Chapter~\ref{chap:lepflav}) would force the mass of such a vector leptoquark to be very large, $m_V\ge 100$~TeV for $\mathcal O(1)$ couplings~\cite{Hung:1981pd,Valencia:1994cj,Smirnov:2007hv,Carpentier:2010ue,Kuznetsov:2012ai,Smirnov:2018ske}, and thus excessively heavy to account for both the charged and neutral current $B$-meson decay anomalies\footnote{As discussed in Chapter~\ref{chap:bphysics}, the inherent New Physics scales implied by the $B$-meson decay anomalies are much lower; $\sim3.3\:\mathrm{TeV}$ for the charged current and $\sim35\:\mathrm{TeV}$ for the neutral current.}. 
In order to understand this, notice that while $V_1$ has a  universal coupling to SM fermions in the unbroken phase, after $SU(2)_L$-breaking a potential misalignment of the quark and lepton eigenstates is generated, leading to LFU-violating $V_1$ couplings. Given the constraints from $\tau$ decays, the $c\nu_\ell$ coupling generated from $b\tau$ through CKM mixing is not sufficiently large to account for $R_{D^{(\ast)}}$ data~\cite{Feruglio:2017rjo}. On the other hand, for a maximal $c\nu_\ell$ coupling (with $\ell\neq \tau$) generated by $d_i\mu$ and $d_i e$ couplings, important constraints arise from $R_{K^{(*)}}$ data for $i=2,3$, and from Kaon decays for $i=1$. Moreover, the $c\nu$ coupling induced by $d\tau$ is heavily CKM-suppressed. Therefore, the only viable possibility is to maximise both $b\tau$ and $s\tau$ couplings, which in turn will induce large couplings between the first two generations of quarks and leptons (given the unitarity of the post-$SU(2)_L$-breaking mixing matrix), thus implying excessive contributions to cLFV.

The question that naturally emerges
is whether one can find a minimal embedding of $V_1$
that successfully allows to overcome the cLFV constraints and address
both $R_{K^{(*)}}$ and $R_{D^{(*)}}$ anomalies. In other words, can
one go beyond the tight constraints arising from a ($3\times 3$)
unitary mixing of quarks and leptons?
A possible way to circumvent the above mentioned constraints is to introduce three ``generations'' of vector leptoquarks, belonging to an identical number of copies of the extended gauge group (e.g. Pati-Salam model based on the gauge group $[SU(4)_c]_i \times  [SU(2)_L]_i \times [SU(2)_R]_i$), with each copy acting on a single SM fermion generation (subject to mixing with additional vector-like fermions), with the largest leptoquark-fermion couplings in association with the third family~\cite{Bordone:2017bld}. 
Another possibility to lower the vector leptoquark mass 
relies in an extended gauge group, $SU(4)\times SU(3)' \times SU(2)_L \times U(1)'$ (often referred to as ``4321''-model), with the third fermion family charged under $SU(4) \times SU(2)_L \times U(1)'$, while the lighter families are only charged under $ SU(3)' \times SU(2)_L \times U(1)'$~\cite{Greljo:2018tuh,Guadagnoli:2020tlx,Fuentes-Martin:2020bnh,Fuentes-Martin:2020bnh,Fuentes-Martin:2020luw,Fuentes-Martin:2020hvc,Cornella:2021sby}.
This leads to an approximate $U(2)$ flavour symmetry, which is softly broken by new vector-like fermions, thus allowing to obtain the desired non-universality in the leptoquark couplings. 

An alternative simplified-model framework, without the need to specify an explicit extended gauge group, was put foward in~\cite{Hati:2019ufv}, which we proceed to describe: working under a single vector leptoquark hypothesis, an effective non-unitary mixing between SM leptons and new vector-like leptons can be used to account for the LFUV structure required to 
simultaneously explain both the charged and the neutral current $B$-meson decay anomalies.
In particular, and motivated
by the phenomenological impact of having non-unitary left-handed
leptonic mixings
in the presence of (heavy) sterile neutral leptons~\cite{Xing:2007zj, Blennow:2016jkn,Fernandez-Martinez:2016lgt, Escrihuela:2015wra} (as extensively discussed in Chapters~\ref{chap:lepflav}-~\ref{chap:ISS}), we have considered the
possibility of non-unitary $V_1$ couplings, as arising from
the presence of $n$ additional vector-like heavy leptons $L$
(also present in the leptoquark construction of~\cite{Bordone:2017bld}).
In the broken phase, the
$V_1$ couplings are then given by a $(3+n)\times (3+n)$
mixing matrix, so that the couplings to SM fermions now correspond to
a $3\times 3$ sub-block, which is no longer
unitary. We hypothesise that
this departure from unitary mixings might indeed hold the key to
simultaneously address $R_{K^{(\ast)}}$ and $R_{D^{(\ast)}}$ data,
while satisfying existing cLFV constraints.

The addition of vector-like heavy charged leptons\footnote{Heavy
  vector-like quarks will not be considered, as they are not required
  for a minimal working model.} (see also Chapter~\ref{sec:g-2paper}) can be seen as an intermediary step
towards a full ultraviolet-complete model, providing a better
framework for the peculiar structure of leptoquark couplings required
by the anomalies. In this framework, the non-unitary mixings will also
lead to the modification of SM-like charged and neutral lepton
currents, establishing an inevitable link to electroweak precision observables, such as lepton flavour violating and/or LFUV $Z$-decays.  The latter
observables will prove to be extremely constraining, ultimately
leading to the exclusion of isosinglet vector-like heavy leptons as a
source of non-universality in $B$-meson decays.

These constraints are much milder for isodoublet heavy leptons:
after arguing that for a single additional heavy charged lepton, cLFV
constraints exclude an explanation of even $R_{D^{(*)}}$
alone, we show that the addition of $n=3$ vector-like
isodoublet leptons allows a simultaneous explanation of both
$R_{K^{(*)}}$ and $R_{D^{(*)}}$ anomalies, while respecting all
available constraints.

\bigskip
Irrespective of the actual New Physics model including (not excessively heavy) vector leptoquarks, the effects can be understood in terms of contributions to the Wilson coefficients.
Following the discussion of the previous chapter (see Table~\ref{tab:1d2d}),
in order to achieve the preferred contributions for the Wilson coefficients, $C_9^{bs\mu\mu} = -C_{10}^{bs\mu\mu}$ and a universal $\Delta C_9^\text{univ.}$, scenarios in which $V_1$ couples 
at the tree level through a left handed ($V-A$) current 
to muons (as well as to down-type quark flavours $b$ and $s$) appear to be favoured the most by the global fits. A nonvanishing $\Delta C_9^{bsee} = -\Delta C_{10}^{bsee}$ along with $C_9^{bs\mu\mu} = -C_{10}^{bs\mu\mu}$ and a universal $\Delta C_9^\text{univ.}$ also provides a reasonable fit but such hypotheses are subject to stringent constraints from cLFV processes. 
Furthermore, and in order to also address the charged current  data ($R_{D^{(\ast)}}$), sizeable tree-level  $\tau$ couplings to second and third generation quarks must also be present, 
and these induce new contributions to the  $C_{V_L}$ Wilson coefficient. Such large $V_1-\tau$ couplings to second and third generation quarks further lead to a large $C_{9(10)}^{bs\tau\tau}$ which then feeds into the muon and electron counterparts (in a universal way) through
RG running\footnote{We further notice that global fits without the \textit{universal} contributions to $C_9^{bs\ell\ell}$ suggest a non-zero \textit{tree-level} contribution to the electron coefficients. However, once the universal contribution is added, the direct \textit{tree-level} contribution is compatible with $0$ at the $1\sigma$ level.}. 
As we have subsequently argued in Section~\ref{sec:smeft}, the required structure of WET Wilson coefficients is naturally generated by a simple combination of SMEFT Wilson coefficients ($(C_1)_{\ell q} = (C_3)_{\ell q}$, see Eq.~\eqref{eqn:smefttowet}).
Interestingly, as previously mentioned, this structure of SMEFT coefficients is in turn naturally given by tree-level matching of $V_1$ interactions to SMEFT.

\bigskip
A possible way to systematically study such $V_1$ extensions relies in considering a simplified-model parametrisation of the vector leptoquark couplings; this allows not only to perform global fits, but also to understand the phenomenological implications of the relevant flavour structure,
which is paramount to establish the current viability of the model, and its prospects for future testability.
In this chapter, we thus pursue this approach 
not only regarding the ``anomalous'' $B$-meson observables, 
but also in what concerns the impact of this BSM construction for a large set of observables (various flavour violating meson decays and cLFV modes) - 
relevant in terms of constraints on the model, or then offering excellent prospects of observation in the near future.

\mathversion{bold}
\section{A simplified-model parametrisation of vector leptoquark $V_1$ couplings}
\mathversion{normal}
\label{sec:simplifiedmodel}
In what follows, we consider a SM extension by a single
vector leptoquark $V_1$, which transforms under the SM
gauge group
${SU}(3)_c\times {SU}(2)_L \times U(1)_Y$ as
$(\mathbf{3},\mathbf{1},2/3)$.
Without loss of generality, we assume that $V_1$ is a gauge boson
of an unspecified gauge extension of $SU(3)_c$ with a
universal (i.e. flavour blind) gauge coupling; without relying
on a specific gauge embedding and/or
Higgs sector, our only working assumption is that
all fermions acquire a mass after EWSB, and that the physical
eigenstates are obtained from the diagonalisation of the corresponding
(generic) mass matrices.
In the weak basis, the interaction of $V_1$ with the SM
matter fields can be written as
\begin{eqnarray}
\label{eq:lagrangian:Vql0}
\mathcal{L} \supset \sum_{i=1}^{3}
V_{1}^\mu \left [
  \frac{\kappa_{L}}{\sqrt{2}}
  \left(\bar{{d}}_{L}^{0,i} \gamma_\mu {\ell}_{L}^{0,i} +
  \bar{{u}}_{L}^{0,i} \gamma^\mu  {\nu}_{L}^{0,i} \right)
  + \frac{\kappa_{R}}{\sqrt{2}}
  \bar{{d}}_{R}^{0,i} \gamma_\mu {\ell}_{R}^{0,i}
  + \frac{\bar{\kappa}_{R}}{\sqrt{2}}
  \bar{{u}}_{R}^{0,i} \gamma_\mu {\nu}_{R}^{0,i}
\right]+\text{H.c.}\, ,
\end{eqnarray}
in which the ``0'' superscript denotes interaction states, and $i=1 - 3$ are
family indices. The couplings $\kappa_{L,R}$ are flavour diagonal, and
universal.

\noindent
In terms of physical fields, and in the absence of right-handed neutrinos\footnote{In principle one can include right-handed neutrinos in the EFT fits on $b\to c\ell\nu$ data~\cite{Mandal:2020htr} and in the leptoquark model. However, couplings to right-handed neutrinos are not necessary to accommodate $R_{D^{(\ast)}}$ and $b\to s\ell\ell$ data, and would significantly increase the number of free parameters. Thus, for simplicity, we do not discuss this possibility here.}, the Lagrangian can be written as
\begin{eqnarray}
\label{eq:lagrangian:Vql_phys3}
\mathcal{L} \supset \sum_{i,j,k=1}^{3}
V_{1}^\mu \left( \bar{d}_{L}^{i}\, \gamma_\mu\,
K_L^{ik}\,\ell_{L}^{k} +\bar{u}_{L}^{j}\, V^{\dagger}_{ji}\,\gamma_\mu\,
K_L^{ik}\, U^{\text P}_{kj}\, \nu_{L}^{j} + \bar{d}_{R}^{i}\, \gamma_\mu\,
K_R^{ik}\,\ell_{R}^{k}\right)
+\text{H.c.}\, ,
\end{eqnarray}
where $V$ is the CKM quark mixing matrix
and $U^{\text P}\equiv U^{\ell\dagger}_L U^{\nu}_L$ the
PMNS leptonic mixing matrix.
We have also introduced $K_{L,R}$ as the ``effective'' leptoquark couplings in the physical mass basis, incorporating possible fermion mixing.
We can then match the ``effective'' leptoquark interactions with SM fermions to a given EFT, such that the Wilson coefficients are parametrised by the effective couplings $K_{L,R}$ and the leptoquark mass $m_{V}$.

For the general vector leptoquark scenario under consideration, the most relevant tree-level Wilson coefficients for $b\to s\ell\ell$ transitions and $R_{D^{(\ast)}}$ observable are given by~\cite{Dorsner:2016wpm}
\begin{eqnarray}
C^{ij;\ell \ell^{\prime}}_{9,10} &=& \mp\frac{\pi}{\sqrt{2}G_F\,\alpha_\text{em}\,V_{3j}\,V_{3i}^{\ast} \,m_V^2}\left(K_L^{i
  \ell^\prime} \,K_L^{j\ell\ast} \right)\,,\nonumber \\
  C^{\,\prime\,ij;\ell \ell^{\prime}}_{9,10} &=& -\frac{\pi}{\sqrt{2}G_F\,\alpha_\text{em} \,V_{3j}\,V_{3i}^{\ast} \,m_V^2}\left(K_R^{i
  \ell^\prime} \,K_R^{j\ell\ast} \right)\,,\nonumber \\
C^{ij;\ell \ell^{\prime}}_{S,P} &=& \pm\frac{\pi}{\sqrt{2}G_F\,\alpha_\text{em} \,V_{3j}\,V_{3i}^{\ast}\, m_V^2}\left(K_L^{i
  \ell^\prime} \,K_R^{j\ell\ast} \right)\,,\nonumber \\
C^{\,\prime\, ij;\ell \ell^{\prime}}_{S,P} &=& \frac{\pi}{\sqrt{2}G_F\,\alpha_\text{em} \,V_{3j}\,V_{3i}^{\ast}\, m_V^2}\left(K_R^{i
  \ell^\prime} \,K_L^{j\ell\ast} \right)\:\text,\nonumber\\
  C_{V_L}^{jk,\ell i} &=& \frac{\sqrt{2}}{4\,G_{F}\,m_V^2}\,
  \frac{1}{V_{jk}}\,
  (V\,K_{L}\, U^P)_{ji}\, K_{L}^{k\ell\ast}\,\text,
  \label{eqn:CV}
\end{eqnarray}
in which $m_V$ is the leptoquark mass. 
We can furthermore estimate the lepton flavour universal leading-log contribution to $C_9^{bs\ell\ell}$ generated via RG running effects as~\cite{Crivellin:2018yvo}
\begin{equation}
    \Delta C_{9}^{ij;\text{univ.}} \approx
-\sum_{\ell=e,\mu,\tau}\frac{\sqrt{2}}{G_F V_{tb}V_{ts}^* m^2_V}\,\frac{1}{6}\,
K^{i\ell}_{L}\,K^{j\ell\ast}_{L}\,\log(m^2_b/m_V^2)\,.
\end{equation}
Variants of the above coefficients (depending on the flavour indices) are responsible for the leading contributions to most of the $b\to s\ell\ell$ and $R_{D^{(\ast)}}$ observables relevant for our analysis. In addition, there are several other observables such as leptonic and semi-leptonic meson decays, as well as cLFV leptonic decays, which 
are important for the analysis.

As discussed in the previous chapter (cf. Section~\ref{sec:bsllfits} and Table~\ref{tab:1d2d}), it is possible to successfully accommodate the anomalies in $b\to s\ell\ell$ transitions by considering left- and right-handed currents.
For the simplified $V_1$ leptoquark moedl here studied, sizeable contributions in the right-handed Wilson coefficients $C_{9,10}^\prime$ (due to large  right-handed couplings) - as can be seen in Eq.~\eqref{eqn:CV}, necessarily lead to large contributions in the scalar and pseudo-scalar coefficients $C_{S,P}^{(\prime)}$.
Contributions to the scalar and pseudo-scalar coefficients are however strongly constrained by $B_s\to\mu\mu$ data, as discussed in Section~\ref{sec:bsllfits}, and in fact preferred to be vanishing.
Moreover, since left-handed couplings are the minimal essential ingredient
frequently called upon to
simultaneously explain the neutral and charged current
anomalies~\cite{Buttazzo:2017ixm}, for simplicity we will henceforth only consider the
latter (i.e., taking $\kappa_{L} \neq 0$ and
$\kappa_{R}=\bar{\kappa}_R=0$). Furthermore, notice that this can be
easily realised in chiral Pati-Salam (PS)
models~\cite{Buttazzo:2017ixm,Balaji:2018zna,Fornal:2018dqn},
and is moreover  phenomenologically well-motivated\footnote{
In the context of PS unification
it has been noted in the literature
that if the vector leptoquark couples to
both left- and right-handed fermion fields with similar gauge
strength, then in the absence of some helicity suppression,
bounds from various searches for lepton flavour
violating mesonic decay modes put a lower limit on the vector
leptoquark mass around 100~TeV~\cite{Valencia:1994cj,Smirnov:2007hv,Carpentier:2010ue,Kuznetsov:2012ai,Smirnov:2018ske}.}.

\begin{figure}[t!]
\centering
\includegraphics[width = 0.65\textwidth]{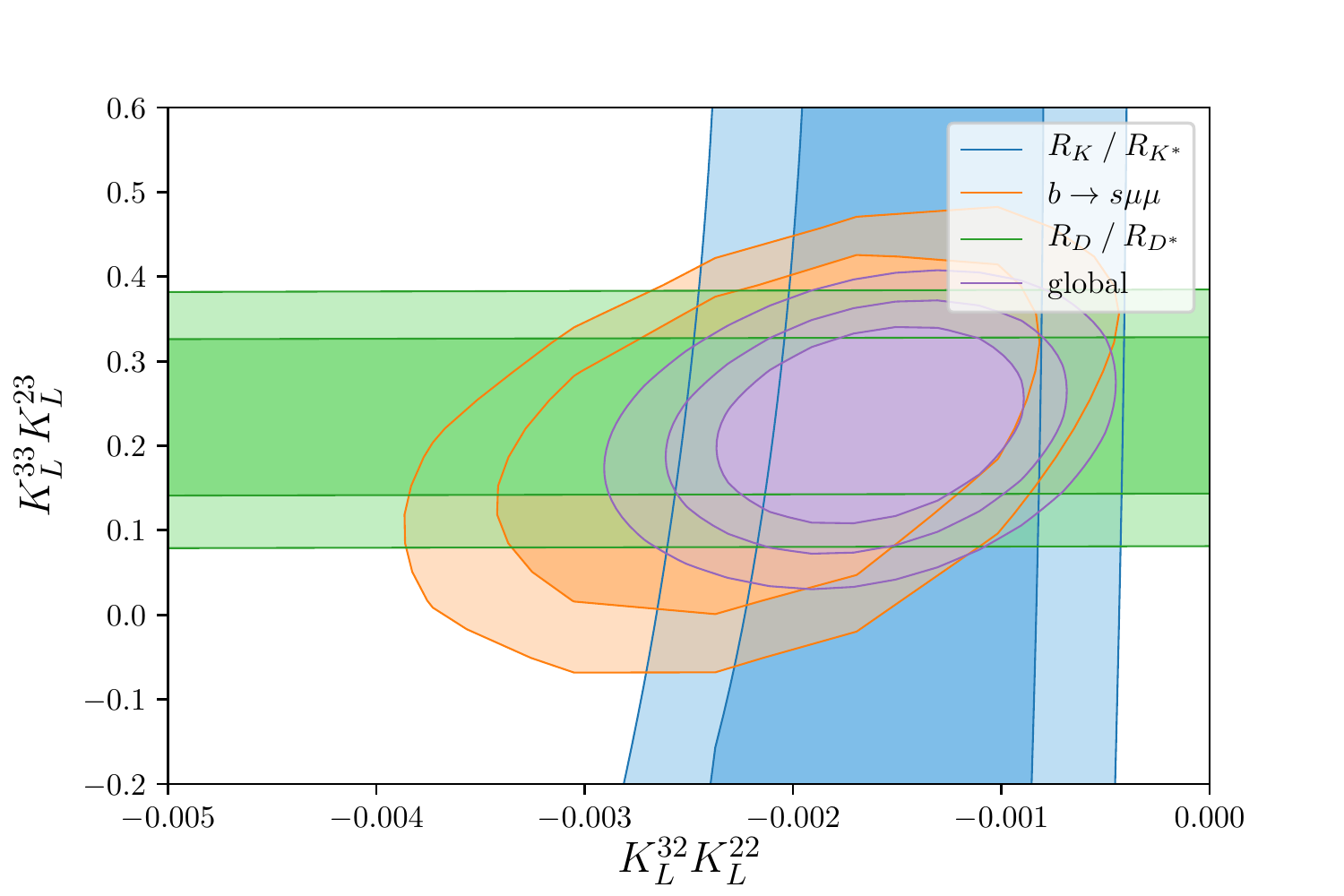}
\caption{Likelihood $1\sigma$ and $2\sigma$ contours from $R_{K^{(*)}}$, $R_{D^{(*)}}$,
  $b\rightarrow s \mu \mu$ observables and the combined global fit
  ($1\sigma$, $2\sigma$ and $3\sigma$) in the plane
  $K_L^{33}K_L^{23}-K_L^{32}K_L^{22}$, defined at the leptoquark mass scale cf.~Eq.~(\ref{eq:lagrangian:Vql_phys3}).  The first generation lepton
  and quark couplings are set to zero. Figure from~\cite{Hati:2019ufv}.}
\label{fig:RK_RD_global}
\end{figure}

In Fig.~\ref{fig:RK_RD_global}, we display
the $1\sigma$ and $2\sigma$ likelihood contours from $R_{K^{(*)}}$, $R_{D^{(*)}}$,
$b\rightarrow s \mu \mu$ observables and the combined global fit
($1\sigma$, $2\sigma$ and $3\sigma$) in the plane
$K_L^{33}K_L^{23}-K_L^{32}K_L^{22}$, where the couplings (see Eq.~(\ref{eq:lagrangian:Vql_phys3})) are varied independently and the
others are set to zero.
As can be seen in Fig. \ref{fig:RK_RD_global}, both
$b\to c\ell\nu$ and $b\to s\ell\ell$ anomalies can be indeed accommodated
simultaneously in such a minimal $V_1$ model.

\section{A hint for the need of non-unitary couplings}
\label{sec:nonuniframework}
The presence of a vector leptoquark, whose interactions with
quarks and leptons are defined in Eqs.~(\ref{eq:lagrangian:Vql0},
\ref{eq:lagrangian:Vql_phys3}), can induce new operators, contributing
to $b$-decays (both neutral and charged currents). In the SM,
$b \to s \ell\ell$ and $b \to c \ell \nu$ decays respectively occur at
one-loop and at tree-level; on the other hand, the new $V_1$-mediated
contributions to both decays arise at the tree-level.  Thus, the new
contributions required to explain $R_{K^{(\ast)}}$ data are
comparatively smaller than those needed to account for the discrepancy
in $R_{D^{(*)}}$ data: in particular, as can be inferred from the inhereent New Physics scales in Chapter~\ref{chap:bphysics}, $R_{D^{(*)}}$ require the mass
scale of $V_1$ to be quite low $\sim \mathcal{O}(1\,\mathrm{TeV})$,
while it is possible to explain $R_{K^{(\ast)}}$ for leptoquark masses
$m_V\sim \mathcal{O}(10\,\mathrm{TeV})$ (taking into account all the
constraints from rare transitions and decays, which will be discussed in detail in the following).  The low mass scale
required to explain the $R_{D^{(\ast)}}$ anomaly effectively precludes
a simultaneous (combined) explanation for both anomalies, due to the
excessive associated contributions to cLFV Kaon decays, in particular
to $K_L \rightarrow e^\pm \mu^\mp$ (which occurs at the tree level).
Consequently, both modes ($K_L \to e^{+} \mu^{-} $ and
$K_L \to e^{-} \mu^{+} $), and therefore the relevant coupling combinations, have to be suppressed separately.  

In the absence of BSM fermions, the effective leptoquark couplings $K_{L}$ (cf. Eq.~\ref{eq:lagrangian:Vql_phys3} are proportional to
an arbitrary unitary matrix (which we hereby denote $V_0$),
 and thus $K_{L}$ can be further cast as
\begin{equation}\label{eq:V0:3x3}
  K_{L}=\frac{\kappa_L}{\sqrt{2}}\, V_0=\frac{\kappa_{L}}{\sqrt{2}}
\begin{pmatrix}
c_{12}c_{13} & s_{12}c_{13} & s_{13}
\\
-s_{12}c_{23}- c_{12}s_{23}s_{13}
&
c_{12}c_{23} - s_{12}s_{23} s_{13}
 & s_{23}c_{13} \\
s_{12} s_{23} - c_{12} c_{23} s_{13} 
&
- c_{12}s_{23} - s_{12} c_{23} s_{13}
 & c_{23} c_{13}
\end{pmatrix}\,,
\end{equation}
in which we used the standard parametrisation of a real $3\times 3$ unitary
matrix in terms of three angles $\theta_{12,23,13}$ (with $c_{ij}$ and
$s_{ij}$ respectively denoting $\cos \theta_{ij}$ and $\sin
\theta_{ij}$), and we restrict ourselves to real parameter space in all our analysis. (We emphasise here that $V_0^L$ does not correspond the PMNS matrix, and that the
above angles are not those associated with neutrino
oscillation data.)

In terms
of the parametrisation of Eq.~\eqref{eq:V0:3x3}, saturating
$R_{D^{(\ast)}}$ requires maximising the $23$ and $33$ entries of
$V_0$ (thus leading to $\theta_{13} \sim 0$ and
$\theta_{23} \sim \frac{\pi}{4}$).  This implies that the branching
fractions of the tree-level Kaon decay modes are proportional to
$\sin^2\theta_{12}$ and $\cos^2\theta_{12}$, respectively.  A
sufficient and simultaneous suppression of contributions to these
modes is then clearly not possible.

\begin{figure}
\centering
\includegraphics[width = 0.65\textwidth]{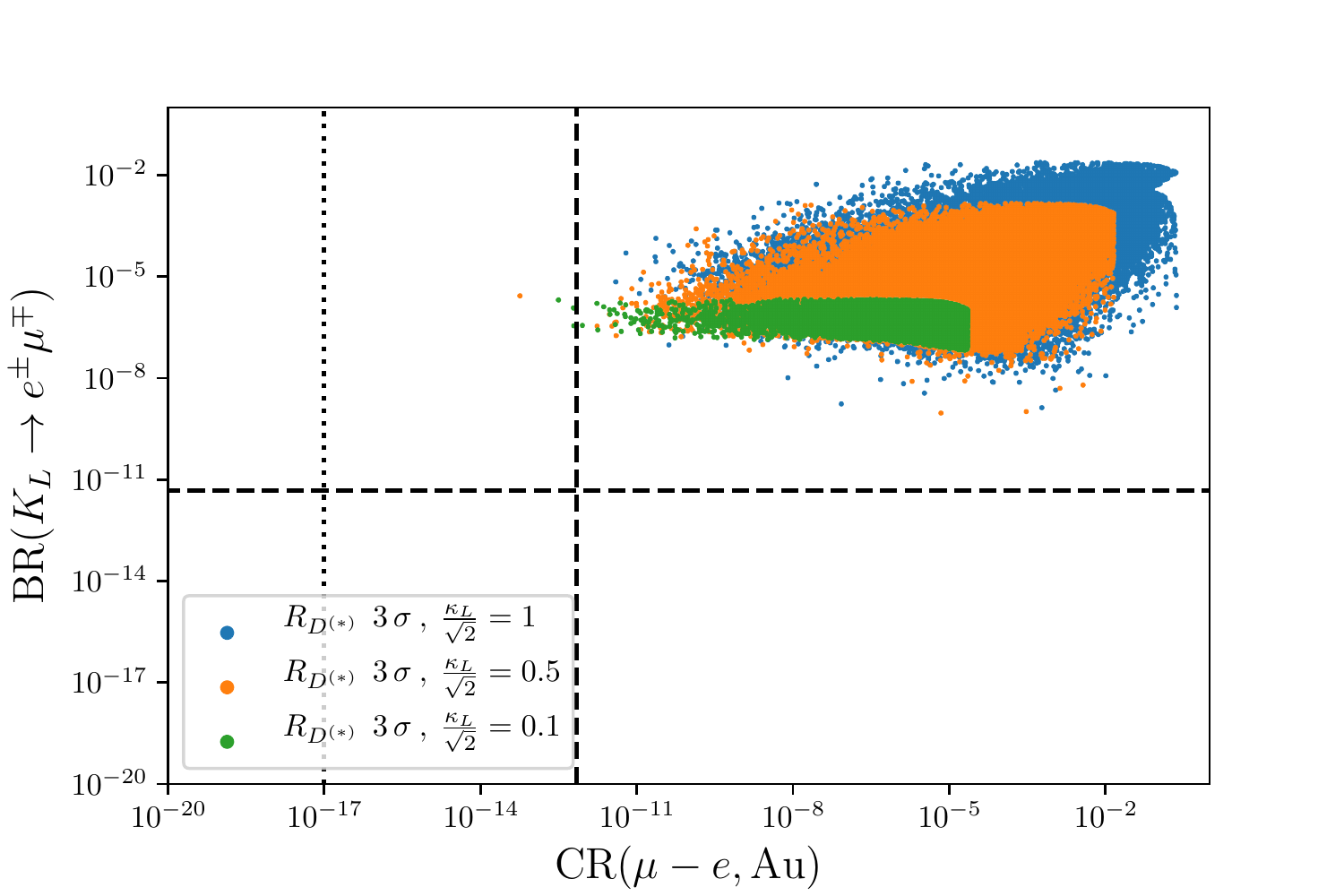}
\caption{Associated predictions for CR($\mu - e$, Au) and
  BR($K_L \to e^\pm \mu^\mp$) for sample points satisfying
  $R_{D^{(\ast)}}$ at the $3\:\sigma$ level, relying on the (unitary)
  parametrisation of Eq.~(\ref{eq:V0:3x3}). The dashed lines represent
  the current experimental upper bounds (see Tables~\ref{tab:cLFVdata}
  and~\ref{tab:semicLFV} in Section~\ref{sec:clfv_intro}), and the dotted
  line a benchmark future sensitivity to CR($\mu - e$, Al) of $\mathcal O(10^{-17})$.
  All mixing angles have been varied randomly between $-\pi$ and $\pi$ and the leptoquark mass is set to $m_V \sim 1.5 \mathrm{TeV}$. The blue, orange and green points respectively correspond to three benchmark choices, $\frac{\kappa_L}{\sqrt{2}}=1,\, 0.5,\, 0.1$. Figure from~\cite{Hati:2019ufv}.}
\label{fig:RDKaon}
\end{figure}

Likewise, excessively large contributions to $\mu-e$ conversion (also
occurring at tree-level) further exclude a low scale realisation, with
$m_V \sim \mathcal{O}(1\,\mathrm{TeV})$.  The above arguments are
illustrated by Fig.~\ref{fig:RDKaon}, in which we display the
predictions for neutrinoless $\mu - e$ conversion and
$K_L \to e^\pm \mu^\mp$ (CP averaged) associated with having contributions to
$R_{D^{(\ast)}}$ within $3 \sigma$ of the current best fit (for
$m_V \sim \mathcal{O}(1\,\mathrm{TeV})$ and three different values of $\frac{\kappa_L}{\sqrt{2}}$).

\bigskip
The above discussion suggests that the minimal flavour structure
encoded in the (unitary) parametrisation of the
leptoquark-quark-lepton currents (Eq.~(\ref{eq:lagrangian:Vql_phys3}))
is insufficient to
account for both anomalies.
A stronger enhancement of LFUV in the leptoquark couplings can be
achieved if one hypothesises that the ``effective'' leptoquark
mixings - i.e. the $3\times 3$ matrix $V_0$ is non-unitary.
As we proceed to discuss, in order to explain
$R_{K^{(*)}}$ and $R_{D^{(*)}}$ data simultaneously, and for
universal gauge couplings, a highly non-unitary flavour misalignment
between quarks and leptons is in fact required.

Such a non-unitary flavour misalignment can be understood in the
presence of heavy vector-like fermions,
${SU}(2)_L$ singlets or doublets, which have non-negligible
mixings with the SM fermions. This can be encoded by generalising the
charged lepton mixing matrix to a $3\times (3+n)$ semi-unitary matrix, so
that SM interaction fields and physical states are related as
$\ell^0_L=U^{\ell}_L \ell_L$ (for $n$ additional heavy states). This is completely analogous to the generalised lepton mixing as it appears in SM extensions via sterile fermions, as discussed in Chapters~\ref{chap:lepflav}-~\ref{chap:ISS}.

\noindent
The Lagrangian of Eq.~(\ref{eq:lagrangian:Vql_phys3}) can thus be
recast as
\begin{eqnarray}
\label{eq:lagrangian:Vql_phys3N}
\mathcal{L} \supset \sum_{i,j=1}^{3} \sum_{k=1}^{3+n}
V_{1}^{\mu} \left(
\bar{d}_{L}^{i} \gamma_\mu  K_L^{ik} \ell_{L}^{k} +
\bar{u}_{L}^{j} V^{\dagger}_{ji} \gamma_\mu K_L^{ik} U^{\text P}_{kj}
\nu_{L}^{j}
\right)
+\text{H.c.}\, .
\end{eqnarray}
Notice that in the above equation, the effective leptoquark coupling
$K_L$ now corresponds to a rectangular
$3\times (3+n)$ matrix, which can be written in terms
of $U_L^\ell$ as $K_L\equiv \frac{\kappa_{L}}{\sqrt{2}} U^{\ell}_L$.

Finally, $K_L$ can be further decomposed as $K_L=\left( K_1 , K_2\right)$,
so that $K_1$ can be now identified with the non-unitary mixings in
the light sectors (contrary to the simple limit of Eq.~(\ref{eq:V0:3x3})). $K_2$ is a $3\times n$ matrix which corresponds to the
$n$ heavy degrees of freedom describing the coupling parameters of the
heavy (vector-like) states.
Inspired by the approach frequently adopted in the context of neutrino
physics, the deviation from unitarity in the $K_1$ block can now be
parametrised as~\cite{Xing:2007zj,Blennow:2016jkn,
Fernandez-Martinez:2016lgt,Escrihuela:2015wra}
\begin{equation}\label{eq:Ndescopm_C1}
  K_1=\frac{\kappa_L}{\sqrt{2}} A\,
  V_0=\frac{\kappa_L}{\sqrt{2}}\begin{pmatrix}\alpha_{11} & 0 & 0\\
\alpha_{21} & \alpha_{22} & 0\\
\alpha_{31} & \alpha_{32} & \alpha_{33}
\end{pmatrix}V_0\,,
\end{equation}
with $V_0$ given in Eq.~(\ref{eq:V0:3x3}).
The left-triangle matrix $A$, characterises the deviation from
unitarity and encodes the effects of the mixings with the heavy states.

As already mentioned, we assume
that the vector leptoquark $V_1$ appears as a gauge boson in an
unspecified $SU(3)_c$ extension. Since neither the gauge embedding nor the
Higgs sector is explicitly specified, our only assumption is that
after EWSB all fermions (SM and vector-like)
are massive, and that the physical eigenstates are obtained from
the diagonalisation of an (effective) generic $(3+n)\times (3+n)$
lepton mass matrix.
For simplicity, we take $n=3$ generations of
heavy leptons in what follows; the $6\times6$ charged lepton mass matrix $\mathcal M_\ell$ can be
diagonalised by a bi-unitary transformation
\begin{equation}
  \mathcal{M_\ell}^{\text{diag}} \,= \,
 U^{\ell\dagger}_L \, \mathcal{M_\ell} \, U^\ell_R\,.
\end{equation}
Being a unitary $6\times6$ matrix, $U^\ell_L$ can be parametrised by
15 real angles and 10 phases, and cast as a the
product of 15 unitary rotations, $\mathcal R_{ij}$. By choosing a
convenient ordering for the products of the complex rotation matrices,
one can establish a parametrisation that allows isolating
the information relative to the heavy leptons in a simple and compact form.
Schematically, this can be described by the following
($2\times2$ block matrix) decomposition~\cite{Xing:2007zj}, to which we adhere for the remainder of our discussion,
\begin{equation}\label{eq:UL:ARBS}
U^\ell_L = \begin{pmatrix}
A & R\\
B & S
\end{pmatrix}
\begin{pmatrix}
V_0 & \mathbb{0}\\
\mathbb{0} & \mathbb{1}
\end{pmatrix}
\end{equation}
further defining
\begin{equation}
\begin{split}
\begin{pmatrix}
A & R\\
B & S
\end{pmatrix}
& =
\mathcal R_{56} \mathcal R_{46} \mathcal R_{36} \mathcal R_{26}
\mathcal R_{16} \mathcal R_{45} \mathcal R_{35}
\mathcal R_{25} \mathcal R_{15} \mathcal R_{34}
\mathcal R_{24} \mathcal R_{14}\:\text,\\
\begin{pmatrix}
V_0 & \mathbb{0}\\
\mathbb{0} & \mathbb{1}
\end{pmatrix}
& = \mathcal R_{23} \mathcal R_{13} \mathcal R_{12}\,.
\end{split}
\end{equation}
Under the above decomposition, one can still identify the SM-like
mixings, given by $V_0$ (cf. Eq.~(\ref{eq:V0:3x3})); the leptoquark
couplings\footnote{Note that this is an identification of the mixing
  elements with the effective leptoquark couplings by choosing the basis
  in which the down-type quarks are diagonal.
} are now parametrised by the $3\times 6$ (rectangular) matrix,
\begin{equation}\label{eq:KL:K1K2}
	K_L\, =\,(K_1, K_2)\, = \,\frac{\kappa_L}{\sqrt{2}}(A \,V_0, R)\,.
\end{equation}
The diagonal elements of the triangular matrix $A$, $\alpha_{ii}$, can be expressed as
\begin{eqnarray}\label{eq:Aalphaii}
\alpha_{ii} \,= \,c_{i6}\,c_{i5}\,c_{i4}\,,
\end{eqnarray}
in terms of the cosines of the mixing angles,
${c}_{ij}=\cos\theta_{ij}$. (The SM-like limit can be recovered for
$A \to \mathbb{1}$.)
The off-diagonal elements can be cast
as~\cite{Xing:2007zj}
\begin{eqnarray}\label{eq:Aalphaij}
\alpha_{21} &=& -c^{}_{14} \,c^{}_{15} \,\hat{s}^{}_{16} \,\hat{s}^\ast_{26} -
c^{}_{14} \,\hat{s}^{}_{15} \,\hat{s}^\ast_{25} \,c^{}_{26}
-\hat{s}^{}_{14}\, \hat{s}^\ast_{24} \,c^{}_{25}\, c^{}_{26}\,, \nonumber\\
\alpha_{32} &=& -c^{}_{24}\, c^{}_{25}\, \hat{s}^{}_{26} \,\hat{s}^\ast_{36} -
c^{}_{24}\, \hat{s}^{}_{25} \,\hat{s}^\ast_{35} \,c^{}_{36} -\hat{s}^{}_{24}
\,\hat{s}^\ast_{34}\, c^{}_{35}\, c^{}_{36}\,, \nonumber\\
\alpha_{31} &=&-c^{}_{14} \,c^{}_{15}\, \hat{s}^{}_{16}\, c^{}_{26}\,
\hat{s}^\ast_{36}
+ c^{}_{14} \,\hat{s}^{}_{15} \,\hat{s}^\ast_{25} \,\hat{s}^{}_{26}
\,\hat{s}^\ast_{36}
- c^{}_{14} \,\hat{s}^{}_{15}\, c^{}_{25} \,\hat{s}^\ast_{35}\,
c^{}_{36}\nonumber\\
&+&\hat{s}^{}_{14} \,\hat{s}^\ast_{24}\, c^{}_{25}\, \hat{s}^{}_{26}\,
\hat{s}^\ast_{36}
+ \hat{s}^{}_{14} \,\hat{s}^\ast_{24} \,\hat{s}^{}_{25} \,
\hat{s}^\ast_{35} \,c^{}_{36}
- \hat{s}^{}_{14} \,c^{}_{24} \,\hat{s}^\ast_{34} \,c^{}_{35}c^{}_{36}\,,
\end{eqnarray}
where
$\hat{s}_{ij} \equiv e^{i\delta^{}_{ij}} \sin\theta^{}_{ij}$,
with $\theta^{}_{ij}$ and
$\delta^{}_{ij}$ respectively being the angles and CP phases associated
with the $\mathcal R_{ij}$ rotation. Finally, it is worth emphasising
that not only the full $6\times6$ matrix $U_L^\ell$ is
unitary,
but its upper $3\times6$ block $(A V_0, R)$ is also semi-unitary on its
own, with $\frac{2}{\kappa_L^2}K_L K_L^\dagger = \mathbb{1}$.

This formalism, which can be easily generalised to $n$ extra
generations, offers the possibility of successfully
separating the information relative to the heavy leptons
in a simple and compact form. Although the couplings (in particular the
$\alpha_{ij}$ entries) can be in general complex, in what follows we
consider a minimal scenario where all couplings are taken to be real.

\section{Constraints from (rare) flavour processes, EWPO and direct searches}
\label{sec:lqconstraints}
The extended framework called upon to address the $B$ meson decay
anomalies - not only the additional vector leptoquark, but also the
presence of extra vector-like fermions, which are the origin of the
non-unitarity of the $V_1$ effective couplings - opens the door to
extensive contributions to numerous observables.

While most of the New Physics contributions occur via higher order (loop)
exchanges, it is important to notice that $V_1$ can also mediate very
rare (or even SM forbidden) processes already at the tree level. As we
proceed to discuss, the
latter observables prove to be particularly constraining, and put
stringent bounds on the degrees of freedom of these leptoquark
realisations.
The ``probing power'' of most of these observables has already been discussed in Chapter~\ref{chap:lepflav}.

In this section we thus begin with a discussion concerning rare and lepton flavour violating semi-leptonic processes, followed by the phenomenological impact of $V_1$ on cLFV observables and electroweak precision observables.
In the end we also give an overview of direct searches for leptoquark states at LHC.
A summary of the current experimental status of LFV searches (current bounds and
future sensitivities) is discussed in Section~\ref{sec:clfv_intro} (see Tables~\ref{tab:cLFVdata} and~\ref{tab:semicLFV}). 
For convenience, the current bounds and future sensitivities for the most important LFV observables are reproduced here in Table~\ref{tab:important_LFV}.

\begin{table}[h!]
	\hspace{-10mm}
	\begin{tabular}{|c|c|c|}
	\hline
	Observable & Current bound & Future sensitivity\\
	\hline
	\hline
	$\text{BR}(\mu\to e \gamma)$	&
 	\quad $<4.2\times 10^{-13}$ \quad (MEG~\cite{TheMEG:2016wtm})	&
 	\quad $<6\times 10^{-14}$ \quad (MEG II~\cite{Baldini:2018nnn}) \\
	$\text{BR}(\tau \to e \gamma)$	&
 	\quad $<3.3\times 10^{-8}$ \quad (BaBar~\cite{Aubert:2009ag})	 &
 	\quad $<3\times10^{-9}$ \quad (Belle II~\cite{Kou:2018nap}) 	 	\\
	$\text{BR}(\tau \to \mu \gamma)$	&
	 \quad $ <4.4\times 10^{-8}$ \quad (BaBar~\cite{Aubert:2009ag})	 &
 	\quad $<10^{-9}$ \quad (Belle II~\cite{Kou:2018nap})		\\
	\hline
	$\text{BR}(\mu \to 3 e)$	&
	 \quad $<1.0\times 10^{-12}$ \quad (SINDRUM~\cite{Bellgardt:1987du}) 	&
	 \quad $<10^{-15(-16)}$ \quad (Mu3e~\cite{Blondel:2013ia})  	\\
	$\text{BR}(\tau \to 3 e)$	&
 	\quad $<2.7\times 10^{-8}$ \quad (Belle~\cite{Hayasaka:2010np})&
 	\quad $<5\times10^{-10}$ \quad (Belle II~\cite{Kou:2018nap})  	\\
	$\text{BR}(\tau \to 3 \mu )$	&
 	\quad $<3.3\times 10^{-8}$ \quad (Belle~\cite{Hayasaka:2010np})	 &
 	\quad $<5\times10^{-10}$ \quad (Belle II~\cite{Kou:2018nap})		\\
	\hline
	$\text{CR}(\mu- e, \text{N})$ &
	 \quad $<7 \times 10^{-13}$ \quad  (Au, SINDRUM~\cite{Bertl:2006up}) &
 	\quad $<10^{-14}$  \quad (SiC, DeeMe~\cite{Nguyen:2015vkk})    \\
	& &  \quad $<2.6\times 10^{-17}$  \quad (Al, COMET~\cite{Krikler:2015msn,KunoESPP19,Adamov:2018vin})  \\
	& &  \quad $<8 \times 10^{-17}$  \quad (Al, Mu2e~\cite{Bartoszek:2014mya})\\
	\hline
	\hline
	$\text{BR}(K_L \to \mu^\pm e^\mp)$ & $< 4.7\times 10^{-12}\quad$~\cite{Tanabashi:2018oca} & ---\\
	\hline
	$\mathrm{BR}(\tau\to\phi\mu)$ & $<8.4\times10^{-8}\quad$~\cite{Tanabashi:2018oca} & $<2\times10^{-9}\quad$ Belle II~\cite{Kou:2018nap}\\
	\hline
	$\mathrm{BR}(B_s\to\mu^\pm\tau^{\mp})$ & $<4.2\times10^{-5}\quad$ LHCb~\cite{Aaij:2019okb} & --- \\
	\hline
	$\mathrm{BR}(B^+\to K^+\tau^+\mu^-)$ & $< 2.8\times 10^{-5}\quad$ BaBar~\cite{Lees:2012zz} & $<3.3\times 10^{-6}\quad$ Belle II~\cite{Kou:2018nap}\\
	\hline
	$\mathrm{BR}(B_s\to\phi\mu^\pm\tau^\mp)$ & $<4.3\times10^{-5}$\cite{Tanabashi:2018oca} & ---\\
	\hline
	\end{tabular}
	\caption{Current experimental bounds and future sensitivities of a selection of the most important cLFV observables which constrain the parameter space of $V_1$ leptoquark models. All upper limits are given at $90\,\%$ confidence level (C.L.). For convenience, the table summarises the most relevant observables as discussed in Chapter~\ref{chap:lepflav}.}
	\label{tab:important_LFV}
\end{table}

\subsection{Lepton flavour violating meson decays}
\label{app:meson}
Here we summarise the different vector leptoquark contributions to
leptonic and semi-leptonic meson decays which arise at tree-level, and to modes with final state neutrinos
(whose new contributions arise at one-loop level). 
Leptoquark SM extensions aiming at addressing the anomalies in $R_{K^{(\ast)}}$
and $R_{D^{(\ast)}}$ data receive strong constraints from
$d_j \to d_i \bar \nu \nu$ transitions (in particular
$s\to d \nu\nu$ and $b\to s \nu\nu$). However, the vector leptoquark
$V_1$ does not generate contributions at tree level,
and the first non-vanishing contribution appears at one loop. Consequently, we find that even with significant uncertainties,
the semi-leptonic decays into charged di-leptons
$d_j \to d_i \ell^-\ell^{\prime +}$ thus always lead to tighter
constraints (both the lepton flavour conserving and the lepton flavour violating modes).
Especially $B \to K \ell \ell^\prime$ and $B_s\to \ell\ell'$ decays with muons and $\tau$-leptons in the final state provide crucial constraints on the parameter space.

We do not include neutral meson oscillations which arise at one-loop level and typically provide much weaker 
constraints if, apart from the leptoquarks, only SM fields are considered. 
However, we notice that this may no longer hold in the presence of additional heavy fermionic states (which might be present in a UV-complete model, as for example heavy vector-like leptons);
in that case, the contributions could be sizeable so that neutral meson oscillations can then lead to important constraints, as discussed in~\cite{Hati:2019ufv,Cornella:2021sby}.

\paragraph{\pmb{$P \rightarrow \ell^- \ell^{\prime +}$}decays}
Vector leptoquarks can induce new contributions to purely leptonic decays of pseudo-scalar mesons, leading to important constraints on the flavour structure of $V_1$ couplings. Here, we provide a brief summary of the formalism for the
computation of the $P \rightarrow \ell^- \ell^{\prime +}$ rates. 
Following the standard decomposition of the hadronic matrix element~\cite{Becirevic:2016zri}
\begin{equation}
  \langle 0\,| \,\bar d_j \,\gamma_\mu\,\gamma_5 \,d_i|\,P(p)\rangle
  \,=\, i \,p_\mu \,f_{P}\,,
\end{equation}
where $f_P$ corresponds to the $P$ meson decay constant,
the branching fraction can be expressed as
\begin{align}
\text{BR}(P &\rightarrow \ell^- \,\ell^{\prime +}) \,=\,
\frac{\tau_{P}}{64 \,\pi^3}\frac{\alpha^{2}
  \,G_F^{2}}{M_P^{3}}\,f_P^{2}\,|V_{3j}\,V_{3i}^{{\ast}}|^2\,
\lambda^{\frac{1}{2}}(M_P, m_{\ell}, m_{\ell^\prime}) \times\nonumber \\
&\times\Bigg\{\left(M_P^{2} - \left(m_{\ell} + m_{\ell^\prime}
\right)^{2} \right)\Bigg|\left(C_9 - C_9^{\prime}\right)\left(m_{\ell}
- m_{\ell^\prime} \right) +\left(C_S - C_S^{\prime}
\right)\frac{M_P^{2}}{m_{d_j} + m_{d_i}}  \Bigg|^{2} + \nonumber \\
&+\left(M_P^{2} - \left(m_{\ell} - m_{\ell^\prime} \right)^{2}
\right)\Bigg|\left(C_{10} - C_{10}^{\prime}\right)\left(m_{\ell} +
m_{\ell^\prime} \right) +\left(C_P -
C_P^{\prime}\right)\frac{M_P^{2}}{m_{d_j} + m_{d_i}}   \Bigg|^{2}
\Bigg\}\,,
\end{align}
where the $\lambda(a,b,c)$ is the standard K\"all\'en-function, defined in Eq.~\eqref{eqn:kallenlambda}. 
Since the vector leptoquarks contribute to the leptonic pseudo-scalar meson decays at the tree level, such processes can provide important and very stringent constraints on the vector leptoquark couplings.

\paragraph{\pmb{$P \rightarrow P^{\prime} \ell^- \ell^{\prime +}$} decays}
The semi-leptonic decays of pseudo-scalar mesons can also be the source of significant constraints on the vector leptoquark couplings. To evaluate the differential branching fractions for these modes, we parametrise
the hadronic matrix elements following the standard convention as
\begin{align}
  &\langle \bar P^{\prime}(p')\,| \,\bar d_i \,\gamma_\mu \,
  d_j\,| \,\bar P (p)\rangle
= \left[(p+p')_\mu - \frac{M_P^{2} - M_{P^{\prime}}^{2}}{q^{2}}q_\mu
  \right]\,f_+(q^{2}) +\frac{M_P^{2} - M_{P^{\prime}}^{2}}{q^{2}}q_\mu
\,f_0(q^{2})\,,\\
&\langle \bar P^{\prime}(p')\,|\,\bar d_i  \sigma_{\mu\nu} \,d_j\,
|\,\bar P(p)\rangle = -i\,\left(p_\mu \,p'_\nu - p_\nu \,p'_\mu
\right)\frac{2}{M_{P}
  + M_{P^{\prime}}}\, f_T(q^{2},\mu)\,,
\end{align}
where the momentum transfer lies in the range
$(m_{\ell} + m_{\ell^\prime})^{2} \leq q^{2}\leq (M_P - M_{P^{\prime}})^{2}$.
For the evaluation of the form factors we closely follow the prescription of~\cite{Khodjamirian:2010vf}. The final differential branching fraction for the decay $P \rightarrow P^{\prime} \ell^- \ell^{\prime +}$  can be expressed in the form
\begin{align}
\frac{d\,\mathrm{BR} (P \rightarrow P^{\prime}
\ell^-\ell^{\prime +})}{d q^{2}} &= |\mathcal
N_{P^{\prime}}(q^{2})|^{2}\times\Big\{\delta_{\ell\ell'}\varphi_7(q^{2})\,|C_7 +
C_7^{\prime}|^{2} + \varphi_9(q^{2})\,|C_9 + C_9^{\prime}|^{2}
+\varphi_{10}(q^{2})\,|C_{10} + C_{10}^{\prime}|^{2} \nonumber\\
&+  \varphi_S(q^{2})\,|C_S + C_S^{\prime}|^{2} + \varphi_P(q^{2})\,|C_P +
C_P^{\prime}|^{2}  + \delta_{\ell\ell'}\varphi_{79}(q^{2})\,\mathrm{Re}\left[(C_7 +
  C_{7}^{\prime})\,(C_9 + C_{9}^{\prime})^{{\ast}} \right]\nonumber\\
&+ \varphi_{9S}(q^{2})\,\mathrm{Re}\left[(C_9 + C_{9}^{\prime})\,(C_S +
  C_{S}^{\prime})^{{\ast}} \right] +
\varphi_{10P}(q^{2})\,\mathrm{Re}\left[(C_{10} + C_{10}^{\prime})\,(C_P +
  C_{P}^{\prime})^{{\ast}} \right] \Big\}\: ,
\end{align}
where
\begin{align}
\varphi_7(q^{2}) &= \frac{2\,m_{d_j}\,|f_T(q^{2})|^{2}}{(M_P +
  M_{P^{\prime}})^{2}}\,\lambda(M_P, M_{P^{\prime}},
\sqrt{q^{2}})\,\left[1 - \frac{(m_{\ell} - m_{\ell^\prime})^{2}}{q^{2}}
  - \frac{\lambda(\sqrt{q^{2}}, m_{\ell}, m_{\ell^\prime})}{3\,q^{4}}
  \right]\:\text, \nonumber\\
\varphi_{9(10)}(q^{2}) &= \frac{1}{2}\,|f_0(q^{2})|^{2}(m_{\ell} \mp
m_{\ell^\prime})^{2}\,\frac{(M_P^{2} -
  M_{P^{\prime}}^{2})^{2}}{q^{2}}\,\left[1 - \frac{(m_{\ell} \pm
    m_{\ell^\prime})^{2}}{q^{2}} \right]\nonumber\\
&+ \frac{1}{2}\,|f_+(q^{2})|^{2}\,\lambda(M_P, M_{P^{\prime}},
\sqrt{q^{2}})\,\left[1 - \frac{(m_{\ell} \mp
    m_{\ell^\prime})^{2}}{q^{2}} - \frac{\lambda(\sqrt{q^{2}},
    m_{\ell}, m_{\ell^\prime})}{3\,q^{4}} \right]\:\text,\nonumber\\
\varphi_{79}(q^{2}) &= \frac{2\,m_{d_j}\,f_+(q^{2})\,f_T(q^{2})}{M_P +
  M_{P^{\prime}}}\,\lambda(M_P, M_{P^{\prime}}, \sqrt{q^{2}})\,\left[1 -
  \frac{(m_{\ell} - m_{\ell^\prime})^{2}}{q^{2}} -
  \frac{\lambda(\sqrt{q^{2}}, m_{\ell}, m_{\ell^\prime})}{3\,q^{4}}
  \right]\:\text,\nonumber\\
\varphi_{S(P)}(q^{2}) &= \frac{q^{2}\,|f_0(q^{2})|^{2}}{2\,(m_{d_j} -
  m_{d_i})^{2}}\,\left(M_P^{2} - M_{P^{\prime}}^{2} \right)^{2}\,\left[1 -
  \frac{(m_{\ell} \pm m_{\ell^\prime})^{2}}{q^{2}} \right]\:\text,\nonumber\\
\varphi_{10P(9S)}(q^{2}) &= \frac{|f_0(q^{2})|^{2}}{m_{d_j} -
  m_{d_i}}\,(m_{\ell} \pm m_{\ell^\prime})(M_P^{2} -
M_{P^{\prime}}^{2})^{2}\,\left[1 - \frac{(m_{\ell} \mp
    m_{\ell^\prime})^{2}}{q^{2}} \right]\,,
\end{align}
and the normalisation factor is given by
\begin{equation}
|\mathcal N_{P^{\prime}}(q^{2})|^{2} = \tau_{P}\,\frac{\alpha^{2}\,
  G_F^{2} \,|V_{3j} \,V_{3i}^{{\ast}}|^{2}}{512 \,\pi^5\,
  M_P^3}\,\frac{\lambda^{\frac{1}{2}}(\sqrt{q^{2}}, m_{\ell},
  m_{\ell^\prime})}{q^{2}}\,\lambda^{\frac{1}{2}}(\sqrt{q^{2}}, M_P,
M_{P^{\prime}})\,.
\end{equation}

\subsection{Charged lepton flavour violation in leptonic processes}
\label{app:cLFV}
The lepton flavour non-universal couplings of vector
leptoquarks (in general non-unitary in the present framework) induce new contributions to cLFV
observables: radiative decays $\ell_i \rightarrow \ell_j \gamma$ and 3-body decays
$\ell_i \rightarrow 3 \ell_j$ at loop level, and neutrinoless
$\mu - e$ conversion in nuclei both at tree and loop level. Further
taking into account the impressive associated experimental
sensitivity, it is clear that these observables lead to important
constraints on the vector leptoquark couplings to SM fermions. It is
important to stress that although the radiative decays are generated
at higher order, relevant anapole contributions can add to the
Wilson coefficients accounting for the tree-level contributions to
neutrinoless $\mu - e$ conversion and $\mu\rightarrow 3e$. The higher
order anapole contributions can have a magnitude comparable to the
tree level ones (or even account for the dominant contribution).  In
addition, dipole operators also contribute significantly to radiative
decays and to neutrinoless $\mu -e$ conversion. 
Although we do take tauonic modes into
account, we notice here that due to the associated current
experimental sensitivity, leptonic tau decays in general lead to
comparatively looser constraints.
Likewise, semi-leptonic lepton flavour conserving meson decays
into final states including tau leptons are typically less
constraining. However, the expected improvements in sensitivity from
dedicated experiments might render the tau modes important probes of
SM extensions via vector leptoquarks\footnote{For a detailed
  discussion regarding semi-leptonic meson decays into final states
  with tau leptons see, for example,~\cite{Capdevila:2017iqn}.}.

Notice that the one-loop dipole and anapole contributions due to
the exchange of vector
bosons generically diverge, and a UV completion must be specified to
obtain a convergent result in a gauge independent manner. We have computed
the anapole and dipole contributions in the Feynman gauge, for which it is
necessary to include the relevant contributions from the Goldstone
modes. To consistently compute these contributions
for vector leptoquarks, we make the minimal working
assumption that the new state corresponds to a (non-abelian)
$SU(3)_c$ gauge extension, whose breaking gives rise to
a would-be Goldstone boson degree of freedom, subsequently absorbed by the
massive vector leptoquark. We thus include this Goldstone mode
(degenerate in mass with $V_1$) to obtain the gauge invariant (finite)
form factors for the dipole and anapole contributions.

\paragraph{Radiative lepton decays \pmb{$\ell_i \rightarrow \ell_j \gamma$}}
Vector leptoquark exchange can induce cLFV $\ell_i \rightarrow \ell_j \gamma$ decays at one-loop level through dipole operators. We parametrise the effective Lagrangian for
radiative leptonic decays $\ell_i \rightarrow \ell_j \gamma$ as
\begin{equation}\label{eqn:radeff}
\mathcal{L}^{\ell_i \to \ell_j \gamma}_\text{eff} \,=\,
-\frac{4G_F}{\sqrt{2}}\,\bar\ell_j\,
\sigma^{\mu\nu}\,F_{\mu\nu}\,
\left(C_L^{\ell_i\ell_j} \,P_L\,  +\,
C_R^{\ell_i\ell_j}\,P_R\right)\,\ell_i \,+\, \text{H.c.}\,,
\end{equation}
where $F_{\mu\nu}$ is the standard electromagnetic field strength tensor. The $\ell_i \rightarrow \ell_j \gamma$ decay width is then given by
\begin{equation}
\Gamma(\ell_i \rightarrow \ell_j\gamma) \,= \,\frac{2 G_F^2\,(m_{\ell_i}^{2}
  - m_{\ell_j}^{2})^{3}}{\pi \,m_{\ell_i}^{3}} \,
\left(|C_L^{\ell_i\ell_j}|^{2} + |C_R^{\ell_i\ell_j}|^2\right)\,.
\end{equation}
The relevant Wilson coefficients $C_{L,R}$ can be obtained in terms of the vector leptoquark couplings\footnote{As discussed in Section~\ref{sec:simplifiedmodel}, we recall that in the current study we work under the assumption that $K_R^{ij}\simeq0$.}, cf. Eq.~(\ref{eq:lagrangian:Vql_phys3}), and are given 
by~\cite{Lavoura:2003xp}
\begin{align}
C_L^{\ell_i\ell_j} \,= \,-\frac{i\,
  N_c}{16\pi^{2}\,M^2} \frac{e}{4\sqrt{2} G_F}\sum_{k}\Bigg\{&\frac{2}{3}
\Big[\left(K_R^{kj{\ast}}\,K_R^{ki}\,m_{\ell_i}
  +K_L^{kj{\ast}}\,K_L^{ki}\,m_{\ell_j} \right) g(t_k) +
  K_R^{kj{\ast}}\,K_L^{ki}\,m_{d_k}\, y(t_k)\Big] \nonumber \\
-&\frac{1}{3}\Big[\left(K_R^{kj{\ast}}\,K_R^{ki}\,m_{\ell_i}
  +K_L^{kj{\ast}}\,K_L^{ki}\,m_{\ell_j} \right)\, f(t_k) +
  K_R^{kj{\ast}}\,K_L^{ki}\,m_{d_k}\, h(t_k)\Big] \Bigg\}\,,
  \label{eqn:radeff1}
\\
C_R^{\ell_i\ell_j} \,= \,-\frac{i \,N_c}{16\pi^{2}\,M^2} \frac{e}{4\sqrt{2} G_F}
\sum_{k}\Bigg\{&\frac{2}{3}
\Big[\left(K_L^{kj{\ast}}\,K_L^{ki}\,m_{\ell_i} +
  K_R^{kj{\ast}}\,K_R^{ki}\,m_{\ell_j} \right) \,g(t_k) +
  K_L^{kj{\ast}}\,K_R^{ki}\,m_{d_k}\, y(t_k)\Big] \nonumber\\
-&\frac{1}{3}\Big[\left(K_L^{kj{\ast}}\,K_L^{ki}\,m_{\ell_i}
 +K_R^{kj{\ast}}\,K_R^{ki}\,m_{\ell_j} \right) \,f(t_k) +
 K_L^{kj{\ast}}\,K_R^{ki}\,m_{d_k} \,h(t_k)\Big] \Bigg\}\,.
 \label{eqn:radeff2}
\end{align}
Here, $t_k = m_{d_k}^{2}/m_{V_1}^2$ and $N_c$ is the number of colours for the internal fermion in the loop. The relevant loop functions are 
\begin{align}
f(t) &= \frac{-5\,t^{3} + 9\,t^{2} - 30\,t + 8}{12\,(t-1)^{3}} +
\frac{3\,t^{2}\,{\ln}(t)}{2\,(t-1)^{4}}\,, \nonumber\\
g(t) &= \frac{-4\,t^{3} + 45\,t^{2} - 33\,t + 10}{12\,(t-1)^{3}} -
\frac{3\,t^{3}\,{\ln}(t)}{2\,(t-1)^{4}}\,, \nonumber\\
h(t) &= \frac{t^{2} + t + 4}{2\,(t-1)^{2}} -
\frac{3t\,{\ln}(t)}{(t-1)^{3}}\,, \nonumber \\
y(t) &= \frac{t^{2} - 11\,t + 4}{2\,(t-1)^2} +
\frac{3\,t^{2}\,{\ln}(t)}{(t-1)^{3}}\,.
\end{align}
\paragraph{Three body decays \pmb{$\ell \to \ell' \ell' \ell'$}}
At the loop level, three body decays can receive contributions from
photon penguins (dipole and off-shell ``anapole''), $Z$ penguins
and box diagrams, arising from flavour violating interactions
involving the vector leptoquark $V_1$ and quarks.
The relevant low-energy effective Lagrangian inducing
the four-fermion operators responsible for $\ell \to \ell' \ell'
\ell'$ decays can be written as~\cite{Okada:1999zk,Kuno:1999jp}
\begin{eqnarray}\label{eq:lto3l}
\mathcal{L}_{\ell \to \ell' \ell' \ell'} &=&
-\frac{4\,G_F}{\sqrt{2}} \left[
g_1 \,(\bar {\ell'}\,P_L\, \ell) (\bar {\ell'}
\,P_L\, \ell')\,+\,g_2 \,(\bar {\ell'} \,P_R\, \ell) (\bar {\ell'}\,
P_R \,\ell') \right. \,+\nonumber\\
&& \left.\,+\,g_3 \,(\bar {\ell'} \,\gamma^\mu \,P_R \,\ell) (\bar {\ell'}
\, \gamma_\mu \,P_R \,\ell')\,+\,
g_4\, (\bar {\ell'} \,\gamma^\mu \,P_L \,\ell) (\bar {\ell'}
\,\gamma_\mu \,P_L\, \ell') \,+\,\right. \nonumber\\
&& \left. \,+\, g_5\, (\bar {\ell'} \,\gamma^\mu \,P_R\, \ell) (\bar {\ell'}
\, \gamma_\mu \,P_L \,\ell')
\,+\, g_6 \,(\bar {\ell'} \,\gamma^\mu \,P_L \,\ell) (\bar {\ell'}
\,  \gamma_\mu \,P_R\, \ell') \right]\, +\,
\text{H.c.}\,,
\end{eqnarray}
to which the photonic dipole terms entering in
$\mathcal{L}^{\ell_i \to \ell_j \gamma}_\text{eff}$,
cf. Eq.~(\ref{eqn:radeff}), must be added; the corresponding
coefficients parametrised by
$C_{L(R)}^{\ell_i\ell_j}$ have already been
discussed in detail in the previous subsection.
Neglecting
Higgs-mediated exchanges, the off-shell anapole photon penguins, $Z$
penguins and box diagrams will give rise to non-vanishing
contributions to the above $g_3$, $g_4$, $g_5$ and $g_6$
coefficients. Note that in the large
$V_1$ mass limit, the off-shell anapole photon-penguin diagrams scale
proportionally to $ |K|^2 \ln(m_q^2/M^2)/M^2$, in contrast to the
contributions from the $Z$-penguins and box diagrams, which are (na\"ively)
proportional to $ |K|^2 m_q^2/M^4$ and  $|K|^4 m_q^2/M^4$
respectively~\cite{Gabrielli:2000te}.
Therefore we only include in our computation the
log-enhanced photonic anapole contributions, in addition to the dipole
ones. Neglecting right-handed couplings of
the leptoquark as before, the only non-vanishing coefficients (at
one-loop) are $g_4 = g_6$.
The relevant amplitude for the anapole contribution can
be written as
\begin{equation}
  i\,\mathcal M_{\text{anapole}} \,= \,i\,e \,\epsilon_\mu^{{\ast}}
  \,M_{\text{anapole}}^{\mu}\,,
\end{equation}
where $M^\mu_{\text{anapole}}$
can be parametrised in terms of a form factor $F^{\gamma \ell\ell^\prime}_L$ as
\begin{equation}
  M_\text{anapole}^\mu\, = \,\frac{1}{(4\pi)^2} \,
  F^{\gamma \ell\ell^\prime}_L\bar\ell^\prime\left(\gamma^\mu \,q^2 - \slashed q \,
  q^\mu\right) P_L\ell\,,
\end{equation}
with $q$ the off-shell photon momentum.
In this convention the $F^{\gamma \ell\ell^\prime}_L$
form factor is independent of $q^2$.
After performing the calculation in the Feynman gauge,
we obtain (in the limit of vanishing external lepton masses)
\begin{equation}
  F^{\gamma \ell\ell^\prime}_L \,=\,
  \frac{N_c}{m^2_V}\sum_i K_L^{i\ell}\,K_L^{i\ell^\prime \ast} \,
  f_a(x_i)\,,
  \label{eqn:anapolelq}
\end{equation}
in which $x_i = m_{d_i}^2/m_V^2$ and $N_c$ is the colour factor
(corresponding to the coloured fields entering in the loop).
Finally, the loop function $f_a(x)$ is given by
\begin{equation}
  f_a(x) \,= \,\frac{4 - 26 \,x + 15 \,x^2 + x^3}{12\,(1-x)^3}
  + \frac{4 - 16\,x - 15\,x^2 + 20\,x^3 - 2\,x^4}{18\,(1-x)^4}\ln(x)\,.
\end{equation}
The leptoquark-induced contributions to the 4-fermion operators are given by
\begin{equation}
  g_4 \,= \,g_6\, = \,
  -\frac{\sqrt{2}}{4 \,G_F}\,\frac{\alpha_e}{4\,\pi}
  \,Q_f \,F^{\gamma \ell\ell^\prime}_L\,.
\end{equation}
In the case of the $\ell \to 3\ell'$ decays, $Q_f=Q_\ell'$ denotes
the charge of the fermion pair at the end of the off-shell photon
(in units of $e$).
As an example, for the case of $\mu \to 3 e$ decays, one obtains the
following branching ratio~\cite{Okada:1999zk,Kuno:1999jp}
\begin{eqnarray}
\text{BR}(\mu \to e e e)&=&
2\,\left(|g_3|^2\,+\,|g_4|^2\right)
\,+\,|g_5|^2+|g_6|^2\,+\nonumber\\
&&+ 8\,e\, \text{Re}\left[C^{\mu e}_R\,
\left(2g_4^*\,+\,g_6^*\right)\,+\,C^{\mu e}_L \,
\left(2g_3^*\,+\,g_5^*\right)\right]\,+\nonumber\\
&&+ \frac{32\,e^2}{m_{\mu}^2}\,
\left\{\ln\frac{m_\mu^2}{m_e^2}\,-\,
\frac{11}{4}\right\}(\left|C_{R}^{\mu e}\right|^2\,+\,
\left|C_{L}^{\mu e}\right|^2)\,,
 \end{eqnarray}
with $C_L$ and $C_R$ as defined in Eqs.~(\ref{eqn:radeff1}, \eqref{eqn:radeff2}).
Similar expressions can be easily inferred for the other cLFV 3-body
decay channels.

\paragraph{Neutrinoless \pmb{$\mu-e$} conversion}
In terms of the relevant effective Wilson coefficients,
the general contribution to the neutrinoless $\mu-e$ conversion
rate is given by~\cite{Dorsner:2016wpm}
\begin{align}
  \Gamma_{\mu -e, \text{N}} \,= \,2\, G_F^{2}
  \,\Big(\,\Big|&\frac{C_R^{\mu e
            \ast}}{m_\mu}\,D + \left(2 \,g_{LV}^{(u)} +
        g_{LV}^{(d)}\right)V^{(p)} + \left(g_{LV}^{(u)} +
        2\,g_{LV}^{(d)} \right)V^{(n)} \nonumber\\
	&+ (G_S^{(u,p)}\,g_{LS}^{(u)} +G_S^{(d,p)}\,g_{LS}^{(d)} +
        G_S^{(s,p)}\,g_{LS}^{(s)})\,S^{(p)} \nonumber \\
	&+ (G_S^{(u,n)}\,g_{LS}^{(u)} +G_S^{(d,n)}\,g_{LS}^{(d)} +
        G_S^{(s,n)}\,g_{LS}^{(s)})\,S^{(n)} \Big|^{2} + (L\leftrightarrow
        R)\Big)\,,
\end{align}
in which the photonic dipole Wilson coefficients
$C_{L(R)}^{\ell_i\ell_j}$ have been introduced in Eq.~(\ref{eqn:radeff}), and defined in Eqs.~(\ref{eqn:radeff1}, \eqref{eqn:radeff2});
the other non-vanishing coefficients,
induced by the tree-level leptoquark exchange or arising from the
photonic anapole contributions, are given by
\begin{align}
g_{LV}^{(d)} &= \frac{\sqrt{2}}{G_F}\left(\frac{1}{\,m_V^2}  \,K_L^{d e} \,
K_L^{d\mu\ast} + \frac{\alpha}{4 \,\pi}  \,Q_d  \,
F^{\gamma \mu e}_L \right) \nonumber\\
g_{LV}^{(u)} &= \frac{\sqrt{2}}{G_F}\left(\frac{\alpha}{4 \,\pi} \, Q_u \, F^{\gamma \mu e}_L
\right)\nonumber\\
g_{RV}^{(d)} &= \frac{\sqrt{2}}{G_F}\left(\frac{\alpha}{4 \,\pi} \, Q_d  \,F^{\gamma \mu e}_L
\right)
\nonumber\\
g_{RV}^{(u)} &= \frac{\sqrt{2}}{G_F}\left(\frac{\alpha}{4 \,\pi}  \,Q_u  \,
F^{\gamma \mu e}_L \right)\,,
\end{align}
with $Q_d = -\frac{1}{3}$ and $Q_u = \frac{2}{3}$, and $F_L^{\gamma\mu e}$ defined in Eq.~\eqref{eqn:anapolelq}.
The values for the overlap integrals ($D, V, S$) are given in
Table~\ref{TabTiAuData}~\cite{Kitano:2002mt},
and the scalar coefficients $G_S^{(d_i,N)}$ can be found
in~\cite{Kosmas:2001mv}.
We again emphasise here that the off-shell anapole contributions,
often neglected in the literature, can have a
contribution comparable to the tree-level leptoquark exchange,
and therefore should be included for a thorough estimation
of the rate of $\mu-e$ conversion in nuclei.

\subsection{Neutral meson mixing: loop effects}
In the presence of $V_1$ leptoquark interactions a contribution to $\left|\Delta F\right|=2$ amplitudes is
generated at one-loop level. Contributions to
neutral meson mixings, $P-\bar P$  with $P= B^0_s, B^0_d ,K^0$,
arise both from SM box diagrams involving top quarks and $W$'s, and from
NP box diagrams involving leptons and
vector leptoquarks.
These contributions can be described
in terms of the following effective Hamiltonian for $|\Delta F| =2$
transitions
\begin{equation}\label{eq:PPmixing}
  \mathcal{H}^{P}_\text{eff} \,= \,
(C_P^\text{SM}+C_P^\text{NP})
\left(\bar d_i\,\gamma^\mu \,P_L \,d_j\right)
\left(\bar d_i \,\gamma_\mu \,P_L\, d_j\right)\,+\text{H.c.},
\end{equation}
with $\{i,j\}$ respectively denoting $\{b,s\}$, $\{b,d\}$ or $\{d,s\}$
for $P=B^0_s$, $B^0_d$ or $K^0$ mesons.
The $|\Delta F| =2$ transitions are sensitive to the
mass scale of the heavy vector-like fermions, and the widths
scale proportionally to $m_V^2$
(similarly to the SM contribution, itself proportional to $m_t^2$).
A complete evaluation of the contributions must further include the
effects of the (physical) scalar field(s); therefore
the computation of these observables requires
specifying a particular UV completion.
Nevertheless, it is possible to draw preliminary conclusions
on the mass scales of the vector leptoquarks and heavy leptonic states
(here denoted by $M$)
based on the New Physics contribution to the diagrams
involving $V_1$.
For example, taking $P=B^0_s$,
one obtains~\cite{Buttazzo:2017ixm,Calibbi:2017qbu}
\begin{equation}
C_{B_s}^{\text{NP}} \,=\,
-\frac{K_L^{2\ell}\,K_L^{3\ell\ast}\,K_L^{2\ell^\prime}\,
  K_L^{3\ell^\prime\ast}}{16\,\pi^2}\,
\left(\frac{D_6}{4M^4} + D_2 - \frac{2\,D_4}{M^2} \right)\,.
\end{equation}
Here $\ell,\ell^\prime=1,...,6$
are the six fermions with the quantum numbers of charged leptons (6 physical eigenstates arising from the mixings of the light SM and heavy vector-like charged leptons).
The loop functions $D_x\equiv D_x\left(M,M,m_s,m_t \right)$
are given by
\begin{equation}
  D_x(m_1,m_2,m_3,m_4)\, =\,\frac{i}{16\,\pi ^2}
  \int{
    \frac{d^d k}{(2\,\pi)^d}
    \frac{(k^2)^{x/2}}{(k^2 - m_1^2)\,(k^2- m_2^2)\,(k^2 - m_3^2)
    \,(k^2 - m_4^2)}}\,\text.
\end{equation}
The $|\Delta F| = 2$ transitions thus lead to a (lower) bound on the heavy leptonic mass scales of around $500$~GeV, while the vector leptoquark mass should lie above the TeV
to keep New Physics contributions to $\Delta M_{B_{s,d}}$ below
$\mathcal{O}(10\%)$, given the experimental constraints.

\subsection{One-loop effects in modes leading to final state neutrinos}
The vector leptoquark can also contribute to $s\to d \nu\nu$ and $b\to s \nu\nu$ transitions at one-loop level.
The $|\Delta S|=1$ rare decays $K^+\,(K_L)\to \pi^+\,(\pi^0)\,\nu_\ell \bar\nu_{\ell^\prime }$ and
$B\to  K^{(\ast)} \nu_\ell\bar \nu_{\ell^\prime}$ correspond to the quark level transition $d_j\to d_i \nu_\ell\bar \nu_{\ell^\prime}$, which can be described by the short-distance effective Hamiltonian~\cite{Buras:2014fpa,Bobeth:2017ecx,Bordone:2017lsy}
\begin{eqnarray}\label{eq:eff-H-Ktopi}
-\mathcal{H}_\text{eff} =
&\frac{4 \,G_F}{\sqrt{2}} \,V_{3i}^\ast \,V_{3j}\,
\frac{\alpha_e}{2\,\pi}\,
\left[C_{L,ij}^{\ell\ell^\prime} \,\left(\bar d_i\,\gamma_\mu \,P_L\,
  d_j\right)\, \left(\bar \nu_\ell\,\gamma^\mu\,
  \,P_L\,\nu_{\ell^\prime}\right) \right. \nonumber\\
&+\left.
C_{R,ij}^{\ell\ell^\prime} \,\left(\bar d_i\,\gamma_\mu \,P_R\,
d_j\right)\, \left(\bar \nu_\ell \,\gamma^\mu
\,P_L\nu_{\ell^\prime}\right) \right] \,+\, \text{H.c.}\, ,
\end{eqnarray}
where $i,j$ corresponds to the down-type quark content of the final and
initial state mesons, respectively.
For vector leptoquarks, the one-loop contributions are a priori divergent; consequently, the corresponding would-be Goldstone modes must be consistently included to obtain the correct result. Following
the prescription of~\cite{Crivellin:2018yvo}, the coefficient $C_{L,fa}^{ij}$ for $d_a \rightarrow d_f\bar\nu_i\nu_j$, due to $V_1$ leptoquark exchange is given by
\begin{align}
C_{L,fa}^{ij} = \sum_{k,l} -\frac{M_W^{2}}{2\,e^{2}\, V_{3a}\,V_{3f}^{\ast}
 \, m_{V_1}^2}\Bigg(&6\,
K_L^{fj}\,K_L^{ai\ast}\,{\ln}\left(\frac{M_W^{2}}{m_{V_1}^2} \right) +
V_{3f}^{\ast}\,V_{3k}\,K_L^{kj}\,V_{3a}\,V_{3l}^{\ast}\,
K_L^{li\ast}\,\frac{m_t^{2}}{M_W^{2}}
\nonumber\\
&+
3\left(V_{3a}\,V_{3k}^{*}\,K_L^{ki\ast}\,K_L^{fj} +
V_{3f}^{\ast}\,V_{3k}\,K_L^{kj}\,K_L^{ai\ast}
\right)\,\frac{m_t^{2}\,{\ln}\left(\frac{m_t^{2}}{M_W^{2}}
  \right)}{m_t^{2} - M_W^{2}}
\Bigg)\:\text,
\end{align}
where $M_W$ and $m_t$ respectively correspond to the masses of the $W$ boson and top quark.
The neutral and charged Kaon decay branching fractions can then be obtained by~\cite{Buras:2004qb,Buras:2015qea}
\begin{align}
&\text{BR}(K^ \pm \to \pi ^ \pm \nu \bar \nu)\, =\,
  \frac{1}{3}\left( 1 + \Delta_\text{EM} \right)\,\eta _\pm \times
  \sum_{f,i = 1}^3 \left\{ \left[
    \frac{\text{Im}\left(\lambda_t\,\tilde X_L^{fi} \right)}{
      \lambda^5}\right]^{2}+ \left[
    \frac{\text{Re} \left(\lambda_c \right)}{
      \lambda} \,P_c\, \delta _{fi}
    + \frac{\text{Re} (
      \lambda_t \,\tilde X_L^{fi})}{ \lambda ^5}
  \right]^{2}\right \},\nonumber\\
  &\text{BR}(K_L \to \pi \nu \bar \nu )\, = \frac{1}{3}{\eta_L}
  \sum_{f,i = 1}^3 \left[\frac{\text{Im}\left(\lambda_t\,
        \tilde X_L^{fi} \right)}{\lambda^5}\right]^{2}\,,
\end{align}
where
\begin{align}
\tilde X_L^{fi} &
  = X_{L}^{\text{SM},fi} - s_W^2\,C_{L,sd}^{fi}\,,\;\quad
P_c = 0.404 \pm 0.024\,, \nonumber\\
\eta_\pm  &=
\left( 5.173 \pm 0.025 \right)\times 10^{-11}
\left[\frac{\lambda}{0.225} \right]^8\,,\nonumber\\
  \eta_L &=\left( 2.231 \pm 0.013 \right)\times 10^{-10}
 \left[\frac{\lambda}{0.225} \right]^8\,,\nonumber\\
\Delta_\text{EM} &=  - 0.003\,,\; \quad
X_{L}^{\text{SM},fi} =
\left(1.481 \pm 0.005 \pm 0.008\right)\,\delta _{fi}\,.
\end{align}
Here, $\lambda$ corresponds to the standard Wolfenstein parametrisation (i.e. the Cabibbo angle), $\lambda_c = V_{cs}^\ast V_{cd}$ and $\lambda_t = V_{ts}^\ast V_{td}$. The decay width for $B\to K^{(*)}\nu\bar{\nu}$ has been derived
in~\cite{Buras:2014fpa}, leading to
$C_{L,sb}^{\text{SM},fi} \approx -1.47/s_W^2\delta_{fi}$, which can be used to normalise the branching ratios as
\begin{equation}
R_{K^{(*)}}^{\nu\bar{\nu}} \,= \,
\frac{1}{3}\sum_{f,i=1}^3
\frac{ \big|C_{L,sb}^{fi} \big|^2}{
  \big|C_{L,sb}^{\text{SM},fi}\big|^2} \,\text.
\end{equation}
%

\subsection{Electroweak precision observables constraining vector-like leptons}
\label{sec:ew}
As extensively discussed in Chapters~\ref{chap:lepflav} and~\ref{sec:massivenu}, non-minimal neutral lepton mixing with additional fermionic states can have drastic phenomenological implications on electroweak precision observables, due to modifications of the tree-level gauge boson interactions.
Likewise, the presence of heavy vector-like fermions (at the source
of the non-unitary couplings of the vector leptoquark to the light
fermions) can have a non-negligible impact on the couplings of SM
fermions to gauge bosons. 
In turn, this can be manifest in new
contributions to several EW precision observables - potentially in
conflict with  SM expectations and precision data -, which will prove to
play a key role in constraining the mixings of the SM charged leptons
with the heavy states.

For the $Z$-couplings, which are modified at the tree level, the most stringent constraints are expected to
arise from leptonic $Z$ decays; in our analysis, we take into account
the LFU ratios and cLFV decay modes of the $Z$ boson. 
For convenience, we summarise in
Table~\ref{tab:EWPO} the EWP observables which are of particular relevance for
the present leptoquark study (experimental measurements and SM predictions).
More details about the observables in general can be found in Chapter~\ref{chap:lepflav}.

\renewcommand{\arraystretch}{1.3}
\begin{table}[h!]
\begin{center}
\begin{tabular}{|c|c|c|}
\hline
Observables  & Experimental data & SM Prediction\\
\hline
$\Gamma_Z$ & $2.4952\pm0.0023\:\mathrm{GeV}$ &
$2.4942\pm0.0008\:\mathrm{GeV}$\\
$\Gamma(Z\to\ell^+\ell^-)$ & $83.984\pm0.086\:\mathrm{MeV} $ &
$83.959\pm0.008 \:\mathrm{MeV}$\\
\hline
$R_e$ & $20.804\pm0.050$ & $ 20.737\pm0.010$\\
$R_\mu$ & $20.785\pm0.033$ & $ 20.737\pm0.010$\\
$R_\tau$ & $20.764\pm0.045$ & $ 20.782\pm0.010$\\
\hline
$\mathrm{BR}(Z\to e^\pm\mu^\mp)$ & $ < 7.5 \times 10^{-7} $ & --\\
$\mathrm{BR}(Z\to e^\pm\tau^\mp)$ & $< 9.8\times 10^{-6} $ & --\\
$\mathrm{BR}(Z\to \mu^\pm\tau^\mp)$ & $ < 1.2\times10^{-5}$ & --\\
\hline
\end{tabular}
\caption{Subset of EWP observables affected by the modified
  $Z$ couplings, with the corresponding experimental measurements and
  SM predictions~\cite{Tanabashi:2018oca}. The ratios $R_\ell$ are
  defined as
  $R_\ell = {\Gamma_\text{had}}/{\Gamma(Z\to\ell^+\ell^-)}$, and
\mbox{$\Gamma(Z\to\ell^+\ell^-)$} denotes an average over $\ell =
  e,\,\mu,\,\tau$.
  For convenience, the most relevant observables discussed in Chapter~\ref{chap:lepflav} are reproduced here.
}\label{tab:EWPO}
\end{center}
\end{table}
\renewcommand{\arraystretch}{1.}

\paragraph{Couplings of the \pmb{$Z$}-boson and photon}

If the heavy vector-like fermion states are $SU(2)_L$ singlets,
mixings with the light $SU(2)_L$ doublets can lead to
modified couplings of the latter to the $Z$ boson ($\bar f f Z$).
For the case of charged leptons,
the relevant couplings can be obtained from the kinetic terms,
\begin{eqnarray}
\mathcal{L}_\text{kin} \,\supset \,
\bar \ell^{0}_{La} \,i \slashed D_a \, \ell^{0}_{La}  \,+ \,
\bar \ell^{0}_{Ra} \,i \slashed D_a \, \ell^{0}_{Ra}   \, =  \,
\bar { \ell}_{L j}  \, (U^\ell_L)^\dagger_{j a} \, i \slashed D_a \,
(U_L^\ell)_{ ak} \,  \ell_{L k}  \,+ \,
\bar {\ell}_{R j} \, (U^\ell_R)^\dagger_{j a} \, i \slashed D_a \,
(U_R^\ell)_{a k} \,  \ell_{R k} \,,
\label{eq:kinE}
\end{eqnarray}
where $\{j, k\}$ denotes the physical fields and
$\{a,b\} = 1 ... 6$ the interaction
states (with $ a \in \{1,2,3\}$ corresponding to SM fermions, and
$a \in \{4,5,6\}$ to the new heavy vector-like states).
The covariant derivative associated with the charges of a given state
$a$ can thus be written
\begin{eqnarray}
D_{\mu,\, a} \,= \, \partial _\mu \,-  \,
i \frac{g_w}{\cos \theta_W} \, \left(T^3_a  \,- \, \sin ^2 \theta_W  \,
Q_a \right) \, Z_\mu  \,-  \,i  \,e  \,Q_a  \,A_\mu \,,
\end{eqnarray}
where $T^3$ and $Q$
respectively denote the weak isospin and the electric charge.
Since the electric charge is the same for all lepton states ($Q_a=-1$),
the couplings of the photon are not modified.

Let us now introduce the ``effective'' $Z$ boson couplings,
\begin{align}\label{eq:gZff:eff}
&(g^{Z\ell_j \ell_k}_L)_\text{eff}\,=\,
   \sum_{a=1}^6 \frac{g_w}{\cos \theta_W} \left(
   T^3_{L\,a}  \,- \, \sin ^2 \theta_W\, Q_a \right) \,
   (U^\ell_L)^\dagger_{j a} (U^\ell_L)_{a k}\,,
\nonumber \\
&(g^{Z\ell_j \ell_k}_R)_\text{eff}\,=\,
   \sum_{a=1}^6 \frac{g_w}{\cos \theta_W}\left(
   T^3_{R\,a}  \,- \, \sin ^2 \theta_W\, Q_a \right) \,
   (U^\ell_R)^\dagger_{j a} (U^\ell_R)_{a k}\,,
\end{align}
where $T^3_{L(R)}$ is the weak isospin of a left-handed
(right-handed) lepton.  Should the SM fermions and the heavy states
belong to the same $SU(2)_L$ representation, universality is trivially
restored (by unitarity) for both $g^{Z\ell_j \ell_k}_{L,R}$ effective
couplings, and one recovers the SM universal couplings. For heavy
isodoublet vector-like states, one has
\begin{equation}
(g^{Z\ell_j \ell_k}_L)_\text{eff}\,=\,\frac{g_w}{\cos \theta_W} \left(
-\frac{1}{2}\,+ \, \sin ^2 \theta_W\, \right) \, \delta_{jk}\,\text,
\end{equation}
However, if the new fields transform differently (have distinct charges)
under $SU(2)_L$,
the $g^{Z\ell_j \ell_k}_{L,R}$ couplings are modified. In
particular, in the presence of \textit{isosinglet heavy states}, one finds
\begin{equation}\label{eq:gZff:eff:singlets}
(g^{Z\ell_j \ell_k}_L)_\text{eff}\,=\,
\,\frac{g_w}{\cos \theta_W} \left(
-\frac{1}{2}\,+ \, \sin ^2 \theta_W\, \right) \, \delta_{jk}
\,+\, \Delta g^{jk}_L\,,
\quad \quad
\text{with} \quad
\Delta g^{jk}_L\, = \,
\sum_{a=4}^6\,  \frac{1}{2}\,\frac{g_w}{\cos \theta_W}
   (U^\ell_L)^\dagger_{j a} (U^\ell_L)_{a k}\,.
\end{equation}
Likewise, vector-like doublets also lead to the modification of
the $g^{Z\ell_j \ell_k}_{R}$ couplings:
\begin{equation}\label{eq:gZff:eff:doublets}
(g^{Z\ell_j \ell_k}_R)_\text{eff}\,=\,
\,\frac{g_w}{\cos \theta_W} \sin ^2 \theta_W \, \delta_{jk}
\,+\, \Delta g^{jk}_R\,,
\quad \quad
\text{with} \quad
\Delta g^{jk}_R\, = \,
\sum_{a=4}^6  \,-\frac{1}{2}\,\frac{g_w}{\cos \theta_W}
   (U^\ell_R)^\dagger_{j a} (U^\ell_R)_{a k}\,.
\end{equation}

\paragraph{Couplings of the \pmb{$W$}-boson}

The possible mixings with the heavy vector-like leptons can also
modify the couplings to the $W$ boson. The charged current interaction
terms can be written
\begin{align}
\mathcal{L}^\text{cc} \, &=\,
\frac{g_w}{\sqrt{2}} \,W_\mu  \, \bar {\nu}^0_{a} \,\gamma^\mu\,
\,\ell^0_{L a}\, + \,
\text{H.c.} \,,\nonumber \\
&=\,
\frac{g}{\sqrt{2}} \,W_\mu  \, \bar {\nu}_{j} \,\gamma^\mu\,
(U^{\nu\dagger}_L)_{ja} \,(U_L)_{ak}\,  \ell_{L k}  \,+\,
\text{H.c.} \,\text,
\end{align}
so that the corresponding charged current couplings are then given by

\begin{equation}
g^{W\nu_j \ell_k}_{L} \,= \,
\frac{g_w}{\sqrt{2}}\, (U^{\nu\dagger}_L)_{ja} \,(U^{\ell}_L)_{ak}
\,=\,
\frac{g_w}{\sqrt{2}}\,U^{\mathrm P\dagger}_{jk}\,,
\end{equation}
where
$U^{\mathrm P}$ denotes the (generalised) PMNS mixing matrix.
A priori, the branching ratios of $W\rightarrow \ell \nu$ can
constrain the mixings of the heavy leptons (see,
e.g.~\cite{Dermisek:2013gta,Poh:2017tfo}). However, these
strongly depend on the neutrino mass generation mechanism (as well
as on the structure of the Higgs sector), and in the present analysis we
will not take them into account;
we nevertheless mention that
for a given Higgs sector,
in the presence of additional isosinglet heavy neutrinos,
the modified charged current vertex can impact
several observables. 
As extensively discussed in Chapters~\ref{chap:lepflav}-~\ref{chap:ISS}, in models with massive neutrinos the non-unitary lepton mixing leads to a modification of the SM gauge boson interactions with SM leptons.
As demonstrated in Chapters~\ref{chap:lepflav}-~\ref{chap:ISS}, this opens the door to a very rich phenomenology on its own.
Therefore, in addition to electroweak
precision measurements of
$\mathrm{BR}(W\rightarrow \ell\nu)$~\cite{Abada:2013aba},
other decays or collider processes with one or
two neutrinos in the final state, as for example
$\tau$ decays, leptonic and semi-leptonic meson
decays~\cite{Atre:2009rg,Abada:2017jjx,Abada:2019bac}, production and
decay of $W$ bosons to di-lepton and two jets at
the LHC~\cite{Bray:2007ru,Dev:2019rxh}, can also lead to interesting
constraints depending on the mass scales of the additional isosinglet
neutral states.

\paragraph{Constraints from EWP observables}
Due to the tree-level modified $Z$-couplings (a consequence of the mixing
of SM fermions with the heavy vector-like fermions),
strong constraints from EWP observables are expected to arise
from the observed lepton universality in charged
leptonic $Z$-decays.

At tree level, the decay width of a massive vector boson to fermions is
given by \cite{Abada:2013aba}
\begin{align}
\label{eqn:Vff}
&\Gamma(V\to f f^\prime) \,= \, \frac{\lambda^{1/2}(m_V, m_f,
  m_{f^\prime})}{48 \pi m_V^3} \nonumber\\
&\times \left[\left(| g_L^{f f^\prime}|^2 + | g_R^{f
      f^\prime}|^2 \right)\left(2m_V^2 - m_f^2 - m_{f^\prime}^2
    - \frac{\left(m_f^2 - m_{f^\prime}^2\right)^2}{m_V^2} \right) + 12
    m_f m_{f^\prime} \,\mathrm{Re} \left(g_L^{f f^\prime} g_R^{f
      f^\prime \ast}\right) \right]\,,
\end{align}
where $g_{L(R)}$ are the chiral couplings and the Käll\'en function is
defined in Eq.~\eqref{eqn:kallenlambda}.
In the case of the $Z$-boson, the relevant couplings have been introduced
in Eq.~(\ref{eq:gZff:eff}). From Eq.~(\ref{eqn:Vff}), and in view of
the very good agreement between the SM predictions and experiment (cf.
Table~\ref{tab:EWPO}), it is clear that any modification of the
tree-level couplings of the $Z$-boson will be subject to very
stringent constraints, which in turn translate into bounds on
the mixing parameters responsible for the $\Delta g_{L(R)}$ terms
(see Eqs.~(\ref{eq:gZff:eff:singlets}, \ref{eq:gZff:eff:doublets})).
Using Eq.~(\ref{eqn:Vff}), one can derive conservative
constraints on $\Delta g_{L(R)}$ from the requirement of compatibility
with the bounds of Table~\ref{tab:EWPO}. As an example, the current
experimental data on $\text{BR}(Z\to e^\pm\mu^\mp)$ and
$\Gamma(\ell^+\ell^-)$ leads to
\begin{align}\label{eq:DeltagLgR}
  |\Delta g^{e\mu}_L |^2 + |\Delta g^{e\mu}_R|^2
  &\lesssim 1.55\times 10^{-6}, \nonumber\\
|\Delta g^{\ell\ell}_L | & \lesssim 5.6 \times 10^{-4},\nonumber\\
|\Delta g^{\ell\ell}_R | & \lesssim 3.5 \times 10^{-4}\,.
\end{align}
The above constraints allow in turn to infer bounds on the elements of
the matrix $A$ (see Eqs.~(\ref{eq:UL:ARBS} - \ref{eq:Aalphaij})).
To illustrate this point, we consider the case of \textit{isosinglet heavy
vector-like leptons} (for which $\Delta g_R=0$)\footnote{Note that in
  the presence of a nontrivial Higgs sector inducing mixings between
  right-handed SM charged leptons and vector-like doublets, one has
  non-vanishing contributions to $ \Delta g_R$, which can
  lead to constraints on the right-handed mixing matrix, parametrised as done for $K_L$, see Eq.~(\ref{eq:Ndescopm_C1}), for a given UV complete
  framework. However, a detailed analysis of such a scenario is beyond
  the scope of our current work. Here, we only include the
  conservative limits for left-handed mixing elements, which are of
  foremost importance to our analysis.}: the limits of
Eq.~(\ref{eq:DeltagLgR}) translate into
\begin{align}
1-|\alpha_{11}| &\lesssim 4 \times 10^{-4}\,, \nonumber\\
1-|\alpha_{22}| &\lesssim 3 \times 10^{-4}\,, \nonumber\\
|\alpha_{21}| &\lesssim 4.6 \times 10^{-4}\, .
\end{align}
These bounds can be well compared to the bounds on $\eta$, measuring the deviation from unitarity of the PMNS, as discussed in Chapter~\ref{sec:massivenu}.

\subsection{Direct searches}
Finally, it is clear that the (negative)
results of direct searches for the exotic states must be taken into account.
At the LHC, pairs of vector leptoquarks can be abundantly produced
in various processes (via $t$-channel lepton exchange and direct
couplings to one or two gluons).
Due to the underlying gauge structure of possible
UV completions, the production cross section
strongly depends on the coupling to gluons which makes the
theoretical predictions for the production of vector leptoquarks
less robust than those for scalar leptoquarks. On the other hand, if the vector leptoquark corresponds to a spontaneously broken non-abelian gauge symmetry, gauge invariance completely fixes the couplings between the vector leptoquark and the gluons, implying lower limits on the vector leptoquark mass (albeit still depending on the branching fractions of the leptoquark) from the negative results in the direct searches for pair production at the LHC, see for instance \cite{Angelescu:2018tyl}.

As a natural consequence of the favoured structure called upon to
maximise the effects on $B$-meson decay anomalies, $V_1$ is expected to
dominantly decay into either $t\bar \nu_\tau$ or $b\bar \tau$. The ATLAS and CMS collaborations have
conducted extensive searches, assuming that
leptoquarks couple exclusively to third generation quarks and
leptons~\cite{Khachatryan:2014ura,Aad:2015caa,Sirunyan:2017yrk,Sirunyan:2018vhk,Sirunyan:2018kzh},
which have led to lower limits on the $V_1$ mass $\sim 1.5$~TeV.
Further important collider signatures are
$pp \to \tau \bar\tau + X$, arising from $t$-channel
leptoquark exchange or from
single leptoquark production in association with a charged lepton.
As argued in~\cite{Cerri:2018ypt,CidVidal:2018eel}, the projected vector
leptoquark reach of HL-LHC with $3$~ab$^{-1}$ is close to
$1.8$~TeV.
Much higher luminosities and/or more sophisticated search strategies
are required to probe the preferred mass scale and couplings of states
at the origin of a combined explanation
of the anomalies (for instance, as suggested by the study
of~\cite{Buttazzo:2017ixm}).
Other potentially interesting search modes could include $b\mu b\tau$ and $b \mu b \mu$ final
states.

In the present analysis, we select (working) benchmark values for the
mass and gauge coupling of the vector leptoquark
allowing to comply with the current available limits. 
In particular, for the following numerical analysis we set $\frac{\kappa_L}{\sqrt{2}} = 1$ as a benchmark choice. Nevertheless, for any other choice consistent with the constraints from direct searches (as discussed above) the qualitative behaviour and the conclusions drawn remain the same. However, for very small values $\kappa_L$ ($\kappa_L\lesssim 0.1$ for $m_V\gtrsim 1.5$~TeV) the number of points in the best fit region for $R_{K^{(*)}}$ and $R_{D^{(*)}}$ anomalies starts to decrease drastically, so that the New Physics effects become negligible with respect to the SM.

\mathversion{bold}
\section{Phenomenological viability of SM extensions via $V_1$ vector leptoquarks and vector-like leptons}
\mathversion{normal}
We now finally address the question of whether a SM extension via vector
leptoquarks can simultaneously provide an explanation to both the
$R_{K^{(\ast)}}$ and $R_{D^{(\ast)}}$ data\footnote{The numerical results presented in this section rely on the experimental status of late 2019 (see~\cite{Hati:2019ufv}), i.e. several updates and new measurements in $b\to s\ell\ell$ are therefore not included. However, this has only a minimal impact on the qualitative conclusions drawn in this section.}, working under the
hypothesis of universal gauge couplings for the vector leptoquark
$V_1$ in the unbroken phase. In this framework, the required flavour non-universality arises
from non-unitary mixings among the SM fermions - a consequence of the
existence of heavy vector-like fermions which have non-negligible
mixings with the light fields.

We first begin by considering the most minimal scenario where the
non-unitary flavour misalignment is due to the presence of a
\textit{single generation of heavy vector-like charged leptons}
(i.e., $n=1$ in Eq.~(\ref{eq:lagrangian:Vql_phys3N})).
Such a minimal field content could already lead to a sufficient amount of
LFUV to account for both $R_{K^{(*)}}$ and $R_{D^{(*)}}$ anomalies.
However, although new contributions to rare meson decays and transitions are
still in good agreement with current experimental bounds, this
scenario is ruled out due to the stringent constraints on cLFV\footnote{
Despite being loop-suppressed in the present $V_1$
  leptoquark  SM extension, $B\to K^{(*)}\nu\bar{\nu}$ decays can in
  general lead to significant constraints; nonetheless, in
  the scenarios here discussed, we find that constraints from LFV meson
  decays, and most importantly cLFV observables,
  provide tighter constraints.} modes.
Excessive contributions to (tree-level) muon-electron conversion
in nuclei play a crucial role in ruling out this realisation, as well
as the closely related radiative decays.
In order to reconcile the model's prediction with
the current bounds on CR($\mu-e$, Au), the photon-penguin contributions
must (at least) partially cancel the sizeable tree-level ones;
however, such large photon-penguin contributions then translate into
unacceptably large $\mu \to e
\gamma$ decay rates, already in conflict with current bounds.

\medskip The required flavour non-universality can be recovered for a
less dramatic unitarity violation; this can be achieved by extending
the particle content by \textit{two or more additional heavy charged
  lepton states}, or formally for $n\geq 2$ in
Eq.~(\ref{eq:lagrangian:Vql_phys3N}).  Although $n=2$ provides more
freedom to evade the constraints found in the case $n=1$, no generic
solution was found in this case, which might then require an extreme fine
tuning to become viable. In the subsequent discussion, we therefore take $n=3$ which
is the minimal case and conveniently replicates the number
of generations in the SM.

\begin{figure}[h!]
   \mbox{\hspace{-8mm}\includegraphics[width=.40\linewidth]{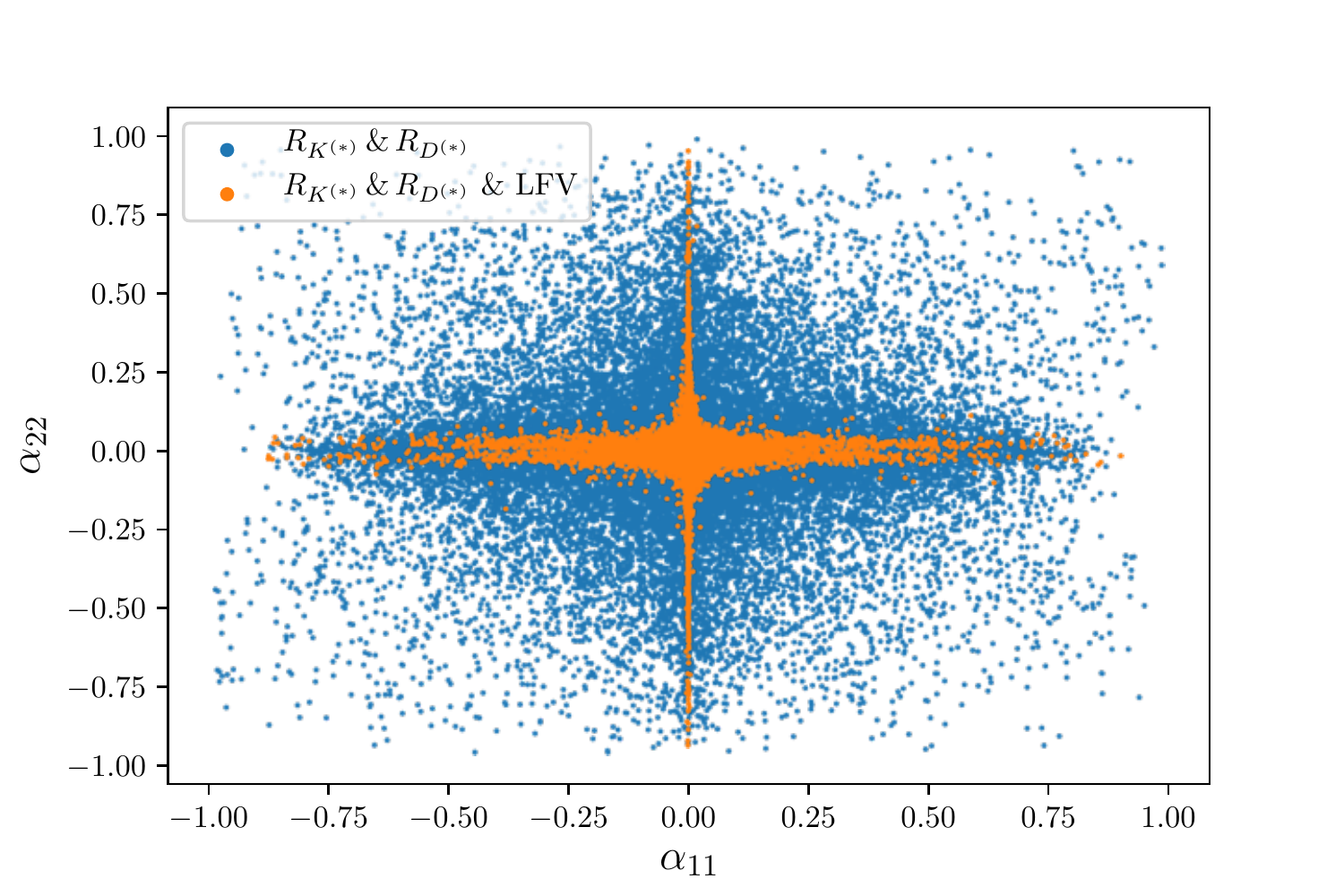}
         \hspace{-6mm}\includegraphics[width=.40\linewidth]{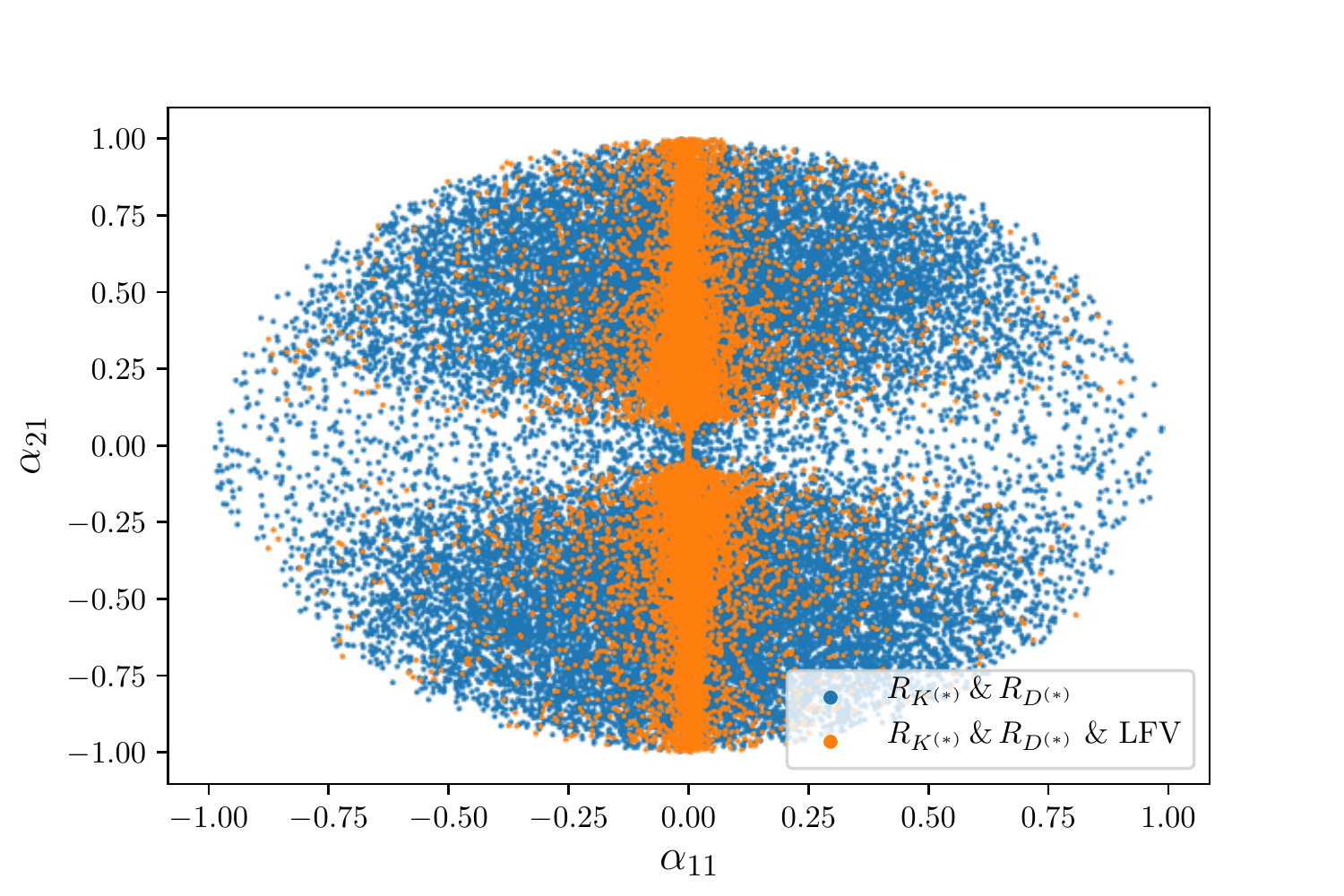}
         \hspace{-6mm}\includegraphics[width=.40\linewidth]{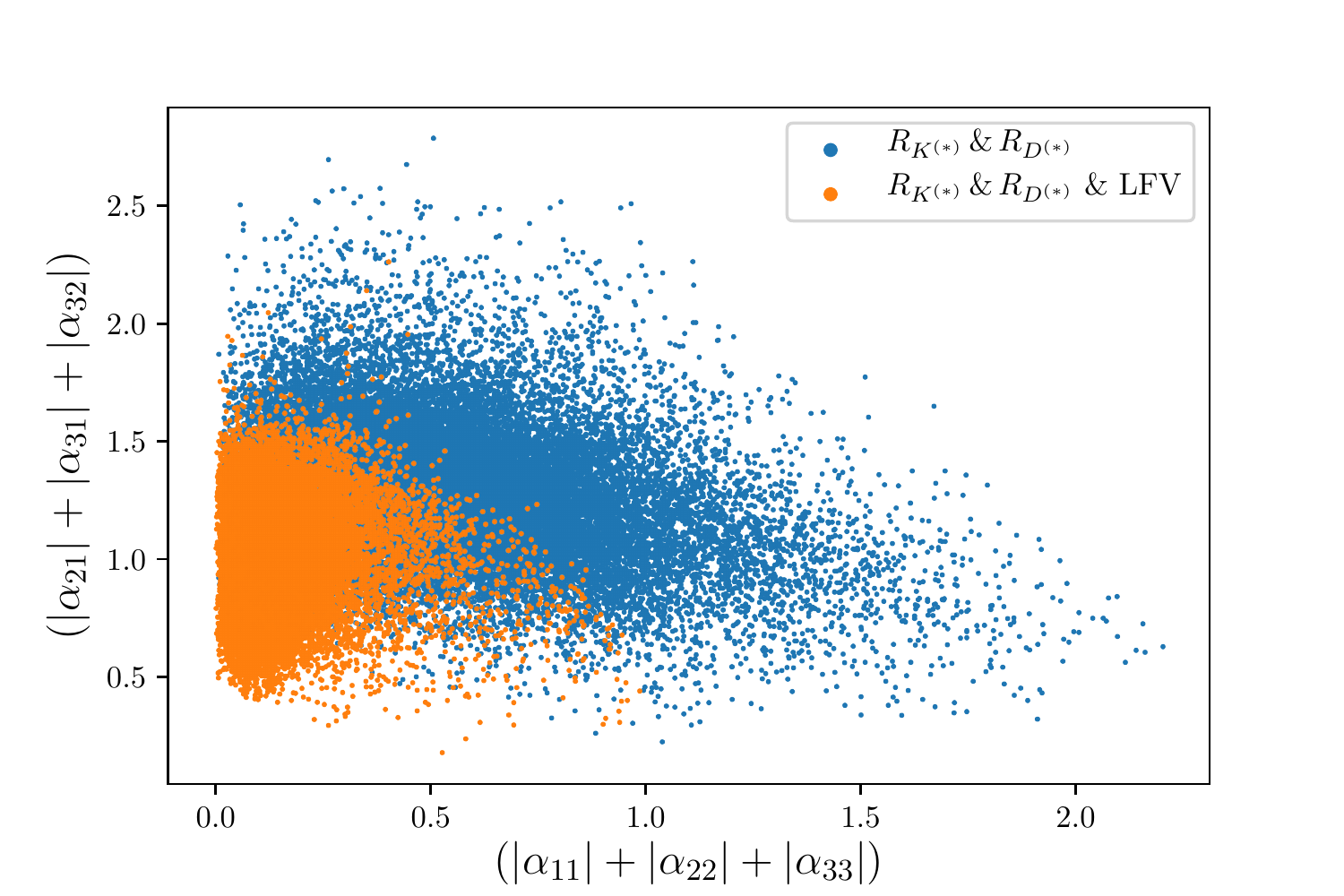}}  \\
    \includegraphics[width=.30\linewidth]{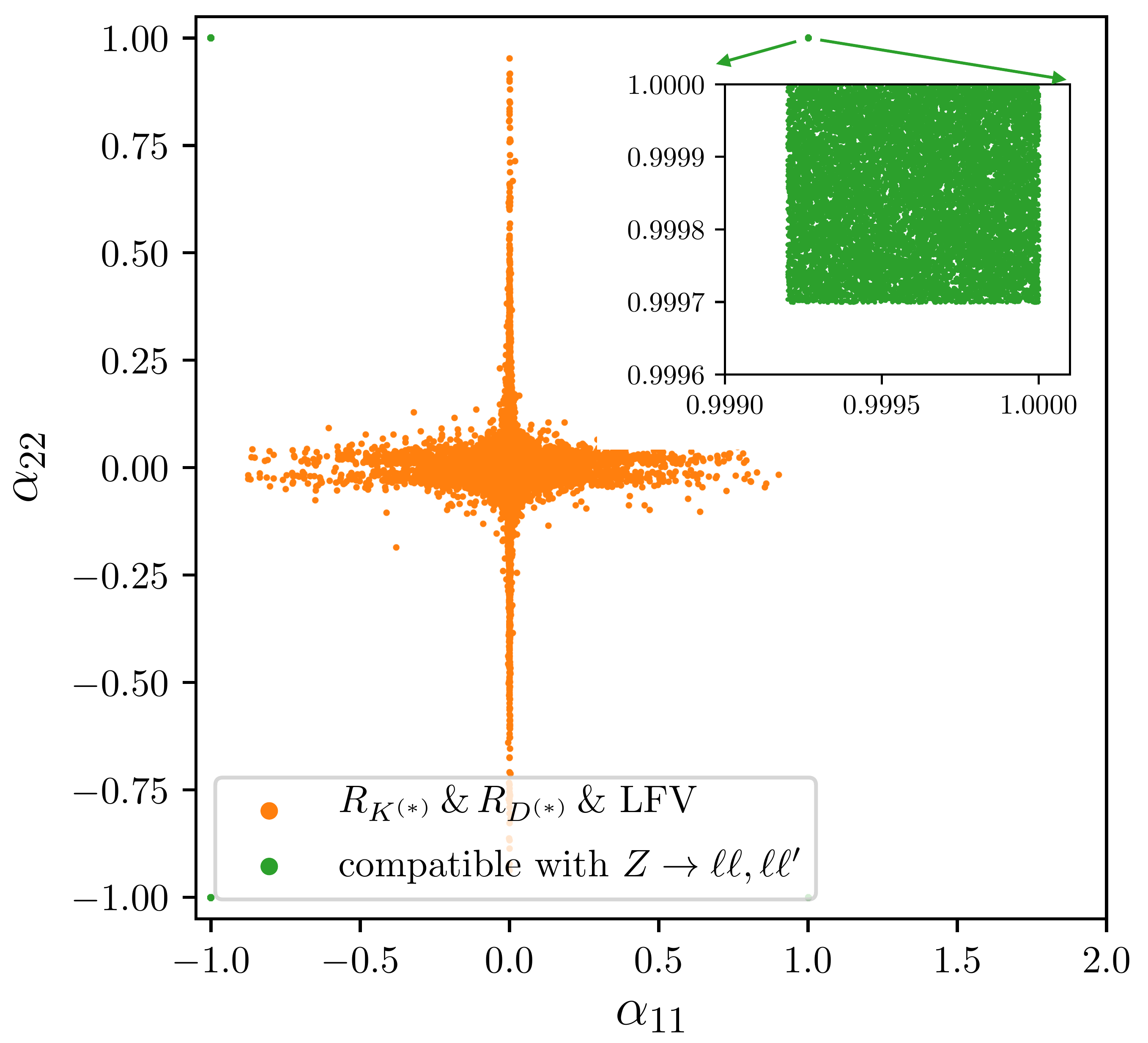}
    \hspace{.5 cm}
    \includegraphics[width=.302\linewidth ]{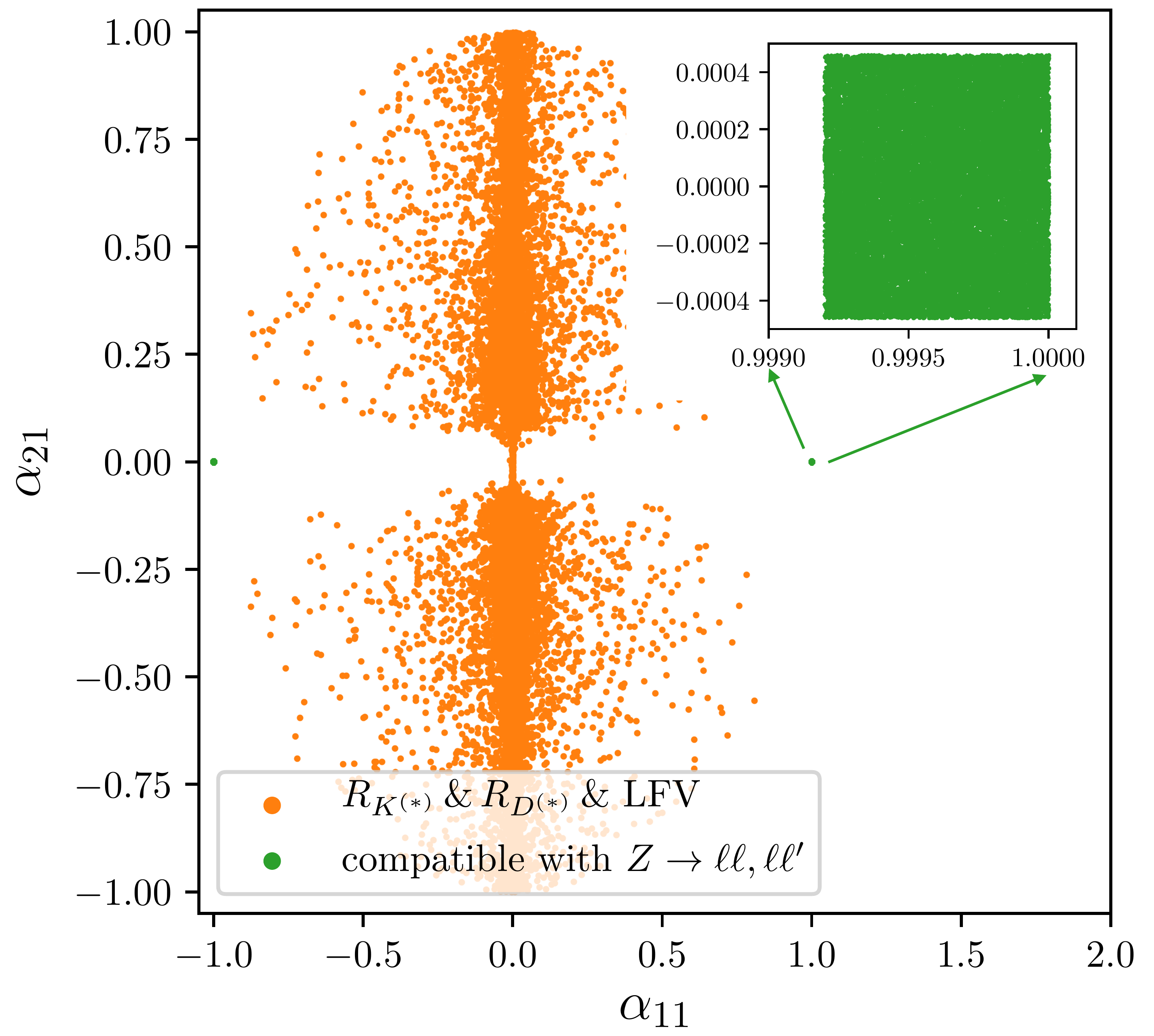}
	 \hspace{.3 cm}
    \includegraphics[width=.305\linewidth]{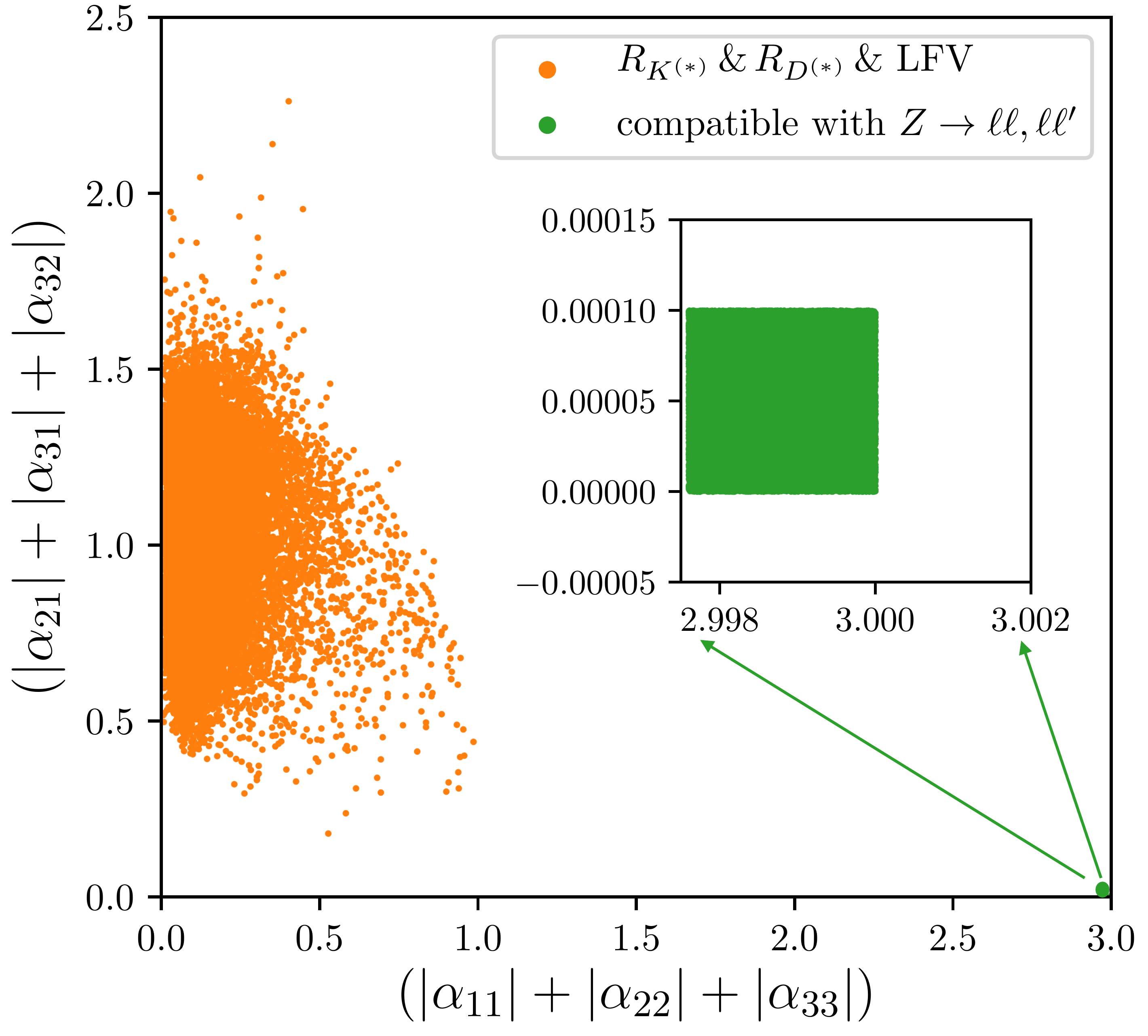}
    \caption{On the
    upper row, $A$-matrix entries complying with
    $R_{K^{(*)}}$ and $R_{D^{(*)}}$ data (blue) and those which in
    addition
    respect all imposed flavour constraints (yellow).
    Lower row: for the case of \textit{isosinglet
    heavy leptons},
    $A$-matrix entries complying with
    $R_{K^{(*)}}$ and $R_{D^{(*)}}$ data as well as LFV bounds (yellow),
    and those which now further comply with bounds from
    $Z$ decays (green). (Notice that the green inset area
    corresponds to a zoom-out of what would otherwise be a tiny region close to
    the border of the parameter space.)
    In all panels we have taken $m_V=1.5$~TeV and all mixing angles have been varied randomly between $-\pi$ and $\pi$. Figures from~\cite{Hati:2019ufv}.}
    \label{fig:alphaii_ij:RDRKcLFV}
\end{figure}

For $n=3$, our study suggests that it is in general possible to
find regions in the parameter space in which the required non-universal
flavour structure to explain both $R_{K^{(*)}}$ and $R_{D^{(*)}}$
can arise in a natural way, while still complying with all constraints
from flavour violating processes (meson and lepton sectors).
This can be seen from the upper row of
Fig.~\ref{fig:alphaii_ij:RDRKcLFV}, in which we display the regimes
for the entries of the matrix $A$ (cf.
Eqs.~(\ref{eq:UL:ARBS} - \ref{eq:Aalphaij})) which account for both $R_{K^{(*)}}$ and
$R_{D^{(*)}}$ data, as well as regions respecting the constraints
arising from the several flavour violating modes considered in our
analysis. Concerning the latter, we find it worth mentioning that
the most stringent constraints arise, as expected, from $K_L \to \mu^\pm e^\mp$,
$\mu \to e \gamma$, and $\mu-e$ conversion in nuclei;
$B$-meson cLFV decays, or  (semi-) leptonic $B$ and
$K$ decays lead to comparatively milder constraints
(or are systematically satisfied).

\begin{figure}[t!]
\hspace*{-4mm}
\includegraphics[width=0.52\textwidth,]{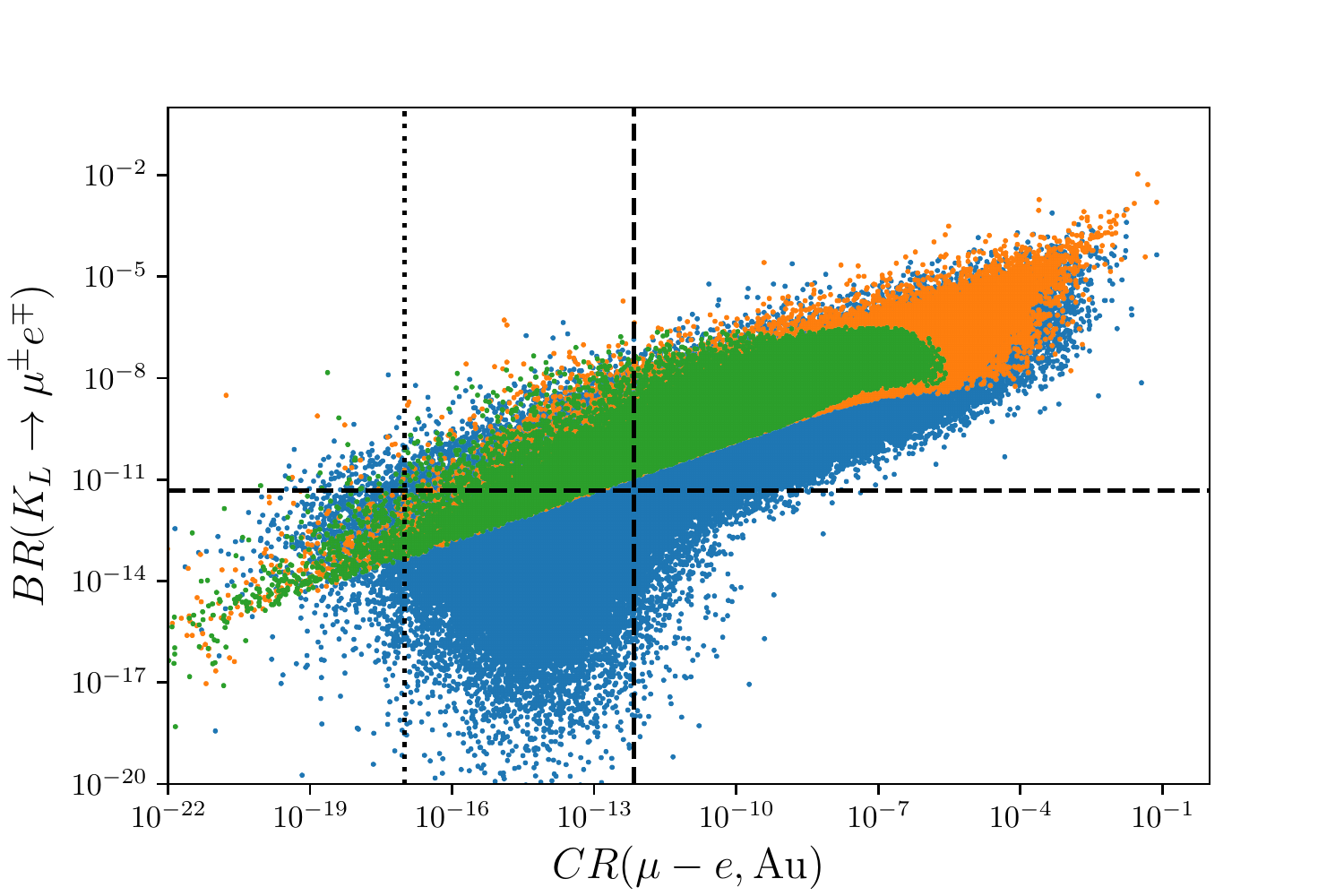}
\hspace*{2mm}
\includegraphics[width=0.52\textwidth,]{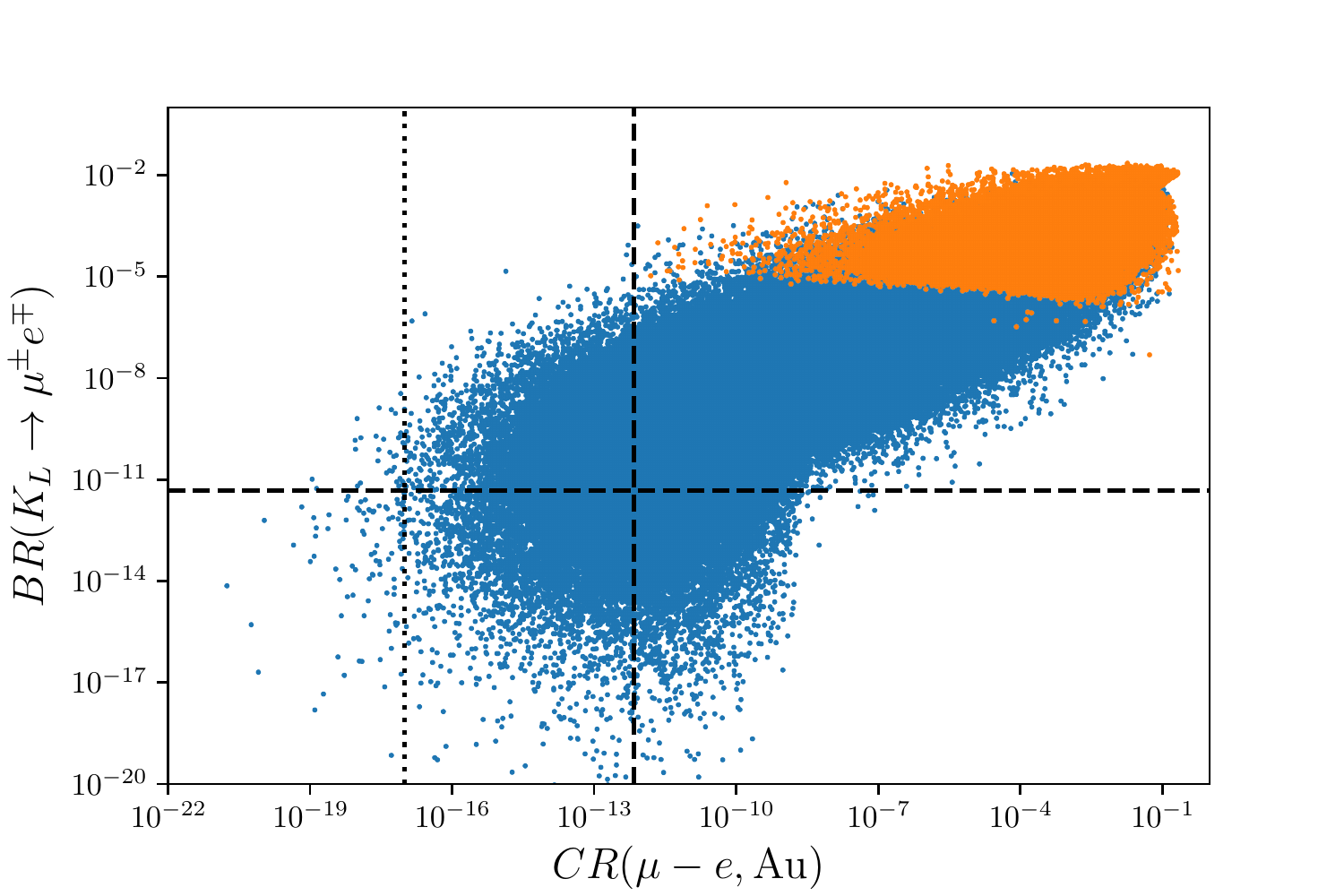}
\caption{Regions in the plane spanned by CR($\mu-e$,N) and
  BR($K_L \to \mu^\pm e^\mp$), accommodating $R_{K^{(*)}}$ (left
  panel) and $R_{D^{(*)}}$ (right panel), respectively for
  leptoquark masses in the intervals
  $m_V \in [15~\text {TeV}, 45~\text {TeV}]$ and
  $m_V \in [1~\text {TeV}, 6~\text {TeV}]$, in the framework of
  non-unitary leptoquark couplings induced by the presence of
  3 generations of \textit{isosinglet heavy leptons}.
  Blue points satisfy  $R_{{K^{(*)}},\,{D^{(*)}}}$ at the 3 $\sigma$ level,
  yellow points are consistent with leptonic $Z$ decays, and
  green points are compatible with all imposed constraints,
  other than those depicted by the corresponding vertical and
  horizontal dashed lines (dotted ones denoting future
  sensitivities).
  In both panels, all mixing angles have been varied randomly between $- \pi$ and $\pi$. Figures from~\cite{Hati:2019ufv}.}
  \label{fig:uni}
\end{figure}

Finally, one should address the compatibility of the
considered SM leptoquark extension with the
constraints arising from EW precision tests; as discussed in the
previous section as well as in other chapters dedicated to lepton flavour phenomenology, non-unitary
mixings can modify the couplings of the $Z$ boson. In particular,
for \textit{isosinglet heavy leptons}, the entries of the matrix $A$ (see Eqs. \eqref{eq:Aalphaii} and \eqref{eq:Aalphaij}) are severely
constrained by the $Z$ width and by bounds on its cLFV decays
($Z \to \ell \ell^\prime$). This is shown on the lower row of
Fig.~\ref{fig:alphaii_ij:RDRKcLFV}, which illustrates the
tension between LFUV and $Z$ bounds - a tension which
ultimately leads to
disfavouring this class of extensions as a phenomenologically viable
NP model to explain both $R_{K^{(*)}}$ and
$R_{D^{(*)}}$ discrepancies.

If one foregoes a solution to the charged current anomalies (i.e.,
$R_{D^{(*)}}$), it is possible to accommodate $R_{K^{(*)}}$
in full agreement with constraints from flavour bounds, relying on
very mild deviations from unitarity, and thus evading constraints from
universality violation in $Z$ decays.
However, one would be led to regions with considerably heavier leptoquarks,
$m_V\gtrsim 15$~TeV. This is depicted on the left panel of
Fig.~\ref{fig:uni}, in which we display regimes complying with $R_{K^{(*)}}$
at the $3\sigma$ level in the plane spanned by two particularly
constraining observables, BR($K_L \to \mu^\pm e^\mp$) and CR($\mu-e$,
N), for $15\text{ TeV}\lesssim m_V\lesssim 45\text{ TeV}$.
As can be verified, a small subset of points (consistent with $R_{K^{(*)}}$ and respecting universality in $Z$ decays) is compatible
with current bounds on the cLFV processes.
This is in agreement with the analyses of various UV-complete models,
such as~\cite{Balaji:2018zna,Fornal:2018dqn}.

For completeness, the right panel of Fig.~\ref{fig:uni} shows a similar study for
$R_{D^{(*)}}$. In order to accommodate $R_{D^{(*)}}$ data, smaller
leptoquark masses are required (in this case we have taken
$1\text{ TeV}\lesssim m_V\lesssim 6\text{ TeV}$), and it is no longer
possible to evade $K_L \to \mu^\pm e^\mp$ and $\mu-e$ conversion
bounds while being consistent with leptonic $Z$-decay
universality. The data displayed in the panels of
Fig.~\ref{fig:alphaii_ij:RDRKcLFV} was obtained for vector leptoquark
masses $m_V\sim 1.5$~TeV; analogous conclusions can be inferred for
$m_V\sim\mathcal(1-3)$~TeV, albeit for different
$\alpha_{ij}$ ranges.

\begin{figure}[h]
  \includegraphics[width=0.52\textwidth, ]{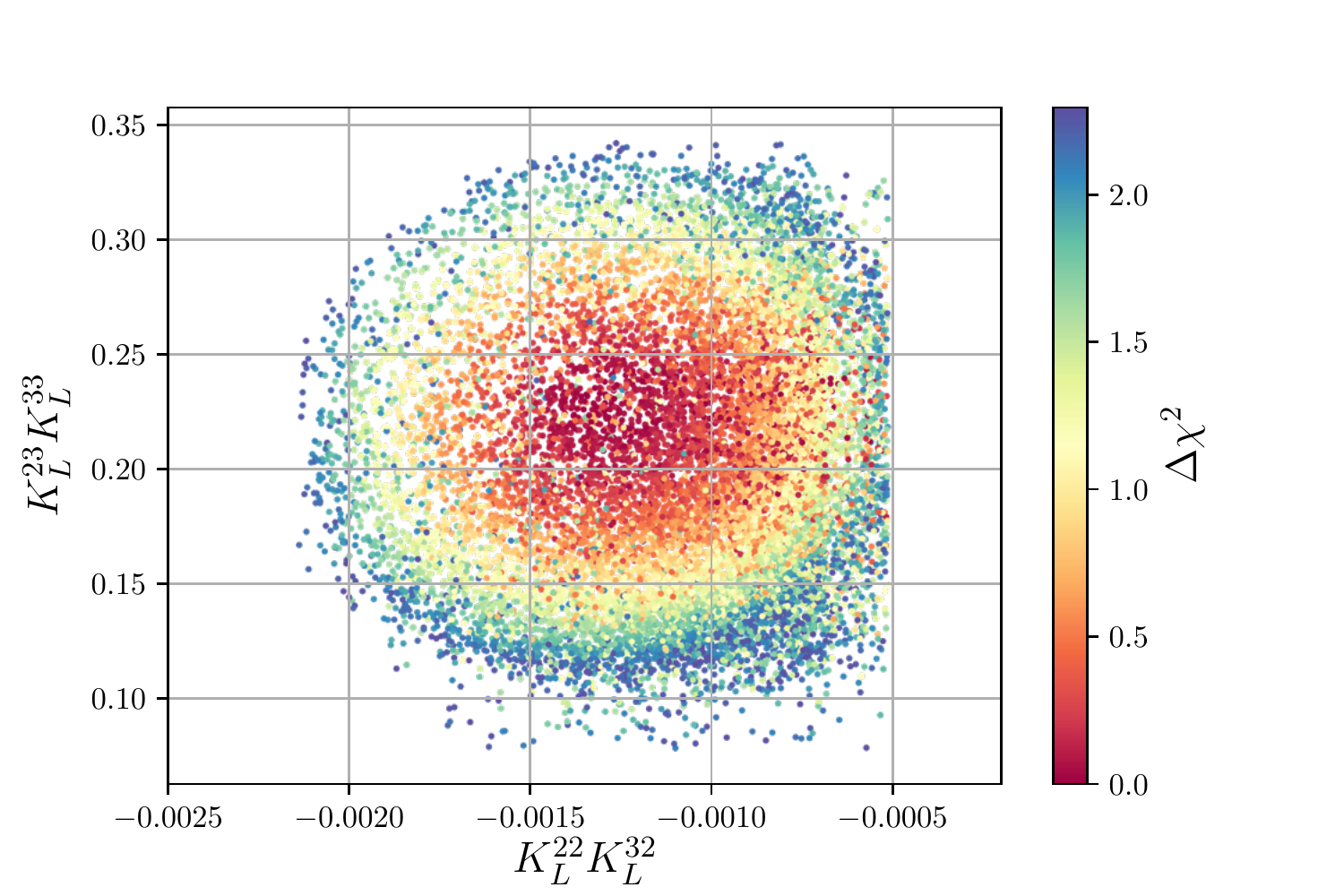}
  \includegraphics[width=0.52\textwidth, ]{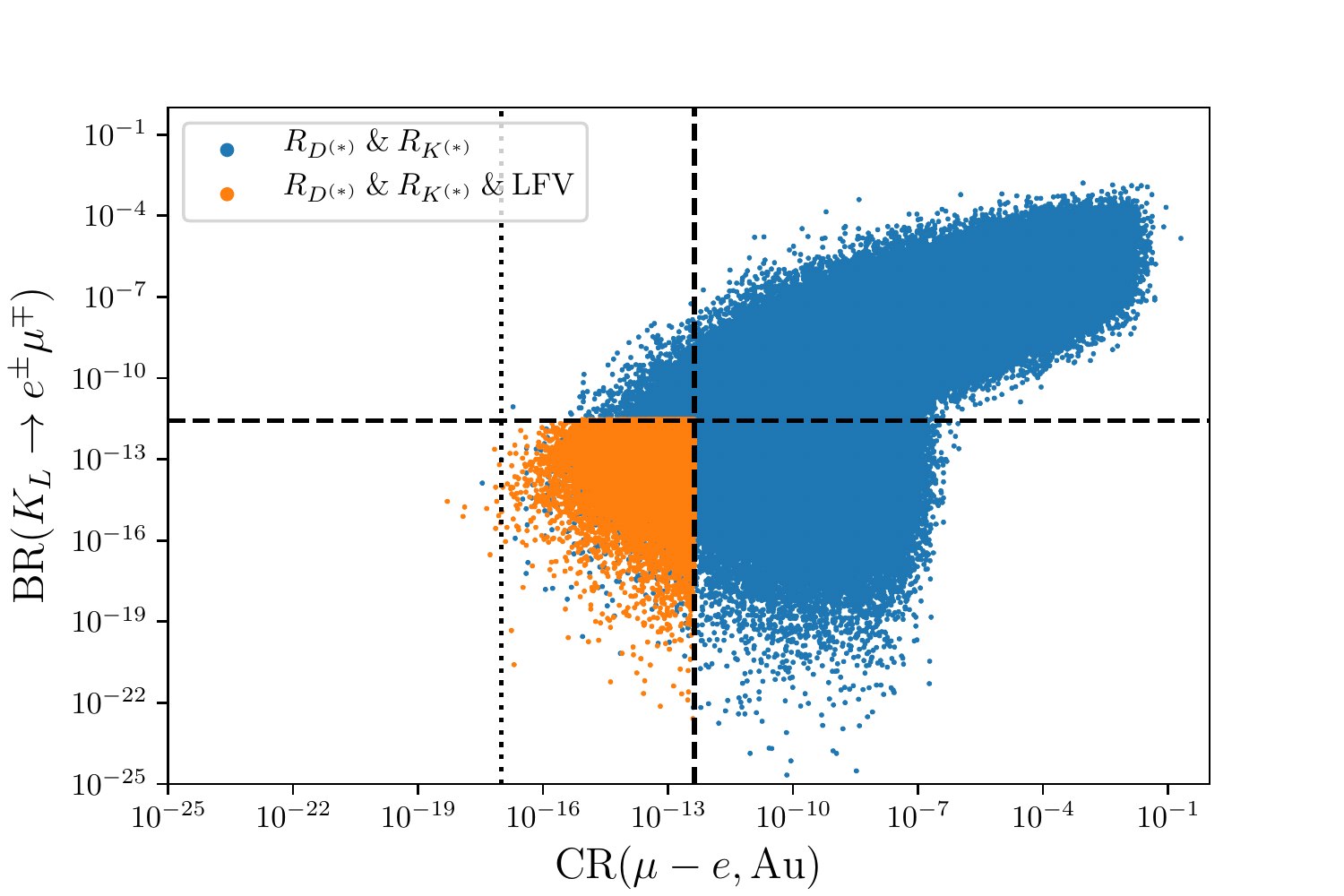}
\caption{On the left, $\Delta\chi^2$ distribution for the fit
  to $R_{K^{(*)}}$ and $R_{D^{(*)}}$ data ($1 \sigma$) in the plane of
  the $(K_L)_{ij}$ couplings.
  All points comply with the different (flavour)
  constraints. On the right,
  regions in the plane spanned by CR($\mu-e$,N) and
  BR($K_L \to \mu^\pm e^\mp$), accommodating both $R_{K^{(*)}}$ and
  $R_{D^{(*)}}$ (blue) and those in addition complying with LFV
  constraints (yellow).
  Both panels correspond to a heavy sector composed
  of \textit{three isodoublet vector-like charged lepton states}, and to having set
  $m_V \sim 1.5$~TeV.
  The $\mathrm \Delta\chi^2$ corresponds to the $1\,\sigma$-region around the best fit point. Figures from~\cite{Hati:2019ufv}.}\label{fig:nuni}
\end{figure}

Since for the case of isosinglet leptons an explanation of $R_{D^{(*)}}$ is excluded by bounds on $Z$ decays, we now consider \textit{isodoublet heavy charged leptons}. The non-unitarity in the couplings of the vector
leptoquark to the SM charged leptons can simultaneously
explain $R_{K^{(*)}}$ and $R_{D^{(*)}}$ data. Moreover, and by
construction, in the case of \textit{isodoublet heavy charged lepton states} the
$Z\ell\ell$ couplings remain universal
(in the absence of mixings between
right-handed SM charged leptons and vector-like doublets,
$\Delta g_R=0$, see Section~\ref{sec:ew}); nevertheless,
flavour observables still play a crucial role, and are (as expected)
responsible for severe constraints on the New Physics degrees of freedom.

The left panel of Fig.~\ref{fig:nuni} offers a global view of this
case, showing the $\Delta\chi^2$ distribution for the fit
to $R_{K^{(*)}}$ and $R_{D^{(*)}}$ data, in the plane spanned by
$(K_L)_{ij}$ ``muon and tau couplings'' ($K_{22}K_{32}-K_{23}K_{33}$),
marginalising over the other couplings. The leptoquark mass is set to
$m_V \sim 1.5$~TeV.
We stress that leading to this plot all couplings
were determined by the underlying non-unitarity parametrisation
(with all mixing angles randomly sampled); in particular, we have not
set the leptoquark couplings to
the first generation of quark and leptons to zero. The displayed points
comply with \textit{all} flavour bounds included in our study,
as described in Section~\ref{sec:lqconstraints}.

The lowest $\Delta\chi^2$ region (dark red ellipsoid) suggests that the
best fit scenario corresponds to New Physics dominantly coupling to
muons and taus. We stress that the patterns emerging from the $\Delta\chi^2$
distribution are not an artefact
of some particular assumption imposed on the couplings, but rather the result
of a very general scan over the full set of (mixing) parameters.

\medskip
The right panel of Fig.~\ref{fig:nuni} offers a projection of the
viable points (displayed on the left panel) in the plane of the most
constraining observables, CR($\mu-e$,N) and BR($K_L \to \mu^\pm e^\mp$).
It is interesting to notice that, to a very good approximation, most
of the currently phenomenologically viable
points lie within future reach of the upcoming muon-electron
conversion dedicated facilities (COMET and Mu2e).

In the near future, and should the $B$-meson decay anomalies be confirmed,
an explanation in terms of such a minimal leptoquark framework
could be probed via its impact for cLFV observables, in particular
$\mu-e$ conversion in nuclei. 

However, the fact that this seems to be the preferred parameter space, might be an artefact of the non-unitary parametrisation of leptoquark couplings.
Therefore, in the next sections, we will explore the interplay of different constraints on the leptoquark couplings, taking them as independent parameters (recall that due to the number of vector-like leptons $n\geq 2$ all entries in $K_L$ can be viewed as independent).
We will further discuss in detail the impact of future negative searches concerning LFV observables.

\mathversion{bold}
\section{Towards a global fit of the vector leptoquark $V_1$ flavour structure}
\mathversion{normal}
\label{sec:lqfit}
Having established that, in order to address the $B$-meson decay anomalies, the flavour structure of the leptoquark couplings is necessarily non-unitary, we now carry out a comprehensive fit of the relevant couplings of the vector leptoquark to the different generations of SM fermions. 
Relying on the simplified-model parametrisation (cf. Section~\ref{sec:simplifiedmodel}), our goal is thus to constrain the entries of the matrix $K_L$ 
(see Eq.~(\ref{eq:lagrangian:Vql_phys3})).
Under the assumption that the relevant couplings are real, a total of nine free parameters will thus be subject to a large number of constraints stemming from data on several SM-allowed leptonic and semi-leptonic meson decays, SM-forbidden cLFV transitions and decays, as well as from an explanation of the (anomalous) observables in the $b \to s \ell \ell$ and $b \to c \tau\nu$ systems.

\noindent
\paragraph{Data relevant for the global fit}
In particular, we take into account the data for the charged current $b\to c\ell\nu$ processes (see Appendix~\ref{app:BFCCC}).
In addition to the LFUV ratios $R_{D^{(\ast)}}$~\cite{Abdesselam:2017kjf, Abdesselam:2018nnh,Aaij:2015yra,Aaij:2017uff,Huschle:2015rga,Hirose:2016wfn,Abdesselam:2019dgh,Lees:2013uzd}, we also include the binned branching fractions of $B\to D^{(\ast)} \ell\nu$ decays~\cite{Aubert:2008yv, Aubert:2007qs,Urquijo:2006wd, Aubert:2009qda}, as listed in Table~\ref{tab:binned_bcellnu}.

Furthermore, we take into account a large array $b\to s\ell\ell$ decays as listed in Appendix~\ref{app:bsll}.
This includes the binned data of the angular observables in the optimised basis~\cite{Descotes-Genon:2013vna} (Table~\ref{tab:ang_data}), the differential branching ratios (Table~\ref{tab:br_bsll}), and the binned LFUV observables (Table~\ref{tab:lfuv_bsll}).
Other than the binned data, we also include the unbinned data of branching ratios in $B_{(s)}\to \ell\ell$~\cite{Chatrchyan:2013bka, Aaij:2017vad, Aaboud:2018mst, Sirunyan:2019xdu,Aaij:2020nol} and inclusive and exclusive branching ratio measurements of $b\to s\gamma$~\cite{Amhis:2014hma,Misiak:2017bgg,Dutta:2014sxo, Aaij:2012ita}.

Other than studying the contributions of the vector leptoquark in the ``anomalous'' channels, we aim to estimate the favoured ranges of all of its couplings to SM fermions. Consequently, we include a large number of additional observables into the likelihoods.
Since most processes only constrain a product of at least two distinct leptoquark couplings, a successful strategy is to include an extensive set of processes, thus allowing to constrain distinct combinations of couplings (as many as possible).

In addition to the $b\to c\ell\nu$ transitions, we also include certain $b\to u\ell\nu$ decays such as $B^0\to\pi\tau\nu$, $B^+\to\tau\nu$ and $B^+\to\mu\nu$, which are listed in Table~\ref{tab:buellnu}.
In many leptoquark models $B\to K^{(\ast)}\nu\bar\nu$ decays provide very stringent constraints. However this is not the case for $V_1$ vector leptoquarks, due to the $SU(2)_L$-structure: the relevant operators for $B\to K^{(\ast)}\nu\bar\nu$ transitions are absent at the tree-level, and are only induced at higher order, thus  leading to weaker constraints. 
Due to the leading operator being generated at the loop-level, a non-linear combination of leptoquark couplings is constrained by this process.
Thus, despite the loop suppression, we include $B\to K\nu\bar\nu$ in the likelihoods, and use the data obtained by Belle~\cite{Grygier:2017tzo,Lutz:2013ftz} and BaBar~\cite{Lees:2013kla,delAmoSanchez:2010bk}. 

To constrain combinations of first and second generation couplings, we further include a large number of binned and unbinned leptonic and semi-leptonic charged current $D$ meson decays, charged and neutral current Kaon decays and SM allowed $\tau$-lepton decays.
The observables and corresponding data-sets can be found in Appendix~\ref{app:sctau} and are listed in Tables~\ref{tab:binned_charm} through \ref{tab:fcnc_strange}.

Finally, as previously discussed, cLFV processes impose severe constraints on the parameter space of vector leptoquark couplings; in particular neutrinoless $\mu-e$ conversion in nuclei and the decay $K_L\to e^\pm\mu^\mp$ provide some of the most stringent constraints for vector leptoquark couplings to the first two generations of leptons~\cite{Hati:2019ufv}. 
Recall that in Table~\ref{tab:important_LFV} we present the current experimental bounds and future sensitivities for various cLFV observables yielding relevant constraints to our analysis.
Depending on the fit set-up, either only a few, or then all of these observables are included in the global likelihood, as explicitly mentioned in the following paragraphs.

\paragraph{Results for the simplified-model fit of the $V_1$ couplings}

Firstly, it is important to emphasise that in our analysis we consider all the entries in the $K_L$ coupling matrix as (real) \textit{free parameters to be determined by the fit}. 
For the leptoquark mass we choose three benchmark-points, $m_{V_1} \in [1.5,\,2.5,\,3.5]\:\mathrm{TeV}$, which allow to illustrate most of the vector leptoquark mass range of interest, while respecting the current bounds from direct searches at colliders~\cite{Khachatryan:2014ura,Aad:2015caa,Aaboud:2016qeg,Aaboud:2019jcc,Aaboud:2019bye,Aad:2020iuy,Sirunyan:2017yrk,Sirunyan:2018vhk,Sirunyan:2018kzh}. 
In particular, notice that masses significantly heavier than a few TeVs preclude a successful explanation of the charged current anomalies, $R_{D^{(*)}}$. For each mass benchmark point we thus obtain best-fit points corresponding to a SM pull around $\sim 6.4\,\sigma$ (with respect to the global likelihood including all lepton flavour conserving observables).

\begin{figure}[h]
	\centering
	\includegraphics[width = 0.6\textwidth]{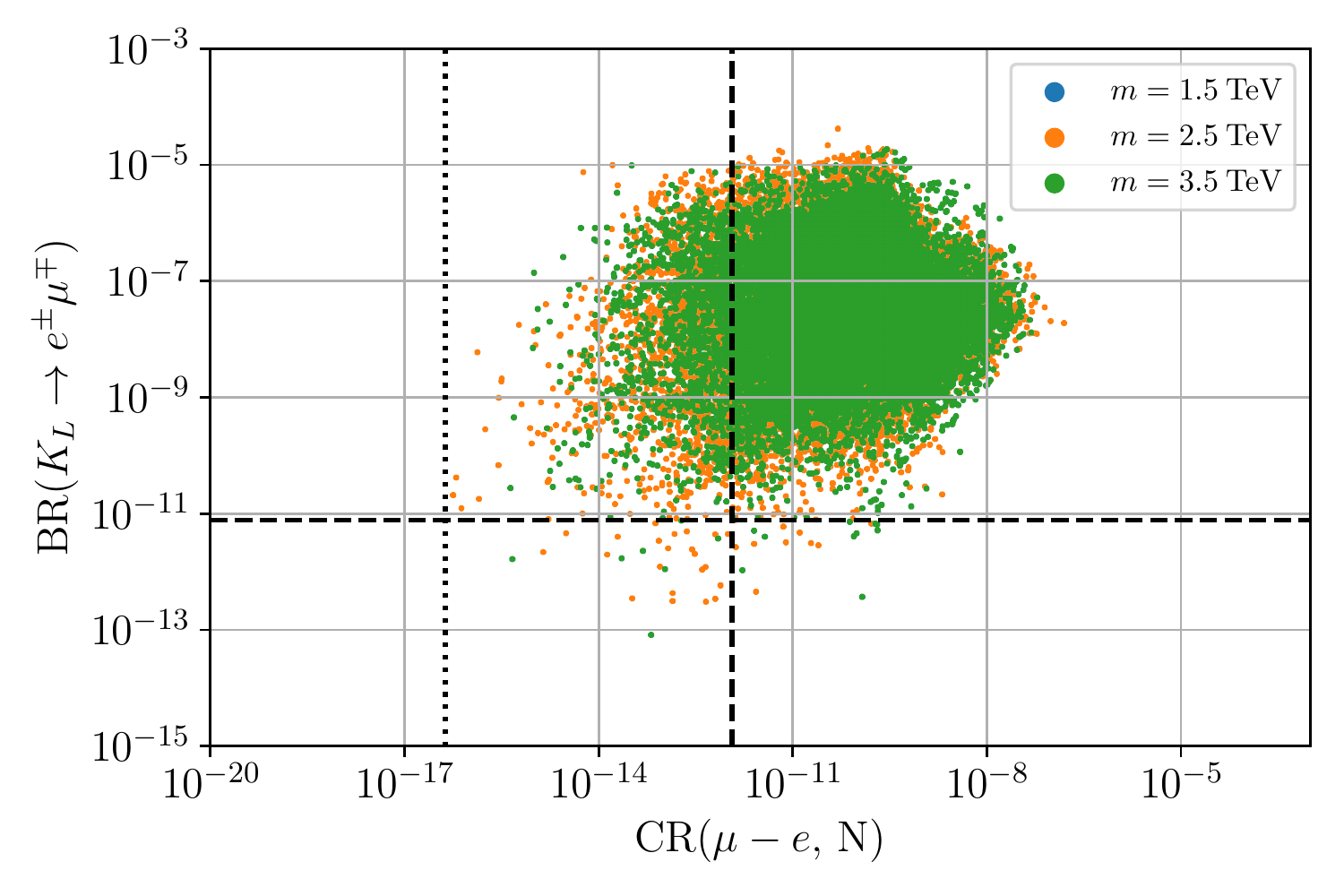}
	\caption{Result of a random scan around the best-fit point (without the inclusion of cLFV bounds on $\mathrm{CR}(\mu - e, \mathrm{Au})$ and $\mathrm{BR}(K_L\to e^\pm\mu^\mp)$ as inputs to the fit). 
	Following a sampling of the global likelihood(s) via MCMC, the sample points shown in the plot are drawn from the posterior distributions of the leptoquark couplings (cf. Appendix~\ref{app:stats}).
	The colour scheme reflects the mass benchmark points: 
	blue, orange and green respectively associated with 
	$m_V$=1.5~TeV, 2.5~TeV and 3.5~TeV.  
	The dashed lines indicate the current bounds at $90\,\%$ C.L., while the dotted line denotes the envisaged future sensitivity of the COMET and Mu2e experiment (for Aluminium nuclei). Figure from~\cite{Hati:2020cyn}.}
	\label{fig:woLFV_KL_CR}
\end{figure}
In Fig.~\ref{fig:woLFV_KL_CR}, we present the results of a random scan around the best-fit points for the vector leptoquark scenario here considered, in the plane spanned by two of the most constraining cLFV observables,  $\mathrm{CR}(\mu - e, \mathrm{N})$ and $\mathrm{BR}(K_L\to e^\pm\mu^\mp)$. The sample points are drawn from the posterior frequency distributions of the leptoquark couplings, following Markov Chain Monte Carlo (MCMC) simulations, as described in Appendix~\ref{app:stats}.
It can be easily seen that for the three mass benchmark choices (corresponding to the different colours in the plot) most of the randomly sampled points are excluded by the strong cLFV constraints. 
Although the involved couplings are compatible with $0$, the constraints on first generation couplings derived from lepton flavour conserving low-energy data (as listed in Appendix~\ref{app:Obs}) are considerably weaker than those from LFV processes. This leads to several ``flat directions'' in the likelihood.
The strongest LFV constraints are from $\mathrm{CR}(\mu - e, \mathrm{Au})$ and $\mathrm{BR}(K_L\to e^\pm\mu^\mp)$, while other LFV constraints on second and third generation couplings are weaker, or on par with constraints from lepton flavour conserving low-energy data.
Therefore, we redefine the strategy of the global fit, and now directly include the upper bounds from $\mathrm{CR}(\mu - e, \mathrm{Au})$ and $\mathrm{BR}(K_L\to e^\pm\mu^\mp)$ as \textit{inputs} in the fitting procedure for the vector leptoquark couplings.

The inclusion of the current upper limits on the observables $\mathrm{CR}(\mu - e, \mathrm{Au})$ and $\mathrm{BR}(K_L\to e^\pm\mu^\mp)$ as input to the fit will consequently shift the best-fit point towards a lower cLFV prediction, also leading to a slightly lower SM pull. However, we find this to be a good compromise in order to identify regimes in the parameter space not yet disfavoured by the current cLFV data. 
In fact, and since 
$\mathrm{CR}(\mu - e, \mathrm{Au})$ and $\mathrm{BR}(K_L\to e^\pm\mu^\mp)$ are indeed two of the most constraining cLFV observables, once the bounds on the latter observables are respected, most of the sample points will be naturally in agreement with   current bounds on most of other cLFV observables (this is a consequence of correlations with other cLFV $\mu-e$ transitions; processes involving  $\tau$-leptons are comparatively less constraining).

In Table~\ref{tab:fits_wKLCRmue} we present our results~\cite{Hati:2020cyn} for the new fits with their corresponding SM pulls. As can be verified, the SM pull is lower, reduced from $\sim6.4\sigma$ to $\sim5.8\sigma$, of which the contributions to the total $\chi^2$ stemming from the charged current $b\to c\ell\nu$ transitions amounts to $\sim 1.5\sigma$, whereas the contributions from the neutral current $b\to s\ell\ell$ transitions amounts to $\sim 4.3\sigma$. 
Furthermore, we show tentative $90\%$ ranges of the posterior (coupling) distributions, obtained by sampling the global likelihood using MCMC.
The ranges, derived  from the histograms of the posterior distributions, are  taken as symmetric intervals between the  $5^\text{th}$ and $95^\text{th}$ percentiles (cf. Appendix~\ref{app:stats}).
We notice here that the vector leptoquark coupling to the first generation SM fermions are consistent with zero, which is an assumption often invoked in literature for simplified analysis. For second and third generation couplings, the quoted ranges of the corresponding fits are in fair agreement with the (order of magnitude) results for the benchmark ranges of second- and third generation couplings quoted in the literature, e.g.~\cite{Cornella:2021sby,Angelescu:2018tyl}. However, given the differences in the coupling parametrisation choices and underlying statistical treatment, the results are not directly comparable.

\begin{table}[h!]

	\hspace*{-8mm}\begin{tabular}{|c|c|c|c|}
		\hline
				$m_{V_1}$ & $K_L$ best-fit & $K_L$ $90\%$ & $\text{pull}_\text{SM}$\\
				\hline
				\hline
				{\footnotesize$1.5\:\mathrm{TeV}$}&
				{\footnotesize$\begin{pmatrix}
									-5.3\times 10^{-6} & 2.6\times 10^{-3} & -0.079\\
									-9.8\times 10^{-4} & -0.03 & 1.1\\
									-3.4\times 10^{-3} & 0.038 & 0.16
								\end{pmatrix}$}&
				{\footnotesize$\begin{pmatrix}
								(-1.2 \to 1.1)\times 10^{-3} & (-1.5\to 9.1)\times 10^{-3}  & -0.11\to 0.009\\
								-0.034 \to 0.036 & -0.063 \to -0.002 & 0.27\to 1.55\\
								-0.050 \to 0.036 & 1.0\times10^{-3} \to 0.11 & 0.08 \to 0.80
								\end{pmatrix}$}&
				{\footnotesize$5.78$}\\
				\hline
				\hline
				{\footnotesize$2.5\:\mathrm{TeV}$}&
				{\footnotesize$\begin{pmatrix}
					-1.9\times 10^{-5} & 4.3\times 10^{-3} & -0.11\\
					2.1\times 10^{-3} & -0.056 & 1.9\\
					-6.9\times 10^{-3} & 0.063 & 0.27
				\end{pmatrix}$}&
				{\footnotesize$\begin{pmatrix}
					(-1.5 \to 2.3) \times 10^{-3} & (-0.26\to1.1)\times 10^{-2}& -0.17 \to 0.014\\
					-0.059\to 0.068 & -0.13 \to -0.009 & 0.43 \to 2.58 \\
					-0.076 \to 0.072 & 0.009 \to 0.21 & 0.13 \to 1.31
				\end{pmatrix}$}&
				{\footnotesize$5.82$}\\
				\hline
				\hline
				{\footnotesize$3.5\:\mathrm{TeV}$}&
				{\footnotesize$\begin{pmatrix}
					2.9\times 10^{-5} & 5.9\times 10^{-3} & -0.14\\
					3.1\times 10^{-3} & -0.078 & 2.6\\
					0.010 & 0.088 & 0.37
				\end{pmatrix}$}&
				{\footnotesize$\begin{pmatrix}
					(-3.6\to2.9)\times 10^{-3} & (-3.7 \to 14.3)\times 10^{-3} & -0.21 \to 0.017\\
					-0.13\to0.078 & -0.18 \to -0.012 & 0.57 \to 3.23\\
					-0.14 \to 0.11 & 0.023 \to 0.32 & 0.22 \to 1.92
				\end{pmatrix}$}&
				{\footnotesize$5.84$}\\
				\hline
	\end{tabular}
	\caption{Results of the fits including the current experimental bounds on $\mathrm{CR}(\mu - e, \mathrm{Au})$ and $\mathrm{BR}(K_L\to e^\pm\mu^\mp)$ in the likelihood: best fit points and symmetric $90\%$ ranges of $K_L^{ij}$. The SM pull is reduced from $\sim6.4\sigma$ to $\sim5.8\sigma$.}
	\label{tab:fits_wKLCRmue}
\end{table}

\begin{figure}[h]
	\mbox{\hspace*{27mm}\includegraphics[width=0.8\textwidth]{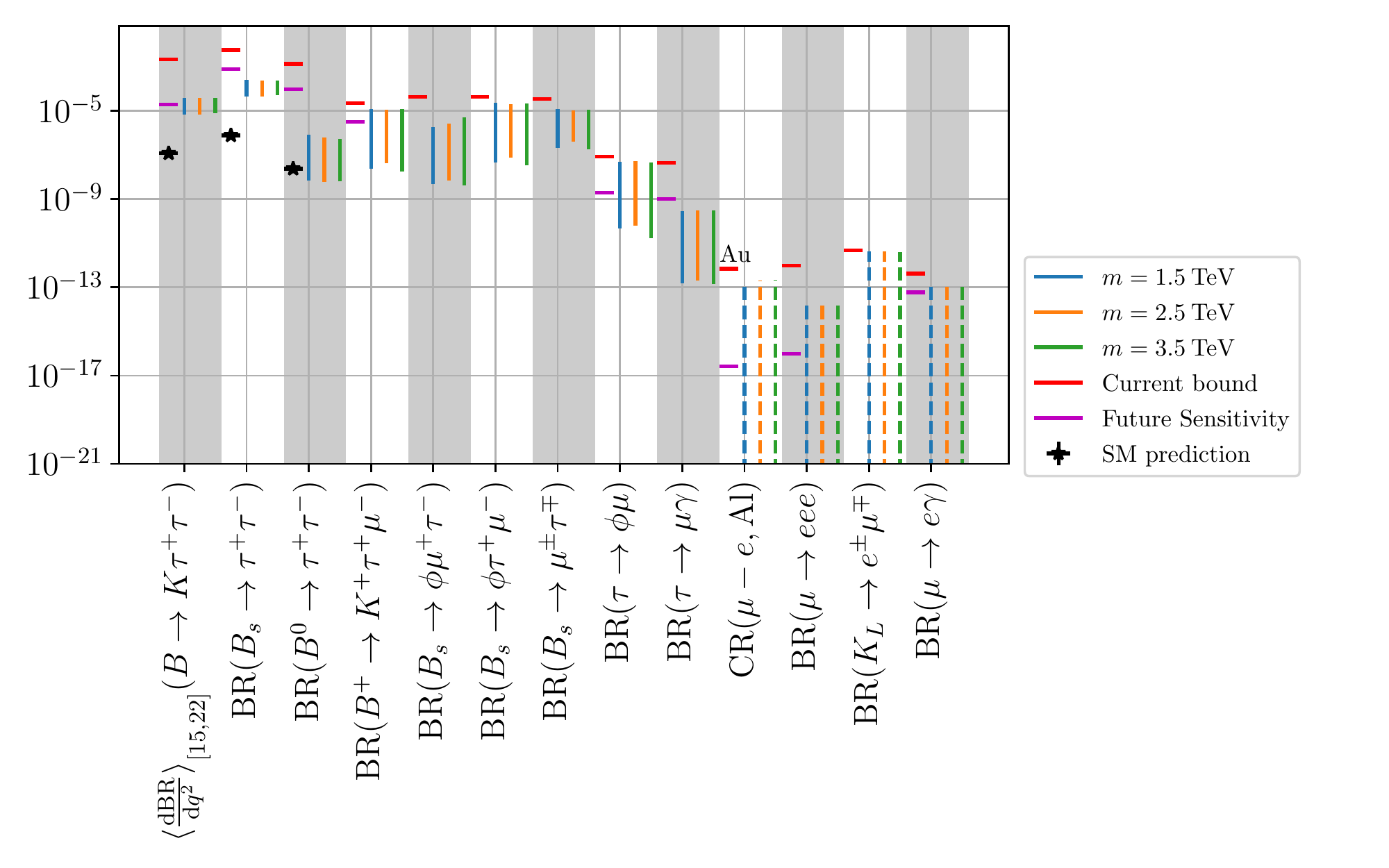}}
	\caption{Predicted ranges for several $\tau$-lepton and LFV observables. The blue, orange and green lines respectively denote the $90\%$ range for leptoquark masses of $1.5,\, 2.5\text{ and } 3.5\:\mathrm{TeV}$ while the horizontal red (purple) lines denote the current (future) bound at $90\,\%$ C.L.;  stars denote SM predictions when appropriate.
	The dashed lines correspond to predictions of observables depending only on couplings that are compatible with $0$ and their top edges correspond to $90\%$ upper limits. (The $90\%$ ranges have been obtained as detailed in Appendix~\ref{app:stats}.) Figure from~\cite{Hati:2020cyn}.
		}
	\label{fig:tau_and_lfv_predictions}
\end{figure}

\noindent
Upon inclusion of the current cLFV constraints we find that the shape of the global likelihood consequently enforces small vector leptoquark couplings to the first two generations of charged leptons, leading to predictions consistent with experimental data.
This thus allows to sample the global likelihood (in terms of the leptoquark couplings) via MCMC techniques (as described in Appendix~\ref{app:stats}).
The posterior distributions of the leptoquark couplings are then used to compute predictions for $B$-meson decays into final states containing $\tau$-leptons, and 
several cLFV observables (including tau decays).
This is presented in Fig.~\ref{fig:tau_and_lfv_predictions} where, for each observable, we depict the current experimental bounds and future sensitivities, the SM predictions (when relevant), as well as the predictions for the three vector leptoquark mass benchmark points - corresponding to the vertical coloured lines.
The dashed lines describe predictions of observables involving only couplings compatible with vanishing values and thus their top edge corresponds to a $90\%$ upper limit, while no lower limit should be implied.

As can be seen from Fig.~\ref{fig:tau_and_lfv_predictions}, a large part of the 
currently allowed parameter space in the $e \mu$ channel (for the three leptoquark mass benchmark points) will
be probed by the upcoming experiments dedicated to searching for neutrinoless $\mu - e$ conversion in Aluminium nuclei, 
Mu2e and COMET, owing to the expected increase in sensitivity. 
In the case of future non-observation of this process, this will lead to strongly improved constraints on the $V_1$ couplings to first first generation fermions.

Moreover, the sensitivity of the lepton flavour violating process $\tau\to \phi \mu$ is expected to be improved by over an order of magnitude at the Belle II experiment, 
which will allow probing a large region of the parameter space associated with the $\mu\tau$ channel. A priori, 
and as can be seen from Fig.~\ref{fig:tau_and_lfv_predictions},
under the current vector leptoquark hypothesis, $\tau\to \phi \mu$ decays have very strong prospects of being observed at Belle II. Conversely, should such a mode not be observed at Belle II, then the $s-\mu$ and $s-\tau$ couplings of the vector leptoquark will be tightly constrained. As a consequence, it might prove extremely challenging to simultaneously address the anomalous neutral and charged current data within the current model.

\section{Impact of future experiments: Belle II and cLFV searches}\label{sec:futureimpact}

Following the overview of the vector leptoquark couplings conducted in the previous section, we now proceed to investigate how our working hypothesis can be effectively probed by the coming future experiments, especially  Belle II and cLFV-dedicated facilities. 

Assuming that the above experiments return only negative search results for the most promising modes, we  
then evaluate how the current $V_1$ hypothesis would still stand as a viable explanation for the LFUV $B$-meson decay anomalies.

\mathversion{bold}
\subsection{Probing the vector leptoquark $V_1$ at coming experiments}
\mathversion{normal}
Concerning the quest for LFUV in  $b\rightarrow s \ell^+\ell^-$ decays,  Belle II is expected to achieve a very high sensitivity for both muon and electron modes, leading to very precise measurements for the ratios $R_K$ and $R_{K^*}$, with the potential to confirm the anomalous LHCb data (if the latter is due to New Physics effects)~\cite{Kou:2018nap}. 
In what concerns $B$-meson decays to $\tau^+ \tau^-$ final states,
Belle II will also provide the first in-depth experimental exploration of these modes. Notice that the latter remain a comparatively less explored set of observables, with relatively weak bounds on the few modes already being searched for: for example, current bounds on $\text{BR}(B^0 \rightarrow \tau^+ \tau^-)<1.3 \times 10^{-3}$ from LHCb~\cite{DeBruyn:2016tiq} and $\text{BR}(B_s \rightarrow \tau^+ \tau^-)< 2.25 \times 10^{-3}$ from Babar~\cite{TheBaBar:2016xwe} are orders of magnitude weaker than the SM predictions. For the purely leptonic decays, the most recent SM computations now include next-to-leading order (NLO) electroweak corrections and next-to-NLO QCD corrections~\cite{Bobeth:2013uxa,Hermann:2013kca, Bobeth:2013tba},
\begin{eqnarray}
\text{BR}(B_s \rightarrow \tau^+ \tau^-)_{\text{SM}} &=& (7.73 \pm 0.49) \times 10^{-7}\,,\nonumber\\
\text{BR}(B^0 \rightarrow \tau^+ \tau^-)_{\text{SM}} &=& (2.22 \pm 0.19) \times 10^{-7}\, .
\end{eqnarray}

\noindent
Within the SM, the exclusive semi-leptonic decays of $B$-mesons to $\tau^+ \tau^-$ final states have been studied by several groups: the modes $B\to K^*\tau^+\tau^-$ and $B_s\to \phi\tau^+\tau^-$ have been computed\footnote{The inclusive $B\to X_s\tau^+\tau^-$ process has been addressed in Refs.~\cite{Guetta:1997fw,Bobeth:2011st}, while indirect constraints on $b\to s\tau^+\tau^-$ operators 
were studied in Ref.~\cite{Bobeth:2011st}.}  in~\cite{Hewett:1995dk,Bouchard:2013mia,Kamenik:2017ghi}. 
To avoid contributions from the resonant decays through the narrow $\psi(2S)$ charmonium resonance (i.e. $B \rightarrow H\psi(2S)$ with $\psi(2S)\rightarrow \tau^+ \tau^-$, where $H=K,K^*, \phi,\cdots$), the relevant SM predictions are typically restricted to an invariant di-tau mass $q^2> 15$ GeV$^2$. 
Taking into account the uncertainties from the relevant form factors and CKM elements, the SM predictions for the branching ratios of the semi-leptonic decays into tau pairs can be determined with an accuracy between 10\% and 15\%. 
Notice that the presence of broad charmonium resonances 
(above the open charm threshold) can further lead to additional subdominant uncertainties, typically of a few percent~\cite{Beylich:2011aq}. 

\noindent 
For the $B \rightarrow K \tau^+ \tau^-$ modes, using the recent lattice $B\rightarrow K$ form factors from the Fermilab/MILC collaboration~\cite{Bailey:2015dka}, the SM predictions for the $q^2\in[15, 22]\, \text{GeV}^2$ have been reported to be~\cite{Du:2015tda},
\begin{eqnarray}
\text{BR}(B^+ \rightarrow K^+ \tau^+ \tau^-)_{\text{SM}} &=& (1.22 \pm 0.10) \times 10^{-7}\,,\nonumber\\
\text{BR}(B^0 \rightarrow K^0 \tau^+ \tau^-)_{\text{SM}} &=& (1.13 \pm 0.09) \times 10^{-7} \, .
\end{eqnarray}
Similar predictions for the $B \rightarrow K^* \tau^+ \tau^-$ modes, with $q^2\in[15, 19]\, \text{GeV}^2$,
have also been reported~\cite{Kou:2018nap,Straub:2018kue} %
\begin{eqnarray}
\text{BR}(B^+ \rightarrow K^{*+} \tau^+ \tau^-)_{\text{SM}} &=& (0.99 \pm 0.12) \times 10^{-7}\,,\nonumber\\
\text{BR}(B^0 \rightarrow K^{*0} \tau^+ \tau^-)_{\text{SM}} &=& (0.91 \pm 0.11) \times 10^{-7} \,.
\end{eqnarray}
The above results rely on the combined fit of lattice QCD and light cone sum rules (LCSR) 
results for $B\rightarrow K$ form 
factors~\cite{Straub:2015ica}.
Finally, the SM prediction for $B_s \rightarrow \phi \tau^+ \tau^-$ mode can also be obtained for the same kinematic region ($q^2 \in [15, 19]\, \text{GeV}^2$)~\cite{Capdevila:2017iqn}
\begin{eqnarray}
\text{BR}(B_s \rightarrow \phi \tau^+ \tau^-)_{\text{SM}} = (0.86 \pm 0.06) \times 10^{-7} \, .
\end{eqnarray}
As already extensively discussed, 
sizeable $b-\tau$ and $s-\tau$ couplings are necessary to explain the charged current anomalous data on $R_{D^{(*)}}$; if $R_{D^{(*)}}$ anomalies are indeed due to New Physics then one expects 
a significant enhancement of the rates of 
$b\to s\tau^+\tau^-$ processes, up to three orders of magnitude from the SM predictions~\cite{Alonso:2015sja,Crivellin:2017zlb,Calibbi:2017qbu,Capdevila:2017iqn}.
This is in line with the corresponding discussion in Chapter~\ref{sec:smeft}, and does not come as a surprise. 
Consequently, this renders searches for $b\to s\tau^+\tau^-$ modes extremely interesting probes of vector leptoquark models aiming at explaining anomalous LFUV data.

Although the LHCb programme includes searches 
for
$B \to {K^{(*)}}{\tau^+\tau^-}$ and $B_s \to {\phi}{\tau^+\tau^-}$ modes, being an $e^+e^-$ experiment Belle II is expected to be more efficient than the LHCb in reconstructing 
$B$ to tau-lepton decays, since many of these modes require reconstructing additional tracks originating from the final state mesons ($K$, $K^*$ or $\phi$). 
Therefore, $b\to s\tau^+\tau^-$ observables will be among the \textit{``golden modes''} aiming at probing the vector leptoquark hypothesis at Belle II. 

\begin{figure}[h!]
	\mbox{\hspace*{27mm}\includegraphics[width=0.8\textwidth]{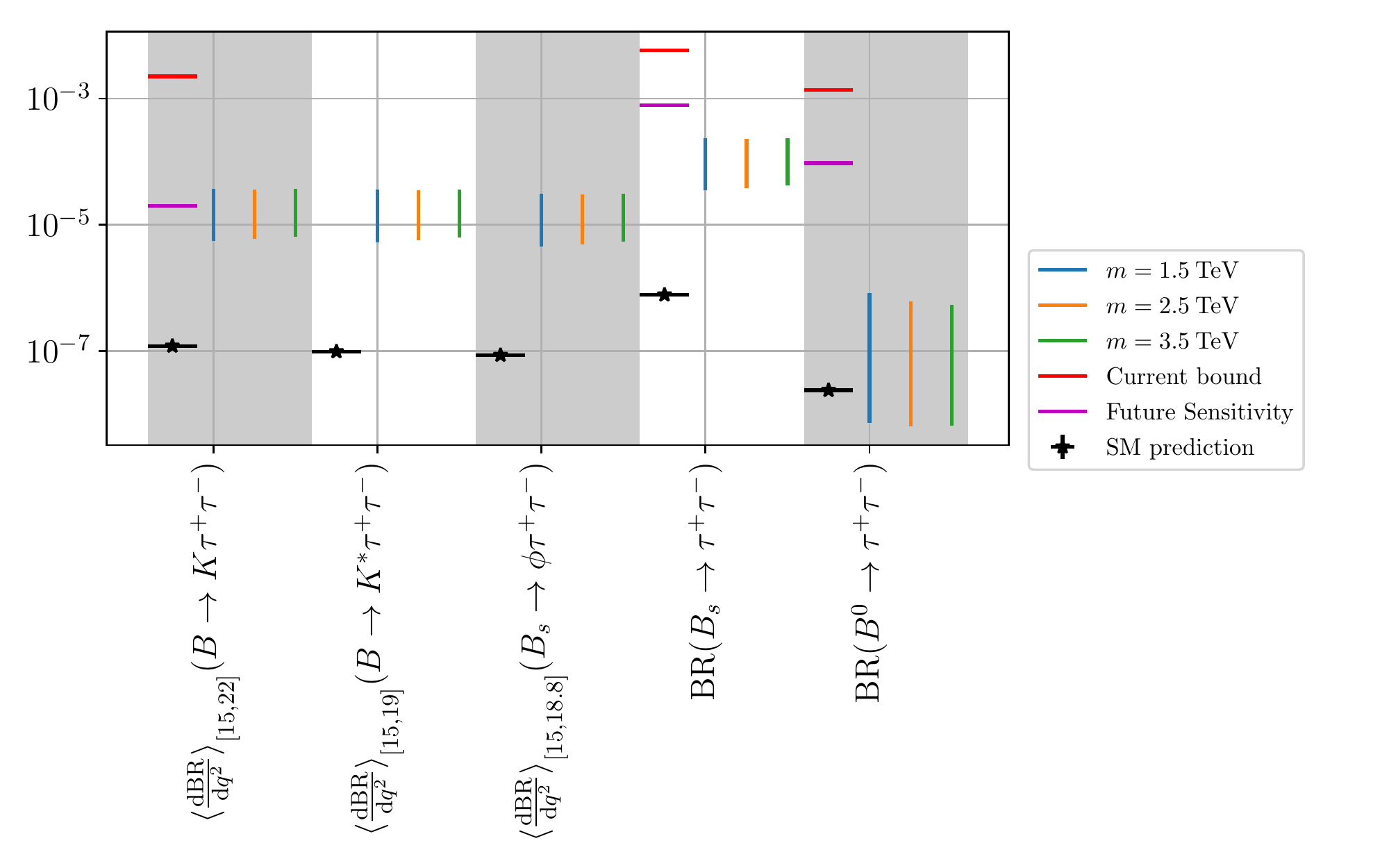}}
	\caption{Predictions for several leptonic and semi-leptonic $B_{(s)}$ to $\tau^+ \tau^-$ decays, for three benchmark values of the vector leptoquark mass (coloured vertical bars). Also displayed, to the left of the different predictions, are the current experimental  limits and the future projected sensitivity from Belle II (horizontal lines), as well as the corresponding SM prediction (black). The ranges correspond to the interval between the $5^\text{th}$ and $95^\text{th}$ percentiles of the posterior distributions, as described in Appendix~\ref{app:stats}. Figure from~\cite{Hati:2020cyn}.}
	\label{fig:tau_sm_allowed_predictions}
\end{figure}

In Fig.~\ref{fig:tau_sm_allowed_predictions} we present
the predictions for several leptonic and semi-leptonic $B_{(s)}$ to $\tau^+ \tau^-$ decays, as arising in the present  
vector leptoquark scenario. 
We display the results for three benchmark leptoquark masses (coloured vertical bars, corresponding to $m_V=$1.5~TeV, 2.5~TeV and 3.5~TeV), together with the current limits and the future projected sensitivity from Belle II, and the corresponding SM predictions. 

As can be clearly observed from Fig.~\ref{fig:tau_sm_allowed_predictions}, all $b\to s\tau\tau$ branching fractions are enhanced with respect to the SM (typically by one to two orders of magnitude). 
This is a direct consequence of accommodating the charged current anomalies (i.e. $R_{D^{(*)}}$), as these call upon sizeable $b-\tau$ and $s(c)-\tau$ couplings. The decay $B^0\to\tau^+\tau^-$ is subject to a milder enhancement due to having the $d-\tau$ coupling already constrained by other observables.

\bigskip
Tau-lepton decays offer powerful probes of vector leptoquark models.
The Belle experiment has searched for 46 distinct cLFV $\tau$ decay modes, using almost its entire data sample of approximately 1000 $\mathrm{fb}^{-1}$;  no evidence for cLFV decays was found, but new 90\% C.L. upper limits on the branching fractions were set, at a level of around  $\mathcal{O}(10^{-8})$. 
At Belle II, if on the one hand the higher beam-induced background will render these searches more challenging, on the other hand its impressive luminosity will 
allow to significantly ameliorate the sensitivities to these modes. As much as 45 billion $\tau$ pairs (in the full dataset) are expected to be produced in $e^+ e^-$ collisions at Belle II, clearly providing very bright prospects for cLFV tau decay searches.
The Belle II experiment is thus expected to improve the sensitivities of the various cLFV decays by more than one order of magnitude, reaching a level of $\mathcal{O}(10^{-9}-10^{-10})$. 

\begin{figure}[h!]
	\mbox{\hspace*{27mm}\includegraphics[width=0.8\textwidth]{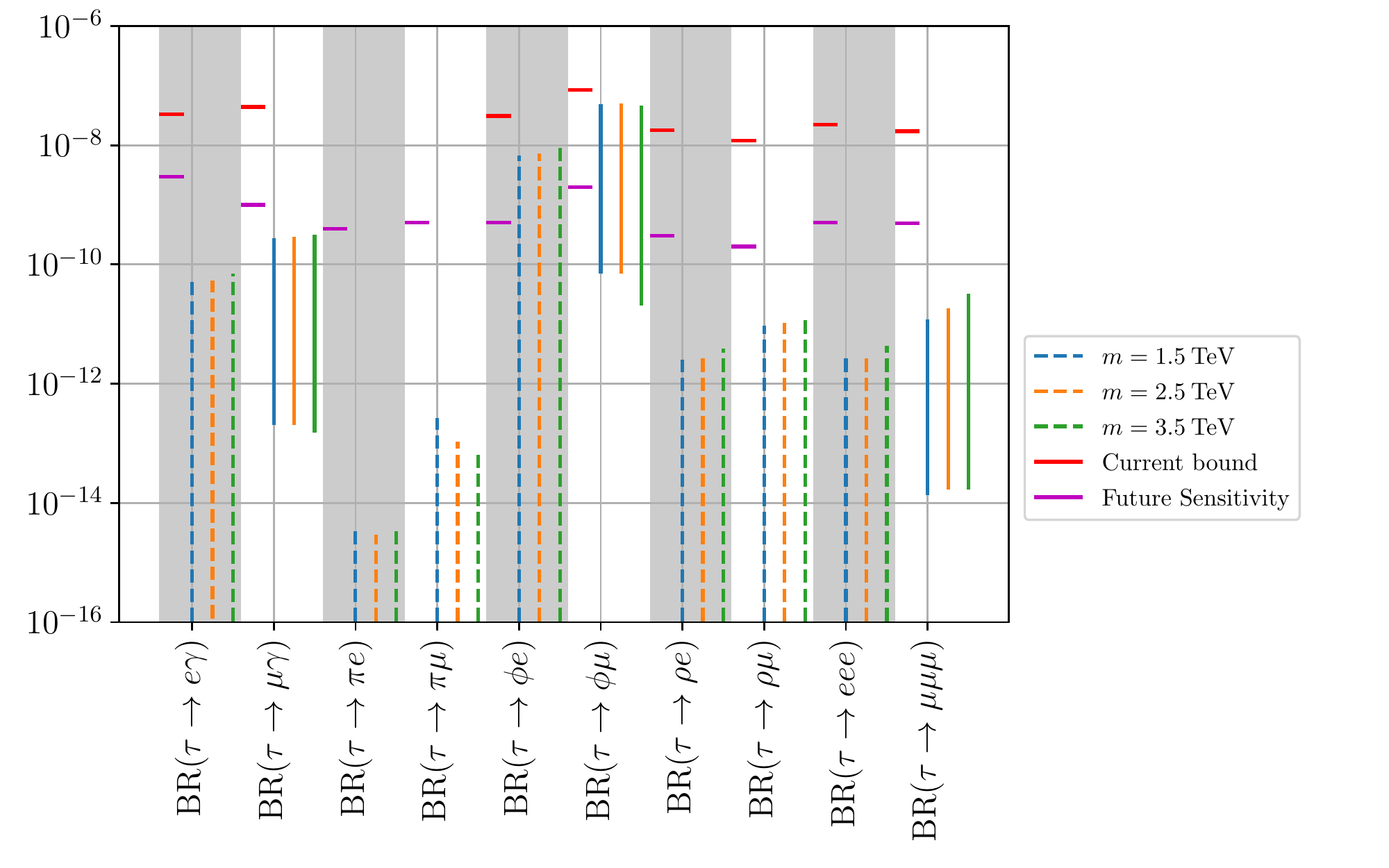}}
    	\caption{Lepton flavour violating $\tau$ decay modes expected to be searched for at the Belle II experiment. The $90\%$ ranges are obtained from sampling points at the around the best-fit point. Line and colour coding as in Fig.~\ref{fig:tau_and_lfv_predictions}. Figure from~\cite{Hati:2020cyn}.}
	\label{fig:taulfv_predictions}
\end{figure}

In Fig.~\ref{fig:taulfv_predictions} we present the predictions of the vector leptoquark scenario for various cLFV tau decay modes which are programmed to be searched for at the Belle II experiment. 
It is interesting to note that among the various observables, the $\tau \to \phi\mu$ decay emerges as the most promising one to probe the vector leptoquark hypothesis  - another ``\textit{golden mode}'', also identified by several other independent groups (see e.g.~\cite{Angelescu:2018tyl}).

\begin{figure}[h!]
	\mbox{\hspace*{27mm}\includegraphics[width=0.8\textwidth]{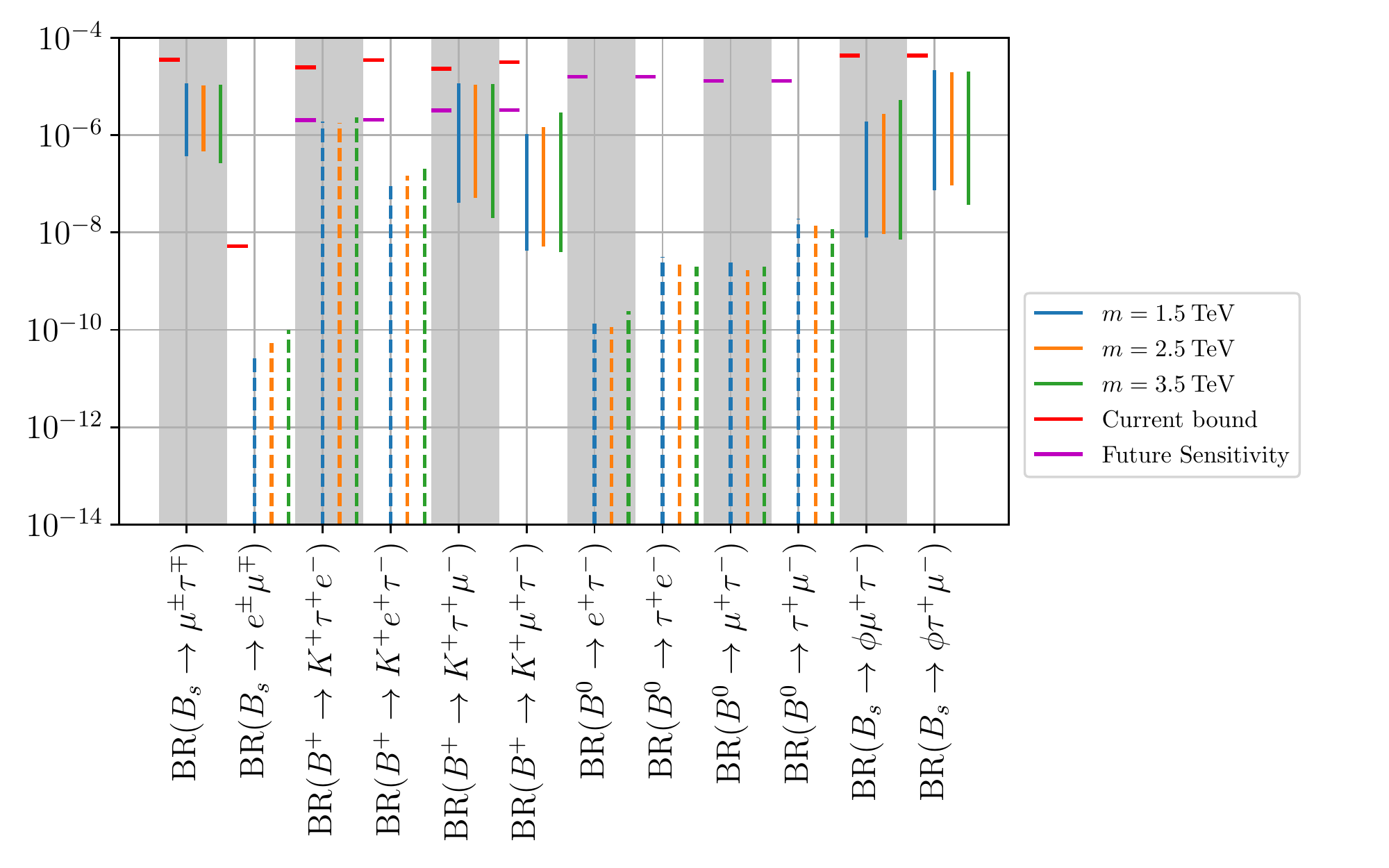}}
	\caption{Lepton flavour violating decay modes of beauty-flavoured mesons to $\tau$-leptons, to be searched for at Belle II. The $90\%$ C.L. ranges are obtained from sampling points around the best-fit point. Line and colour coding as in Fig.~\ref{fig:tau_and_lfv_predictions}. Figure from~\cite{Hati:2020cyn}.}
	\label{fig:bfv_predictions}
\end{figure}

The Belle II experiment will also search for a number of cLFV leptonic and semi-leptonic $B$-meson decays (some into final state $\tau$s). In Fig.~\ref{fig:bfv_predictions} we present our findings for these cLFV processes.
In the context of the present vector leptoquark model, one thus expects sizeable contributions for $B_s \rightarrow \tau^+ \mu^-$,
$B^+ \rightarrow K^+ \tau^+ e^-$, $B^+ \rightarrow K^+ \tau^+ \mu^-$ and $B_s \rightarrow \phi \tau^+ \mu^-$ (for the different benchmark masses considered), close to  current bounds, and clearly within reach of future sensitivities\footnote{Notice that the rates for $B_s$ decays into $\phi \tau^- \mu^+$ are typically less enhanced than those for the (opposite charge) $\phi \tau^+ \mu^-$ mode: this is a consequence of the leptoquark couplings involved, with the combination $K_L^{22}K_L^{33}$ (entering the former) in general smaller than $K_L^{23}K_L^{32}$ (appearing in the latter), as can be inferred, for example, from Table~\ref{tab:fits_wKLCRmue}.}. Together with the decay channels identified following the results displayed in Fig.~\ref{fig:tau_sm_allowed_predictions}, these cLFV 
modes appear particularly promising to observe a signal of a vector leptoquark New Physics scenario explaining the $B$-meson decay anomalies.

\subsection{Impact of future negative searches}
\label{sec:future_Belle}
A final point to be addressed concerns the impact of 
future null results from Belle II and other experiments searching for cLFV: if no cLFV signal is found, and no enhancement of $B$-meson decay rates 
is observed, to which extent will this affect the prospects of a vector leptoquark hypothesis as a viable explanation of the $B$-meson decay anomalies?  
To assess the implication of such a scenario we re-conduct the fit whose results were summarised in  Table~\ref{tab:fits_wKLCRmue}, now including the projected future sensitivities from Belle II  and cLFV-dedicated experiments (COMET, Mu2e, MEG II and Mu3e).
Recall that the Belle II observables taken into account in this fit are listed in Appendix~\ref{app:Obs}  (Table~\ref{tab:belleii}), with the future sensitivities always corresponding to the assumption of the full anticipated luminosity of $50\:\mathrm{ab}^{-1}$; the future sensitivities for the cLFV dedicated experiments have been summarised in the first part of Table~\ref{tab:important_LFV}.

The results of this new fit~\cite{Hati:2020cyn} (corresponding to null results in the several ``golden modes'' previously discussed) are presented in Table~\ref{tab:fits_wCOMET_Belle}. 
A comparison of these results with those of  Table~\ref{tab:fits_wKLCRmue} suggests that all leptoquark couplings would be well constrained (with the exception of the $d-\tau$ one). We again notice here that the vector leptoquark coupling to the first generation SM fermions remain consistent with zero.

\begin{table}[h!]
	\hspace{-1.1cm}
    \begin{tabular}{|c|c|c|c|}
	\hline
	{$m_{V_1}$} & $K_L$ {best-fit} & $K_L$ $90\%$ & {$\text{pull}_\text{SM}$}\\
	\hline
	\hline
	{\footnotesize$1.5\:\mathrm{TeV}$} &
	{\footnotesize$\begin{pmatrix}
		-1.9\times 10^{-6} & -9.5\times10^{-3} & -0.011\\
		6.4\times10^{-6} & -0.021 & 0.31\\
		-3.2\times10^{-6} & 0.061 & 0.49
	\end{pmatrix}$}&
	{\footnotesize$\begin{pmatrix}
		(-6.6\to 8.6)\times10^{-4} & (-2.3\to 8.8)\times10^{-3} & -0.056\to0.008\\
		-0.012\to 0.011 & -0.037 \to -0.009  & 0.13 \to 0.59\\
		(-3.1 \to 2.5)\times 10^{-3} & 0.030 \to 0.12  & 0.19\to1.02
	\end{pmatrix}$}&
	{\footnotesize$5.52$}\\
	\hline
	\hline
	{\footnotesize$2.5\:\mathrm{TeV}$}&
	{\footnotesize$\begin{pmatrix}
		3.8\times10^{-6} & -8.7\times10^{-3} & -0.031\\
		3.9\times10^{-5} & -0.032 & 0.53\\
		2.7\times10^{-5} & 0.11 & 0.81
	\end{pmatrix}$}&
	{\footnotesize$\begin{pmatrix}
		(-8.5\to9.0)\times10^{-4} & (-2.9\to13.9)\times10^{-3} & -0.062\to 0.013\\
		-0.017\to 0.017 & -0.077 \to -0.018 & 0.13 \to 0.92\\
		(-3.3\to 5.8)\times 10^{-3} & 0.041 \to 0.18 & 0.23\to 1.79
	\end{pmatrix}$}&
	{\footnotesize$5.58$}\\
	\hline
	\hline
	{\footnotesize$3.5\:\mathrm{TeV}$}&
	{\footnotesize$\begin{pmatrix}
		-1.2\times10^{-5} & 0.012 & -0.012\\
		3.1\times10^{-4} & -0.044 & 0.71\\
		-4.0\times10^{-5} & 0.16 & 1.19
	\end{pmatrix}$}&
	{\footnotesize$\begin{pmatrix}
		(-1.4\to 1.4)\times10^{-3} & (-6.5\to14.6)\times 10^{-3} & -0.10\to 0.011\\
		-0.025\to0.024 & -0.10\to -0.02 & 0.23\to1.39 \\
		(-7.9\to4.8)\times 10^{-3} & 0.063 \to 0.36 & 0.32\to2.41
	\end{pmatrix}$}&
	{\footnotesize$5.61$}\\
	\hline
	\end{tabular}
	\caption{Best-fit points, symmetric $90\%$ ranges and SM pulls of the fits containing the envisaged sensitivities of the Belle II, COMET, Mu2e, Mu3e and MEG II experiments where the non-observation of all included cLFV observables is assumed.}
	\label{tab:fits_wCOMET_Belle}
\end{table}

One can now re-project
the new fit results onto the plane of the anomalous $B$-meson decay observables, by randomly sampling around the best fit points presented in Table~\ref{tab:fits_wCOMET_Belle}.  
For the $V_1$ scenario under consideration, the strongest impact of a non-observation of cLFV processes and non-enhanced rates for $B$-meson decays to $\tau^+ \tau^-$ final states occurs for the fit of the charged current anomalies $R_{D}$ and $R_{D^{*}}$. 
This is a consequence of having significantly stronger constraints on the vector leptoquark couplings to $\tau$-leptons following the negative search results from Belle II and future cLFV experiments, and will render $V_1$ less efficient in contributing to both $R_{D^{(*)}}$.

\begin{figure}[h!] 
	\hspace*{30mm}
	\includegraphics[width= 0.8 \textwidth]{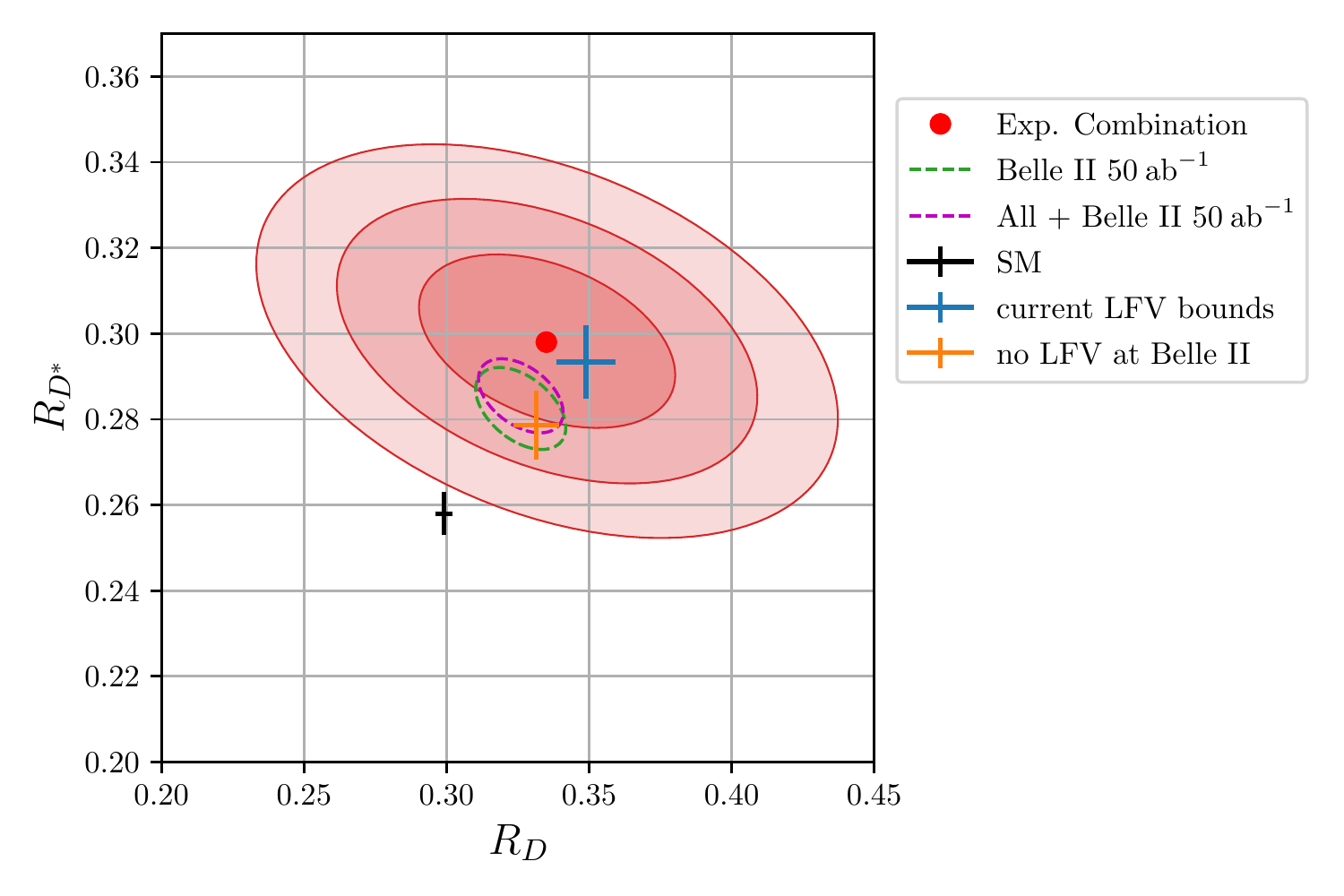}
	\caption{Likelihood contours and vector leptoquark predictions for $R_D$ and $R_{D^\ast}$. Red regions correspond to different likelihood contours obtained from a na\"ive combination of the experimental likelihoods. The blue cross denotes the predictions at the best-fit point to current LFV data. The orange cross denotes the predictions at the best-fit point with assumed null results of LFV processes at Belle II, Mu2e and COMET.
	The black cross denotes the SM prediction~\cite{Amhis:2019ckw}.
	The green dashed contour line describes the na\"ive extrapolation of the current combination of Belle data~\cite{Huschle:2015rga,Hirose:2016wfn,Abdesselam:2019dgh} to the anticipated future precision of the Belle II experiment, while the purple dashed contour line is a na\"ive combination of the Belle II projection with the current data.
	Figure from~\cite{Hati:2020cyn}
	}\label{fig:rdrds_after_belleii}
\end{figure}

We present in Fig.~\ref{fig:rdrds_after_belleii} the different likelihood contours and leptoquark predictions, for different benchmark masses\footnote{The central values and uncertainties of the predictions at the best-fit points are almost identical for all mass benchmark points.} and fit set-ups, as well as best-fit points for the distinct experimental scenarios. 
The  impact for the $b\to c\ell\nu$ fit can be observed in the $R_{D}-R_{D^{*}}$ plane depicted in Fig.~\ref{fig:rdrds_after_belleii}, as the preferred ``region'' (orange cross) is pulled towards the SM prediction, and away from the current experimental best fit point (red circle).

Notice however that potential negative results from Belle II and future cLFV experiments do not significantly affect the fit to anomalous $b\rightarrow s \ell\ell$ observables. 

\bigskip
The above discussion clearly emphasises the key role played by Belle II and future cLFV experiments in probing the vector leptoquark scenario as a unified explanation to the $B$-decay anomalies, especially in view of a new determination of $R_{D^{(*)}}$ (central value and associated uncertainties).
Scenarios can be envisaged in which future experimental data 
corroborates current $R_{D^{(*)}}$ values (no change in the central value, corresponding to the red ``dot'' in Fig.~\ref{fig:rdrds_after_belleii}), but accompanied by a reduction of the associated errors (implying tighter likelihood contours): this could then potentially contribute to disfavour $V_1$ as a viable explanation to the charged current $B$-meson decay anomalies.
However, if future Belle II data (dashed contours in Fig.~\ref{fig:rdrds_after_belleii}, see also Fig.~\ref{fig:rdrds_proj} and related discussion in Chapter~\ref{chap:bphysics}) evolves along current Belle data, vector leptoquarks would still remain exceptional candidates to explain the $B$-meson decay anomalies, while avoiding detection in cLFV processes in the future.

\section{Summary and outlook}

Being a well-motivated New Physics candidate, leptoquark extensions of the SM have been increasingly investigated, in view of their potential for a simple, minimalistic scenario to explain the current hints of LFUV arising from $B$-meson decay data. Vector leptoquarks transforming as $(\mathbf{3},\mathbf{1},2/3)$ are particularly appealing, as they offer a simultaneous explanation for both charged and neutral current $B$-meson  decay anomalies, parametrised by the $R_{K^{(\ast)}}$ and $R_{D^{(\ast)}}$ observables.

In this chapter, we have thus investigated how minimal constructions, containing the vector leptoquark $V_1$, successfully account for 
the anomalies in both  $R_{K^{(\ast)}}$ and $R_{D^{(\ast)}}$. 
Minimal extensions by a single $V_1$ leptoquark are in general
disfavoured due to the strong cLFV constraints on the (unitary) quark-lepton-$V_1$ couplings. 
In~\cite{Hati:2019ufv} we have suggested that
the pattern of mixings required to simultaneously address
$R_{K^{(*)}}$ and $R_{D^{(*)}}$ with a single $V_1$
could be interpreted within a
framework of non-unitary $V_1 \ell q$ couplings:
the mixings of the
SM charged leptons with the additional vector-like heavy
leptons can lead to effectively non-unitary $V_1 \ell q$ couplings,
offering the required amount of LFUV to  account for both anomalies~\cite{Hati:2019ufv}.
As we have argued, the most
minimal non-unitary scenario (i.e. $n=1$) consistent with
both $R_{K^{(*)}}$ and
$R_{D^{(*)}}$, is ruled out as it leads to excessive contributions to cLFV
observables such as muon-electron conversion in nuclei.
We have thus considered
three families of vector-like heavy
leptons, and we have carried out a detailed analysis of the impact for
an extensive array of flavour violation and EW precision
observables.
Our findings revealed that the $SU(2)_L$ charges of the heavy charged
leptons are of paramount importance for the
model's viability:
for isosinglet heavy leptons, the mass of the leptoquark must be
sufficiently large to avoid excessive contributions to $Z$
decays, which then prevents an explanation of $R_{D^{(*)}}$.
This is expected to happen for heavy leptons of any $SU(2)_L$
representation except for isodoublets. In this particular case, the
$Z\ell \ell$ couplings remain universal, and we have shown that the
non-unitarity in the couplings allows to successfully explain both
sets of anomalies, while complying with all considered current bounds.

Furthermore, after having established that the couplings of the vector leptoquark to SM quarks and leptons are necessarily non-unitary,  we have extended the phenomenological analysis: we emphasise that starting from a completely general simplified-model parametrisation, 
we presented a fit for the full $3\times 3$ 
matrix ($K_L^{ij}$) encoding the $V_1 \ell q$ couplings,  
taking into account various relevant flavour observables and the anomalous LFUV data.
Not only this approach can provide a better guidance towards the probable UV completions for the vector leptoquark scenario to adress the LFUV anomalies but it can also reveal prospects for the relevant new observables which can be used to probe the vector leptoquark scenario (often missed in analysis of vector leptoquark coupling matrices with ad-hoc vanishing couplings to the first generation of SM fermions). 
Relying on this alternative formalism for the phenomenological fitting of the vector leptoquark couplings, we thoroughly investigated the impact of such a New Physics scenario: 
we considered the prospects for an extensive array of observables, including (in addition to the anomalous $B$-meson decay observables) leptonic cLFV transitions, several $B$ decay modes to final states including $\tau^+ \tau^-$ pairs, flavour violating $\tau$ decays as well as cLFV (semi-) leptonic decays of $B$-mesons. In view of the excellent experimental prospects, we have investigated several very promising ``\textit{golden modes}'' to (indirectly) test the $V_1$ scenario. 
Among these channels one finds $\tau\to\phi\mu$ decays, $b\to s\tau\tau$ and $b\to s\tau\mu$ transitions, as well as $\mu - e$ conversion in nuclei.  
These modes, searched for at Belle II and coming cLFV experiments, will play a crucial role in testing the vector leptoquark hypothesis as a single explanation to the $R_{K^{(\ast)}}$ and $R_{D^{(\ast)}}$ anomalies.

As we have discussed, the confirmation of LFUV in $B$-meson decays,
(strongly) enhanced rates for $B$-meson decays to $\tau^+ \tau^-$ final states, as well as an observation of cLFV transitions in certain channels (by itself a massive discovery!), would all contribute to substantiate a vector leptoquark New Physics scenario - although some of the latter signals could indeed arise from other BSM constructions. Conversely, the non-observation of such signals at Belle II and future cLFV experiments has the potential to falsify the vector leptoquark scenario as a solution to the anomalous $R_{D^{(*)}}$ data, if the latter anomaly persists in future measurements with reduced uncertainty (without significant changes in the central values). Should this be the case, and although New Physics models containing vector leptoquarks could still address the neutral current $B$-decay anomalies (i.e. $R_{K^{(*)}}$), 
a common explanation of both sets of anomalies would be certainly more challenging. 

\medskip
Finally, we want to stress that even in the event of accumulating more and more indirect signals consistent with $V_1$ leptoquark model predictions, be it $b\to s\mu\tau$ LFV decays or significantly enhanced $b\to s\tau\tau$ transitions, by no means this would imply a ``discovery'' of leptoquarks.
Low-scale $V_1$  leptoquarks can only ultimately be discovered and directly probed at colliders.
Particularly in the context of future hadron colliders, the high-luminosity LHC, the high-energy LHC (running at $27\:\mathrm{TeV}$) and of a planned FCC-hh, potential signals of simplified vector leptoquark models ($V_1$ and $V_3$) accommodating $b\to s\ell\ell$ data were recently pursued in~\cite{Hiller:2021pul}. It is shown that direct searches at these future colliders would be able to directly probe leptoquark masses up to $\sim 20\:\mathrm{TeV}$.
More ambitiously from the experimental side, but even more appealing from the phenomenological point of view, would be direct searches at a high-energy ($\mathrm{TeV}$-scale) muon collider.
It has been recently shown~\cite{Asadi:2021gah} that a $3\:\mathrm{TeV}$ muon collider would be sufficient to entirely probe the currently preferred parameter space of simplified $V_1$ models that accommodate $b\to s\ell\ell$ data
(see also~\cite{Huang:2021biu} for $b\to s\ell\ell$ inspired scalar leptoquarks $S_3$ models and $Z'$-boson constructions).
Hopefully, in the near future, direct searches will either unveil the existence of these New Physics stats, or then provide stringent bounds allowing to constrain (and potentially rule out) their viability range.
\chapter{Final remarks and future prospects}
\label{chap:concs}
\minitoc
During the past decades, the endeavour of uncovering New Physics beyond the Standard Model has relied of multiple approaches, both on the experimental and theoretical sides.
Experimentally, there are dedicated efforts to directly search for New Physics at \textit{high energies} and indirectly search for New Physics effects at \textit{high intensities}.
Motivated by New Physics models attempting at solving (or at least ameliorating) the theoretical caveats of the SM (e.g. the hierarchy problem), for instance supersymmetric or extra-dimensional models, most experiments were devoted to searching directly for the new (heavy) states at the high-energy facilities, such as LEP, Tevatron, or LHC.  

Being complementary to high energy searches, precision measurements of electroweak and flavour observables at both low and high energies have always paved the way to direct discoveries of states present in the SM (e.g. the electroweak gauge bosons or the Higgs).
Prior to the present theoretical formulation of the SM, indirect hints on ``New Physics'' effects also provided guide lines for model building, as it is the case of $\beta$ decays contradicting the two-body decay picture and eventually leading to the introduction of neutrinos, the discovery of P violation leading to the introduction of $V-A$ interactions stemming from the $SU(2)_L$ gauge symmetry, the discovery of CP violation leading to the hypothesis of the third quark generation, and many more.

Despite the discovery of the Higgs boson at the LHC, direct signals for new states have so far eluded experimental observation. 
In turn, negative search results continuously increase the energy scale at which New Physics could be present, which is in many cases already above the TeV-scale.
Thus, theoretical objectives of naturalness that often necessitate New Physics to be present at the TeV scale (in order to solve the theoretical issues of the SM) are challenged and one should re-evaluate the guiding principles for model-building.

With the discovery of neutrino oscillations, one clear guideline for New Physics is given. 
Being the first laboratory evidence for New Physics, neutrino oscillations urgently call for extensions of the SM, in order to offer a viable mechanism of neutrino mass generation.
Interestingly, due to offering a new source of CP violation and calling upon weakly interacting states, New Physics extensions aiming at providing an explanation for neutrino masses can often be connected to the baryon asymmetry of the universe and the dark matter problem.
As of today, neutrino physics has entered its precision era, and a world-wide experimental and theoretical effort is devoted to resolve the many open questions related to it.
Additionally, the door to an immensely rich phenomenology connected to the lepton sector has been opened, since due to the presence of neutrino masses many accidental symmetries of the SM appear to be broken in Nature.
Consequently, the interest in high-intensity searches dedicated to the lepton sector has steadily increased.

The violation of accidental (lepton) symmetries of the SM, such as charged lepton flavour conservation and lepton flavour universality (both violated due to the presence of neutrino masses) opens many possible paths to search for New Physics.
While massive neutrinos consist of only one possible source of lepton flavour and lepton flavour universality violation, indirect signals indicating the breaking of these symmetries in synergy with possible other indirect signals of New Physics will provide crucial guidelines for both experimental direct searches and theoretical efforts to describe New Physics interactions.
Clearly, the lepton sector is emerging as a powerful laboratory to search for New Physics.

In addition to the discovery of neutrino oscillations, other ``lepton-flavoured'' observables exhibit significant deviations from their respective SM predictions. Among them are the anomalous magnetic moment of the muon, the LFUV observables in semi-leptonic $B$-meson decays and numerous deviations in the $b\to s\ell\ell$ system.
Can these anomalies suggest a path to the underlying  New Physics model, not only capable of explaining  them, but also allowing for the many other shortcomings of the SM?

A first starting point is given by data-driven ``bottom-up'' approaches fuelled by EFT analyses, to find requirements at low energies on a potential New Physics candidate. Although the main goal should always be to aim for a complete UV  description  of Nature (i.e. a full theoretical construction accounting for all the SM observational and theoretical caveats), early avenues can be inferred from minimal BSM realisations, as identified from the results of the bottom-up approach. Such  ad-hoc extensions, in which the SM is minimally enlarged by the strictly necessary ingredients to address individual problems (be them scalar fields, vectors, neutral fermions...) might offer guidelines towards the construction of more complete frameworks.
For this purpose, it is thus  of paramount importance to devote resources to fully understand the low-energy implications of such minimal constructions.
Likewise, in order to clarify the presence of New Physics in low-energy observables ($B$-anomalies, $(g-2)_\ell$ etc.), and in all cases to reduce the (sometimes still significant) theoretical uncertainties, the SM contributions, in particular (non-perturbative) QCD and long-distance effects, need to be mastered.

On the experimental front, the future looks extremely bright.
Concerning neutrino physics, there are several upcoming facilities dedicated to precisely measure the PMNS parameters, leptonic CP violation and determine the absolute mass scale of neutrinos.
Furthermore, a steadily increasing amount of data in neutrino-nucleus scattering is accumulated, allowing to indirectly constrain non-standard neutrino interactions and mixing scenarios.
At the high intensity frontier in  what regards the charged lepton sector, numerous upcoming muon dedicated facilities will greatly improve the searches for charged lepton flavour violation,
while Belle II will allow to significantly improve bounds on a large number of flavour violating $\tau$-lepton decays.
Moreover, the ``$B$-anomalies'' and numerous other interesting processes, including tauonic final states and (semi-) leptonic LFV decays, will be probed and potentially discovered.
As complementary tests of the Standard Model and its symmetries,  programmes on rare charm decays and CP violation in the charm sector have just started.
Finally, at the high energy frontier, run 3 of LHC is expected to start soon, with the high-luminosity upgrade coming in the foreseeable future.
In the farther future, albeit ambitious, very promising collider projects are planned,
including a high-energy electron-positron collider and possibly even a TeV-scale muon collider.
High-energy lepton machines will ultimately allow to push further the energy and intensity frontier in the quest to unravel New Physics.

\appendix
\chapter{Loop functions in general neutrino mixing scenarios}
\label{app:loopfunctions_neutrinos}
\minitoc
Below we summarise the loop functions for the cLFV transitions mediated by massive Majorana neutrinos, as well as their relevant limits\footnote{Note that in
  Ref.~\cite{Ilakovac:1994kj} the loop function $F_\text{Xbox}$ is
  named $F_\text{box}$ and has an opposite global sign when compared to
  Ref.~\cite{Alonso:2012ji}, which also reflects in the form factor
  $F_\text{box}^{\beta 3\alpha}$.}, as taken from
Refs.~\cite{Alonso:2012ji,Ilakovac:1994kj}. 
The photon dipole and anapole functions and relevant asymptotic limits, are given by
\begin{eqnarray}
    F_\gamma(x) &=& \frac{7 x^3 - x^2 - 12x}{12(1-x)^3} - \frac{x^4 -
      10x^3 + 12x^2}{6(1-x)^4}\log x\,,\nonumber\\ 
    F_\gamma(x) &\xrightarrow[x\gg1]{}& -\frac{7}{12} -
    \frac{1}{6}\log x\,,\nonumber\\ 
    F_\gamma(0) &=& 0\,,\label{eqn:lfun:fgamma}\\
    G_\gamma(x) &=& -\frac{x(2x^2 + 5x - 1)}{4(1-x)^3} -
    \frac{3x^3}{2(1-x)^4}\log x\,,\nonumber\\ 
    G_\gamma(x) &\xrightarrow[x\gg1]{}& \frac{1}{2}\,,\nonumber\\
    G_\gamma(0) &=& 0\,.\label{eqn:lfun:ggamma}
\end{eqnarray}
The loop functions of the $Z$-penguins are given by a
two-point 
function
\begin{eqnarray}
    F_Z(x) &=& -\frac{5 x}{2(1 - x)} - \frac{5x^2}{2(1-x)^2}\log x\,,\nonumber\\
    F_Z(x) &\xrightarrow[x\gg 1]{}& \frac{5}{2} - \frac{5}{2}\log
    x\,,\nonumber\\ 
    F_Z(0) &=& 0\,,\label{eqn:lfun:fz}
\end{eqnarray}
and two three-point functions which are symmetric under interchange
of the arguments: 
\begin{eqnarray}
    G_Z(x,y) &=& -\frac{1}{2(x-y)}\left[\frac{x^2(1-y)}{1-x}\log x -
      \frac{y^2(1-x)}{1-y}\log y \right]\,,\nonumber\\ 
    G_Z(x, x) &=& -\frac{x}{2} - \frac{x\log x}{1-x}\,,\nonumber\\
    G_Z(0,x) &=& -\frac{x\log x}{2(1-x)}\,,\nonumber\\
    G_Z(0,x)&\xrightarrow[x\gg 1]{}& \frac{1}{2}\log x\,,\nonumber\\
    G_Z(0,0) &=& 0\,,\label{eqn:lfun:gz}\\
    H_Z(x,y) &=& \frac{\sqrt{xy}}{4(x-y)}\left[\frac{x^2 - 4x}{1 -
        x}\log x - \frac{y^2 - 4y}{1 - y}\log
      y\ \right]\,,\nonumber\\ 
    H_Z(x,x) &=& \frac{(3 - x)(1-x) - 3}{4(1-x)} - \frac{x^3 - 2x^2 +
      4x}{4(1-x)^2}\log x\,,\nonumber\\ 
    H_Z(0,x) &=& 0\,.\label{eqn:lfun:hz}
\end{eqnarray}
The (symmetric) box-loop-functions and their limits are given by
\begin{eqnarray}
    F_\text{box}(x,y) &=& \frac{1}{x-y}\left\{\left(4 +
    \frac{xy}{4}\right)\left[\frac{1}{1-x} + \frac{x^2}{(1-x)^2} \log
      x - \frac{1}{1-y} - \frac{y^2}{(1-y)^2}\log
      y\right]\right.\nonumber\\  
    &\phantom{=}& \left. -2xy\left[\frac{1}{1-x} + \frac{x}{(1-x)^2}
      \log x - \frac{1}{1-y} - \frac{y}{(1-y)^2}\log y
      \right]\right\}\,,\nonumber\\ 
    F_\text{box}(x,x) &=& -\frac{1}{4(1-x)^3}\left[x^4 - 16x^3 + 31x^2
      - 16 + 2x\left(3x^2 + 4x - 16\right)\log x\right]\,,\nonumber\\ 
    F_\text{box}(0,x) &=& \frac{4}{1 - x} + \frac{4x}{(1-x)^2}\log
    x\,,\nonumber\\ 
    F_\text{box}(0,x)&\xrightarrow[x\gg 1]{}& 0\,,\nonumber\\
    F_\text{box}(0,0) &=& 4\,,\label{eqn:lfun:fbox}\\
    F_\text{Xbox}(x,y) &=& -\frac{1}{x-y}\left\{\left(1 + \frac{xy}{4}
    \right)\left[\frac{1}{1-x} + \frac{x^2}{(1-x)^2} \log x -
      \frac{1}{1-y} - \frac{y^2}{(1-y)^2}\log
      y\right]\right.\nonumber\\  
    &\phantom{=}& \left. -2xy\left[\frac{1}{1-x} + \frac{x}{(1-x)^2}
      \log x - \frac{1}{1-y} - \frac{y}{(1-y)^2}\log y
      \right]\right\}\,,\nonumber\\ 
    F_\text{Xbox}(x,x) &=& \frac{x^4 - 16x^3 + 19x^2 - 4}{4(1-x)^3} +
    \frac{3x^3 + 4x^2 - 4x}{2(1-x)^3}\log x\,,\nonumber\\ 
    F_\text{Xbox}(0,x) &=& -\frac{1}{1-x} - \frac{x}{(1 - x)^2}\log
    x\,,\nonumber\\ 
    F_\text{Xbox}(0,x)&\xrightarrow[x\gg 1]{}& 0\,,\nonumber\\
    F_\text{Xbox}(0,0) &=& -1\,,\label{eqn:lfun:fxbox}\\
    G_\text{box}(x,y) &=& -\frac{\sqrt{xy}}{x-y}\left\{(4 +
    xy)\left[\frac{1}{1-x} + \frac{x}{(1-x)^2} \log x - \frac{1}{1-y}
      - \frac{y}{(1-y)^2}\log y\right]\right.\nonumber\\  
    &\phantom{=}& \left. -2\left[\frac{1}{1-x} + \frac{x^2}{(1-x)^2}
      \log x - \frac{1}{1-y} - \frac{y^2}{(1-y)^2}\log y
      \right]\right\}\,,\nonumber\\ 
    G_\text{box}(x,x) &=& \frac{2x^4 - 4x^3 + 8x^2 - 6x}{(1-x)^3} -
    \frac{x^4 + x^3 + 4x}{(1-x)^3}\log x\,,\nonumber\\ 
    G_\text{box}(0,x) &=& 0\,.\label{eqn:lfun:gbox}
\end{eqnarray}

\chapter{Phase dependence of cLFV observables - full analytical expressions}
\label{app:analytic.phase.observables}
\minitoc

\noindent
In order to analytically study the phase dependence of the cLFV form factors (leading to the results discussed in Chapter~\ref{chap:CPV}, we work under the approximation $\sin\theta_{\alpha 4}\approx\sin\theta_{\alpha5} \ll 1$. Let us further assume that the masses of the heavy states are close to each other 
and of the order of a few $\mathrm{TeV}$, $m_4\approx m_5 \gtrsim \Lambda_\text{EW}$.
\section{Photon penguins}
The photon penguin form factors exhibit a common structure. From inspection of the relevant loop functions (cf. Appendix~\ref{app:loopfunctions_neutrinos}), it can be seen that contributions from the light neutrino mass eigenstates are negligible. Furthermore, since the heavy states are assumed to be close in mass, the loop functions are approximately equal and can be thus factored out of the sum.
The form factor is then given by 
\begin{eqnarray}
    G_\gamma^{\beta \alpha} &=& \sum_{i =1}^{5}
    \mathcal{U}_{\alpha i}\,\mathcal{U}_{\beta i}^\ast\, G_\gamma(x_i)\:,\nonumber \\ 
    &\approx& (\mathcal U_{\alpha 4}\,\mathcal U_{\beta 4}^\ast + \mathcal U_{\alpha 5}\,\mathcal U_{\beta 5}^\ast) \,G_\gamma(x_{4,5}) \ ,\label{eq:cLFV:FF:Ggamma:approx}
\end{eqnarray}
and inserting now the entries of the mixing matrix of Eq.~\eqref{eqn:allrot} 
(in the limit in which $\varphi_{i}=0$) 
yields
\begin{equation}
    G_\gamma^{\beta \alpha} \approx e^{-\frac{i}{2}(\Delta_4^{\alpha\beta}+\Delta_5^{\alpha\beta})}\left( s_{\alpha 4} s_{\beta_4} e^{-\frac{i}{2}(\Delta_4^{\alpha\beta}-\Delta_5^{\alpha\beta})}  + s_{\alpha 5} s_{\beta 5}e^{\frac{i}{2}(\Delta_4^{\alpha\beta}-\Delta_5^{\alpha\beta})}\right) G_\gamma(x_{4,5})\,,
    \label{eqn:Ggaphase}
\end{equation}
with $s_{ij} = \sin\theta_{ij}$ and $x_{4,5} = m_4^2/M_W^2 = m_5^2/M_W^2$, and where we have defined
\begin{equation}
    \Delta_i^{\alpha\beta} = \delta_{\alpha i} - \delta_{\beta i}\,,
\end{equation}
with the properties $\Delta_i^{\alpha\beta} = - \Delta_i^{\beta\alpha}$, $\Delta_i^{\alpha\rho} + \Delta_i^{\rho\beta} = \Delta_i^{\alpha\beta}$ and $\Delta_i^{\alpha\alpha} = 0$.
Further assuming that the mixing between the active neutrinos and $\nu_4$ and $\nu_5$ is approximately equal, i.e. $\sin\theta_{\alpha 4}\approx \sin\theta_{\alpha 5}$, we can simplify $G_\gamma^{\beta\alpha}$ to 
\begin{equation}
    G_\gamma^{\beta \alpha} \approx s_{\alpha 4}s_{\beta 4} e^{-\frac{i}{2}(\Delta_4^{\alpha\beta}+\Delta_5^{\alpha\beta})} 2 \cos\left(\frac{\Delta_4^{\alpha\beta}-\Delta_5^{\alpha\beta}}{2}\right)  G_\gamma(x_{4,5})\,,
\end{equation}
such that the branching fraction for the radiative decays is given by
\begin{equation}
    \mathrm{BR}(\ell_\beta\to \ell_\alpha\gamma)\propto |G_\gamma^{\beta\alpha}|^2 \approx 4 s_{\alpha 4}^2 s_{\beta 4}^2 \cos^2\left(\frac{\Delta_4^{\alpha\beta}-\Delta_5^{\alpha\beta}}{2}\right) \, G^2_\gamma(x_{4,5})\,.
\end{equation}
Similar results can be obtained for $F_\gamma^{\beta\alpha}$.

\mathversion{bold}
\section{$Z$ penguins}
\mathversion{normal}
The form factor generated by $Z$ penguin diagrams can be split into three different parts. It consists of bubble diagrams and triangle diagrams with or without Majorana mass insertions, and is given in the ``$3+2$ toy model'' by
\begin{eqnarray}
    F_Z^{\beta \alpha} &=& F_Z^{(1)} + F_Z^{(2)} + F_Z^{(3)}\\
    &=&\sum_{i,j =1}^{5}
    \mathcal{U}_{\alpha i}\,\mathcal{U}_{\beta j}^\ast \left[\delta_{ij} \,F_Z(x_j) +
    C_{ij}\, G_Z(x_i, x_j) + C_{ij}^\ast \,H_Z(x_i, x_j)\right]\:.
    \label{eq:fz3}
\end{eqnarray}
The first term which is proportional to the function $F_Z$ can be rewritten in the same way as the photon penguins, and is thus given by
\begin{eqnarray}
    F_Z^{(1)} &=& \sum_{i=1}^{5}\, \mathcal U_{\alpha i}\, \mathcal U_{\beta i}^\ast \, F_Z(x_i)\nonumber\\
    &\approx& (\mathcal U_{\alpha 4}\, \mathcal U_{\beta 4}^\ast + \mathcal U_{\alpha 5}\, \mathcal U_{\beta 5}^\ast) \tilde F_Z(x_{4,5})\nonumber\\
    &\approx& 2 s_{\alpha 4}s_{\beta 4}\,e^{-\frac{i}{2}(\Delta_4^{\alpha\beta}+\Delta_5^{\alpha\beta})}\: \cos\left(\frac{\Delta_4^{\alpha\beta}-\Delta_5^{\alpha\beta}}{2}\right)  F_Z(x_{4,5})\,,
\end{eqnarray}
in analogy to $G_\gamma^{\beta\alpha}$, see Eq.~\eqref{eqn:Ggaphase}.
The second and the third terms are more involved due the presence of $C_{ij}$.
The second term can be written as
\begin{eqnarray}
    F_Z^{(2)} &\approx& \sum_{\rho \in \{e, \mu, \tau\}} \left[(\mathcal U_{\alpha 4} U_{\rho 4}^\ast +\mathcal U_{\alpha 5} \,\mathcal U_{\rho 5}^\ast)(\mathcal U_{\beta 4}^\ast \,\mathcal U_{\rho 4} +\mathcal U_{\beta 5}^\ast \,\mathcal U_{\rho 5})\right]\tilde G_Z(x_{4,5})\nonumber\\
    &=& \sum_{\rho \in \{e, \mu, \tau\}} \left[(s_{\alpha 4}s_{\rho 4} e^{-i \Delta_4^{\alpha\rho}} + s_{\alpha 5}s_{\rho 5} e^{-i \Delta_5^{\alpha\rho}})(s_{\beta 4}s_{\rho 4} e^{i \Delta_4^{\beta\rho}} + s_{\beta 5}s_{\rho 5} e^{i \Delta_5^{\beta\rho}})\right] \tilde G_Z\nonumber\\
    &\approx& \sum_{\rho \in \{e, \mu, \tau\}} 4s_{\alpha 4}s_{\beta 4} s_{\rho 4}^2\,e^{-\frac{i}{2}(\Delta_4^{\alpha \beta} + \Delta_5^{\alpha\beta})}\:\cos\left(\frac{\Delta_4^{\alpha \rho} - \Delta_5^{\alpha\rho}}{2}\right) \cos\left(\frac{\Delta_4^{\beta \rho} - \Delta_5^{\beta\rho}}{2}\right)\tilde G_Z\,,
\end{eqnarray}
where we introduced $\tilde G_Z = \tilde G_Z(x_{4,5}) \equiv  G_Z(x_{4,5}, x_{4,5})$, which is also used in the following for loop functions that depend on 2 parameters, in the limit of degenerate masses (cf. Appendix~\ref{app:loopfunctions_neutrinos}).
Due to the appearance of $C_{ij}^\ast$, in the third  term of Eq.~(\ref{eq:fz3}), the Majorana phases will also be present. This corresponds to a Majorana mass insertion in the corresponding triangle diagram.
The last term can be cast as
\begin{eqnarray}
    F_Z^{(3)} &\approx& \sum_{\rho \in \{e, \mu, \tau\}}\left[(\mathcal U_{\alpha 4} \,\mathcal U_{\rho 4} +\mathcal U_{\alpha 5} \,\mathcal U_{\rho 5})(\mathcal U_{\beta 4}^\ast \,\mathcal U_{\rho 4}^\ast +\mathcal U_{\beta 5}^\ast \,\mathcal U_{\rho 5}^\ast)\right]\tilde H_Z(x_{4,5})\nonumber\\
    &=&e^{-\frac{i}{2}(\Delta_4^{\alpha\beta} + \Delta_5^{\alpha\beta})}\sum_{\rho \in \{e, \mu, \tau\}}\left[\left(s_{\alpha 4}s_{\rho 4}e^{-\frac{i}{2}(\Delta_\alpha^{45} + \Delta_\rho^{45} - 2(\varphi_4 - \varphi_5))} + s_{\alpha 5}s_{\rho 5}e^{\frac{i}{2}(\Delta_\alpha^{45} + \Delta_\rho^{45} - 2(\varphi_4 - \varphi_5))}\right)\right.\nonumber\\
    &\phantom{=}&\hspace{2cm}
    \times\left.\left(s_{\beta 4}s_{\rho 4}e^{\frac{i}{2}(\Delta_\beta^{45} + \Delta_\rho^{45} - 2(\varphi_4 - \varphi_5))} + s_{\beta 5}s_{\rho 5}e^{-\frac{i}{2}(\Delta_\beta^{45} + \Delta_\rho^{45} - 2(\varphi_4 - \varphi_5))}\right)\right]\tilde H_Z\nonumber
\end{eqnarray}
\begin{eqnarray}
    &\approx& 4e^{-\frac{i}{2}(\Delta_4^{\alpha\beta} + \Delta_5^{\alpha\beta})} \tilde H_Z\,s_{\alpha 4}s_{\beta 4}\times\nonumber\\
    &\phantom{=}&\times\,\sum_{\rho \in \{e, \mu, \tau\}}\left[s_{\rho 4}^2\cos\left(\frac{\Delta_\alpha^{45} + \Delta_\rho^{45}}{2} - (\varphi_4 - \varphi_5)\right)\cos\left(\frac{\Delta_\beta^{45} + \Delta_\rho^{45}}{2} - (\varphi_4 - \varphi_5)\right)\right]\,.
\end{eqnarray}
In the case of vanishing Dirac phases this can be further simplified to 
\begin{eqnarray}
    F_Z^{(3)} \approx 4s_{\alpha 4}s_{\beta 4} \tilde H_Z(x_{4,5}, x_{4,5})\,\sum_\rho \left[s_{\rho 4}^2 \cos^2(\varphi_4 - \varphi_5)\right]\,.
\end{eqnarray}

Moreover, and unique to the $Z$-penguin, there are non-negligible contributions to the form factor stemming from light and heavy virtual neutrinos in the loop.
We write the corresponding limit of the loop function as $G_Z(0, x_{4,5}) = \overline{G}_Z$.
As usual, using the above approximations, the corresponding part of the form factor can be cast as
\begin{eqnarray}
    F_Z^{(2)}(0,x_{4,5}) &\approx&\left[ \sum_{i = 1}^3\sum_{j=4,5}\sum_{\rho=e,\mu,\tau} \mathcal U_{\alpha i}\,\mathcal U_{\beta j}^\ast\,\mathcal U_{\rho i}^\ast\,\mathcal U_{\rho j} + \sum_{i = 4,5}\sum_{j=1}^3\sum_{\rho=e,\mu,\tau} \mathcal U_{\alpha i}\,\mathcal U_{\beta j}^\ast\,\mathcal U_{\rho i}^\ast\,\mathcal U_{\rho j}\right] \overline{G}_Z\nonumber\\
    &=& \sum_\rho \left[\left(\mathcal U_{\alpha 1}\,\mathcal U_{\rho 1}^\ast + \mathcal U_{\alpha 2}\,\mathcal U_{\rho 2}^\ast + \mathcal U_{\alpha 3}\,\mathcal U_{\rho 3}^\ast \right) \left( \mathcal U_{\beta 4}^\ast\,\mathcal U_{\rho 4} + \mathcal U_{\beta 5}^\ast\,\mathcal U_{\rho 5}\right) \right. + \nonumber\\
    &\phantom{=}&\hspace{0.5cm} +\left. \left(\mathcal U_{\alpha 4}\,\mathcal U_{\rho 4}^\ast + \mathcal U_{\alpha 5}\,\mathcal U_{\rho 5}^\ast \right) \left(\mathcal U_{\beta 1}^\ast\,\mathcal U_{\rho 1} + \mathcal U_{\beta 2}^\ast\,\mathcal U_{\rho 2} + \mathcal U_{\beta 3}^\ast\,\mathcal U_{\rho 3}\right)\right]\,\overline{G}_Z\,.
\end{eqnarray}
Making use of the unitarity of the full $5\times5$ mixing matrix, i.e. 
\begin{equation}
    \sum_{i = 1}^{5} \mathcal U_{\alpha i}\, \mathcal U_{\rho i}^\ast = \delta_{\alpha\rho}\:\:\Rightarrow\:\: \sum_{i = 1}^{3} \mathcal U_{\alpha i}\, \mathcal U_{\rho i}^\ast = \delta_{\alpha\rho} - \mathcal U_{\alpha 4}\, \mathcal U_{\rho 4}^\ast - \mathcal U_{\alpha 5}\, \mathcal U_{\rho 5}^\ast\,,
\end{equation}
one finally has
\begin{eqnarray}
    F_Z^{(2)}(0,x_{4,5}) &\approx& \sum_\rho \left[\left(\delta_{\alpha\rho} - \mathcal U_{\alpha 4}\, \mathcal U_{\rho 4}^\ast - \mathcal U_{\alpha 5}\, \mathcal U_{\rho 5}^\ast\,\right)\left( \mathcal U_{\beta 4}^\ast\,\mathcal U_{\rho 4} + \mathcal U_{\beta 5}^\ast\,\mathcal U_{\rho 5} \right)\right.\nonumber\\
    &\phantom{=}&\left. + \left(\mathcal U_{\alpha 4}\,\mathcal U_{\rho 4}^\ast + \mathcal U_{\alpha 5}\,\mathcal U_{\rho 5}^\ast \right)\left(\delta_{\beta\rho} -  \mathcal U_{\beta 4}^\ast\,\mathcal U_{\rho 4} - \mathcal U_{\beta 5}^\ast\,\mathcal U_{\rho 5} \right)\right]\overline{G}_Z\nonumber\\
    &=& 2\,\overline{G}_Z \sum_\rho\left[\delta_{\alpha\rho} s_{\beta 4}s_{\rho 4} e^{-\frac{i}{2}(\Delta_4^{\rho\beta} + \Delta_5^{\rho\beta})}\cos\left(\frac{\Delta_4^{\rho\beta} - \Delta_5^{\rho\beta}}{2}\right) + \right.\nonumber\\
    &\phantom{=}& \left.+\delta_{\beta\rho} s_{\alpha 4}s_{\rho 4} e^{-\frac{i}{2}(\Delta_4^{\alpha\rho} + \Delta_5^{\alpha\rho})}\cos\left(\frac{\Delta_4^{\alpha\rho} - \Delta_5^{\alpha\rho}}{2}\right)\right.\nonumber\\
    &\phantom{=}& \left. - 4s_{\alpha 4}s_{\beta 4}s_{\rho 4}^2 e^{-\frac{i}{2}(\Delta_4^{\alpha\beta} + \Delta_5^{\alpha\beta})}\cos\left(\frac{\Delta_4^{\alpha\rho} - \Delta_5^{\alpha\rho}}{2}\right)\cos\left(\frac{\Delta_4^{\beta\rho} - \Delta_5^{\beta\rho}}{2}\right)\right]\,.
\end{eqnarray}

\section{Box diagrams}
The form factor generated by box diagrams is given by
\begin{eqnarray}
    F_\text{box}^{\beta 3 \alpha} &=& F_\text{box}^{(1)} + F_\text{box}^{(2)}\nonumber\\
    &=&\sum_{i,j = 1}^{3+k} \mathcal{U}_{\alpha i}\,\mathcal{U}_{\beta j}^\ast\left[\mathcal{U}_{\alpha i} \,\mathcal{U}_{\alpha j}^\ast\, G_\text{box}(x_i, x_j) - 2 \,\mathcal{U}_{\alpha i}^\ast \,\mathcal{U}_{\alpha j}\, F_\text{Xbox}(x_i, x_j) \right]\:,
\end{eqnarray}
where the first term corresponds to a diagram with a possible Majorana mass insertion, thus depending on the Majorana phases.
The first term $F_\text{box}^{(1)}$ can then be written as
\begin{eqnarray}
    F_\text{box}^{(1)} &\approx& (\mathcal U_{\alpha 4}^2 + \mathcal U_{\alpha 5}^2)(\mathcal U_{\beta 4}^\ast \,\mathcal U_{\alpha 4}^\ast + \mathcal U_{\beta 5}^\ast \,\mathcal U_{\alpha 5}^\ast )\, \tilde G_\text{box}(x_{4,5})\nonumber\\
    &=& \left(s_{\alpha 4}^2 e^{-2i(\delta_{\alpha 4} - \varphi_4)} + s_{\alpha 5}^2 e^{-2i(\delta_{\alpha 5} - \varphi_5)}\right)\left(s_{\alpha 4}s_{\beta 4} e^{i(\delta_{\alpha 4} + \delta_{\beta 4} - 2\varphi_4)} + s_{\alpha 5}s_{\beta 5} e^{i(\delta_{\alpha 5} + \delta_{\beta 5} - 2\varphi_5)}\right) \tilde G_\text{box}\nonumber\\
    &=& e^{-\frac{i}{2}(\Delta_4^{\alpha\beta} + \Delta_5^{\alpha \beta})}\left(s_{\alpha 4}^2e^{-i(\Delta_\alpha^{45} - (\varphi_4 - \varphi_5))} + s_{\alpha 5}^2e^{i(\Delta_\alpha^{45} - (\varphi_4 - \varphi_5)}\right)\nonumber\\
    &\phantom{=}&\hspace{2cm} \times \left(s_{\alpha 4}s_{\beta 4}e^{\frac{i}{2}(\Delta_\alpha^{45} + \Delta_{\beta}^{45} - 2(\varphi_4 - \varphi_5))} + s_{\alpha 5}s_{\beta 5}e^{-\frac{i}{2}(\Delta_\alpha^{45} + \Delta_{\beta}^{45} - 2(\varphi_4 - \varphi_5))}\right)\tilde G_\text{box}\nonumber\\
    &\approx& 4 e^{-\frac{i}{2}(\Delta_4^{\alpha\beta} + \Delta_5^{\alpha \beta})} s_{\alpha 4}^3s_{\beta 4}\cos\left(\Delta_{\alpha}^{45} - (\varphi_4 - \varphi_5)\right)\cos\left(\frac{\Delta_{\alpha}^{45} + \Delta_{\beta}^{45}}{2} - (\varphi_4 - \varphi_5)\right)\tilde G_\text{box}\,,
\end{eqnarray}
which again can be further simplified in the case of vanishing Dirac phases to
\begin{equation}
    F_\text{box}^{(1)} \approx 4 s_{\alpha 4}^3 s_{\beta 4}\cos^2(\varphi_4 - \varphi_5)\,\tilde G_\text{box}\,.
\end{equation}
The second term is independent of the Majorana phases and can be written as
\begin{eqnarray}
    F_\text{box}^{(2)} &\approx& -2 (|\mathcal U_{\alpha 4}|^2 + |\mathcal U_{\alpha 5}^2|)(\mathcal U_{\beta 4}^\ast \,\mathcal U_{\alpha 4} + \mathcal U_{\beta 5}^\ast \,\mathcal U_{\alpha 5})\,\tilde F_\text{Xbox}(x_{4,5})\nonumber\\
    &=& -2 (s_{\alpha 4}^2 + s_{\alpha 5}^2)\left(s_{\alpha 4}s_{\beta 4} e^{-i(\delta_{\alpha 4} - \delta_{\beta 4})} + s_{\alpha 5}s_{\beta 5} e^{-i(\delta_{\alpha 5} - \delta_{\beta 5})}\right) \tilde F_\text{Xbox}\nonumber\\
    &=& -2(s_{\alpha 4}^2 + s_{\alpha 5}^2)e^{-\frac{i}{2}(\Delta_4^{\alpha \beta} + \Delta_5^{\alpha\beta})}\left(s_{\alpha 4}s_{\beta 4}e^{-\frac{i}{2}(\Delta_4^{\alpha\beta} -\Delta_5^{\alpha\beta})} + s_{\alpha 5}s_{\beta 5}e^{\frac{i}{2}(\Delta_4^{\alpha\beta} -\Delta_5^{\alpha\beta})}\right)\tilde F_\text{Xbox}\nonumber\\
    &\approx& - 8 e^{-\frac{i}{2}(\Delta_4^{\alpha \beta} + \Delta_5^{\alpha\beta})} s_{\alpha 4}^3s_{\beta 4}\cos\left(\frac{\Delta_4^{\alpha\beta} - \Delta_5^{\alpha\beta}}{2}\right)\tilde F_\text{Xbox}\,.
\end{eqnarray}

\medskip
The box diagrams contributing to neutrinoless muon-electron conversion show a similar behaviour as that of photon- and $Z$-penguin diagrams with one neutrino in the loop.

\chapter{Additional information on lepton mixing with flavour and CP symmetries}
\minitoc
\section{Lepton mixing in the model-independent scenario}
\label{app:leptonmixing}
In this appendix, we revisit the four different types of lepton mixing patterns, Case~1) through Case 3~b.1), that have been identified in the study of~\cite{Hagedorn:2014wha}.
We mention for each case the generator $Z$, the CP transformation $X$ and the expressions for $\sin^2\theta_{ij}$, $J_{\mathrm{CP}}$, $I_1$ and $I_2$
and, where available, (approximate) formulae for the sines of the CP phases as well as (approximate) sum rules among the lepton mixing parameters.
We remind that the residual symmetry in the charged lepton sector, $G_\ell$,  is always chosen as the $Z_3$ group which corresponds to the diagonal 
subgroup of the $Z_3$ symmetry, contained in $G_f$ and arising from the generator $a$, and the $Z_3$ symmetry $Z_3^{(\mathrm{aux})}$. As discussed,
this leads to a diagonal charged lepton mass matrix and, consequently, to no contribution to lepton mixing from the charged lepton sector, see Eq.~(\ref{eq:Ul}).

For the lepton mixing parameters extracted from the PMNS, we follow the conventions of the PDG in the parametrisation of a unitary mixing matrix ($W$) in terms of 
the lepton mixing angles and the Dirac phase $\delta$~\cite{ParticleDataGroup:2020ssz}
\begin{equation}
\label{eq:defW}
W=
\begin{pmatrix}
c_{12} c_{13} & s_{12} c_{13} & s_{13} e^{- i \delta} \\
-s_{12} c_{23} - c_{12} s_{23} s_{13} e^{i \delta} & c_{12} c_{23} - s_{12} s_{23} s_{13} e^{i \delta} & s_{23} c_{13} \\
s_{12} s_{23} - c_{12} c_{23} s_{13} e^{i \delta} & -c_{12} s_{23} - s_{12} c_{23} s_{13} e^{i \delta} & c_{23} c_{13}
\end{pmatrix}
\end{equation}
with $s_{ij}=\sin \theta_{ij}$ and $c_{ij}=\cos \theta_{ij}$,
while we define the Majorana phases $\alpha$ and $\beta$ through
\begin{equation}
\label{eq:defP}
P=\left( \begin{array}{ccc}
 1  & 0 & 0\\
 0  &  e^{i \alpha/2}  & 0\\
 0 & 0  & e^{i (\beta/2 + \delta)}
\end{array}
\right)
\end{equation}
so that 
\begin{equation}
\label{eq:UPMNSWP}
U_{\mathrm{PMNS}} = \left( \begin{array}{ccc}
U_{e1} & U_{e2} & U_{e3}\\
U_{\mu1} & U_{\mu2} & U_{\mu3}\\
U_{\tau1} & U_{\tau2} & U_{\tau3}
\end{array}
\right)
= W \, P
\end{equation}
with $0 \leq \theta_{ij} \leq \pi/2$ and $0 \leq \alpha, \beta, \delta \leq 2 \, \pi$. We extract the sine squares of the lepton mixing angles as follows
\begin{equation}
\label{eq:sin2thij}
\sin^2 \theta_{13} = |U_{e3}|^2 \; , \;\; \sin^2 \theta_{12} = \frac{|U_{e2}|^2}{1-|U_{e3}|^2} \; , \;\; \sin^2 \theta_{23} = \frac{|U_{\mu3}|^2}{1-|U_{e3}|^2} \, .
\end{equation}
The CP phases are most conveniently extracted with the help of the CP invariants $J_{\mathrm{CP}}$~\cite{Jarlskog:1985ht}, $I_1$ and $I_2$~\cite{Jenkins:2007ip}
\begin{equation}
\label{eq:Jarlskog:1985ht}
J_{\mathrm{CP}} = \mathrm{Im} \left[ U_{e1} U_{e3}^\ast U_{\tau 1}^\ast U_{\tau 3}  \right]
= \frac 18 \, \sin 2 \theta_{12} \, \sin 2 \theta_{23} \, \sin 2 \theta_{13} \, c_{13} \, \sin \delta 
\end{equation}
and 
\begin{equation}
\label{eq:I1I2}
I_1 = \mathrm{Im} [U_{e2}^2 (U_{e1}^\ast)^2] = s^2_{12} \, c^2 _{12} \, c^4_{13} \, \sin \alpha \; , \;\;
I_2 = \mathrm{Im} [U_{e3}^2 (U_{e1}^\ast)^2] = s^2 _{13} \, c^2 _{12} \, c^2_{13} \, \sin \beta \; .
\end{equation}
From these, $\sin\delta$, $\sin\alpha$ and $\sin\beta$ can be computed.

\subsection{Case 1)}
\label{sec31}

For Case 1), the generator $Z$ of the residual $Z_2$ symmetry and the CP transformation $X$ are given by
\begin{equation}
\label{eq:ZXCase1}
Z=c^{n/2} \;\; \mbox{and} \;\; X=a\, b \, c^s \, d^{2 \, s} \, X_0 \;\; \mbox{with} \;\; 0 \leq s \leq n-1 \; .
\end{equation}
Note that the index $n$ has to be even. 
The matrix $\Omega (\mathbf{3})$ and the indices $f$ and $h$ of the rotation $R_{fh} (\theta)$, appearing in Eq.~(\ref{eq:Unumodind}),
are
\begin{equation}
\label{eq:OmegaR13Case1}
\Omega (\mathbf{3}) = e^{i \, \phi_s} \, U_{\mathrm{TB}} \, \left(
\begin{array}{ccc}
1 & 0 & 0\\
0 & e^{-3 \, i \, \phi_s} & 0\\
0 & 0 & -1
\end{array}
\right)
\;\; \mbox{and} \;\; 
R_{13} (\theta) = \left(
\begin{array}{ccc}
\cos \theta & 0 & \sin\theta\\
0 & 1 & 0\\
-\sin\theta & 0 & \cos\theta
\end{array}
\right)
\end{equation}
with
\begin{equation}
\label{eq:UTB}
U_{\mathrm{TB}} = \left(
\begin{array}{ccc}
\sqrt{\frac 23} & \frac{1}{\sqrt{3}} & 0\\
-\frac{1}{\sqrt{6}} & \frac{1}{\sqrt{3}} & \frac{1}{\sqrt{2}}\\
-\frac{1}{\sqrt{6}} & \frac{1}{\sqrt{3}} & -\frac{1}{\sqrt{2}}
\end{array}
\right) 
\end{equation}
and
\begin{equation}
\label{eq:phisdef}
\phi_s=\frac{\pi \, s}{n} \; .
\end{equation}
The matrix $K_\nu$, present in Eq.~(\ref{eq:Unumodind}), is set to the identity matrix for concreteness.

The main results of Case 1) are the following:\\ 
$a)$ the solar mixing angle is constrained by 
\begin{equation}
\label{eq:sin2th12limitCase1}
\sin^2\theta_{12} \gtrsim \frac 13 \; ,
\end{equation}
$b)$ none of the mixing angles depends on the parameters $n$ and $s$
\begin{equation}
\label{eq:sin2thetaijCase1}
\sin^2\theta_{13}= \frac 23 \, \sin^2\theta \; , \;\; \sin^2\theta_{12} = \frac{1}{2+\cos 2 \theta} \; , \;\; \sin^2 \theta_{23} = \frac 12 \, \Big( 1+ \frac{\sqrt{3} \, \sin 2 \theta}{2+\cos 2 \theta} \Big) \; ,
\end{equation}
$c)$ the size of the free angle $\theta$ is (mainly) fixed by the measured value of the reactor mixing angle $(\theta_{13})$ and $\theta$ takes two different values in the interval between $0$ and $\pi$,
\begin{equation}
\label{eq:thetaCase1}
\theta \approx 0.18 \;\; \mbox{and} \;\; \theta \approx 2.96 \; ,
\end{equation}
$d)$ two approximate sum rules among the mixing angles can be established
\begin{equation}
\label{eq:sumrulesCase1}
\sin^2 \theta_{12} \approx \frac 13 \, \left( 1 + \sin^2\theta_{13} \right) \;\; \mbox{and} \;\; \sin^2\theta_{23} \approx \frac 12 \, \left( 1 \pm \sqrt{2} \, \sin \theta_{13} \right) 
\end{equation}
with $\pm$ depending on $\theta \lessgtr \pi/2$,\\
$e)$ the Dirac phase $\delta$ and the Majorana phase $\beta$ are both trivial, $\sin\delta=0$ and $\sin\beta=0$,\\
$f)$ the Majorana phase $\alpha$ only depends on the parameter $s$ (the ratio $s/n$) and its sine reads
\begin{equation}
\label{eq:sinalphaCase1}
\sin \alpha = - \sin 6 \, \phi_s \; ,
\end{equation}
$g)$ for $s=0$ and $s=\frac n2$, CP is not violated.

\subsection{Case 2)}
\label{sec32}

The residual $Z_2$ symmetry in the sector of the neutral states is the same as in Case 1), while the CP transformation $X$ depends on two different parameters
\begin{equation}
\label{eq:ZXCase2}
Z=c^{n/2} \;\; \mbox{and} \;\; X=c^s \, d^t \, X_0\,, \;\; \mbox{with} \;\; 0 \leq s, t \leq n-1 \; .
\end{equation}
Like for Case 1), the index $n$ of $G_f$ has to be even.
A more convenient choice of parameters than $s$ and $t$ are $u$ and $v$, which are related to the former by
\begin{equation}
\label{eq:uvdeffromst}
u=2 \, s-t \;\; \mbox{and} \;\; v=3 \, t \; .
\end{equation}
The matrix $\Omega (\mathbf{3})$ and the indices $f$ and $h$ of the rotation matrix $R_{fh} (\theta)$ in Eq.~(\ref{eq:Unumodind}) read
\begin{equation}
\label{eq:OmegaR13Case2}
\Omega (\mathbf{3}) = e^{i \, \phi_v/6} \, U_{\mathrm{TB}} \, R_{13} \, \left( -\frac{\phi_u}{2} \right) \, \left(
\begin{array}{ccc}
1 & 0 & 0\\
0 & e^{-i \, \phi_v/2} & 0\\
0 & 0 & -i
\end{array}
\right)
\;\; \mbox{and} \;\; 
R_{13} (\theta) = \left(
\begin{array}{ccc}
\cos \theta & 0 & \sin\theta\\
0 & 1 & 0\\
-\sin\theta & 0 & \cos\theta
\end{array}
\right) 
\end{equation}
with
\begin{equation}
\label{eq:phiuphivdef}
\phi_u=\frac{\pi \, u}{n} \;\; \mbox{and} \;\; \phi_v=\frac{\pi \, v}{n} \; .
\end{equation}
For the definition of $U_{\mathrm{TB}}$ see Eq.~(\ref{eq:UTB}). 
Like for Case 1) we set $K_\nu$ to the identity matrix.

The main features of the mixing pattern of Case 2) are:\\
$a)$ the solar mixing angle has a lower limit identical to the one of Case 1), see Eq.~(\ref{eq:sin2th12limitCase1}),\\
$b)$ the lepton mixing angles depend on the parameters $u$ and $n$ as well as the free angle $\theta$
\begin{equation}
\label{eq:sin2thetaijCase2}
\!\!\!\sin^2 \theta_{13} = \frac 13 \, (1-\cos \phi_u \, \cos 2 \theta) \; , \;\; \sin^2 \theta_{12} = \frac{1}{2+\cos \phi_u \cos 2 \theta} \; , \;\;
\sin^2 \theta_{23} = \frac 12 \, \left( 1 + \frac{\sqrt{3} \, \sin \phi_u \cos 2\theta}{2+\cos\phi_u \cos 2 \theta} \right) \; ,
\end{equation}
$c)$ the size of $\cos \phi_u \, \cos 2 \theta$ (and thus of $u/n$ and $\theta$) is constrained by the measured value of the reactor mixing angle.
Taking into account symmetries of the formulae in $(u,\theta)$, discussed in~\cite{Hagedorn:2014wha}, it is sufficient to consider small values of $u/n$ and $\cos 2\theta \approx 1$.
The choice $u=0$ is associated with distinctive features (see point $g)$ below).\\
$d)$ the mixing angles fulfil two (approximate) sum rules: the one already found for Case 1), see first approximate equality in Eq.~(\ref{eq:sumrulesCase1}), and 
\begin{equation}
\label{eq:sumruleCase2}
6 \, \sin^2 \theta_{23} \, (1-\sin^2 \theta_{13}) = 3+\sqrt{3} \, \tan\phi_u - 3 \, \left( 1+\sqrt{3} \, \tan\phi_u \right) \, \sin^2 \theta_{13} \; ,
\end{equation}
$e)$ the Dirac phase $\delta$ and the Majorana phase $\beta$ depend on the parameters $u$ and $n$ as well as on the free angle $\theta$. Information on them
is most conveniently given in terms of the CP invariants $J_{\mathrm{CP}}$ and $I_2$ (see Section~\ref{sec2})
\begin{equation}
\label{eq:Jarlskog:1985htI2Case2}
J_{\mathrm{CP}} = - \frac{\sin 2\theta}{6 \, \sqrt{3}} \;\; \mbox{and} \;\; I_2= \frac 19 \, \sin 2 \,\phi_u \, \sin 2 \theta \; ,
\end{equation}
$f)$ the Majorana phase $\alpha$ depends, to very good accuracy, only on the parameters $v$ and $n$ (through the ratio $v/n$)
\begin{equation}
\label{eq:sinalphaCase2}
\sin\alpha \approx -\sin\phi_v \; ,
\end{equation}
$g)$ for the choice $u=0$, the atmospheric mixing angle and the Dirac phase are both maximal, $\sin^2 \theta_{23}=1/2$ and $|\sin\delta|=1$, while
the Majorana phase $\beta$ is trivial, $\sin\beta=0$, and the Majorana phase $\alpha$ exactly fulfils the approximate equality in Eq.~(\ref{eq:sinalphaCase2}).\\
$h)$ if $v=0$ is permitted, this leads to a trivial Majorana phase $\alpha$, $\sin\alpha=0$,\\
$i)$ furthermore, three symmetry transformations of the formulae of the lepton mixing parameters (in the parameters $u$ and $\theta$) have been found in~\cite{Hagedorn:2014wha}. Two of them
are independent, e.g.~
\begin{equation}
\label{eq:symmtrafosCase2}
\begin{array}{lll}
u \; \rightarrow \; u+n\, ,& \theta \; \rightarrow \; \frac{\pi}{2} - \theta \, :& \sin^2 \theta_{ij}, \, J_{\mathrm{CP}}, \, I_2 \; \mbox{are invariant and} \;\; I_1 \; \mbox{changes sign;}
\\ 
u \; \rightarrow \; 2\, n-u\, ,& \theta \; \rightarrow \; \pi - \theta \, :& \sin^2 \theta_{13}, \, \sin^2 \theta_{12}, \, I_1, \, I_2 \; \mbox{are invariant,} \;\; J_{\mathrm{CP}} \; \mbox{changes sign} 
\\
&&\mathrm{and} \; \sin^2\theta_{23} \; \rightarrow \;\; 1-\sin^2\theta_{23} \; .
\end{array}
\end{equation}

\subsection{Case 3 a) and Case 3 b.1)}
\label{sec33}

\noindent Case 3 a) and Case 3 b.1) are based on a different residual $Z_2$ symmetry in the sector of the neutral states than that of Case 1) and Case 2). This $Z_2$ symmetry depends on the parameter $m$.
Similarly, the CP transformation $X$ depends on the parameter $s$.
The explicit form of the generator $Z$ and of $X$ is
\begin{equation}
\label{eq:ZXCase3}
Z=b \, c^m \, d^m \;\; \mbox{and} \;\; X=b \, c^s \, d^{n-s} \, X_0 \;\; \mbox{and} \;\; 0 \leq m, \, s \leq n-1 \; .
\end{equation}
Since $Z$ contains the generator $b$, Case 3 a) and Case 3 b.1) can only be realised with the flavour symmetry $G_f=\Delta (6 \, n^2)$. 
The value of the parameter $m$ and, consequently, the choice of the residual $Z_2$ symmetry are strongly constrained by the measured values of the lepton mixing angles.

A possible form of $\Omega (\mathbf{3})$ and the matrix $R_{fh} (\theta)$ are given by 
\begin{equation}
\label{eq:OmegaR12Case3}
\!\!\!\Omega (\mathbf{3}) = e^{i \, \phi_s}\, \left( \begin{array}{ccc}
1 & 0 & 0\\
0 & \omega & 0\\
0 & 0 & \omega^2 
\end{array}
\right) \, U_{\mathrm{TB}} \, \left(
\begin{array}{ccc}
1 & 0 & 0\\
0 & e^{-3 \, i \, \phi_s} & 0\\
0 & 0 & -1
\end{array}
\right) \, R_{13} (\phi_m) \;\; \mbox{and} \;\; R_{12} (\theta) = \left(
\begin{array}{ccc}
\cos\theta & \sin\theta & 0\\
-\sin\theta & \cos\theta & 0\\
0 & 0 & 1
\end{array}
\right) 
\end{equation}
with
\begin{equation}
\label{eq:phimdef}
\phi_m=\frac{\pi \, m}{n} \; . 
\end{equation}
Again, the matrix $K_\nu$ is set to the identity matrix.

Two viable types of mixing patterns are found~\cite{Hagedorn:2014wha}: in Case 3 a) the parameter $m$ fixes the values of the atmospheric and reactor mixing 
angles, while in Case 3 b.1) the parameter $m$ is around $n/2$ in order to successfully accommodate the solar mixing angle. We first recapitulate the results for Case~3~a)
and then those for Case 3 b.1).

\subsubsection{Case 3 a)}
\label{sec331}

The relevant properties of the mixing pattern of Case 3 a) are:\\
$a)$ the value of $m/n$ is strongly constrained by the measured value of the reactor mixing angle. This value has to be either close to $0$ or to $1$.
Not only $\sin^2\theta_{13}$ is fixed by $m/n$, but also the value of the atmospheric mixing angle
\begin{equation}
\label{eq:sin2theta1323Case3a}
\sin^2\theta_{13} = \frac 23 \, \sin^2 \phi_m \;\; \mbox{and} \;\; \sin^2 \theta_{23} = \frac 12 \, \left( 1+ \frac{\sqrt{3} \, \sin 2 \, \phi_m}{2+\cos 2 \, \phi_m} \right) \; ,
\end{equation}
$b)$ due to this strong correlation a sum rule can be derived for $\sin^2 \theta_{13}$ and $\sin^2 \theta_{23}$
\begin{equation}
\label{eq:sumruleCase3a}
\sin^2\theta_{23} \approx \frac 12 \, \left( 1 \pm \sqrt{2} \, \sin\theta_{13} \right) 
\end{equation}
with $\pm$ for $m/n$ close to $0$ or $1$, respectively,\\
$c)$ the solar mixing angle depends on the parameter $s$ and on the free angle $\theta$ as well
\begin{equation}
\label{eq:sin2theta12Case3a}
\sin^2\theta_{12} = \frac{1+\cos 2\,\phi_m\,\sin^2\theta+\sqrt{2}\,\cos\phi_m\,\cos 3\,\phi_s \, \sin 2\theta}{2+\cos 2 \, \phi_m}.
\end{equation}
Note that the solar mixing angle can be accommodated to its measured best-fit value for most of the choices of the parameter $s$. In particular, $\sin^2\theta_{12}$
is no longer constrained to be larger than $1/3$, as for Case 1) and Case 2). For most choices of $s$ two values of the free angle $\theta$, one close to $0$ or $\pi$ and another depending on the parameter $s$, permit an acceptable fit
to the measured value of $\sin^2\theta_{12}$.\\
$d)$ the CP invariants $J_{\mathrm{CP}}$, $I_1$ and $I_2$ depend in general on all parameters, $n$, $m$, $s$ and $\theta$,
\begin{eqnarray}
\label{eq:Jarlskog:1985htI1I2Case3a}
&&J_{\mathrm{CP}} = - \frac{1}{6 \, \sqrt{6}} \, \sin 3 \, \phi_m \, \sin 3 \, \phi_s \, \sin 2 \theta \; ,
\\ \nonumber
&&I_1=-\frac 19 \, \cos \phi_m \, \sin 3 \, \phi_s \, \left( 4 \, \cos \phi_m \, \cos 3 \, \phi_s \, \cos 2 \theta + \sqrt{2} \, \cos 2 \, \phi_m \, \sin 2 \theta \right) \; , 
\\ \nonumber
&&I_2=\frac 49 \, \sin^2\phi_m \, \sin 3 \, \phi_s \, \sin\theta \, \left( \cos 3\, \phi_s \, \sin\theta -\sqrt{2} \, \cos\phi_m \, \cos\theta \right) \; , 
\end{eqnarray}
$e)$ approximate values can be found for the sines of the CP phases when the constraints on $m/n$ and $\theta$, arising from accommodating the lepton mixing angles, are used. These 
are
\begin{equation}
\label{eq:sinalphaCase3a}
|\sin\alpha| \approx |\sin 6 \, \phi_s| \; ,
\end{equation}
and
\begin{eqnarray}
\label{eq:sinbetasindeltaCase3a}
\mbox{for} \; \theta \approx 0, \, \pi&&\sin\delta \approx 0 \;\; \mbox{and} \;\; \sin\beta \approx 0 \; ,
\\ 
\nonumber
\mbox{for} \; \theta \not\approx 0, \, \pi&&|\sin\delta| \approx \left| \frac{3 \, \sin 6 \,\phi_s}{5+ 4 \, \cos 6 \, \phi_s}\right| \;\; \mbox{and} \;\; |\sin\beta| \approx 2 \, |\sin 6 \, \phi_s| \, \left| \frac{2+\cos 6 \, \phi_s}{5+4 \, \cos 6 \, \phi_s} \right| \; .
\end{eqnarray}
Note that the magnitude of $\sin\beta$ has an upper limit, $|\sin\beta| \lesssim 0.87$.\\
$f)$ if two values of the free angle $\theta$ permit an acceptable fit to the measured lepton mixing angles for a certain choice of $s$, the sine of the Majorana phase $\alpha$ for the two different values of $\theta$ has the same magnitude,
but opposite sign. If only one value of $\theta$ leads to a good fit to the experimental data, the Majorana phase $\alpha$ is trivial, $\sin\alpha=0$,\\
$g)$ for $s=0$, all CP phases are trivial, i.e.~$\sin\alpha=\sin\beta=\sin\delta=0$.\\
$h)$ for the choice $s=\frac n2$, the free angle $\theta$, that leads to the best accommodation of the measured values of the lepton mixing angles, is $\theta=0$. Consequently, 
the solar mixing angle is bounded from below, i.e.~$\sin^2\theta_{12} \gtrsim 1/3$, and all CP phases are trivial,\\ 
$i)$ like for Case 2), three symmetry transformations of the formulae of the lepton mixing parameters in the parameters $m$, $s$ and the free angle $\theta$ have been found in~\cite{Hagedorn:2014wha}. Two of them
are independent, e.g.~
\begin{equation}
\label{eq:symmtrafosCase3}
\begin{array}{lll}
s\; \rightarrow \; n-s \; ,&\theta \; \rightarrow \; \pi-\theta \; :&\sin^2\theta_{ij} \; \mbox{are invariant and} \;  J_{\mathrm{CP}}, \, I_1, \, I_2 \; \mbox{change sign;} 
\\ 
m \; \rightarrow \; n-m \; ,&\theta \; \rightarrow \; \pi-\theta \; :&\sin^2\theta_{13}, \, \sin^2\theta_{12}, \, I_1, \, I_2 \; \mbox{are invariant,} \; J_{\mathrm{CP}} \; \mbox{changes sign} 
\\ 
&&\mbox{and} \; \sin^2\theta_{23} \; \rightarrow 1-\sin^2\theta_{23} \; .
\end{array}
\end{equation}

\subsubsection{Case 3 b.1)}
\label{sec34}

The lepton mixing pattern of Case 3 b.1) arises from the matrices $\Omega (\mathbf{3})$ and $R_{12} (\theta)$ in Eq.~(\ref{eq:OmegaR12Case3}), if these are multiplied from the 
right with the cyclic permutation matrix $P_{\mathrm{cyc}}$
\begin{equation}
\label{eq:PcycCase3b1}
P_{\mathrm{cyc}} = \left( \begin{array}{ccc}
0 & 1 & 0\\
0 & 0 & 1\\
1 & 0 & 0
\end{array}
\right) \; \mbox{, i.e.~} \;\; U_{\mathrm{PMNS}}=U_\nu=\Omega (\mathbf{3}) \, R_{12} (\theta) \, P_{\mathrm{cyc}} \; .
\end{equation}
This cyclic permutation corresponds to a re-ordering of the columns of the PMNS mixing matrix. 
The properties of the lepton mixing pattern of Case 3 b.1) can be summarised as follows:\\
$a)$ all lepton mixing parameters depend on $n$, $m$, $s$ and the free angle $\theta$,
\begin{eqnarray}
\label{eq:sin2thetaijCase3b1}
\sin^2 \theta_{13} &=& \frac 13 \, \left( 1+ \cos 2 \, \phi_m \, \sin^2\theta +\sqrt{2} \, \cos\phi_m \, \cos 3 \, \phi_s \, \sin 2 \theta \right) \; ,
\\ \nonumber
\sin^2 \theta_{23} &=& \frac 12 \, \left( 1+ \frac{2 \, \sqrt{3}\, \sin\phi_m\,\sin\theta \, (\sqrt{2}\,\cos 3 \, \phi_s \, \cos\theta-\cos\phi_m \, \sin\theta)}{2-\cos 2 \, \phi_m \,\sin^2\theta-\sqrt{2}\,\cos\phi_m\,\cos3 \, \phi_s \, \sin2 \theta} \right) \; ,
\\ \nonumber
\sin^2\theta_{12} &=& 1 - \frac{2 \, \sin^2 \phi_m}{2-\cos 2 \, \phi_m \,\sin^2\theta-\sqrt{2}\,\cos\phi_m \, \cos 3 \, \phi_s \, \sin 2\theta}
\end{eqnarray}
and
\begin{eqnarray}
\label{eq:Jarlskog:1985htI1I2Case3b1}
J_{\mathrm{CP}} &=& -\frac{1}{6 \, \sqrt{6}} \, \sin 3 \, \phi_m \, \sin 3 \, \phi_s \, \sin 2 \theta \; ,
\\ \nonumber
I_1 &=& -\frac 49 \, \sin^2 \phi_m \, \sin 3 \, \phi_s \, \sin\theta \, \left( \cos 3 \, \phi_s \, \sin\theta-\sqrt{2} \, \cos\phi_m \, \cos\theta \right) \; ,
\\ \nonumber
I_2 &=& -\frac 49 \, \sin^2 \phi_m \, \sin 3 \, \phi_s \, \cos\theta \, \left( \cos 3 \, \phi_s \, \cos\theta+\sqrt{2} \, \cos \phi_m \, \sin\theta \right) \; ,
\end{eqnarray}
$b)$ the parameter $m$ is strongly constrained by the measured value of $\sin^2\theta_{12}$, i.e.~$m\approx \frac n2$,\\
$c)$ for $m=\frac n2$, two approximate sum rules among the lepton mixing angles are found
\begin{equation}
\label{eq:sumrulesCase3b1}
\sin^2\theta_{12} \approx \frac 13 \, \left( 1-2 \, \sin^2\theta_{13} \right) \;\; \mbox{and} \;\; \sin^2\theta_{23} \approx \frac 12 \left( 1+\sqrt{\frac23} \, \frac{\cos 3 \, \phi_s \, \sin2 \theta_0}{1-\sin^2 \theta_{13}} \right)
\end{equation}
with $\theta_0 \approx 1.31$ or $\theta_0 \approx 1.83$, constrained by the measured value of the reactor mixing angle,\\
$d)$ for $m= \frac n2$ and $s= \frac n2$, the atmospheric mixing angle is maximal, $\sin^2\theta_{23}= \frac 12$,\\
$e)$ for $m= \frac n2$, the Majorana phases only depend on the parameter $s$ (the ratio $s/n$), and have the same magnitude, 
\begin{equation}
\label{eq:sinaplhabetaCase3b1}
\sin\alpha=\sin\beta=-\sin 6 \, \phi_s
\end{equation}
and the Dirac phase fulfils the approximate relation 
\begin{equation}
\label{eq:sindeltaCase3b1}
\sin\delta \approx \pm \sin 3 \,\phi_s \;\; \mbox{with} \; \pm \; \mbox{referring to} \;\; \theta \lessgtr \pi/2\; .
\end{equation}
Taking into account the constraints on the free angle $\theta$ and the parameter $s$, arising from the experimental data on lepton mixing angles, the magnitude of the sine of the Dirac phase is bounded from below, $|\sin\delta| \gtrsim 0.71$,\\ 
$f)$ for $m= \frac n2$ and $s= \frac n2$, the Dirac phase is maximal, $|\sin\delta|=1$, while both Majorana phases are trivial, $\sin\alpha=0$ and $\sin\beta=0$,\\ 
$g)$ for $s=0$, CP is not violated,\\ 
$h)$ for Case 3 b.1) the same symmetry transformations hold as for Case 3 a), see point $i)$ in Section~\ref{sec331}, Eq.~(\ref{eq:symmtrafosCase3}).

\subsection{Impact of the ISS embedding on lepton mixing}
\label{sec:ISSimpact}
\vspace{0.11in}
\noindent Furthermore, we can estimate the deviations in the (approximate) sum rules induced by effects of non-unitarity of the lepton mixing matrix, such as the ones in Eq.~(\ref{eq:sumrulesCase1}). 
These are discussed in turn for each of the cases, Case~1) through Case 3 b.1).

\subsection*{Case 1)}

Two approximate sum rules have been found for Case 1), see Eq.~(\ref{eq:sumrulesCase1}). The effects of non-unitarity of the lepton mixing matrix on these are expected to be as follows: for the first sum rule,
relating the solar and the reactor mixing angle, using the best-fit value $|U_{e 3}|^2\approx 0.022$~\cite{Esteban:2020cvm}, we have
\begin{equation}
\label{eq:sumrule1Case1nonuni}
\Delta \Sigma_1 \approx -2 \, \eta_0 \, \left( \frac{1+|U_{e3}|^4}{1-|U_{e3}|^4} \right) \approx -2 \, \eta_0
\end{equation}
with $\Delta\Sigma_1$ corresponding to the relative deviation of the non-unitary result from the unitary one and defined as

\begin{equation}
\Delta\Sigma_1 = \frac{\left( \frac{3 \, \sin^2 \theta_{12}}{1 + \sin^2\theta_{13}} \right)_{\mathrm{ISS}} - \left( \frac{3 \, \sin^2 \theta_{12}}{1 + \sin^2\theta_{13}} \right)_{\mathrm{MIS}}}{\left( \frac{3 \, \sin^2 \theta_{12}}{1 + \sin^2\theta_{13}} \right)_{\mathrm{MIS}}} \;\; \mbox{with} \;\; \left( \frac{3 \, \sin^2 \theta_{12}}{1 + \sin^2\theta_{13}} \right)_{\mathrm{MIS}} \approx 1 \;\; \mbox{from Eq.~(\ref{eq:sumrulesCase1}),}
\end{equation}
while the deviation for the second sum rule, the one involving the atmospheric and the reactor mixing angle, is of the form
\begin{equation}
\label{eq:sumrule2Case1nonuni}
\Delta\Sigma_2 \approx - \sqrt{2} \, \eta_0 \, \left( \frac{\sqrt{2} \pm |U_{e3}|\, (1+|U_{e3}|^2)}{(1\pm \sqrt{2} \, |U_{e3}|)\, (1-|U_{e3}|^2)} \right) \approx -1.87 \, (-2.31) \, \eta_0
\end{equation}
for $+ (-)$.
$\Delta\Sigma_2$ is defined analogously to $\Delta\Sigma_1$ with the help of the second approximate sum rule in Eq.~(\ref{eq:sumrulesCase1}).
The different signs refer to the different signs in the sum rule.
We note that none of the relative deviations, $\Delta\Sigma_1$ and $\Delta\Sigma_2$, depends on the parameters $n$, $s$ 
or on the precise value of the free angle $\theta_S$, up to the sign in $\Delta\Sigma_2$. 
For $y_0 \sim 1$ and $M_0 \sim 1000$~GeV we expect these to be 
\begin{equation}
\label{eq:estimateDSigma12y01}
\Delta\Sigma_1 \approx -0.03 \;\; \mbox{and} \;\; \Delta\Sigma_2 \approx -0.028 \, (-0.035) 
\end{equation}
for $+ (-)$ from the expression for $\Delta\Sigma_2$ in Eq.~(\ref{eq:sumrule2Case1nonuni}).

\subsection*{Case 2)}

For Case 2) we also have two approximate sum rules: one which coincides with the first sum rule of Case 1) and another one, relating the atmospheric and the reactor mixing angles, shown in Eq.~(\ref{eq:sumruleCase2}).
The effects of non-unitarity (of the PMNS mixing matrix) on the latter one are estimated to be of the order of 
\begin{equation}
\label{eq:sumruleCase2nonuni}
\Delta\Sigma_3 \approx - 2 \, \eta_0 \, \left( \frac{\sqrt{3}+\tan\phi_u}{\sqrt{3}\, (1-|U_{e3}|^2) + (1-3\, |U_{e3}|^2) \, \tan\phi_u} \right) \; ,
\end{equation}
where $\Delta\Sigma_3$ is defined in the analogous way as $\Delta\Sigma_1$. The form of $\Delta\Sigma_3$ can be simplified by remembering that $u/n$ is required to be small and thus we expand in $\phi_u=\frac{\pi \, u}{n}$ up to the linear order.
At the same time, we use the best-fit value for $|U_{e3}|^2 \approx 0.022$~\cite{Esteban:2020cvm} so that we have
\begin{equation}
\label{eq:sumruleCase2nonunisimp}
\Delta\Sigma_3 \approx  -2.05 \, \eta_0 \, (1+ 0.026 \, \phi_u) \; .
\end{equation}
This shows that there is only a very mild dependence of $\Delta\Sigma_3$ on $\phi_u = \frac{\pi\, u}{n}$. 
Furthermore, there is no explicit dependence of $\Delta\Sigma_3$ on the parameter $v$ and the free angle $\theta_S$.
Numerically we find for $y_0 \sim 1$ and $M_0 \sim 1000$~GeV that 
\begin{equation}
\label{eq:estimateDSigma3y01}
\Delta\Sigma_3 \approx -0.031 \; ,
\end{equation}
which is of a size very similar to the other relative deviations.

\subsection*{Case 3 a) and Case 3 b.1)}

For Case 3 a) the approximate sum rule, found in Eq.~(\ref{eq:sumruleCase3a}), is actually identical to the second sum rule for Case 1), see Eq.~(\ref{eq:sumrulesCase1}), taking into account the different signs in both of them.
We thus expect very similar results also for Case 3 a).

For Case 3 b.1) two approximate sum rules are derived for $m=\frac n2$, see Eq.~(\ref{eq:sumrulesCase3b1}). For the first of these two, we find as relative deviation of the non-unitary result from the unitary one
\begin{equation}
\label{eq:sumrule1Case3b1nonuni}
\Delta\Sigma_4 \approx -2 \, \eta_0 \, \left( \frac{1-2 \, |U_{e3}|^4}{1-3 \, |U_{e3}|^2+2 \, |U_{e3}|^4} \right) \approx -2.14 \, \eta_0 \; ,
\end{equation}
while for the second one we have
\begin{equation}
\label{eq:sumrule2Case3b1nonuni}
\Delta\Sigma_5 \approx - 2 \, \eta_0 \, \left( \frac{\sqrt{3}+ \sqrt{2} \, \cos 3 \, \phi_s \, \sin 2 \, \theta_0}{\sqrt{3} \, (1-|U_{e3}|^2) + \sqrt{2} \, \cos 3 \, \phi_s \, \sin 2 \, \theta_0} \right) \approx -2.05 \, \eta_0 \mp 0.019 \, \eta_0 \, \cos 3 \, \phi_s \; ,
\end{equation}
where we have again used $|U_{e3}|^2 \approx 0.022$~\cite{Esteban:2020cvm} and $\theta_0 \approx \frac{\pi}{2} \pm \epsilon$ with $\epsilon \approx 0.26$, cf. text below Eq.~(\ref{eq:sumrulesCase3b1}). We thus see that the relative deviation $\Delta\Sigma_5$ only weakly depends on the value
of the parameter $s$, related to the chosen CP transformation $X$. 
Furthermore, we infer that neither $\Delta\Sigma_4$ nor $\Delta\Sigma_5$ depends strongly on the parameter $n$ or the free angle $\theta_S$.
Using $y_0 \sim 1$ and $M_0 \sim 1000$~GeV, we have for the two relative deviations 
\begin{equation}
\label{eq:estimateDSigma45y01}
\Delta\Sigma_4 \approx -0.032 \;\; \mbox{and} \;\; \Delta\Sigma_5 \approx -0.031 \; .
\end{equation}

\section{Numerical analysis of lepton mixing in the symmetry endowed ISS}
\label{app:numericalISS}
In this Appendix, we study numerically the impact of the heavy sterile states of the $(3,3)$ ISS framework on the results for the lepton mixing parameters, and if available, on the approximate sum rules among these.
We do so for each of the different cases, Case 1) through Case 3 b.1), for some viable choices of the group theory parameters, e.g.~the index $n$ of the flavour symmetry $G_f$. 
We also compare these findings to the analytical estimates, presented in Section~\ref{sec4}.

Before detailing results for the different cases in sections~\ref{sec51}-\ref{sec54}, we present the viable ranges the Dirac neutrino Yukawa coupling $y_0$, the mass scale $M_0$ of the heavy sterile states, as well as on the parameters $\mu_i$, due to the bounds on the unitarity of the PMNS mixing matrix.

In view of the discussion in Section~\ref{sec:unitarityISS}, 
we will in general
assume that the mass scale $M_0$ varies in the range 
\begin{equation}
\label{eq:M0range}
500 \, \mathrm{GeV} \lesssim M_0 \lesssim 5000 \, \mathrm{GeV} \; .
\end{equation}
Although mostly lying beyond future collider reach~\cite{Antusch:2016ejd}, 
the chosen range for $M_0$ (and thus for $M_{NS}$ and the heavy mass 
spectrum) is motivated by its 
phenomenological interest, as it is in general associated 
with extensive observational imprints, being thus indirectly
accessible in numerous dedicated facilities~\cite{Abada:2014vea,Arganda:2014dta,Abada:2014kba,Abada:2014cca,Arganda:2015naa,Abada:2015oba,DeRomeri:2016gum}.  

Concerning the Yukawa coupling $y_0$, and following the results
displayed in Fig.~\ref{fig:Constraints_eta},
we will in general illustrate our results for 
two different values of the Yukawa coupling $y_0$,
\begin{equation} 
\label{eq:y0choice}
y_0= 0.5 \;\; \mbox{and} \;\; y_0=0.1 \; .
\end{equation}
Nevertheless, we will exceptionally consider larger values of 
the Yukawa coupling $y_0=1$, in order to better illustrate 
the effects of the deviations from unitarity of the PMNS mixing matrix. 
These cases will be clearly 
identified in the discussion; 
unless otherwise stated,
disfavoured regimes associated with bounds on $\eta_{\alpha\beta}$ will be indicated by a grey-shaded
area in the corresponding plots.

Finally, we consider the free parameters $\mu_i$.
As can be seen from Eq.~(\ref{eq:m123LOopt1}), in the case of option 1,
$\mu_i$ are directly proportional to the light neutrino masses
$m_i$. Thus, they are experimentally constrained 
by the measured mass squared differences and by the bound on the sum
of the light neutrino masses coming from cosmology. The 
latest experimental data are collected in
Table~\ref{tab:nufit} in Section~\ref{sec:osci}. 
We notice that in our numerical study, the two mass squared differences are always adjusted to their experimental best-fit value~\cite{Esteban:2020cvm}.

A few comments are still in order concerning the light neutrino mass 
spectrum - the value of the lightest neutrino mass $m_0$, and the ordering
(NO vs. IO).  
Regarding $m_0$,
we have verified that the results 
for the lepton mixing parameters are always independent of its
choice. 
Throughout this section, we have thus fixed its value to
\begin{equation}
\label{eq:m0fixed}
m_0 = 0.001 \, \mathrm{eV}.
\end{equation}
Furthermore, we note that we have performed the numerical analysis for
both NO and IO, 
and no (numerically significant) differences were found, neither for the relative deviations of the lepton mixing parameters, nor for the (approximate) sum rules.
Accordingly, all the
results of this section will be only illustrated for the case of a NO 
light neutrino mass spectrum. 
However, notice that upon discussion of the prospects of the current framework 
concerning $0\nu\beta\beta$ decay in Section~\ref{sec6}, 
we will consider both orderings of the mass spectrum, and also vary $m_0$.

\bigskip
Leading to the fits presented in the following subsections, 
we only consider experimental constraints on the lepton mixing angles 
and the two mass squared differences, but not on the CP phase
$\delta$, since the latter 
is only very mildly experimentally constrained (a summary of the
relevant neutrino oscillation data is given in Section~\ref{sec:osci}). 

Predicting the CP phases for the different choices of the CP symmetry
requires identifying the values of the free angle $\theta_S$
which lead to a set of lepton mixing parameters in agreement with
experimental data. 
The free angle $\theta_S$ is fit by maximising the joint likelihood of the
predictions for $\sin^2\theta_{ij}$. 
We refer to the combination of
experimental data (latest update on global fits) provided by NuFIT
5.0~\cite{Esteban:2020cvm} (see Section~\ref{sec:osci}). 
In order to fit $\theta_S$ with high accuracy, and especially to take
into account the ``double well'' structure in $\sin^2\theta_{23}$, we
interpolate the numerical $\Delta\chi^2$ values (available
at~\cite{Esteban:2020cvm}), and linearly extrapolate beyond the provided ranges to
ensure a smooth behaviour for arbitrary input values. 
The interpolated $\chi^2$ functions are then transformed into 
probability distributions so that a global
joint likelihood of all relevant parameters ($\sin^2\theta_{ij}$ and,
in the case of the (3,3) ISS framework, also $\Delta m_{ij}^2$) can be
constructed. 
To ensure that the cosmological bound on the sum of light neutrino
masses is respected, a half-normal distribution (as a gaussian upper
limit) is further included.
We first fit predictions for $U_S$ and therefore $\theta_S$
on the data for $\sin^2\theta_{ij}$,  
from which we proceed to consider the model-dependent, i.e. (3,3) ISS
framework. 
This is done by maximising the joint likelihood function using the
\texttt{migrad} algorithm of the \texttt{iminuit} library~\cite{iminuit}. 
Local maximum likelihood estimators lying outside of the global
$3\,\sigma$ region around the (experimental) best-fit point are
rejected. 

Due to the peculiar structure and almost degenerate heavy states in
the ISS mass matrix, a large numerical precision ($\sim 100$ digits)
is needed for a reliable matrix diagonalisation. This is achieved
using the \texttt{mpmath} python library~\cite{mpmath} and algorithms within.  
To study the effects of the heavy sterile states on the
predictions for lepton mixing parameters, we use
``\textit{effective}'' mixing angles (and phases) which we define as in Eqs.~(\ref{eq:sin2thij},\ref{eq:Jarlskog:1985ht},\ref{eq:I1I2}), but with $U_\text{PMNS}$ replaced by $\tilde{U}_\nu$. 
The free angle $\theta_S$ then needs to be re-fitted, using the
results obtained within the model-independent approach as starting
values for the fit, thus allowing to study deviations from those
predictions. 

Keeping the lightest neutrino mass $m_0$ fixed - and thus the
lightest Majorana mass - 
($\mu_1$ or $\mu_3$ depending on the ordering of the light neutrino mass spectrum),  
the remaining two Majorana masses $\mu_i$ are treated as free
parameters to be determined by a fit to $\sin^2\theta_{ij}$ and
$\Delta m_{ij}^2$ data. 
The starting values for $\mu_i$ are determined by
inverting the leading order expression given in Eq.~\eqref{eq:mnuLO}
and a modified Casas-Ibarra parametrisation~\cite{Casas:2001sr}  
\begin{equation}
    \mu_S \,\simeq \,M_{NS}^T \,m_D^{-1} \,U^\ast_S \, m^\text{diag}_\nu
    \,U^\dagger_S \,{m_D^T}^{-1} \,M_{NS}\,, 
\end{equation}
where $U_S$, the matrix which diagonalises $\mu_S$ and at leading order also the light neutrino mass matrix, is determined by the flavour symmetry $G_f$ and CP and the residual symmetry $G_\nu$.

\subsection{Case 1)}
\label{sec51}

In order to scrutinise the effects of the $(3,3)$ ISS framework and its heavy states on the lepton mixing parameters,
we choose a value of the index $n$ that allows studying several different values of the parameter $s$ (and thus CP transformations $X$) for Case 1). In this way, 
the behaviour of the Majorana phase $\alpha$, see Eq.~(\ref{eq:sinalphaCase1}), can be studied systematically. Concretely, in the following we use
\begin{equation}
\label{eq:numchoiceparaCase1}
n=26 \;\; \mbox{and} \;\; 0 \leq s \leq 25 \,.
\end{equation}

\begin{figure}[t!]
\begin{center}
\parbox{3in}{\includegraphics*[scale=0.5]{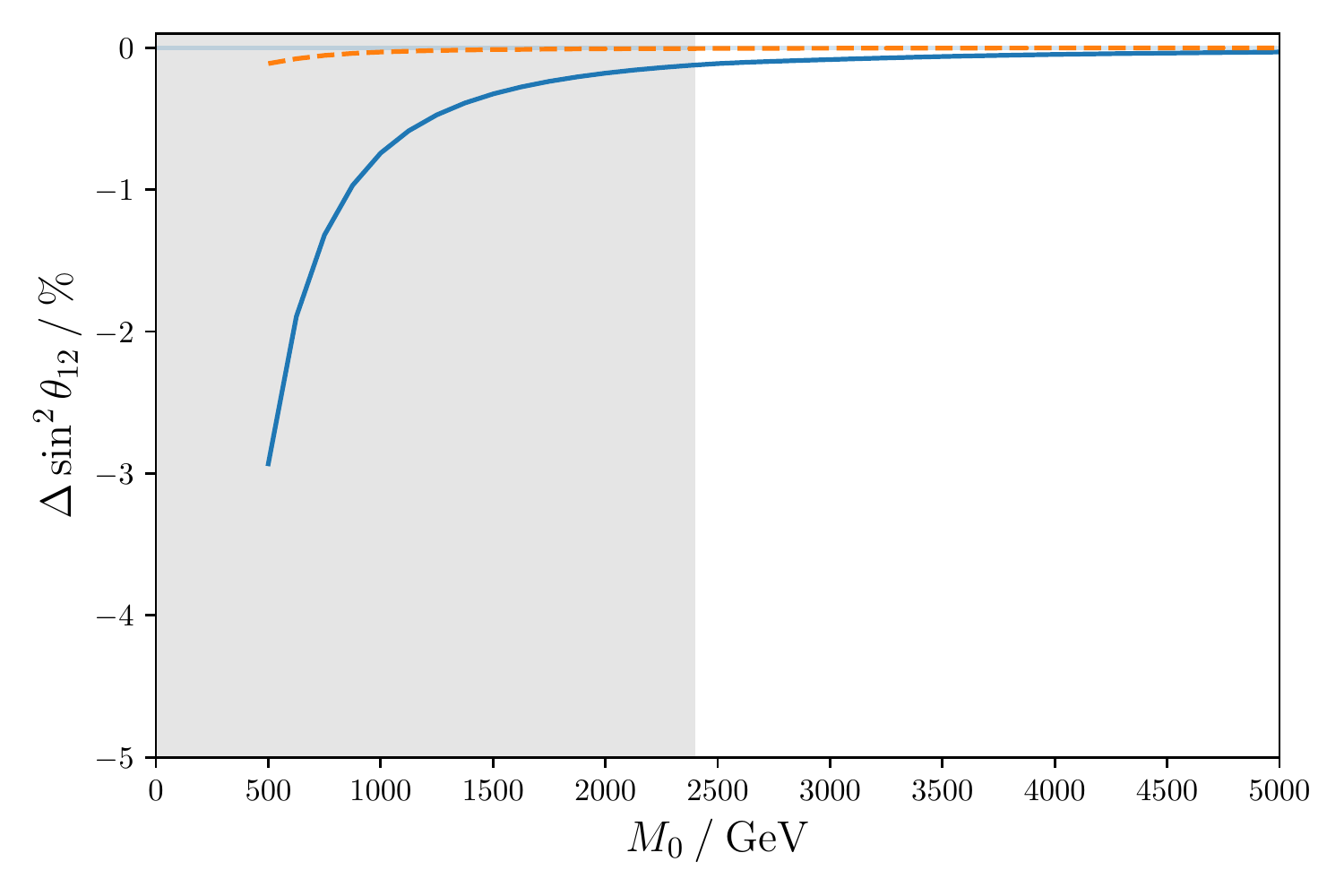}}
\hspace{0.2in}
\parbox{3in}{\includegraphics*[scale=0.5]{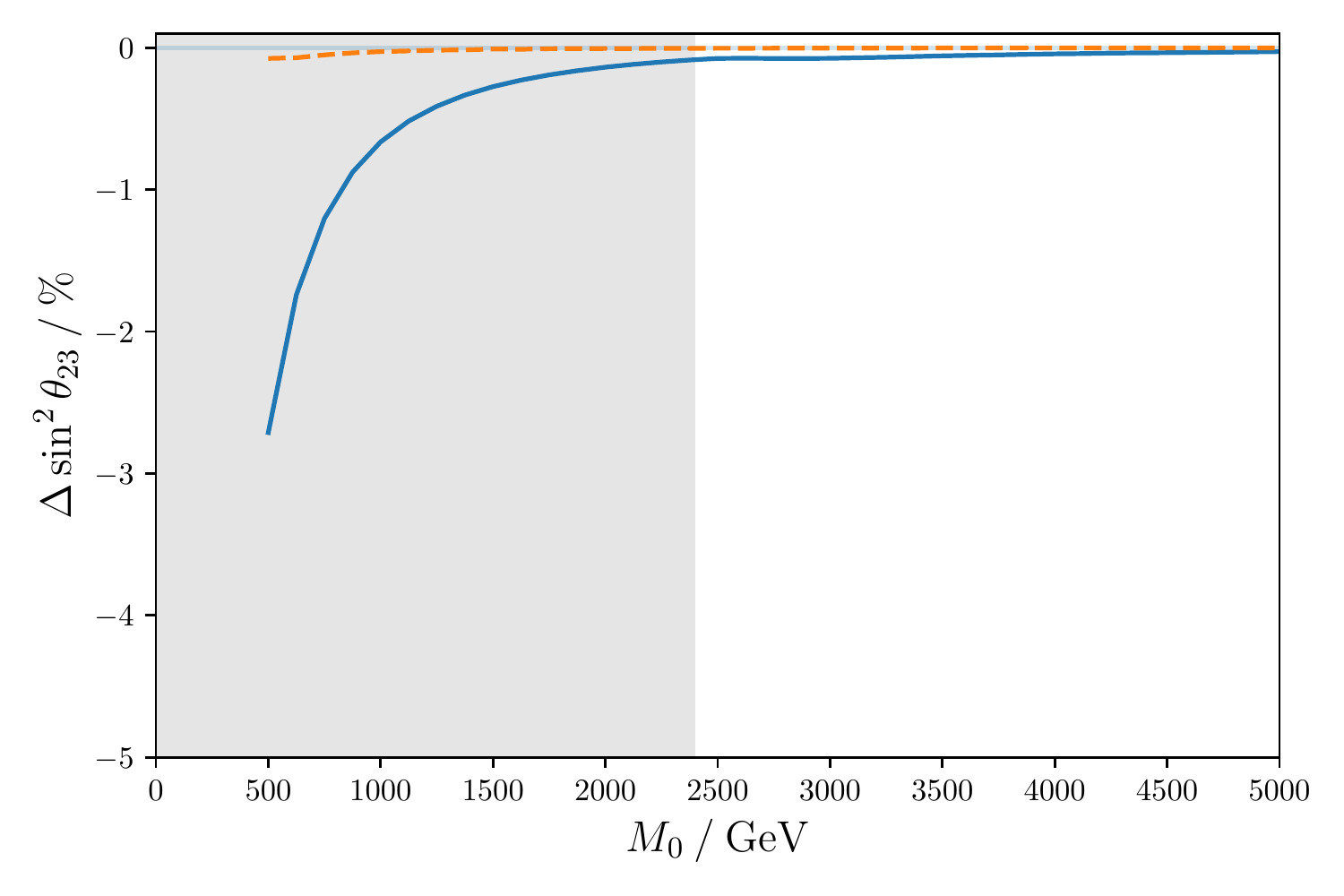}}
\end{center}
 \caption{{\small \textbf{Case 1)} 
 Relative deviation of $\sin^2\theta_{12}$ (left) and
$\sin^2\theta_{23}$ (right) as obtained for option 1 of the $(3,3)$ ISS
from the corresponding values derived in the 
model-independent scenario, as a function of 
$M_0$ (in GeV).  
For concreteness, we have fixed $s=1$ and $n=26$.
The curves are associated with distinct values of the Yukawa coupling $y_0$:
the orange (dashed) curve corresponds to $y_0=0.1$ and the blue
(solid) one to $y_0=0.5$. A grey-shaded area denotes regimes
disfavoured due to conflict with experimental bounds (see detailed discussion in Section~\ref{sec:unitarityISS}). 
Figures from~\cite{Hagedorn:2021ldq}.
\label{fig:Case1_s2thij}}}
\end{figure}

Based on the results obtained in the model-independent scenario (see Section~\ref{sec31}), and the analytical estimates of the effects due to the heavy sterile states
of the $(3,3)$ ISS framework carried in Section~\ref{sec4}, only the CP phase $\alpha$ is expected to show a dependence on the parameters $n$ and $s$ (through the ratio $s/n$). 
This is confirmed by our numerical analysis. 
Without loss of generality we thus set  $s=1$ to study the relative deviations $\Delta\sin^2\theta_{12}$ and $\Delta\sin^2\theta_{23}$. 
These are shown in Fig.~\ref{fig:Case1_s2thij}, respectively in the left and right plots, as a function of $M_0$, which determines the scale of the heavy mass spectrum. 
We notice that their sign and size is consistent with the estimate found in 
Eq.~(\ref{eq:estimateDsy01}).\footnote{Notice that following Eq.~(\ref{eq:etaopt1}), $y_0 \sim 1$ and $M_0 \sim 1000$~GeV lead to the same result for the quantity $\eta_0$ as $y_0 \sim 0.5$ and $M_0 \sim 500$~GeV.}
The relative deviation of the reactor mixing angle, $\Delta\sin^2\theta_{13}$, is not shown and does not fulfil the expectations from the analytical estimate, since it turns out to be positive and always below
$0.5\%$ for values of $y_0 \lesssim 0.5$ and $M_0 \gtrsim 500$~GeV. 
This is a consequence of having $\theta_{13}$ driving the fit to determine $\theta_S$, due to its associated experimental precision, see Table~\ref{tab:nufit} in Section~\ref{sec:osci}.
Consequently, we find for $\theta_S$ values around $0.19$, which are slightly larger than those obtained in the model-independent scenario, see Eq.~(\ref{eq:thetaCase1}).
We note that in the plots shown here, we always have $\theta_S < \pi/2$, since this leads to a much better agreement with the experimentally preferred value of the atmospheric mixing angle: $\sin^2 \theta_{23} \approx 0.604$ to be compared to the experimental values 
$\sin^2 \theta_{23} = 0.570^{+0.018} _{-0.024}$ for light neutrinos with NO and $\sin^2 \theta_{23} = 0.575^{+0.017} _{-0.021}$ for light neutrinos with IO~\cite{Esteban:2020cvm}.
\begin{figure}[t!]
\begin{center}
\parbox{3in}{\includegraphics*[scale=0.5]{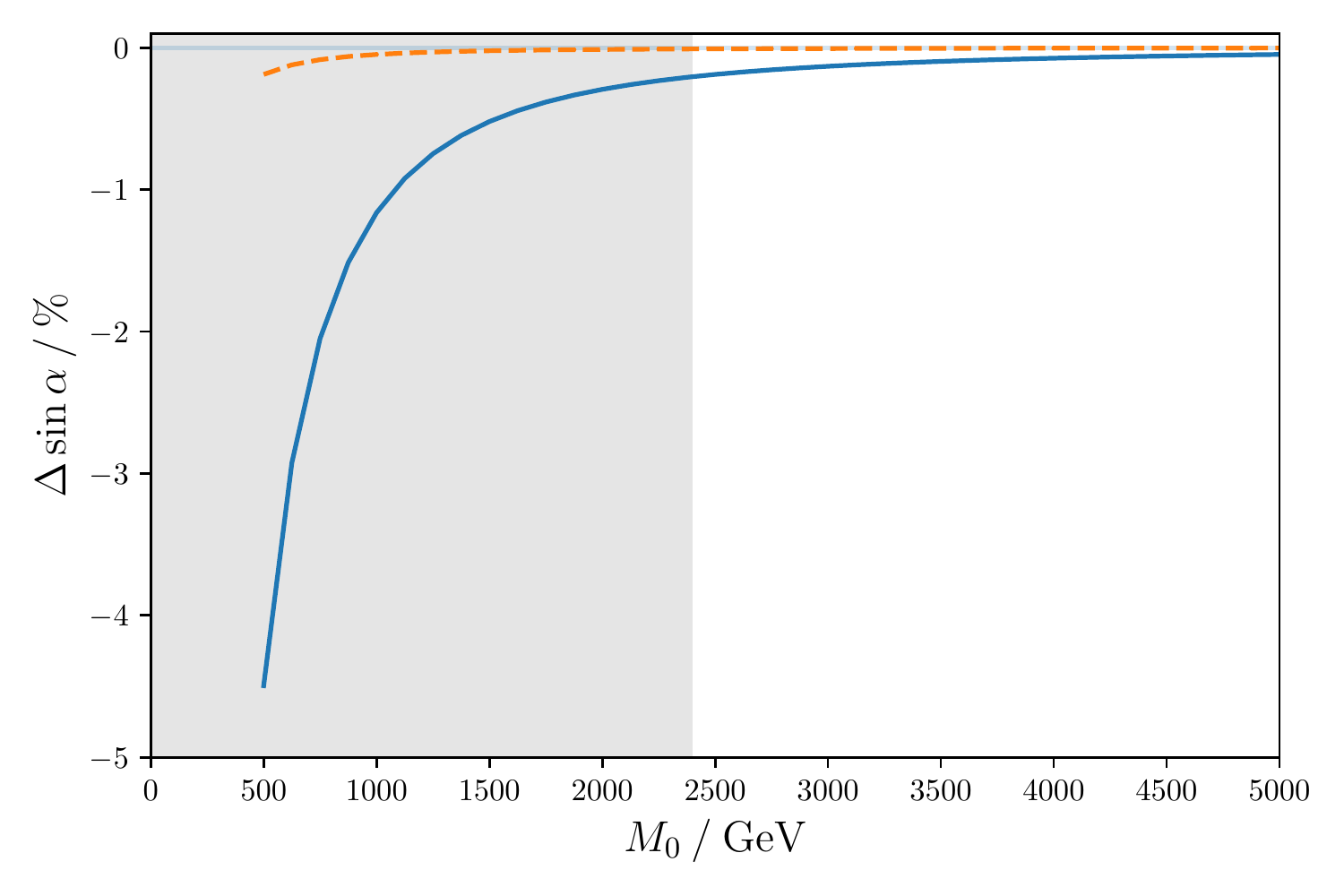}}
\hspace{0.2in}
\parbox{3in}{\includegraphics*[scale=0.5]{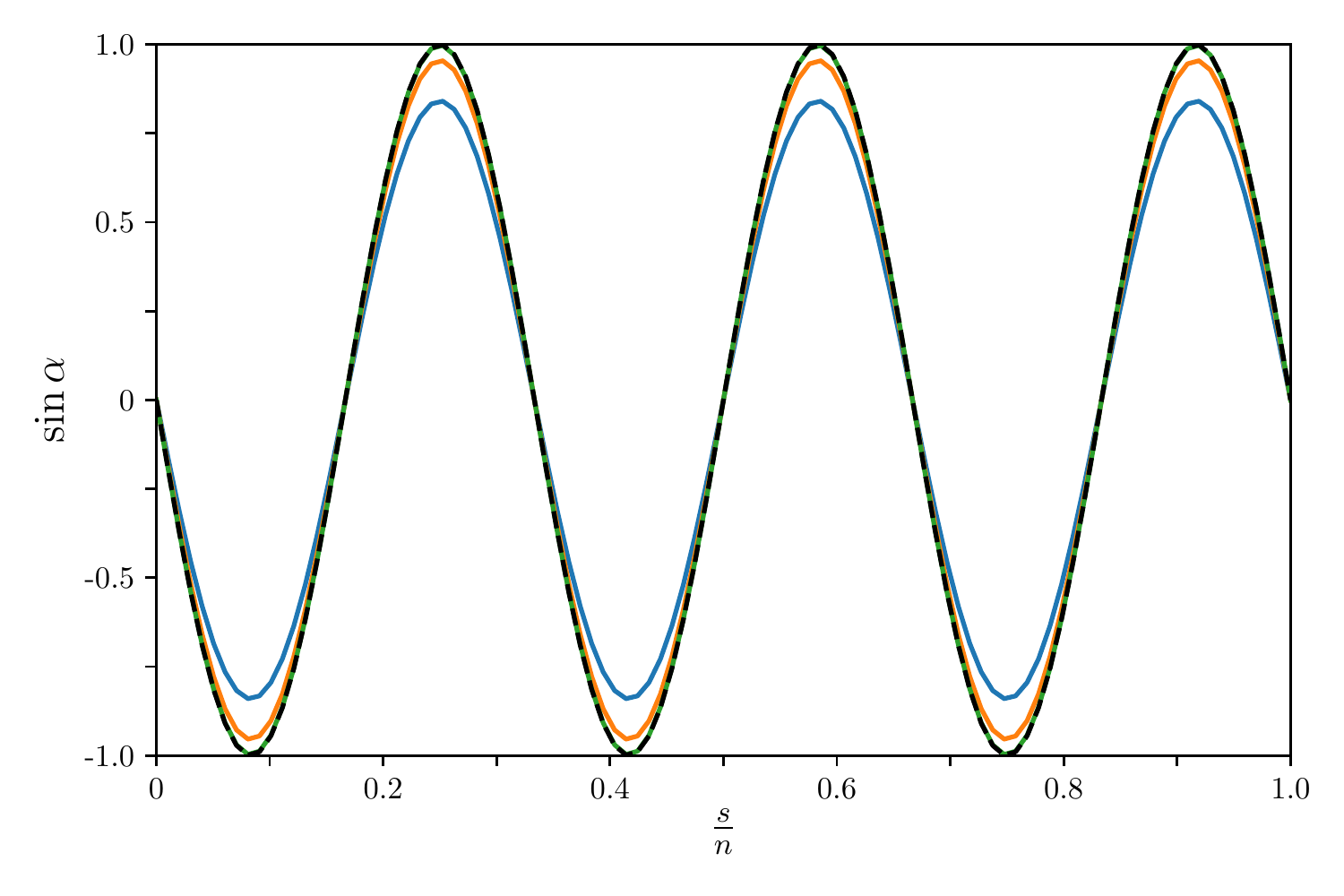}}
\end{center}
\caption{\small {\textbf{Case 1)} \textbf{Left plot}: 
Relative deviation of $\sin\alpha$ as obtained for option 1 of the $(3,3)$ ISS framework from the corresponding model-independent
prediction, with respect to $M_0$ (in GeV).
Line and colour code as in Fig.~\ref{fig:Case1_s2thij}. 
\textbf{  Right plot}:  
$\sin\alpha$ with respect to $s/n$ (fixing $n=26$ and continuously varying $0\leq  s < 26$). 
The black (dashed) curve
displays the result for $\sin\alpha$ obtained in the model-independent
scenario, see Eq.~(\ref{eq:sinalphaCase1}).
The coloured (solid) curves refer to distinct values of $M_0$:
blue for $M_0=500$~GeV, orange for $M_0=1000$~GeV and green for $M_0=5000$~GeV.  
We have chosen $y_0=1$ in order to better display the
deviation from the model-independent scenario (notice that such a
value of $y_0$ requires 
$M_0 \gtrsim 4800$~GeV to comply with the experimental bounds on $\eta_{\alpha\alpha}$
at the $3\, \sigma$ level, cf. Section~\ref{sec:unitarityISS}).
Figures from~\cite{Hagedorn:2021ldq}.
\label{fig:Case1_salpha}}}
\end{figure}

Moving on to the relative deviation of the Majorana phase $\alpha$, we note that also in this case
the size, sign and behaviour of the relative deviation $\Delta\sin\alpha$ (depending on $y_0$ and $M_0$) does not depend on the actual choice of the parameter $s$.
Thus, we have again taken $s=1$. 
In the left plot in Fig.~\ref{fig:Case1_salpha}, we present the relative deviation of $\sin\alpha$ as obtained for option 1 of the $(3,3)$ ISS framework from the corresponding model-independent
prediction, with respect to $M_0$ (in GeV).
Comparing the maximal size of the relative deviation of $\sin\alpha$ ($\Delta\sin\alpha$) with the ones of the solar and the atmospheric mixing angles, $\Delta\sin^2\theta_{12}$
and $\Delta\sin^2\theta_{23}$, previously displayed in Fig.~\ref{fig:Case1_s2thij}, we confirm that the latter are slightly smaller than the former, as expected from the analytical estimate in Eq.~(\ref{eq:estimateDsy01}). 
The right plot in Fig.~\ref{fig:Case1_salpha} illustrates the suppression of the value of $\sin\alpha$ depending on $s/n$
for three different values of $M_0$, 
$M_0=500$~GeV, $1000$~GeV and $5000$~GeV, and these are compared to the result expected in the model-independent scenario, see Eq.~(\ref{eq:sinalphaCase1}).
We have chosen here $y_0=1$ in order to enhance the visibility 
of the deviations between the model-independent scenario and the (3,3) ISS presented in this plot, although such a large value of the Yukawa coupling requires $M_0$ to be at least $M_0 \gtrsim 4800$~GeV in order to comply with the experimental bounds on
the quantities $\eta_{\alpha\beta}$, see Section~\ref{sec:unitarityISS}. Beyond this suppression of the value of $\sin\alpha$, we note that the periodicity in $s/n$ is still the same, independently of the effects of non-unitarity of $\tilde{U}_\nu$, confirming the analytical estimates of Section~\ref{sec42}. We have also numerically verified the analytical expectation that the Dirac phase $\delta$ as well as the Majorana phase $\beta$
remain trivial, i.e.~$\sin\delta=0$ and $\sin\beta=0$.

\begin{figure}[t!]
\begin{center}
\parbox{3in}{\includegraphics*[scale=0.5]{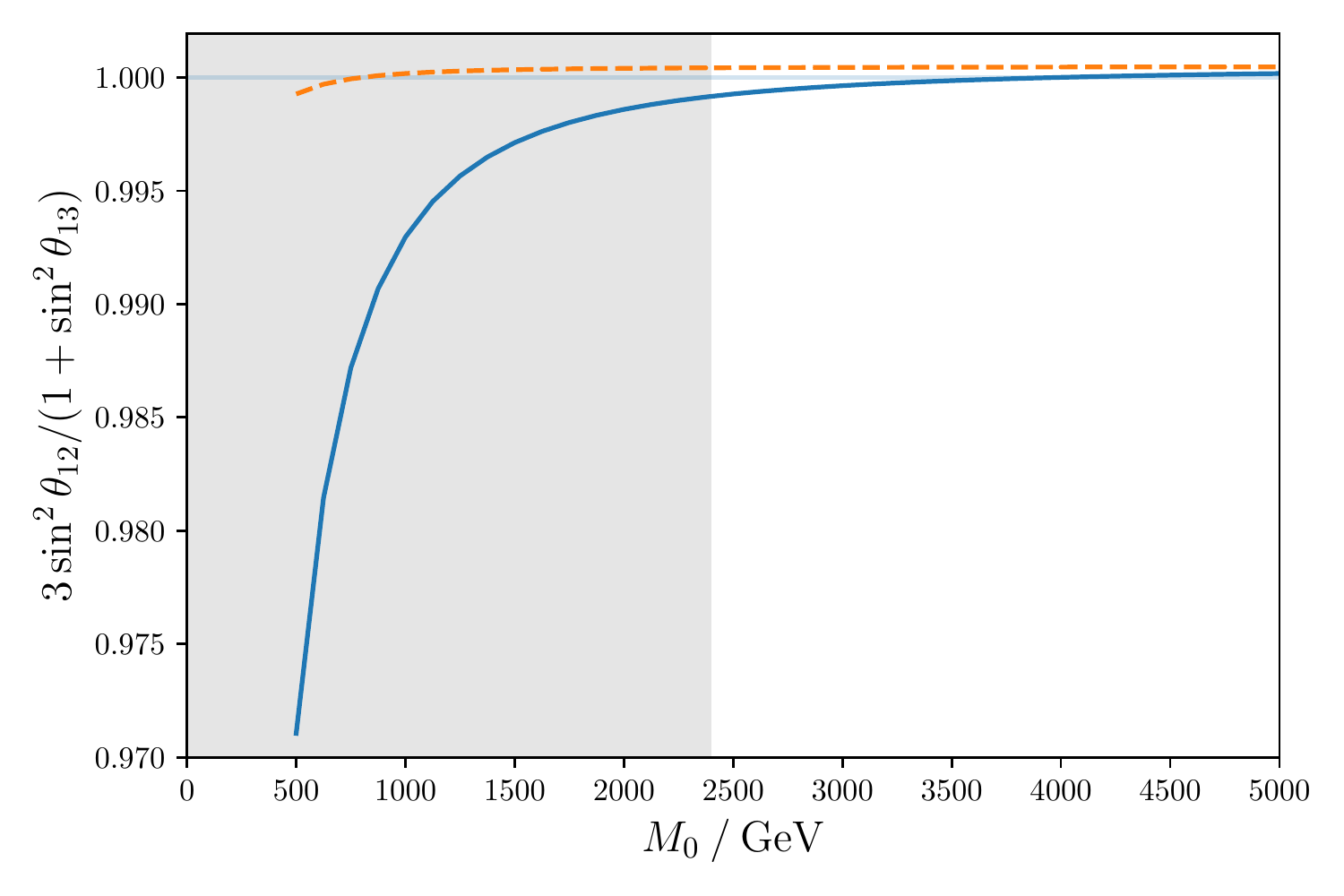}}
\hspace{0.2in}
\parbox{3in}{\includegraphics*[scale=0.5]{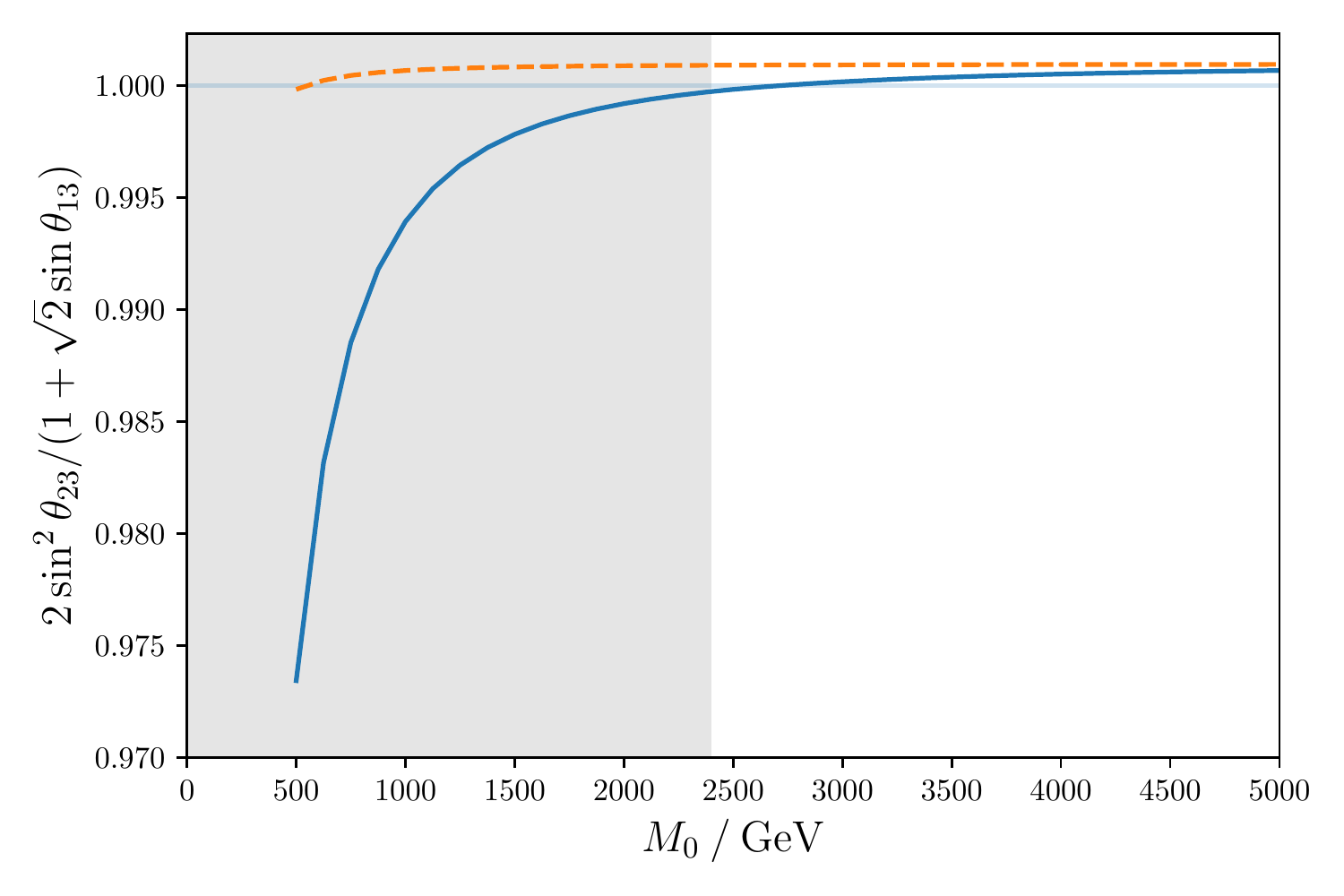}}
\end{center}
\caption{{\small \textbf{Case 1)} 
Validity check of approximate sum rules
for option 1 of the $(3,3)$ ISS framework 
with respect to the mass $M_0$ (in GeV).
 Line and colour code as in Fig.~\ref{fig:Case1_s2thij}.
 We note that for the second sum rule (right plot) we focus on the approximate sum rule with a plus sign, since we present results for $\theta_S < \pi/2$, see Eq.~(\ref{eq:sumrulesCase1})
and below. 
Figures from~\cite{Hagedorn:2021ldq}.
\label{fig:Case1_Sigma12}}}
\end{figure}

Finally, we address the validity of the two approximate sum rules, see Eq.~(\ref{eq:sumrulesCase1}). As can be seen from the plots in Fig.~\ref{fig:Case1_Sigma12},
 deviations do not exceed the level of $-3\%$, in agreement with the analytical estimate. Furthermore, we numerically confirm that the maximally achieved relative deviation is slightly
larger for the first sum rule than for the second, for $\theta_S < \pi/2$. 
We also note that for large values of $M_0$, where effects of the non-unitarity of $\tilde{U}_\nu$ should be suppressed, both ratios related to the two different sum rules become slightly larger than one. This is consistent with the fact that
these sum rules only hold approximately.

\subsection{Case 2)}
\label{sec52}

  %
 \begin{figure}[t!]
\begin{center}
\parbox{3in}{\includegraphics*[scale=0.5]{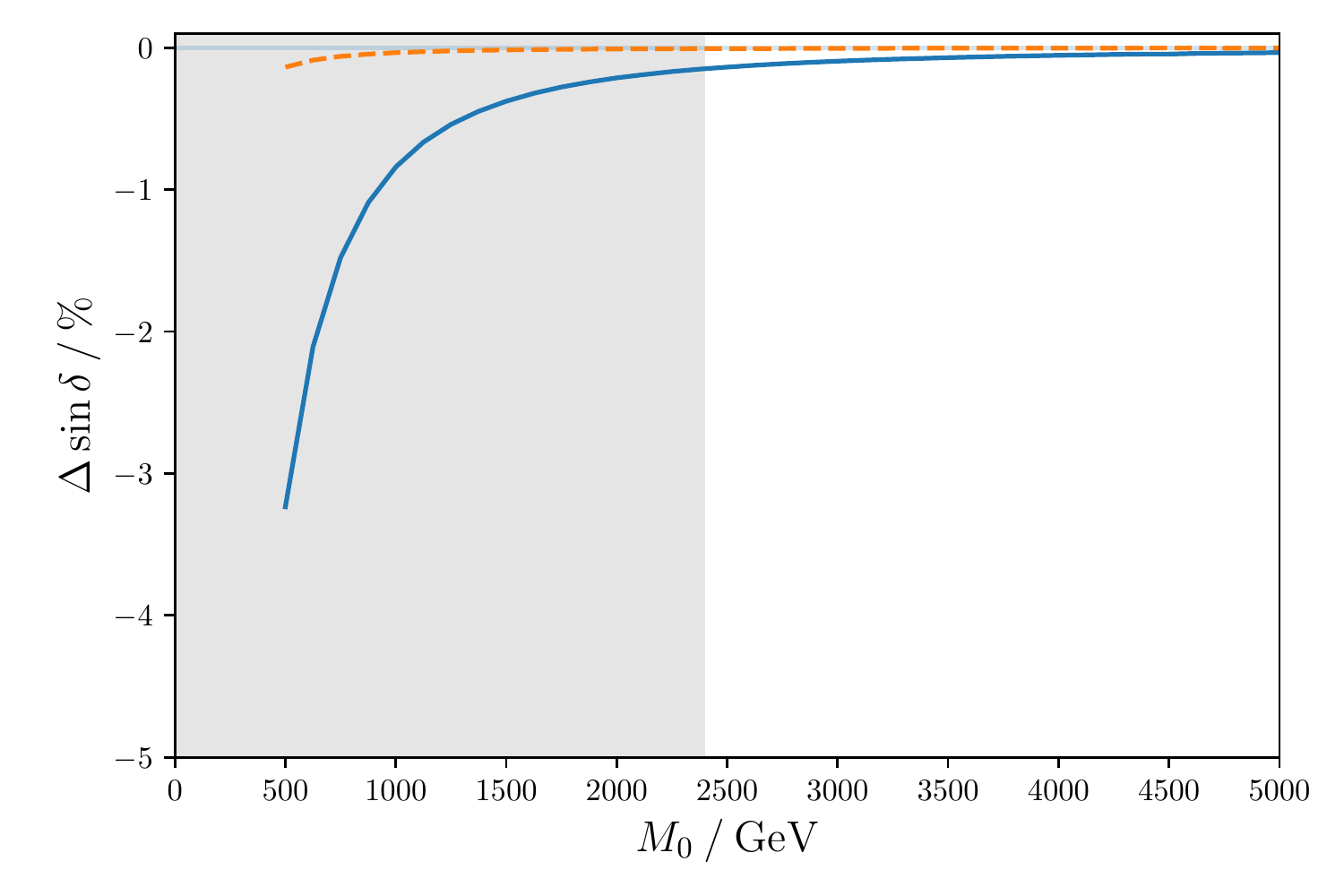}}
\hspace{0.2in}
\parbox{3in}{\includegraphics*[scale=0.5]{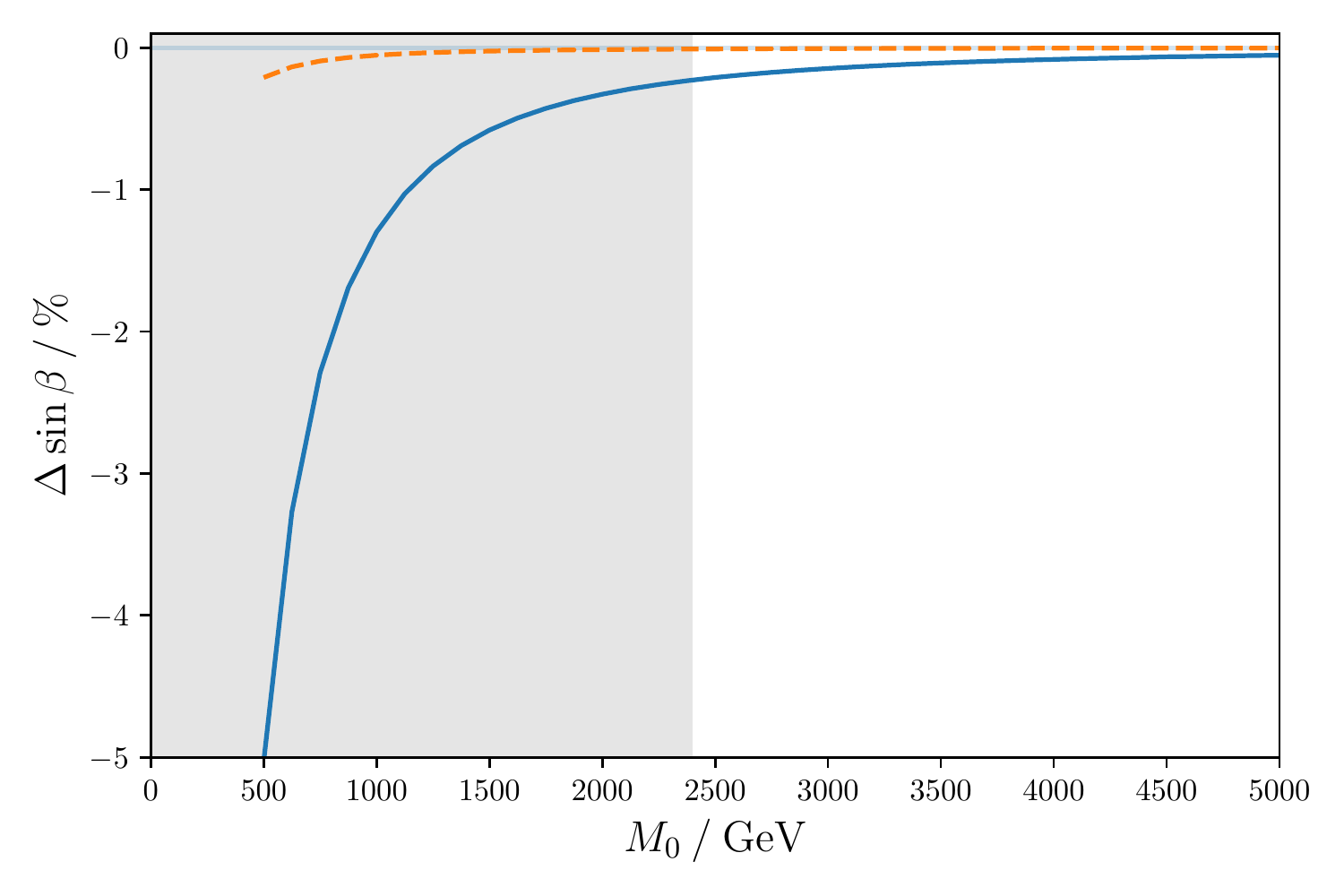}}
\end{center}
\caption{{\small \textbf{Case 2)} Relative deviations $\Delta\sin\delta$ (left plot) and $\Delta\sin\beta$ (right plot), as obtained for option 1 of the $(3,3)$ ISS framework, from the values obtained in the model-independent scenario, 
 with respect to the mass $M_0$ (in GeV). 
 The concrete choice of $v$ is irrelevant and thus we 
 have set $v=3$. Line and colour code as in Fig.~\ref{fig:Case1_s2thij}.
 Figures from~\cite{Hagedorn:2021ldq}.
\label{fig:Case2_sdelta_sbeta}}}
\end{figure}
In our numerical study, we choose as representative values of the index $n$ and of the parameter $u$
\begin{equation}
\label{eq:numchoiceparaCase2}
n=14 \;\; \mbox{and} \;\; u=1,
\end{equation}
also commenting on results for the choices $u=-1$, $u=15$ as well as $u=0$ in order to comprehensively analyse the features of Case 2). 
For the parameter $v$, we consider all permitted values according to the relations in Eqs.~(\ref{eq:ZXCase2},\ref{eq:uvdeffromst}) and the chosen value of $u$, e.g.~for $u=1$ we have
\begin{equation}
\label{eq:numchoiceparavCase2u1}
v= 3, \, 9, \, 15, \, 21, \, 27, \, 33, \, 39  \; .
\end{equation}
We start by discussing the relative deviations of $\sin^2\theta_{ij}$. The results for $\Delta\sin^2\theta_{12}$ and $\Delta\sin^2\theta_{23}$ are consistent with the analytical expectations, see Eq.~(\ref{eq:Deltasin2thetaij_gen}).
 Indeed, the plots for $\Delta\sin^2\theta_{12}$ and $\Delta\sin^2\theta_{23}$ look very similar to those presented in Fig.~\ref{fig:Case1_s2thij} for Case 1).
However, the relative deviation $\Delta\sin^2\theta_{13}$ does not agree with the analytical expectations and instead is always very small, showing that like in Case 1),  $\sin^2\theta_{13}$
is typically adjusted to its experimental best-fit value (since it also drives the fit for the present case).
 
We confirm numerically that the deviations of $\sin^2 \theta_{ij}$ 
do not depend on the choice of the parameter $v$ and we
have thus fixed  $v=3$. 
As regards the dependence of $\Delta\sin^2\theta_{ij}$ on the parameter $u$, we have also checked that the
 aforementioned different choices of $u$ all lead to the same result. 
 
\begin{figure}[t!]
\begin{center}
\parbox{3in}{\includegraphics*[scale=0.5]{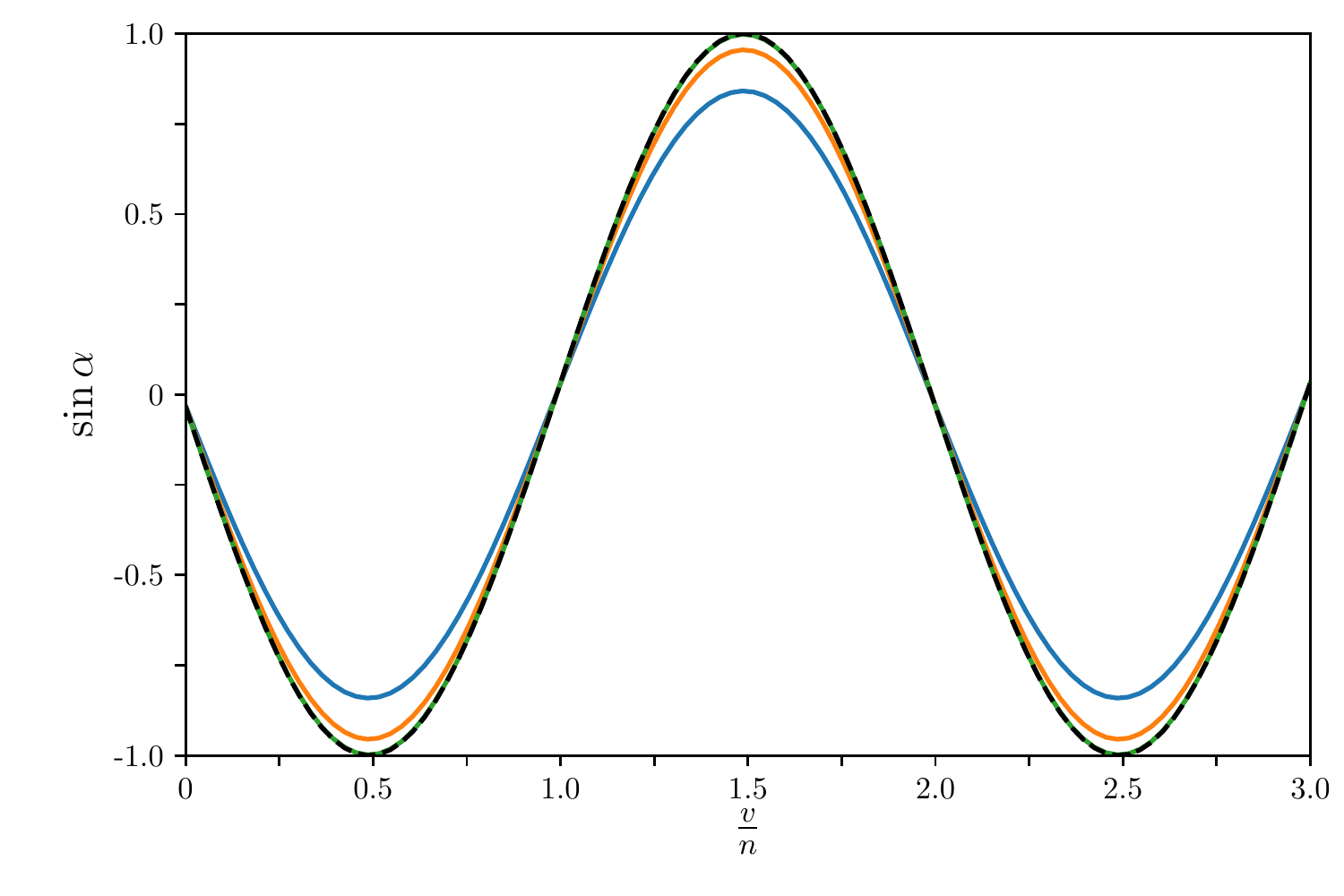}}
\hspace{0.2in}
\parbox{3in}{\includegraphics*[scale=0.5]{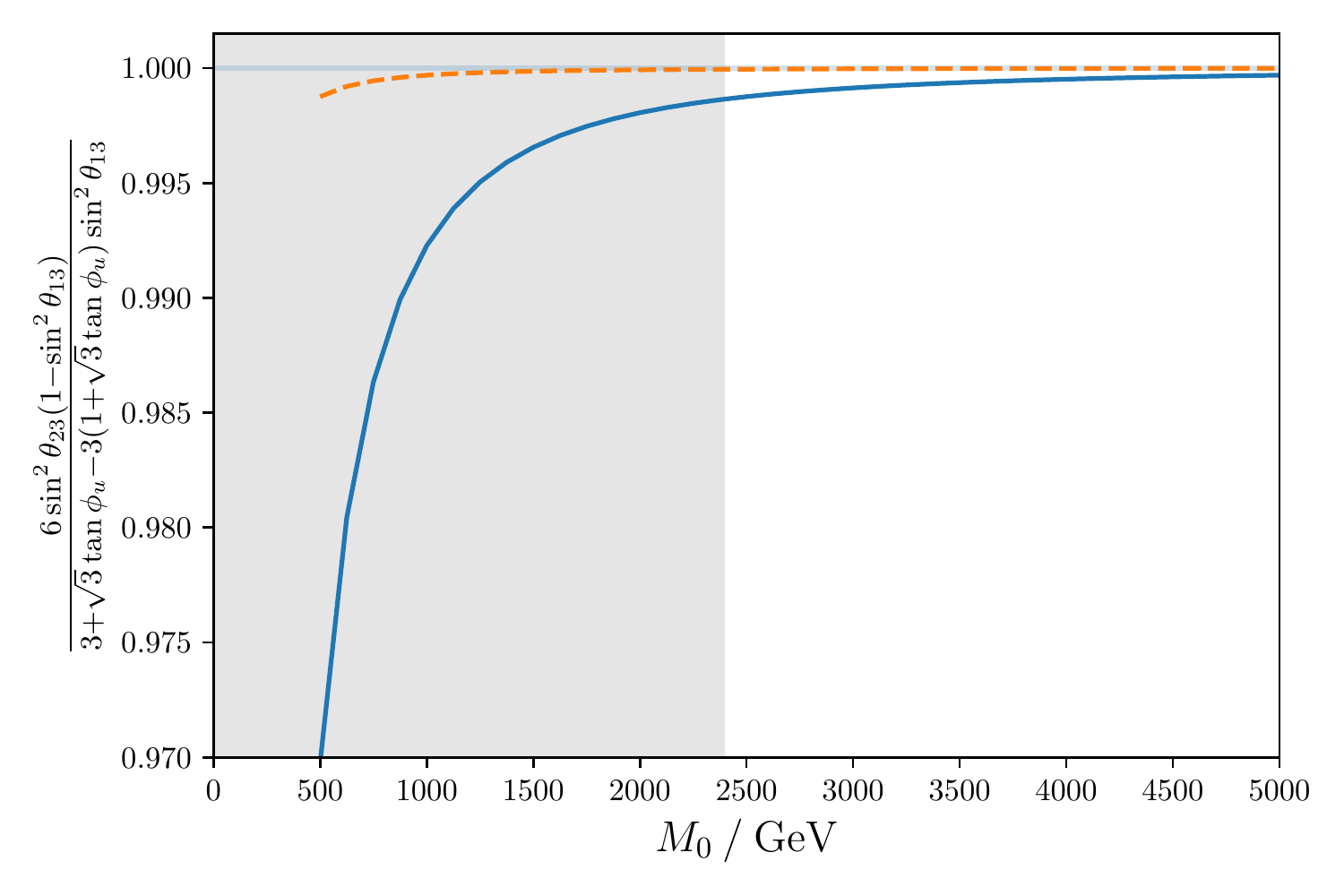}}
\end{center}
\caption{\small {\textbf{Case 2)} \textbf{Left plot}: $\sin\alpha$ with respect to $v/n$ (fixing $n=14$ and continuously varying $0 \leq v < 3n = 42$). 
The different (coloured) curves refer to three different masses $M_0$ like in Fig.~\ref{fig:Case1_salpha}, also setting $y_0=1$.
 The black (dashed) curve displays the result for $\sin\alpha$, obtained in the model-independent scenario, see Eq.~(\ref{eq:sinalphaCase2}).
 \textbf{Right~plot}: Validity check of the exact sum rule in Eq.~(\ref{eq:sumruleCase2}) for option 1 of the $(3,3)$ ISS framework
 with respect to the mass $M_0$ (in GeV). Line and colour code as in Fig.~\ref{fig:Case2_sdelta_sbeta}.
 We have chosen $n=14$ and $u=1$ so that $\tan\phi_u \approx 0.23$. 
 Figures from~\cite{Hagedorn:2021ldq}.
\label{fig:Case2_salpha_Sigma3}}}
\end{figure}

 For the relative deviations of the CP phases $\delta$ and $\beta$, $\Delta\sin\delta$ and $\Delta\sin\beta$, we present our findings in Fig.~\ref{fig:Case2_sdelta_sbeta}.
  Since these deviations are also independent of the choice of $v$, we choose $v=3$ for concreteness. 
The plot for $\Delta\sin\alpha$ looks very similar to the corresponding one of Case 1), see left plot in Fig.~\ref{fig:Case1_salpha}.
The sign and size of the deviations are in accordance with the analytical expectations, see 
 Eqs.~(\ref{eq:DeltasinCPphases_gen},\ref{eq:estimateDsy01}). We note that both Majorana phases $\alpha$ and $\beta$ experience slightly larger effects from the non-unitarity of the lepton mixing matrix (i.e., the presence of the heavy sterile states) than the Dirac phase $\delta$.
 The effects of the non-unitarity of $\tilde{U}_\nu$ on the behaviour of $\sin\alpha$ with respect to $v/n$, shown in the left plot of Fig.~\ref{fig:Case2_salpha_Sigma3}, are very similar to those
 encountered when studying $\sin\alpha$ with respect to $s/n$ for Case 1), see the right plot in Fig.~\ref{fig:Case1_salpha}.
 Again, we emphasise that the periodicity of $\sin\alpha$ in $v/n$ is not altered by the effects of the non-unitarity of the PMNS mixing matrix.

Next, we detail our numerical results for the relative deviations of the two (approximate) sum rules found for Case 2), see Section~\ref{sec32}. We have checked that for the sum rule which is common for Case 1) and Case 2)
(see first approximate equality in Eq.~(\ref{eq:sumrulesCase1})), the results do coincide with those shown in the left plot in Fig.~\ref{fig:Case1_Sigma12}.
Concerning the exact sum rule, shown in Eq.~(\ref{eq:sumruleCase2}), the numerical results are given in the right plot in Fig.~\ref{fig:Case2_salpha_Sigma3}. We see that the size and sign 
of the relative deviation agree with the analytical estimate shown in Eq.~(\ref{eq:estimateDSigma3y01}). We have also checked numerically that the results do not depend on the choice of $u$ and $v$; while the plot presented relies on
$u=1$ and $v=3$, similar results have been found for the other mentioned choices of $u$ and the admitted values of $v$. 

We comment on the choice $u=0$ that predicts maximal atmospheric mixing and maximal Dirac phase $\delta$, $\sin\beta=0$ and the exact equality in Eq.~(\ref{eq:sinalphaCase2}):
the relative deviations $\Delta\sin^2\theta_{23}$ and $\Delta\sin\delta$ are of the same sign and size,  
and exhibit the same dependence on the Yukawa coupling $y_0$ and on the mass scale $M_0$ as occurs for the choice $u=1$.
Furthermore, the fact that the Majorana phase $\beta$ is trivial is not altered by the effects of non-unitarity of $\tilde{U}_\nu$, as expected from the analytical estimates, see Section~\ref{sec4}.
 The results for the Majorana phase $\alpha$ look very similar
to those displayed in Fig.~\ref{fig:Case1_salpha} (left plot) and Fig.~\ref{fig:Case2_salpha_Sigma3} (left plot). 
Moreover, we confirm that whenever the choice $v=0$ is permitted,  the Majorana phase $\alpha$ vanishes 
independently of the deviations of $\tilde{U}_\nu$
from unitarity. 

Finally, we notice that we have performed a numerical check to confirm that  the symmetry transformations in the parameters $u$ and $\theta$, see Eq.~(\ref{eq:symmtrafosCase2}) under point $i)$ in Section~\ref{sec32}, are still valid.

\subsection{Case 3 a)}
\label{sec53}

As representative values for $n$ and $m$, we take
\begin{equation}
\label{eq:numchoiceparaCase3a}
n=17 \;\; \mbox{and} \;\; m=1 \; ,
\end{equation}
since these can  satisfactorily accommodate the experimental data on the reactor and the atmospheric mixing angles
for light neutrinos with NO/IO~\cite{Esteban:2020cvm},
according to the expectations from the model-independent scenario, see Section~\ref{sec331} and, especially, Eq.~(\ref{eq:sin2theta1323Case3a}).
We consider all possible values of the parameter $s$. In addition to $m=1$, we also study the results on lepton mixing for the choice $m=16$. 
The rather large value of the index $n$ of the flavour symmetry is needed in order to achieve a sufficiently small value of $m/n$ (or $1-m/n$).

 \begin{figure}[t!]
\begin{center}
\parbox{3in}{\includegraphics*[scale=0.55]{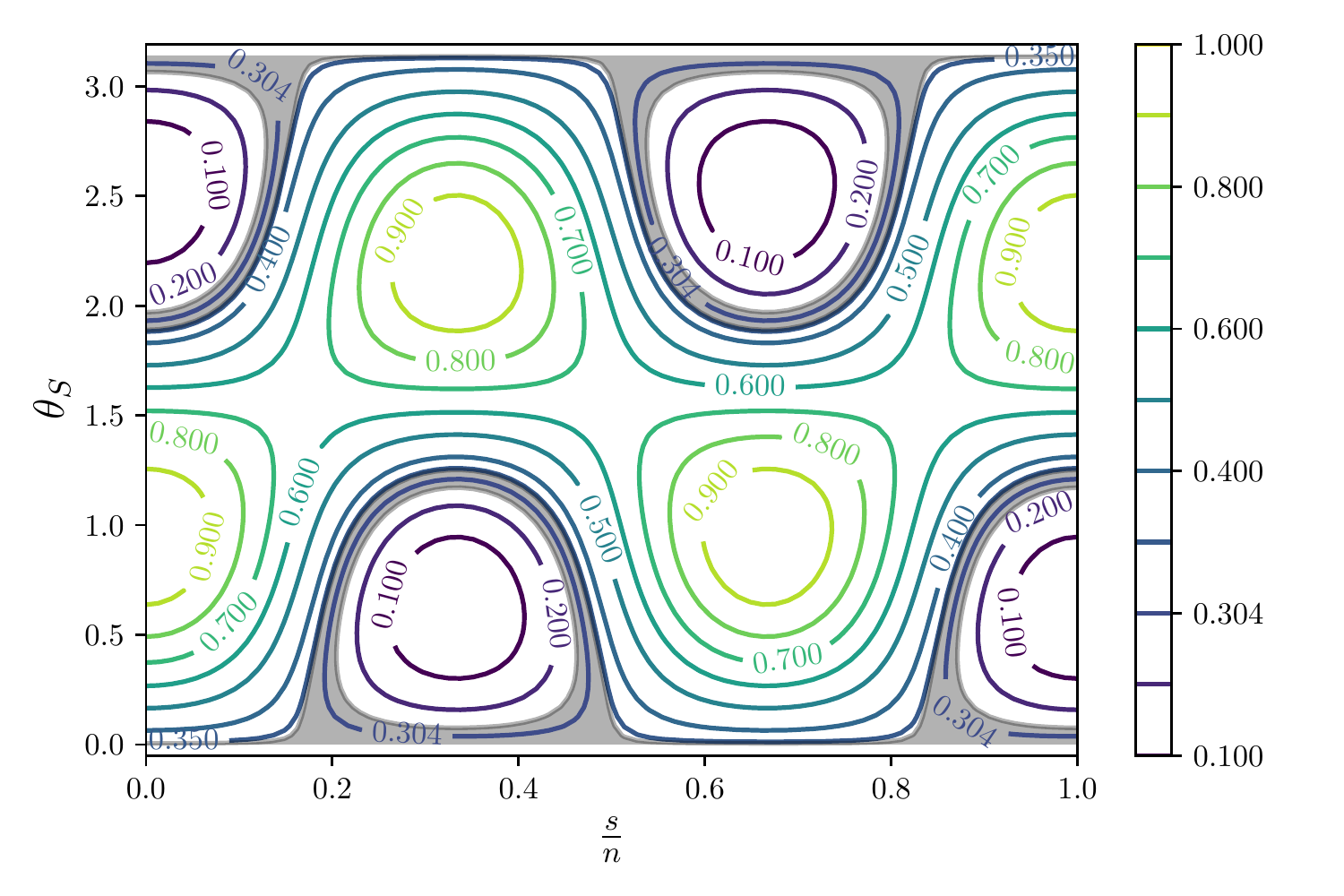}}
\hspace{0.2in}
\parbox{3in}{\includegraphics*[scale=0.55]{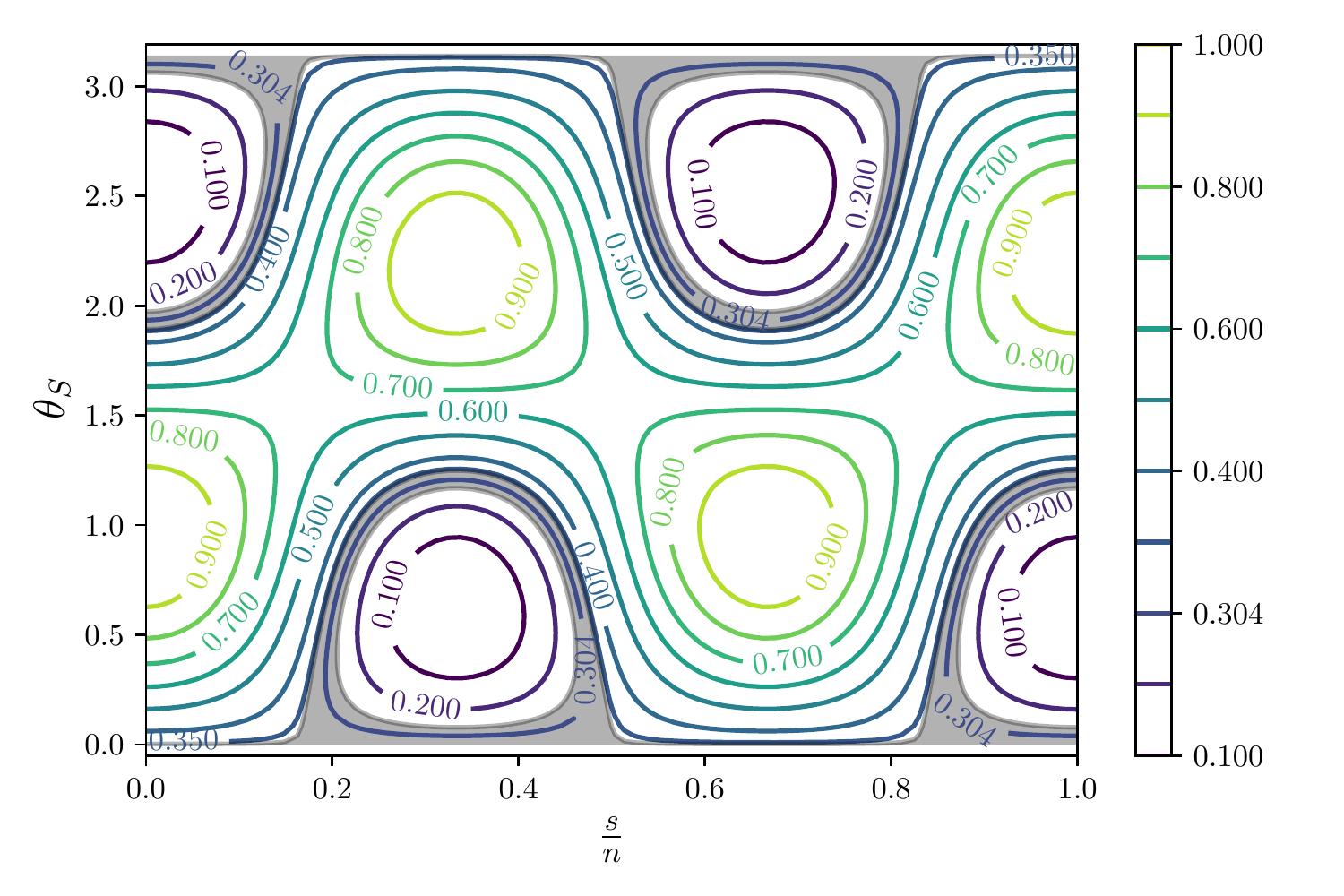}}
\end{center}
\caption{{\small \textbf{Case 3 a)} Contour plots for $\sin^2\theta_{12}$ in the $(s/n-\theta_S)$ plane, obtained for option 1 of the $(3,3)$ ISS framework.
The left plot is for $M_0=1000$~GeV and the right one for $M_0=5000$~GeV. We fix $y_0=0.5$ in order to amplify the differences between the two plots.
Here the grey-shaded areas denote 
values of $\sin^2\theta_{12}$ which are 
experimentally favoured at the $3 \, \sigma$ level~\cite{Esteban:2020cvm}.
Figures from~\cite{Hagedorn:2021ldq}.
\label{fig:Case3a_s2th12_contour}}}
\end{figure}
Since fixing $n$ and $m$ determines completely the value of the reactor and the atmospheric mixing angles, we only consider, like for Case 1) and Case 2), 
the relative deviations $\Delta\sin^2\theta_{13}$ and $\Delta\sin^2\theta_{23}$. We note that their size and sign do agree with the analytical expectations, see Eq.~(\ref{eq:estimateDsy01}).
 Furthermore, we confirm numerically that there is no dependence of these results on the parameter $s$ and the free angle $\theta_S$.
Since in Case 3 a) $\theta_{12}$ is the only lepton mixing angle that depends on the free angle $\theta_S$, $\sin^2\theta_{12}$ naturally drives the fit, and thus the relative deviation $\Delta\sin^2\theta_{12}$ is always very small.
 Given that $\sin^2\theta_{12}$ further depends on the parameter $s$, we present
 in Fig.~\ref{fig:Case3a_s2th12_contour}
 plots for $\sin^2\theta_{12}$ in the $(s/n-\theta_S)$ plane
for two different values of the mass scale $M_0=1000$~GeV (left plot) and $M_0=5000$~GeV (right plot). 
We fix the Yukawa coupling to $y_0=0.5$ in order to better perceive the differences in the plots for the two different values of $M_0$,
although such a large value of $y_0$ does require $M_0 \gtrsim 2400$~GeV in order to comply with the experimental constraints on $\eta_{\alpha\beta}$, see Section~\ref{sec:unitarityISS}. 
As one observes in Fig.~\ref{fig:Case3a_s2th12_contour}, the visible differences are still very small. 
We stress that here the grey-shaded areas indicate the values of $\sin^2\theta_{12}$ that are experimentally favoured at the $3 \, \sigma$ level~\cite{Esteban:2020cvm}.
As can be clearly seen from Fig.~\ref{fig:Case3a_s2th12_contour}, for most values of $s$
a successful accommodation of the experimental data can be obtained for 
two different values of the free angle $\theta_S$.
 One of these values is close to $\theta_S \approx 0$ or $\theta_S \approx \pi$.
These plots can be compared with a very similar one shown in the original analysis of the different mixing patterns,  see~\cite{Hagedorn:2014wha}.

 \begin{figure}[t!]
\begin{center}
\parbox{3in}{\includegraphics*[scale=0.55]{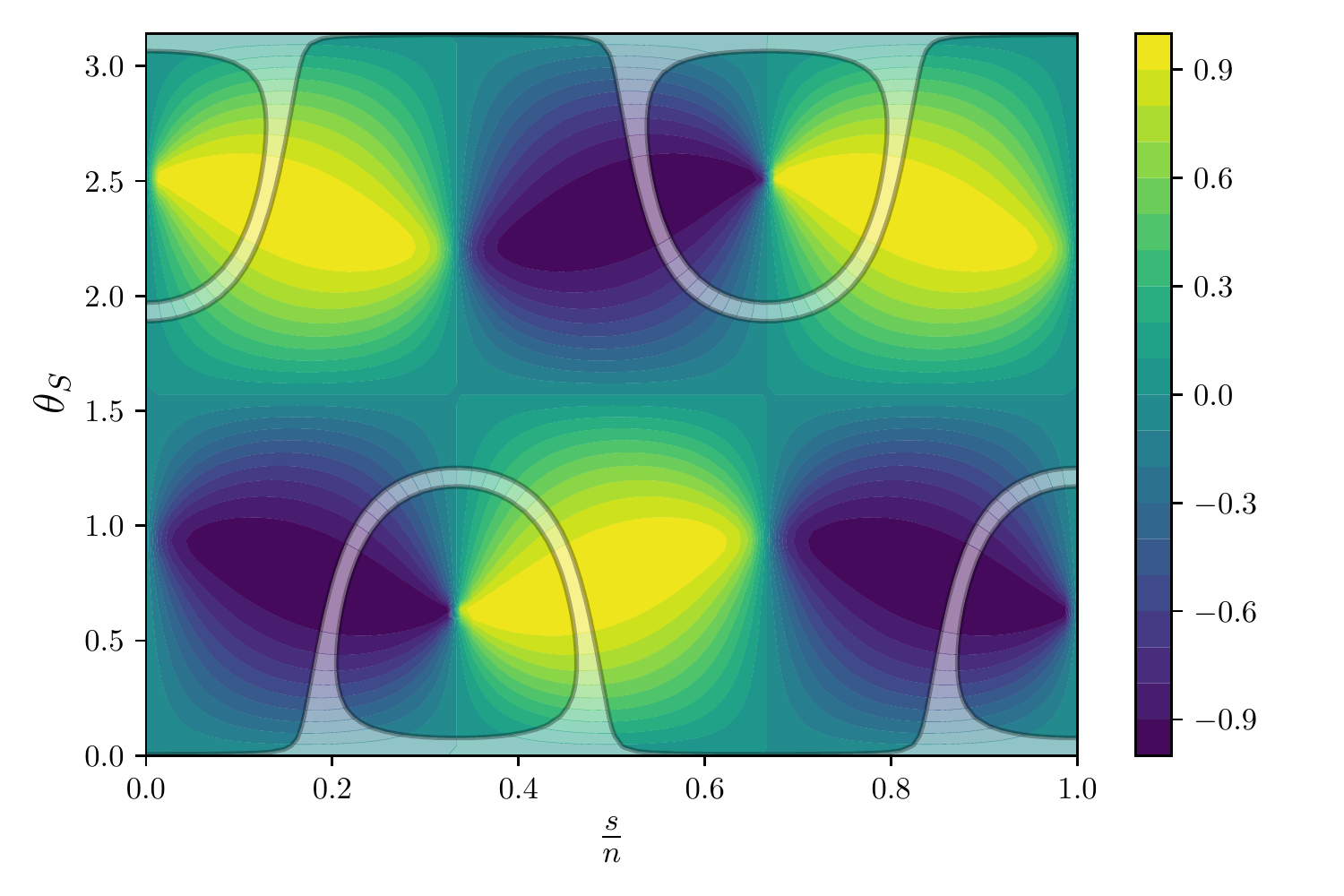}}
\hspace{0.2in}
\parbox{3in}{\includegraphics*[scale=0.55]{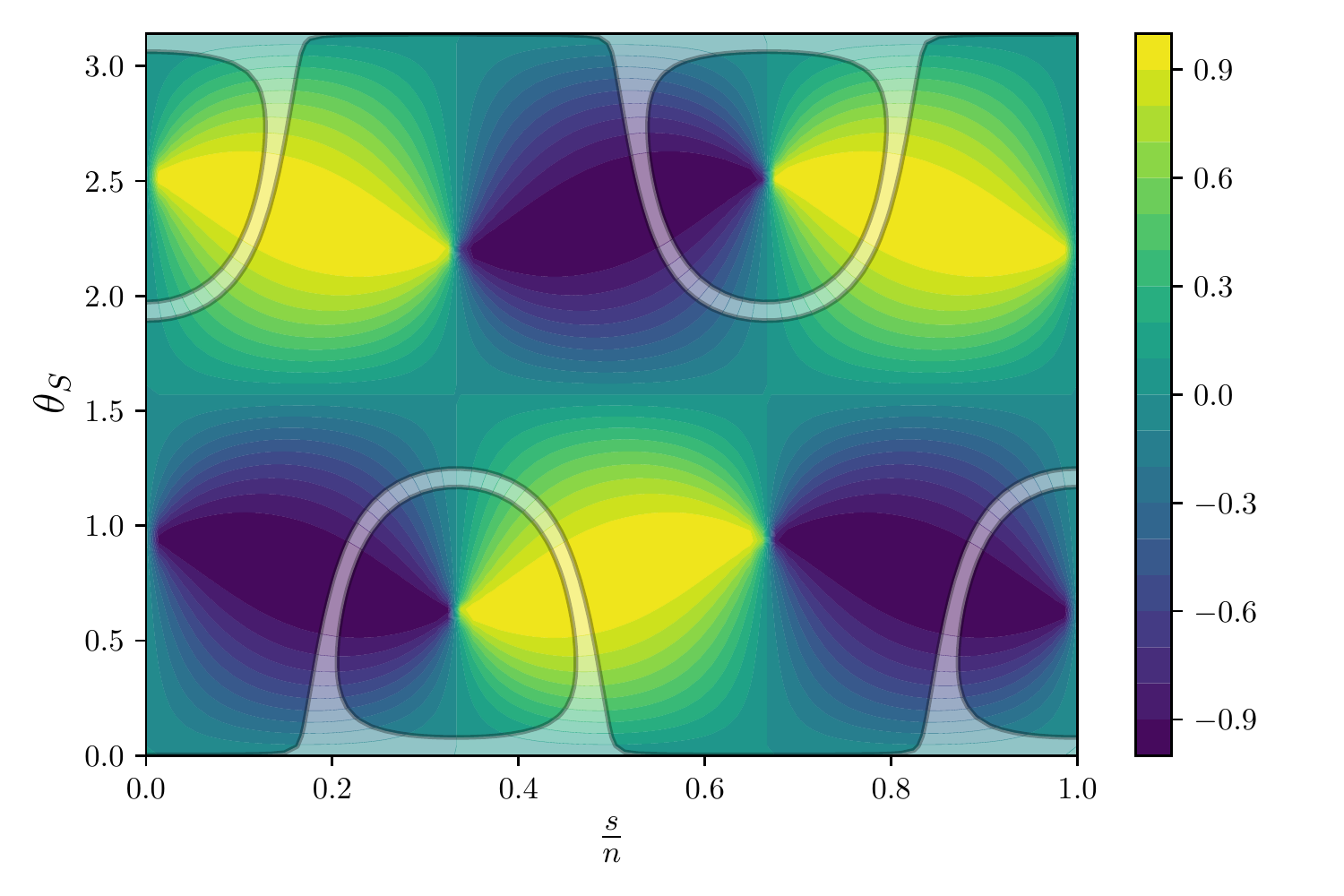}}
\vspace{0.1in}
\parbox{3in}{\includegraphics*[scale=0.55]{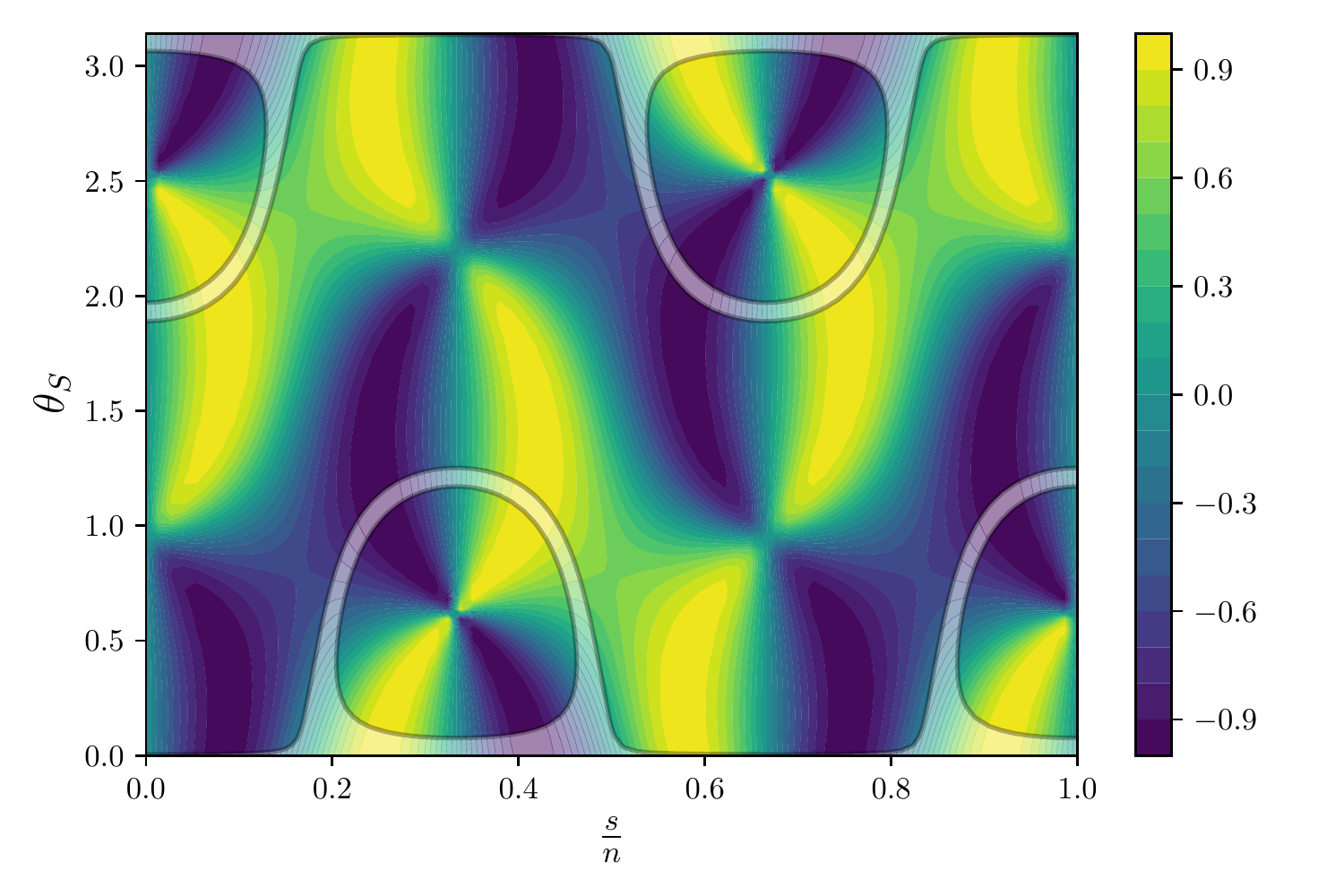}}
\hspace{0.2in}
\parbox{3in}{\includegraphics*[scale=0.55]{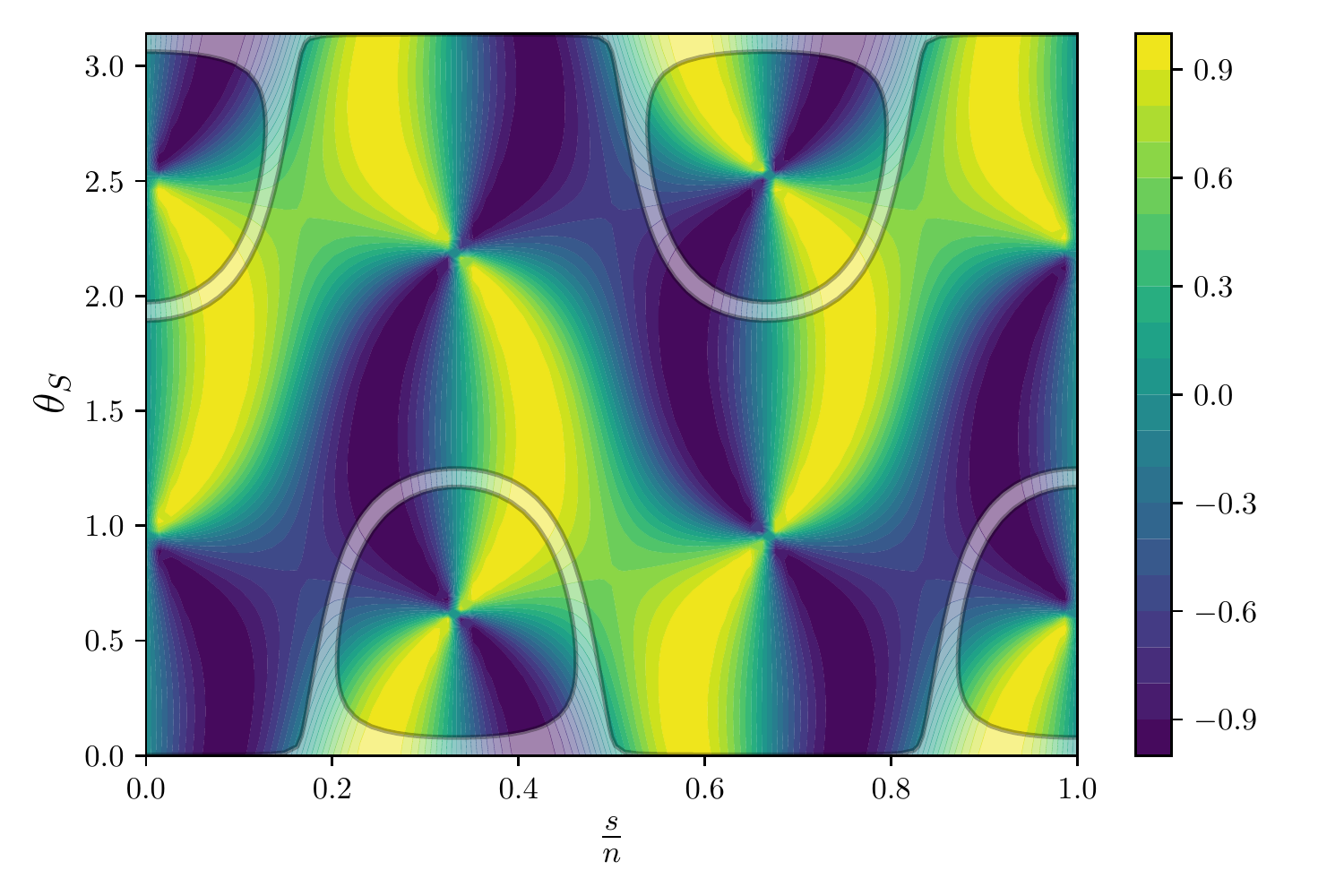}}
\vspace{0.1in}
\parbox{3in}{\includegraphics*[scale=0.55]{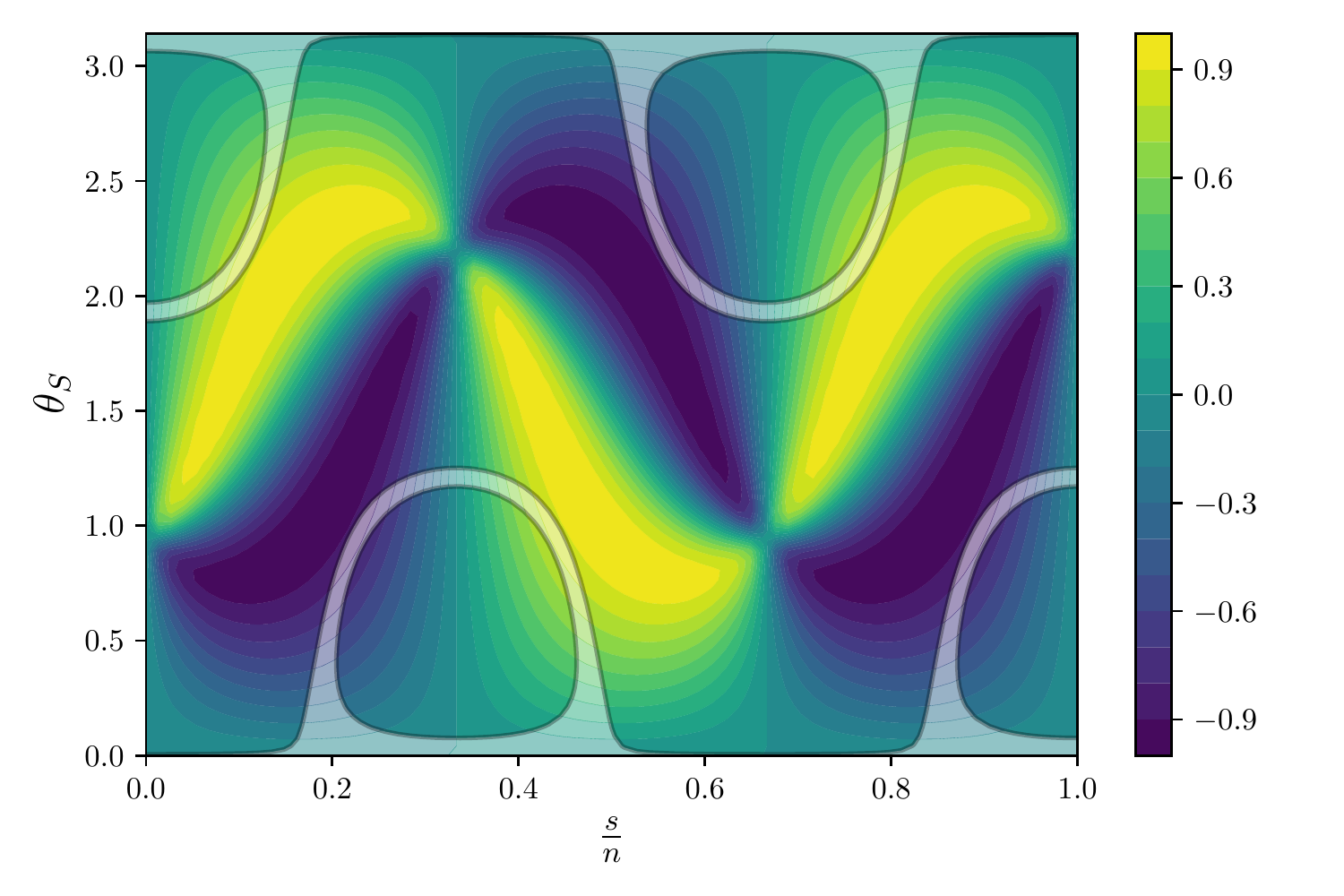}}
\hspace{0.2in}
\parbox{3in}{\includegraphics*[scale=0.55]{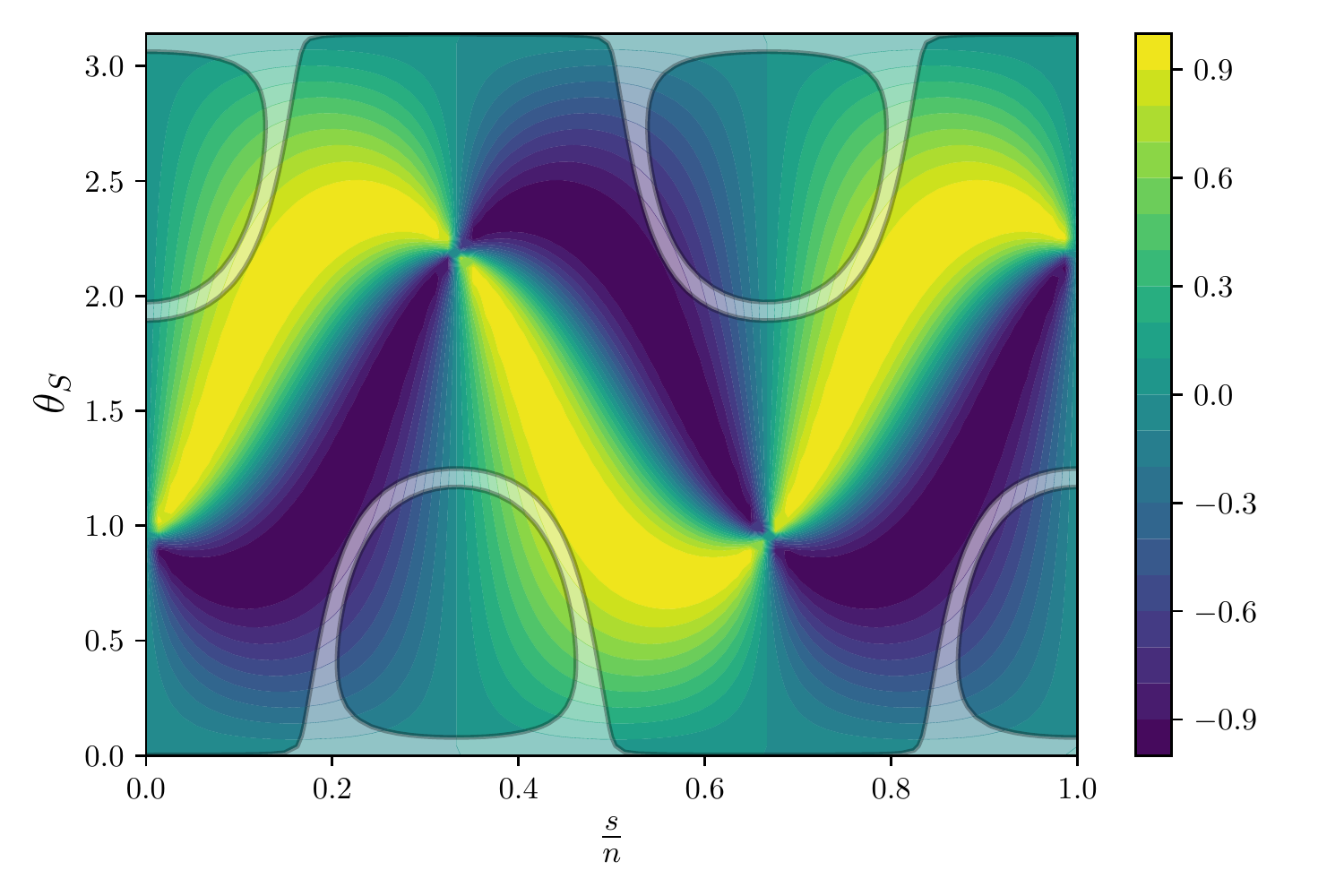}}
\end{center}
\caption{{\small \textbf{Case 3 a)} Contour plots for the sines of the CP phases in the $(s/n-\theta_S)$ plane, 
obtained for option 1 of the $(3,3)$ ISS framework. 
From top to bottom (first to third row), $\sin\delta$, $\sin\alpha$ and $\sin\beta$. 
On the left column plots, $M_0=1000$~GeV while on the right $M_0=5000$~GeV. We again fix $y_0=0.5$ (see Fig.~\ref{fig:Case3a_s2th12_contour}). 
The colour scheme denotes the values of the sines, from $-1$ (dark blue) to $+1$ (light yellow), as indicated by the colour bar on the right of each plot.
The white/grey-shaded areas correspond here to those of Fig.~\ref{fig:Case3a_s2th12_contour}, and indicate the values of the solar mixing angle that are experimentally preferred at the $3 \, \sigma$ level.
Figures from~\cite{Hagedorn:2021ldq}.
\label{fig:Case3a_sdelta_salpha_sbeta_contour}}}
\end{figure}
The results for the relative deviations of the CP phases, $\Delta\sin\delta$, $\Delta\sin\alpha$ and $\Delta\sin\beta$, look similar to those obtained for the already presented cases, Case 1) and Case 2). 
For this reason, we prefer to show contour plots for the sines of all three CP phases in the $(s/n-\theta_S)$ plane.
These can be found in Fig.~\ref{fig:Case3a_sdelta_salpha_sbeta_contour}, 
where we display $\sin\delta$, $\sin\alpha$ and $\sin\beta$, for two different values of $M_0$, $M_0=1000$~GeV (left plots) and $M_0=5000$~GeV (right plots). The colour scheme denotes the values of the sines (indicated by the colour bar on the right of each plot). We again take $y_0=0.5$ in order to enhance the visibility of differences in the plots. The white/grey-shaded areas indicate the values of the solar mixing angle that are experimentally preferred at the $3 \, \sigma$ level. It turns out that visible differences between the plots for $M_0=1000$~GeV and 
$M_0=5000$~GeV are (mainly) found in regions of the $(s/n-\theta_S)$ plane that
are not compatible with the experimental value of $\sin^2\theta_{12}$ at the $3 \, \sigma$ level. Nevertheless, the results presented in these plots are interesting, since the validity of the approximate formulae for the sines of the CP phases
(found in Eqs.~(\ref{eq:sinalphaCase3a},\ref{eq:sinbetasindeltaCase3a}) under point $e)$ for the model-independent scenario) as well as the fact that the absolute value of $\sin\beta$ is bounded to be smaller than $\sim 0.9$, can be checked.
Furthermore, they can be directly compared with the results for the model-independent scenario presented in~\cite{Hagedorn:2014wha}.
Again, we confirm numerically that the effects of non-unitarity of the PMNS mixing matrix do not affect the 
vanishing of $\sin\delta$, $\sin\alpha$ and/or $\sin\beta$ (occurring for certain choices of group theory parameters).   
The approximate sum rule, quoted in Eq.~(\ref{eq:sumruleCase3a}), is valid with a plus sign for the choice $n=17$ and $m=1$. Studying its behaviour depending on the Yukawa coupling $y_0$ and on the mass scale $M_0$ thus
leads to results 
very similar to those obtained for the second approximate sum rule (see second approximate equality in Eq.~(\ref{eq:sumrulesCase1})), for values of the free angle $\theta_S <\pi/2$, as shown
in the right plot of Fig.~\ref{fig:Case1_Sigma12}. 

In the end, we note that we have numerically confirmed that the symmetry transformations, given under point $i)$ in Section~\ref{sec331}, are valid.

\subsection{Case 3 b.1)}
\label{sec54}

For the last case, we focus on
\begin{equation}
\label{eq:numchoiceparaCase3b1}
n=20 \;\; \mbox{and} \;\; m=11 \; .
\end{equation}
All viable values of the parameter $s$ are studied. We choose the index $n$ of the flavour symmetry to be rather large\footnote{As shown in~\cite{Hagedorn:2014wha}, values 
of $n$ as small as $n=2$ are sufficient in order to successfully accommodate the experimental data on lepton mixing angles.} in order to allow studying different values of $m$, while achieving  good agreement with
experimental data on the solar mixing angle. In addition to the value $m=11$ we also perform a numerical analysis for $m=9$
and $m=10$.

\begin{figure}[t!]
\begin{center}
\parbox{3in}{\hspace{-0.3in}\includegraphics*[scale=0.55]{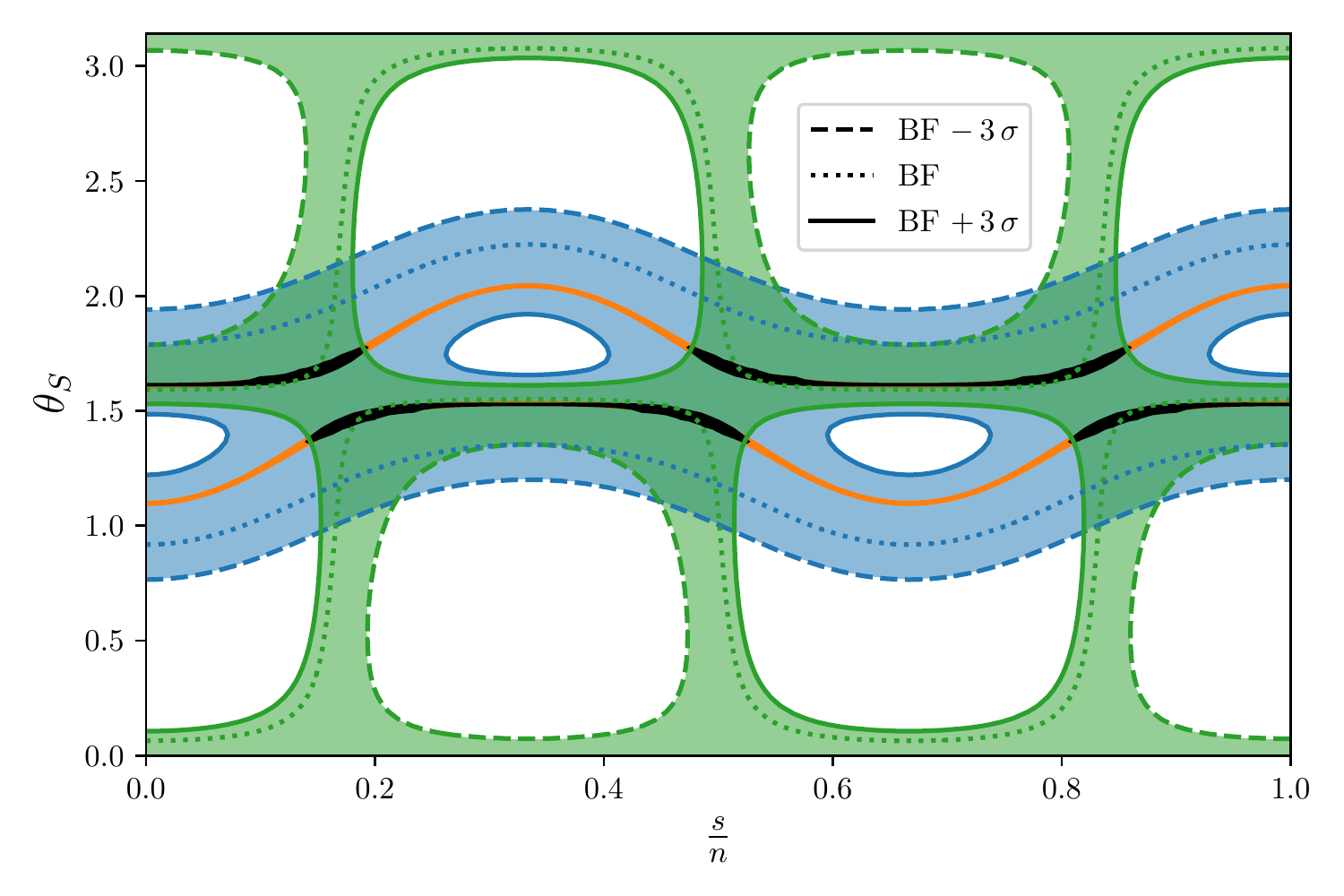}}
\hspace{-0.1in}
\parbox{3in}{\includegraphics*[scale=0.55]{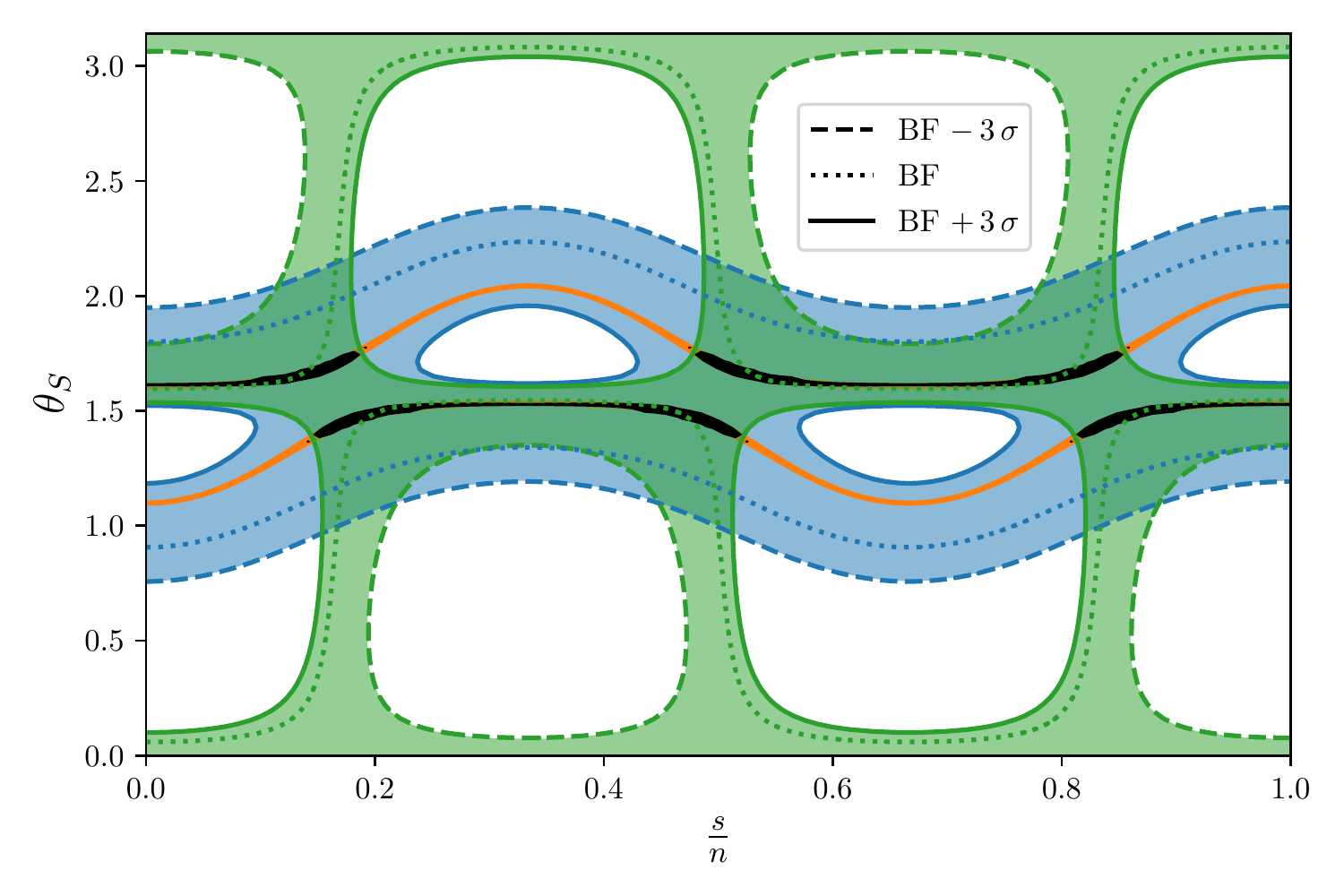}}
\end{center}
\caption{{\small \textbf{Case 3 b.1)} Contour plots for $\sin^2\theta_{ij}$, obtained for option 1 of the $(3,3)$ ISS framework, in the $(s/n-\theta_S)$ plane. Blue, green and orange respectively correspond to $\sin^2\theta_{12}$, $\sin^2\theta_{23}$ and $\sin^2\theta_{13}$.
Dotted lines indicate the experimental best-fit (BF) value for each $\sin^2\theta_{ij}$, while the coloured surfaces correspond to 
a $3 \sigma$ interval: dashed (solid) lines respectively define the BF$\mp 3 \sigma$ boundaries.  
Their overlap is highlighted in black. On the left, $M_0=1000$~GeV, while on the right $M_0=5000$~GeV. We again fix $y_0=0.5$ in order to amplify the differences between the two plots.
Figures from~\cite{Hagedorn:2021ldq}.
\label{fig:Case3b1_s2thij_contour}}}
\end{figure}

For Case 3 b.1) all mixing angles turn out to depend on the parameter $s$ and the free angle $\theta_S$, in addition to the two parameters $n$ and $m$ which we have fixed, see  Eq.~(\ref{eq:sin2thetaijCase3b1}).
In what follows we identify the areas in the $(s/n-\theta_S)$ plane in which the three mixing angles (individually and simultaneously) are in agreement with the experimental data at the $3 \, \sigma$ level~\cite{Esteban:2020cvm}.
This is shown in the contour plots in Fig.~\ref{fig:Case3b1_s2thij_contour}, for $\sin^2\theta_{12}$ (blue), $\sin^2\theta_{23}$ (green) and $\sin^2\theta_{13}$ (orange) and their combination (black), for two different values of the mass 
scale $M_0$, $M_0=1000$~GeV (left plot) and $M_0=5000$~GeV (right plot).
We note that we have again chosen $y_0=0.5$ for better visibility of the differences in the plots, although in this case $M_0=1000$~GeV leads to conflict with the experimental constraints on the quantities $\eta_{\alpha\beta}$, see Section~\ref{sec:unitarityISS}. We see that the areas of agreement with experimental data at the $3 \,\sigma$ level slightly differ between $M_0=1000$~GeV and $M_0=5000$~GeV. 
However, their overlap (shown in black in the two plots) is not visibly affected, and thus the parameter space in the $(s/n-\theta_S)$ plane compatible with the experimental data on lepton mixing angles hardly depends on 
the mass scale $M_0$. Indeed, comparing these two plots to a similar one, presented in the original analysis of the mixing pattern Case 3 b.1) for the model-independent scenario~\cite{Hagedorn:2014wha}, we confirm that all agree very well.
We note that the by far strongest constraint on the allowed parameter space in the $(s/n-\theta_S)$ plane is imposed by the reactor mixing angle $\sin^2\theta_{13}$.
The results shown in the plots in Fig.~\ref{fig:Case3b1_s2thij_contour} also confirm that all values of the parameter $s$ lead to a successful accommodation of the experimental data of the mixing angles for $n=20$ and $m=11$. 
The values of the free angle $\theta_S$
are then close to $\pi/2$.
 Regarding the size and sign of the relative deviations $\Delta\sin^2\theta_{12}$ and $\Delta\sin^2\theta_{23}$, we note that these are consistent with the analytical estimates, see Eq.~(\ref{eq:estimateDsy01}), whereas for $\Delta\sin^2\theta_{13}$ 
we always find it to be very small due to the pull in the fit that drives the adjustment of the free angle $\theta_S$ to match the best-fit value of the reactor mixing angle. This is analogous to what has been observed for Case 1) 
and Case 2).  
 \begin{figure}[t!]
\begin{center}
\parbox{3in}{\includegraphics*[scale=0.55]{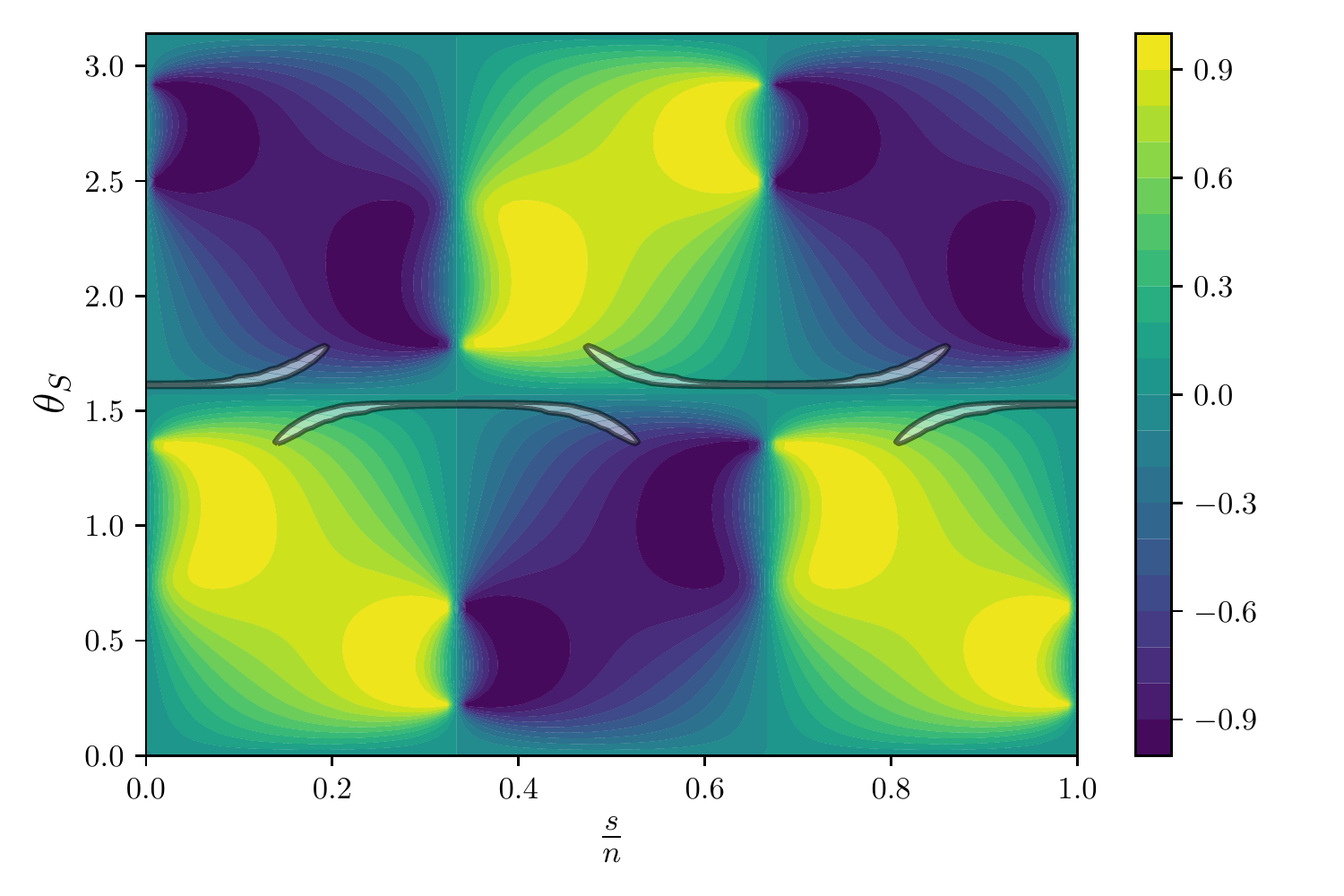}}
\hspace{0.2in}
\parbox{3in}{\includegraphics*[scale=0.55]{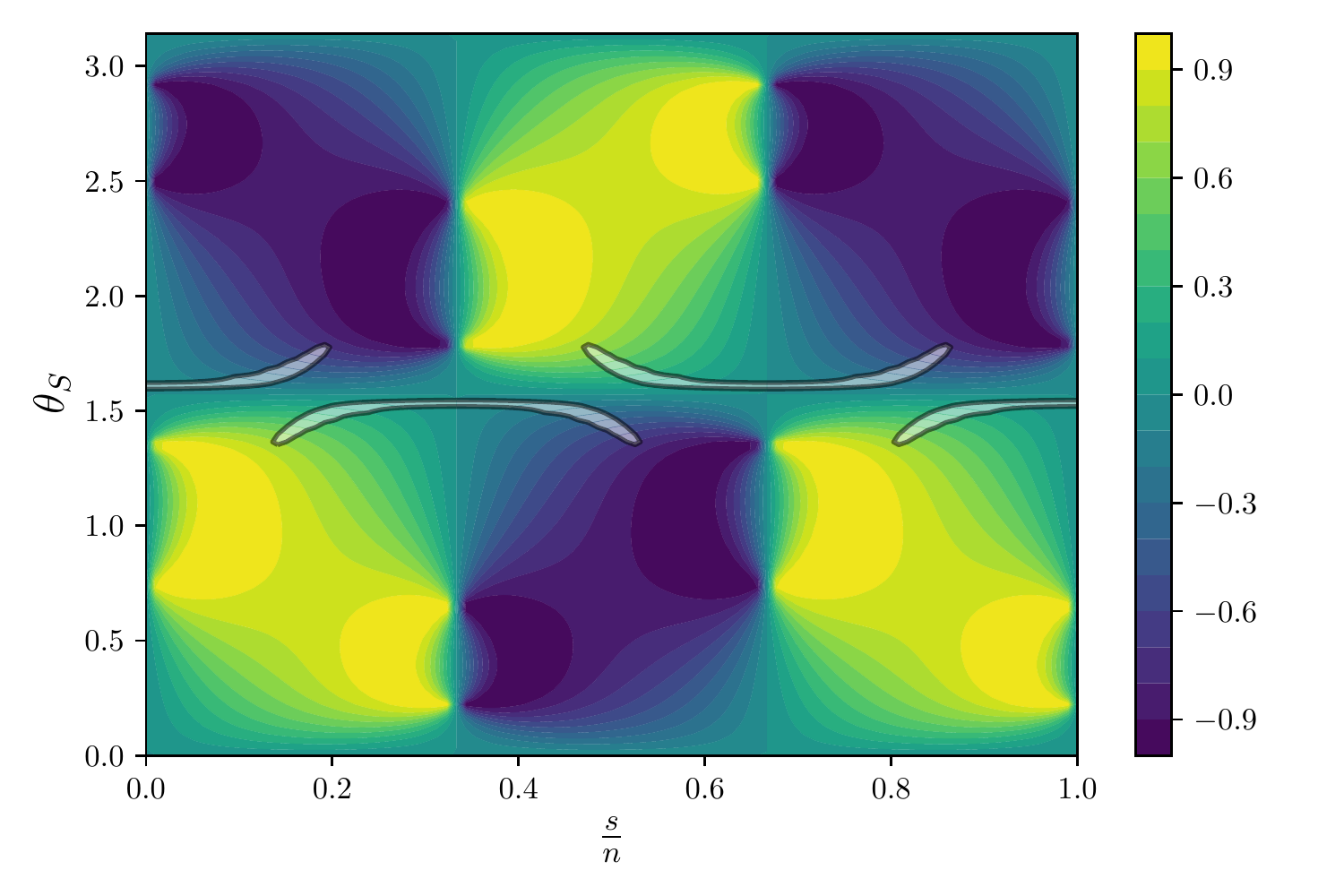}}
\vspace{0.1in}
\parbox{3in}{\includegraphics*[scale=0.55]{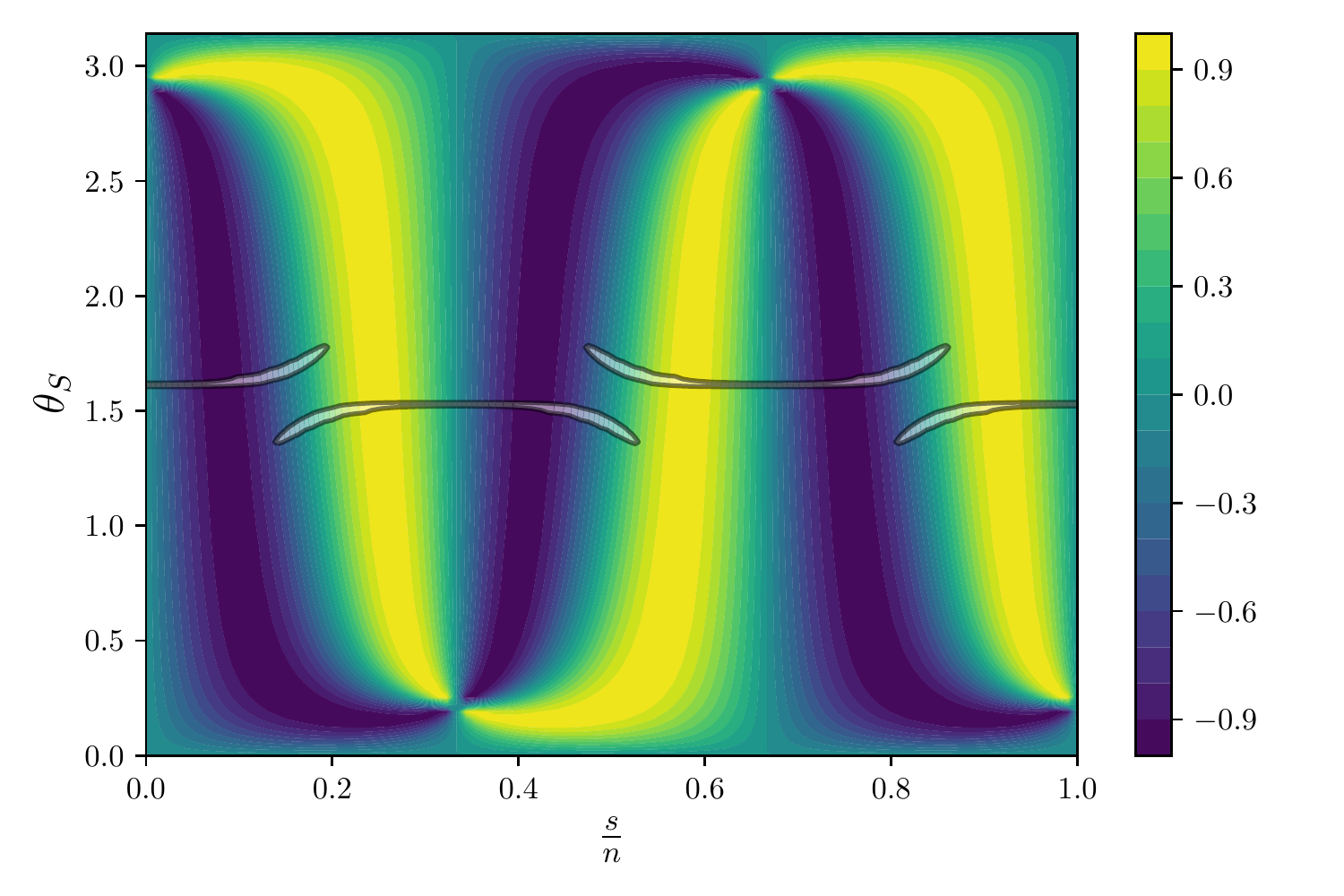}}
\hspace{0.2in}
\parbox{3in}{\includegraphics*[scale=0.55]{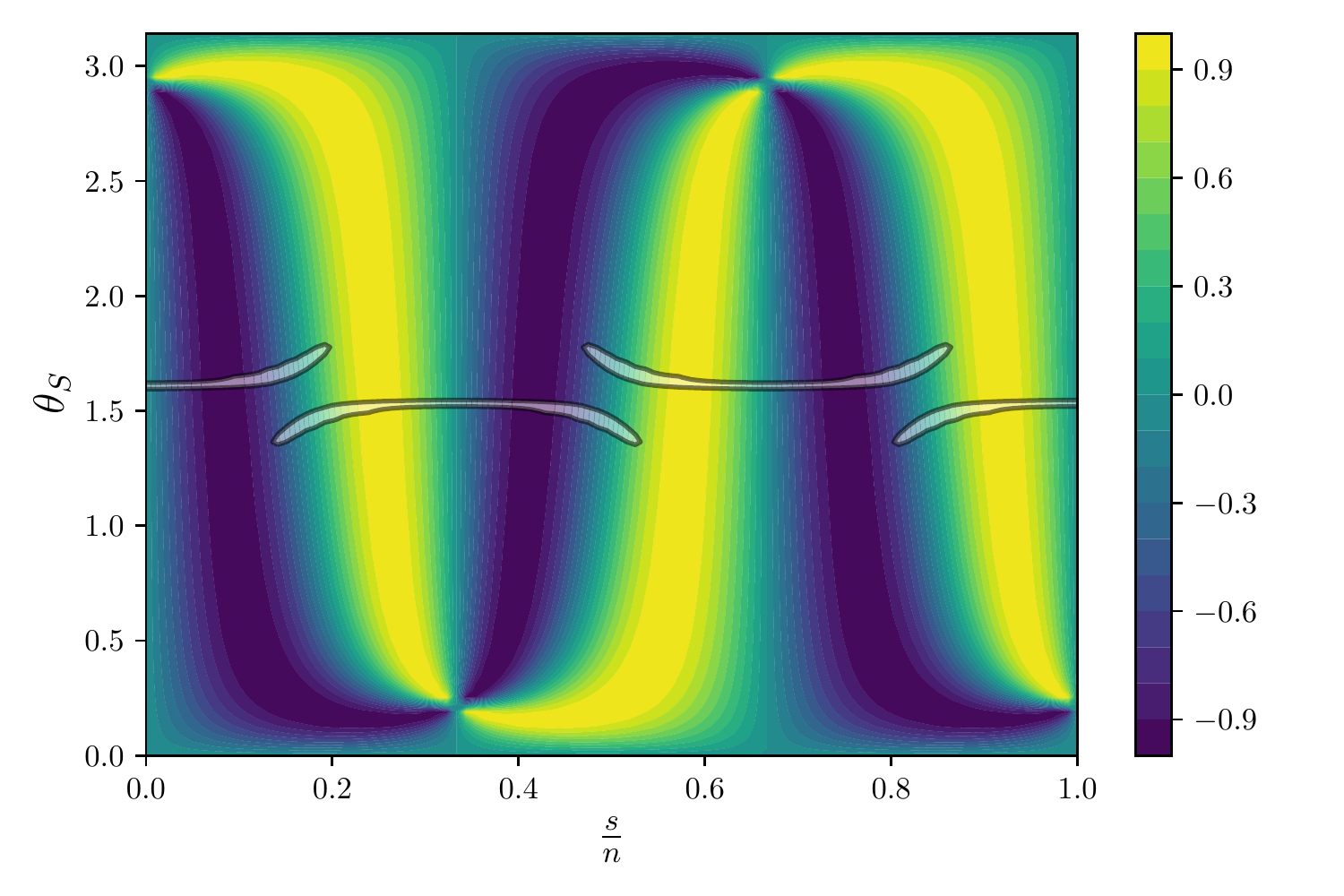}}
\vspace{0.1in}
\parbox{3in}{\includegraphics*[scale=0.55]{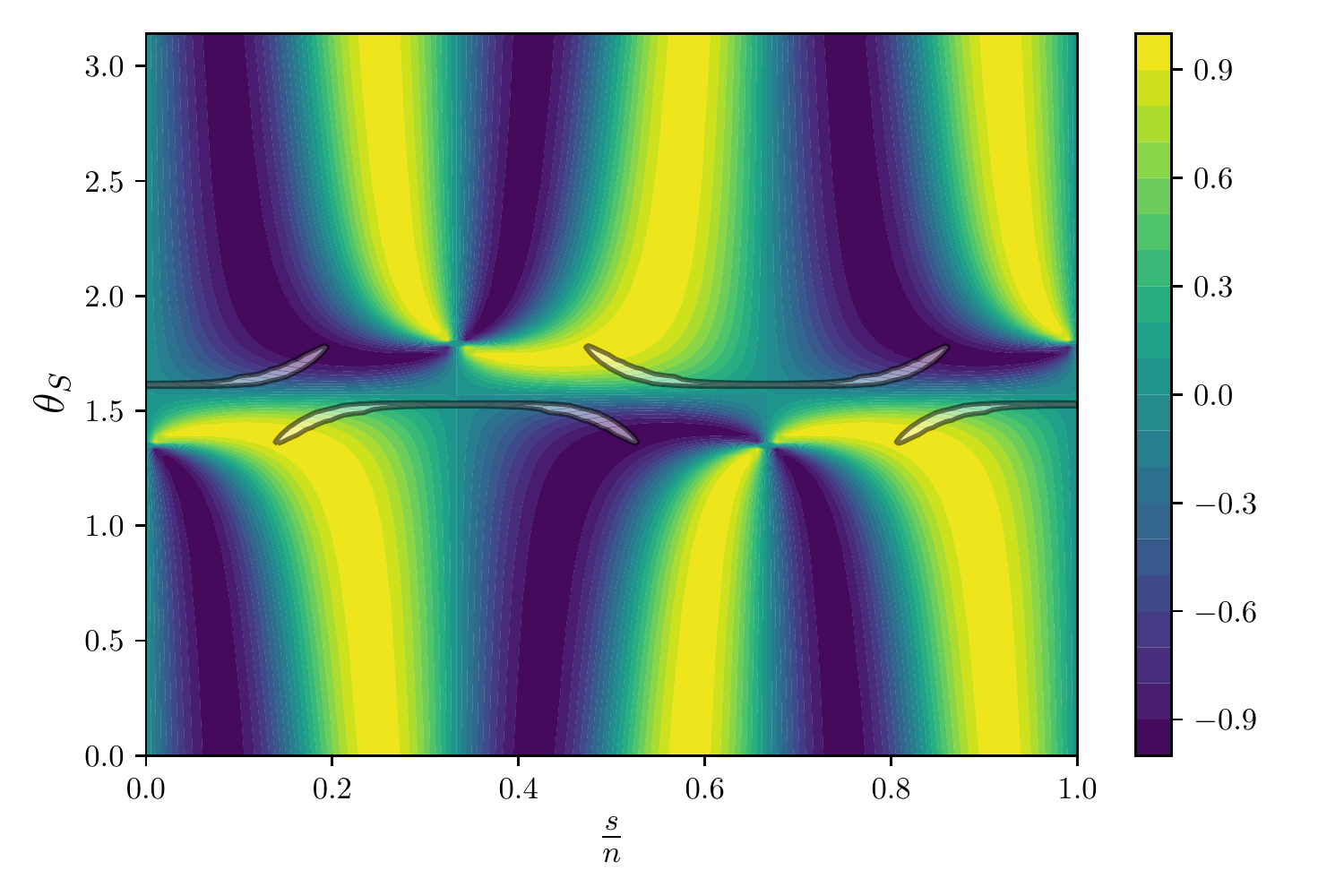}}
\hspace{0.2in}
\parbox{3in}{\includegraphics*[scale=0.55]{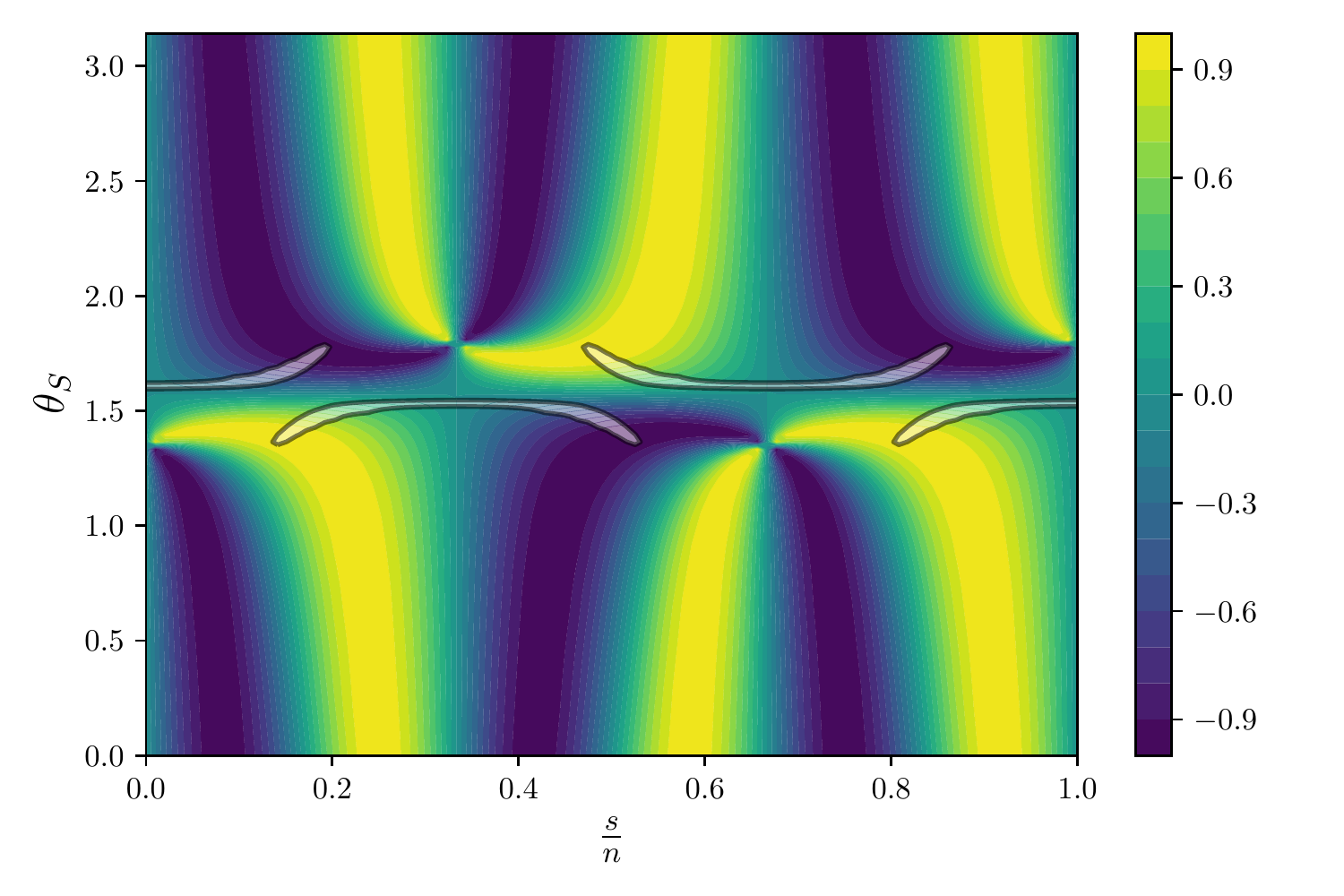}}
\end{center}
\caption{{\small \textbf{Case 3 b.1)} Contour plots for the sines of the CP phases, obtained for option 1 of the $(3,3)$ ISS framework, in the $(s/n-\theta_S)$ plane. 
From top to bottom (first to third row), $\sin\delta$, $\sin\alpha$ and $\sin\beta$.
The white/grey-shaded areas correspond to the black regions in the plots in Fig.~\ref{fig:Case3b1_s2thij_contour}, and indicate the regions in which all three mixing angles agree with experimental data at the $3 \, \sigma$ level.
Input parameters ($M_0$ and $y_0$) and colour coding as in Fig.~\ref{fig:Case3a_sdelta_salpha_sbeta_contour}.
Figures from~\cite{Hagedorn:2021ldq}.
\label{fig:Case3b1_sdelta_salpha_sbeta_contour}}}
\end{figure}
In what concerns the CP phases, we proceed in the same way as for the three mixing angles, and show in Fig.~\ref{fig:Case3b1_sdelta_salpha_sbeta_contour} several contour plots in the $(s/n-\theta_S)$ plane. 
We choose the same values of $M_0$ and $y_0$ as for the analogous study done for Case 3 a); conventions and colour-coding are identical to Fig.~\ref{fig:Case3a_sdelta_salpha_sbeta_contour}.
Like in Case 3 a), the visible differences for the different values of $M_0$ are mostly found in regions of the $(s/n-\theta_S)$ plane that disagree with experimental data on the three mixing angles by more than $3\, \sigma$.
We can observe that the absolute value of $\sin\delta$ has an upper bound $\sim 0.8$ for the choice $n=20$ and $m=11$, whereas the sines of both Majorana phases 
are a priori not constrained.
 Comparing the relative deviations of the sines of the CP phases, $\Delta\sin\alpha$, $\Delta\sin\beta$ and $\Delta\sin\delta$, with the analytical estimates, see Eq.~(\ref{eq:estimateDsy01}), we find agreement in the size; notice however that 
 the sign of the relative deviations $\Delta\sin\beta$ and $\Delta\sin\delta$ is positive. 

\begin{figure}[t!]
\begin{center}
\parbox{3in}{\includegraphics*[scale=0.5]{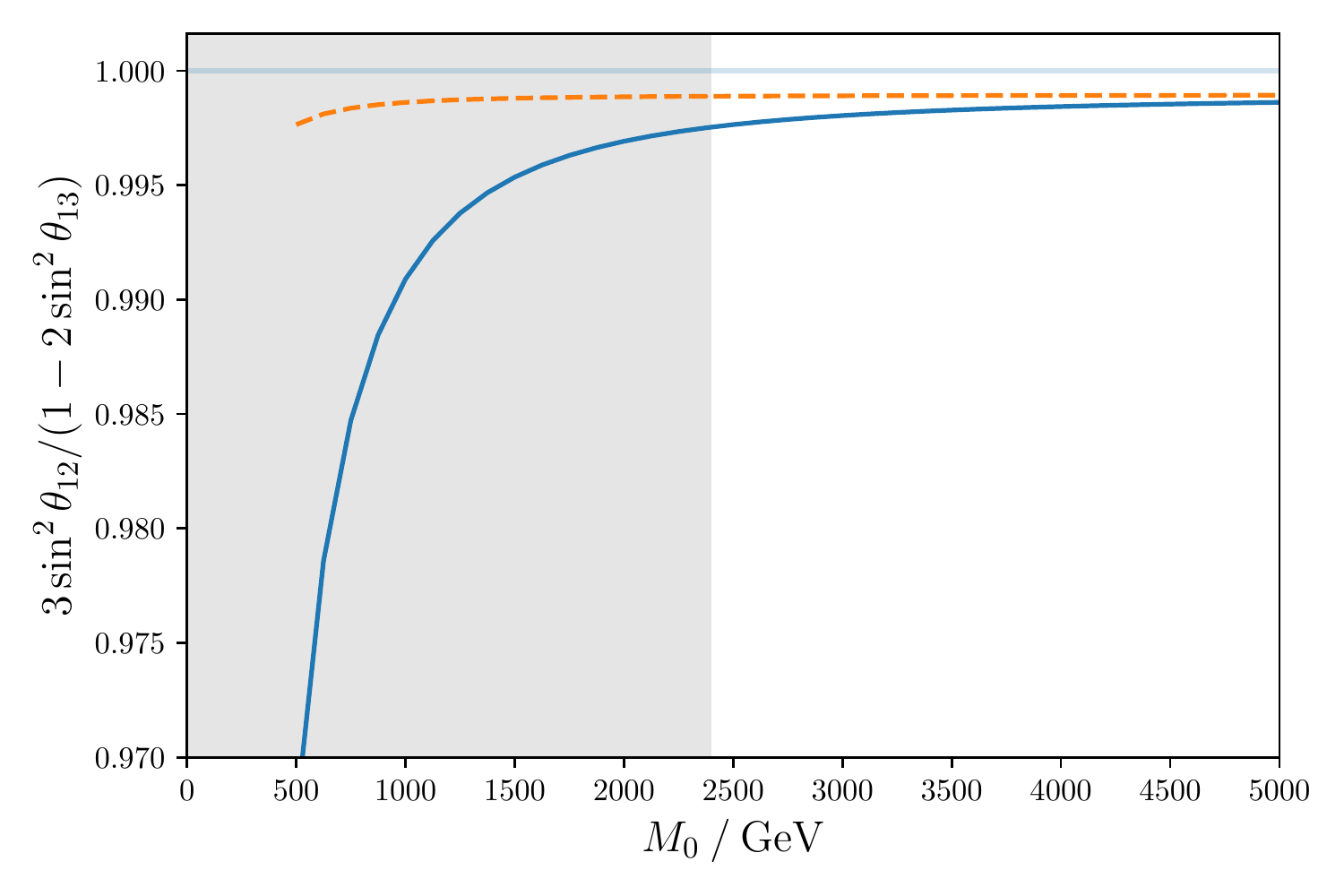}}
\hspace{0.2in}
\parbox{3in}{\includegraphics*[scale=0.5]{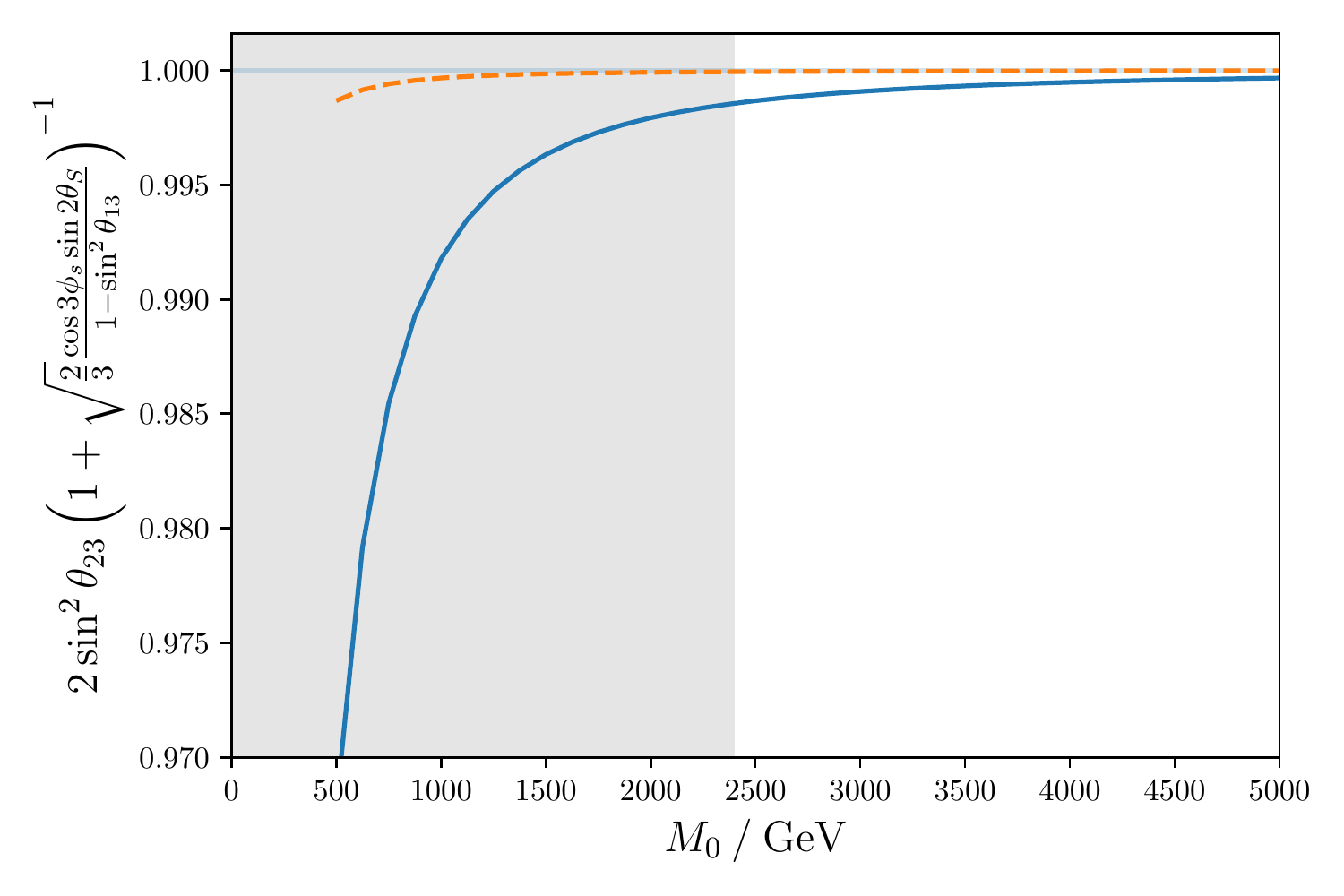}}
\end{center}
\caption{{\small \textbf{Case 3 b.1)} 
 Validity check of approximate sum rules in Eq.~(\ref{eq:sumrulesCase3b1}) for option 1 of the $(3,3)$ ISS framework 
 with respect to the mass $M_0$ (in GeV).
 In addition to $n=20$ and $m=10$, we fix $s=4$ ($\cos3 \, \phi_s \approx -0.31$) and $\theta_S \approx 1.83$ for the evaluation of both approximate sum rules.
Otherwise, same conventions and colour-coding as in Fig.~\ref{fig:Case1_s2thij}.
Figures from~\cite{Hagedorn:2021ldq}.
\label{fig:Case3b1_Sigma45}}}
\end{figure}
As shown in the model-independent scenario, several simplifications of the formulae in Eqs.~(\ref{eq:sin2thetaijCase3b1}, \ref{eq:Jarlskog:1985htI1I2Case3b1}) can be made for $m=\frac n2$ (corresponding to $m=10$ for the present case). 
In particular, two approximate sum rules are found, see Eq.~(\ref{eq:sumrulesCase3b1}). In the following, we investigate how these are affected by the presence of the ISS heavy sterile states. We proceed in an analogous way as done for the (approximate) sum rules found for the other cases. Our results are displayed in Fig.~\ref{fig:Case3b1_Sigma45} for two different values of the Yukawa coupling, $y_0=0.1$ and $y_0=0.5$, and can be compared to the analytical estimates for the relative deviations 
$\Delta\Sigma_4$ and $\Delta\Sigma_5$, see Eqs.~(\ref{eq:sumrule1Case3b1nonuni},\ref{eq:sumrule2Case3b1nonuni},\ref{eq:estimateDSigma45y01})
in Section~\ref{sec:ISSimpact}. We note that the results have been obtained for the choice $s=4$ ($\cos3 \, \phi_s \approx -0.31$). This choice has been made since it leads to a value of the atmospheric mixing angle which agrees best with current experimental data~\cite{Esteban:2020cvm}. Furthermore, we remark that we have replaced $\theta_0$ by $\theta_S$ in the second approximate sum rule in Eq.~(\ref{eq:sumrulesCase3b1})
 which, however, turns out to be very close to $\theta_0\approx 1.83$. 
 As can be seen in Fig.~\ref{fig:Case3b1_Sigma45}, for $y_0 \sim 0.5$ and $M_0 \sim 500$~GeV we find a deviation of about $-3 \%$ with respect to the results obtained in the model-independent approach. We thus confirm the analytical expectation (see Eq.~\eqref{eq:estimateDSigma45y01}), which was obtained for $y_0 = 1$ and $M_0 = 1000$~GeV (leading to the same value of $\eta_0$, cf. Eq.~(\ref{eq:etaopt1})).
For large $M_0$ the displayed ratios may not lead to exactly one, since the two sum rules only hold approximately.

For the choice of $m = \frac{n}{2} = 10$,
we can also check the (approximate) validity of the statements made for the sines of the CP phases and for the lower bound on the absolute value of the sine of the CP phase $\delta$, as observed in the  
model-independent scenario (compare to point $e)$ in Section~\ref{sec34}). Indeed, these hold, up to the expected deviations due to the effects of non-unitarity of $\tilde{U}_\nu$; moreover,  the equality of the sines of the two Majorana phases $\alpha$ and 
$\beta$ still holds exactly (see first equality in Eq.~(\ref{eq:sinaplhabetaCase3b1})).

For the choice $m=\frac n2=10$ and additionally $s=\frac n2=10$, one expects from the model-independent scenario that the atmospheric mixing angle and the Dirac phase are maximal, while both Majorana phases are trivial.
This also holds to a very good degree for option 1 of the $(3,3)$ ISS framework, for values of $M_0 \gtrsim 500$~GeV and $y_0 \lesssim 1$. 
In general, in all occasions in which a trivial CP phase is expected in the model-independent scenario, the same is obtained for option 1 of the $(3,3)$ ISS framework.

For Case 3 b.1) we numerically confirm that the symmetry transformations, given in Eq.~(\ref{eq:symmtrafosCase3}) under point $i)$ for Case 3 a), also hold.

\chapter{Statistics and treatment of experimental data}
\label{app:stats}
\minitoc

A proper statistical treatment of the experimental data and of the theoretical uncertainties is imperative for a precision analysis of flavour observables.
In general, the goal is to find a set of theoretical predictions for the observables of interest ($\mathcal O_i^\text{th}$) which agrees best with the experimental data on the observables ($\vec{\mathcal O}_i^\text{exp}$).
In order to determine the agreement with data, one builds a likelihood comprising the probability distributions of experimental data, evaluated at the theoretical predictions.
Schematically, we multiply the probability distribution functions (pdf) provided by the experimental data
\begin{equation}
	\mathcal L = \prod_i \mathrm{pdf}_i\left(O_i^\text{exp}, O_i^\text{th}(\vec p)\right)\,,
\end{equation}
in which the theoretical predictions depend on a set of given input parameters $\vec p$, all associated with additional sources of uncertainty.
Maximising this likelihood function then leads to the maximum likelihood estimator -- i.e. ``best-fit point'' -- as the point of highest probability.
In practice, one is only interested in a subset of the theoretical input parameters, or fit parameters ($\vec\theta$), leaving the remaining input parameters as nuisance parameters ($\vec\xi$) to be ``integrated out''. 
To do this, one in general follows either the \textit{Bayesian} or the \textit{Frequentist} approach, both computationally very expensive.

Another much faster approach which is used throughout this work is a \textit{gaussian approximation of the likelihood}, which can be written as
\begin{equation}
	- 2\Delta \mathrm{log} \mathcal L(\vec \theta) \approx \chi^2 = (\vec{\mathcal O}_\text{th}(\vec \theta) - \vec{\mathcal O}_\text{exp})^T \times \left(\mathcal C_\text{th} + \mathcal C_\text{exp}\right)^{-1}\times\left(\vec{\mathcal O}_\text{th}(\vec \theta)- \mathcal{\vec O}_\text{exp}\right)\,\text.
	\label{eqn:loglikelihood}
\end{equation}
In the above, $\vec{\mathcal O}_\text{exp}$ are the central values of the observables as measured by experiments, $\vec{\mathcal O}_\text{th}(\vec \theta)$ the central values of the theoretical predictions with respect to the nuisance parameters (but dependent on the fit parameters $\theta_i$), $\mathcal C_\text{exp}$ the covariance matrix of the measurements of \textit{all} included observables and $\mathcal C_\text{th}$ the covariance matrix of the predictions of \textit{all} included observables.
The theoretical covariance matrix now contains all theoretical uncertainties of the observables (and their correlations) and is obtained by randomly sampling the nuisance parameters according to their probability distributions.
Note that in this way the nuisance parameters $\vec \xi$ are ``effectively integrated out'' and the likelihood function to be optimised only depends on the parameters of interest, $\vec\theta$.
This approach was first employed in~\cite{Altmannshofer:2014rta}.

The experimental covariance matrix is estimated by first sampling  all experimental probability distributions (with a sample size of $10^6$ random values), including the effects of correlations among them.
In a second step, the mean values and the combined covariance matrix are estimated from the random samples. 
This however leads to an incorrect inclusion of strict upper limits, for instance a half-normal distribution, since mean values of samples drawn from a half-normal distribution (or related distributions) do not correspond to the true central values, which are $0$.
To circumvent this problem, all observables that only have experimental upper bounds are not included in Eq.~\eqref{eqn:loglikelihood}. Their likelihood is evaluated using their specific probability distributions (as provided by the experiments), at the expense of neglecting theoretical uncertainties. The probability distributions are then subsequently added to the global likelihood.

To take into account the theoretical uncertainties and correlations we use a similar Monte-Carlo method - all input parameters are randomly sampled ($N_\text{MC SM} = 10^4$) according to their probability distributions.
Then all observables are computed for each sample, to estimate the theoretical covariance matrix, which then also includes the theoretical correlations between observables.

The resulting approximate log-likelihood (or $\chi^2$) is then minimised using the \texttt{MIGRAD} algorithm implemented in the \texttt{minuit}~\cite{James:1975dr,iminuit} library. For the fits of the Wilson coefficients we compute the asymmetric errors with the \texttt{MINOS} algorithm. For leptoquark fits this however requires excessively large computation times. 
Therefore, we sample the likelihoods depending on the leptoquark couplings employing MCMC-simulations using the \texttt{emcee} python package~\cite{Foreman-Mackey:2012any}. This results in posterior distributions of the couplings and observables of interest.
The given best fit values are the maximum likelihood estimators as derived from the fit, in good agreement with the maximum likelihood estimators of the posterior distributions.
The quoted $90\%$ ranges are derived from the histograms of the posterior distributions. Here we take symmetric intervals between the $5^\text{th}$ and $95^\text{th}$ percentiles, while predicted upper limits (denoted as dashed lines) correspond to the $90^\text{th}$ percentile.

\chapter{Experimental data used in leptoquark fits}
\label{app:Obs}
\minitoc

\noindent
In this appendix we list the observables taken into account in the different fit set-ups (model-independent, see Chapter~\ref{chap:bphysics}, and for simplified leptoquark models, see Chapter~\ref{chap:lq}), as well as the datasets used for the fits. The observables (and datasets) are sorted according to the different hadronic and leptonic systems.
\mathversion{bold}
\section{Charged current $B$-decays}
\label{app:BFCCC}
\paragraph{Observables in $b\to c\ell\nu$}
\mathversion{normal}
First and foremost we include the very relevant LFUV ratios $R_{D^{(\ast)}}^{\tau\ell}$, commonly denoted $R_{D^{(\ast)}}$, into the global likelihoods. Analogously, a ratio comparing muons and electrons in the final state ($R_{D^{\ast}}^{\mu e}$) can be defined, which shows excellent agreement with the SM~\cite{Abdesselam:2017kjf, Abdesselam:2018nnh}.
For $R_{D^\ast}^{\tau\ell}$ we use the uncorrelated measurements by LHCb~\cite{Aaij:2015yra,Aaij:2017uff} and Belle~\cite{Hirose:2016wfn}, whereas for $R_D^{\tau\ell}$ there are several measurements,  obtained by BaBar~\cite{Lees:2013uzd} and Belle~\cite{Huschle:2015rga,Hirose:2016wfn,Abdesselam:2019dgh}, always in correlation with $R_{D^\ast}^{\tau\ell}$.

Numerous other observables are taken into account in addition to the anomalous ratios $R_{D^{(\ast)}}$. The extensive array of experimental data (in binned branching fractions of the decay $B\to D^{(\ast)}\ell\nu$) used in our fits is presented in Table~\ref{tab:binned_bcellnu}.

\begin{table}[h!]
\begin{center}
	\begin{tabular}{|c|c|c|}
	\hline
	Observables & $q^2$-bins in $\mathrm{GeV}^2$ & Datasets\\
	\hline
	\hline
	$\braket{\mathrm{BR}}(B^+\to D\tau\nu)$ & $[4, 4.53]$, $[4.53, 5.07]$, $[5.07, 5.6]$, $[5.6, 6.13]$ & Belle'15~\cite{Huschle:2015rga}\\
    $\braket{\mathrm{BR}}(B^0\to D\tau\nu)$	& $[6.13, 6.67]$, $[6.67, 7.2]$, $[7.2, 7.73]$, $[7.73, 8.27]$ & \\
	& $[8.27, 8.8]$, $[8.8, 9.33]$, $[9.33, 9.86]$, $[9.86, 10.4]$ & \\
	& $[10.4, 10.93]$, $[10.93, 11.47]$, $[11.47, 12.0]$ &\\
	\hline
	\hline
	$\braket{\mathrm{BR}}(B^+\to D^\ast\tau\nu)$ & $[4, 4.53]$, $[4.53, 5.07]$, $[5.07, 5.6]$, $[5.6, 6.13]$ & Belle'15~\cite{Huschle:2015rga}\\
    $\braket{\mathrm{BR}}(B^0\to D^\ast\tau\nu)$ & $[6.13, 6.67]$, $[6.67, 7.2]$, $[7.2, 7.73]$, $[7.73, 8.27]$ & \\
	& $[8.27, 8.8]$, $[8.8, 9.33]$, $[9.33, 9.86]$, $[9.86, 10.4]$ & \\
	& $[10.4, 10.93]$ & \\
	\hline
	\hline
	$\braket{\mathrm{BR}}(B^+\to D\mu\nu)$ & $[0.0, 1.03]$, $[1.03, 2.21]$, $[2.21, 3.39]$, $[3.39, 4.57]$ & Belle'15~\cite{Glattauer:2015teq}\\
	$\braket{\mathrm{BR}}(B^+\to D e\nu)$ & $[4.57, 5.75]$, $[5.75, 6.93]$, $[6.93, 8.11]$, $[8.11, 9.3]$ &\\
	&  $[9.3, 10.48]$, $[10.48, 11.66]$ &\\
	\hline
	\hline
	$\braket{\mathrm{BR}}(B^0\to D\mu\nu)$ & $[0.0, 0.97]$, $[0.97, 2.15]$, $[2.15, 3.34]$, $[3.34, 4.52]$ & Belle'15~\cite{Glattauer:2015teq}\\
	$\braket{\mathrm{BR}}(B^0\to D e\nu)$ & $[4.52, 5.71]$, $[5.71, 6.89]$, $[6.89, 8.07]$, $[8.07, 9.26]$ &\\
	& $[9.26, 10.44]$, $[10.44, 11.63]$ &\\
	\hline
	\end{tabular}
	\caption{Datasets of binned branching fractions in $B\to D^{(\ast)}\ell\nu$.}
	\label{tab:binned_bcellnu}
\end{center}
\end{table}

\noindent
Furthermore, we include the unbinned branching fractions $\mathrm{BR}(B^+\to D^{(\ast)} \mu\nu)$, $\mathrm{BR}(B^+\to D^{(\ast)} e\nu)$~\cite{Aubert:2008yv, Aubert:2007qs} and the inclusive branching fraction $\mathrm{BR}(B\to X_c e\nu)$~\cite{Urquijo:2006wd, Aubert:2009qda}.

\mathversion{bold}
\paragraph{Other charged current $B$-decays}
\mathversion{normal}
In addition to charged current $b\to c\ell\nu$ decays, we also include certain $b\to u\ell\nu$ decays to obtain further constraints on the leptoquark couplings to the first quark generation. These can be found in Table~\ref{tab:buellnu}.
\begin{table}[h!]
\begin{center}
	\begin{tabular}{|c|c|c|}
	\hline
	Observable & SM prediction & Measurement/Limit\\
	\hline
	\hline
	$\mathrm{BR}(B^0\to\pi\tau\nu)$ & $(8.4\pm1.1)\times10^{-5}$ & $(1.52 \pm 0.72 \pm 0.13) \times 10^{-4}$ Belle'15~\cite{Hamer:2015jsa}\\
	\hline
	$\mathrm{BR}(B^+\to\tau\nu)$ & $(8.8\pm0.6)\times10^{-5}$ & $(1.09 \pm 0.24) \times 10^{-4}$ PDG~\cite{Tanabashi:2018oca}\\
	\hline
	$\mathrm{BR}(B^+\to\mu\nu)$ & $(4.0\pm0.3)\times10^{-7}$ & $<1\times10^{-6}$ HFLAV'18~\cite{Amhis:2019ckw}\\
	\hline
	\end{tabular}
	\caption{Datasets on further charged current $B$-meson decays. The SM predictions are obtained using \texttt{flavio}~\cite{Straub:2018kue}.}
	\label{tab:buellnu}
\end{center}
\end{table}
\mathversion{bold}
\section{Observables from $b\to s\ell\ell$ transitions}
\mathversion{normal}
\label{app:bsll}
Leading to the fits of Sections~\ref{sec:bsll} and~\ref{sec:lqfit}, we include a large number of different binned and unbinned observables into the respective likelihoods. These play a crucial r\^ole in efficiently constraining the $b\to s$ transition FCNC operators and subsequently the leptoquark couplings involved.

\mathversion{bold}
\paragraph{Binned observables in $b\to s\ell\ell$}
\mathversion{normal}
We take into account all available data for the angular observables in the optimised basis \cite{Descotes-Genon:2013vna}. Depending on the experiment providing the data, the (sub)sets of observables and bins vary. The datasets for the angular observables taken into account is summarised in Table \ref{tab:ang_data}, whereas the data on the differential branching fractions is shown in Table \ref{tab:br_bsll}.
We notice that in all cases we neglect the bin between $6$ and $8\:\mathrm{GeV}^2$ as, due to the $c\bar c$ resonances,
QCD factorisation is no longer a good approximation in this region~\cite{Beneke:2001at}.
Furthermore, we do not take into account the bin $[0.1, 0.98]\:\mathrm{GeV}^2$: 
the different form factor treatments in \texttt{flavio}~\cite{Straub:2018kue} and Ref.~\cite{Descotes-Genon:2013vna} lead to significant discrepancies in the associated theoretical uncertainties in this bin, while for all other bins there is a good agreement.
Moreover, in the region of large hadronic recoil, we always take into account the narrow bins, whereas at low hadronic recoil we average over the kinematic region above the resonances.

\begin{table}[h!]
\begin{center}
	\begin{tabular}{|c|c|c|}
	\hline
	Observables & $q^2$-bins in $\mathrm{GeV}^2$ & Datasets\\
	\hline
	\hline
	$\braket{\mathcal O}(B^{0}\to K^\ast\mu^+\mu^-)$ & &\\
	\hline
	$\braket{F_L}$, $\braket{P_1}$, $\braket{P_2}$, $\braket{P_3}$, & $[1.1, 2.5]$, $[2.5, 4]$, &  LHCb'15\cite{Aaij:2015oid}, LHCb'20\cite{Aaij:2020nrf}\\
	$\braket{P_4^\prime}$, $\braket{P_5^\prime}$, $\braket{P_6^\prime}$, $\braket{P_8^\prime}$  &$[4,6]$, $[15, 19]$ & \\
	\hline
	$\braket{F_L}$, $\braket{P_1}$, $\braket{P_4^\prime}$ & $[0.04, 2]$, $[2,4]$, $[4, 6]$ & ATLAS'17\cite{Aaboud:2018krd}\\
	$\braket{P_5^\prime}$, $\braket{P_6^\prime}$, $\braket{P_8^\prime}$ & & \\
	\hline
	$\braket{F_L}$, $\braket{A_{FB}}$, & $[1,2]$, $[2, 4.3]$ & CMS'17\cite{CMS:2017ivg} \\
	 $\braket{P_1}$, $\braket{P_5^\prime}$ & $[4.3, 6]$, $[16,19]$ & \\
	\hline
	$\braket{F_L}$, $\braket{A_{FB}}$ & $[0,2]$, $[2,4.3]$, $[16, 19.3]$ & CDF'12\cite{CDF:2012qwd}\\
	\hline
    \hline
	$\braket{\mathcal O}(B^{+}\to K^\ast\mu^+\mu^-)$ & &\\
	\hline
	$\braket{F_L}$, $\braket{P_1}$, $\braket{P_2}$, $\braket{P_3}$, & $[1.1, 2.5]$, $[2.5, 4]$, &  LHCb'20\cite{Aaij:2020ruw}\\
	$\braket{P_4^\prime}$, $\braket{P_5^\prime}$, $\braket{P_6^\prime}$, $\braket{P_8^\prime}$  &$[4,6]$, $[15, 19]$ & \\
	\hline
	\hline
	$\braket{\mathcal O}(B^0\to K^\ast e^+ e^-)$ & &\\
	\hline
	$\braket{F_L}$, $\braket{P_1}$, $\braket{P_2}$, $\braket{\mathrm{Im}(A_T)}$ & $[0.0008, 0.257]$ & LHCb'20~\cite{Aaij:2020umj}\\
	\hline
	\hline
	$\braket{\mathcal O}(B_s\to \phi \mu^+ \mu^-)$ & &\\
	\hline
	$\braket{F_L}$, $\braket{S_3}$, $\braket{S_4}$, $\braket{S_7}$ & $[0.1, 2]$, $[2, 5]$, $[15, 19]$ & LHCb'15~\cite{Aaij:2015esa}\\
	\hline
	\end{tabular}
	\caption{Datasets on angular $b\to s\mu\mu$ observables taken into account in the analysis. The 2 digits appearing after each collaborations' name denote the years of the respective publications.}
	\label{tab:ang_data}
\end{center}
\end{table}

\begin{table}[h!]
\begin{center}
	\begin{tabular}{|c|c|c|}
	\hline
	Observables & $q^2$-bins in $\mathrm{GeV}^2$ & Datasets\\
	\hline
	\hline
	$\braket{\frac{\mathrm{dBR}}{\mathrm d q^2}}(B^+\to K^+\mu^+\mu^-)$ & $[1.1, 2]$, $[2,3]$, $[3,4]$ & LHCb'14~\cite{Aaij:2014pli}\\
	& $[4, 5]$, $[5, 6]$, $[15, 22]$ & \\
	\hline
	$\braket{\frac{\mathrm{dBR}}{\mathrm d q^2}}(B^0\to K^0\mu^+\mu^-)$ & $[0.1, 2]$, $[2,4]$, $[4, 6]$, $[15, 22]$ & LHCb'14~\cite{Aaij:2014pli}\\
	\hline
	$\braket{\frac{\mathrm{dBR}}{\mathrm d q^2}}(B^+\to K^*\mu^+\mu^-)$  & 
	$[0.1, 2]$, $[2, 4]$, $[4, 6]$, $[15, 19]$ & LHCb'14~\cite{Aaij:2014pli}\\
	\hline
	$\braket{\frac{\mathrm{dBR}}{\mathrm d q^2}}(B^0\to K^*\mu^+\mu^-)$ &
	$[1.1, 2.5]$, $[2.5, 4]$, $[4, 6]$, $[15, 19]$ & LHCb'16~\cite{Aaij:2016flj}\\
	\hline
	\hline
	$\braket{\frac{\mathrm{dBR}}{\mathrm d q^2}}(B_s\to \phi\mu^+\mu^-)$ & 
	$[0.1, 2]$, $[2, 5]$, $[15, 19]$ & LHCb'15~\cite{Aaij:2015esa}\\
	& $[1,2.5], [2.5,4], [4,6]$ & LHCb'21~\cite{LHCb:2021zwz}\\
	\hline
	\end{tabular}
	\caption{Datasets on binned differential branching ratios in $B\to K^{(*)}\mu\mu$ decays taken into account in the analysis.}
	\label{tab:br_bsll}
\end{center}
\end{table}

In addition to the binned observables in $b\to s\mu\mu$, we also include the $b\to s\ell\ell$ LFUV observables  into the likelihoods.
The bins and datasets of the ratios of (differential) branching fractions $R_{K^{(\ast)}}$, as well as differences of angular observables between electrons and muons in the final state,
\begin{equation}
    Q_{4,5} \equiv P_{4,5}^{'\mu\mu} - P_{4,5}^{'ee} 
\end{equation}
are listed in Table~\ref{tab:lfuv_bsll}.

\begin{table}[h!]
\begin{center}
	\begin{tabular}{|c|c|c|}
	\hline
	Observables & $q^2$-bins in $\mathrm{GeV}^2$ & Datasets\\
	\hline
	\hline
	$\braket{R_K}$ & $[1.1, 6.0]$, $[0.1, 4.0]$, $[1.0,6.0]$, $[14.18, 19.0]$ & (LHCb'19~\cite{Aaij:2019wad}), LHCb'21~\cite{LHCb:2021trn}, Belle'19~\cite{Abdesselam:2019lab}\\
	\hline
	$\braket{R_{K^\ast}}$ & $[0.045, 1.1]$, $[1.1, 6.0]$, $[15,19]$ & LHCb'17~\cite{Aaij:2017vbb}, Belle'19~\cite{Abdesselam:2019wac}\\
	\hline
	$\braket{Q_4}$, $\braket{Q_5}$ & $[0.1, 4]$, $[1.0, 6.0]$, $[14.18, 19.0]$ & Belle'16~\cite{Wehle:2016yoi}\\
	\hline
	\end{tabular}
	\caption{Datasets of observables in $B\to K^{(*)}\ell\ell$ decays sensitive to LFU violation.}
	\label{tab:lfuv_bsll}
\end{center}
\end{table}

\paragraph{Leptonic FCNC decays}
Having sizeable new physics effects in $B\to K^{(\ast)}\mu\mu$ (as required to fit the anomalous data) opens the possibility of having new contributions to other rare $b\to s\ell\ell$ decays, which have either been found to be consistent with the SM, or are yet to be observed. 

Meson decay modes without a hadron in the final state suffer from significantly smaller hadronic uncertainties, since QCD corrections can be absorbed into a redefinition of the decay constant, and all QED and electroweak corrections remain fully perturbative.
Consequently, these decays provide very clean probes for NP effects especially in $C_{7, 10}^{(')}$, but also in $C_{S,P}^{(')}$ Wilson Coefficients.
A recent LHCb analysis~\cite{Aaij:2020nol} of $B_{(s)}\to ee$ yields upper bounds at the $\mathcal{O}(10^{-9})$ level. 
For $B_{(s)}\to \mu\mu$, the situation is more complicated, since the decays are always measured in correlation to each other.
While the decay $B_s\to \mu\mu$ has been observed and measured by several experiments~\cite{LHCb:2021qbv,LHCb:2021awg,Chatrchyan:2013bka, Aaij:2017vad, Aaboud:2018mst, Sirunyan:2019xdu}, as of today only upper limits on the decay $B^0\to\mu\mu$ are available (at the $10^{-10}$ level), due to insufficient statistics. 
In order to avoid losing important correlations in the measurements, we use the 2-dimensional likelihoods (including negative values for $\mathrm{BR}(B^0\to\mu\mu)$) and sample them to obtain a naïve combination, following the prescription of Ref.~\cite{Aebischer:2019mlg,Altmannshofer:2021qrr}. 

\paragraph{Other observables}
To constrain contributions to $C^{(')\:bs\gamma}_7$ in the dipole operator, we also include the branching fractions $\mathrm{BR}(B\to K^\ast\gamma)$~\cite{Amhis:2014hma}, $\mathrm{BR}(B\to X_s\gamma)$~\cite{Misiak:2017bgg} and $\mathrm{BR}(B_s\to\phi\gamma)$\cite{Dutta:2014sxo, Aaij:2012ita}. Notice that all these observables 
correspond to the full branching fractions, implying that they are calculated and measured over the full kinematic region.

\mathversion{bold}
\section{Strange, charm and $\tau$-lepton decays}
\mathversion{normal}
\label{app:sctau}
The above listed data mostly allows to constrain combinations of second and third generation quark leptoquark couplings (to all leptons). To achieve more precise constraints for the second and first generation quarks, we further include numerous decays of strange and charm flavoured mesons.
Since the light mesons cannot decay into $\tau$-leptons, we also use data on SM allowed $\tau$-lepton decays, as a complementary source of information.

\paragraph{Binned charm decays}
In addition to the precise measurements of the full branching fractions of several charmed meson decay modes, there are also precise measurements of the $q^2$ distributions for several charged current decay modes in semi-leptonic charm decays with an electron in the final state. The datasets used are presented in Table~\ref{tab:binned_charm}.
\begin{table}[h!]
\begin{center}
	\begin{tabular}{|c|c|c|}
	\hline
	Observables & $q^2$-bins in $\mathrm{GeV}^2$ & Datasets\\
	\hline
	\hline
	$\braket{\mathrm{BR}}(D^{+,\,0}\to K e\nu)$ & $[0.0,0.2]$, $[0.2,0.4]$, $[0.4,0.6]$, $[0.6,0.8]$ & CLEO~\cite{Besson:2009uv}, BESIII~\cite{Ablikim:2015ixa,Ablikim:2017lks}\\
	& $[0.8, 1.0]$, $[1.2, 1.4]$, $[1.4, 1.6]$, $[1.6, 1.88]$ & \\
	\hline
	$\braket{\mathrm{BR}}(D^{0}\to \pi e\nu)$ & $[0.0,0.2]$, $[0.2,0.4]$, $[0.4,0.6]$, $[0.6,0.8]$ &  BESIII~\cite{Ablikim:2015ixa}\\
	& $[0.8, 1.0]$, $[1.2, 1.4]$, $[1.4, 1.6]$, $[1.6, 1.8]$ & \\
	& $[1.8, 2.0]$, $[2.0, 2.2]$, $[2.2, 2.4]$, $[2.4, 2.6]$ & \\
	& $[2.6, 2.98]$ & \\
	\hline
	$\braket{\mathrm{BR}}(D^{+}\to \pi e\nu)$ & $[0.0,0.3]$, $[0.3,0.6]$, $[0.6,0.9]$, $[0.9,1.2]$ &  CLEO~\cite{Besson:2009uv}, BESIII~\cite{Ablikim:2017lks}\\
	& $[1.2, 1.5]$, $[1.5, 2.0]$, $[2.0, 2.98]$ & \\
	\hline
	\end{tabular}
	\caption{Datasets on binned branching fractions in charged current charm decays.}
	\label{tab:binned_charm}
\end{center}
\end{table}

\paragraph{Unbinned observables}
Besides the binned semi-leptonic charm decays, we also include the full branching fractions for charged current leptonic and semi-leptonic charm decays, charged and neutral current decays of strange flavoured mesons, and charged current semi-leptonic $\tau$-lepton decays. The charged current decays are listed in Table~\ref{tab:charged_charm_strange_tau} and the neutral current ones in Table~\ref{tab:fcnc_strange}.
\begin{table}[h!]
\begin{center}
	\begin{tabular}{|c|c|c|}
	\hline
	Observable & SM prediction & Measurement/Limit\\
	\hline
	\hline
	$\mathrm{BR}(D^0\to K\mu\nu)$ & $(3.54\pm0.25)\times10^{-2}$ & $(3.31\pm0.13)\times10^{-2}\quad$~\cite{Tanabashi:2018oca}\\
	\hline
	$\mathrm{BR}(D^0\to Ke\nu)$ & $(3.55\pm0.25)\times10^{-2}$ & $(3.53\pm0.028)\times10^{-2}\quad$~\cite{Tanabashi:2018oca}\\
	\hline
	$\mathrm{BR}(D^+\to K\mu\nu)$ & $(9.04\pm0.55)\times10^{-2}$ & $(8.74\pm0.19)\times10^{-2}\quad$~\cite{Tanabashi:2018oca}\\
	\hline
	$\mathrm{BR}(D^+\to Ke\nu)$ & $(9.08\pm0.64)\times10^{-2}$ & $(8.73\pm0.0)\times10^{-2}\quad$~\cite{Tanabashi:2018oca}\\
	\hline
	$\mathrm{BR}(D^0\to\pi\mu\nu)$ & $(2.67\pm0.16)\times10^{-3}$ & $(2.37\pm0.24)\times10^{-3}\quad$~\cite{Tanabashi:2018oca}\\
	\hline
	$\mathrm{BR}(D^0\to\pi e\nu)$ & $(2.68\pm0.15)\times10^{-3}$ & $(2.91\pm0.04)\times10^{-3}\quad$~\cite{Tanabashi:2018oca}\\
	\hline
	$\mathrm{BR}(D^+\to\pi e\nu)$ & $(3.48\pm0.22)\times10^{-3}$ & $(3.72\pm0.17)\times10^{-3}\quad$~\cite{Tanabashi:2018oca}\\
	\hline
	\hline
	$\mathrm{BR}(D^+\to \tau\nu)$ & $(1.09\pm0.01)\times10^{-3}$ & $<1.2\times10^{-3}\quad$~\cite{Tanabashi:2018oca}\\
	\hline
	$\mathrm{BR}(D^+\to \mu\nu)$ & $(4.10\pm0.05)\times10^{-4}$ & $(3.74\pm0.17)\times10^{-4}\quad$~\cite{Tanabashi:2018oca}\\
	\hline
	$\mathrm{BR}(D^+\to e\nu)$ & $(9.64\pm0.12)\times10^{-9}$ & $<8.8\times10^{-6}\quad$~\cite{Tanabashi:2018oca}\\
	\hline
	\hline
	$\mathrm{BR}(D_s\to \tau\nu)$ & $(5.32\pm0.05)\times10^{-2}$ & $(5.48\pm0.23)\times10^{-2}\quad$~\cite{Tanabashi:2018oca}\\
	\hline
	$\mathrm{BR}(D_s\to \mu\nu)$ & $(5.46\pm0.05)\times10^{-3}$ & $(5.50\pm0.23)\times10^{-3}\quad$~\cite{Tanabashi:2018oca}\\
	\hline
	$\mathrm{BR}(D_s\to e\nu)$ & $(1.28\pm0.01)\times10^{-7}$ & $<8.3\times10^{-5}\quad$~\cite{Tanabashi:2018oca}\\
	\hline
	\hline
	$\mathrm{BR}(K^+\to\pi\mu\nu)$ & $(3.39\pm0.04)\times10^{-2}$ & $(3.35\pm0.03)\times10^{-2}\quad$~\cite{Tanabashi:2018oca}\\
	\hline
	$\mathrm{BR}(K^+\to\pi e\nu)$ & $(5.13\pm0.05)\times10^{-2}$ & $(5.07\pm0.04)\times10^{-2}\quad$~\cite{Tanabashi:2018oca}\\
	\hline
	$\mathrm{BR}(K_L\to\pi\mu\nu)$ & $(27.11\pm0.26)\times10^{-2}$ & $(27.04\pm0.07)\times10^{-2}\quad$~\cite{Tanabashi:2018oca}\\
	\hline
	$\mathrm{BR}(K_L\to\pi e\nu)$ & $(40.93\pm0.46)\times10^{-2}$ & $(40.55\pm0.11)\times10^{-2}\quad$~\cite{Tanabashi:2018oca}\\
	\hline
	\hline
	$\mathrm{BR}(K^+\to\mu\nu)$ & $(63.08\pm0.83)\times10^{-2}$ & $(63.56\pm0.11)\times10^{-2}\quad$~\cite{Tanabashi:2018oca}\\
	\hline
	$\mathrm{BR}(K^+\to e\nu)$ & $(1.561\pm0.023)\times10^{-5}$ & $(1.582 \pm 0.007) \times10^{-5}\quad$~\cite{Tanabashi:2018oca}\\
	\hline
	\hline
	$\mathrm{BR}(\tau\to K\nu)$ & $(7.09\pm0.11)\times10^{-3}$ & $(6.96\pm0.10)\times10^{-3}\quad$~\cite{Tanabashi:2018oca}\\
	\hline
	$\mathrm{BR}(\tau\to \pi\nu)$ & $(10.84\pm0.14)\times10^{-2}$ & $(10.82\pm0.05)\times10^{-3}\quad$~\cite{Tanabashi:2018oca}\\
	\hline
	\end{tabular}
	\caption{Data on charged current charm and strange flavoured meson decays. The SM predictions are obtained using \texttt{flavio}~\cite{Straub:2018kue}.}
	\label{tab:charged_charm_strange_tau}
\end{center}
\end{table}

\begin{table}[h!]
\begin{center}
	\begin{tabular}{|c|c|c|}
	\hline
	Observable & SM prediction & Measurement/Limit\\
	\hline
	\hline
	$\mathrm{BR}(K_L\to \mu^+\mu^-)$ & $(7.45\pm1.24)\times10^{-9}$ & $(6.84\pm0.11)\times10^{-9}\quad$~\cite{Tanabashi:2018oca}\\
	\hline
	$\text{BR}(K^+ \to \pi^+ \nu\bar \nu)$ &
	$(8.4 \pm 1.0) \times 10^{-11}
	\phantom{|}^{\phantom{|}}_{\phantom{|}}$\cite{Buras:2015qea} &
	\begin{tabular}{l}
	$17.3^{+11.5}_{-10.5} \times 10^{-11}
	\quad$\cite{Artamonov:2008qb}
	\\
	$< 1.78 \times 10^{-10}
	\quad$\cite{CortinaGil:2020vlo}
	\end{tabular}
	\\
	\hline
	$\text{BR}(K_L \to \pi^0 \nu\bar \nu)$ &
	$(3.4 \pm 0.6) \times 10^{-11}\quad$\cite{Buras:2015qea} &
	$<  2.6 \times 10^{-8}
	\quad$\cite{Ahn:2009gb}\\
	\hline
	\end{tabular}
	\caption{Data on FCNC kaon decays. The SM predictions are obtained using \texttt{flavio}~\cite{Straub:2018kue} if not otherwise stated.}
	\label{tab:fcnc_strange}
\end{center}
\end{table}

\section{Belle II Observables}
As discussed in Section~\ref{sec:future_Belle}, we use specific fit set-ups which allow for an extrapolation of the current situation into the near future.
The future sensitivities, taken into account as data, are listed in Table~\ref{tab:belleii}; these always correspond to the full anticipated luminosity of $50\:\mathrm{ab}^{-1}$.

\begin{table}[h!]
\begin{center}
	\begin{tabular}{|c|c|c|}
	\hline
	Observable & Current bound & Belle II Sensitivity\\
	\hline
	\hline
	$\text{BR}(\tau \to e \gamma)$	&
 	\quad $<3.3\times 10^{-8}$ \quad BaBar~\cite{Aubert:2009ag}	 &
 	\quad $<3\times10^{-9}$ \quad  	 	\\
	\hline
	$\text{BR}(\tau \to \mu \gamma)$	&
	 \quad $ <4.4\times 10^{-8}$ \quad BaBar~\cite{Aubert:2009ag}	 &
 	\quad $<10^{-9}$ \quad 		\\
	\hline
	$\text{BR}(\tau \to 3 e)$	&
 	\quad $<2.7\times 10^{-8}$ \quad Belle~\cite{Hayasaka:2010np}&
 	\quad $<5\times10^{-10}$ \quad  	\\
 	\hline
	$\text{BR}(\tau \to 3 \mu )$	&
 	\quad $<3.3\times 10^{-8}$ \quad Belle~\cite{Hayasaka:2010np}	 &
 	\quad $<5\times10^{-10}$ \quad 		\\
	\hline
	\hline
	$\mathrm{BR}(\tau\to\pi e)$ & $<8\times10^{-8}$\quad Belle~\cite{Miyazaki:2007jp} & $<4\times10^{-10}$\\
	\hline
	$\mathrm{BR}(\tau\to\pi \mu)$ & $<1.1\times10^{-7}$\quad Belle~\cite{Miyazaki:2007jp} & $<5\times10^{-10}$\\
	\hline
	$\mathrm{BR}(\tau\to\phi e)$ & $<3.1\times10^{-8}$\quad Belle~\cite{Miyazaki:2011xe} & $<5\times10^{-10}$\\
	\hline
	$\mathrm{BR}(\tau\to\phi \mu)$ & $<8.4\times10^{-8}$\quad Belle~\cite{Miyazaki:2011xe} & $<2\times10^{-9}$\\
	\hline
	$\mathrm{BR}(\tau\to\rho e)$ & $<1.8\times10^{-8}$\quad Belle~\cite{Miyazaki:2011xe} & $<3\times10^{-10}$\\
	\hline
	$\mathrm{BR}(\tau\to\rho \mu)$ & $<1.2\times10^{-8}$\quad Belle~\cite{Miyazaki:2011xe} & $<2\times10^{-10}$\\
	\hline
	\hline
	$\mathrm{BR}(B^+\to K^+\tau^+e^-)$ & $<1.5\times10^{-5}$\quad BaBar~\cite{Lees:2012zz} & $<2.1\times10^{-6}$\\
	$\mathrm{BR}(B^+\to K^+\tau^-e^+)$ & $<4.3\times10^{-5}$\quad BaBar~\cite{Lees:2012zz} &\\
	\hline
	$\mathrm{BR}(B^+\to K^+\tau^+\mu^-)$ & $<2.8\times10^{-5}$\quad BaBar~\cite{Lees:2012zz} & $<3.3\times10^{-6}$\\
	$\mathrm{BR}(B^+\to K^+\tau^-\mu^+)$ & $<4.5\times10^{-5}$\quad BaBar~\cite{Lees:2012zz} &\\
	\hline
	\hline
	$\mathrm{BR}(B^0\to e^\pm\tau^{\mp})$ & $<2.8\times10^{-5}$\quad BaBar~\cite{Aubert:2008cu} & $ <1.6\times10^{-5}$\\
	\hline
	$\mathrm{BR}(B^0\to \mu^\pm\tau^{\mp})$ & $<1.4\times10^{-5}$\quad LHCb~\cite{Aaij:2019okb} & $ <1.3\times10^{-5}$\\
	\hline
	\end{tabular}

	\vspace{0.5cm}
	\begin{tabular}{|c|c|c|}
	\hline
	Observable & SM prediction & Belle II Sensitivity\\
	\hline
	\hline
	$\mathrm{BR}(B^0\to\tau\tau)$ & $(2.22 \pm 0.19)\times10^{-8}$\quad~\cite{Bobeth:2013uxa,Hermann:2013kca, Bobeth:2013tba} & $<9.6\times10^{-5}$\\
	\hline
	$\mathrm{BR}(B_s\to\tau\tau)$ & $(7.73 \pm 0.49)\times10^{-7}$\quad~\cite{Bobeth:2013uxa,Hermann:2013kca, Bobeth:2013tba} & $<8.1\times10^{-4}$\\
	\hline
	$\braket{\mathrm{BR}}(B\to K\tau^+\tau^-)_{[15, 22]}$ & $(1.20\pm0.12)\times10^{-7}$\quad~\cite{Capdevila:2017iqn} & $<2\times10^{-5}$\\
	\hline
	\end{tabular}
	\caption{Observables for which Belle II will improve on current experimental sensitivities. The SM predictions are obtained using \texttt{flavio}~\cite{Straub:2018kue}, unless otherwise stated.}
	\label{tab:belleii}
\end{center}
\end{table}
\backmatter

\addchap{Résumé en français}
\section{Introduction}
Suite \`a la découverte en 2012 d'un boson scalaire au LHC, ayant les propriétés d'un boson de Higgs, le secteur électrofaible du modèle standard (SM) a finalement été complété~\cite{ATLAS:2015yey}.
Bien que s'agissant d'une percée massive, cette découverte était bien anticipée, car les résultats des mesures de précision électrofaibles du LEP et du Tevatron indiquaient que, si le modèle standard était une description exacte de la nature, le LHC devrait découvrir un boson scalaire avec une masse autour de $\sim 100\:\mathrm{GeV}$.
Dans le passé, les mesures de précision des désintégrations électrofaibles, comme par exemple la désintégration du muon, ont conduit à des limites inférieures fortes sur les masses des bosons de jauge électrofaibles bien avant leur découverte directe. La construction du secteur des saveurs du modèle standard a connu une évolution similaire :
après la découverte du quark étrange, le ``modèle à trois quarks" avec une symétrie de saveur $SU(3)$ a d'abord conduit à la prédiction de nombreux nouveaux états liés sous la forme de baryons et de mésons, qui ont ensuite été découverts.
Ce ``modèle" avait cependant le problème frappant de prédire des courants neutres changeant de saveur (FCNC) au niveau de l'arbre, ce qui n'était pas confirmé par les données expérimentales.
Cela a ainsi conduit à l'hypothèse du quark charme, afin de supprimer les transitions FCNC via un mécanisme généralisé de Glashow-Iliopoulos-Maiani (GIM)~\cite{Glashow:1970gm}.
De plus, la découverte de la violation de CP dans les désintégrations de kaons a conduit à l'hypothèse d'une troisième génération de quarks, car la violation de CP n'est possible que s'il y a au moins trois familles~\cite{Kobayashi:1973fv}.
Les mesures de précision ultérieures du mélange de mésons neutres $K^0 - \bar K^0$ ont permis d'établir des limites inférieures strictes sur la masse du quark top, qui serait beaucoup plus lourd que les autres quarks.
Le vaste effort combiné des recherches expérimentales directes et indirectes, ainsi que des études phénoménologiques visant à interpréter les données au cours des soixante dernières années a mené la physique des particules dans une ère sans précédent.  
Dans presque tous les secteurs de la physique des hautes énergiesLes, les mesures de précision  corroborent les prédictions du modèle standard avec une grande exactitude.

En dépit de ce qui est devenu une réussite théorique et phénoménologique évidente, le modèle standard a été confronté avec certains échecs, suggérant que celui ci ne pouvait pas  complètement décrire la nature. La raison la plus évidente est liée aux fait que le MS ne tient pas compte d'une théorie quantique de la gravité, et ne peut donc pas décrire toutes les interactions fondamentales connues.
En outre, la description du secteur de Higgs est loin d'être satisfaisante, car elle ne repose pas sur un principe fondamental.
La compréhension du mécanisme exact de la brisure de symétrie électrofaible est également liée au problème des saveurs ; pourquoi les masses des fermions sont-elles si hiérarchisées ? Pourquoi y a-t-il trois générations de fermions ?
Au-delà des questions théoriques (et esthétiques), le modèle standard manque d'un candidat viable pour la matière noire et ne peut pas expliquer l'asymétrie baryonique de l'Univers.
En outre, et de manière plus frappante, la découverte des oscillations des neutrinos et leur description réussie via le mécanisme ``Pontecorvo-Maki-Nakagawa-Sakata"~\cite{Pontecorvo:1957cp,Pontecorvo:1957qd,Maki:1962mu}, implique nécessairement que les neutrinos sont massifs, contrairement à ce qui est prédit par le MS ; la première preuve irréfutable de  l'existence de Nouvelle Physique est ainsi découverte en laboratoire .

Pour résoudre les problèmes mentionnés, de nombreux modèles et cadres ont été proposés, incluant fréquemment des états de nouvelle physique présents à l'échelle du TeV.
Jusqu'à présent, aucun signal de ces nouveaux états n'a  été directement découvert au LHC.
Cependant, les mesures des observables des saveurs et les tests de précision électrofaibles imposent indirectement des contraintes strictes sur l'espace des paramètres et sur l'échelle de masse des modèles de nouvelle physique.
En particulier, les mesures de précision des observables de la saveur des hadrons effectuées au cours des vingt dernières années ont permis d'imposer des contraintes strictes à la matrice de mélange des quarks de Cabibbo-Kobayashi-Maskawa, mettant fortement en évidence
l'unitarité de cette dernière.
Par conséquent, tout contenu supplémentaire de fermions qui interagit avec les quarks du modèle standard ne peut avoir de grands mélanges, et l'existence d'une  quatrième génération de quarks a été exclue.
En outre, les mesures de précision de désintégrations rares (telles que $B_s\to \mu\mu$, en excellent accord avec la prédiction du modèle standard) ont permis d'écarter presque tous les modèles visant à résoudre l'énigme de la rupture de la symétrie électrofaible ; en d'autres termes, ``la saveur est le cimetière habituel des théories électrofaibles au-delà du modèle standard".

Contrairement au secteur des quark, le secteur  des leptons, ainsi que leur transitions de saveur, est loin d'être maîtrisé.
Alors que les entrées de la matrice de mélange des quarks ont été déterminées avec une grande précision, l'effort expérimental pour mesurer les paramètres de mélange des leptons vient juste d'atteindre son ``ère de précision".
En soi, le secteur des neutrinos est à l'origine de nombreuses questions ouvertes ; on ne connaît actuellement ni l'échelle absolue ni le mécanisme à l'origine des masses des neutrinos.
De plus, alors que le modèle de mélange des quark est très hiérarchisé (tout comme le spectre des masses des quark), la PMNS ne l'est pas, ce qui aggrave encore le ``problème (global) de la saveur".
Comme les phénomènes d'oscillation des neutrinos impliquent nécessairement que ces derniers sont massifs, on s'attend à ce que les symétries de saveur leptoniques accidentelles du modèle standard soient violées dans la nature.
Il s'agit de la conservation de la saveur leptonique individuelle et de l'universalité de la saveur leptonique.
Ainsi, les tests expérimentaux de ces symétries semblent particulièrement intéressants pour parvenir à une meilleure compréhension du secteur leptonique, pour contraindre les contributions de la Nouvelle Physique et éventuellement pour découvrir des indices indirects de ses effets pouvant se manifester à basse énergie dans les phénomènes leptoniques.

S'il est clair que la saveur leptonique neutre est violée dans la nature, les recherches de processus violant la saveur leptonique chargée n'ont jusqu'à présent donné que des résultats négatifs.
Cela permet d'imposer des contraintes strictes aux modèles visant à fournir un mécanisme viable de génération de la masse des neutrinos.
De la même manière, les mesures des observables de précision sensibles à la violation de l'universalité de la saveur des leptons à haute énergie (désintégrations $W\to\ell\nu$ et $Z\to \ell\ell$) et à basse énergie (désintégrations (semi-)leptoniques des $K$ et des $\pi$) semblent être cohérentes avec les prédictions du modèle standard, ce qui conduit à des limites strictes sur l'unicité de la matrice de mélange des leptons et donc sur la présence (hypothétique) de fermions neutres supplémentaires  qui pourraient se mélanger avec leptons neutres du modèle standard.
Il est cependant important de noter que la violation de l'universalité de la saveur des leptons, et la violation de la saveur des leptons chargés, peuvent également se produire dans des modèles de la Nouvelle Physique, sans avoir aucun lien (direct) avec le mécanisme de  génération de masse des neutrinos.

Bien que jusqu'à présent l'écrasante majorité des observables de la saveur mesurées semblent être en accord avec le paradigme de la saveur du modèle standard, ces dernières années, plusieurs observables liées à la saveur des leptons ont commencé à présenter des écarts significatifs par rapport à leurs prédictions respectives du modèle standard.
Parmi ces dernières figurent les moments magnétiques anormaux de l'électron et du muon.
En particulier, les mesures du moment magnétique anormal du muon restent toujours en contradiction avec la prédiction du modèle standard\footnote{Les évaluations récentes des contributions de la 
``hadronic vacuum polarisation", obtenues par la QCD sur réseau, conduisent à des tensions beaucoup plus faibles entre la prédiction du modèle standard et les mesures}.
En combinant les mesures effectuées à Brookhaven et à Fermilab, la tension s'élève actuellement à $+4.2 \,\sigma$ (écarts types).
Plus récemment, en raison de la disponibilité de mesures indépendantes de la constante de structure fine électromagnétique $\alpha_e$ (à l'aide d'atomes de césium\footnote{Une mesure récente de $\alpha_e$ à l'aide d'atomes de rubidium présente une tension avec le résultat du césium, autour de $ >5 \,\sigma$,  et conduit à une tension moins important pour le moment magnétique anormal de l'électron.}),  une tension avec la prédiction du modèle standard a été aussi découverte concernant  le moment magnétique anormal de l'électron, conduisant à une déviation de $2.5\,\sigma$ entre théorie et expérience.
Il est intéressant de noter que le signe (et la grandeur) des écarts respectifs pourrait indiquer la présence d'interactions de Nouvelle Physique violant l'universalité de la saveur leptonique.

Les rapports des largeurs de désintégration semi-leptoniques des mésons (courants chargés et neutres) sont directement sensibles à la violation de l'universalité de la saveur leptonique.
Au cours de la dernière décennie les mesures des rapports $R_{D^{(\ast)}}\equiv B\to D^{(\ast)}\tau\nu / B\to D^{(\ast)}\ell\nu$ et $R_{K^{(\ast)}}\equiv B\to K^{(\ast)}\mu\mu/B\to K^{(\ast)}ee$ présentent des tensions persistantes avec leurs prédictions respectives du modèle standard, atteignant plus récemment $3.1\,\sigma$ pour la mesure de $R_K$.
De plus, les mesures des fractions d'embranchement différentielles de $B\to K^\ast\mu\mu$ et $B_s\to \phi\mu\mu$, ainsi que les mesures des coefficients angulaires dans la désintégration $B\to K^\ast (\to K\pi)\mu\mu$ montrent des déviations (locales) atteignant $>3\,\sigma$.
Si elles sont interprétées en termes de présence de nouvelle physique, les anomalies dites des mesons ``$B$", en particulier dans les transitions de courants neutres $b\to s\ell\ell$, semblent dessiner une image cohérente : il existerai une ``force qui éloigne les muons".
Alors que la découverte et les mesures des oscillations des neutrinos sont les premières preuves irréfutables de la Nouvelle Physique, les anomalies de saveur dans les moments magnétiques anormaux et les désintégrations des mésons $B$ sont certainement des indices \textit{indirectes} intéressants  sur la Nouvelle Physique.

\section{Le modèle standard}
Le modèle standard de la physique des particules~\cite{Weinberg:1967tq,Glashow:1961tr,Salam:1968rm} offre une description extraordinairement réussie et pourtant simple de la nature à ses plus petites échelles ; il offre un cadre commun pour décrire les particules élémentaires et leurs interactions électrofaible et forte.
Malgré son succès exceptionnel, il est maintenant fermement établi que le modèle standard (SM) ne peut rendre compte d'un certain nombre d'observations, et il faut donc envisager des constructions théoriques - incluant de nouveaux degrés de liberté (nouvelles particules et/ou nouvelles interactions), capables d'expliquer certaines données expérimentales.
De plus, un fort intérêt théorique alimente également l'étude de la ``Nouvelle Physique au-delà du SM (BSM)", car cette dernière pourrait fournir une solution, ou du moins améliorer, certaines des énigmes théoriques du SM.

\medskip
Le modèle standard est une théorie quantique des champs, renormalisable, invariante sous le groupe de Poincaré et le groupe de jauge semi-simple (local) $SU(3)_c\times SU(2)_L\times U(1)_Y$.
En plus des bosons de jauge associés, le SM comprend trois familles de quarks et de leptons, ainsi qu'un seul champ scalaire fondamental.
Leurs représentations sous les groupes non-abéliens $SU(3)_c$ et $SU(2)_L$, ainsi que leur charge sous le groupe de jauge abélien $U(1)_Y$, sont listées dans le tableau~\ref{res_tab:SM_res}. La convention de la (hyper)charge $U(1)_Y$ est telle que $Q_f^\text{em} = Y_f^{U(1)}+T_{3\,f}^{SU(2)}$.
\renewcommand{\arraystretch}{1.3}
\begin{table}[ht!]
\begin{center}
  \begin{tabular}{|c|c|c|c|}
  \hline
      Champ & $SU(3)_c$ & $SU(2)_L$ & $U(1)_Y$\\
  \hline
  \hline
  $Q = \left(u_L, \, d_L\right)^T$ & $ \mathbf{3} $ & $ \mathbf{2} $ &
  $ \frac{1}{6} $ \\ 
  $ \ell = \left(\nu_L, \, e_L\right)^T $ & $ \mathbf{1} $ & $
  \mathbf{2} $ & $ -\frac{1}{2} $ \\ 
  $ u_R $ & $ \mathbf{3} $ & $ \mathbf{1} $ & $ \frac{2}{3} $ \\ 
  $ d_R $ & $ \mathbf{3} $ & $ \mathbf{1} $ & $ -\frac{1}{3} $\\ 
   $ e_R $ & $ \mathbf{1} $ & $ \mathbf{1} $ & $ -1 $ \\
  \hline
  $ H = (H^+, H^0)^T$ & $ \mathbf{1} $ & $ \mathbf{2} $ & $ \frac{1}{2} $ \\
  \hline
    $G$ & $\mathbf{8}$ & $\mathbf{1}$ & 0\\
    $W$ & $\mathbf{1}$ & $\mathbf{3}$ & 0\\
    $B$ & $\mathbf{1}$ & $\mathbf{1}$ & 0\\ 
  \hline
  \end{tabular}
  \end{center}
\caption{Contenu en champs du Modèle Standard et
  les représentations correspondantes sous le groupe de jauge $SU(3)_c\times SU(2)_L$, ainsi que leur charge sous le groupe de jauge abélien $U(1)_Y$.} 
  \label{res_tab:SM_res}
\end{table}
\renewcommand{\arraystretch}{1.}

Une fois que le groupe de jauge et le contenu en matière ont été définis, le Lagrangien (renormalisable) est entièrement déterminé,
\begin{eqnarray}
    \mathcal L_\text{SM} &=&  -\frac{1}{4} B_{\mu\nu}B^{\mu\nu} - \frac{1}{4}W_{\mu\nu}^a W_a^{\mu\nu} - \frac{1}{4} G_{\mu\nu}^a G_a^{\mu\nu}\nonumber\\
    &\phantom{=}& + i \bar Q_L^i \slashed D Q_L^i + i \bar u_R^i \slashed D u_R^i + i \bar L_L^i \slashed D L_L^i + i \bar e_R^i \slashed D e_R^i\nonumber\\
    &\phantom{=}& + Y_{ij}^u \bar Q_L^i \tilde H u_R^j + Y_{ij}^d \bar Q_L^i H d_R^j + Y_{ij}^\ell \bar L_L^i H e_R^j + \text{H.c.}\nonumber\\
    &\phantom{=}& + \left|D_\mu H\right|^2 + \mu^2 \left|H\right|^2 - \lambda \left|H\right|^4\,.
\end{eqnarray}
Dans ce qui précède, $i,j = 1,2,3$ sont des indices de famille, $\slashed D = D_\mu \gamma^\mu$, et $\tilde H = i \sigma_2 H$ ; $Y^f$ désigne les couplages de Yukawa, $\lambda$ l'auto-couplage quartique du Higgs et $\mu$ le terme de masse du Higgs. À l'exception de $\mu$, tous les couplages précédents sont sans dimension, de sorte que, théoriquement, le SM peut être extrapolé à une large gamme d'énergies (ou d'échelles). 
De plus, les interactions entre les champs de jauge et les fermions sont codées dans la dérivée covariante de jauge, donnée par
\begin{equation}
   D_\mu = \partial_{\mu} + i g_s G_\mu^a T^{SU(3)}_a + i g_w W_\mu^a T^{SU(2)}_a + i g' Y B_\mu\,,
\end{equation}
où les couplages $g_s, g_w, g'$ désignent les différents couplages de jauge de $SU(3)_c$, $SU(2)_L$ et $U(1)_Y$, et $T^{(\mathcal G)}_a$ sont les générateurs du groupe de jauge (non-abélien) $\mathcal G$ dans la représentation du fermion (ou boson) sur laquelle agit la dérivée.
Les termes cinétiques des champs de jauge $F$ s'écrivent en fonction de leurs tenseurs de champ, qui sont définis comme suit
\begin{equation}
    F_{\mu\nu}^a \equiv \partial_\mu F_\nu^a - \partial_\nu F_\mu^a + i g_{(\mathcal{G})} f^{abc}_{(\mathcal G)} F_\mu^b F_\nu^c\,,
\end{equation}
où $g_{(\mathcal{G})}$ est le couplage de jauge associé et $f^{abc}_{(\mathcal G)}$ sont les constantes de structure de l'algèbre de liaison correspondante couverte par le groupe de jauge $\mathcal G$ (pour le groupe abélien, tous les $f_{U(1)}^{abc}=0$).

Pour autant que les paramètres du secteur de Higgs (qui sont nécessairement introduits à la main) remplissent certaines conditions ($\mu^2,\lambda > 0$), le champ de Higgs développe une valeur d'attente du vide (vev) $\langle H\rangle = (0, \frac{v}{\sqrt{2}})^T$ avec $v= \sqrt{\mu^2/\lambda}\simeq 246~\mathrm{GeV}$, et brise (spontanément) le groupe de jauge SM en $SU(3)_c\times U(1)_\text{em}$. Ce phénomène est le mécanisme dit de Brout-Englert-Higgs (BEH)~\cite{Englert:1964et,Higgs:1964ia,Higgs:1964pj}.
Après la brisure de symétrie électrofaible (EWSB), trois bosons de jauge massifs émergent, les $Z^0$ et $W^\pm$, tandis que les gluons et une combinaison linéaire de $B$ et $W$, le photon $A$ (ou $\gamma$), restent sans masse.
Dans la phase brisée, les champs de jauge physiques (électrofaibles) peuvent être écrits en termes de champs de jauge originaux $W$ et $B$ comme suit
\begin{align}
    W_\mu^\pm &= \frac{1}{\sqrt{2}}(W_\mu^1 \mp i W_\mu^2)\quad&\text{avec masse}\quad &M_W = \frac{g_w v}{2}\,,\nonumber\\
    Z_\mu^0 &= \frac{1}{\sqrt{g_w^2 + g^{\prime 2}}}(g W_\mu^3 - g' B_\mu)\quad&\text{avec masse}\quad &M_Z = \frac{v}{2}\sqrt{g_w^2 + g^{\prime 2}}\,,\nonumber\\
    A_\mu &= \frac{1}{\sqrt{g_w^2 + g^{\prime 2}}} (g' W_\mu^3 + g B_\mu)\quad&\text{avec masse}\quad &M_A = 0\,.
\end{align}
En outre, les fermions acquièrent des masses 
\begin{equation}
    m^f = Y^f \langle H\rangle
\end{equation}
via leurs couplages avec le boson de Higgs.

Dans le SM, la dynamique des fermions est régie par les interactions avec les bosons de jauge et par les couplages de Yukawa.
En général, les couplages de Yukawa (et donc les matrices de masse $m^f$) ne sont pas diagonaux.
Afin d'obtenir les champs de fermions physiques (massifs), les couplages de Yukawa des fermions doivent être diagonalisés.
Après le EWSB, les termes de masse des quark ($q=u,d$) dans le Lagrangien peuvent être reformulés comme suit
\begin{equation}
    \mathcal L_\text{mass}^q \sim \bar q_{L}^i M_{ij}^q q_R^j = \bar q_{L}^i V_{L}^{q\dagger}\, V_L^q\, M_{ij}^q\, V_{R}^{q\dagger}\, V_R^q\, q_R^j = \hat{\bar q}_L^i m_i^q \hat{q}_R^i\,,
\end{equation}
où la base physique (de masse) est désignée par $\hat{\phantom{q}}$.
Les matrices unitaires $V_{L,R}^q$ diagonalisent les matrices de masse et relient les bases d'interaction aux bases de masse par les relations suivantes
\begin{equation}
    m_\text{diag}^q = V_L^q\, M_{ij}^q\, V_R^{q\dagger}\,,\quad\text{et}\quad\hat{q}_{L,R} = V_{L,R}^q q_{L,R}\,,
\end{equation}
et de forme équivalente pour les leptons chargés.

Dans le secteur des quarks\footnote{Pour le secteur leptonique, puisque les neutrinos sont sans masse en raison de l'absence de neutrinos droitiers et/ou de triplets de Higgs, on peut sans perte de généralité choisir de travailler dans une base dans laquelle les couplages d'Yukawa des leptons chargés sont diagonaux, ce qui conduit à des interactions de courant chargé conservant strictement la saveur leptonique.}
l'insertion des transformations ci-dessus dans le lagrangien d'interaction conduit à une violation de saveur dans les courants chargés (cc), paramétré par la matrice de mélange des quarks, dite de Cabibbo-Kobayashi-Maskawa (CKM) ($V_\text{CKM}$)
\begin{equation}
    \mathcal L_\text{cc}^q \sim - \frac{g_w}{\sqrt{2}} V_\text{CKM}^{ij} W_\mu^+ \bar u_{L i} \gamma^\mu d_{L j}\,,\quad V_\text{CKM} = V_L^u V_L^{d\dagger}\,.
\end{equation}
La matrice CKM est une matrice complexe et (spéciale) unitaire $3\times3$, et donc entièrement paramétrée par 4 paramètres réels (paramètres physiques)\footnote{Une matrice spéciale unitaire $3\times3$ possède en général huit paramètres libres. Cependant, quatre des phases complexes peuvent être réabsorbées dans les redéfinitions des champs de quark (qui sont des particules de Dirac) et sont donc non physiques, ce qui laisse quatre paramètres physiques.}.
Ces paramètres sont généralement exprimés, dans la paramétrisation dite standard, en termes de trois angles de mélange réels et d'une phase.
En raison du désalignement entre les bases de masse des quarks de type up et down, la matrice CKM est en général non triviale et conduit donc à une violation de saveur hadronique, qui a été observée expérimentalement dans un certain nombre de désintégrations de mésons et de baryons.
De plus, le mécanisme de Kobayashi-Maskawa~\cite{Kobayashi:1973fv}, via sa phase unique (physique), fournit naturellement une source de violation de CP.
Les interactions FCNC (flavour changing neutral currents) restent absentes au niveau arbre et sont, à un ordre supérieur, naturellement supprimées par un mécanisme de GIM généralisé~\cite{Glashow:1970gm}.

En tout, le SM possède 18 paramètres libres : les trois couplages de jauge ($g_s, g_w, g'$), les deux paramètres du potentiel de Higgs, 10 paramètres dans le secteur des quark (masses des quark, trois angles CKM et une phase violant CP), et les trois masses des leptons chargés.

Bien que ne pas imposé a priori, le Lagrangien du SM présente un certain nombre de symétries ``accidentelles", qui permettent de comprendre et d'expliquer certaines propriétés et certains phénomènes. Les symétries accidentelles exactes du SM correspondent à : 
\begin{itemize}
    \item Conservation du nombre de baryons ($U(1)_B$ global) : chaque quark porte $B_q = 1/3$, alors que les leptons ne le portent pas ($B_\ell = 0$). Cette symétrie n'est brisée au niveau quantique que dans les processus dits de sphaléron, qui conservent toutefois $B-L$, $L$ étant le nombre total de leptons.
    \item Conservation du nombre leptonique individuel (par saveur - $U(1)_{L_e}\times U(1)_{L_\mu}\times U(1)_{L_\tau}$ global) : entre autres conséquences, ces symétries interdisent les processus violant la saveur des leptons, tels que les désintégrations violant la saveur des leptons chargés (par exemple, $\mu\to e\gamma$) ainsi que les oscillations de neutrinos. En outre, la conservation du nombre de leptons individuels rend tous les couplages des bosons de jauge SM aux leptons universels en termes de saveur.
    Cela implique naturellement la conservation globale du nombre leptonique total, une symétrie globale $U(1)_L$, qui n'est également brisée qu'au niveau quantique dans les processus sphalériens déjà mentionnés.
\end{itemize}
Il est intéressant de noter que ces symétries accidentelles pourraient également indiquer des pistes vers des extensions (préférées) du SM.

Le Lagrangien du SM possède également plusieurs symétries accidentelles approximatives, correspondant à des symétries globales exactes, uniquement brisées par de petits couplages.
Dans la limite de la disparition des couplages de Yukawa et de $g' = 0$, le SM possède une symétrie globale supplémentaire $SU(2)$ (sous laquelle le Higgs se transforme en doublet).
Le vev de Higgs brise cette symétrie, en préservant néanmoins la symétrie dite ``custodiale" $SU(2)_C$, sous laquelle les bosons massifs $W^\pm$ et $Z^0$ forment un triplet de masse dégénérée $M_C = M_W = M_Z$.
Le petit couplage $g'$ brise alors cette symétrie, et dans la limite de la disparition des couplages de Yukawa on trouve
\begin{equation}
    \frac{M_W^2}{M_Z^2\cos^2\theta_w} \equiv \rho = 1\,,
\end{equation}
avec l'angle de mélange faible $\tan\theta_w = g'/g_w$.
Les corrections dues aux couplages de Yukawa (dominés par le Yukawa du quark top $y_t = \frac{m_t \sqrt{2}}{v}\simeq 1$) n'apparaissent qu'à des ordres supérieures (boucle) ; cette symétrie, elle aussi  accidentelle, donne donc lieu à une prédiction non triviale du SM, $\rho \simeq 1$, qui a été très bien testée expérimentalement.

Dans la limite où les couplages de Yukawa sont zero, $Y^f = 0$, le SM possède cinq symétries globales supplémentaires $U(3)$, associées aux trois familles de $Q_L, u_R, d_R, L_L, e_R$.
Ces dernières symétries (et leurs sous-groupes) permettent de comprendre de nombreuses propriétés en physique des saveurs (quark), et ouvrent la voie à l'étude de modèles de Nouvelle Physique des saveurs (par exemple avec des symétries de saveurs, dont nous parlerons plus tard).
Nous notons ici que toutes ces symétries accidentelles sont une conséquence directe du fait que seuls les opérateurs renormalisables (dimension $\leq 4$) sont inclus dans le Lagrangien du SM.

\medskip
Le SM constitue l'une des théories les plus abouties de la physique moderne : fondé sur un cadre théorique élégant, il décrit \textit{presque toutes} les observations expérimentales en physique des particules avec une grande précision.
Après la découverte des bosons de jauge $Z$ et $W$, le nouveau boson découvert au LHC répond de plus en plus aux exigences SM d'un ``boson de Higgs" (en particulier concernant son spin/parité).
Les mesures des collaborations ATLAS et CMS,  donnent la valeur moyenne de~\cite{ATLAS:2015yey}
\begin{equation}
    m_H = (125.09 \pm 0.24)\:\mathrm{GeV}\,.
    \label{res_eqn:higgsdirect}
\end{equation}
Le secteur électrofaible du SM a été testé par une quantité impressionnante de mesures.
Les observables de précision électrofaibles (EWPO) ont été systématiquement utilisés pour vérifier les prédictions du SM et pour contraindre ses paramètres inconnus (liés au secteur de Higgs).
Bien que les EWPO comprennent un énorme ensemble de mesures possibles, celles-ci ont été réduites par les groupes de travail du LEP et du Tevatron, et comprennent par exemple la masse et la largeur du boson $W$, et diverses observables du pôle du $Z$ telles que l'angle de mélange faible $\sin^2\theta_w$, les largeurs de désintégration des fermions SM, parmi beaucoup d'autres.
En plus de déterminer les propriétés électrofaibles du SM, ces mesures permettent d'effectuer des contrôles de cohérence approfondis du SM\footnote{Pour un aperçu complet de l'état actuel des expériences, voir par 
exemple~\cite{ParticleDataGroup:2020ssz}.}.
Par exemple, un ajustement global de tous les EWPO aux données expérimentales conduit à une masse de Higgs de~\cite{ParticleDataGroup:2020ssz}. 
\begin{equation}
    m_H^\text{EWPO} = 90^{+18}_{-16}\:\mathrm{GeV}\,,
\end{equation}
en accord avec la mesure directe de la masse (cf. Eq.~\eqref{res_eqn:higgsdirect}) à $1.8\,\sigma$.

\begin{figure}
    \centering
    \includegraphics[width=0.6\textwidth]{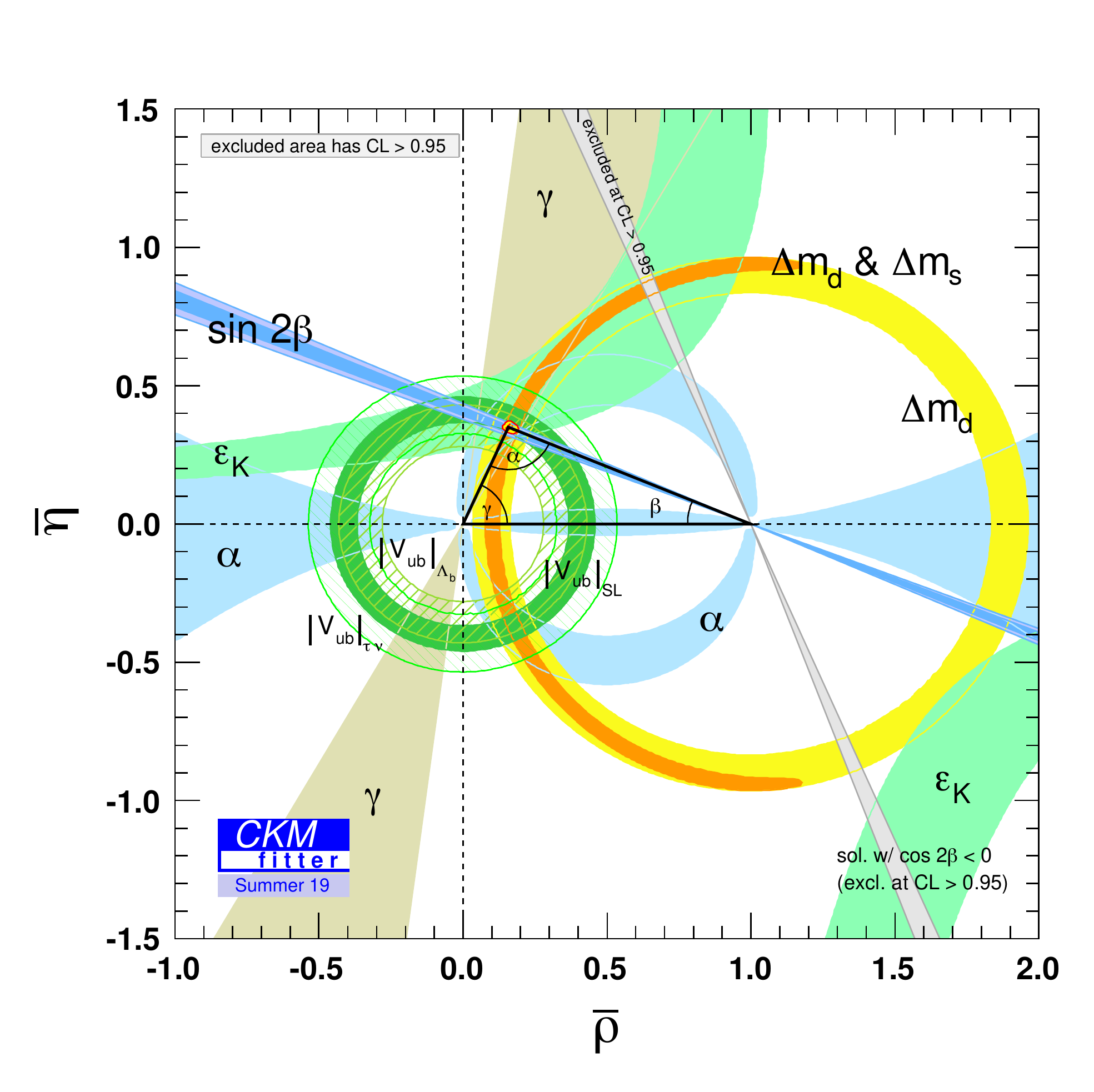}
    \caption{Contraintes sur l'unitarité de la matrice CKM déduites d'un grand nombre de mesures expérimentales de processus hadroniques violant la saveur (quark). Figure extraite de~\cite{Charles:2004jd}.}
    \label{res_fig:UTCKM_res}
\end{figure}

Cette ``success story" se poursuit dans le domaine de la physique des saveurs (quark).
Les observables de mélange de quark et de violation de CP mesurées expérimentalement sont en général bien expliquées par le paradigme CKM de la saveur, enraciné dans la matrice CKM unitaire et le mécanisme GIM ; la matrice CKM présente une structure fortement hiérarchique, et la détermination de ses éléments est bien en accord avec l'unitarité.
Les paramètres fondamentaux du paradigme CKM, y compris les déviations possibles de l'unitarité, ont été contraints par une grande série d'observables avec une précision impressionnante~\cite{Charles:2004jd}, comme l'on peut voir sur la Fig.~\ref{res_fig:UTCKM_res}.

Enfin, l'étude des transitions et des désintégrations de saveurs (hadroniques) a connu des développements importants au cours des dernières années.
Si, du côté expérimental, de plus en plus de données ont été accumulées grâce  à de nombreuses expériences dédiées, du côté théorique, des progrès importants ont été aussi réalisés, les effets d'ordre supérieur étant de mieux en mieux maîtrisés dans les calculs permettant la preediction des observables.

Les interactions fortes entre quarks, médiées par les gluons, sont décrites par la chromodynamique quantique (QCD), dont la nature non perturbatrice à basse énergie (grandes distances) peut être traitée avec succès par la QCD sur réseau (LQCD).

À ce jour, l'ensemble des expériences passées et actuelles a permis de dresser un tableau très clair et cohérent de l'état du SM : à quelques exceptions près, la nature au niveau élémentaire peut être décrite par la chromodynamique quantique et les interactions électrofaibles, qui peuvent être modélisées par un mécanisme de Higgs pour le EWSB.
En outre, le paradigme du CKM décrit précisément les transitions de saveur des quarks et la violation de CP (avec un mécanisme GIM généralisé supprimant les FCNC indésirables).
En d'autres termes, le modèle standard fournit une description exceptionnelle de la nature et fonctionne bien mieux que ce à quoi on aurait pu s'attendre au départ.

\section{Neutrinos massifs}
La découverte des oscillations de neutrinos a constitué la première preuve irréfutable NP observée en laboratoire.
De plus, les neutrinos massifs ouvrent la porte à la violation de la saveur leptonique :  les oscillations des neutrinos signalent la violation de la saveur leptonique neutre et, en l'absence d'un principe fondamental (symétrie imposée), toute extension du SM permettant incluant des neutrinos massifs et des mélanges leptoniques devrait également permettre des processus de violation de la saveur leptonique chargée, tels que les désintégrations $\mu\to e\gamma$.

Malgré le vaste effort expérimental mondial pour déterminer les paramètres d'oscillation des neutrinos massifs, plusieurs questions restent ouvertes, ce qui suggère que notre compréhension du secteur leptonique neutre est loin d'être complète.
Tout d'abord, étant électriquement neutres, les neutrinos peuvent être décrits comme des fermions de Dirac ou de Majorana, ce qui signifie qu'ils pourraient être leur propre antiparticule.
En outre, l'échelle de masse absolue et l'odre du spectre des neutrinos légers demeurent  inconnus. 

Les mesures de la largeur de désintégration invisible du boson $Z$ confirment l'existence de $3$ états neutres légers (avec des masses inférieures à la moitié de la masse du boson $Z$), les 3 neutrinos actifs. Néanmoins, l'existence de fermions neutres supplémentaires, des états dits stériles (sans interactions de jauge SM), reste une possibilité viable et attrayante.
En particulier, les leptons neutres lourds (HNL) sont souvent invoqués dans les extensions du SM qui visent à prendre en compte les données d'oscillation, et offrent un mécanisme intéressant pour la génération de la masse des neutrinos.

Afin d'accommoder les données d'oscillation des neutrinos, le SM doit être étendu.
Si l'on impose la conservation du nombre leptonique total, les neutrinos sont des fermions de Dirac et nous pouvons étendre de façon minimale le contenu du champ du SM par trois neutrinos droitiers $\nu_R$, ce qui permet d'écrire directement un terme d'interaction de Yukawa entre le doublet de leptons du SM et $\nu_R$, $Y^\nu H \bar L_L \nu_R$, en totale analogie avec les autres fermions (quarks et leptons chargés).
Après le EWSB, cela conduit à un terme de masse de la forme
\begin{equation}
    \mathcal L_\text{mass}^\text{Dirac} = \bar \ell_L m_\ell \ell_R + \bar \nu_L m_D\nu_R + \text{H.c.}\,,
\end{equation}
dans laquelle $m_\ell$ est la matrice de masse des leptons chargés, $m_D = Y^\nu v/\sqrt{2}$ et $v$ est le vev du Higgs du SM.
Comme dans le secteur des quarks, les termes de masse des leptons chargés et des neutrinos peuvent alors être diagonalisés par des transformations bi-unitaires
\begin{equation}
    m_\nu^\text{diag} = V_L^\nu m_D V_R^{\nu\dagger}\,,\quad m_\ell^\text{diag} = V_L^\ell m_\ell V_R^{\ell\dagger}\,.
\end{equation}
Les transformations entre la base d'interaction et la base de masse (désignée par $\hat{\phantom\ell}$) sont ainsi données par
\begin{equation}
    \hat\nu_{L,R} = V_{L,R}^\nu \nu_{L,R}\,,\quad \hat\ell_{L,R} = V_{L,R}^\ell \ell_{L,R}\,.
\end{equation}
Dans la base de masse, nous pouvons alors définir le spineur de Dirac physique $\psi_\nu = \nu_L + \nu_R$ qui satisfait l'équation de Dirac.
La matrice PMNS est alors donnée par $U_{\text{PMNS}} = V_L^{\ell\dagger}V_L^\nu$, ou si nous choisissons de travailler dans la base faible dans laquelle les couplages d'Yukawa des leptons chargés sont diagonaux, simplement par $U_\text{PMNS} = V_L^\nu$.  

Bien que cette extension ad hoc fournisse une explication fonctionnelle des données d'oscillation, pour assurer la compatibilité avec les limites expérimentales de l'échelle de masse absolue des neutrinos ($m_\nu\lesssim 0.1\,\mathrm{eV}$), il faudrait que les couplages de Yukawa $Y^\nu$ soient extrêmement petits, $Y^\nu\lesssim 10^{-12}$. Cela soulève la question de savoir pourquoi il existe une si grande hiérarchie dans les couplages de Yukawa entre les secteurs des leptons chargés et neutres (ou pire encore, si l'on considère tous les fermions) ; par conséquent, cette hiérarche soulève la question de la naturalité des couplages d'Yukawa.
Cependant, un aspect plus problématique c'est qu'en raison de la nature de singlet de $\nu_R$ (pas de charge électrique ni de couleur, et aussi singlet $SU(2)_L$), la symétrie de jauge du SM permet en principe d'écrire un terme de masse de  Majorana de la forme $m_{RR}\bar\nu_R\nu_R^c$. À moins qu'une symétrie ne soit appliquée, un tel terme conduirait à la violation du nombre total  leptonique $L$ (par deux unités).

Malgré ses défauts, cette extension ad-hoc du SM par  des neutrinos de Dirac peut être défendue comme un simple ajout  d'états de spin supplémentaires au contenu en champs du SM ; elle est donc attrayante en raison de sa minimalité.

\medskip
Au prix d'une rupture de l'invariance de jauge ou d'une perte de renormalisabilité, le contenu en champs du SM autorise un terme de masse de Majorana de la forme $m_{LL}\overline{\nu_L}\nu_L^c$.
Contrairement aux fermions portant une charge de jauge, les spineurs $\psi$ et $\psi^c$ d'un fermion neutre (auquel aucune charge conservée globalement n'est associée) ne correspondent pas nécessairement à des champs différents, mais plutôt à des états d'hélicité différents, et obéissent donc à la même équation de mouvement.
Cela implique que l'on pourrait avoir $\psi = \psi^c$, ce qui est communément appelé la ``condition de Majorana".
Un bispineur de Majorana peut alors être construit à partir d'une seule composante chirale, $\psi_\text{M} = \psi_L + C\bar\psi_L^T$, donnant lieu à un terme de masse de la forme
\begin{equation}
    \mathcal L_\text{mass}^\text{Majorana} = \frac{1}{2} m_\text{M} \left(\overline{\psi_L^c}\psi_L + \overline{\psi_L} \psi_L^c\right)\,.
\end{equation}
En principe, ce type de terme de masse pourrait être réalisé dans la nature pour les neutrinos, puisque ce sont des particules neutres ; les désintégrations double-bêta sans neutrinos et autres interactions LNV seraient donc  possibles. 

Cependant, avec les neutrinos du SM, un terme de masse de la forme $m_\text{M} \overline{\nu_L}\nu_L^c$ viole l'invariance de jauge $SU(2)_L$, car il se transforme comme un triplet $SU(2)_L$.
L'invariance de jauge peut être récupérée si l'on suppose que ce terme provient d'un opérateur non renormalisable de dimension 5, appelé opérateur de Weinberg, qui est le seul opérateur de dimension 5 invariant de jauge qui peut être construit à partir des champs du SM (cf. Chapitre~\ref{sec:eft}. 
Il est donné par
\begin{equation}
    \mathcal L_{d=5} = \frac{C_{ij}}{2\Lambda}(\overline{L_i^c}\widetilde H^\ast)(\widetilde H^\dagger L_j)\,,
\end{equation}
$\Lambda$ est l'échelle de la Nouvelle Physique associée à la brisure du nombre  leptonique total.
Ici, l'opérateur de Weinberg se transforme sous $SU(2)_L$ comme un singlet  fermionique, ce qui suggère qu'il peut être généré au niveau arbre par des fermions singlets tels que les neutrinos RH $\nu_R$, ce qui est le cas d'un ``type I seesaw mechanism''.

\begin{figure}
    \centering
\mbox{    
\includegraphics[width=0.25\textwidth]{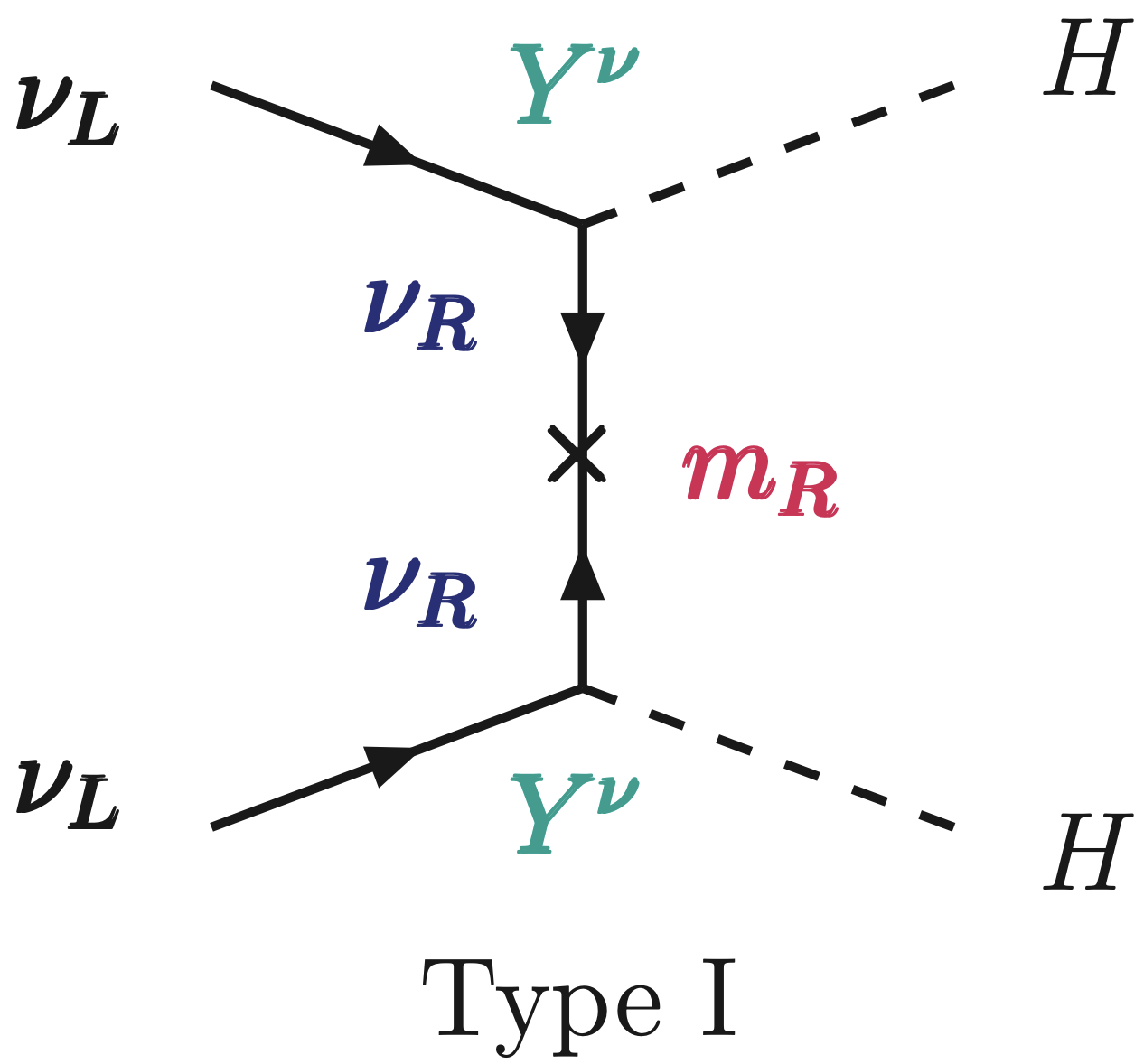}
\hspace{5mm}\includegraphics[width=0.35\textwidth]{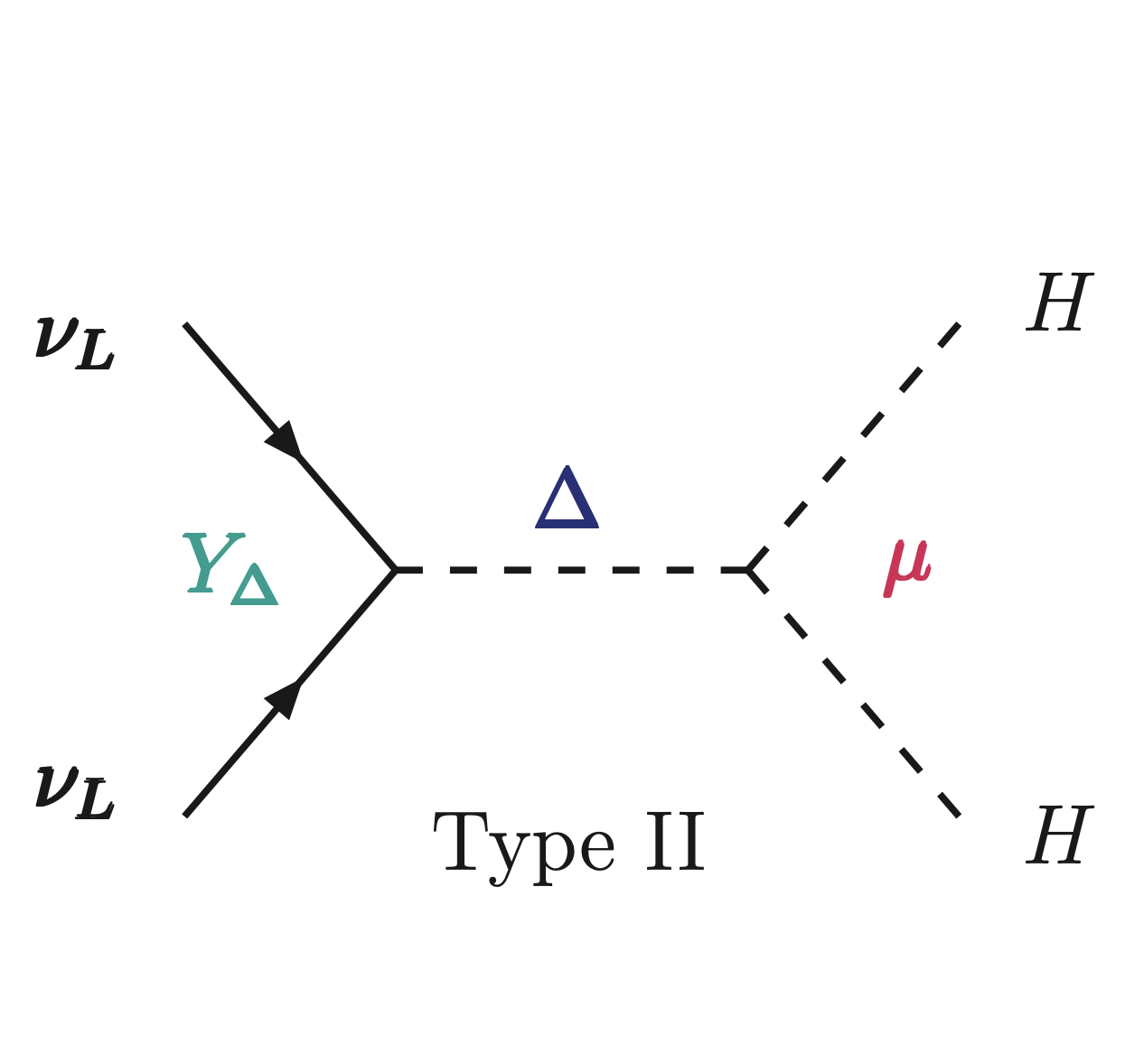}
\hspace{5mm}\includegraphics[width=0.28\textwidth]{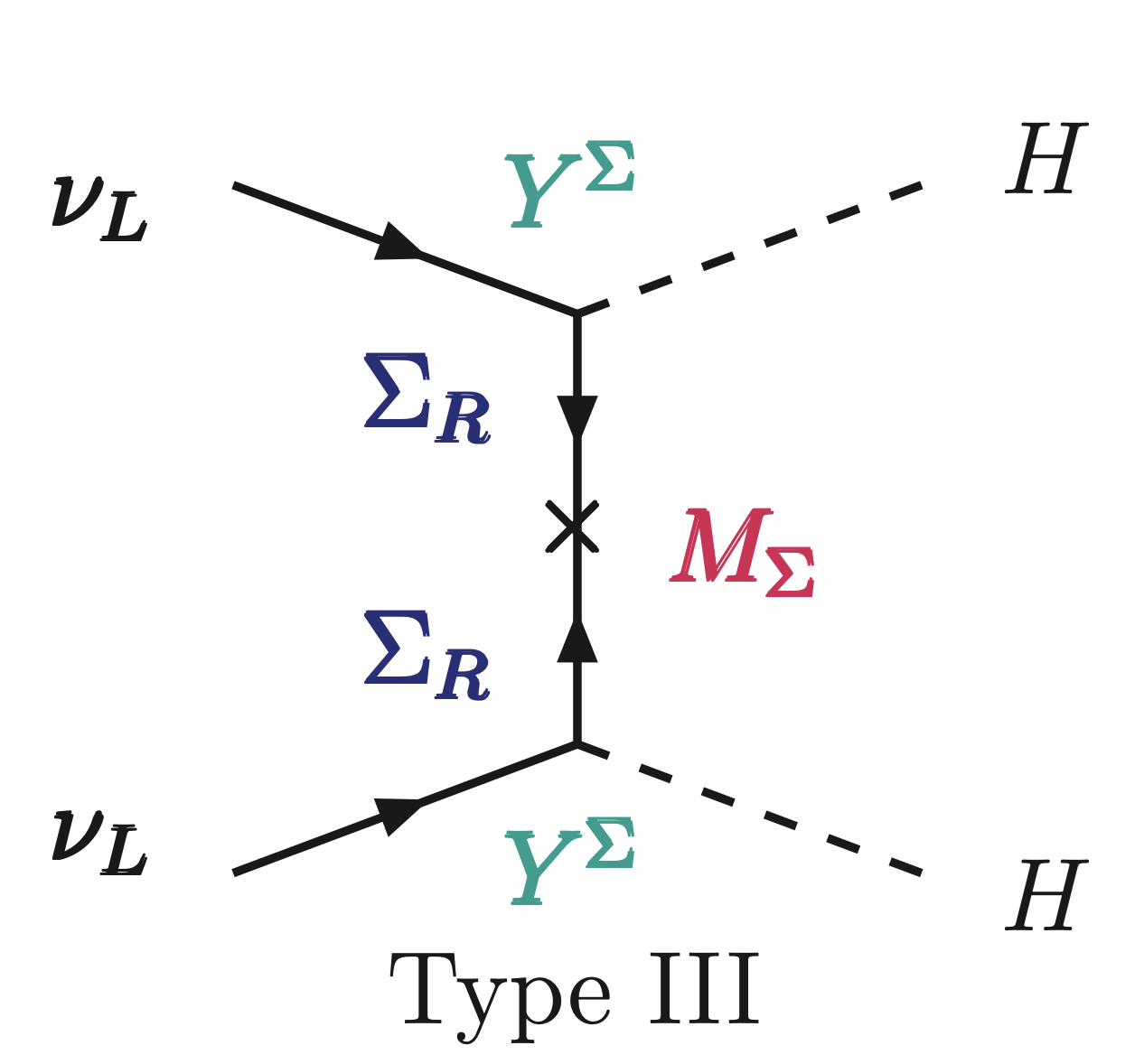}
}
    \caption{Réalisations au niveau arbre de l'opérateur de Weinberg, représentées par des diagrammes de Feynman.
    De gauche à droite, les ``seesaw mechanisms'' de type I, de type II et de type III.}
    \label{fig:seesaws_res}
\end{figure}

L'invariance de jauge permet deux réalisations supplémentaires de l'opérateur de Weinberg, via des triplets scalaires ou des triplets de fermions.
Dans la Fig.~\ref{fig:seesaws_res}, nous illustrons les trois types de ``seesaw mechanism'' par les diagrammes associés au niveau arbre, donnant lieu à des réalisations de l'opérateur de Weinberg :
en supposant une réalisation au niveau arbre, cela conduit respectivement au type II~\cite{Barbieri:1979ag,Cheng:1980qt,Magg:1980ut,Lazarides:1980nt,Schechter:1980gr,Mohapatra:1980yp} (extensions SM via un triplet scalaire) et au type III~\cite{Foot:1988aq,Ma:1998dn} (extensions SM via un triplet fermionique).
Indépendamment de la réalisation, après le EWSB, l'opérateur de Weinberg donne lieu à un terme de masse de Majorana effectif pour les neutrinos gauches, ayant la forme suivante
\begin{equation}
    \mathcal L_{d=5} = \frac{v^2 C_{ij}}{2 \Lambda} (\overline{\nu_{i L}^c} \nu_{L j}) + \text{H.c.}\,,
\end{equation}
où la suppression $\Lambda_\text{EW}/\Lambda$ est manifeste. 
Selon la réalisation spécifique, les coefficients $C_{ij}$  (sans dimension) contiennent une combinaison de couplages, de facteurs de boucle, etc..
Si $C_{ij}\sim \mathcal O(1)$, la compatibilité avec les données expérimentales actuelles impliquerait pour l'échelle de la Nouvelle Physique $\Lambda \sim \mathcal O(10^{16})\:\mathrm{GeV}$, proche de l'échelle GUT. C'est le cas du ``vanilla type I seesaw mechanism".
En fait, les premières propositions d'un ``seesaw" de type I ont été faites dans le cadre de modèles GUT $SO(10)$~\cite{Minkowski:1977sc,Yanagida:1979as,Glashow:1979nm,Gell-Mann:1979vob,Mohapatra:1979ia}.

Cependant, selon le modèle UV sous-jacent, les couplages effectifs $C_{ij}$ peuvent également être (très) petits, que ce soit en raison de la suppression d'une possible réalisation au niveau boucle (si l'opérateur de Weinberg n'est pas réalisé au niveau arbre), ou en raison d'arguments basés sur des symétries, ce qui est le cas de nombreuses variantes de ``seesaw" réalisées à basse échelle, comme le Inverse Seesaw (ISS)~\cite{Schechter:1980gr,Gronau:1984ct,Mohapatra:1986bd},  
le Linear Seesaw (LSS)~\cite{Barr:2003nn,Malinsky:2005bi} et le 
$\nu$-MSM~\cite{Asaka:2005an,Asaka:2005pn,Shaposhnikov:2008pf}.
En outre, les masses des neutrinos peuvent également être générées par des opérateurs de dimension supérieure. Pour une classification exhaustive, voir par exemple ~\cite{Gargalionis:2020xvt,Gargalionis:2019drk}.

\section{Le rôle des phases leptoniques violant CP dans les observables cLFV}
Comme déjà mentionné, les oscillations de neutrinos 
impliquent que les leptons neutres sont massifs et que les saveurs des leptons ne sont pas conservées, ce qui ouvre à son tour la possibilité d'avoir de processus tels que la violation de la saveur des leptons chargés (cLFV) et la violation de CP leptonique (CPV), interdits dans le SM.


De nombreux processus LNV (y compris les doubles désintégrations bêta sans émission de neutrinos, ou les désintégrations de mésons (semi-)leptoniques) sont connus pour présenter une forte dépendance des phases CPV leptoniques~\cite{Abada:2019bac}. In~\cite{Abada:2021zcm}, 
une étude approfondie des effets des phases de Dirac et de Majorana en ce qui concerne les transitions et les désintégrations cLFV leptoniques a été réalisée, et dans ce qui suit nous mettons en évidence les résultats les plus pertinents.

\subsection{Le rôle des phases : une première approche}
Nous avons considéré un ``3+2 toy model'' effectif, dans lequel 2 leptons neutres lourds (HNL) sont ajoutés au contenu du SM. Aucune hypothèse n'est faite sur le mécanisme de génération de la masse des neutrinos. Le spectre contient 5 états massifs de Majorana, et les mélanges leptoniques sont encodés dans une matrice $5\times5$, paramétrée via 10 angles de mélange $\theta_{\alpha j}$ et 10 phases violant  CP - 6 de Dirac $\delta_{\alpha j}$ et 4 de Majorana $\varphi_j$. Dans la limite de petits angles de mélange, les mélanges actifs-stériles sont donnés par 
\begin{equation}
\mathcal{U}_{\alpha (4,5)} \approx 
\left (\begin{array}{cc}
s_{14} e^{-i(\delta_{14}-\varphi_4)} &
s_{15} e^{-i(\delta_{15}-\varphi_5)} \\
s_{24} e^{-i(\delta_{24}-\varphi_4)} &
s_{25} e^{-i(\delta_{25}-\varphi_5)} \\
s_{34} e^{-i(\delta_{34}-\varphi_4)} &
s_{35} e^{-i(\delta_{35}-\varphi_5)} 
\end{array}
\right)\,,
\end{equation}
avec $s_{\alpha i} = \sin \theta_{\alpha i}$.
C'est important de remarquer que la matrice PMNS (mélanges leptoniques d'états ``gauches")
n'est plus unitaire, ce qui conduit à des courants leptoniques (chargés et neutres) modifiés, et donc à des contributions significatives à plusieurs observables interdites dans le cadre du SM.

\begin{figure}
    \centering
\mbox{\hspace*{-5mm}    \includegraphics[width=0.51\textwidth]{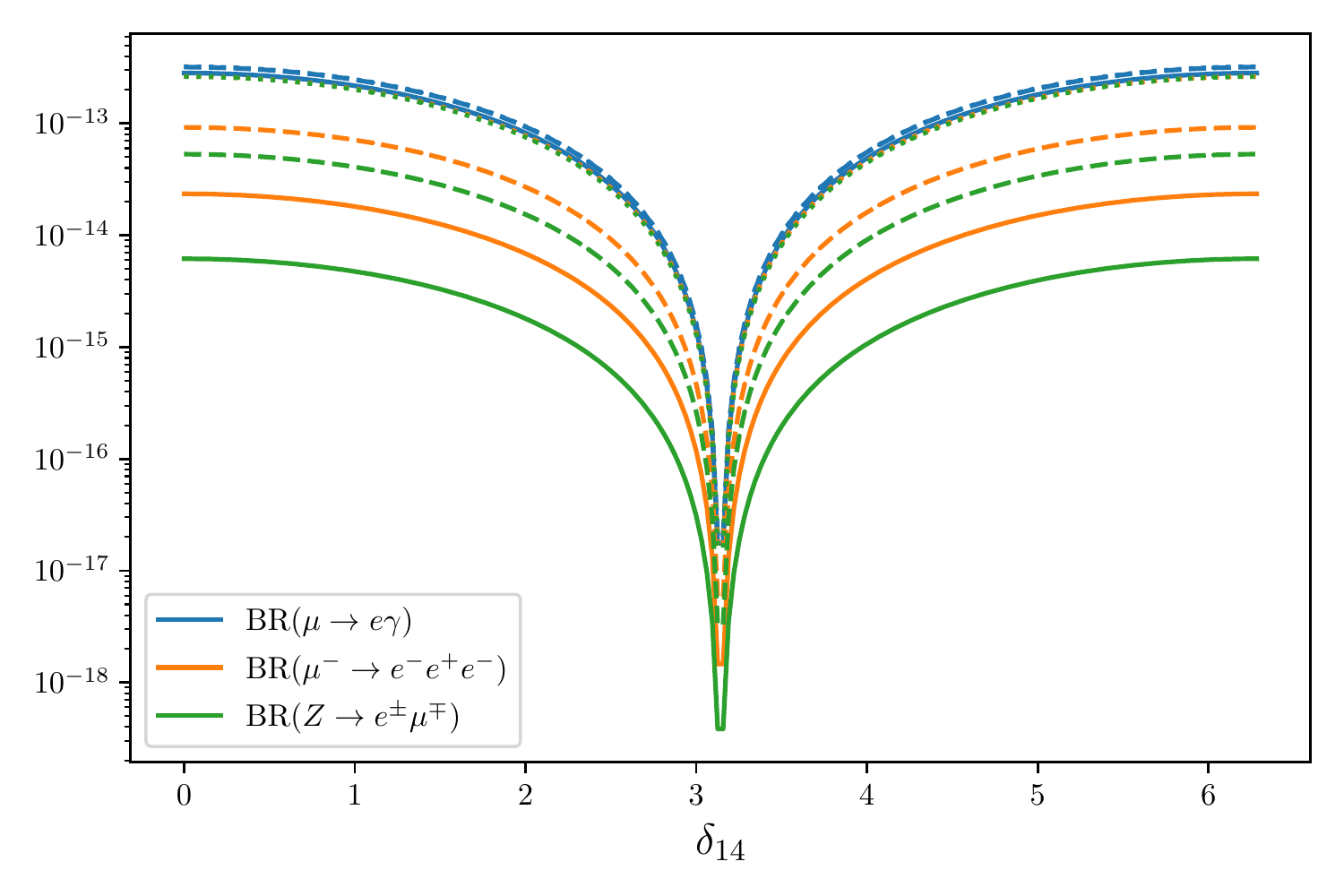}\hspace*{2mm}
\includegraphics[width=0.51\textwidth]{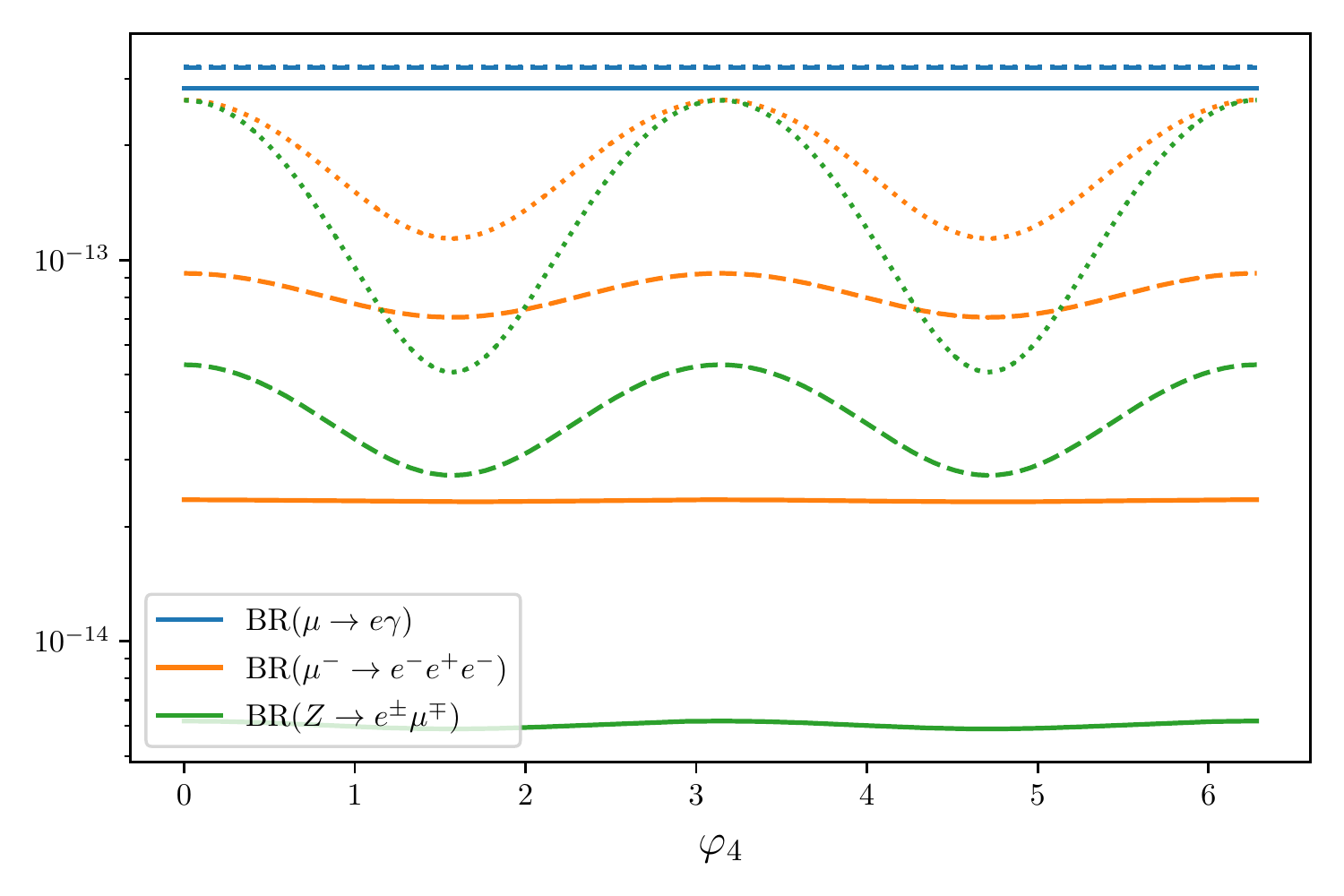}}
    \caption{ Dépendance des observables cLFV sur la phase de Dirac violant CP $\delta_{14}$ (à gauche) et la phase de Majorana $\varphi_4$ (à droite). Les lignes pleines, pointillées et en pointillé correspondent respectivement à $m_4=m_5=$ 1, 5, 10~TeV. D'après~\cite{Abada:2021zcm}.}
    \label{res_fig:1}
\end{figure}

Afin d'illustrer le rôle des phases CPV concernant les observables cLFV, considérons le cas des désintégrations $\mu \to e \gamma$, médiées par des bosons $W$, ainsi que par des neutrinos légers et lourds. 
Le raport d'embranchement associé (voir~\cite{Abada:2021zcm}) est donné par 
\begin{equation}
\text{BR}(\mu \to e \gamma)\propto 
       |G_\gamma^{\mu e}|^2\,, \text{with} \quad 
    G_\gamma^{\mu e} \, =\, \sum_{i=4,5} 
    \mathcal{U}_{e i}\, \mathcal{U}_{\mu i}^* \, 
    G_\gamma(m^2_{N_i}/M^2_W)\,.
\end{equation}
Dans la limite de $m_4 \approx m_5$ et pour $\sin \theta_{\alpha 4} \approx \sin \theta_{\alpha 5} \ll 1$    
le facteur de forme est donné par
\begin{equation}
|G_\gamma^{\mu e}|^2 \approx 4 s_{14}^2 s_{24}^2 \cos^2\left(\frac{\delta_{14}+\delta_{25}-\delta_{15}-\delta_{24}}{2}\right) G_\gamma^2(x_{4,5})\,.
\end{equation}
Le taux des transitions cLFV dépend clairement des phases de Dirac, avec une annulation totale obtenue dans le cas $\delta_{14}+\delta_{25}-\delta_{15}-\delta_{24} = \pi$. 
D'autres facteurs de forme (par exemple ceux associés aux diagrammes de type pingouin et boîte $Z$, pertinents pour les désintégrations à trois corps et pour la conversion muon-électron, par exemple) dépendent également des phases de Dirac et de Majorana, mais ont des expressions associées plus complexes. 
La dépendance de plusieurs transitions $\mu-e$  par rapport aux phases de Dirac est illustrée sur le graphique de gauche de la Fig.~\ref{res_fig:1}, pour $\delta_{14}$ ; sous l'hypothèse simple 
$\sin \theta_{\alpha 4} =\sin \theta_{\alpha 5}$, et pour $m_4=m_5=1$~TeV, on retrouve le comportement identifié ci-dessus (et l'annulation formelle, pour $\delta_{14}=\pi$), présent pour les désintégrations $\mu \to e \gamma$, $\mu \to 3 e$ et $Z\to e \mu$. 
Une dépendance similaire est trouvée pour les phases de Majorana dans les observables considérées (sauf pour les désintégrations radiatives, auxquelles les phases de Majorana CPV ne contribuent pas). Ceci est illustré sur le panneau droit de la Fig.~\ref{res_fig:1}, pour le même ensemble d'observables et d'hypothèses sous-jacentes.

Les processus s'appuyant sur des topologies différentes (boîtes, $Z$ et penguoins photon, ...) peuvent présenter un degré significatif d'interférence (destructive ou constructive) des contributions distinctes, de sorte que les phases CPV de Dirac et de Majorana peuvent conduire à des annulations ou à des augmentations des taux associés. 
Il est également important de mentionner que chaque fois que des interactions $Z\nu\nu$ sont présentes, toutes les saveurs (et donc toutes les phases) contribuent.

\subsection{Vers des scénarios réalistes}
Après la première approche simple mentionnée ci-dessus, nous effectuons maintenant une étude réaliste de l'impact des phases CPV sur les observables cLFV ; des balayages complets de l'espace des paramètres ont été effectués (à la fois pour les angles de mélange et aussi pour toutes les phases), et toutes les contraintes disponibles (pertinentes) furent appliquées. En ce qui concerne ces dernières, et en plus des différentes contraintes cLFV, nous prenons en compte les résultats expérimentaux et les limites sur des extensions SM via des HNL à l'échelle du TeV\footnote{Nous considérons les contraintes provenant des observables de précision électrofaible ($M_W$, $G_F$, largeur invisible de $Z$, ...), les tests d'universalité leptonique (désintégrations leptoniques de $W$ et $Z$, rapports des désintégrations leptoniques de mésons, rapports des désintégrations (semi)leptoniques de tau, ...), les doubles désintégrations bêta sans émission de neutrinos, et enfin les contraintes d'unitarité perturbative ($\Gamma_{N_{4,5}}/m_{4,5} \leq 1/2$) ; pour une description détaillée et les références correspondantes, voir~\cite{Abada:2021zcm}.}.

\begin{figure}[ht!]
    \centering
\mbox{\hspace*{-5mm}    \includegraphics[width=0.51\textwidth]{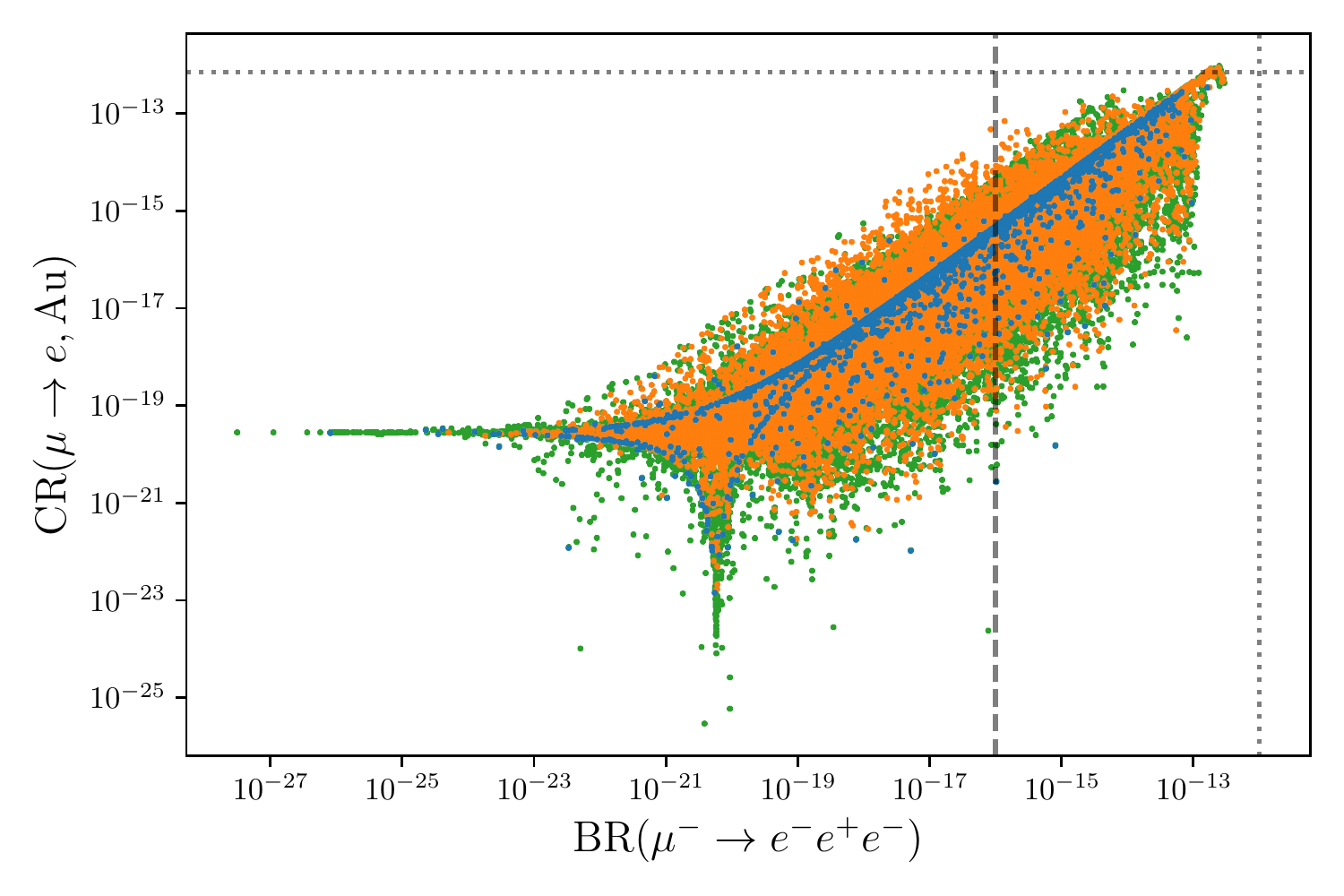}\hspace*{2mm}
\includegraphics[width=0.51\textwidth]{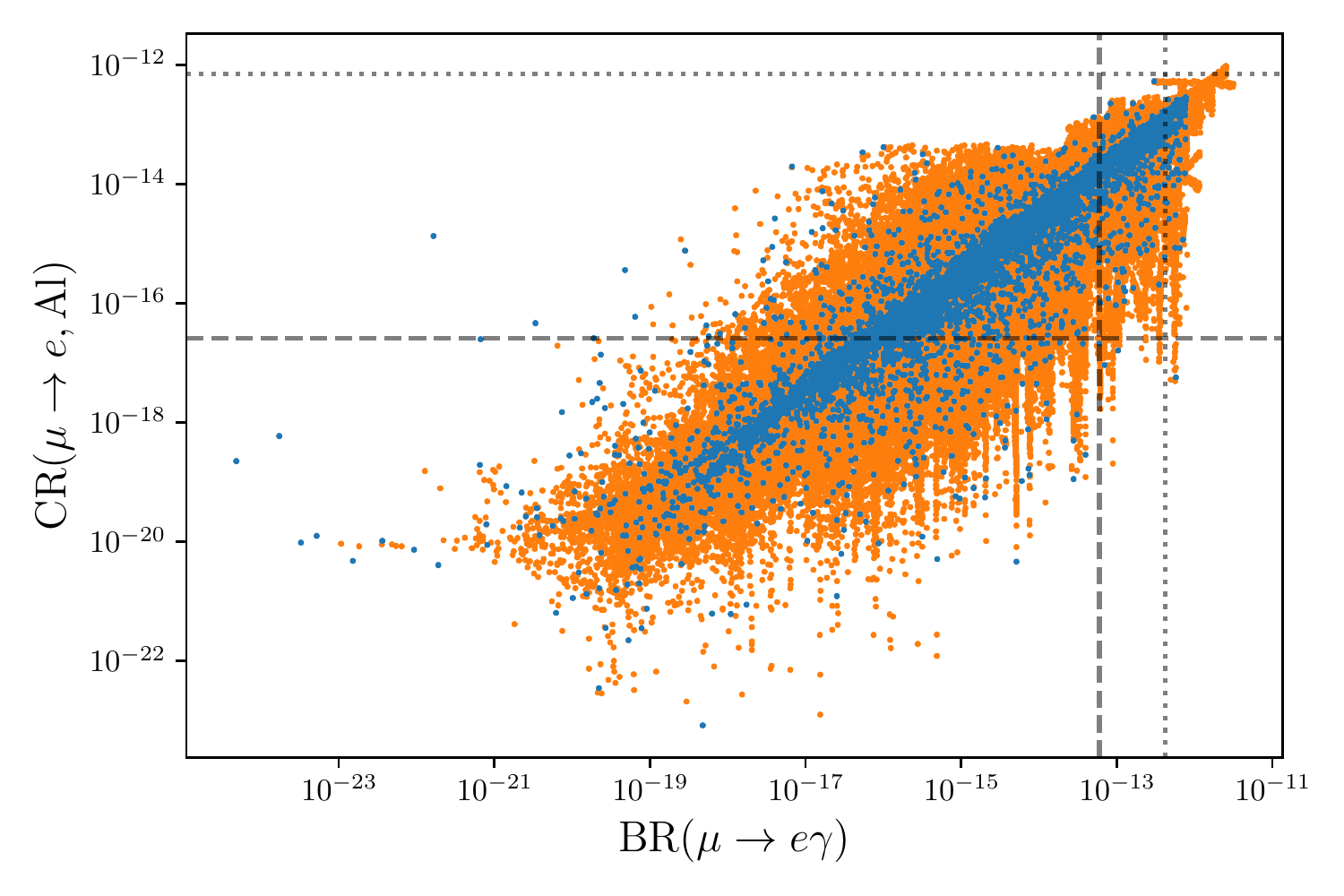}}
\caption{
    Corrélation des observables $\mu-e$ cLFV, pour des valeurs variables des phases CPV de Dirac et de Majorana : nulles (bleu), non nulles (orange), ``grille spéciale" (vert), cf. description dans le texte. De~\cite{Abada:2021zcm}.
        }
    \label{res_fig:2}
\end{figure}

Sur le graphique de gauche de la Fig.~\ref{res_fig:2}, nous présentons les effets des phases CPV sur la corrélation entre les taux de deux observables du secteur $\mu-e$, CR($\mu-e$, N) et BR($\mu \to 3e$). Pour aboutir aux résultats présentés, un balayage aléatoire a été effectué sur un espace de paramètres semi-contraint : en particulier, on n'impose plus que $\theta_{\alpha 4} \approx \pm \theta_{\alpha 5}$. Nous avons pris des états lourds dégénérés ($m_4=m_5=1$~TeV), et pour chaque point, les phases CPV $\delta_{\alpha 4}$ et $\varphi_4$ ont été fixées à zéro (points bleus), ont été variées de façon aléatoire (orange) et ont aussi été variées sur une grille (vert), cette dernière possibilité visant à s'assurer que les cas spéciaux d'``annulation" sont inclus.
Étant donné que dans le régime consideré pour les masses des fermions lourds, les deux observables reçoivent des contributions dominantes des pingouins $Z$, on s'attend à ce que les taux associés soient corrélés ; un tel comportement est effectivement observé - cf. la ligne bleue épaisse du graphique CR($\mu-e$, N) vs. BR($\mu \to 3e$). Cependant, et dès que les phases CPV sont non-nulles, on observe une perte de corrélation, d'autant plus frappante pour les valeurs ``spéciales" des phases $\{0, \frac{\pi}{4}, \frac{\pi}{2}, \frac{3\pi}{4}, \pi\}$ - ceux ci correspondant aux points verts.   
Compte tenu de ce comportement, il est important de souligner que les extensions de type HNL du SM ne doivent pas être écartées lors de l'observation d'un seul signal cLFV ; par exemple, 
si les futures recherches dans les collisionneurs suggèrent fortement la présence d'états stériles avec des masses proches de 1~TeV, et si BR($\mu \to 3 e$)$\approx 10^{-15}$ est mesuré, on ne doit pas forcément s'attendre à une observation de CR($\mu-e$, Al). Alors que pour les phases CPV nulles on s'attendrait à ce que cette dernière BR soit de~$\mathcal{O}(10^{-14})$, en présence de phases violant CP 
la plage attendue pour la conversion muon-électron est vaste, avec CR($\mu-e$, Al) potentiellement aussi bas que $10^{-18}$. 

Enfin, les résultats d'une vue d'ensemble de l'espace des paramètres du ``3+2 toy model" sont affichés sur le panneau de droite de la Fig.~\ref{res_fig:2}, dans lequel nous présentons 
CR($\mu-e$, N) vs. BR($\mu \to e \gamma$). Les résultats proviennent d'un balayage complet des paramètres de mélange (tous les angles $\theta_{\alpha 4}$ et $\theta_{\alpha 5}$ variés indépendamment, et toutes les phases de Dirac et de Majorana aussi variées aléatoirement\footnote{Les états lourds ne sont plus dégénérés, mais leurs masses sont prises suffisamment proches pour permettre des effets d'interférence ($m_4=1$~TeV, avec $m_5-m_4 \sim \Gamma_{N_{4,5}} \in [40~\text{Mev}, 210~\text{GeV}]$).}).
Encore une fois, on observe une perte de corrélation (qui serait autrement présente) pour les phases CPV non-nulles ; de plus, et comme mentionné ci-dessus, on peut maintenant avoir des taux importants pour une seule des observables. L'observation expérimentale de la conversion de $\mu \to e \gamma$ ne doit pas nécessairement être accompagnée de l'observation de la conversion de $\mu-e$ dans l'aluminium (et vice-versa).

\subsection{Discussion}
Comme discuté ici, la présence de phases CPV de Dirac et/ou de Majorana peut avoir un fort impact sur les taux d'observables cLFV, conduisant à une suppression ou à une augmentation des taux. 

La présence éventuelle de phases leptoniques - qui sont une caractéristique générique des mécanismes de génération de masse de neutrinos - devrait également être prise en compte lors de l'interprétation de données futures. Les phases CPV jouent un rôle crucial dans l'évaluation de la viabilité des (régimes des) extensions du SM via HNL. Plusieurs exemples sont fournis dans le tableau~\ref{res_table:1}, dans lequel nous résumons les prédictions de certains points de référence P$_i$ (pour des choix distincts des angles de mélange actif-stérile) concernant les observables cLFV, ainsi que les prédictions associées aux valeurs non-nulles des phases (P$^\prime_i$) :
\begin{eqnarray}
\label{res_eq:Pi:angles}
& \text{P}_1: &
s_{14} = 0.0023\,, \:s_{15} = -0.0024\,,\:
s_{24} = 0.0035\,, \:s_{25} = 0.0037\,, \:
s_{34} = 0.0670\,, \:s_{35} = -0.0654\,, \nonumber \\
& \text{P}_2: &
s_{14} = 0.0006\,, \: s_{15} = -0.0006\,, \: 
s_{24} = 0.008\,, \: s_{25} = 0.008\,, \: 
s_{34} = 0.038\,, \: s_{35} = 0.038\,, \nonumber \\
& \text{P}_3: &
s_{14} = 0.003\,, \: s_{15} = 0.003\,, \:  
s_{24} = 0.023\,, \:  s_{25} = 0.023\,, \:  
s_{34} = 0.068\,, \:  s_{35} = 0.068\,. 
\end{eqnarray}
Les variantes P$^\prime_i$ ont des angles de mélange identiques, mais en association avec les configurations de phase suivantes :
\begin{eqnarray}\label{res_eq:Pi:phases}
\text{P}^\prime_1:    
\delta_{14} = \frac{\pi}{2}\,, \:  
\varphi_4 = \frac{3\pi}{4}\,;\quad  
\text{P}^\prime_2:    
\delta_{24}=\frac{3\pi}{4}\,, \:  
\delta_{34} = \frac{\pi}{2}\,, \:  
\varphi_4 = \frac{\pi}{\sqrt{8}}\,; \quad  
\text{P}^\prime_3: 
\delta_{14}\approx \pi\,, \:  
\varphi_4\approx \frac{\pi}{2}\,.
\end{eqnarray}
\noindent Nous avons choisi $m_4=m_5=5$ TeV pour les trois points de référence.

À titre d'exemple, notons que le régime de grands angles de mélange associé à P$_3$ serait exclu en raison d'un conflit avec les limites actuelles pour les observables cLFV; cependant, la présence de phases CPV permet de concilier facilement les prédictions avec l'observation (P$_3^\prime$), et donc de rendre viable le régime de mélange associé.

En résumé, la présence de phases CPV leptoniques (à la fois de Dirac et de Majorana) devrait être systématiquement prise en compte dans l'analyse phénoménologique des extensions HNL du SM en ce qui concerne les prospectives pour la cLFV.

\renewcommand{\arraystretch}{1.}
\begin{table}[h!]
\caption{Prédictions pour les observables cLFV en association avec P$_i$, et variantes avec des phases de violation de CP non-nulles, P$_i'$. Les symboles 
({\small\XSolidBrush}, $\checkmark$, $\circ$) désignent 
les taux en conflit avec les limites expérimentales actuelles, les prédictions dans la limite de la sensibilité future et celles hors de portée future.
}
\vspace*{2mm} \hspace*{10mm}
    \begin{tabular}{|l|c|c|c|c|c|}
    \hline
 & BR($\mu \to e\gamma$) & BR($\mu \to 3e$) & CR($\mu - e$, Al) & BR($\tau \to 3\mu$)& BR($Z \to \mu \tau$)\\
 \hline\hline
$\text P_1$ & $ 3\times 10^{-16}$ \:\:$\circ$ & 
$ 1\times 10^{-15}$ \:\:$\checkmark$& 
$ 9\times 10^{-15}$ \:\:$\checkmark$& 
$ 2\times 10^{-13}$ \:\:$\circ$&
$ 3\times 10^{-12}$ \:\:$\circ$\\
$\text P_1^\prime$
& $ 1\times 10^{-13}$ \:\:$\checkmark$& 
$2\times 10^{-14}$ \:\:$\checkmark$& 
$1\times 10^{-16}$ \:\:$\checkmark$& 
$1\times 10^{-10}$ \:\:$\checkmark$& 
$2\times 10^{-9}$ \:\:$\checkmark$\\
\hline
\hline
$\text P_2$ 
& $2\times 10^{-23}$ \:\:$\circ$
& $2\times 10^{-20}$ \:\:$\circ$ 
& $2\times 10^{-19}$ \:\:$\circ$ 
&  $1\times 10^{-10}$ \:\:$\checkmark$
& $3\times 10^{-9}$ \:\:$\checkmark$\\
$\text P_2^\prime$
& $6\times 10^{-14}$ \:\:$\checkmark$
& $4\times 10^{-14}$ \:\:$\checkmark$
& $9\times 10^{-14}$ \:\:$\checkmark$
&  $8\times 10^{-11}$ \:\:$\checkmark$
& $1\times 10^{-9}$ \:\:$\checkmark$\\
 \hline
 \hline
 $\text{P}_3$ 
 & $2\times 10^{-11}$ \:\:{\footnotesize \XSolidBrush}
 & $3\times 10^{-10}$ \:\:{\footnotesize \XSolidBrush} 
 & $3\times 10^{-9}$ \:\:{\footnotesize \XSolidBrush}
 & $2\times 10^{-8}$ \:\:$\checkmark$
 & $8\times 10^{-7}$ \:\:$\checkmark$\\
$\text P_3^\prime$
& $8\times 10^{-15}$ \:\:$\circ$
  & $1\times 10^{-14}$ \:\:$\checkmark$
  & $6\times 10^{-14}$ \:\:$\checkmark$
  & $2\times 10^{-9}$ \:\:$\checkmark$
  & $1\times 10^{-8}$ \:\:$\checkmark$\\
 \hline
\end{tabular}\label{res_table:1}
\end{table}
\renewcommand{\arraystretch}{1.}

\section{Moments magnétiques anormaux des leptons chargés}
Le moment magnétique (dipôle) d'une particule chargée est une mesure de la tendance de cette particule à s'aligner avec un champ magnétique.
Pour un fermion, ou en particulier un lepton chargé de spin $\vec S$ et de masse $m_\ell$, le moment magnétique est donné par
\begin{equation}
    \vec M = g_\ell \,\frac{e}{2 m_\ell}\, \vec S\,,
\end{equation}
dans laquelle $g_\ell$ est la ``force de couplage" du lepton à un champ magnétique, appelée ``facteur de Landé".
L'équation de Dirac implique $g_\ell = 2$, mais ce résultat est sensible aux corrections quantiques.
En électrodynamique quantique (QED), un lepton chargé couplé à un champ magnétique externe est décrit par un courant leptonique invariant de jauge couplé à un photon 
hors couche.
Le courant leptonique électromagnétique invariant de jauge peut en général être paramétré comme suit
\begin{equation}
    \mathcal J_\mu = \bar \ell (p')\left[F_1(q^2) \gamma_\mu + \frac{i}{2m_\ell} F_2(q^2) \,\sigma_{\mu\nu} q^\nu - F_3(q^2)\gamma_5 \,\sigma_{\mu\nu}q^\nu + F_4(q^2)(q^2\gamma_\mu - 2 m_\ell q_\mu)\gamma_5\right]\ell(p)\,,
\end{equation}
dans laquelle $q$ est le momentum du photon et $F_i$ sont les facteurs de forme électromagnétiques.
Le facteur de Landé est alors donné par
\begin{equation}
    g_\ell = 2(F_1(0) + F_2(0))\,.
\end{equation}
Au niveau arbre dans le SM, nous avons $F_1(0) = 1$ et $F_{2,3,4}(0) = 0$, ce qui conduit à $g_\ell = 2 = g_\text{Dirac}$.
Dans la théorie des perturbations, les corrections d'ordre supérieur apportées à $F_1$ ne modifient que le couplage original au photon et donnent donc la dépendance d'échelle de la charge électronique $e$, de sorte que les corrections de $g_\ell$ ne peuvent provenir que de contributions d'ordre supérieur à $F_2(0)$.
L'autre facteur de forme $F_3(0)$ induit le moment dipolaire électrique $d_\ell$, tandis que $F_4$ n'est pertinent que pour les échanges de photons virtuels à courte distance, souvent appelés ``anapole".

Les corrections d'ordre supérieur contribuant à $F_2(0)$ et donc à $g_\ell$, sont aisément capturées dans le soi-disant \textit{moment magnétique anormal} défini comme suit
\begin{equation}
    a_\ell \equiv \frac{g_\ell - g_\text{Dirac}}{g_\text{Dirac}} = \frac{g_\ell - 2}{2} = F_2(0)\,,
\end{equation}
communément appelé $(g-2)_\ell$.
La première correction à l'ordre suivant (NLO) en QED a été calculée pour la première fois en 1948, ce qui a donné $a_\ell = \frac{\alpha_e}{2\pi}$, où $\alpha_e = \frac{e^2}{4\pi}$ est la constante de structure fine électromagnétique.
Depuis, de nombreux progrès ont été réalisés.
En général, les corrections quantiques du moment magnétique anormal peuvent être divisées en trois catégories. Il y a les contributions de la QED pure, qui ne dépendent que de la masse du lepton chargé et de $\alpha_e$, et qui ont été entièrement calculées perturbativement jusqu'à une précision de 5 boucles. 
Pour le moment magnétique anormal du muon $a_\mu$, les corrections provenant des interactions faibles ont également été calculées jusqu'à la précision NLO (2 boucles).
De plus, les corrections QCD dues à la ``light-by-light scattering'' hadronique~\cite{Colangelo:2015ama,Green:2015sra,Gerardin:2016cqj,Blum:2016lnc,Colangelo:2017qdm,Colangelo:2017fiz,Blum:2017cer,Hoferichter:2018dmo,Hoferichter:2018kwz}, dues à la polarisation du vide hadronique
hadroniques~\cite{Chakraborty:2016mwy,Jegerlehner:2017lbd,DellaMorte:2017dyu,Davier:2017zfy,Borsanyi:2017zdw,Blum:2018mom,Keshavarzi:2018mgv,Colangelo:2018mtw,Davier:2019can}, 
et les corrections hadroniques d'ordre supérieur~\cite{Kurz:2014wya,Colangelo:2014qya} doivent être prises en compte pour obtenir une prédiction SM suffisamment précise.
Avant le calcul le plus récent, basé sur la QCD sur réseau\footnote{En raison des incertitudes relativement importantes dans les calculs antérieurs de QCD sur réseau, une autre méthode pour déterminer le LO HVP repose sur une approche basée sur les données de production de hadrons à partir de photons virtuels dans la diffusion $e^+e^-$. Pour une revue de ces évaluations, voir~\cite{Aoyama:2020ynm}.}, de la contribution de la polarisation hadronique du vide d'ordre supérieur (LO HVP)
par la collaboration BMW~\cite{Borsanyi:2020mff}, la prédiction SM 
compilée récemment par le ``Muon $g-2$ Theory Inititative''~\cite{Aoyama:2020ynm} s'est avérée être la suivante
\begin{equation}\label{res_eq:amu:SMwhite}
    a_\mu^\text{SM}\, =\, 116\, 591\, 810 \, (43) \times 10^{-11}\,,
\end{equation}
où l'incertitude est dominée par les contributions hadroniques.

Suite aux premiers résultats récemment divulgués de l'expérience ``g-2" E989 à  FNAL~\cite{Muong-2:2021ojo}, qui sont en accord avec les résultats précédents de l'expérience BNL E821~\cite{Bennett:2006fi}, la moyenne expérimentale actuelle du moment magnétique anormal du muon~\cite{Muong-2:2021ojo} est donnée par
\begin{equation}\label{res_eq:amu:exp}
    a_\mu^^text{exp}\, =\, 116\, 592\, 061\, (41) \times 10^{-11}\,,
\end{equation}
une valeur qui doit être comparée à la prediction du SM (cf. Eq.~\eqref{res_eq:amu:SMwhite}), ce qui conduit à une tension de $4.2\,\sigma$ entre  
la théorie et l'observation
\begin{equation}\label{res_eq:amu:delta}
    \Delta a_\mu\, \equiv a_\mu^\text{SM}\,- \, 
    a_\mu^\text{exp}\,
    \, =\, 251 \, (59) \times 10^{-11}\,.
\end{equation}
La précision impressionnante de la prédiction théorique et des mesures expérimentales rend $a_\mu$ une observable de haute précision, extrêmement sensible aux contributions de la Nouvelle Physique.

Si on prend en compte le calcul de la collaboration BMW, 
la valeur obtenue ($a_\mu^\text{SM}=116\,591\,954\, (57)\times 10^{-11}$) suggérerait $\Delta a_\mu= 107\, (70)\times 10^{-11}$, correspondant à une tension de $1.5\,\sigma$ entre  
la théorie et l'observation. 
En attendant une confirmation\footnote{Dans~\cite{Crivellin:2020zul}, il a été souligné que de telles contributions de polarisation hadronique du vide pourraient potentiellement conduire à des conflits avec les analyses électrofaibles, induisant des tensions dans d'autres
  observables pertinentes (jusqu'à présent en bon accord avec le SM).} 
indépendante
 des calculs des contributions des LO HVP basés sur la QCD sur réseau, nous nous baserons dans ce qui suit sur $\Delta a_\mu$ obtenu à partir de la valeur SM telle que donnée dans Eq.~(\ref{res_eq:amu:SMwhite}).
 Une vue d'ensemble des moyennes des prédictions SM et des mesures expérimentales est présentée dans la Fig.~\ref{fig:g-2overview}~\cite{Lellouchplot}.
 \begin{figure}
    \centering
    \includegraphics[width=0.6\textwidth]{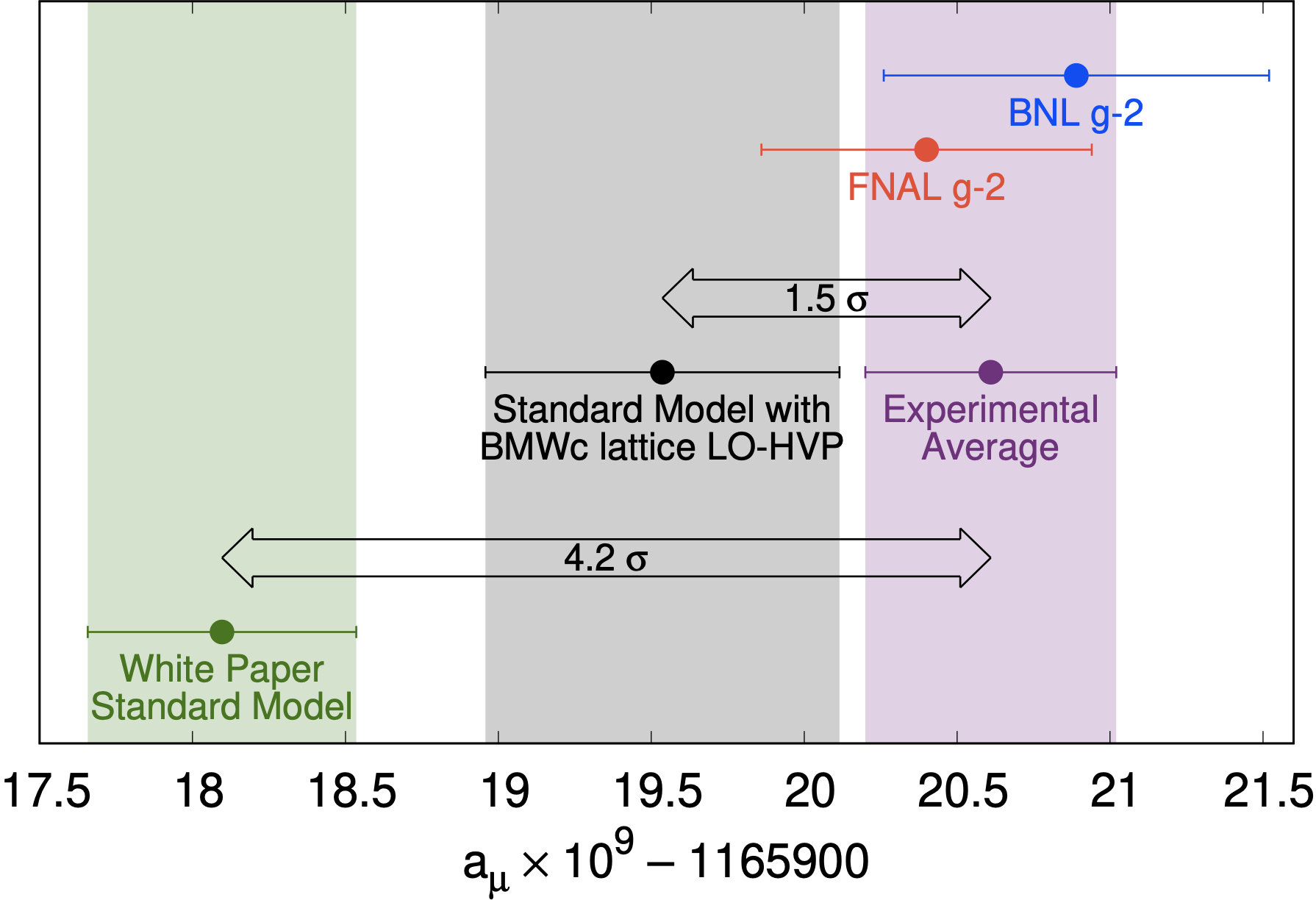}
    \caption{Aperçu des moyennes actuelles des prédictions SM et des mesures expérimentales de $a_\mu$. La région verte représente la prédiction SM compilée dans~\cite{Aoyama:2020ynm}, la région grise représente la prédiction SM prenant en compte la détermination de HVP par QCD sur réseau telle qu'obtenue dans~\cite{Borsanyi:2020mff}, et la région violette représente la moyenne expérimentale des mesures de BNL~\cite{Bennett:2006fi} et FNAL~\cite{Muong-2:2021ojo}.
    Figure tirée de~\cite{Lellouchplot}.}
    \label{res_fig:g-2overview}
\end{figure}

Dans l'hypothèse d'une tension significative entre la théorie et l'observation, telle que donnée par 
l'équation~(\ref{res_eq:amu:delta}), il est évident
que l'on a besoin d'une Nouvelle Physique capable de combler un tel écart ; 
plusieurs modèles minimaux, ainsi que des modèles de NP plus complets, ont été explorés en profondeur à la lumière des récents résultats expérimentaux (pour une revue récente, voir, par exemple,~\cite{Athron:2021iuf} et ses références).  

\smallskip
Afin d'accommoder la tension dans $a_\mu$, on s'attend à ce que des contributions de Nouvelle Physique apparaissent au niveau d'une boucle.
Selon le rapport des masses des champs BSM (scalaire $S$, vecteur $V$, ou fermions $F$) se propageant à l'intérieur de la boucle $m_F/m_S$ ou $m_F/m_V$, et selon la grandeur des couplages (chiraux) pertinents, d'importantes contributions de Nouvelle Physique sont possibles.
En général, la tension $\Delta a_\mu$ peut être expliquée avec des champs BSM comparativement légers et des couplages importants menant à des améliorations chirales.
Les explications de la nouvelle physique impliquant des champs BSM légers sont toutefois soumises à un large éventail d'autres contraintes indirectes provenant du LHC, des usines de saveurs et des recherches de matière noire.
Pour une étude complète des modèles candidats impliquant jusqu'à trois champs BSM, qui expliquent la tension dans $a_\mu$ (avec des liens possibles avec, par exemple, la matière noire), voir~\cite{Athron:2021iuf}.

\bigskip
Le moment magnétique anormal de l'électron $a_e$ a été calculé en QED avec une précision impressionnante (à 4 boucles). 
Du point de vue expérimental, jusqu'à récemment, les mesures de $a_e$ ont été utilisées pour déduire la valeur de $\alpha_e$ à basse énergie.
Il est intéressant de noter qu'une mesure précise de $\alpha_e$ utilisant des atomes de $\mathrm{Cs}$~\cite{Parker:2018vye,Yu:2019gdh}, est à l'origine d'une autre divergence, cette fois-ci
concernant le moment magnétique anormal de l'électron.
La mesure expérimentale du moment magnétique anormal de l'électron
$a_e$~\cite{Hanneke:2008tm} 
\begin{equation}
a_e^\text{exp}\, =\, 1\,159\,652\,180.73(28)\times 10^{-12}\,
\end{equation}
présente actuellement une déviation de $2.5\,\sigma$ par rapport à la prédiction du SM (en se basant sur la valeur de $\alpha_e$ déduite des atomes de césium), 
\begin{equation}
\label{res_Delta_aeCS}
\Delta a_e^\text{Cs}\,=\,a_e^\text{exp} - a_e^\text{SM}\,\sim\, 
-0.88 (0.36)\times 10^{-12}\,.
\end{equation}
Dans~\cite{Morel:2020dww}, une estimation plus récente de $\alpha_e$ a été obtenue, cette fois-ci en utilisant des atomes de rubidium ; la nouvelle détermination de $\alpha_e$ (impliquant une déviation globale au-dessus du niveau de $5\,\sigma$ pour $\alpha_e$) suggère maintenant des tensions plus petites entre l'observation et la prédiction théorique, \begin{equation}\label{res_eq:ae:deltaRb}
    \Delta a_e^\text{Rb} \, =\, 0.48 \, (0.30)\times 10^{-12}\,,
\end{equation}
correspondant à une déviation de $\mathcal{O}(1.7\,\sigma)$.
En plus de signaler des déviations par rapport à l'attente SM, 
il est intéressant de noter l'impact potentiel de \textit{tous deux}
$\Delta a_e$ et $\Delta a_\mu$ : en plus d'avoir un signe opposé, le rapport $\Delta a_\mu/\Delta a_e$ n'obéit pas  
aux estimations naïves $\sim m^2_\mu/ m^2_\mu$ (attendues de l'opérateur dipôle magnétique, dans lequel une insertion de masse du lepton SM est responsable de l'inversion de chiralité requise~\cite{Giudice:2012ms}). Ce comportement rend une explication commune des deux tensions assez difficile, faisant appel à un départ de l'hypothèse de violation minimale de la saveur (MFV), ou à aller au-delà des extensions simples 
du SM 
(par une seule particules de NP qui se couple aux leptons chargés~\cite{Davoudiasl:2018fbb,Kahn:2016vjr,Crivellin:2018qmi,Dorsner:2020aaz}). 
Remarqueons que le schéma émergeant de deux $\Delta a_e$ et $\Delta a_\mu$ pourrait également être perçu comme suggérant une violation de l'universalité des saveurs.

Enfin, en ce qui concerne le moment magnétique anormal du $\tau$-lepton, la précision expérimentale~\cite{DELPHI:2003nah} est encore très faible par rapport à l'incertitude théorique~\cite{Eidelman:2007sb},
\begin{eqnarray}
    a_\tau^\text{SM} &=& (117721\pm 5)\times 10^{-8}\,,\nonumber\\
    -0.052 < a_\tau^\text{exp} &<& 0.013\,,
\end{eqnarray}
de sorte que, malheureusement, cette observable ne peut pas encore être utilisée pour déduire des informations utiles sur les éventuelles contributions de la nouvelle physique.

\section{Anomalies de désintégration du méson $B$ et leptoquarks}
Dans le SM, les leptons chargés ne sont distinguables que par leur masse. En particulier, tous les  couplages électrofaibles avec les bosons de jauge ne dépendent pas 
de la saveur des leptons, ce qui conduit à une symétrie accidentelle appelée universalité de la saveur des leptons (LFU), dont la validité a été déterminée avec une très grande précision, par exemple dans les désintégrations de $Z\to \ell^+\ell^-$ et de $W^\pm\to \ell^\pm \nu$ ($\ell = e, \mu, \tau$)~\cite{ParticleDataGroup:2020ssz}.

Cependant, au cours de la dernière décennie, des indices de la violation de l'LFU dans les désintégrations de $b\to c\ell\nu$ et de $b\to s\ell\ell$ ont commencé à émerger, avec une tension croissante par rapport aux attentes du SM.
En particulier, les mesures des rapports de branchement ``théoriquement propres" ${R_{D^{(\ast)}}} = \mathrm{BR}(B\to D^{(\ast)}\tau\nu)/\mathrm{BR}(B\to D^{(\ast)}\ell\nu)$~\cite{Abdesselam:2019dgh} et ${R_{K^{(\ast)}}} = \mathrm{BR}(B\to K^{(\ast)}\mu\mu)/\mathrm{BR}(B\to K^{(\ast)}ee)$~\cite{Aaij:2019wad,Aaij:2017vbb} s'écartent d'environ $2 - 3\,\sigma$ de leurs prédictions théoriques qui, jusqu'à des corrections dûes à l'espace de phase, devraient être 1 dans le SM.
Les moyennes des mesures expérimentales actuelles  et les prédictions du SM peuvent être trouvées dans le tableau~\ref{res_tab:data}.
\begin{table}[h!]

\begin{center}
{\small
\begin{tabular}{|c|c|c|c|c|}
\hline
     &  $R_{K}$ & $R_{K^{*}}$ & $R_{D}$ & $R_{D^{*}}$\\
     \hline
Prédictions du SM  &  $\simeq 1$ & $\simeq 1$ 
& $0.299 \pm 0.003$ & $0.258 \pm 0.005$ \\
\hline
Mesures expérimentales     & $0.845 \pm 0.06$ & $0.69 \pm 0.12$  
& $0.340 \pm 0.030$ & $0.295 \pm 0.014$\\
\hline
\end{tabular}
}
\label{res_tab:data}
\caption{Prédictions du SM et (moyennes des) mesures expérimentales des observables ``théoriquement propres" de LFU.}
\end{center}

\end{table}

En outre, les mesures des observables angulaires dans les désintégrations de $B^{0,+}\to K^\ast \mu^+\mu^-$ présentent des déviations (locales) de $2-3\,\sigma$ dans plusieurs intervalles (``bins") de $q^2$.
Ces mesures~\cite{LHCb:2020lmf,LHCb:2020gog} ont été récemment mises à jour, confirmant et renforçant les hypothèses NP actuellement privilégiées.

Très récemment, la collaboration LHCb a mis à jour sa mesure de $R_{K} = 0.846^{+0.044}_{-0.041}$~\cite{LHCb:2021trn} avec un écart par rapport à la prédiction du SM atteignant désormais $3.1 \sigma$, fournissant ainsi la première \textit{évidence} de la violation du LFU\footnote{Pour une vue d'ensemble (animée) de l'évolution de l'ajustement aux données de $b\to sell\ell$ lors de l'inclusion des nouvelles mesures, voir \href{http://moriond.in2p3.fr/2021/EW/slides/ani_fit_evo.mp4}{\textcolor{blue}{http://moriond.in2p3.fr/2021/EW/slides/ani\_fit\_evo.mp4}}.}.

Ces tensions persistantes avec le SM semblent indiquer indirectement la présence d'une nouvelle physique, probablement à l'échelle du TeV. 
De nombreuses approches différentes ont été explorées pour identifier lesquels
modèles (minimaux) de NP réussissent le mieux à réconcilier les prédictions théoriques avec les données expérimentales. 
Avant d'aborder les perspectives de différentes extensions du SM par des leptoquarks (LQ) vecteurs du (l'un des scénarios les plus prometteurs et motivés pour expliquer simultanément les deux anomalies), nous considérerons une approche basée sur la théorie effective des champs (EFT), 
qui est indépendante du modèle. Cela permettra d'identifier de manière générique lesquelles classes de modèles de NP offrent le contenu et les interactions les plus appropriés pour expliquer les données. 

\subsection{EFT et ``global fits"}
L'approche EFT repose sur la paramétrisation des effets de NP en termes d'opérateurs d'ordre supérieur non-renormalisables (traces résiduelles d'états plus lourds dans la théorie de basse énergie ). En partant de sous-ensembles pertinents du Lagrangien effectif, exprimés en termes de coefficients de Wilson (WC) semileptoniques $C^{q q^\prime ; \ell \ell^\prime }$ et d'opérateurs effectifs, nous
commentons comment des scénarios bien motivés pour (des ensembles de) $C^{q q^\prime ; \ell \ell^\prime }$ deviennent significativement favorisés par les données actuelles.

\begin{figure}
    \centering
    \includegraphics[width=0.48\textwidth]{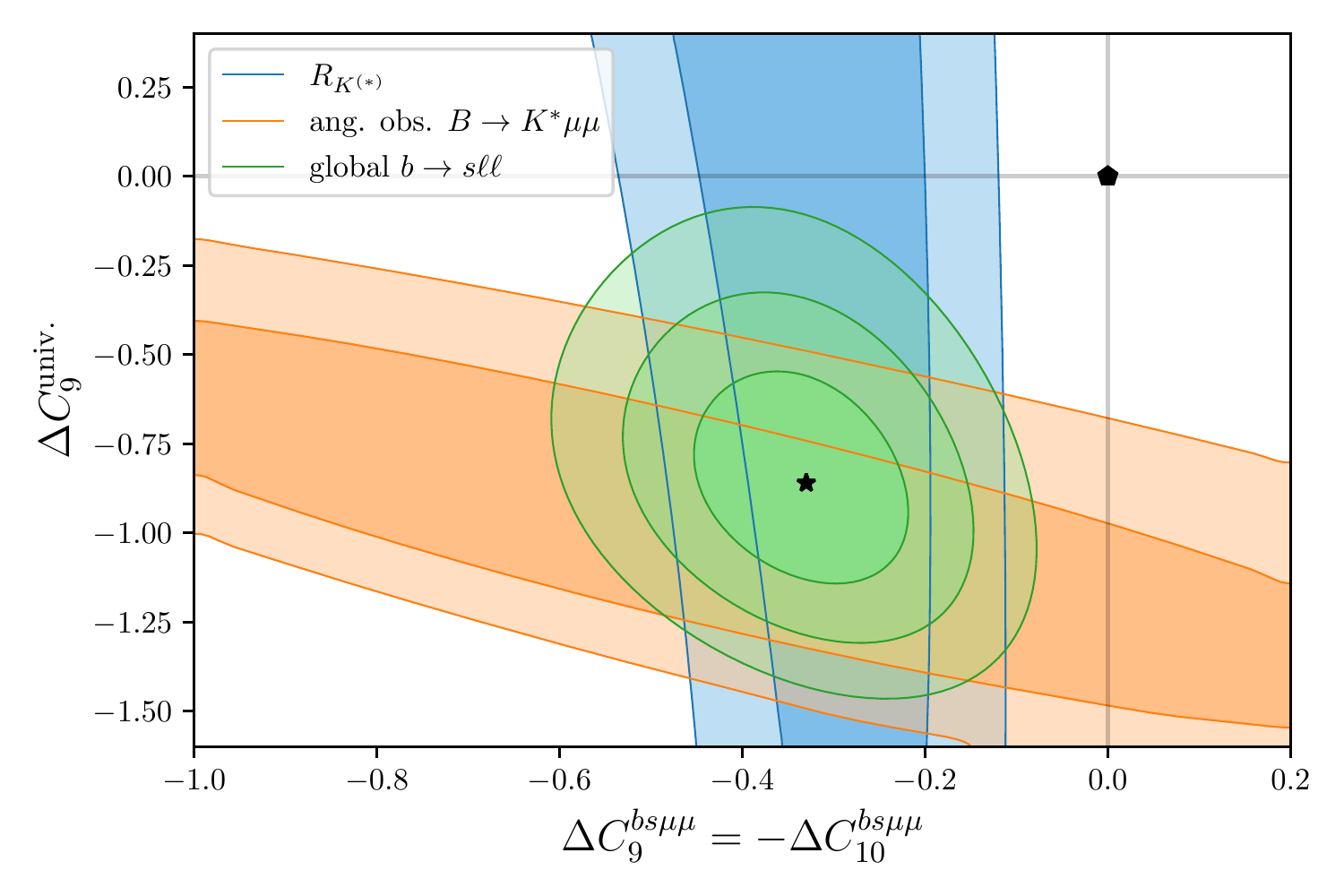}\includegraphics[width=0.48\textwidth]{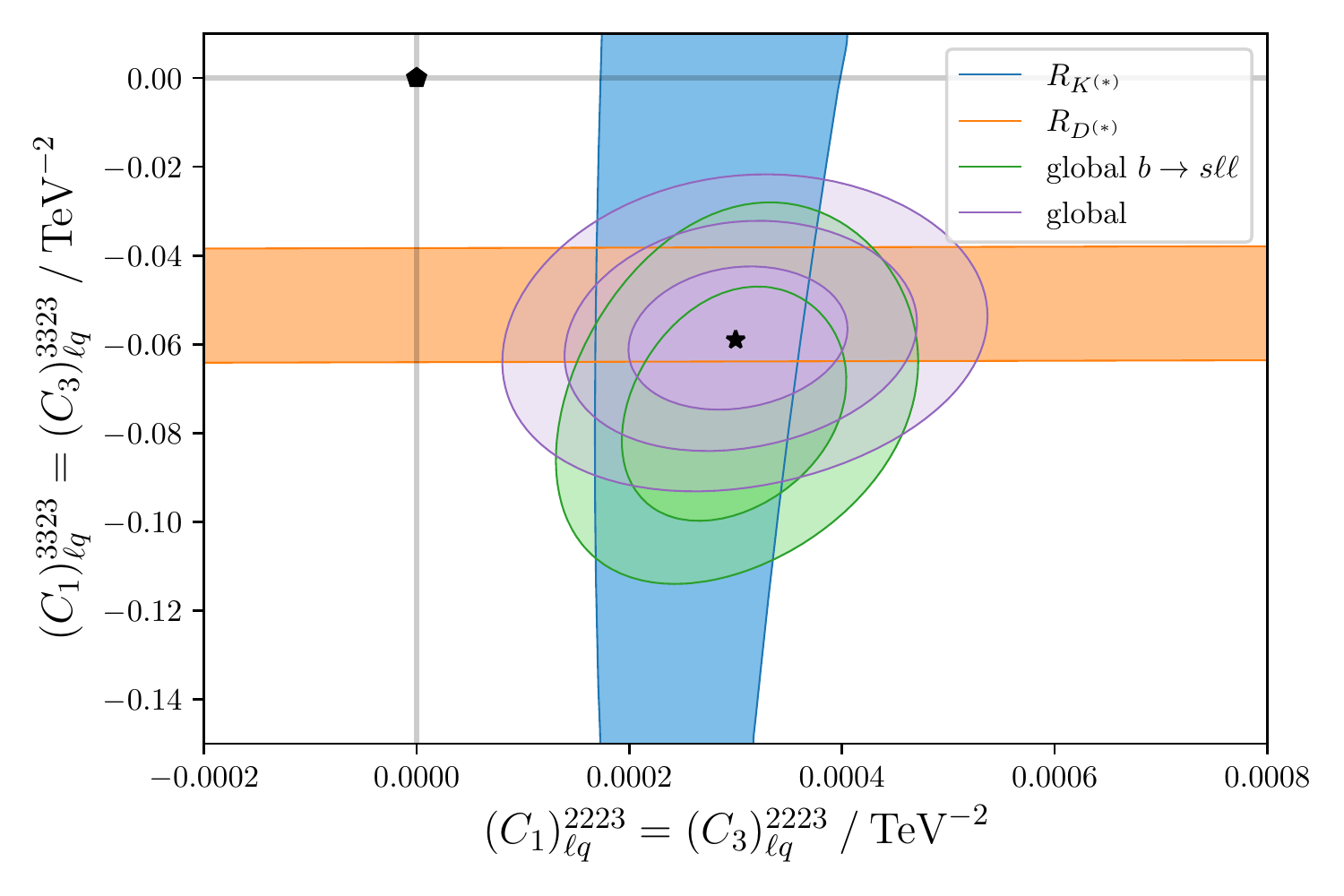}
    \caption{Likelihood contours (1$\sigma$, 2$\sigma$ (et 3$\sigma$ pour l'ajustement global)) dans le plan des coefficients de Wilson WET et SM-EFT. Les lignes de contour en pointillé indiquent la situation avant la mesure de $R_{K}$ de 2021, un pentagone la prédiction du SM, un losange l``ancien" point de meilleur ajustement et une étoile le nouveau point de meilleur ajustement après inclusion des données actualisées de $R_{K}$. \textbf{Gauche:} Ajustement des coefficients WET à $4.8\,\mathrm{GeV}$ avec le point de meilleur ajustement $\Delta C_9^{bs\mu\mu} = -0.34^{+0.08}_{-0.08}$ et $\Delta C_9^\mathrm{univ.} = -0.74^{+0.19}_{-0.17}$. \textbf{Droite:} Ajustement des coefficients SM-EFT à $2\,\mathrm{TeV}$ avec le meilleur point d'ajustement de $(C_1)_{\ell q}^{2223} = (2.9^{+0.6}_{-0. 6})\times 10^{-4}\,\mathrm{TeV}^{-2}$ et $(C_1)_{\ell q}^{3323} = 0.056^{+0.01}_{-0.01}\,\mathrm{TeV}^{-2}$ ;   l'écart est de $7.9\sigma$ par rapport à la prédiction du SM.}
    \label{res_fig:fits}
\end{figure}

Le sous-ensemble du Lagrangien effectif pour les transitions de courant chargé $d_k\to u_j\ell\nu_i$ est donné par
\begin{equation}
    \mathcal L_{\mathrm{eff}} \simeq -\frac{4 G_F}{\sqrt{2}}V_{jk}\left[(1 + C_{V_L}^{jk\ell i})(\bar u_j \gamma_\mu d_k)(\bar \ell \gamma^\mu P_L \nu^i) \right] + \mathrm{H.c.}\,,
\end{equation}
où $V_{jk}$ sont des éléments de la matrice de mélange CKM.
Alors que les anomalies des courants chargés $R_{D^{(\ast)}}$ peuvent être expliquées par les contributions des NP au coefficient vectoriel gauche $\mathcal C_{V_L}^{c b \tau \nu}$, celles des courants neutres -- en particulier en raison des déviations des observables angulaires -- nécessitent une analyse EFT spécifique pour identifier la structure d'opérateur (ou la combinaison de structures) préférée par les données expérimentales. 
Un sous-ensemble du Lagrangien effectif à basse énergie pour les transitions de $b\to sell\ell$ peut être exprimé comme suit
\begin{equation}
    \mathcal L_\mathrm{eff} \simeq \frac{4 G_F}{\sqrt{2}}V_{t d_j}V_{t d_i}^\ast\left[\frac{\alpha_e}{4\pi} C_9^{ij\ell\ell'}(\bar d_i\gamma^\mu P_L d_j)(\bar \ell \gamma_\mu \ell') + \frac{\alpha_e}{4\pi} C_{10}^{ij\ell\ell'}(\bar d_i\gamma^\mu P_L d_j)(\bar \ell \gamma_\mu\gamma_5 \ell') \right]\, .
\end{equation}
Il s'avère qu'un scénario de NP très intéressant est donné en prenant en compte de contributions de nouvelle physique $V-A$ violant le LFU dans $\Delta C_9^{bs\mu\mu} = - \Delta C_{10}^{bs\mu\mu}$, en plus d'une contribution de NP vectorielle de LFU désignée par $\Delta C_9^\mathrm{univ.}$ qui est de universelle et ajoutée à $C_9^{bs\mu\mu}$ et $C_9^{bsee}$. 
Un ajustement de toutes les données disponibles de $b\to s\ell\ell$
y compris la mesure actualisée de $R_{K}$, conduit à une amélioration de $\sim 6.6 \sigma$ par rapport à la prédiction du SM. Le meilleur point d'ajustement est donné par $\Delta C_9^{bs\mu\mu} = -0.34^{+0.08}_{-0.08}$ et $\Delta C_9^{\mathrm{univ.}} = -0.74^{+0.19}_{-0.17}$, ce qui montre une préférence de $\sim 3\sigma$ pour les contributions non-nulles de NP aux WCs, comme manifeste par les contours  
dans le plan des théories effectives faibles (WET) et des coefficients de Wilson SM-EFT, montrés dans le panneau gauche de la Fig.~\ref{res_fig:fits}.

De manière intéressante, et comme il a été souligné dans la Réf.~\cite{Crivellin:2018yvo}, une contribution universelle à $C_9^{bs\ell\ell}$ peut être générée à partir 
d'effets
du groupe de renormalisation (RG), dûs à des operateurs semi-tauoniques.
Dans le panneau de droite de la 
Fig.~\ref{res_fig:fits} nous affichons les
``likelihood contours"\footnote{En plus des données $b\to s\ell\ell$ et $R_{D^{(\ast)}}$, la likelihood globale contient également les fractions de branchement échantillonnées
dans les désintégrations exclusives $B\to D^{(\ast)}\ell\nu$.}
dans le plan des coefficients 
de Wilson SM-EFT de type singlets et triplets de $SU(2)_L$ semi-muoniques et semi-tauoniques  dans $b\to s\ell\ell$. Comme on peut le voir, les contours de $b\to s\ell\ell$ ne sont pas indépendants des opérateurs semi-tauoniques, un effet des contributions universelles induites par RGE à $C_9^{bs\ell\ell}$ à l'échelle de masse des quarks $b$.

Cette corrélation entre les anomalies des courants chargés et neutres à hautes énergies suggère fortement une explication combinée des deux tensions dans un modèle minimal.

Il est intéressant de noter que la structure préférée des opérateurs EFT (à la fois dans SM-EFT et WET) est naturellement générée par un LQ vecteur singlet de $SU(2)_L$, $V_1$, contribuant aux transitions anomales des courants chargés et neutres au niveau arbre, tout en échappant aux contraintes strictes des désintégrations $d_i\to d_j \nu\bar\nu$.

\subsection{Leptoquarks vecteurs}
\begin{figure}
    \centering
    \mbox{\raisebox{-2mm}{\includegraphics[width=0.48\textwidth]{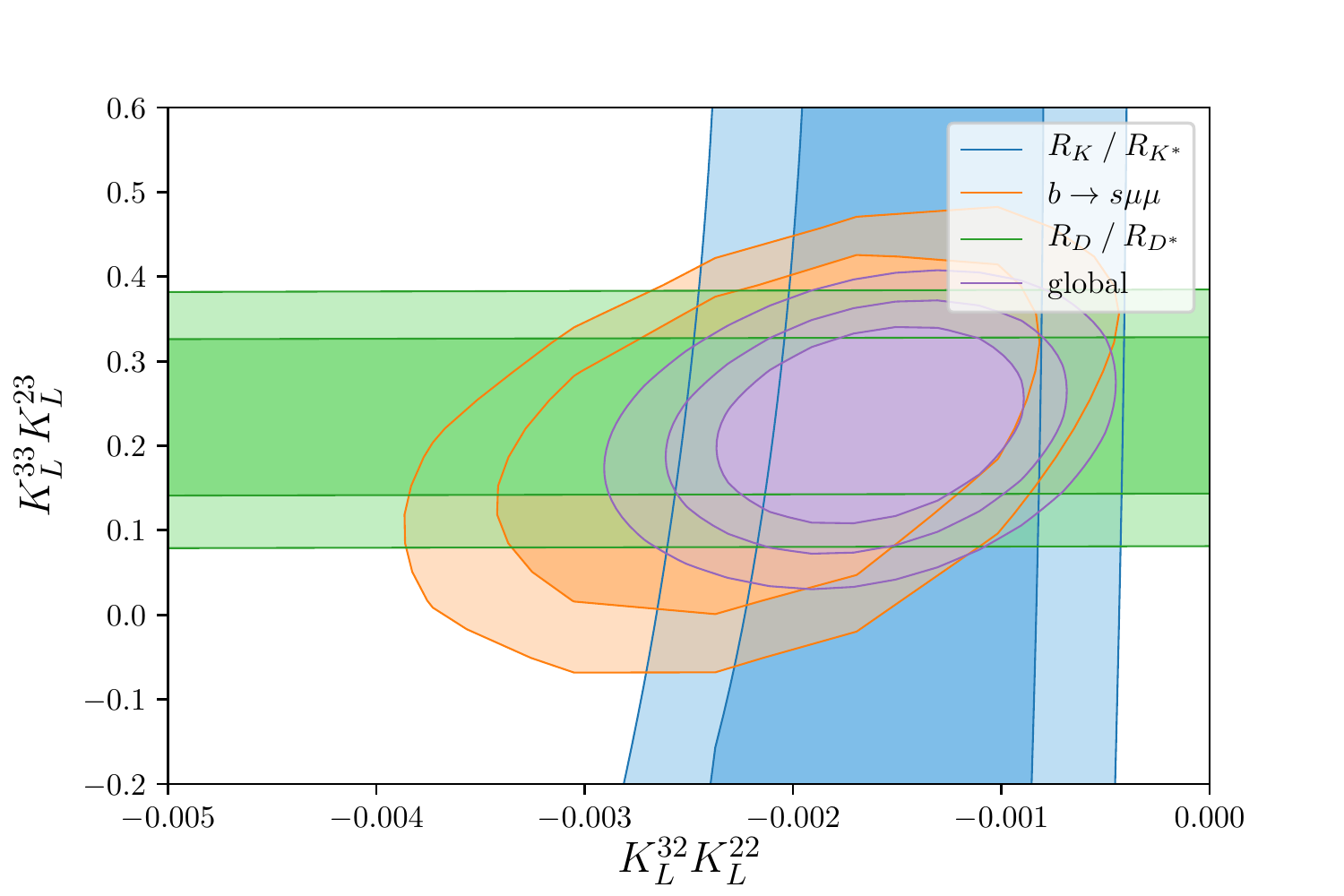}}
    \raisebox{-8mm}{\includegraphics[width=0.48\textwidth]{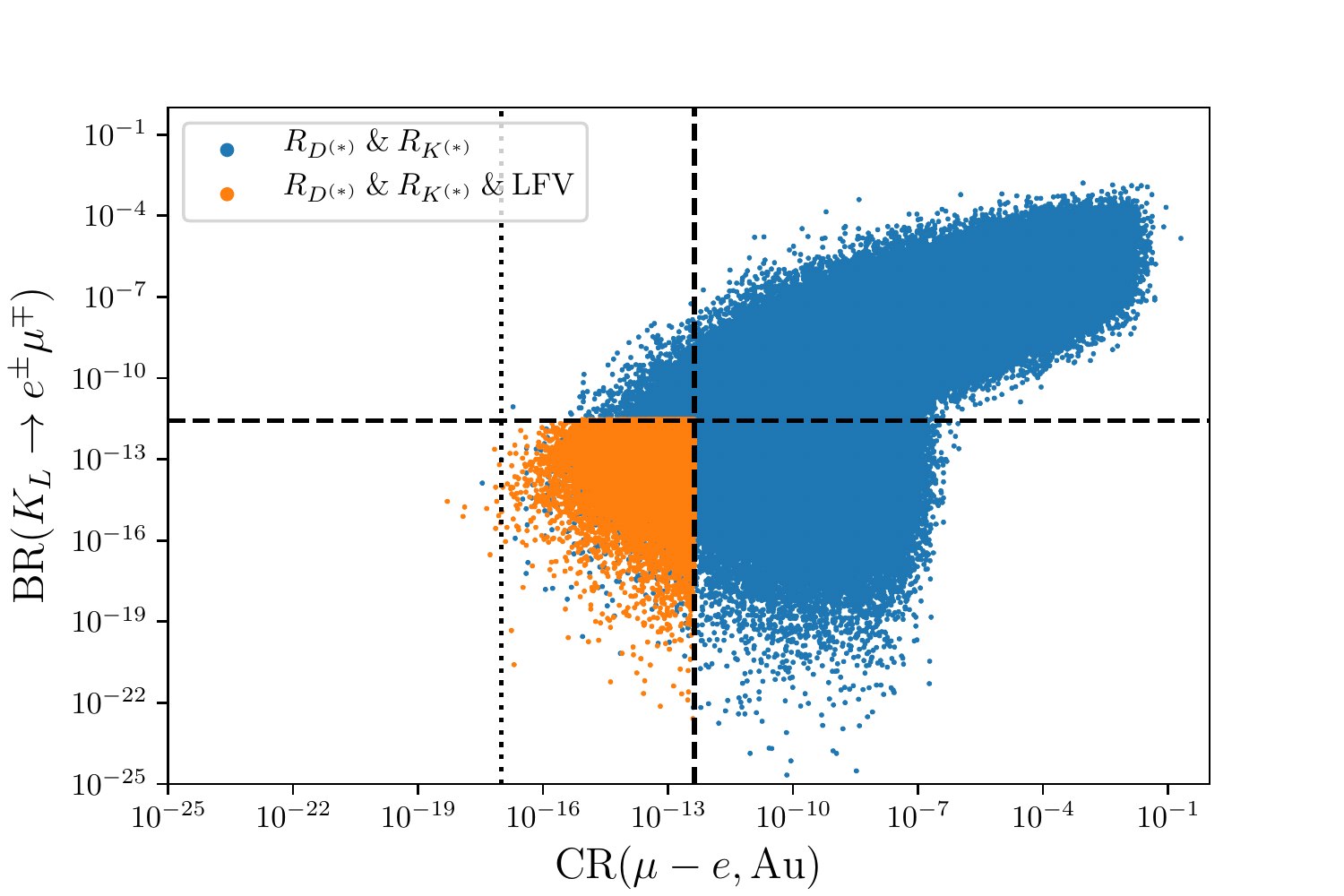}}}
    \caption{\textbf{Gauche:} Ajustement des combinaisons des couplages dominants des LQ $V_1$ ($K_L^{3i}K_L^{2i}$) aux données anormales. Les contours de correspondent à 1 $\sigma$, 2 $\sigma$ et 3 $\sigma$. \textbf{Droit:} Échantillons de Monte-Carlo générés tenant compte des anomalies $B$, affichés dans le plan des deux observables les plus contraignantes, CR$(\mu-e, \mathrm{Au})$ et $K_L \to e^\pm \mu^\mp$. Les points bleus violent au moins une limite LFV tandis que les points orange respectent toutes les contraintes imposées. Figures tirées de Ref.~\cite{Hati:2019ufv}.}
    \label{res_fig:LQfit}
\end{figure}

Des leptoquarks qui sont des vecteurs de jauge, tels que $V_1$, apparaissent naturellement dans les (grandes) théories unifiées, spécifiquement à partir de l'unification quark-lepton, comme dans les modèles Pati-Salam.
Dans notre étude de Ref.~\cite{Hati:2019ufv}, et  
au lieu d'explorer une complétion UV spécifique pour les leptoquarks $V_1$, nous avons choisi de trouver des exigences sur les couplages de $V_1$ aux fermions du SM d'une manière efficace.
Le sous-ensemble d'éléments pertinents 
(gauches~\footnote{En raison de l'absence d'indices forts 
suggérant des contributions droitières  aux opérateurs WET non-nulles, nous nous limitons à des couplages LQ gauches.}) Les couplages LQ aux fermions du SM 
peuvent être paramétrés de manière générale comme suit
\begin{equation}
    \mathcal L \simeq \sum_{i,j,k,l = 1}^{3} V_1^\mu \left(\bar d_L^i\gamma_\mu K_L^{ik}\ell_L^k + \bar u_L^j V_{ji}^\dagger \gamma_\mu K_L^{ik} U_{kl}^P\nu_L^l \right) + \mathrm{H.c.}\,,
\end{equation}
où $K_L^{ij}$ désigne les couplages LQ effectifs et $U^P$ est la matrice de mélange leptonique PMNS.
La correspondance entre les couplages LQ (à l'échelle de masse du LQ, $m_{V_1} \simeq 1.5\,\mathrm{TeV}$) et les WC identifiés par l'analyse EFT peut être effectuée comme suit :
\begin{equation}
C^{ij;\ell \ell^{\prime}}_{9,10} = \mp\frac{\pi}{\sqrt{2}G_F\,\alpha_\mathrm{em}\,V_{3j}\,V_{3i}^{\ast} \,m_{V_1}^2}\left(K_L^{i
  \ell^\prime} \,K_L^{j\ast} \right)\,, 
  C_{jk,\ell i}^{V_L} = \frac{\sqrt{2}}{4\,G_{F}\,m_{V_1}^2}\,
  \frac{1}{V_{jk}}\,
  (V\,K_{L}\, U^P)_{ji}\, K_{L}^{k\ell\ast}\,.
  \label{res_eqn:CV}
\end{equation}
Dans le panneau gauche de la Fig.~\ref{res_fig:LQfit}, nous montrons des contours dans le plan des couplages dominants des LQ. Remarqueons que l'on trouve la même corrélation entre les couplages semi-muoniques et semi-tauoniques dans les contours préférés par les données $b\to s\ell\ell$ que celle rencontrée précédemment dans les ``fits" EFT (cf. Fig.~\ref{res_fig:fits}).

Comme nous l'avons montré dans notre analyse dans~\cite{Hati:2019ufv}, afin d'échapper à ces limites, des couplages effectivement non-unitaires aux fermions du SM sont nécessaires. Par exemple, cela peut être réalisé en introduisant des fermions de type vectoriel, plus spécifiquement des leptons de type vectoriel, qui se mélangent avec les fermions du SM.
Dans notre approche, nous avons donc paramétré les couplages des LQ via 12 rotations unitaires incorporant le mélange entre les leptons du SM et les 3 générations supplémentaires de doublets leptoniques $SU(2)_L$ de type vectoriel\footnote{D'autres représentations de leptons vectoriels sont 
exclues, car le modèle de mélange requis aurait un impact considérable sur les couplages $Z\ell\ell$, déjà au niveau arbre, violant ainsi les limites expérimentales de ces quantités, mesurées avec grande précision.}.
Dans le panneau de droite de la Fig.~\ref{res_fig:LQfit}, nous montrons les résultats d'un balayage aléatoire où nous faisons varier les 12 angles de mélange dans l'intervalle $[-\pi,\pi]$, présentant nos résultats dans le plan des deux observables LFV les plus contraignants, à savoir les désintégrations $K_L\to \mu^\pm e^\mp$ et la conversion $\mu-e$ sans neutrinos dans les noyaux. Les points affichés (en bleu) sont en accord avec les données anormales au niveau de $3\,\sigma$ ; cependant, la plupart de ces derniers points conduisent à la violation d'au moins une limite expérimentale du LFV. Les points respectant toutes les contraintes imposées (expliquant $R_{K,D}^{(*)}$ et respectant toutes les limites expérimentales) sont marqués en orange. Notez que la majeure partie de l'espace de paramètres actuellement privilégié peut être sondée par les prochaines expériences COMET et Mu2e~\cite{Adamov:2018vin,Bartoszek:2014mya}, toutes deux dédiées à la recherche de conversions $\mu-e$ sans neutrinos dans l'aluminium.

Dans une seconde analyse phénoménologique actualisée, nous avons pris les 9 couplages des LQ gauches comme des paramètres indépendants et nous les avons ajustés à plus de 350 observables~\footnote{Ces derniers comprennent les données $b\to s\ell\ell$ et $b\to c\ell\nu$, un grand nombre de désintégrations de mésons et de $\tau$ (semi-)leptoniques ($b$, $c$ et $s$) violant et conservant la saveur leptonique et plusieurs processus LFV purement leptoniques. Une liste complète peut être trouvée dans la Réf.~\cite{Hati:2020cyn}.}, pour trois points de référence pour la masse du $V_1$ - $m_{V_1}\in[1.5, 2.5, 3.5]\,\mathrm{TeV}$ ; ceci permet de trouver une région dans l'espace des paramètres à $9$ dimensions dans laquelle les anomalies des $B$ peuvent être expliquées tout en échappant aux contraintes des processus LFV~\cite{Hati:2020cyn}.
Nous avons fait l'hypothèse de que la distribution postérieure des couplages était approximativement gaussienne et nous les avons échantillonnés selon leur distribution. A partir des échantillons de Monte-Carlo, nous avons ensuite calculé des plages postérieures pour les observables autour du (des) point(s) de meilleur ajustement, en nous basant sur l'espace des paramètres préférés actuellement par les données expérimentales.

\begin{figure}
    \centering
    \mbox{\hspace{-10mm}\raisebox{0mm}{\includegraphics[width=0.65\textwidth]{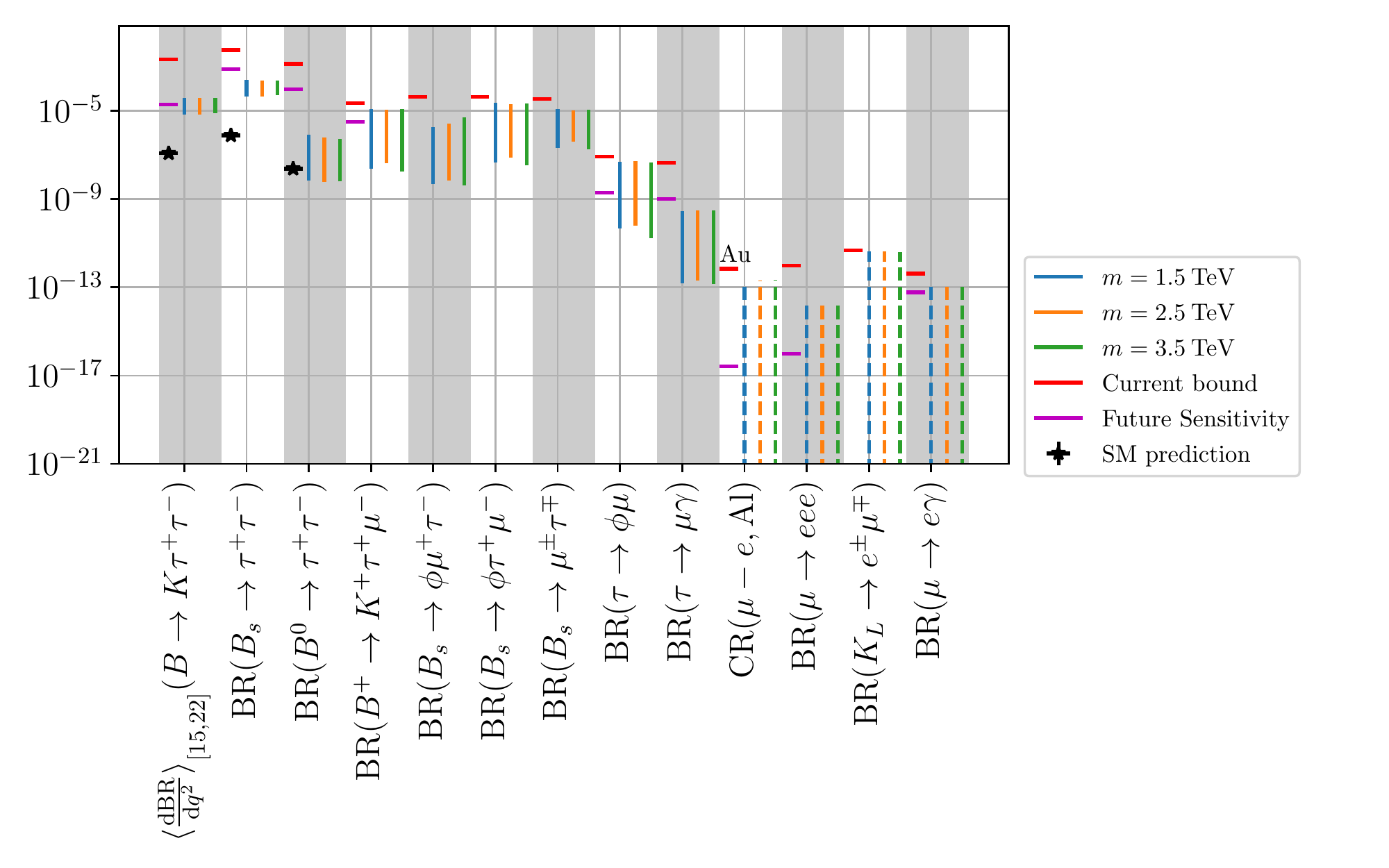}}
    {\hspace{-4mm}\raisebox{6mm}{\includegraphics[width=0.55\textwidth]{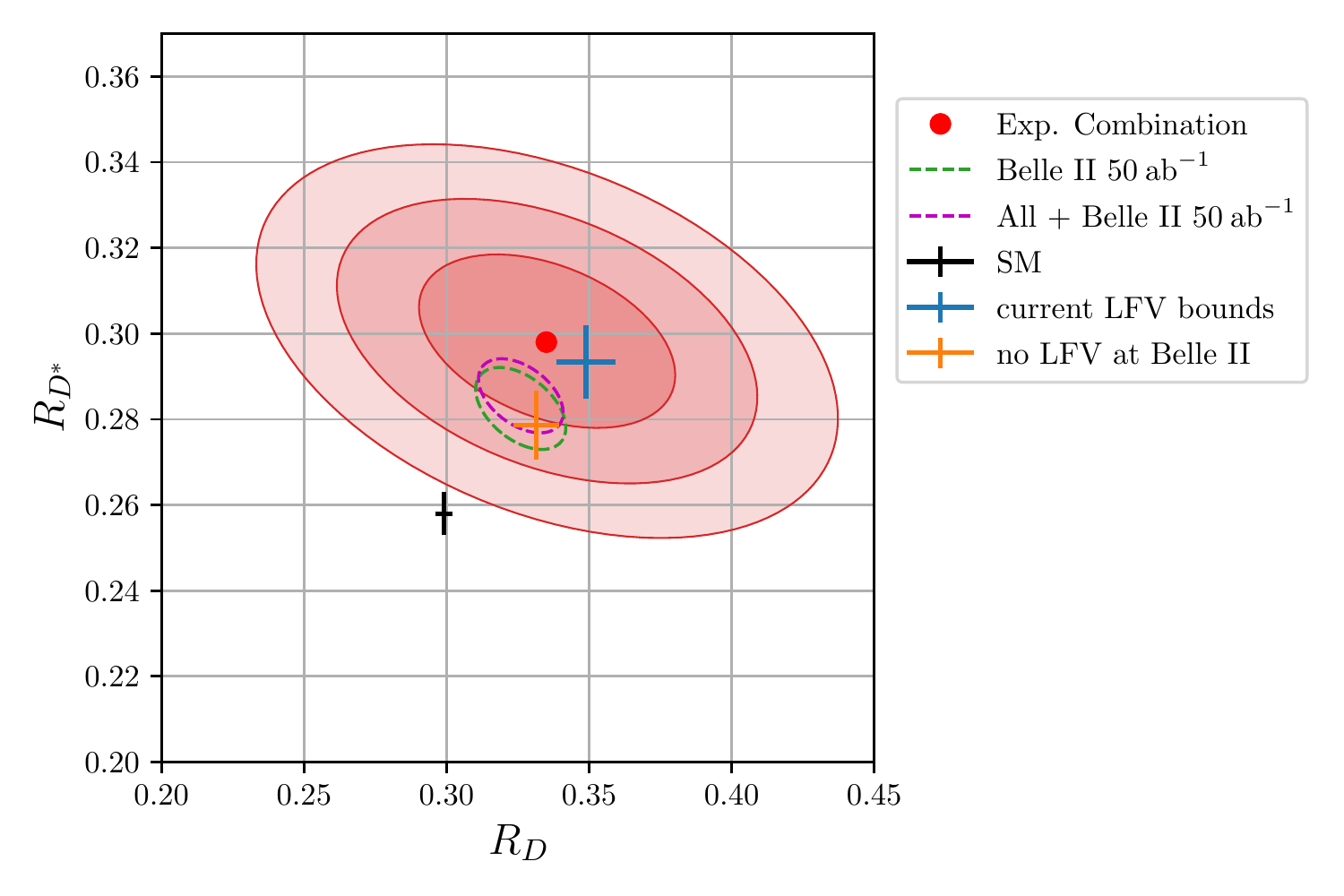}}}}
    \caption{\textbf{Gauche:} Plages postérieures des prédictions du LQ vecteur pour plusieurs observables recherchées par Belle II, COMET et Mu2e. \textbf{Droite:} Plages postérieures des prédictions du LQ pour $R_{D^{(\ast)}}$ basées sur les limites actuelles (sensibilités futures) des processus LFV pertinents pour le test de l'hypothèse d'un LQ vectorielle, présentée en bleu (orange). Les prédictions pour des différentes masses coïncident. Sont également présentées les extrapolations de la mesure actuelle de Belle à la sensibilité de Belle II (lignes de contour en pointillés). Tirés de Ref.~\cite{Hati:2020cyn}.}
    \label{res_fig:predictions}
\end{figure}

Plusieurs désintégrations rares de mesons $B$, impliquant des taus, et de nombreuses désintégrations LFV du tau seront recherchées par l'expérience Belle II~\cite{Kou:2018nap}, avec des sensibilités améliorées.
En raison des couplages importants du LQ aux quarks $b, s$- et $c$, ainsi qu'aux leptons chargés, nous nous attendons a priori à des augmentations importantes des processus $b\to s \tau^+\tau^-$ et processus LFV.
Dans le panneau gauche de la Fig.~\ref{res_fig:predictions}, nous montrons les plages de $1\sigma$ des distributions postérieures pour plusieurs de ces observables, ainsi que les limites expérimentales actuelles et les sensibilités futures.
En particulier, les désintégrations $B\to K\tau\mu$ et $\tau\to\phi\mu$ sont des canaux prometteurs à Belle II. De plus, l'intervalle prédit de $\mu-e$ sans neutrinos dans l'aluminium sera (presque) entièrement sondée par COMET et par Mu2e~\cite{Adamov:2018vin,Bartoszek:2014mya}.

Les résultats de cette analyse ne sont en aucun cas une garantie de découvrir des signaux LFV dans ces canaux. 
Nous avons donc étudié l'impact des résultats \textit{nulls} dans les canaux LFV de Belle II, Mu2e et COMET - remplaçant ainsi les limites LFV actuelles par des sensibilités futures dans notre ajustement.
Les résultats d'un tel ajustement futur hypothétique sont présentés dans le panneau de droite de la Fig.~\ref{res_fig:predictions}, où nous montrons les prédictions du modèle LQ pour $R_{D^{(\ast)}}$ sur la base des limites LFV actuelles et futures (en bleu et orange respectivement).
De plus, nous avons extrapolé la mesure actuelle de $R_{D^{(\ast)}}$ par Belle à la précision anticipée de Belle II avec une luminosité intégrée de $50,\mathrm{ab}^{-1}$, et nous l'avons combinée avec toutes les autres mesures disponibles, indiquées par les ellipses vertes et violettes (en pointillés)~\cite{Hati:2020cyn}.
Curieusement, en l'absence de signaux LFV, le point de meilleur ajustement prévu pour les LQ (en orange) se rapproche de la prédiction du SM, bien qu'il chevauche le contour de $1\,\sigma$ de la sensibilité extrapolée de Belle II.
D'autre part, si les mesures futures coïncident avec la valeur centrale de la moyenne globale actuelle (avec une meilleure précision), une explication $V_1$-LQ des anomalies $R_{D^{(\ast)}}$ serait en conflit avec les limites futures des processus LFV (encore une fois en l'absence de toute découverte de LFV à Belle II).
Ainsi, l'évolution des mesures futures de $R_{D^{(\ast)}}$ s'avérera déterminante pour falsifier l'hypothèse du vecteur LQ. 

\subsection{Perspectives}
Ces dernières années, de nombreux indices de la présence d'une violation de LFU dans les désintégrations semi-leptoniques du méson $B$ de courant chargé et neutre ont émergé en association avec plusieurs observables.
Les analyses EFT actuelles semblent favoriser les modèles minimaux qui peuvent traiter simultanément les tensions dans les deux canaux, en raison d'une préférence pour les contributions \textit{universelles} à $C_9^{bs\ell\ell}$ à l'échelle de masse du quark $b$ (qui peuvent être induites par des effets de RG).
Cette interprétation est encore renforcée par la mesure très récemment mise à jour de $R_{K}$.
Suite à la récente mesure de LHCb~\cite{LHCb:2021trn}, nous avons mis à jour les ajustements de plusieurs hypothèses de NP conduisant à un bon accord avec les données.

Compte tenu de l'absence de couplages au niveau arbre entre les quarks de type down et les neutrinos, 
les LQ de type vecteur singlet de $SU(2)_L$ sont d'excellents candidats pour une explication combinée des anomalies de désintégration des mésons $B$, 
bien que soumis à des contraintes strictes provenant des observables de LFV.
Nous avons montré qu'une structure de saveur non-unitaire des couplages des LQ à la matière du SM est nécessaire afin de respecter les nombreuses limites des observables de saveur, qui ont été mesurées comme étant en accord avec le SM ; une telle structure peut être générée par des mélanges de leptons du SM avec des états doublets lourds de type vecteur.
Nous avons exploré plus en profondeur la phénoménologie des saveurs de ce modèle simple de LQ vecteurs, en effectuant une analyse statistique spécifique ; cela a permis d'identifier plusieurs ``modes dorés" qui ont d'excellentes chances d'être observés par les expériences à venir dans un futur proche.
Enfin, nous avons souligné l'importance des futures mesures de $R_{D^{(\ast)}}$ pour la viabilité du modèle.

\section{Conclusions}
Au cours des dernières décennies, l'effort de découverte de la nouvelle physique au-delà du modèle standard s'est appuyé sur de multiples approches, tant du point de vue expérimental que théorique.
Sur le plan expérimental, des efforts sont déployés pour rechercher directement la nouvelle physique à \textit{hautes énergies} ainsi que pour rechercher indirectement les effets de la nouvelle physique à \textit{hautes intensités}.
Motivées par les modèles de nouvelle physique qui tentent de résoudre (ou du moins d'améliorer) les problèmes théoriques du SM (par exemple le problème de la hiérarchie), comme c'est le cas des modèles supersymétriques ou extradimensionnels, la plupart des expériences ont été consacrées à la recherche directe des nouveaux états (lourds) dans les collisionneurs à haute énergie, comme le LEP, le Tevatron ou le LHC.  

Complémentaires aux recherches à haute énergie, les mesures de précision des observables électrofaibles et de saveur, à basse et hautes énergies ont toujours précédé et ouvert la voie à des découvertes directes d'états du SM (par exemple les bosons de jauge électrofaibles ou le boson de Higgs).
Avant la formulation théorique actuelle du SM, des indices indirects sur les effets de la ``Nouvelle Physique" ont également fourni des lignes directrices pour la construction de modèles, comme c'est le cas des désintégrations $\beta$ contredisant l'image de désintégration à deux corps et conduisant finalement à l'introduction des neutrinos, la découverte de la violation P conduisant à l'introduction des interactions $V-A$ provenant de la symétrie de jauge $SU(2)_L$, la découverte de la violation CP conduisant à l'hypothèse d'une troisième génération de quarks, parmi
d'autres.

Malgré la découverte du boson de Higgs au LHC, les signaux directs de nouveaux états ont jusqu'à présent échappé à l'observation expérimentale. 
En retour, les résultats négatifs des recherches repoussent
l'échelle d'énergie à laquelle la nouvelle physique pourrait être présente, laquelle est dans de nombreux cas déjà supérieure au TeV.
Ainsi, les objectifs théoriques de naturalité qui motivent la présence de la nouvelle physique à l'échelle du TeV (afin de résoudre les problèmes théoriques du SM) sont remis en question et il convient de réévaluer les principes directeurs de la construction de modèles.

Avec la découverte des oscillations de neutrinos, une ligne directrice pour la Nouvelle Physique a été clairement identifiée. 
En tant que première preuve en laboratoire de la nouvelle physique, les oscillations de neutrinos forcent l'extension du SM, de façon à inclure un mécanisme viable pour la de génération de la masse des leptons neutres.
Il est intéressant de noter qu'en offrant une nouvelle source de violation de CP et en faisant appel à des états 
qui interagissent faiblement, les extensions de la nouvelle physique visant à fournir une explication des masses de neutrinos ouvre aussi des connections avec  d'autres problèmes du MS, tels que l'asymétrie baryonique de l'univers et le problème de la matière noire.
Aujourd'hui, la physique des neutrinos est entrée dans un ère de précision, et un effort expérimental et théorique mondial est consacré à la résolution des nombreuses questions ouvertes qui lui sont liées.
En outre, une phénoménologie immensément riche liée au secteur leptonique s'est ouverte  à present :
en raison de la présence de masses de neutrinos, de nombreuses symétries accidentelles du SM semblent être brisées dans la nature.
Par conséquent, l'intérêt pour les recherches à haute intensité dédiées au secteur leptonique n'a cessé de croître.

La violation des symétries (leptoniques) accidentelles du SM, telles que la conservation de la saveur leptonique chargée et l'universalité de la saveur leptonique (toutes deux violées en raison de la présence de masses de neutrinos), ouvre de nombreuses voies possibles pour la recherche de la nouvelle physique.
Alors que les neutrinos massifs ne constituent qu'une source possible de violation de la saveur leptonique et de l'universalité de la saveur leptonique, les signaux indirects indiquant la rupture de ces symétries, en synergie avec d'autres signaux indirects possibles de nouvelle physique, fourniront des orientations cruciales pour les recherches expérimentales directes ainsi que pour les efforts théoriques visant à décrire les interactions de nouvelle physique.
Il est clair que le secteur leptonique est en train de devenir un outil très puissant en ce qui concerne la recherche de la nouvelle physique.

En plus de la découverte des oscillations des neutrinos, d'autres observables ``à saveur leptonique" présentent des écarts significatifs par rapport à leurs prédictions SM respectives. Parmi elles, on peut citer le moment magnétique anormal du muon, les observables LFU dans les désintégrations semi-leptoniques des mésons $B$ et de nombreuses déviations dans le système $b\to s\ell\ell$.
Ces anomalies peuvent-elles suggérer une voie vers le modèle sous-jacent de la Nouvelle Physique, non seulement capable de les expliquer, mais aussi de prendre en compte les nombreuses autres lacunes du SM ?

Un premier point de départ est donné par les approches ``bottom-up" basées sur les données et alimentées par les analyses EFT, afin de trouver les exigences à basse énergie d'un candidat potentiel à la nouvelle physique. Bien que l'objectif principal doit toujours être de viser une description complète de la nature au niveau UV (c'est-à-dire une construction théorique complète tenant compte de toutes les réserves observationnelles et théoriques du SM), les premières pistes peuvent être déduites de réalisations BSM minimales, telles qu'identifiées à partir des résultats de l'approche ``bottom-up". De telles extensions ad hoc, dans lesquelles le SM est élargi de façon minimale par des ingrédients strictement nécessaires pour résoudre des problèmes individuels (qu'il s'agisse de champs scalaires, de vecteurs, de fermions neutres...), pourraient offrir des lignes directrices pour la construction de cadres plus complets. Il est donc primordial de consacrer des ressources pour comprendre pleinement les implications à basse énergie de ces constructions minimales.
De même, afin de clarifier la présence de la Nouvelle Physique dans les observables à basse énergie (anomalies $B$, $(g-2)_\ell$ etc.), et dans tous les cas pour réduire les incertitudes théoriques (parfois encore importantes), les contributions du SM, en particulier de la QCD (non-perturbative) et les effets à longue distance, doivent être maîtrisées.

Sur le plan expérimental, l'avenir semble extrêmement prometteur.
En ce qui concerne la physique des neutrinos, plusieurs expériences sont planifiées, destinées à mesurer avec précision les paramètres de la PMNS, la violation de CP leptonique et à déterminer l'échelle de masse absolue des neutrinos.
En outre, une quantité croissante de données sur la diffusion neutrino-nucleon est accumulée, ce qui permet de contraindre indirectement les interactions et les scénarios de mélange non-standard des neutrinos.
À la frontière des haute intensités, en ce qui concerne le secteur des leptons chargés, les nombreuses futures expériences  dédiées aux muons amélioreront considérablement les recherches de violation de la saveur des leptons chargés,
tandis que Belle II permettra d'améliorer de manière significative les limites d'un grand nombre de désintégrations de leptons violant la saveur du $\tau$.
En outre, les ``anomalies $B$" et de nombreux autres processus intéressants, y compris les états finaux tauoniques et les désintégrations LFV (semi-) leptoniques, seront sondés et potentiellement découverts.
En tant que tests complémentaires du modèle standard et de ses symétries, les programmes sur les désintégrations rares du charme et la violation de CP dans le secteur du charme viennent aussi de commencer.
Enfin, à hautes énergies, le troisième cycle du LHC devrait bientôt commencer, avec la mise à niveau à haute luminosité dans un avenir proche.
Dans un avenir plus lointain, des projets de collisionneurs très prometteurs, bien qu'ambitieux, sont prévus,
notamment un collisionneur électron-positron à haute énergie et peut-être même un collisionneur de muons à l'échelle du TeV.
Les machines leptoniques à haute énergie permettront finalement de repousser les frontières de l'énergie et de l'intensité dans la quête de la nouvelle physique.

\end{document}